# Title

Infector characteristics exposed by spatial analysis of SARS-CoV-2 sequence and demographic data analysed at fine geographical scales.

# Authors

Anna Gamża[1], Samantha Lycett[1], Will Harvey[1], Joseph Hughes[3], Sema Nickbakhsh[4], David L Robertson[3], Alison Smith Palmer[4], Anthony Wood[1], Rowland Kao[1,2]

Affiliations:

1) The Roslin Institute, University of Edinburgh, Edinburgh, UK

2) School of Physics and Astronomy, University of Edinburgh, Edinburgh, UK

3) MRC-University of Glasgow Centre for Virus Research, Glasgow, UK

4) Public Health Scotland, Glasgow, UK

# Abstract

Characterising drivers of SARS-CoV-2 circulation is crucial for understanding COVID-19 because of the severity of control measures adopted during the pandemic. Whole genome sequence data augmented with demographic metadata provides the best opportunity to do this. We use Random Forest Decision Tree models to analyse a combination of over 4000 SARS-CoV2 sequences from a densely sampled, mixed urban and rural population (Tayside) in Scotland in the period from August 2020 to July 2021, with fine scale geographical and socio-demographic metadata. Comparing periods in versus out of "lockdown" restrictions, we show using genetic distance relationships that individuals from more deprived areas are more likely to get infected during lockdown but less likely to spread the infection further. As disadvantaged communities were the most affected by both COVID-19 and its restrictions, our finding has important implications for informing future approaches to control future pandemics driven by similar respiratory infections.

# Introduction

As of August 2024, the SARS-CoV-2 pandemic has resulted in over 776 million reported cases and 7 million confirmed COVID-19-associated deaths worldwide (World Health Organization, 2024). While the pandemic is now declared by WHO to not be a Public Health Emergency of International Concern (World Health Organization, 2023), SARS-CoV-2 continues to circulate worldwide, with ongoing waves of infections and increased death rates still a possibility (Wang et al., 2021) especially in more vulnerable populations. The likely occurrence of pandemics of other respiratory pathogens, such as influenza, also remain high (Marani et al., 2021).

For SARS-CoV-2, there is a considerable literature on the risk of becoming infected and symptomatic. An increased risk of infection has been associated with individuals from lower socio-economic backgrounds (high deprivation) (Carrión et al., 2021; Garcia-Morata et al., 2022; Green et al., 2021; Gu et al., 2021; Harris et al., 2023; Mongin et al., 2022). Some association was also found between the age and risk of infection, with older people likely to be more susceptible (Ehlert, 2021; Harris et al., 2023). While population density was not consistently associated with risk of acquiring the infection (Carozzi et al., 2022; Khavarian-Garmsir et al., 2021; Lak et al., 2021) there was a positive association with how well connected different areas of residence are determined for example by links to public transport (Gu et al., 2022; Lu et al., 2021; Wang et al., 2023). Substantial uncertainty still remains over risk factors for transmitting the infection. Existing methods, mostly based on contact tracing, provided useful insights, however have serious limitations due to the requirement for substantial investigative effort, the impact of biases and subjectivity, poor resolution, or being limited to conditions of low density contacts (Hossain et al., 2022; Müller & Kretzschmar, 2021). Additionally, for more detailed contact tracing, like monitoring movements with the mobile data, questions about its ethics due to private data collection and protection were raised (Dar et al., 2020; Jung et al., 2020).

Here, we use the genetic sequences of virus samples obtained from confirmed SARS-CoV-2 cases to study patterns of viral circulation in a mixed urban and rural population and spanning low-to-high levels of deprivation. Of substantial interest is the question of how periods of extremely strict restrictions on people's activity, known as "lockdowns", affected this circulation, i.e. the risk of spreading SARS-CoV-2 further, and in particular whether its impact differed across demographic groups. This is a question of particular importance given its known differential impact on health and well-being (Dorn et al., 2020; Gama et al., 2021; Gray et al., 2020; Green et al., 2021).

Much of the data gathered during the pandemic period is unprecedented in scope, and their analysis can provide insight into the efficacy of past control measures that inform targeted control measures in response to both future waves of COVID-19 and future pandemics of similar emergent respiratory infections. Methods to integrate epidemiological and genetic data are promising but usually require either extremely dense datasets, sophisticated bespoke analytical techniques or both, and can otherwise be difficult to implement (Kao et al., 2014). We follow previous analyses (Crispell et al., 2019; Rossi et al., 2022) that used decision tree-based machine learning approaches designed to account for such methodological issues. For the SARS-CoV-2 we consider the simple measure of genetic distance between virus sequences and use much larger datasets at broader geographical scales showing greater genetic variation than has been analysed in previous studies. As the data was provided by NHS Scotland and Scottish Government statistics in standardised very fine geographical resolution areas (data zones, with approx. 500-1000 residents) (Scottish Government, 2024b), we adopt a geographical epidemiology perspective (Vallée, 2023), focusing on the demographic characteristics of this areas, expanding the previous analyses of case incidence only (Lak et al., 2021; Wood et al., 2023).

We demonstrate that analysing genetic distance between pairs can be used to identify useful relationships that account for infector and infectee roles in transmission process. By quantify the relative importance of demographic and geographical relationships correlated with genetic distance we therefore gain insight into patterns that drive fine scale geographical transmission.

# Results

To reveal transmission dynamics in the community focusing on individuals that spread the infection further, we made use of the SARS-CoV-2 sequences from one of the 14 NHS Health Boards in Scotland (NHS Tayside), obtained from surveillance samples from the community, together with socio-demographic data. All data were recorded at the level of data zones, small administrative units standardised by number of residents (approx. 500-1000 individuals). The Tayside region was chosen due to its mix of rural and urban areas (see Supplementary Figure 4.2), a broad representation of low-to-high deprivation, and large number of available sequences coming from its robust testing program (NHS Tayside, 2022). During the three year period of SARS-CoV-2 pandemic in the UK, around 3 million virus genomes were sequenced and published by COVID-19 Genomics UK consortium and UK Public Health Agencies (COG-UK consortium, 2023). Of these 342,372 sequences were recorded from Scotland between 01 July 2020 and 13 February 2023, representing 12.81% of the 2,672,172 registered positive SARS-CoV-2 cases. Here, we focus on the pandemic period from 04 August 2020 to 30 July 2021, for which there are 4880 sequences from NHS Tayside that are retained in our analysis. This subset of sequenced cases shows no substantial biases compared to all positive case data (see Supplementary Note 1).

To study the characteristics of infectors, we compared the analysed sequences to the reference strain (the first published SARS-CoV-2 sequence: isolate Wuhan-Hu-1; MN908947.3 (Wu et al., 2020a, 2020b)) to make an estimate of directionality of infection route between any two sequences by identifying which sequence was genetically closer to the origin of the whole epidemic (i.e. closer to the Wuhan reference strain), and thus it is more likely to be 'infector' for purposes of statistical analyses. Using the ordered pairs of real SARS-CoV-2 sequences we then analysed the genetic distance relationships between the sequences and socio-demographic variables using a statistical machine learning model to examine the transmission dynamics under different social movement restriction regimes in the course of the pandemic.

*Identification of the sub-population with overrepresentation of infectors*

First, we demonstrate that ordering sequences in a pair according to their similarity to the ancestor sequence can be used to identify sets of individuals that are more likely than average to be infectors and that this is associated with greater representation in the group we label "Sequence 1" (i.e. genetically closer to the reference strain). We analysed sets of transmission trees generated in simulations of a stochastic SIS individual-based model. We considered the extreme scenario where only half the population is able to pass on infection, but all hosts are equally susceptible. Statistics for paired data for analysis, where the sequences were ordered based on their patristic distance to first infector (Case 0), were compared to a negative control where sequences were ordered randomly (see Methods section for further details). Statistics were calculated from 54 "major" (bigger than 200 cases) simulated outbreaks in a population of 500 hosts. Figure 1 shows that ordering sequences by their distance to the SARS-CoV-2 reference strain separates the two types of hosts (one being non-infectious) successfully, but only if the pairs separated by long genetic distances are excluded by implementing genetic distance cutoff. Bigger cutoffs introduce too much noise from past transmission chains to distinguish between these host types.

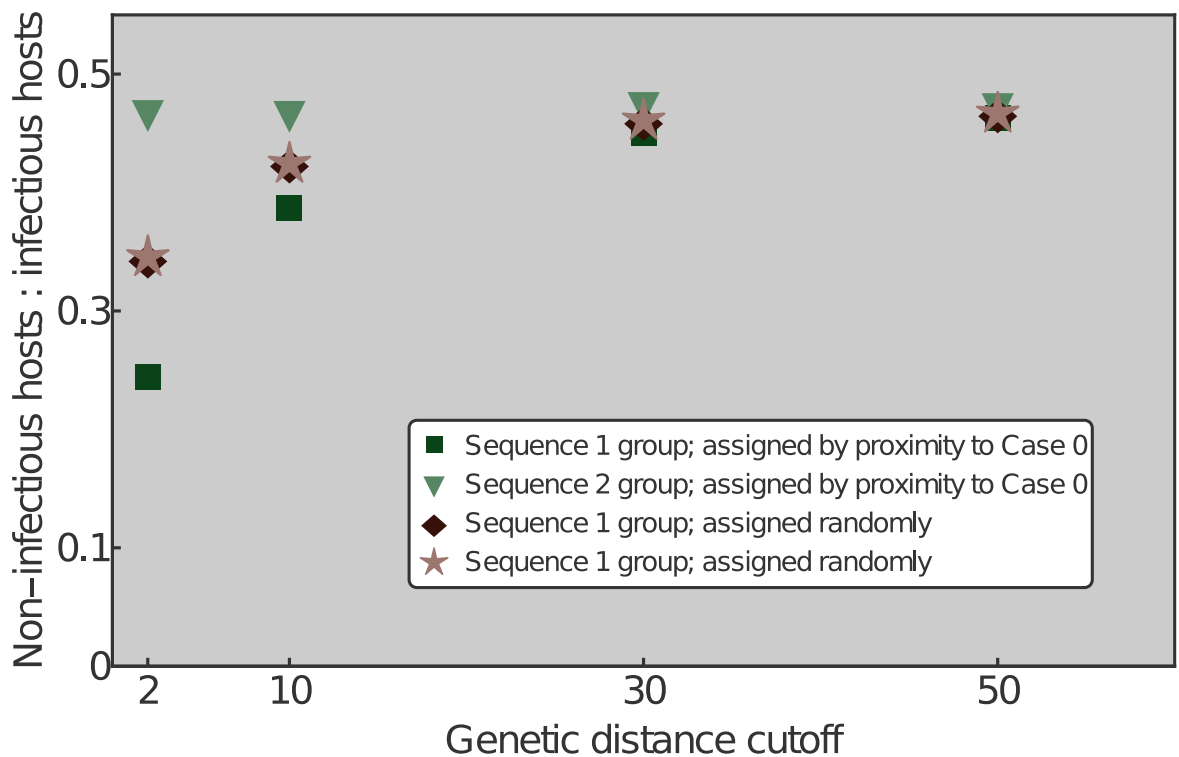

Figure 1. Results of a agent-based SIS simulation of transmission trees, where all individuals are equally susceptible, but only 50% are able to transmit infection. The y-axis shows the proportion of non-infectious hosts (50% of total susceptible population) when sequencs are assigned to one of two groups; in the first set (assiged by proximity), sequences are assigned to Group 1 if they are genetically more closely related to the source infection (i.e. proximity on a transmission tree to case 0), and group two otherwise (if equally close, assigned at random); in the second set, individuals are assigned randomly to each group.

For the real life sequence data we generated pairs of sequences that were registered within four weeks from each other (rolling interval, see Methods section for details) for which we calculated genetic distance and assigned properties describing the cases (such as sex or age band) and geographical area they come from, such as population size or "deprivation rank" based on the Scottish Index of Multiple Deprivation that describes the deprivation of areas across seven domains: income, employment, education, health, access to services, crime and housing (Scottish Government, 2024a).

As the analysis of simulated transmission trees confirmed that both ordering of sequences within a pair and the genetic cutoff are essential to identify groups with overrepresentation of infectors, we ordered the sequences in a pair such that the sequence more similar to the wild type SARS-CoV-2 (first published SARS-CoV-2 sequence isolate Wuhan-Hu-1; MN908947.3 (Wu et al., 2020a, 2020b)) is marked as "Sequence 1" group and the less similar one as "Sequence 2". We implemented a genetic cutoff by rejecting pairs that were separated by more than approx. 40 SNPs (see Methods section for details).

*Effects of lockdown*

Random Forest Decision Trees were used to analyse the paired sequences supplemented with metadata in order to explore the relationship between the genetic distance between ordered pairs and 18 potential statistical predictor variables for the whole analysed timeline. Out of 18 variables (Table 2 in Methods section) we identified 8 as the most significant for the model fit (see Supplementary Note 2 and Supplementary Dataset 2 for details) for the genetic distance and included them in the final model: age category, sum of contemporaneous SARS-CoV-2 cases, sum of work-related geographical connections, geographical distance, total population, deprivation rank (Scottish Index of Multiple Deprivation (Scottish Government, 2024a)) and temporal distance (see Methods section for details).

The level of restriction (Containment Health Index (Hale et al., 2021)) was the most important variable describing the observed patterns of transmission (See Supplementary Figure 2.11). Strict lockdown was implemented in the very beginning of 2021 in Scotland when the number of SARS-CoV-2 infections was increasing (second pandemic wave (Public Health Scotland, 2024)) to mitigate the effect of the pandemic and protect vulnerable individuals (Scottish Government, 2021). To examine how the transmission patterns were affected by this lockdown period we ran separate Random Forest models for the two periods 04$^{th}$ August 2020 to 01$^{st}$ January 2021 & 01$^{st}$ May 2021 to 30$^{th}$ July 2021 (out of lockdown lockdown) and 02$^{nd}$ January 2021 to 30$^{th}$ April 2021 (in lockdown). The feature importance plots presented on Figure 1 show that the genetic distance relationship with other variables was substantially different during the lockdown period than outside of it. The model fit was better for the within lockdown period ($R^2$=0.80 vs $R^2$=0.74). In lockdown, the variables assigned to Sequence 1 group were consistently and markedly less important than the same variables assigned to Sequence 2 group.

To gain further insight into the direct relationship between selected variables and genetic distance we generated Accumulate Local Effects (ALE) plots (see Methods for details) for all the variables in the final Random Forest models. The summary of the relationship between genetic distance and the selected variables is presented below (see Table 1), full analysis of and the dependence for these and auxiliary variables is provided in Supplementary Note 2.

Table 1. Summary statistics of ALE plots

| Variable | Relative importance | | ALE effect size on the genetic distance prediction | | ALE effect direction | |
|---|---|---|---|---|---|---|
| | Out of lockdown* | In lockdown | Out of lockdown | In lockdown | Out of lockdown | In lockdown |
| Commuting Connections | 0.21 (0.47) | 0.45 | 2.2E-5 | 8.9E-5 | convex | negative |
| Contemporaneous cases Seq 1 | 0.45 (0.98) | 0.39 | 3.1E-4 | 6.3E-5 | negative | unclear |
| Contemporaneous cases Seq 2 | 0.46 (1.00) | 0.75 | 2.5E-4 | 2.8E-4 | negative | unclear |
| Deprivation Seq 1 | 0.40 (0.87) | 0.40 | 3.6E-4 | 1.5E-4 | negative | negative |
| Deprivation Seq 2 | 0.40 (0.88) | 1.00 | 2.0E-4 | 5.2E-4 | negative | positive |
| Geographical distance | 0.45 (0.98) | 0.68 | 8.9E-5 | 2.0E-4 | positive | positive |
| Population size Seq 1 | 0.33 (0.72) | 0.41 | 6.0E-4 | 1.6E-4 | negative | negative |
| Population size Seq 2 | 0.34 (0.76) | 1.00 | 2.2E-4 | 2.1E-3 | negative | unclear |
| Temporal distance | 0.25 (0.55) | 0.27 | 2.4E-5 | 3.1E-5 | negative | convex |

\* in brackets importance relative to out of lockdown period only

In Figure 2 we compare the importance of model variables, and the dependence between variable (as indicated by Accumulated Local Effects (ALE) plots), in and out of lockdown periods using the Random Forest modelling results.

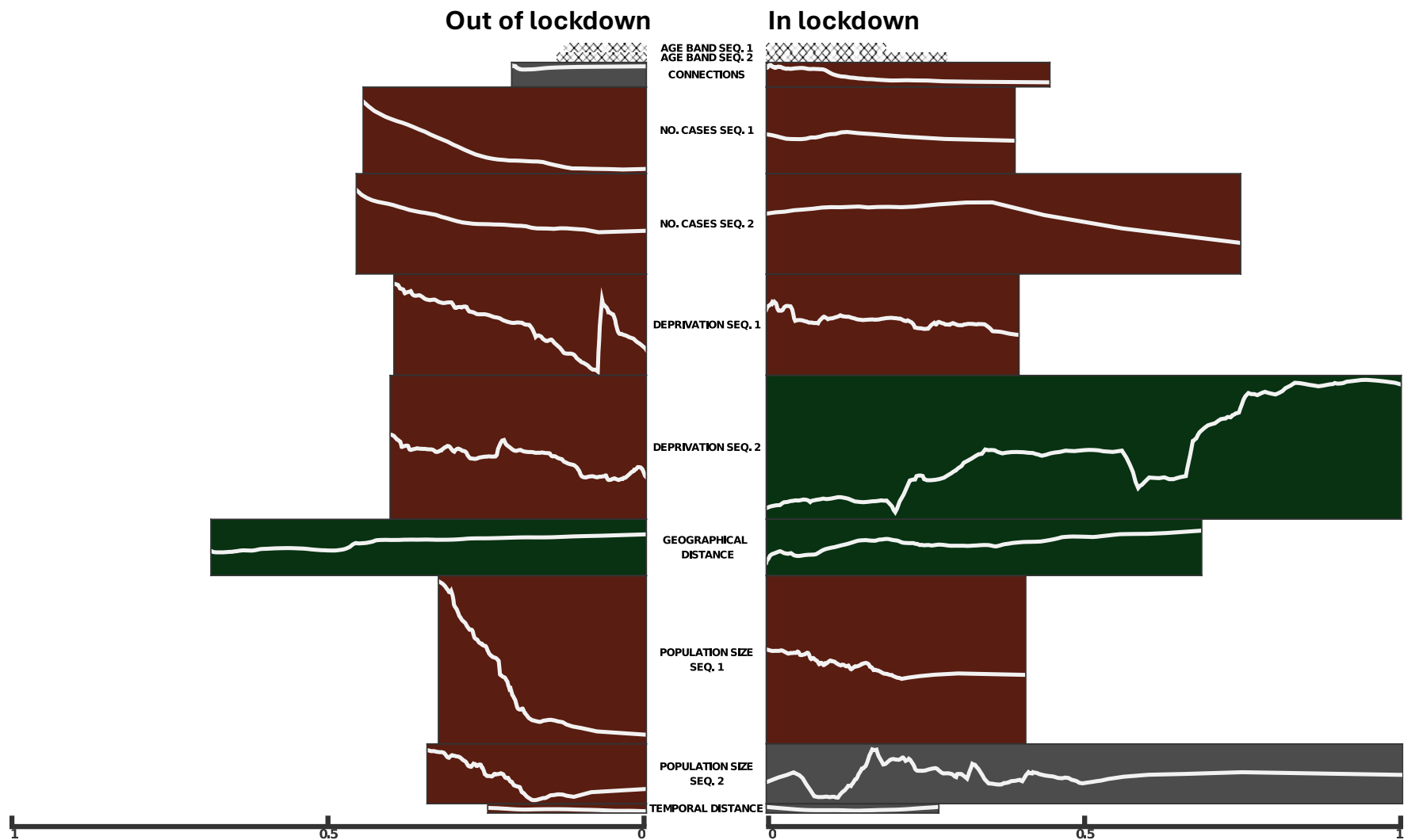

Figure 2. Relative feature importance plot (x axis) of variables generated from the Random Forest model fit describing the genetic distance between pairs of SARS-COV-2 sequences (with sequence 1 being genetically more similar to the wild type than sequence 2) where genetic distance is less than approx. 40 SNPs; sequences were taken from the NHS Tayside in Scotland for two periods: in lockdown (02nd January 2021 to 30th April 2021) and out of lockdown (04th August 2020 to 01st January 2021 and 01st May 2021 to 30th July 2021); the vertical height of the bars is scaled by the magnitude of the effect that each variable has on the genetic distance prediction (from Accumulated Local Effects -ALE plots) and ALE curves are added for every variable rescaled on their x axes to fit the importance plot and smoothed (rolling mean, t=5) to indicate the shape of relationship between the variable and genetic distance; detailed ALE plots with full x scales are provided in Supplementary Note 2, the bar colour indicates the shape of the ALE curve (green for positive slope, red for negative, grey for other and patterned where no ALE plot was generated); all variables increase in value from left to right (both in and out of lockdown); increasing deprivation rank indicates decreasing level of deprivation.

More detailed results of the importance analysis and all ALE plots (and their description) are provided in Supplementary Note 2. Both feature importance and ALE dependence suggest that deprivation, number of contemporaneous cases, population size and geographical distance are the best indicators of genetic distance relationships. Genetic distance is negatively correlated with number of contemporaneous cases and population size for both in and out of lockdown periods and for both Sequence 1 and Sequence 2 groups; it is also positively correlated with geographical distance for all analysed periods. Notably, deprivation and population size of the data zones for Sequence 2 group, show different trends in and out of lockdown.

*Deprivation specific infector characteristics*

To further investigate whether the different patterns observed for deprivation are an effect of the infectors being more likely to be classified as Sequence 1, we generated Random Forest model fits and consequent ALE plots for models supplied with sequence pair data selected with whole range of genetic distance cutoffs values considering the out of lockdown (Figure 3) and within lockdown (Figure 4) periods. The full results are presented in Supplementary Note 2.

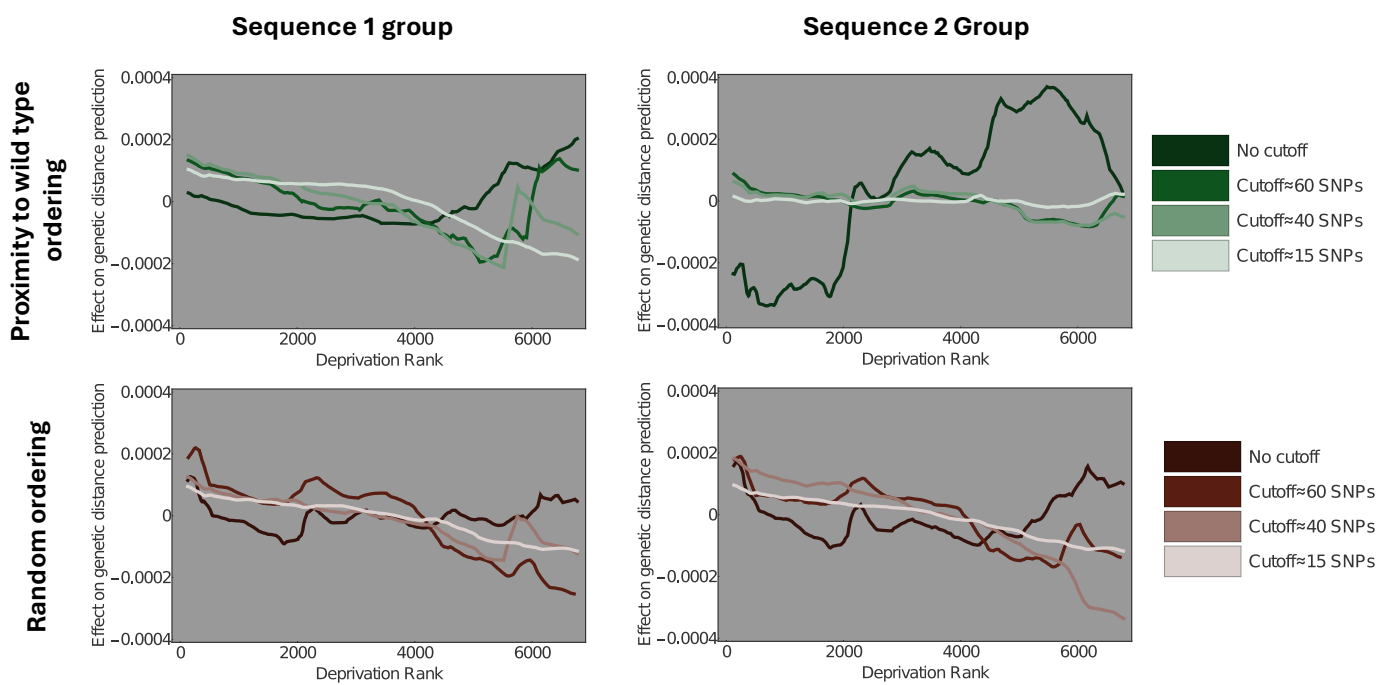

Figure 3. ALE plots based on Random Forest model fits generated for out of lockdown periods (04[th] August 2020-01[st] January 2021 & 01[st] May 2021-30[th] July 2021) for paired SARS-CoV-2 sequence data from NHS Tayside, Scotland selected with genetic distance cutoff for data where pairs were either ordered by their similarity to wild type SARS-CoV-2 sequence (Sequence 1 more similar; top row) or randomly ordered (bottom row); The Scottish Index of Multiple Deprivation ranks areas from the most to least deprived (such that bigger rank indicates lower deprivation); the correlation curves were smoothed for readability (moving average; t=10), full data are provided in Supplementary Dataset 1.

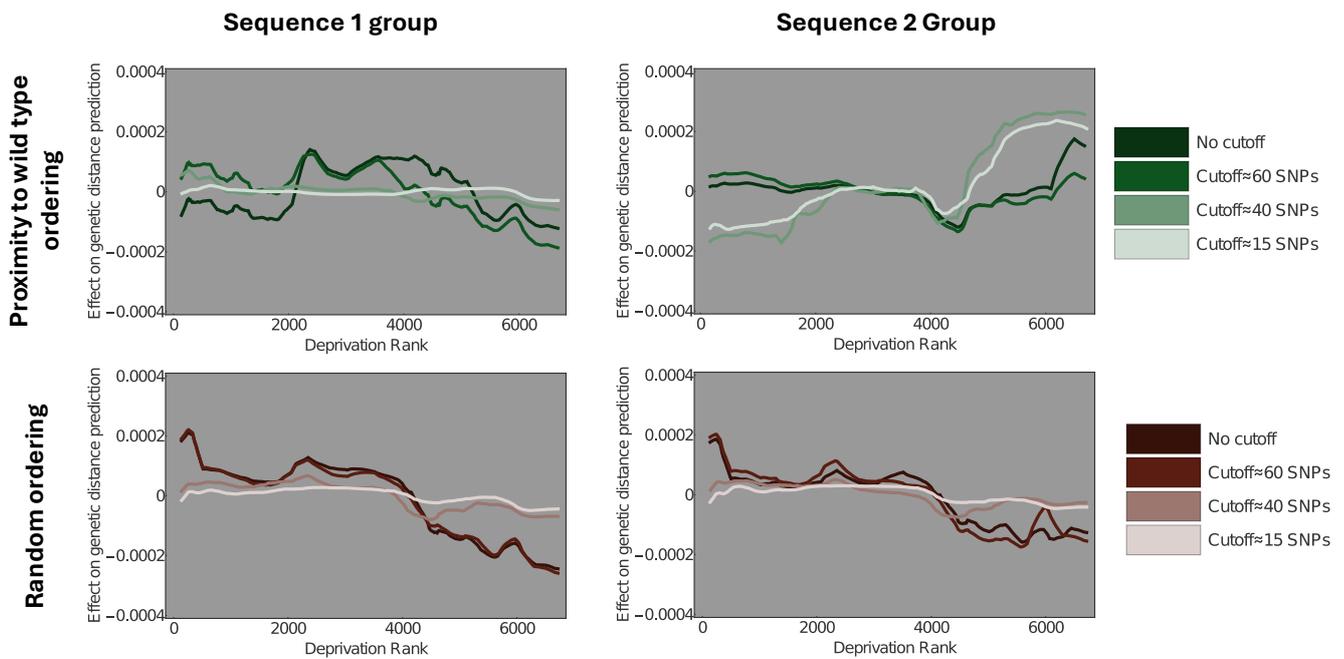

Figure 4. ALE plots based on Random Forest model fits generated for of lockdown period (02nd January 2021 to 30th April 2021) for paired SARS-CoV-2 sequence data from Tayside, Scotland selected with varied genetic distance cutoff for data where pairs were either ordered by their similarity to wild type SARS-CoV-2 sequence (Sequence 1 more similar; top row) or randomly ordered (bottom row); The Scottish Index of Multiple Deprivation ranks areas from the most to least deprived (such that bigger rank indicates lower deprivation); the correlation curves were smoothed for readability (moving average; t=10), full data are provided in Supplementary Dataset 1.

The ALE plots presented on Figures 3 and 4 indicate that the effect of lockdown was strongly deprivation dependent. The change in correlation curves generated for a range of genetic cutoffs between sequence pairs, when ordered by their similarity to wild type virus, compared to random ordering of pairs, confirms that the difference in patterns associated with sequence 2 only persists in the ordered pairing and is thus consistent with an infector/infectee effect. This finding corroborates simulation results (Figure 1).

## Discussion

Here we have analysed the Scottish SARS-CoV-2 healthcare data registered between 04th August 2020 and 30th July 2021 (mid pandemic period) from a particular health board area of Scotland, NHS Tayside, comparing the in and out of lockdown periods. With the data we explored patterns of communal transmission, in particular the infector characteristics, implicated by relationships between genetic distance and various variables recorded for small geographical areas (data zones) each described by, among others, their location, deprivation, number of cases at the time of detection, and population size. This area of Scotland was chosen due to its more tractable size, the availability of large numbers of sequences, and substantial urban and rural populations (see Supplementary Figure 4.2). Tayside also has substantial areas of deprivation (see Supplementary Figures 4.5 & 4.6), making it a useful point for this initial study. Nevertheless, previous analyses of cases suggest that drivers of case distributions in Tayside are no different from elsewhere in Scotland (Wood et al. 2023) and so are results are likely to be more broadly applicable.

Patterns of transmission were markedly different for the analysed lockdown period compared to the out of lockdown periods as demonstrated by the significant difference in relationships between genetic distance and deprivation, number of contemporaneous cases, population size and genetic distance between areas. Our study also shows that geographical proximity was an indicator of closer genetic connection. Our data analysis was limited to the first year of the SARS-CoV-2 epidemic in Scotland, during which various movement restrictions, including the full lockdown, were implemented (The Scottish Parliament, 2023) which influenced natural connectivity of the areas. Non-essential long travel and commute was banned and/or discouraged which weakens the connections to distant areas. Our analysis confirms that reductions in mobility during lockdown did have a substantial impact on the ability of the virus to spread geographically, with more infections occurring within, rather than between, the areas of residence compared to periods out of lockdown (see Supplementary Figure 2.4). The analysis of Dundee city data (See Supplementary Figure 6.1) indicates that while homogenous mixing may be a valid assumption when modelling small, urban areas, when whole regions are considered, the mixing depends strongly on distance between the areas.

We have shown that for deprivation the patterns are different for the two groups considered (Sequence 1 and Sequence 2) and we hypothesize that this is driven by infectors/infectee relationships that alter in impact depending on the severity of the restrictions at the time. The negative correlation between genetic distance and deprivation for out of lockdown period is much stronger for the Sequence 1 group than for Sequence 2 group, but interestingly, in the lockdown period the trend is reversed- the dependence between genetic distance and deprivation for Sequence 1 sample is almost flat, while for Sequence 2 sample there is a strong positive correlation. In simulations, we have shown that, where some individuals are more likely to cause infections than others, these individuals are more often found in the Sequence 1 group than in the Sequence 2 group, when the genetic distance cutoff is implemented. As the sequences are otherwise treated in an identical fashion, we claim that differences in the correlations between our risk factors and the genetic distance observed across these two groups are due to differences in the drivers of SARS-CoV-2 circulation – i.e. who the infectors are. Notably these patterns are different in and out of lockdown; we posit that this is due to a shift in activity patterns, such that both infector and infectee relationships become more pronounced in lockdown, indicating that, while lockdown relationships protected more individuals from less deprived areas from becoming infected, it was less effective amongst the more deprived. This finding is consistent with observations that individuals with lower socio-economic background may be less able to reduce physical distancing measures during lockdowns (Carrión et al., 2021; Shushtari et al., 2021), e.g. because they are essential workers (Gama et al., 2021). Moreover, survey data from the UK suggest that adherence to measures was negatively associated with lower income and positively associated with the level of measures (Kale et al., 2022). Both associations may contribute to the patterns we observed in our study. Perhaps more surprisingly, the evidence that individuals residing in more affluent areas were more responsible for the geographical spread of SARS-CoV-2, as indicated by the less deprived areas having shorter genetic distances during the out of lockdown period, which in contact networks would be interpreted as having a greater role in cross community spread (i.e. higher betweenness centrality). While there are some indications that persons residing in more affluent areas were more likely to be international travellers responsible for introducing COVID-19 from overseas locations (Sturrock et al., 2021) this is the first observation that we are aware of, showing a role for residents of affluent areas in circulating virus within sub-national regions. This may be linked to observations of the role of social gatherings as hotspots of infection (Hass & Jokar Arsanjani, 2021; Liu et al., 2021; Wang et al., 2023) if persons from more affluent areas are more likely to travel farther for this purpose; however confirming this would require further work.

It has been shown consistently in the literature that individuals from a lower socio-economic background are less likely to be tested for COVID-19 while at the same time being more likely to be infected and suffer more severe health outcomes as a consequence of the infection (Carrión et al., 2021; Garcia-Morata et al., 2022; Green et al., 2021; Gu et al., 2021; Harris et al., 2023; Landier et al., 2023; Mongin et al., 2022). While it is possible that the patterns we see are influenced by detection bias, our analysis of the data biases (Supplementary Note 1) shows that the sequenced and PCR tests from whole Scotland are biased compared to all tests combined (i.e. including antigen tests) for areas of higher health deprivation (Supplementary Figure 1.2). This may be because people from a more deprived background are more likely to develop severe symptoms of the disease and the testing bias is introduced due to more severe cases being PCR tested while seeking medical attention and less severe cases self-testing using antigen test or no test at all. Therefore, there are two opposing testing biases at play: on one hand people from more deprived areas are less likely to get tested, on the other hand if they seek testing their sample is more likely to be sequenced. Future studies should be aimed at quantifying which testing bias is stronger and how that can influence our findings.

We confirm here the expectation that the greater the number of cases, or the population of the area the greater the potential for infectious connectivity. More cases or increased population usually indicates ongoing community transmission, while places and periods that have less cases or smaller population are more likely to have cases imported from outside. Our analysis also confirms that, as intended, the lockdown was successful at scaling down contact networks in communities.

While here we compared periods with and without the enforcement of a strict, throughout whole analysed period measures of some degree were implemented (The Scottish Parliament, 2023). As a next step more detailed analysis of comparing the timeline of various restrictions and genetic patterns should be carried to inform how these restrictions may be reflected in patterns of transmission we observed. Focus of our study was on detection of transmission patterns on spatial meso-scale, describing local interactions between areas housed by small populations. As SARS-CoV-2 virus sequences does not always mutate after transmitted from one host to the other (De Maio et al., 2021) resulting in zero length genetic distance between many pairs. In such case the nuanced relationship between two sequences cannot be established from genetic data alone and as a result close interaction patterns such as household transmission events are lost.

We have demonstrated that genetic relationships between sequenced virus samples can be used in geo-epidemiological analysis to explore patterns of transmission determined by infectors. Such knowledge may be used to generate signatures that could be exploited to quantify the impact of differing levels of nonpharmaceutical interventions, which varied during the COVID-19 pandemic in Scotland as elsewhere. We show that our results, based on robust machine learning analyses, help us to understand the role of demographic factors in maintaining the geographical mid-pandemic circulation of SARS-CoV-2. Should similar data be available in the future, these approaches could be used in real time to aid in future epidemic control

## Acknowledgments

The study was funded by Chief Scientist Office (grant HIPS/21/63), and Medical Research Council (MC_UU_00034/5 ).

The SARS-CoV-2 genomics data was initially generated through a network of laboratories as part of the COVID-19 Genomics UK Consortium (COVID-19 Genomics UK Consortium (cogconsortium.uk)). The COVID-19 Genomics UK Consortium is supported by funding from the UK Medical Research Council, which is part of UK Research and Innovation, the UK National Institute of Health and Care Research (grant code MC_PC_19027) and Genome Research Limited, operating as the Wellcome Sanger Institute.

More recent SARS-CoV-2 genomics data were generated by the NHS Sequencing Service in Scotland (COVID-19 Whole Genome Sequencing (WGS) - Public health microbiology - Services - Public Health Scotland).

The SARS-CoV-2 test data was supplied by PHS via eDRIS (electronic Data Research and Innovation Service).

# Methods

*Simulation of transmission trees*

To simulate transmission trees we constructed a stochastic SIS individual based model, in which the new cases as well as the source of their infection were tracked. Each simulation started with a population of total 500 hosts, equally distributed into two groups: 250 in Group 1 and 250 in Group 2; 499 of the hosts were susceptible, and 1 host from Group 2 was infectious (Case 0). Homogeneous mixing is assumed for both groups.

Transmission was modelled with constant 5-day infectious period. When an individual recovered it was able to be infected again and the next infection was registered as a separate case. Group 2 had constant infectivity and susceptibility for all scenarios, while Group 1 had proportionally smaller or bigger infectivity and/or susceptibility depending on the scenario at hand.

The transmission rate was dependent on the groups two individuals at contact come from and was described by following matrix:

|  | Susceptible host | |
|---|---|---|
| Infectious host | **Group 1** | **Group 2** |
| **Group 1** | $\beta_{(1,1)}=xy\beta$ | $\beta_{(1,2)}=x\beta$ |
| **Group 2** | $\beta_{(2,1)}=y\beta$ | $\beta_{(2,2)}=\beta$ |

where β is a base transmission rate, x is proportional change in Group 1 infectivity and y is proportional change in Group 1 susceptibility.

For each simulation step each host could contact one other host with the probability of meeting a host of a particular type (susceptible host from either Group 1 or 2, infectious host from group 1 or infectious host from group 2) as proportionate to the numbers of hosts of a relevant type in the whole population. The probability of infection per one contact with an infectious host was:

$P_{inf} = 1 - e^{-\beta_{(i,s)}}$,

where $\beta_{(i,s)}$ is the transmission rate attributable to each susceptible-infectious pair that are in contact (see transmission matrix above). Each simulation was run for 300 steps. We simulated 12 scenarios with various x and y value combination. Each scenario was repeated 100 times; the details of the simulation scenarios are described in Table 2.

Table 2. Simulated scenarios for a two-group transmission system, with basic transmission rate (β), Group 1 infectivity and susceptibility modifiers (x and y respectively), number of simulated repetitions and sum of major (> 200 cases) and minor (≤ 200 cases) outbreaks.

|  | β | x | y | Repetitions | Major outbreaks | No/Minor outbreaks |
|---|---|---|---|---|---|---|
| **Scenario 1** | 0.12 | 1 | 1 | 100 | 84 | 16 |
| **Scenario 2** | 0.12 | 1 | 1.5 | 100 | 91 | 9 |
| **Scenario 3** | 0.12 | 1 | 0.5 | 100 | 74 | 26 |
| **Scenario 4** | 0.12 | 0 | 1 | 100 | 54 | 46 |
| **Scenario 5** | 0.12 | 0 | 1.5 | 100 | 64 | 36 |
| **Scenario 6** | 0.12 | 0 | 0.5 | 100 | 30 | 70 |
| **Scenario 7** | 0.12 | 0.5 | 1 | 100 | 80 | 20 |
| **Scenario 8** | 0.12 | 0.5 | 1.5 | 100 | 89 | 11 |
| **Scenario 9** | 0.12 | 0.5 | 0.5 | 100 | 52 | 48 |
| **Scenario 10** | 0.12 | 1.5 | 1 | 100 | 88 | 12 |
| **Scenario 11** | 0.12 | 1.5 | 1.5 | 100 | 91 | 9 |
| **Scenario 12** | 0.12 | 1.5 | 0.5 | 100 | 83 | 17 |

For each scenario we selected all major outbreaks (outbreaks of size bigger than 200 hosts) and created transmission trees from tracked infector-infectee pairs. Next, using the transmission trees we calculated patristic distances (counting each branch of a tree as one unit) for all the pairs of cases present on a tree. Cases in each pair were either ordered according to their proximity to Case 0 (Sequence 1 being case with a shorter patristic distance to Case 0 and Sequence 2 with a longer distance) or alternatively they were ordered randomly (each pair generated from a tree either stayed in the same configuration or was swapped, in both cases with probability 0.5). To account for various genetic distance cutoff values, we selected from each dataset the pairs that were separated by distance lower than particular cutoff, and next we calculated the proportion of individuals from Group 1 to individuals of Group 2 for both the Sequence 1 and Sequence 2 groups, considering two ordering variants: first, for pairs ordered by proximity to Case 0 and second, as a control, pairs ordered randomly. In the main manuscript we presented summary results for Scenario 4. Detailed results for all the scenarios are presented in Supplementary Dataset 1.

*SARS-CoV-2 Scottish Data*

We use data in spatial resolution of Scottish data zones, small areas or residence housed by approx. 500-1000 individuals (Scottish Government, 2024c). The SARS-CoV-2 PCR and antigen tests and whole genome sequence data were gathered as a part of public health pandemics monitoring program in years 2020-2023. For the period we analysed here (04[th] August 2020 to 30[th] July 2021), of 390,025 of positive cases (361,218 PCR positive and 28,807 antigen positive) we were able to match 62,796 (approx. 16.10%) with sequence data that was obtained from PCR test samples (COG-UK consortium, 2023).

NHS Tayside is a region in central Scotland housing around 400,000 people; its main cities are Dundee and Perth (Scottish Government, 2018). To ensure that the analysis is computationally feasible we focused on one region of Scotland. We choose Tayside as our case region as during the pandemic it was a region with persistent SARS-COV-2 circulation and it has a robust testing program (NHS Tayside, 2022); it consists of rural as well as urban areas of variable population density and deprivation (see maps in Supplementary Note 4). In NHS Tayside, out of 29,672 of recorded positive cases (27,058 PCR positive and 2,614 antigen positive), we were able to match 4,880 (approx. 16.45 %) to sequence data.

*Data sources*

For our analysis we combined data from eight datasets, the detailed description of the raw datasets with the source information is provided in Table 3. Two of the datasets (dataset I-II in Table 3) are confidential, one dataset can be accessed with restrictions (dataset III in Table 3), and five of the datasets are public and can be accessed without licence (datasets IV-VIII below in Table 3).

Table 3. Datasets used to assess the dependencies between SARS-Cov-2 genetic distance and various factors describing the areas of residence (data zones) of the tested individuals in NHS Tayside, Scotland in period from theperiod 04th August 2020 to 30th July 2021.

| | | | |
|---|---|---|---|
| **Dataset I** | Scottish SARS-CoV-2 test data | Record of all SARS-CoV-2 positive PCR and antigen tests in Scotland recorded from 01st July 2020 to 25 July 2023; each case was described with the following variables: Patient ID, care home ID (if relevant), collection date, age band (each of 5 years), sex, test type, test reason, NHS board of residence, Data Zone of residence, test results of s gene status, specimen ID. | Protected access provided by NHS Scotland per data agreement; |
| **Dataset II** | Sequence and case IDs (from Public Health Scotland) | Dataset that links specimen ID (used in Dataset I) with sequence ID (Used in dataset III), additionally it also contains collection date. | Protected access provided by Public Health Scotland per data agreement; |
| **Dataset III** | Scottish SARS-CoV-2 sequences | Fasta files with sequence nucleotide code from trimmed sequenced with masked alignment and metadata for these. | Public access via COVID-19 Genomics UK Consortium website (COG-UK consortium, 2023); accesses in 2022, currently can be used as a version archived on 5 May 2023) |
| **Dataset IV** | Commuting flow data | 2011 Census Special Workplace Statistics in Output Zone level for location of usual residence and place of work from 2011 census flow data (dataset census code WF01BUK_oa); each output zone describes with three variables: 1) area of usual residence; 2) Area of workplace; 3) Number of persons. | Access for non-profit and educational use provided via UK Data service portal (UK Data Service) |
| **Dataset V** | Scottish Index of Multiple Deprivation | Data from Scottish Index of Multiple Deprivation 2020v2 (SIMD) that ranks data zones from the most deprived (rank 1) to the least deprived (rank 6976) based on assessment of seven sub-domains: income, employment, education, health, access to services, crime and housing; in lookup file each Data Zone was described with following variables: Data Zone code, Data Zone Name, SIMD Rank, SIMD Vigintile, SIMD Decile, SIMD Quintile, Income domain rank, Employment domain rank, Education domain rank, Health domain rank, Access domain rank, Crime domain rank, Housing domain rank, Population (as off 2017), Working_Age_Population (as off 2017) and 12 variables with various geographical area codes and names as used for small area statistics in Scotland (Scottish Government, 2024c). | Public access via SIMD website (Scottish Government, 2024b) |
| **Dataset VI** | Data Zone Centroids | Point coordinates describing the population weighted centre of 2011 Data Zones in Scotland defined in 2011 by Scottish Government | Publicly available via GOV.UK (Scottish Government, 2024a) |
| **Dataset VII** | Geographical codes | Connected codes for Output Area, Data Zones and Intermittent Zones ("Output Area 2011 to Data Zones and Intermittent Zones 2011") as used for small area statistics in Scotland (Scottish Government, 2024c) | Publicly available via National Records of Scotland website with census indexes coding (National Records of Scotland) |
| **Dataset VIII** | Restriction level data | Dataset prepared for Oxford Covid-19 Government Response Tracker (OxCGRT)(Thomas Hale et al., 2021) ("OxCGRT compact subnational v1") describing, on country level, various factors systematically measuring changes in global response to covid in time; variables relevant to out study describe: country, date, average Stringency Index, average Government Response Index, and average Containment Health Index (see (Thomas Hale et al., 2023) for description of indices) | Publicly available via OxCGRT GitHub (Hale et al., 2021) |

*Data processing*

Sequence and case data

We used SARS-CoV-2 whole genome sequence data derived from Scotland (Dataset III in Table 3) that already were trimmed and had masked alignment. Due to memory problems the fasta file containing all the sequences was divided into 9 separate files and further analysis of sequences was from the separated files. Out of 2,985,080 sequences for the UK provided by the COG Consortium (COG-UK consortium, 2023) we selected 355,126 sequences coming from Scotland (labelled with "Scotland"). For these we selected sequences with IDs that matched the IDs that were present in the Sequence and case IDs data (Dataset II).

Out of 355,126 Scottish sequences 342,360 (approx. 96.41 %) were connected to a unique case identifier (ID). Additionally, 37 549 (out of 379,909; approx. 9.88%) sequence IDs from Dataset II were not connected to any sequence data and thus were excluded. Fifty eight cases were connected to more than one sequence ID. We retained these sequences under the assumption these are correct (e.g. when someone is concurrently infected with two strains). As the number of rejected entries after matching (either sequences from Dataset III or sequenced cases IDs from Dataset II) was small, we did not perform analysis exploring the reasons for the data mismatch. As date of collection was reported in both Dataset I and Dataset II we cross checked if the dates were the same for respective IDs. 2231/342360 (approx. 0.65%) of matched Scottish sequences had mismatched collection dates. In all cases we selected the earlier date reported; the most plausible explanation being that the first was the recorded sampling date and the second a different laboratory entry date. As the number of mismatched dates was small and the difference in dates was short in the majority of cases (see Supplementary Table 5.1 in Supplementary Note 5) we assumed that the effect of the detected discrepancy on our final results will be negligible and therefore we have not performed any additional analysis.

For this analysis we selected sequence data (connected to case IDs) from the NHS Tayside only (NHS board Tayside; codes S08000027 and S08000030 as assigned by Office for National Statistics (Office for National Statistics, 2024) from period 04$^{th}$ August 2020 to 30$^{th}$ July 2021, which means in the final analysis we included 4880 sequences.

To explore connectivity of sampled sequences we studied the correlation between the genetic distance of pairs of sequences and a set of factors describing the case characteristics (age category, sex, detection time and place of living), and the demographic characteristics associated with the Data Zone of residence (e.g. deprivation, population size, number of SARS-CoV-2 positive cases in the area and commuting connections). The final data set (combined Dataset I and II; NHS Tayside only; period 04th August 2020 to 30th July 2021) was used to generate pairs of sequences that we subsequently used to calculate genetic distance between them. To avoid matching together sequences that were distant in time (and therefore not relevant for studying relatively close relationships), the sequence pairs were generated only within a time widows of 4 weeks (28 days) with a rolling interval of 2 weeks (14 days)- see Fig. 5. When the pair was present in two windows the record from earlier window was kept.

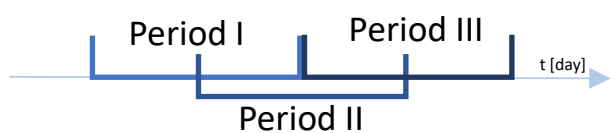

Figure 5. Schematic representation of 28 day time windows with 14 day rolling interval.

The genetic distance between paired sequences was calculated for each pair using the Tamura and Nei 1993 model (Tamura & Nei, 1993) (R; function: dist.dna, package: APE (Emmanuel Paradis, 2024)).

To decide which sequence within a pair is more likely to have emerged earlier (e.g. is closer on the transmission tree to the first ever SARS-CoV-2 case), sequences in each pair were ordered after comparison to the first SARS-COV-2 complete genome sequence published in GenBank (first published SARS-CoV-2 sequence isolate Wuhan-Hu-1; MN908947.3 (Wu et al., 2020a, 2020b))- assumed to be a wild type virus. For each sequence in the pair, the distance to the wild type was calculated using Tamura and Nei 1993 model (R; function: dist.dna, package: APE); the sequence with the shortest distance (more similar) to the wild type was named Sequence 1 and sequence with the longest distance (less similar) was named Sequence 2.

Each sequenced case was provided with the sex of the tested individual (binary: female/male) and 5 year age band (classified in 16 categories, see Table 4). To reduce the number of age categories we assigned cases to 4 new age categories- each covering 20 years; see Table 4 for details. Additionally, as needed for some of the fitted models, we implemented binary variables describing if paired cases were within the same age group/sex or not.

Table 4. New 20-year age classes assigned to cases originally described with 5-year age classes.

| Original age class (5 years span) [years old] | New age class (20 years span) [years old] |
|---|---|
| 00-04 | 0-19 |
| 05-09 | |
| 10-14 | |
| 15-19 | |
| 20-24 | 20-39 |
| 25-29 | |
| 30-34 | |
| 35-39 | |
| 40-44 | 40-59 |
| 45-49 | |
| 50-54 | |
| 55-59 | |
| 60-64 | 60+ |
| 65-69 | |
| 70-74 | |
| 75+ | |

The temporal distance was calculated as a difference (in days) between Sequence 1 and Sequence 2. The order of sequences was preserved, with negative temporal distance meaning that the Sequence 1 was detected later than Sequence 2.

Geographical data

In addition to case demographic data we studied how various factors describing data zones of residence (but not necessarily tested individuals) are connected to SARS-CoV-2 transmission. Each case in Dataset I and subsequently each sequence matched to (some of) the cases was provided with the data zone (2011 version) of the residence of the tested individual. The version we used following the 2011 Census to have population of approximately 500-1000 inhabitants at that time, with compact shape and "contain households with similar social characteristics", as a unit that reflects the properties of local communities ensuring the anonymity of the people (Scottish Government, 2024c). The 2011 version of the data as at the time of analysis, the 2022 Census data were not yet available. These data include a total of 6,976 data zones described with geographical boundaries and population weighted centroids (here Dataset VI). As they are the standardised geography, many population data in Scotland are published at data zone resolution or in units that can easily be aggregated/transformed to them (National Records of Scotland, 2024) (e.g. see Dataset VII that describes the relations between various geographies). These include datasets we used together with SARS-CoV-2 case data in our analysis, namely Commuting flow data (here Dataset IV) and Scottish Index of Multiple Deprivation (here Dataset V).

Data zone centroids coordinates were provided by SpatialData.gov.scot (Scottish Government, 2024a) the centroid position is a population weighted centre of the all Scottish Data Zones defined in 2011. As the centroids were provided in British National Grid coordinates (BNG, ESPG:27700) we converted them in R (function: st_transform; package: sf (Edzer Pebesma, 2024)) into latitude and longitude format (lat/long, WGS84; EPSG:4326). In all analysis we used the WGS84 system. The centroids were used to calculate the geographical distance between Data Zones assigned to each case within a pair (R; function: st_distance; package: sf).

Commuting flow data were provided from 2011 census (UK Data Service) and describe the origin output zone (output zone of the residence) and destination output zone (output zone of work) together with a number of people that commute from the origin to the destination. Output zones are smaller sub-geographies lying within data zones (Scotland's Census, 2021); for consistency we aggregated any data from the output zone level to the data zone level using the coding provided in Dataset VII. Next, we created the "Connections" variable; for every pair we summed up the number of people exchanged between the two data zones forming a pair and for each one also number of people exchanged with the data zones that both of the paired areas have connections with (have either non-zero number of people coming to work from each paired data zone or non-zero number of people coming to work to each paired data zone). The algorithm and Connections variable definition is illustrated on Supplementary Figure 5.1 (Supplementary Note 5).

Additionally, we linked to the cases data the Scottish Index of Multiple Deprivation indices (Scottish Government, 2024b) describing deprivation in the data zone of residence (Dataset V). Deprivation rank based on (SIMD) describes the deprivation of areas accounting for various factors (housing, health, income, crime, education and access deprivation); the smaller the rank the bigger the deprivations of the data zone (Scottish Government, 2024b). We included a variable describing overall/main deprivation index ("Rank"), it's 8 sub-domains (Income, Employment, Education, Health, Access, Crime and Housing) as well as population estimates (as of year 2017). Apart from SIMD indices we also added the mean and difference of the main deprivation index and population size to the data for each sequence pair.

To consider the local circulation of virus within data zones, for each pair we also calculated the number of SARS-CoV-2 cases present in each of their data zones of residence in the period the pair was formed in (using the same time widows of 4 weeks with rolling interval of 2 weeks as for the pair formation; Fig. 5). Apart from the sum of the cases we also calculated the mean and difference of cases.

Other

Additionally, to account for level of restrictions applied in Scotland in 2020/2021 we used the Oxford Covid-19 Government Response Tracker that systematically describes restrictions used in various countries during the pandemics and how they change in time (Thomas Hale et al., 2023; Thomas Hale et al., 2021). The three indices we used (Stringency Index, Government Response Index, and Containment Health Index) were reported for the whole Scotland (no spatial resolution) and were variable in time. We used the index together with official government communications (The Scottish Parliament, 2023) to define in lockdown (02nd January 2021 to 30th April 2021) and out of lockdown (04th August 2020 to 01st January 2021 and 01st May 2021 to 30th July 2021) periods for the timeline we analysed as relevant to Tayside.

*Random Forest model*

To describe the relationship between genetic distance for paired sequenced cases and other variables we considered many versions of the Random Forest model for which we estimated $R^2$ and feature importance (permutation importance) (See Supplementary Dataset 2 with summary of all the model versions and Supplementary Note 2 for model results). We selected the final version presented on Fig. 2 in the Results section for which we generated Accumulated Local Effects (ALE) plots to gain additional insight into selected variables (see Supplementary Note 2 for full results).

Technical information

Random Forest analysis was implemented in R using package "ranger" (Marvin N. Wright, 2023) that allowed us to perform multicore calculations that substantially reduce computing time. Each model run was performed with the same set of parameters. Number of trees was set at 500 and number of variables to possibly split at in each node (mtry) was set at 5. The output was generated for permutation importance. Other parameters regarding Random Forest fitting were set at default values as defined by the package authors. Each model version was run for all the data first (i.e. 100% data used for training) to get the best insight into patterns that can be captured by the algorithm and subsequently for chosen model we assessed the model performance by considering versions with only portions of data used as a training sample (see the "Model performance assessment" section below). ALE plots were generated by ALEplot function (R; package: ALEPlot (Apley, 2018)) with "predict" function from ranger package (R; package: ranger) using Random Forest models trained on the 100% data and keeping number of intervals ("k") set to be 500 (or the maximum number of intervals if less than 500 data points available).

Model selection

For the model selections we used data from both the in and out of lockdown periods together (to obtain one Random Forest fit to all analysed timeline from 04th August 2020 and 30th July 2021). The in and out of lockdown period analysis was conducted using only the variables of the selected final model. During the model selection process we ran a total of 15 versions of the Random Forest with various combinations of variables (see Supplementary Dataset 2 for summary description of all tested versions). Based on the feature importance and influence on $R^2$ (see Supplementary Figure 2.8 in Supplementary note 2) the following variables were selected for the final model: age category, sum of all positive cases, sum of connections, geographical distance, total population for Data Zone, deprivation rank for Data Zone, temporal distance. All the variables we considered are described in Table 5.

Table 5. Variables used in Random Forest model to study genetic distance dependence for SARS-COV-2 cases in NHS Tayside, Scotland in period 04.08.2020 to 30.07.2021.

| Variable | | Description | Data source |
|---|---|---|---|
| **Genetic distance** | GeneticDistance | Genetic distance between two SARS_COV-2 sequences in a particular pair, calculated from full sequence dataset using TN93 (Tammura and Nei 1993) model (Tamura & Nei, 1993) with 0.5 gamma correction and pairwise deletion for missing data | Dataset III |
| **Age category** | Age_cat_Seq1 | Age category for sequence 1; the 5-year categories have been aggregated into 20-year categories | Dataset I |
| | Age_cat_Seq2 | Age category for sequence 2; the 5-year categories have been aggregated into 20-year categories | Dataset I |
| **Sum of SARS-COV-2 cases** | Cases_Total_Seq1 | Total number of positive PCR and antigen tests in time window of the particular pair for Sequence 1 | Dataset I |
| | Cases_Total_Seq2 | Total number of positive PCR and antigen tests in time window of the particular pair for Sequence 2 | Dataset I |
| **Sum of work related connections between datazones** | Connections | Number of persons from data zone of Sequence 1 and data zone of Sequence 2 that work in the same data zone elsewhere and number of persons that live elsewhere and work in data zone of sequence 1 or data zone of sequence 2 (see Supplementary Figure 5.1) | Dataset IV |
| **Geographical distance between datazones** | GeographicalDistance | Distance (in meters) between centroids that represent the population weighted centre of data zones for particular pair | Dataset VI |
| **Total population for datazone** | Population_Seq1 | Total number of people living in data zone of Sequence 1 estimated as of 2017 | Dataset V |
| | Population_Seq2 | Total number of people living in data zone of Sequence 2 estimated as of 2017 | Dataset V |
| **Deprivation rank for datazone** | SIMD_Rank_Seq1 | Scottish Index of Multiple Deprivation rank for data zone of Sequence1 | Dataset V |
| | SIMD_Rank_Seq2 | Scottish Index of Multiple Deprivation rank for data zone of Sequence1 | Dataset V |
| **Temporal distance between cases** | Temporal distance | Time between sampling for Sequence 1 and Sequence 2, based on sample collection dates provided in case detest and PHS sequence dataset; if date in PHS sequence dataset was different than date in case dataset, the earlier date was taken; negative is when Sequence 2 was sampled earlier than Sequence 1. | Dataset I / Dataset II |
| **Rejected and additional variables** | | | |
| | Cases_Ant_Seq1 | Total number of positive antigen tests in time window of the particular pair for Sequence 1 | Dataset I |
| **Sum of antigen positive cases** | Cases_Ant_Seq2 | Total number of antigen tests in time window of the particular pair for Sequence 2 | Dataset I |
| **Sum of PCR positive cases** | Cases_PCR_Seq1 | Total number of positive PCR tests in time window of the particular pair for Sequence 1 | Dataset I |
| | Cases_PCR_Seq2 | Total number of positive PCR tests in time window of the particular pair for Sequence 2 | Dataset I |
| **Residence to work flow** | Flow_Seq1_Seq2 | Number of persons traveling to work from data zone of Sequence 1 to data zone of Sequence 2 | Dataset IV |
| | Flow_Seq2_Seq1 | Number of persons traveling to work from data zone of Sequence 2 to data zone of Sequence 1 | Dataset IV |
| **Sex** | Sex_Seq1 | Sex of the sampled case for Sequence 1 | Dataset I |
| | Sex_Seq2 | Sex of the sampled case for Sequence 2 | Dataset I |
| **Mean sum of SARS-COV-2 cases** | Case_Total_Mean | Mean number of cases | |
| **Difference in total population for paired datazones** | Population_Difference | Difference between the total numbers of positive PCR and antigen tests in time window for the particular pair (absolute value) | |
| **Mean of total population for paired datazones** | Population_Mean | Mean of the total number of positive PCR and antigen tests in time window of the particular pair | |
| **Difference in deprivation rank for paired datazones** | SIMD_Rank_Mean | Mean of the Scottish Index of Multiple Deprivation ranks for the particular pair | |
| **Mean in deprivation rank for paired datazones** | SIMD_Rank_Difference | Difference between the Scottish Index of Multiple Deprivation ranks for the particular pair (absolute value) | |

We considered three variants of the number of contemporaneous cases variable: the total number of all positive tests, total number of all PCR tests and total number of all antigen tests. In the Random Forest model runs with all three of these variables the total number of all tests had the biggest importance, and metrices representing only one test type (PCR or antigen) have similar but slightly lower importance (see Supplementary Figure 2.10). As data for both tests were available in our dataset we used the most informative metrics: the total number of all tests.

In the models defined to study the level of restriction, we tested three indices: average Stringency Index, average Government Response Index, and average Containment Health Index. The feature importance plot (Supplementary Figure 2.11 from Supplementary Note 2) indicates that Containment Health Index has the biggest average importance and as such we used it to decide on lockdown period dates (Supplementary Figure 2.10).

Model performance assessment

To assess implementation of the Random Forest models, for the selected model versions (Supplementary Dataset 2, Model Version: RF6Y1) we added randomly generated variables (R; function: runif, package: stats, part of R 3.6.2 (R, 1970); generated numbers between 1 and 1000). As expected, the random variable was returned as being unimportant (permutation importance close to zero (see Supplementary Figure 2.13 from Supplementary Note 2).

To assess the prediction error for selected Random Forest models, we randomly divided the datasets into two training and testing sub-sets in a 75:25 ratio (R; function: sample_n, package: dplyr (Hadley Wickham, 2023)). The model trained on the randomly selected 75% (2,077,588/2,770,117) cases is presented on Supplementary Figure 2.13 (Supplementary Note 2). Both the $R^2$ and feature importance are similar to the original model that was trained on the whole dataset (Supplementary Figure 2.6 from Supplementary Note 2)). The remaining 25% of sequences (666,418/2,770,117) were used as a test dataset to estimate the prediction error of the trained model. The plot of residuals for both training and testing datasets is presented on Fig. 6.

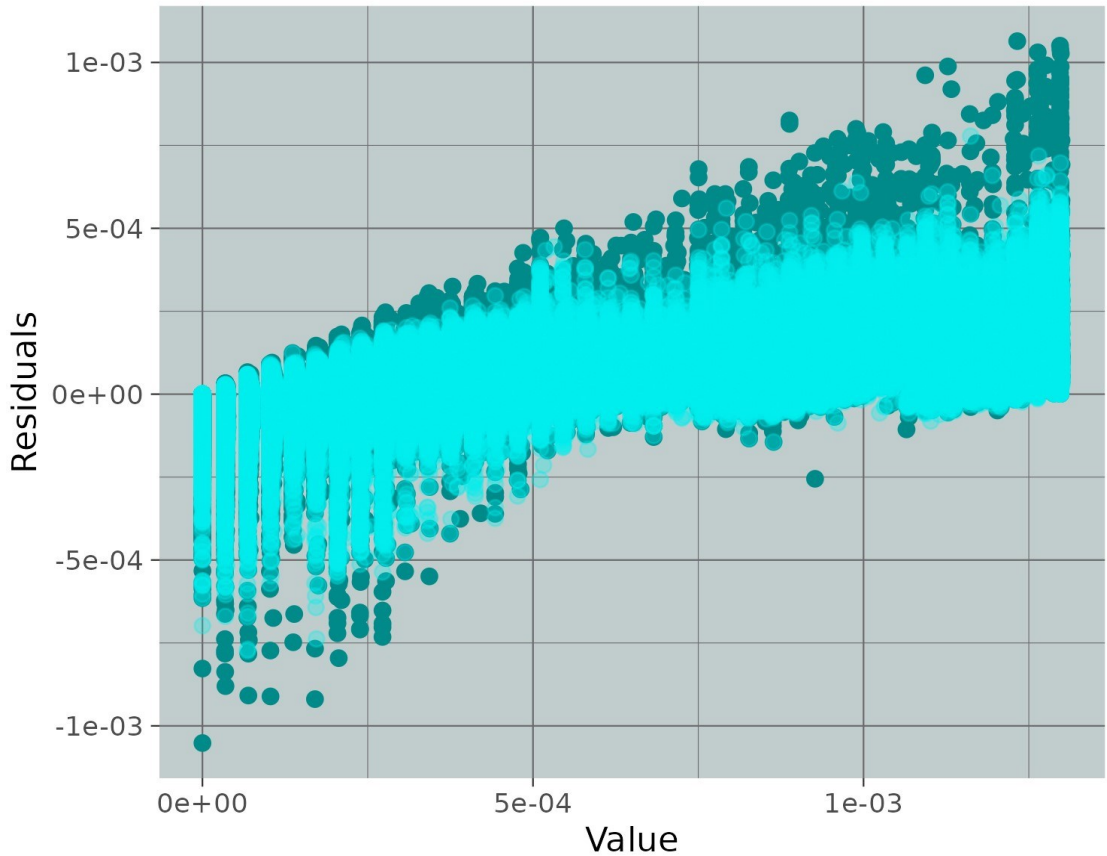

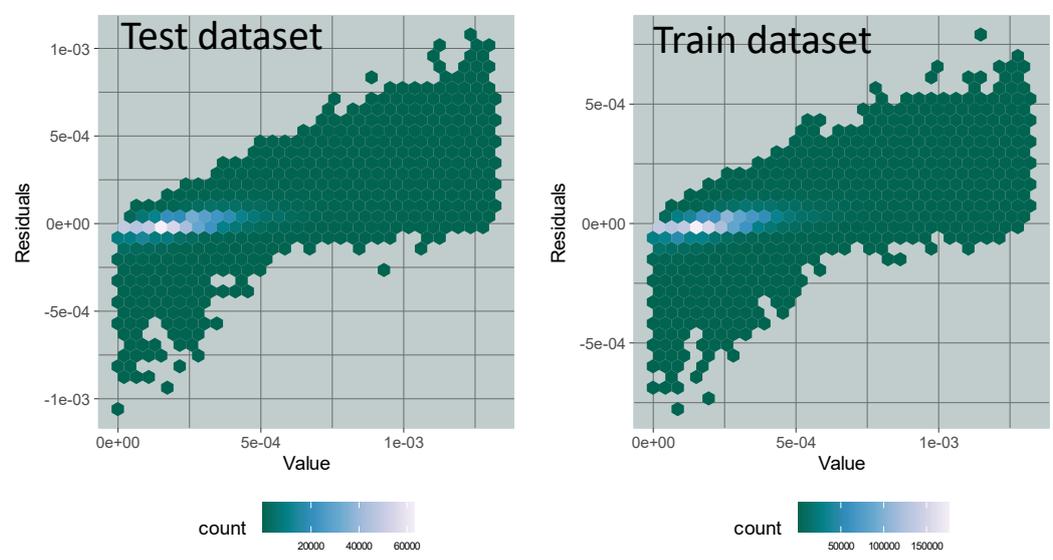

Figure 6. Residuals model predictions for "Test dataset" and "Train dataset" from Random Forest on trained "Train dataset"; upper panel shows direct comparison of the residuals for both datasets, while lower panel shows density plot of the residuals; train dataset was comprised of 75% (2,077 588/2,770,117) of randomly selected pairs of SARS-COV-2 sequences from NHS Tayside, Scotland in period from 04[th] August 2020 to 30[th] July 202.

Models with varied genetic cutoff value

To explore the difference between ordering the pairs according to their similarity to wild type and random ordering, we generated RF model fits for paired sequence data without implementing any genetic cutoff and for data selected with 5 cutoff values (genetic distance cutoff of 0.002, 0.0015, 0.0013, 0.001, 0.0005, 0.0001, using the Tamura and Nei 1993 model (Tamura & Nei, 1993)) for two sorting scenarios: 1) sequences within pair ordered by proximity to wild type SARS-CoV-2 virus (first published SARS-CoV-2 sequence isolate Wuhan-Hu-1; MN908947.3 (Wu et al., 2020a, 2020b)) and 2) sequences within pair ordered randomly (sequences kept the same ordering or swapped both with probability 0.5). In the Figure 3 we present smoothed lines for only three distances: 0.0005, 0.0013, 0.0020. The summary description of all models is provided in Supplementary Dataset 2, the summary ALE plots are provided in Supplementary Dataset 1, and the importance plots are presented in Supplementary Note 2 (Supplementary Figures 2.15-2.21).

**Methods' references**

# Title

Infector characteristics exposed by spatial analysis of SARS-CoV-2 sequence and demographic data analysed at fine geographical scales.

## Authors

Anna Gamża[1], Samantha Lycett[1], Will Harvey[1], Joseph Hughes[3], Sema Nickbakhsh[4], David L Robertson[3], Alison Smith Palmer[4], Anthony Wood[1], Rowland Kao[1,2]

Affiliations:

1) The Roslin Institute, University of Edinburgh, Edinburgh, UK

2) School of Physics and Astronomy, University of Edinburgh, Edinburgh, UK

3) MRC-University of Glasgow Centre for Virus Research, Glasgow, UK

4) Public Health Scotland, Glasgow, UK

# Supplementary Information

**Supplementary Note 1.  Data biases**
The initial analysis comparing sequenced PCR tests, and all positive tests was conducted using all Scottish case data reported from period 04[th] August 2020 to 30[th] July 2021. We compared the basic statistics of the cases namely: sex and age of the cases, number of positive cases in time, NHS board the cases came from and Data Zone they inhabit.

The sequenced cases data for both the whole of Scotland and only the Tayside region is shown to be representative of all positive (detected) cases in terms of sex, age structure, geographical coverage (see Supplementary Fig. 1.1, panels A,B,D,E, F). We detected some temporal biases, showing that for some periods a smaller proportion of detected cases was sequenced than for the others (Supplementary Fig. 1.1, panel C). Both sequenced (Supplementary Figure 1.2 C) and PCR positive cases (Supplementary Figure 1.2 E)  are biased towards areas with higher health deprivation (subdomain accounting for illness and mortality, alcohol and drug related hospitalisation, emergency hospital stays, mental health drugs prescriptions and low birth weight (Scottish Government, 2024)) comparing to LFD positive cases and to all positive cases (PCR and LFD together) (Supplementary Figure 1.2 C); this pattern was not observed when considering the Tayside data only (Supplementary Figure 1.2 B, D, F, H)

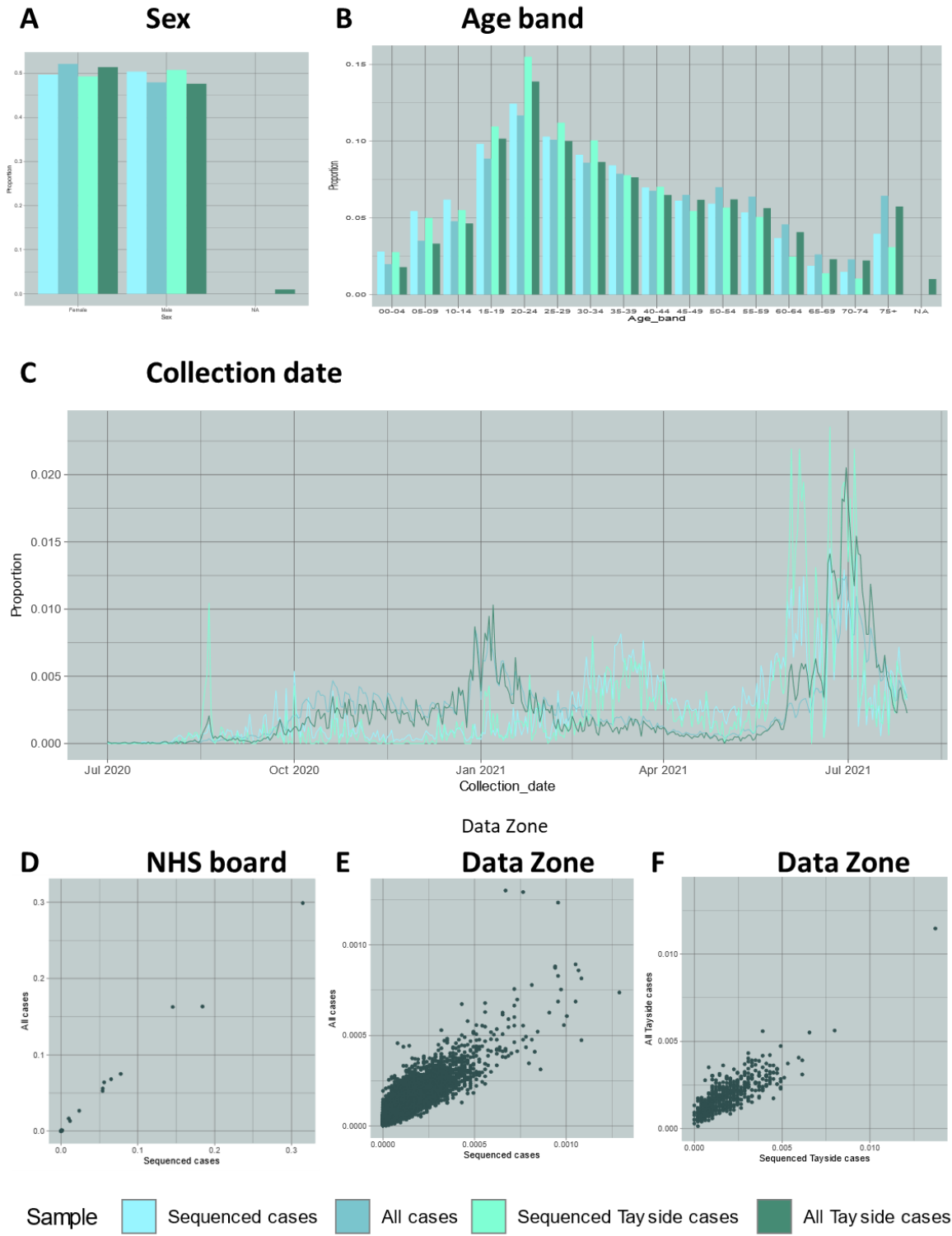

Supplementary Figure 1.1 Comparison of various characteristics for all detected positive and all sequenced SARS-CoV-2 cases in all regions of Scotland and in Tayside regions only, region from period 04[th] August 2020 to 30[th] July 2021; A) comparison of the cases sex; B) comparison of the cases age band; C) comparison of the rolling mean of proportions of cases observed every day; D) comparison of the cases NHS board; E) comparison of the cases Data Zone for whole Scotland; F) comparison of the cases Data Zone for whole Tayside only.

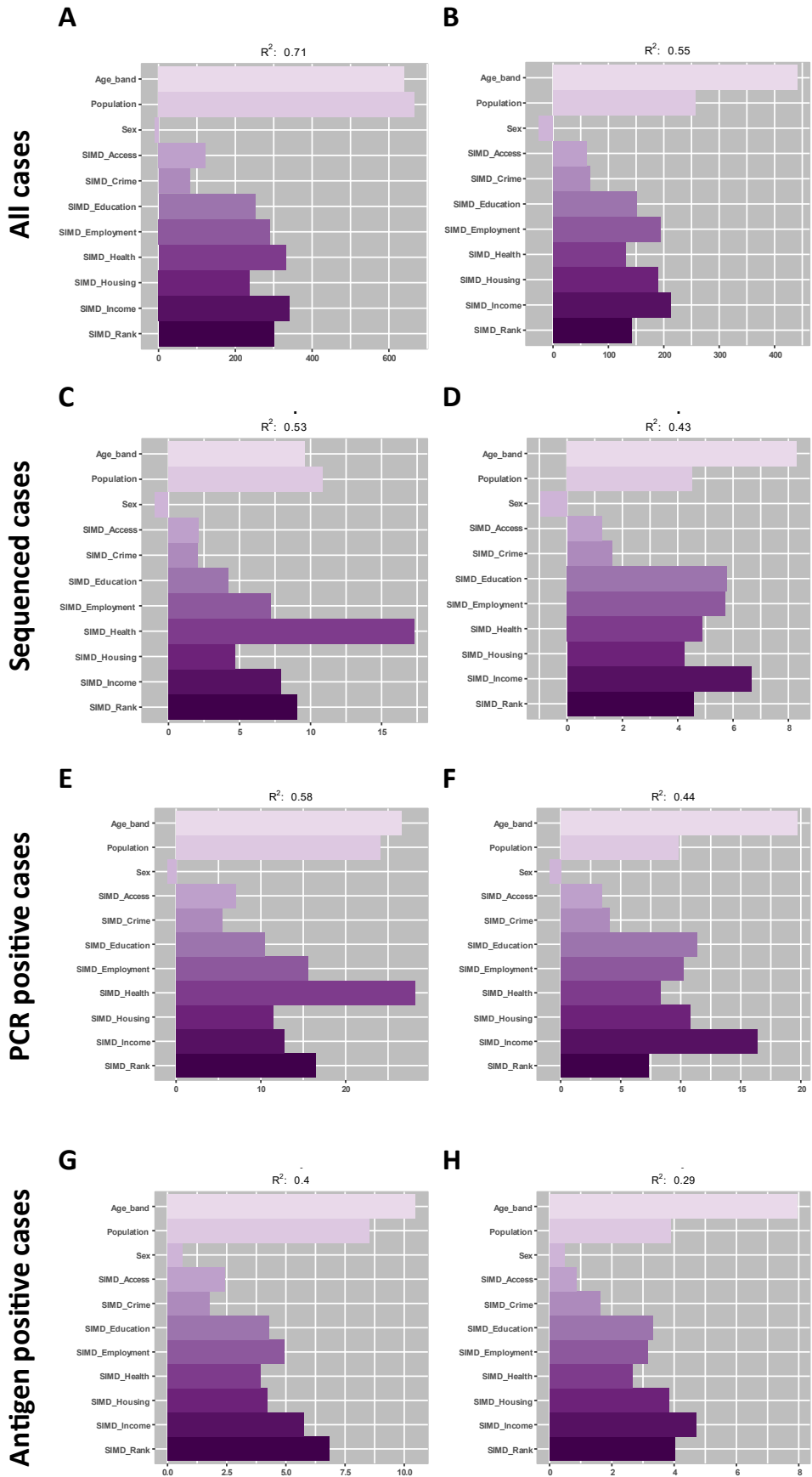

Supplementary Figure 1.2. Feature important of selected analysed variables for a Random Forest model describing the total number of SARS-COV-2 cases in each data zone separately for each age band (of 5 years) and sex in Scotland from period 04[th] August 2020 to 30[th] July 2021.

**Supplementary Note 2. Full Random Forest results (ALE plots, model selection and alternative model fits)**

*Full ALE plots for in and out of lockdown periods*

Deprivation (Scottish Index of Multiple Deprivation Rank)

Deprivation is measured in Scotland by Scottish Index of Multiple Deprivation that ranks Data Zones from the most to least deprived (such that higher rank indicates lower deprivation) (Scottish Government, 2024).

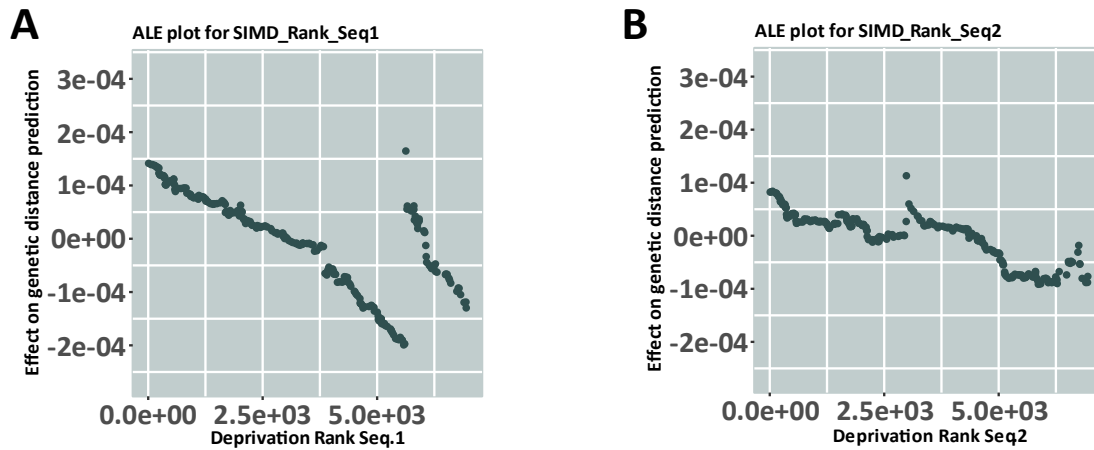

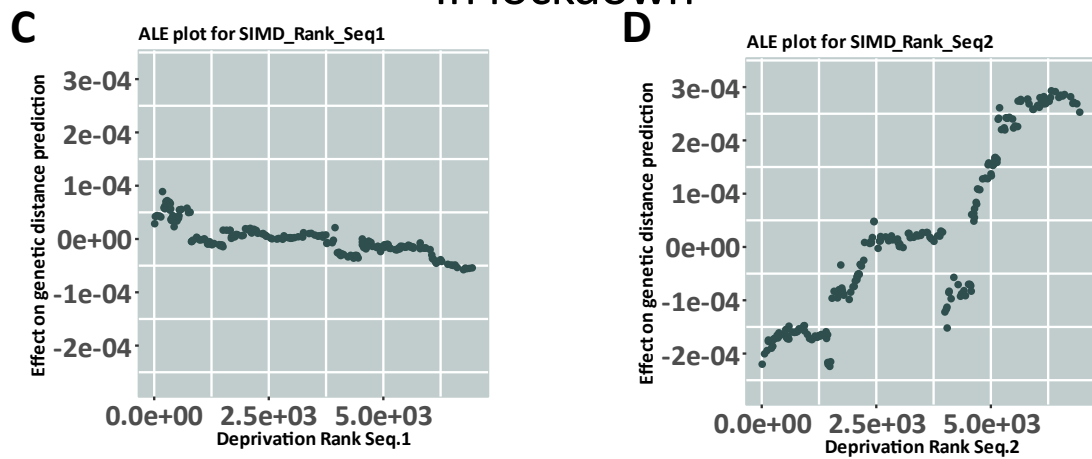

Supplementary Figure 2.1. Accumulated Local Effects plots for the deprivation (SIMD) rank in data zones of residence of sampled individuals generated from Random Forest fit to sequenced cases from Tayside, from period 04$^{th}$ August 2020 to 30$^{th}$ July 2021

Number of contemporaneous cases and population size

The total number of contemporaneous cases is an indicator of the severity of the epidemics in the Data Zones housed by paired cases in the 30 day period these cases were tested. Population size describes the number of people registered as living in each data zone (as of 2017) recorded in SIMD dataset (Scottish Government, 2024).

For both total number of contemporaneous cases and population size variables trends observed in ALE plots show similar patterns (Supplementary Figure 2.2 and Supplementary Figure 2.1).

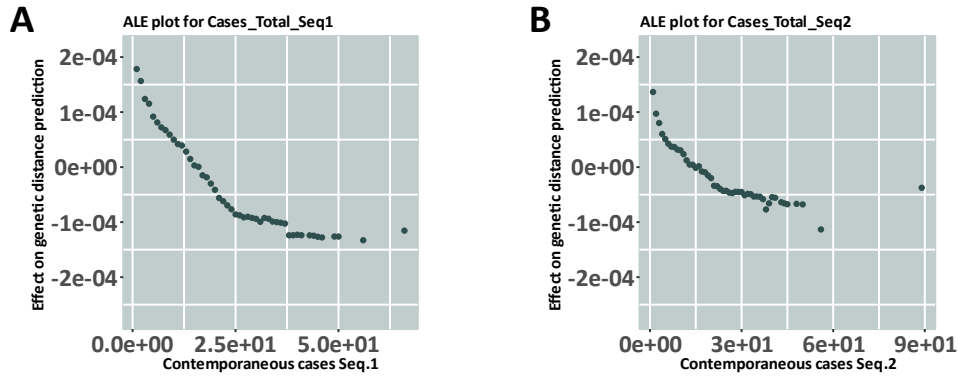
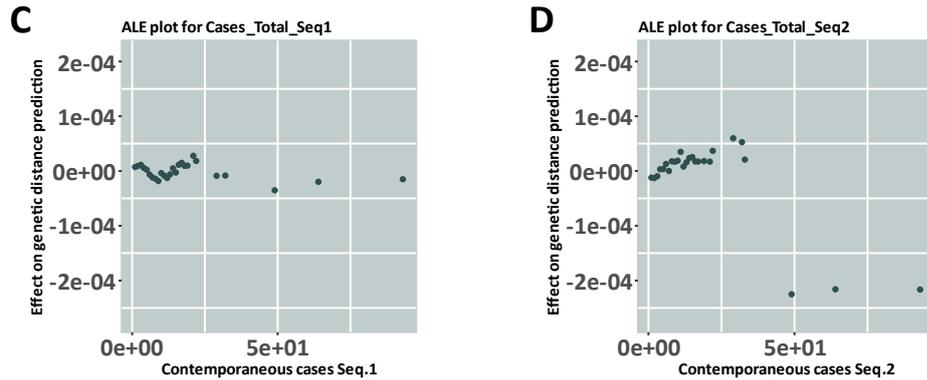

Supplementary Figure 2.2. Accumulated Local Effects plots for the number of contemporaneous cases in data zones of residence of sampled individuals generated from Random Forest fit to sequenced cases from Tayside, from period 04[th] August 2020 to 30[th] July 2021.

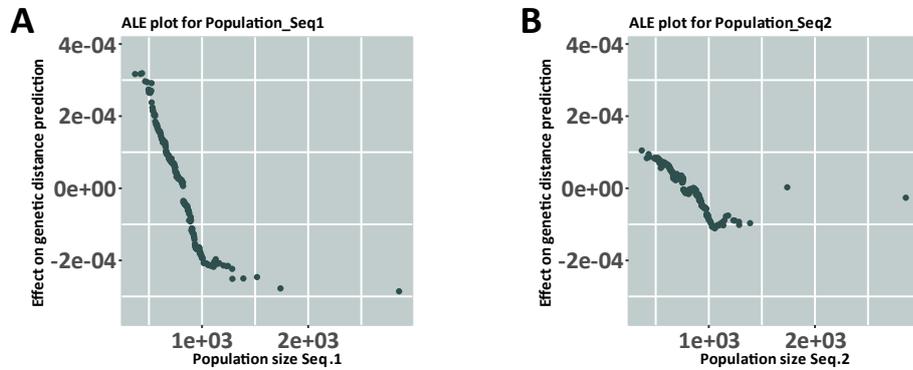
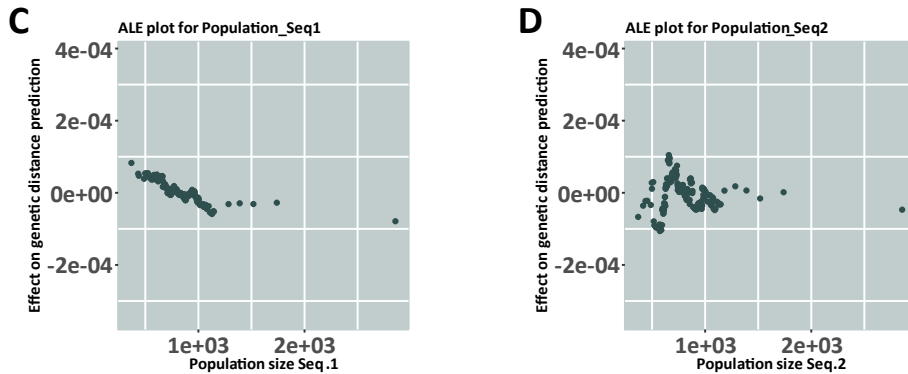

Supplementary Figure 2.3. Accumulated Local Effects plots for the population size in data zones of residence of sampled individuals generated from Random Forest fit to sequenced cases from Tayside, from period 04[th] August 2020 to 30[th] July 2021

Distance dependence of transmission

Geographical distance is the shortest distance between the population weighted centres of two data zones in the paired sequences.

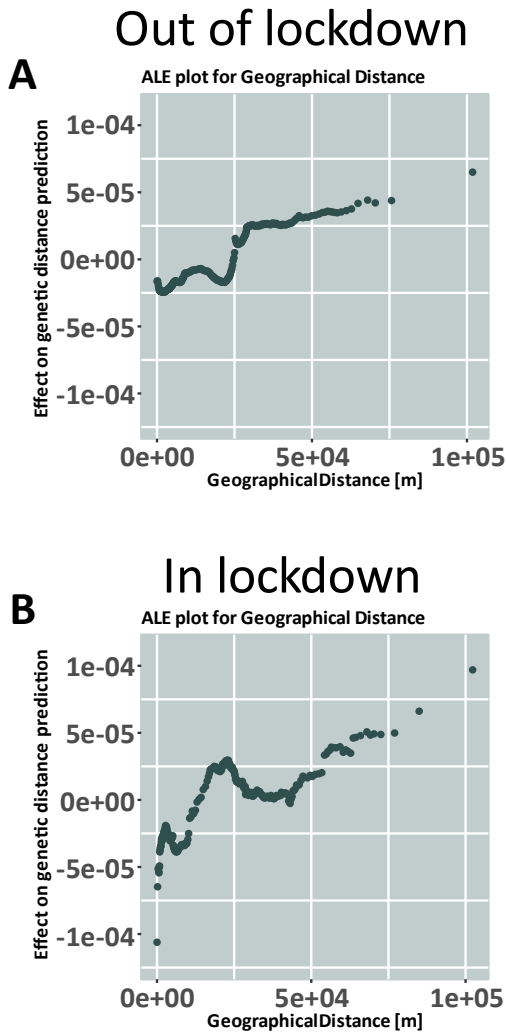

Supplementary Figure 2.4. Accumulated Local Effects plots for the geographical distance of the data zones of residence generated from Random Forest fit to sequenced cases from Tayside, from period 04$^{th}$ August 2020 to 30$^{th}$ July 2021

Commuting connections

The commuting connections variable describes all the connections between two data zones made by commuting workers: workers exchanged between two data zones but also workers from these zones going to work to the same (third) data zone and workers from this third data zone going to work in either of the two paired data zones.

## In lockdown

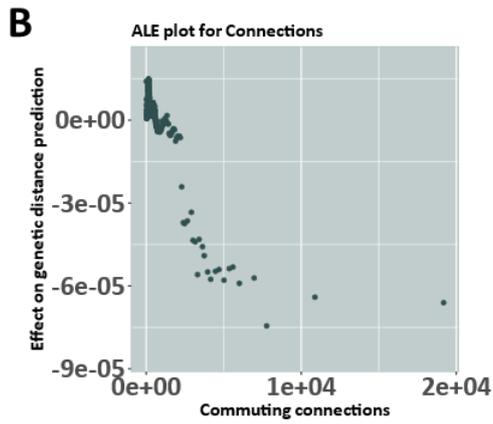

## Out of lockdown

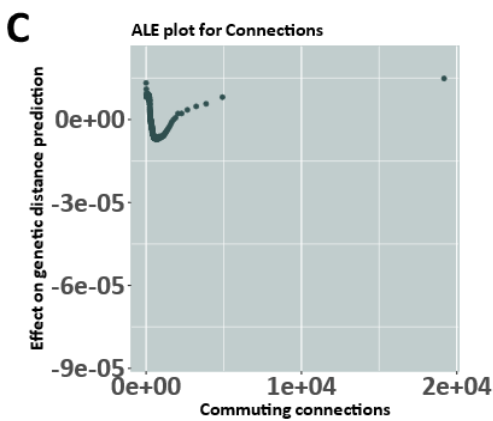

Supplementary Figure 2.5. Accumulated Local Effects plots for commuting connections in data zones of residence of sampled individuals generated from Random Forest fit to sequenced cases from Tayside, from period 04[th] August 2020 to 30[th] July 2021

Temporal distance

Temporal distance is the number of days that passed before sequence 1 and sequence 2; if sequence 2 was registered earlier temporal distance is negative. As we implemented a temporal cut off, pairs were formed only within sample windows of 30 days with rolling intervals. Thus the temporal distance is always between -30 and 30 days.

### A Out of lockdown

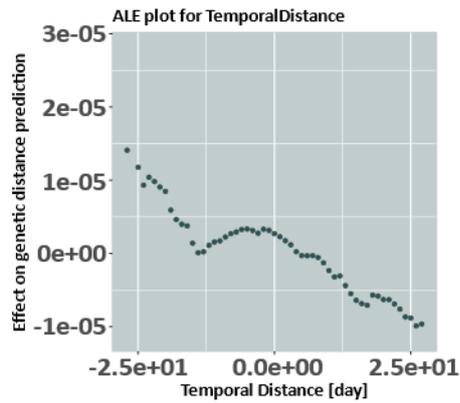

### B In lockdown

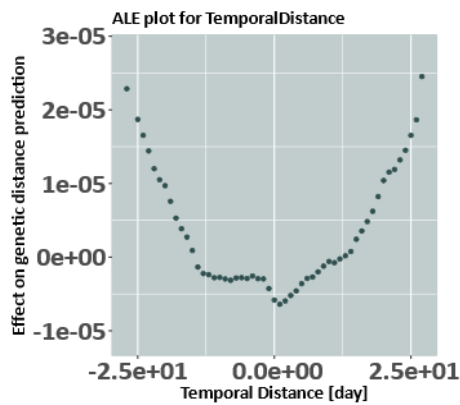

Supplementary Figure 2.6. Accumulated Local Effects plots for time distance in data zones of residence of sampled individuals generated from Random Forest fit to sequenced cases from Tayside, for in lockdown period () and out of lockdown period 04[th] August 2020 to 30[th] July 2021

Age class

Age class is the 20 years age interval assigned to the cases. The RF feature importance analysis (Fig. 1) shows that it was the least important variable in order of importance we included in the final model, and it was the only categorical variable. As a categorical variable, we did not implement ALE plots for it.

*Importance and ALE plots for the whole analysis period (Tayside, 04th August 2020 to 30th July 2021)*

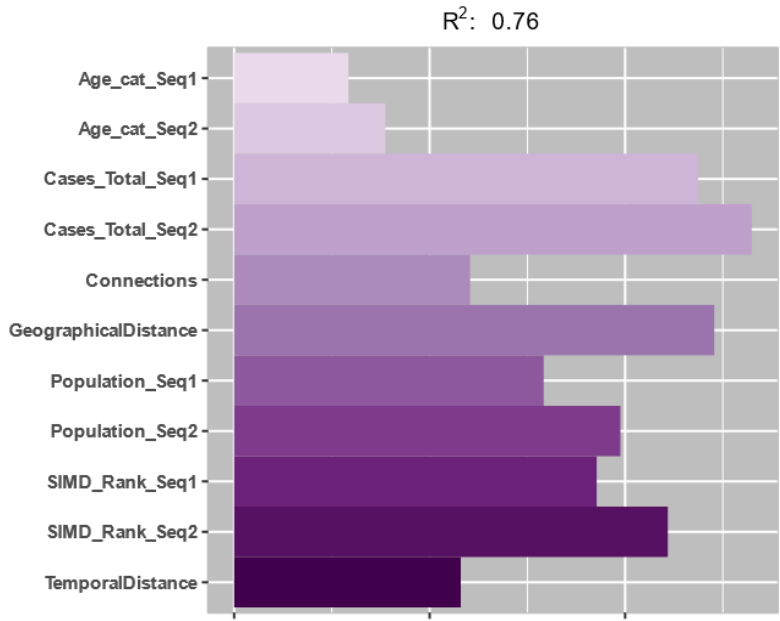
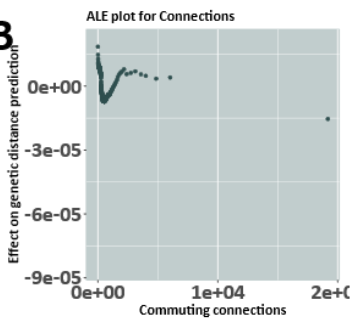
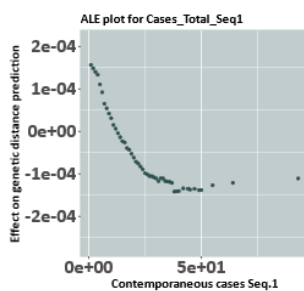
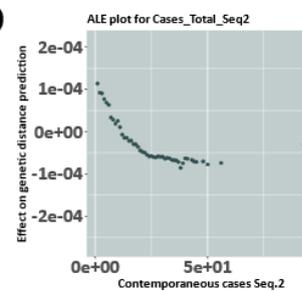
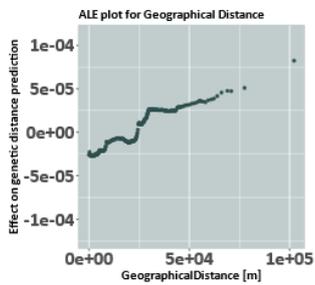
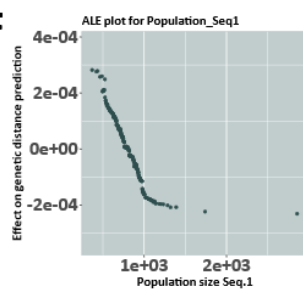
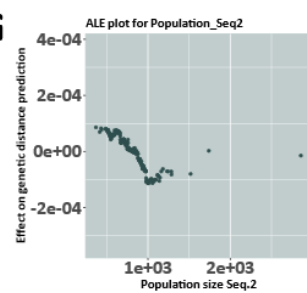
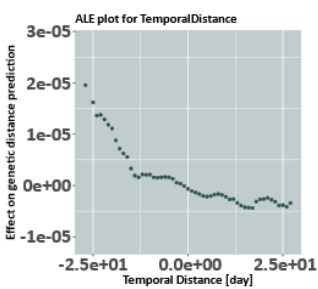
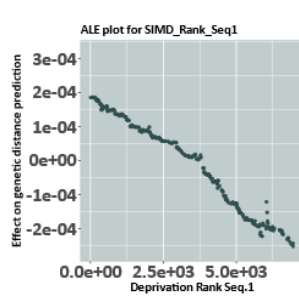
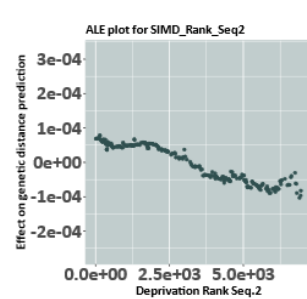

Supplementary Figure 2.7. Feature important (A) and ALE plots (B-J) of analysed variables for a Random Forest model describing the genetic distance between pairs of SARS-COV-2 sequences (with sequence 1 being genetically more similar to the wild type than sequence 2) for the Tayside region in the period from 04th August 2020 to 30th July 2021; B) ALE plot for the Commuting Connections variable; C) ALE plot for the Contemporaneous Cases Sequence 1 variable; D) ALE plot for the Contemporaneous Cases Sequence 2 variable; E) ALE plot for the Geographical Distance variable; F) ALE plot for the Population Size Sequence 1 variable; G) ALE plot for the Population Size Sequence 2 variable; H) ALE plot for the Temporal Distance variable; I) ALE plot for the Deprivation (SIMD Rank) Sequence 1 variable; J) ALE plot for the Deprivation (SIMD Rank) Sequence 2 variable;

*Model version selection*

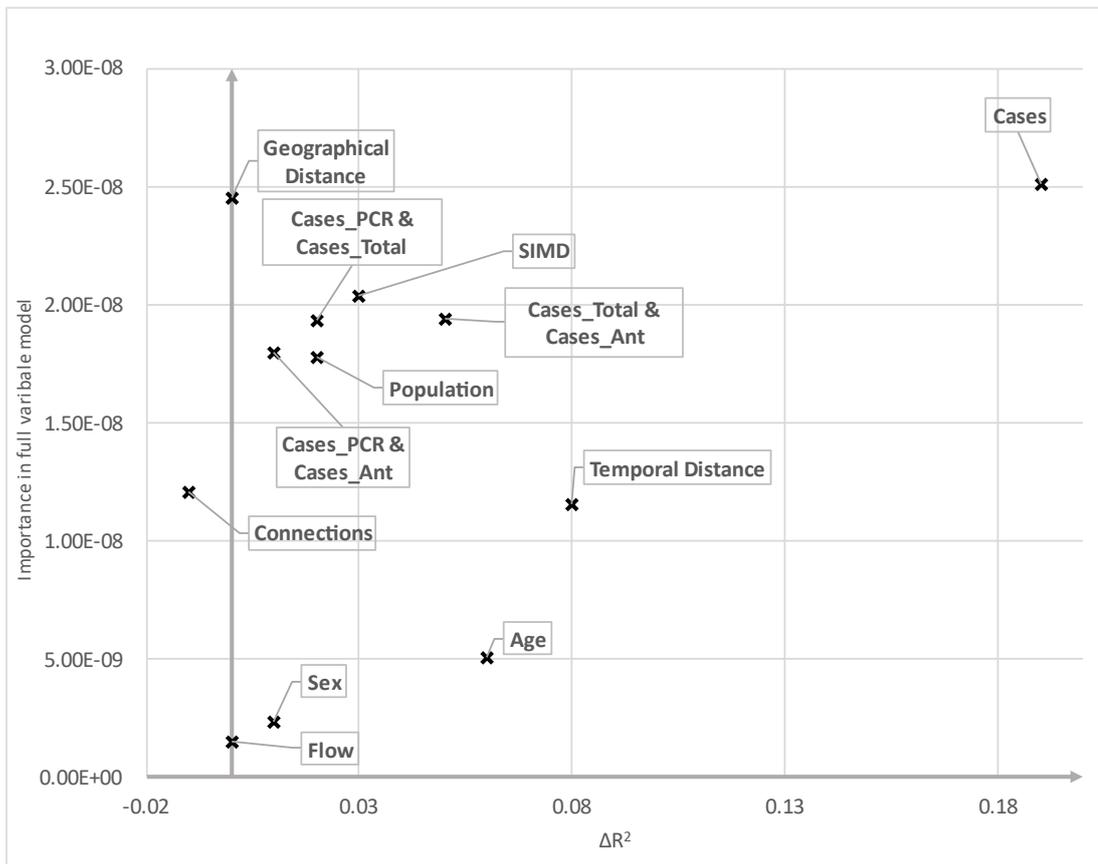

Supplementary Figure 2.8. Change in $R^2$ between all variables model (RF1Y1) and models without plotted variables plotted with the importance of every variable for the "all variables" model.

*Summary variables*

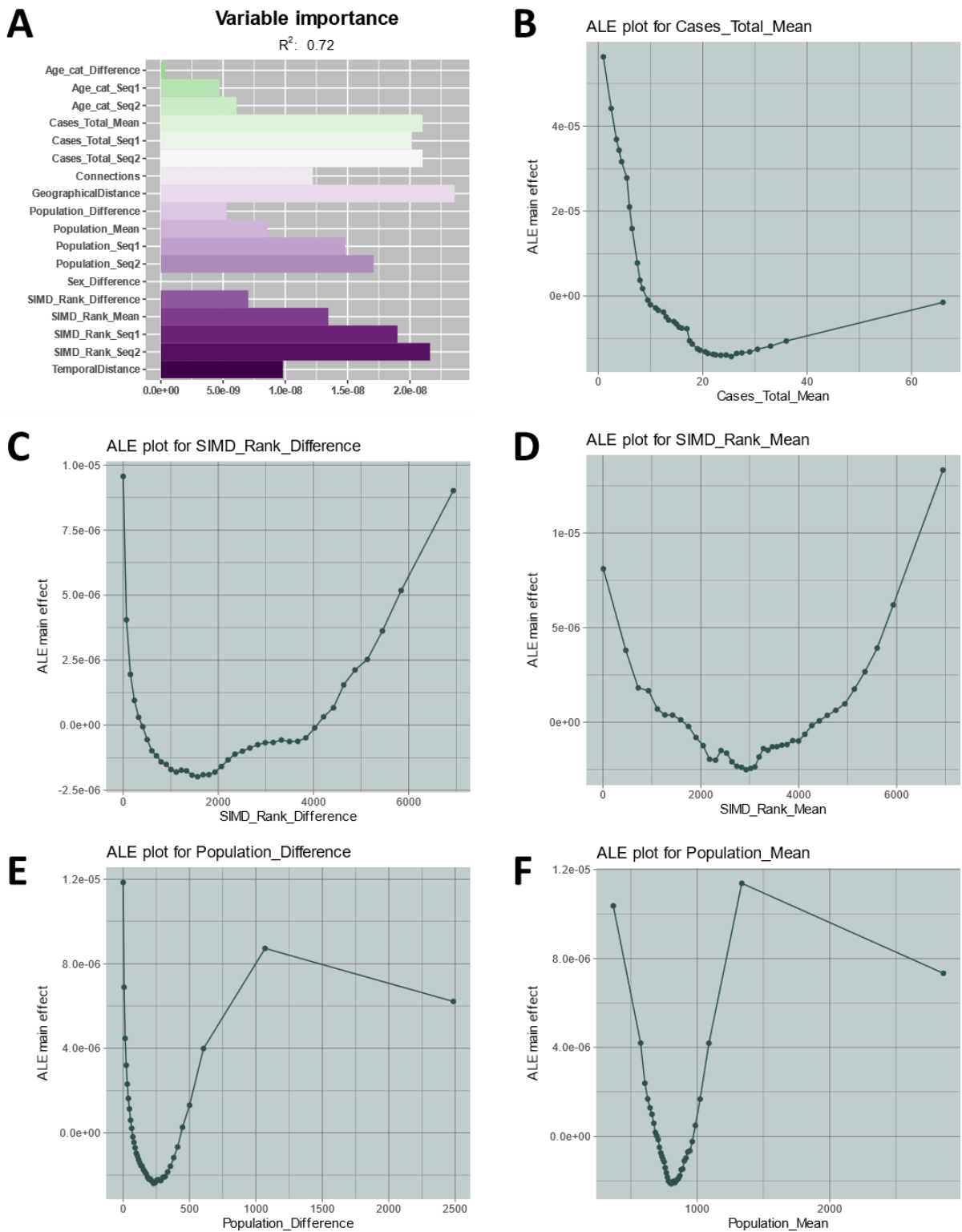

Supplementary Figure 2.9. Feature importance (A) and ALE plots of selected analysed variables for a Random Forest model with summary variables (means and differences of selected variables) describing the genetic distance between pairs of SARS-COV-2 sequences (with sequence 1 being genetically more similar to the wild type than sequence 2) for the Tayside region in the period from 04.08.2020 to 30.07.2021; B) ALE plot for the Total Cases Mean variable; C) ALE plot for the Deprivation (SIMD Rank)- Difference variable; D) ALE plot for the Deprivation (SIMD Rank)- Mean variable; E) ALE plot for the Population Size- Difference variable; F) ALE plot for the Population Size- Mean variable.

*Separated antigen and PCR positive cases*

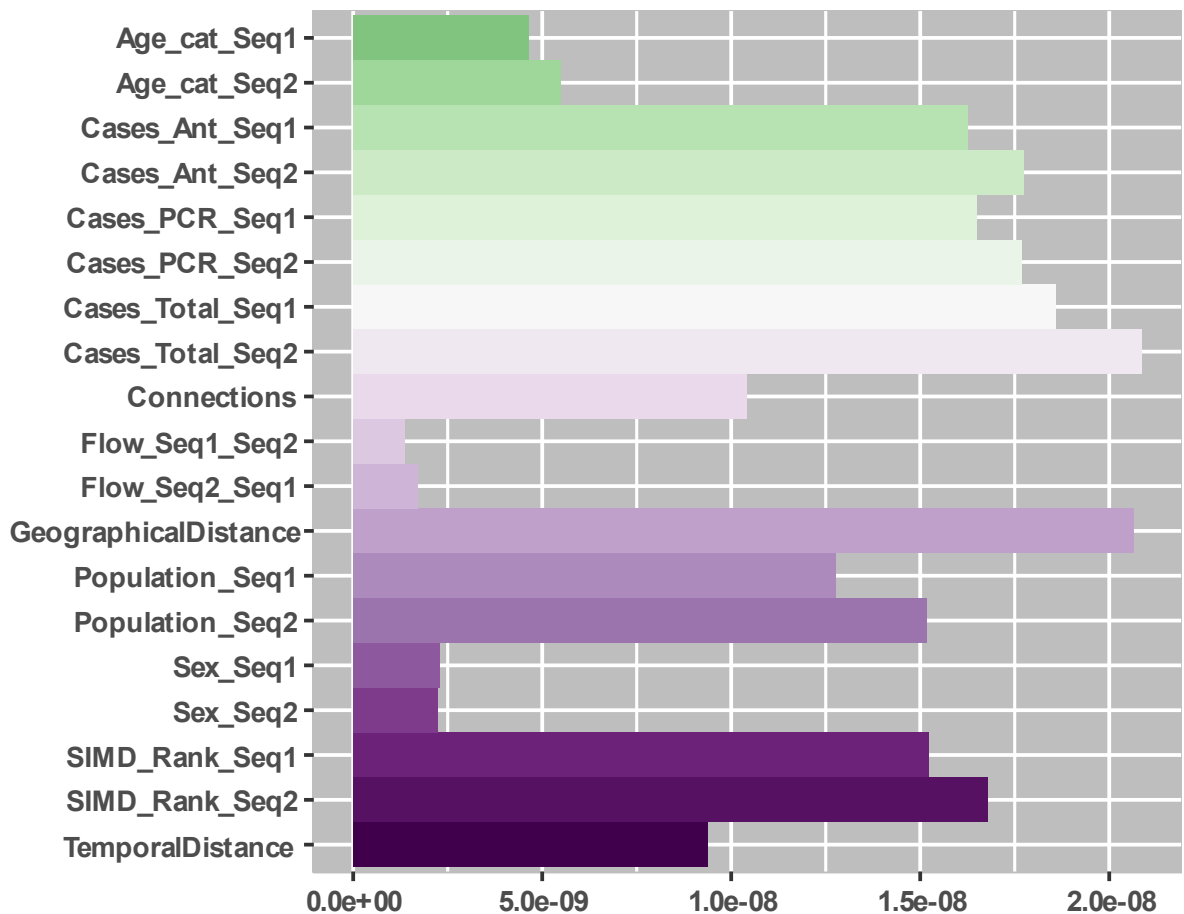

Supplementary Figure 2.10. Feature importance (A) of analysed variables for a Random Forest model with additional number of contemporaneous cases variables separated for PCR and antigen (LFD) tests describing the genetic distance between pairs of SARS-COV-2 sequences (with sequence 1 being genetically more similar to the wild type than sequence 2) for the Tayside region in the period from 04[th] August 2020 to 30[th] July 2021;

*Health Containment index*

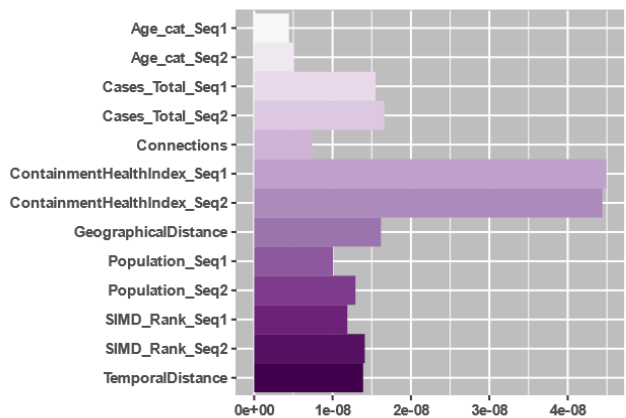

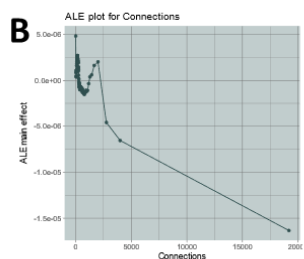
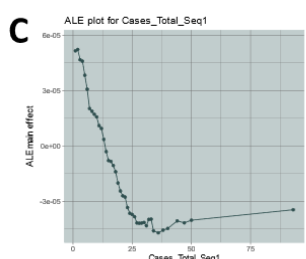
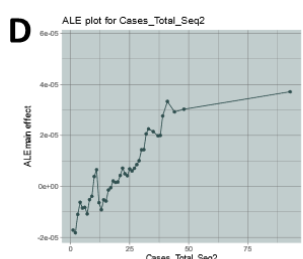

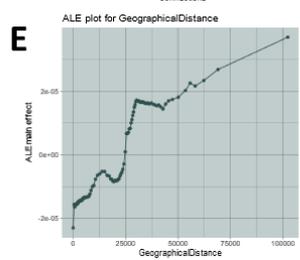
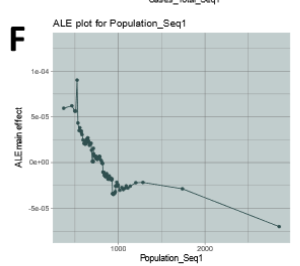
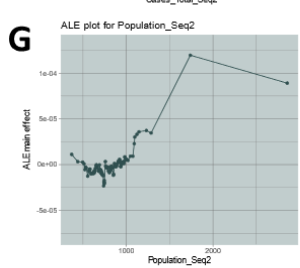

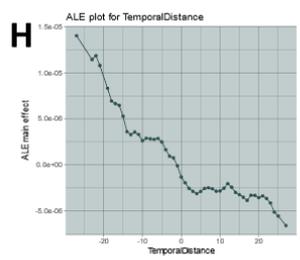
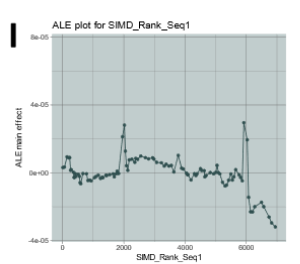
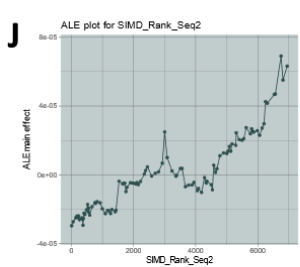

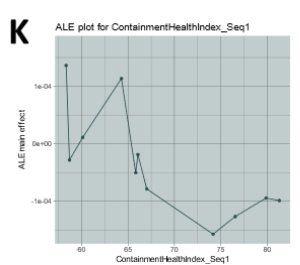
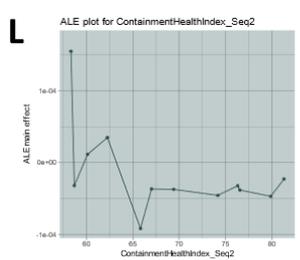

Supplementary Figure 2.11. Feature important (A) and ALE plots (B-J) of analysed variables for a Random Forest model with added Containment Health Index variable describing the genetic distance between pairs of SARS-COV-2 sequences (with sequence 1 being genetically more similar to the wild type than sequence 2) for the Tayside region in the period from 04th August 2020 to 30th July 2021; B) ALE plot for the Commuting Connections variable; C) ALE plot for the Contemporaneous Cases Sequence 1 variable; D) ALE plot for the Contemporaneous Cases Sequence 2 variable; E) ALE plot for the Geographical Distance variable; F) ALE plot for the Population Size Sequence 1 variable; G) ALE plot for the Population Size Sequence 2 variable; H) ALE plot for the Temporal Distance variable; I) ALE plot for the Deprivation (SIMD Rank) Sequence 1 variable; J) ALE plot for the Deprivation (SIMD Rank) Sequence 2 variable; K) ALE plot for the Containment Health Index Sequence 1 variable; L) ALE plot for the Containment Health Index Sequence 2 variable; model RFR2Y1 in Supplementary Dataset 2.

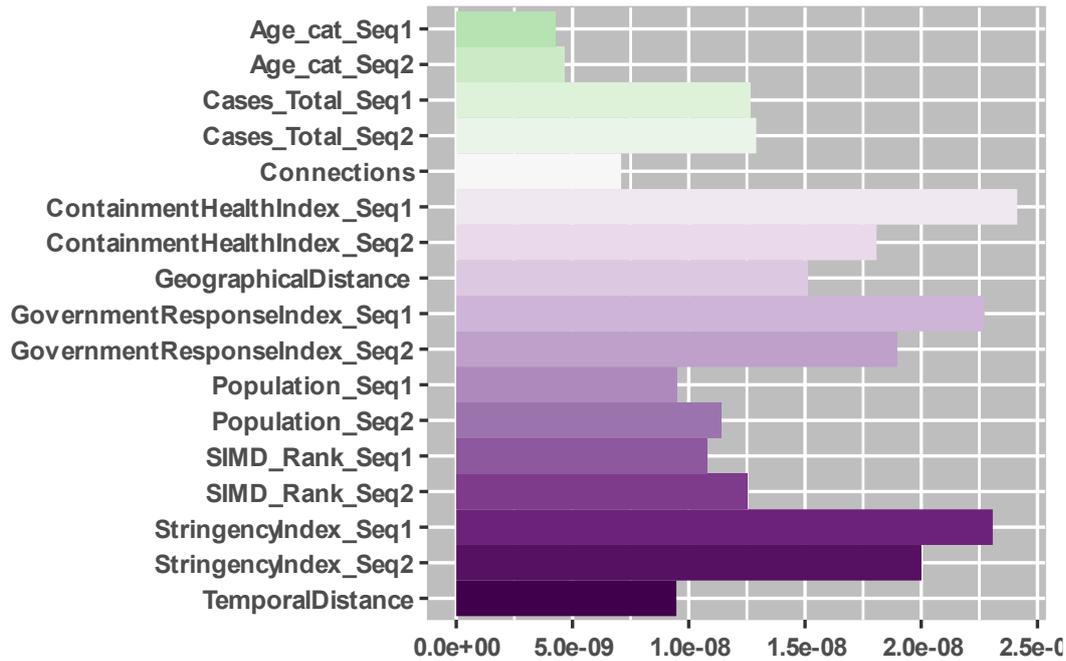

Supplementary Figure 2.12. Feature important (A) of analysed variables for a Random Forest model with added Containment Health Index variable describing the genetic distance between pairs of SARS-COV-2 sequences (with sequence 1 being genetically more similar to the wild type than sequence 2) for the Tayside region in the period from 04[th] August 2020 to 30[th] July 2021; B) ALE plot for the Commuting Connections variable; C) ALE plot for the Contemporaneous Cases Sequence 1 variable; D) ALE plot for the Contemporaneous Cases Sequence 2 variable; E) ALE plot for the Geographical Distance variable; F) ALE plot for the Population Size Sequence 1 variable; G) ALE plot for the Population Size Sequence 2 variable; H) ALE plot for the Temporal Distance variable; I) ALE plot for the Deprivation (SIMD Rank) Sequence 1 variable; J) ALE plot for the Deprivation (SIMD Rank) Sequence 2 variable; K) ALE plot for the Containment Health Index Sequence 1 variable; L) ALE plot for the Containment Health Index Sequence 2 variable; model RFR1Y1 in Supplementary Dataset 2.

*Model with additional random variable*

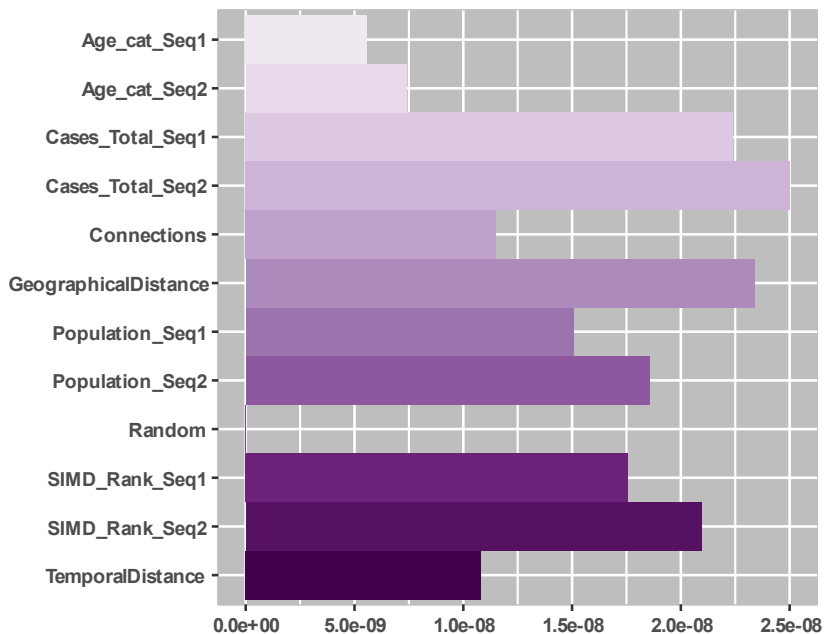

Supplementary Figure 2.13. Feature importance of analysed variables for a Random Forest model with additional random variable (numbers between 1 and 1000 drawn from uniform distribution) describing the genetic distance between pairs of SARS-COV-2 sequences (with sequence 1 being genetically more similar to the wild type than sequence 2) for the Tayside region in the period from 04[th] August 2020 to 30[th] July 2021; this is model RF13Y1 in Supplementary Dataset 2

*Model trained with 75% of data*

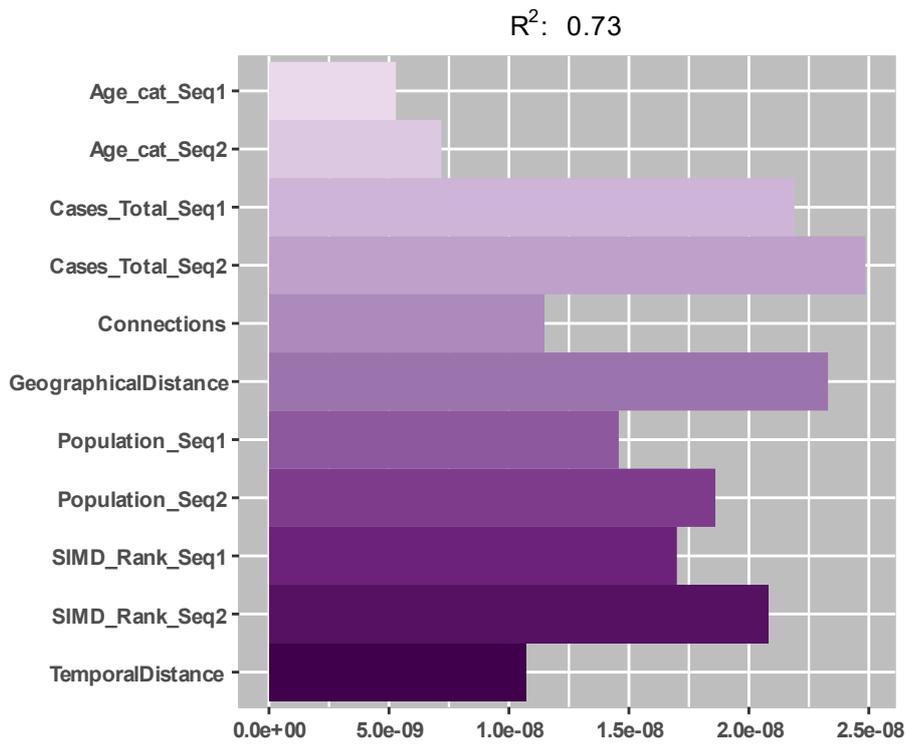

Supplementary Figure 2.14. Feature importance of analysed variables for a Random Forest model trained on 75% of randomly selected data describing the genetic distance between pairs of SARS-COV-2 sequences (with sequence 1 being genetically more similar to the wild type than sequence 2) for the Tayside region in the period from 04[th] August 2020 to 30[th] July 2021; this is model RF6Y1Train in Supplementary Dataset 2

*Models with variable genetic distance cutoff*

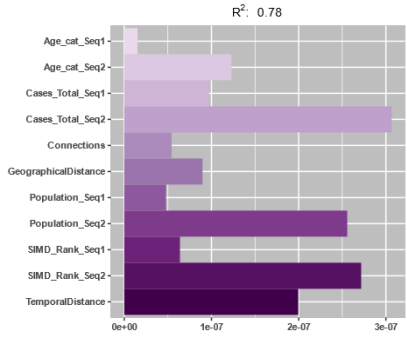
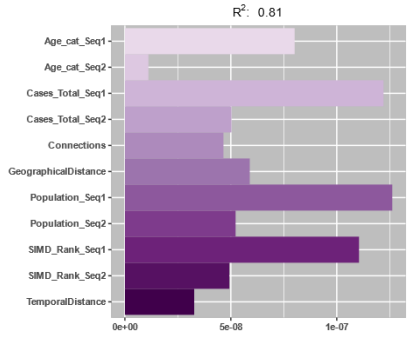
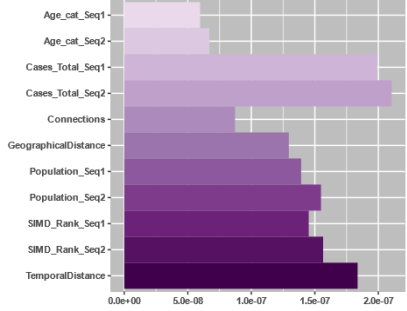
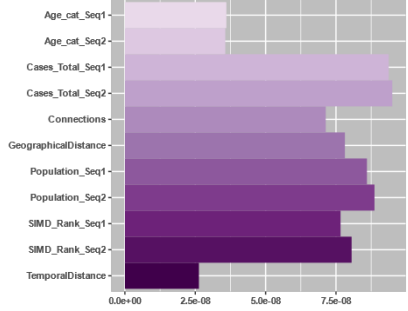

Supplementary Figure 2.15. Feature importance of analysed variables for Random Forest describing the genetic distance between pairs of SARS-COV-2 sequences (with either sequence 1 being genetically more similar to the wild type than sequence 2) or sequences ordered randomly) with no genetic distance cutoff, for Tayside region in period from 04[th] August 2020 to 30[th] July 2021; this is model RFNTY1 in Supplementary Dataset 2

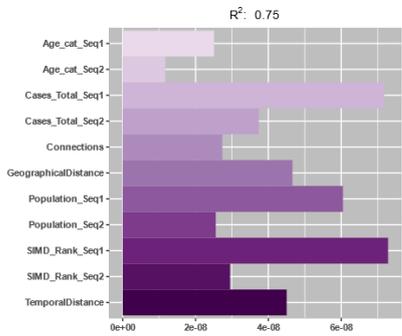
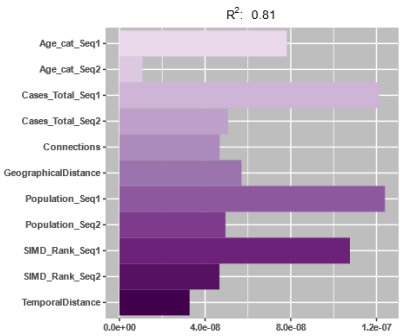
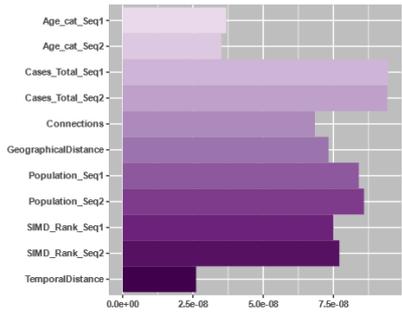
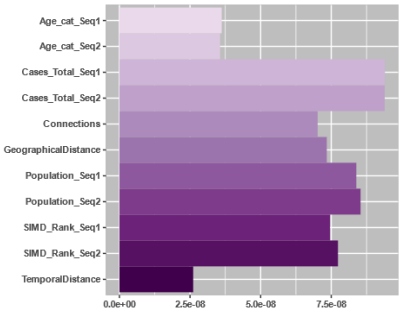

Supplementary Figure 2.16. Feature importance of analysed variables for Random Forest describing the genetic distance between pairs of SARS-COV-2 sequences (with either sequence 1 being genetically more similar to the wild type than sequence 2) or sequences ordered randomly) with genetic distance cutoff=0.0020, for Tayside region in period from 04th August 2020 to 30th July 2021; this is model RFGT1Y1 in Supplementary Dataset 2

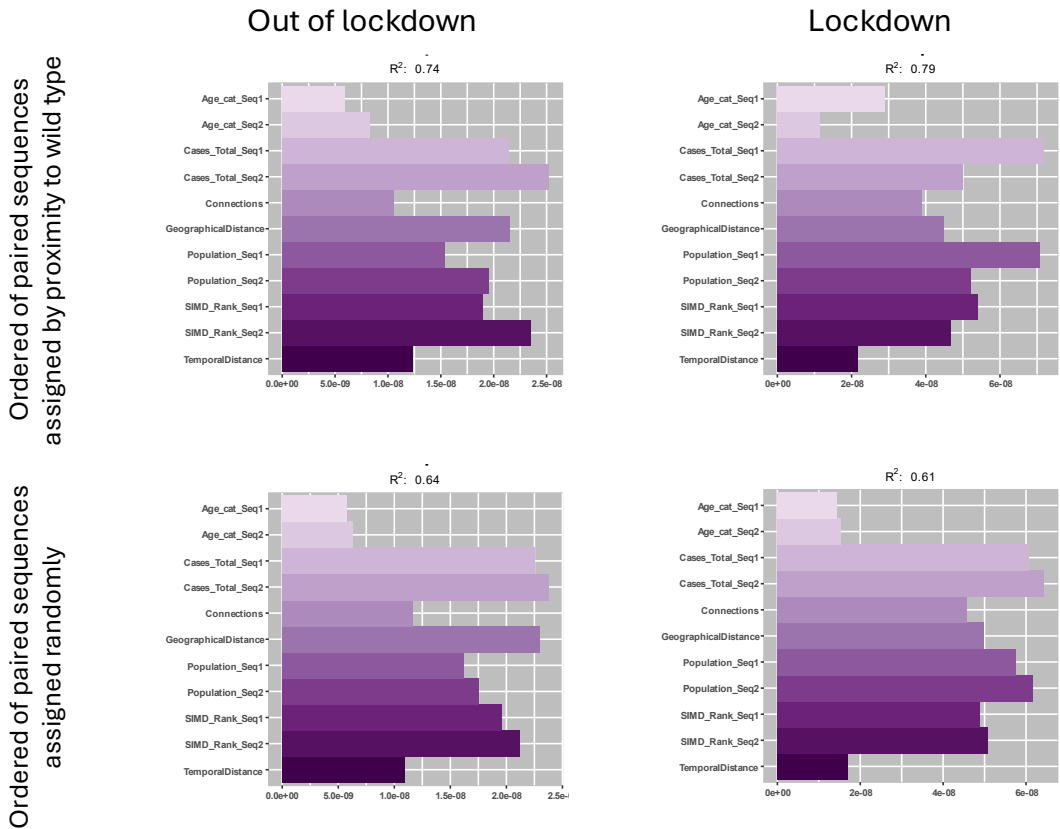

Supplementary Figure 2.17. Feature importance of analysed variables for Random Forest describing the genetic distance between pairs of SARS-COV-2 sequences (with either sequence 1 being genetically more similar to the wild type than sequence 2) or sequences ordered randomly) with genetic distance cutoff=0.0015, for Tayside region in period from 04[th] August 2020 to 30[th] July 2021; this is model RFGT2Y1 in Supplementary Dataset 2

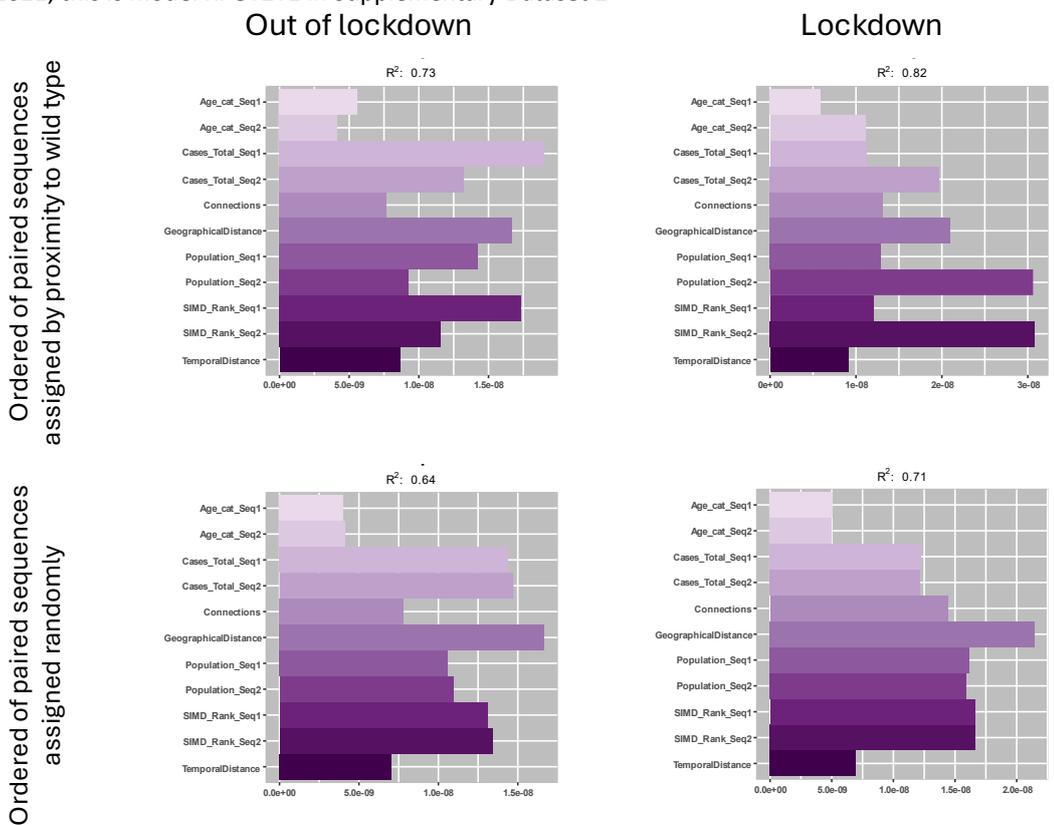

Supplementary Figure 2.18. Feature importance of analysed variables for Random Forest describing the genetic distance between pairs of SARS-COV-2 sequences (with either sequence 1 being genetically more similar to the wild type than sequence 2) or sequences ordered randomly) with genetic distance cutoff=0.0010, for Tayside region in period from 04th August 2020 to 30th July 2021; this is model RFGT3Y1 in Supplementary Dataset 2

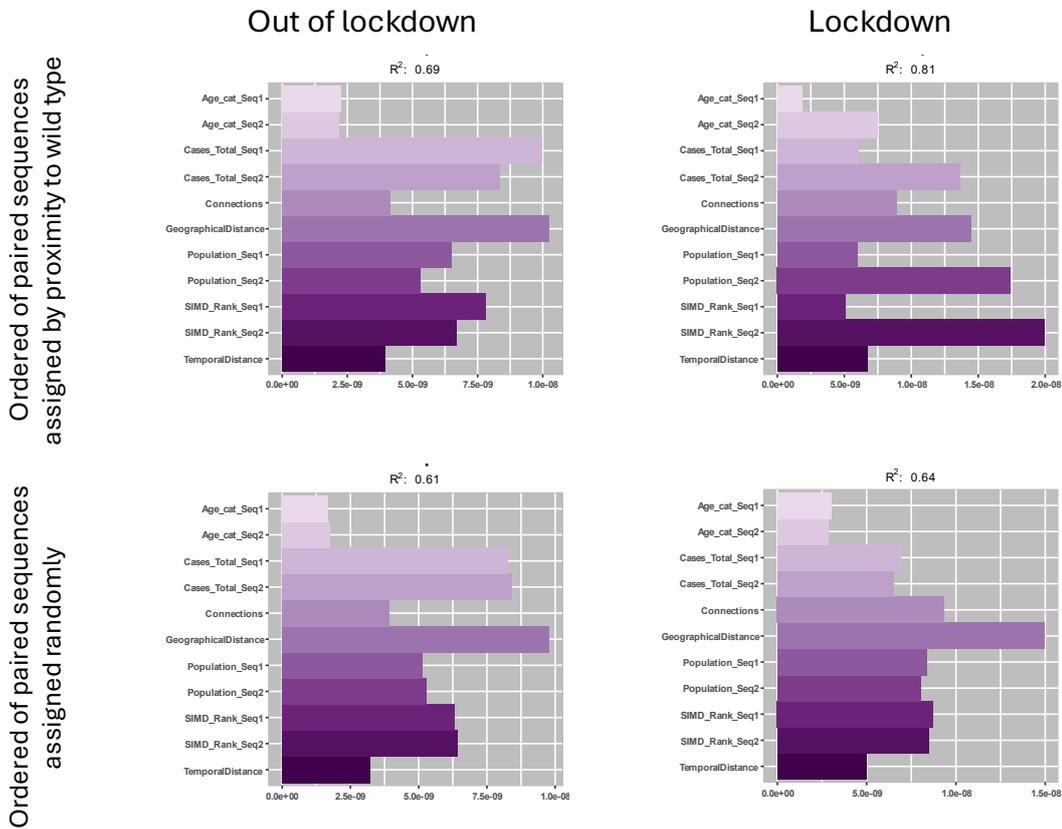

Supplementary Figure 2.19. Feature importance of analysed variables for Random Forest describing the genetic distance between pairs of SARS-COV-2 sequences (with either sequence 1 being genetically more similar to the wild type than sequence 2) or sequences ordered randomly) with genetic distance cutoff=0.0005, for Tayside region in period from 04[th] August 2020 to 30[th] July 2021; this is model RFGT4Y1 in Supplementary Dataset 2

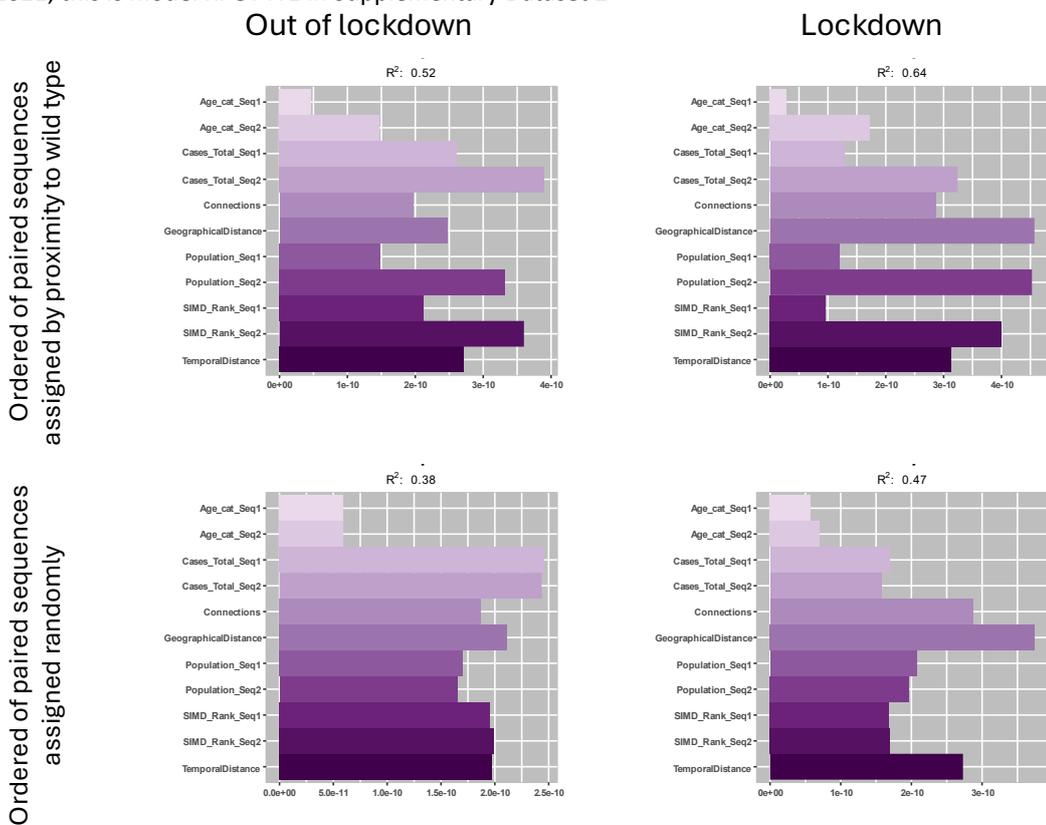

Supplementary Figure 2.20. Feature importance of analysed variables for Random Forest describing the genetic distance between pairs of SARS-COV-2 sequences (with either sequence 1 being genetically more similar to the wild type than sequence 2) or sequences ordered randomly) with genetic distance cutoff=0.0001, for Tayside region in period from 04th August 2020 to 30th July 2021; this is model RFGT5Y1 in Supplementary Dataset 2

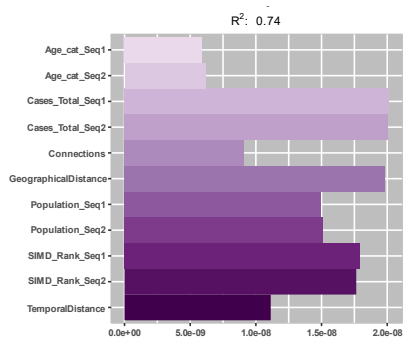 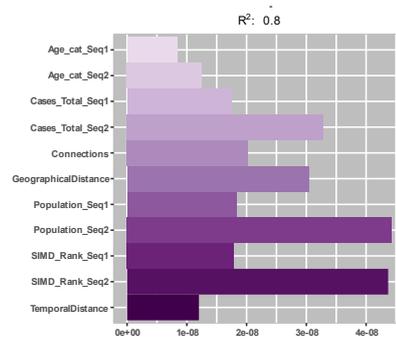
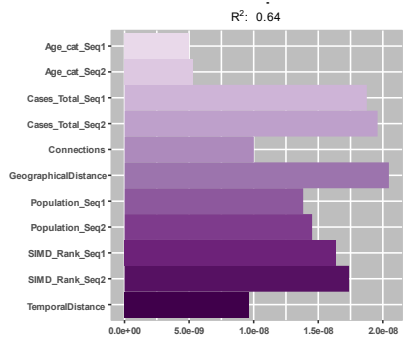 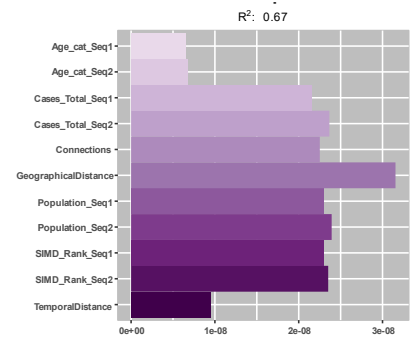

Supplementary Figure 2.21. Feature importance of analysed variables for Random Forest describing the genetic distance between pairs of SARS-COV-2 sequences (with either sequence 1 being genetically more similar to the wild type than sequence 2) or sequences ordered randomly) with genetic distance cutoff=0.0013, for Tayside region in period from 04th August 2020 to 30th July 2021; this is model RFMTY1 in Supplementary Dataset 2

# Supplementary Note 3. Correlation analysis

## 3.1. Total number of cases

Total cases Seq.1

Total cases Seq. 2

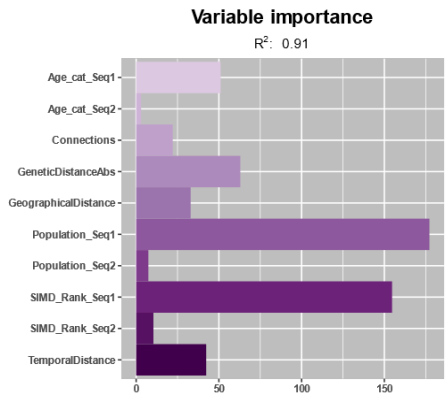
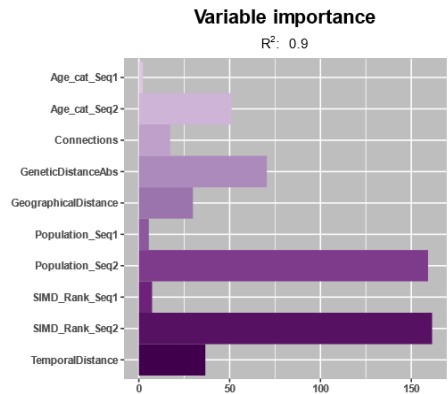
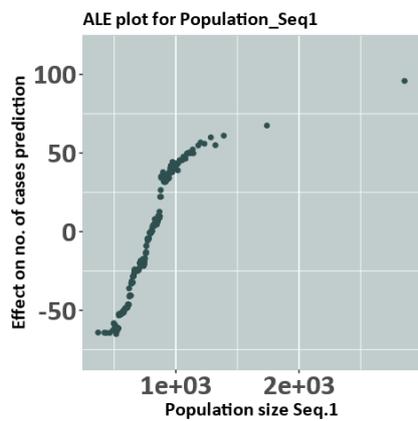
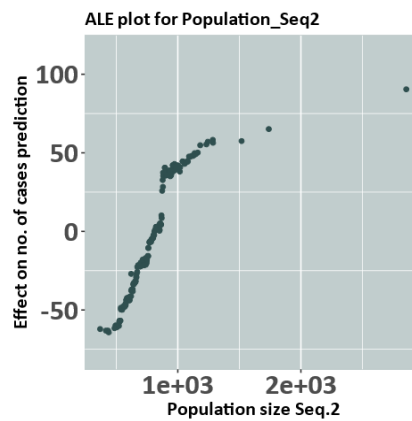
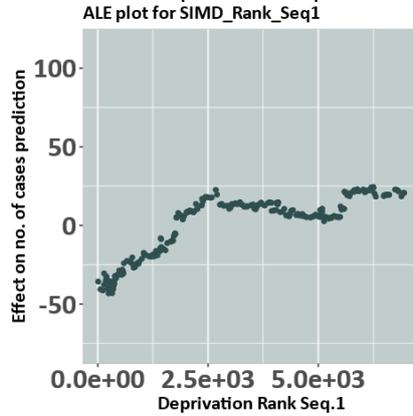
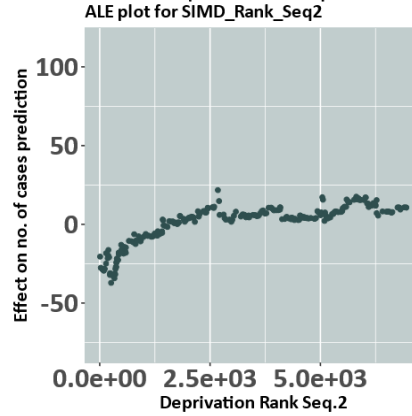

Supplementary Figure 3.1. Feature importance of analysed variables for a Random Forest model describing the correlation between total number of cases and other model variables for the Tayside region in the period from 04th August 2020 to 30th July 2021; with additional ALE plots for the most importance correlates.

*3.2. Geographical distance*

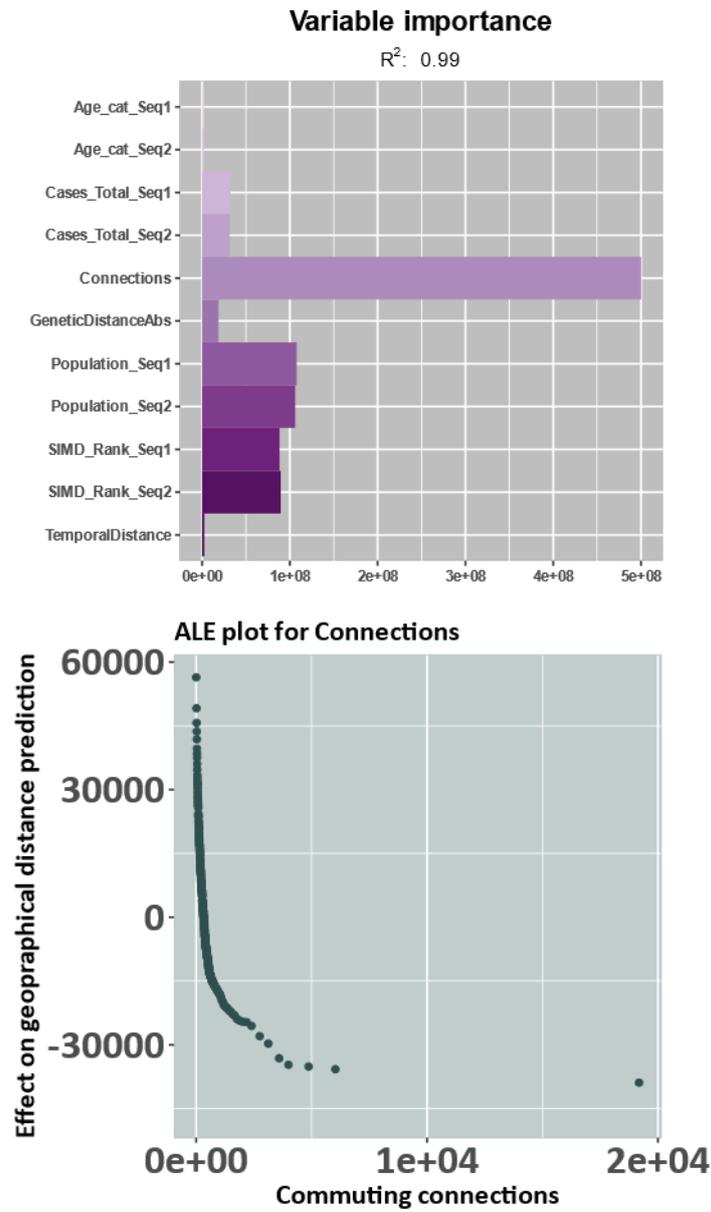

Supplementary Figure 3.2. Feature importance of analysed variables for a Random Forest model describing the correlation between geographical distance and other model variables for the Tayside region in the period from 04th August 2020 to 30th July 2021; with additional ALE plots for the most importance correlates.

## 3.3. Deprivation rank

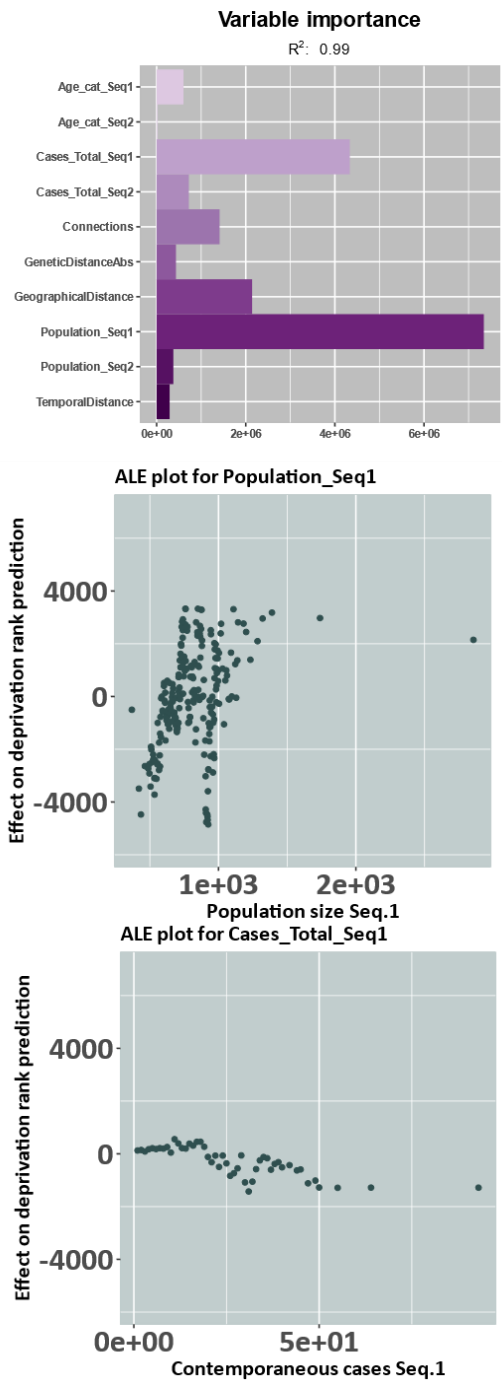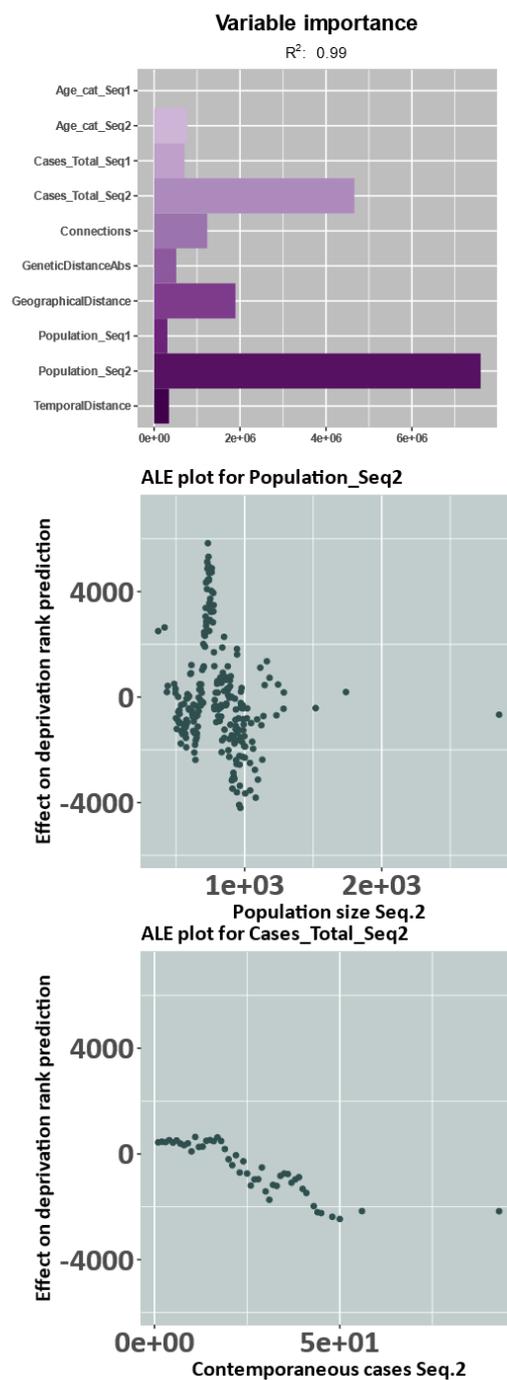

Supplementary Figure 3.3. Feature importance of analysed variables for a Random Forest model describing the correlation between deprivation (SIMD) rank and other model variables for the Tayside region in the period from 04th August 2020 to 30th July 2021; with additional ALE plots for the most importance correlates.

*3.4. Population size*

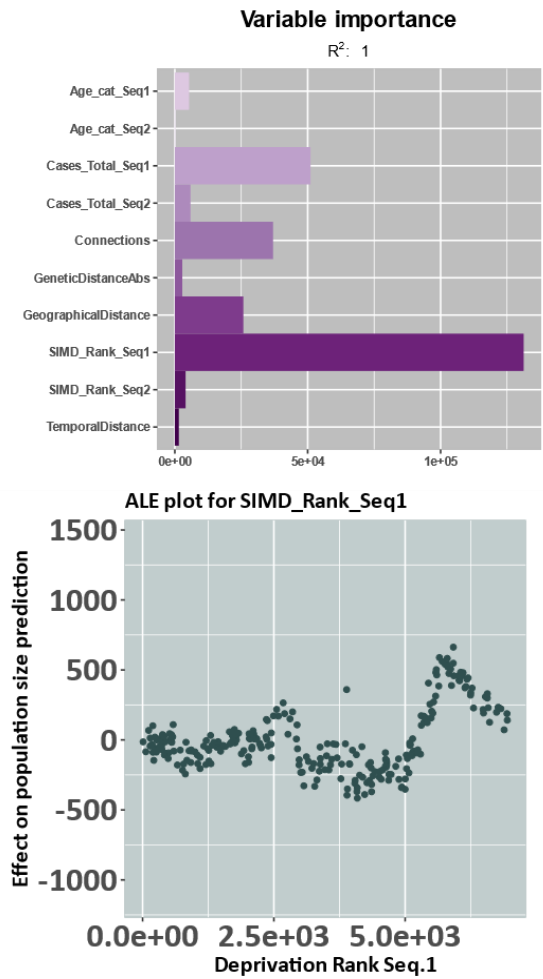
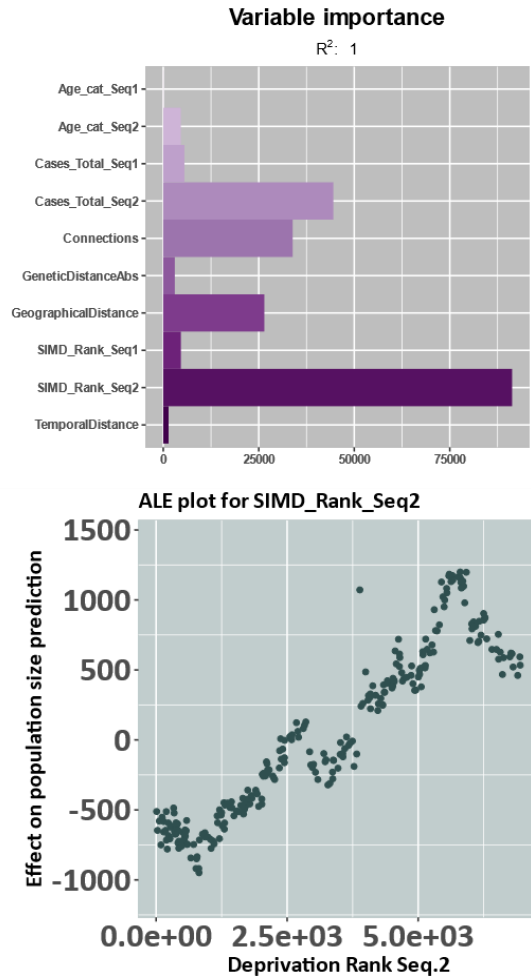

Supplementary Figure 3.4. Feature importance of analysed variables for a Random Forest model describing the correlation between population size (as of 2017) and other model variables for the Tayside region in the period from 04th August 2020 to 30th July 2021; with additional ALE plots for the most importance correlates.

## 3.5. Commuting connections

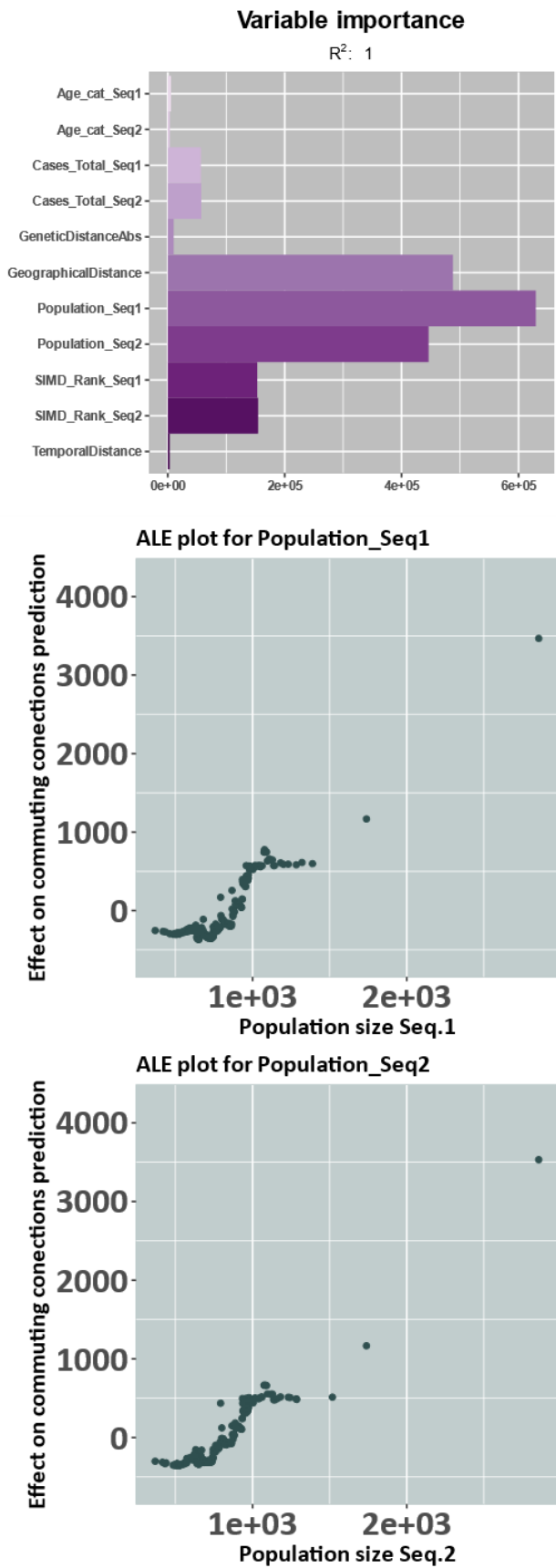

Supplementary Figure 3.5. Feature importance of analysed variables for a Random Forest model describing the correlation between commuting connections and other model variables for the Tayside region in the period from 04th August 2020 to 30th July 2021; with additional ALE plots for the most importance correlates.

*3.6. Temporal distance*

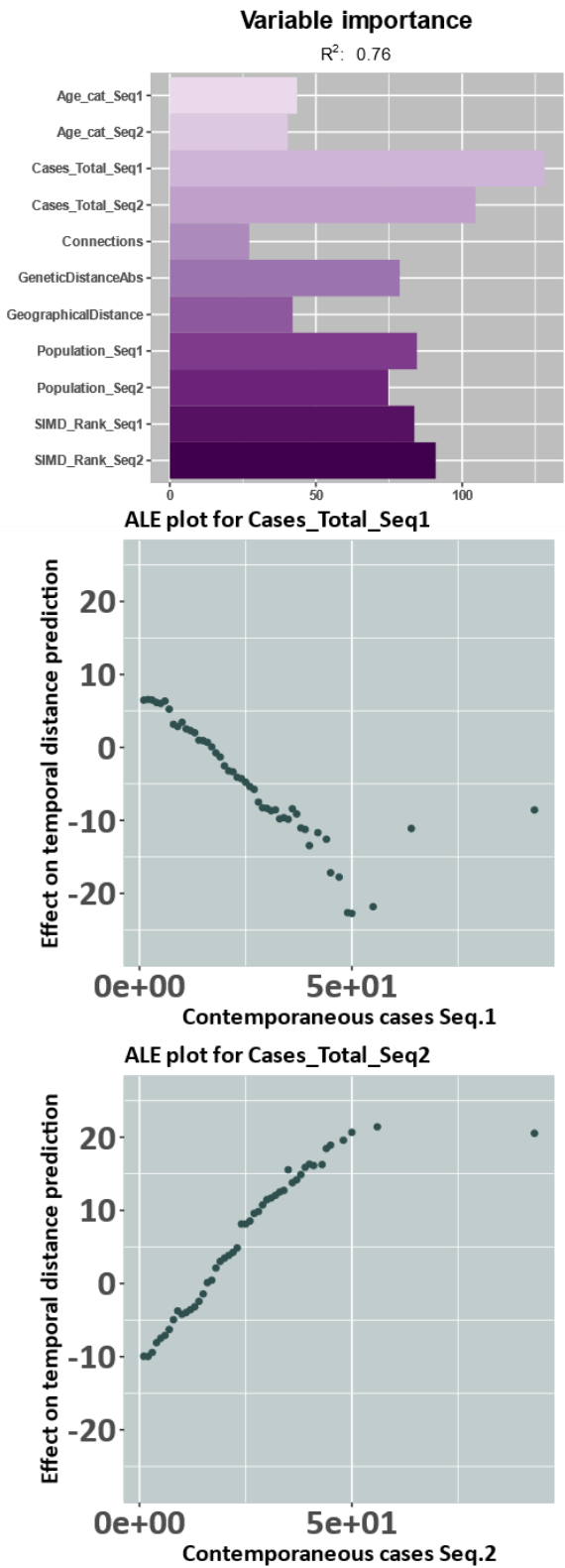

Supplementary Figure 3.6. Feature importance of analysed variables for a Random Forest model describing the correlation between temporal distance and other model variables for the Tayside region in the period from 04th August 2020 to 30th July 2021; with additional ALE plots for the most importance correlates.

## 3.6. Additional figures

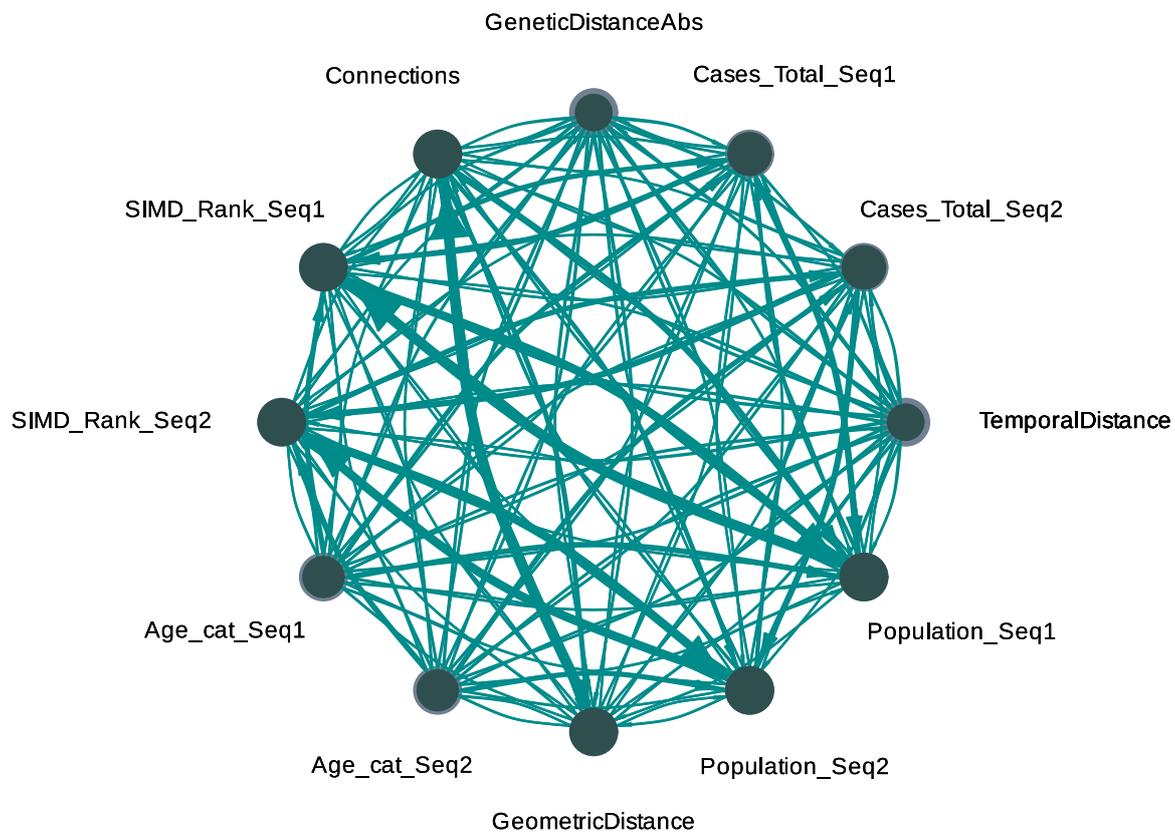

Supplementary Figure 3.6.1. Correlation graph of the selected variables

## Supplementary Note 4. Tayside maps

For illustration purposes we generated an interactive html widget with Tayside map marked with marked data zone boundaries with layers describing the urban-rural classification of the areas, their deprivation (SIMD) rank and population size as of 2017. The widget is provided with the published code and screenshots of the layers are presented below.

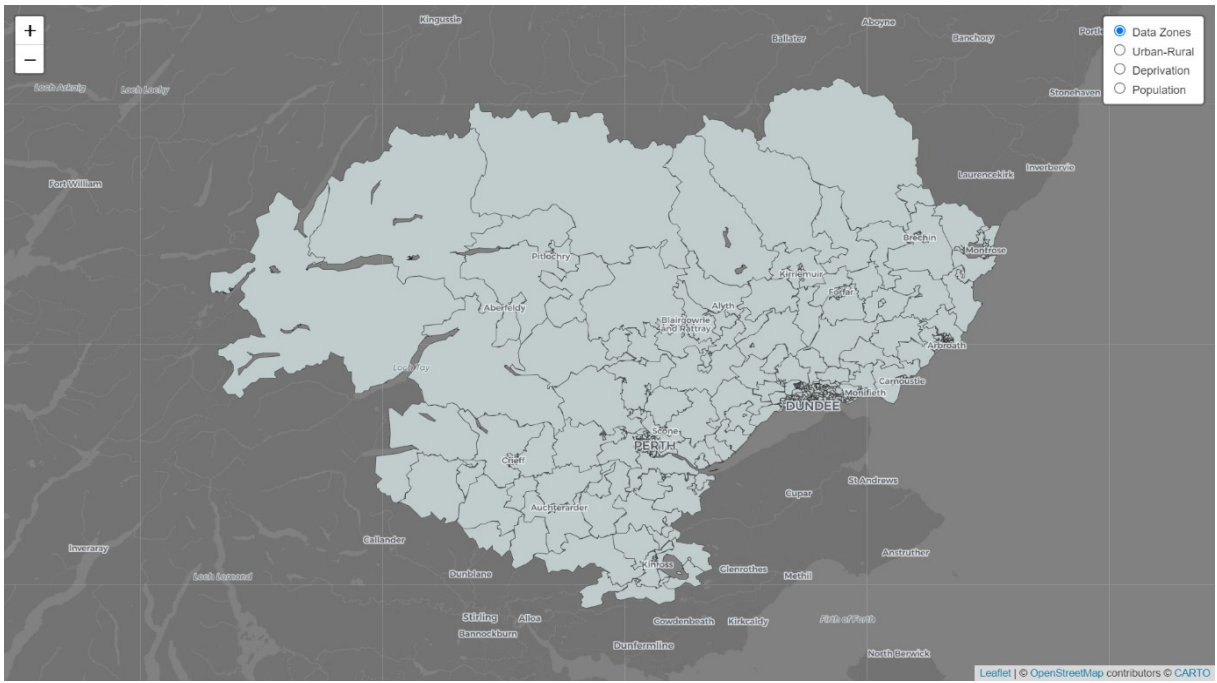

Supplementary Figure 4.2. Tayside map divided into Data Zones

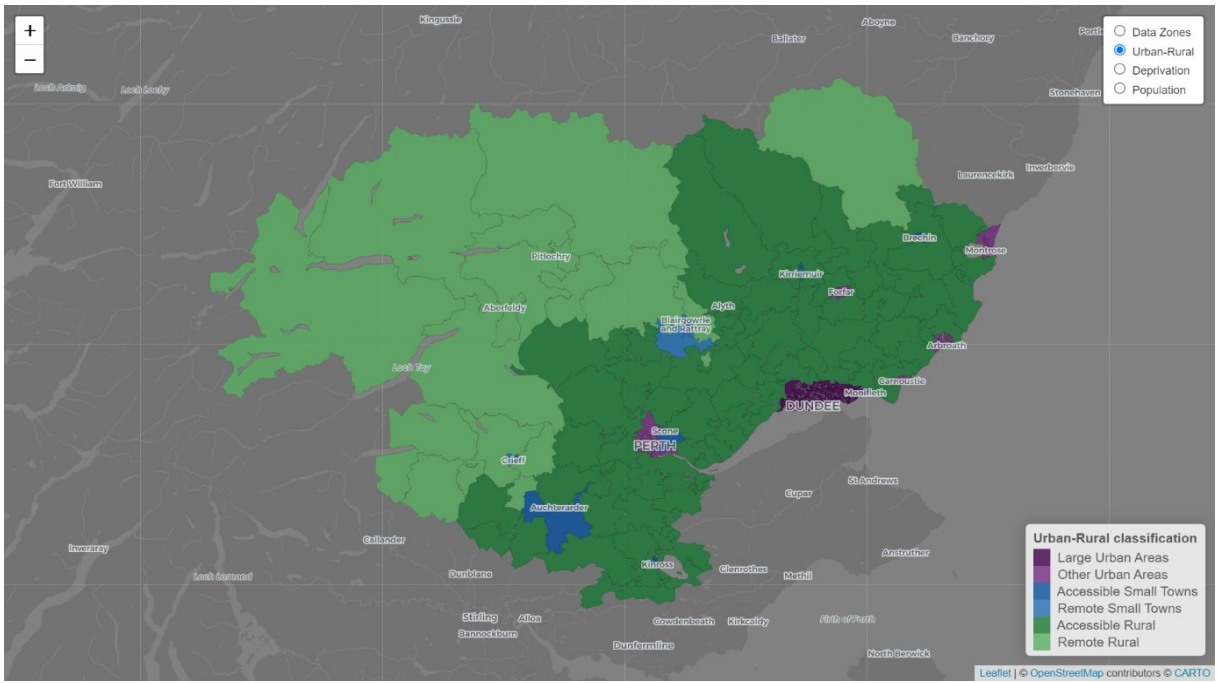

Supplementary Figure 4.2. Tayside map with marked urban and rural areas on Data Zone level

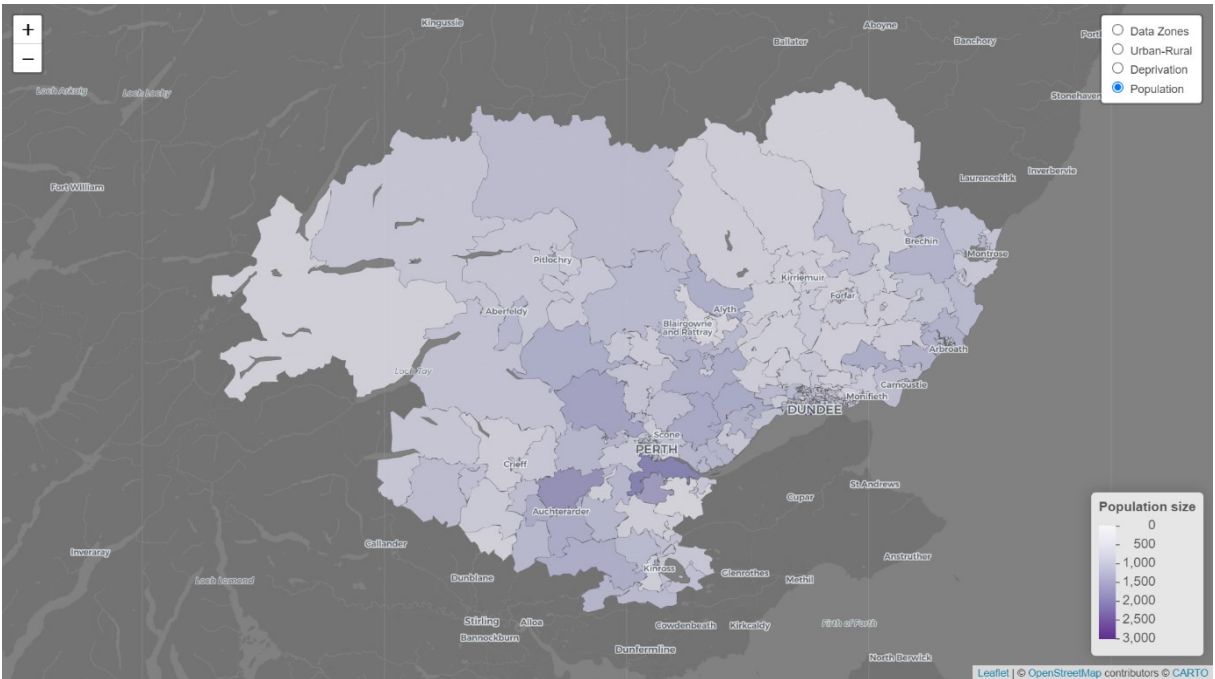
Supplementary Figure 4.3. Tayside map with marked population density (as of 207) at the Data Zone level

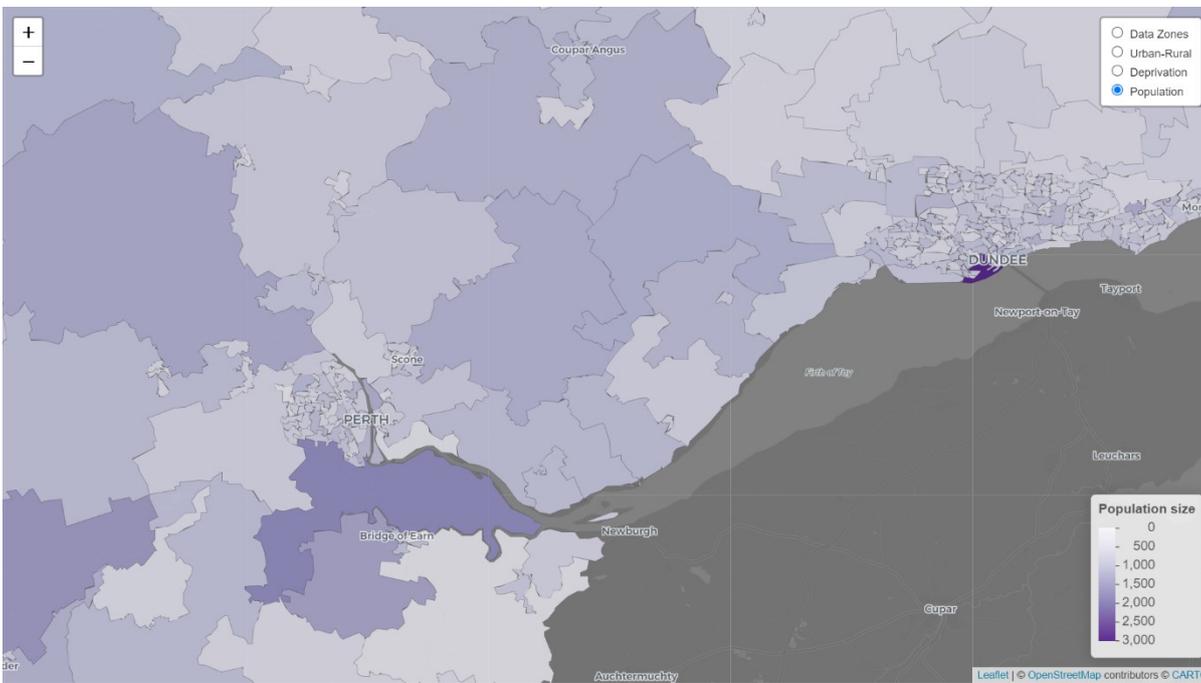
Supplementary Figure 4.4. Zoom into part of Tayside with cities of Perth and Dundee with marked population density (as of 2017) at the Data Zone level

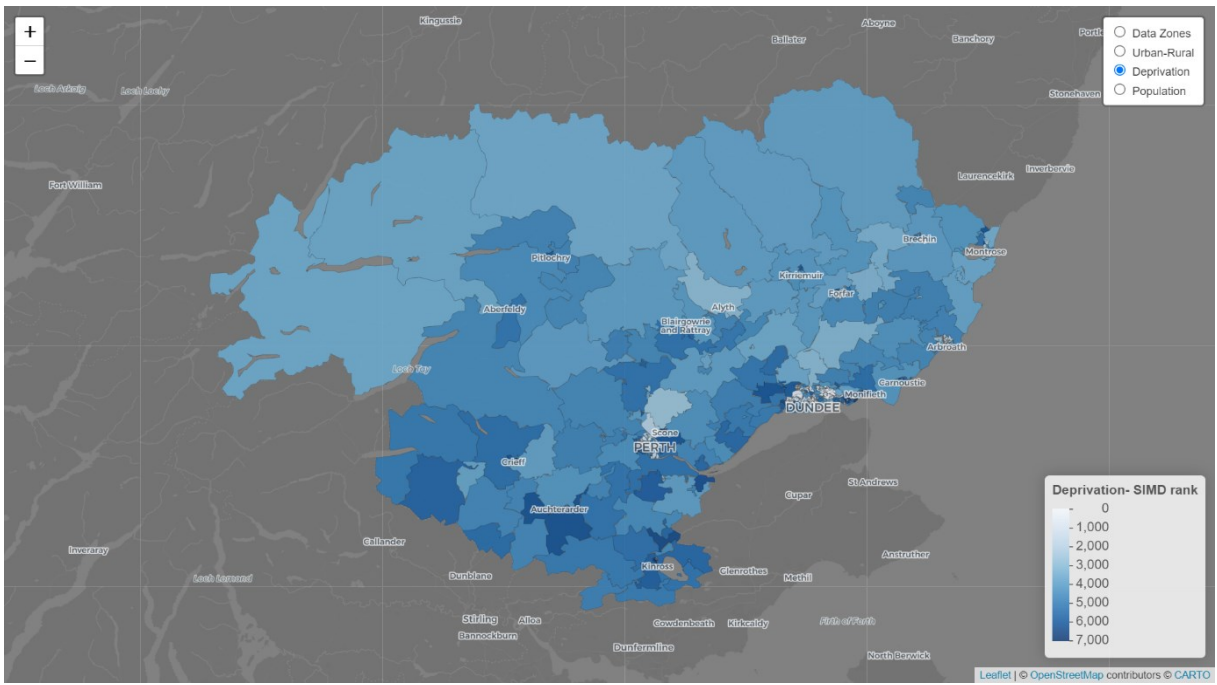

Supplementary Figure 4.5. Tayside map with marked deprivation at the Data Zone level

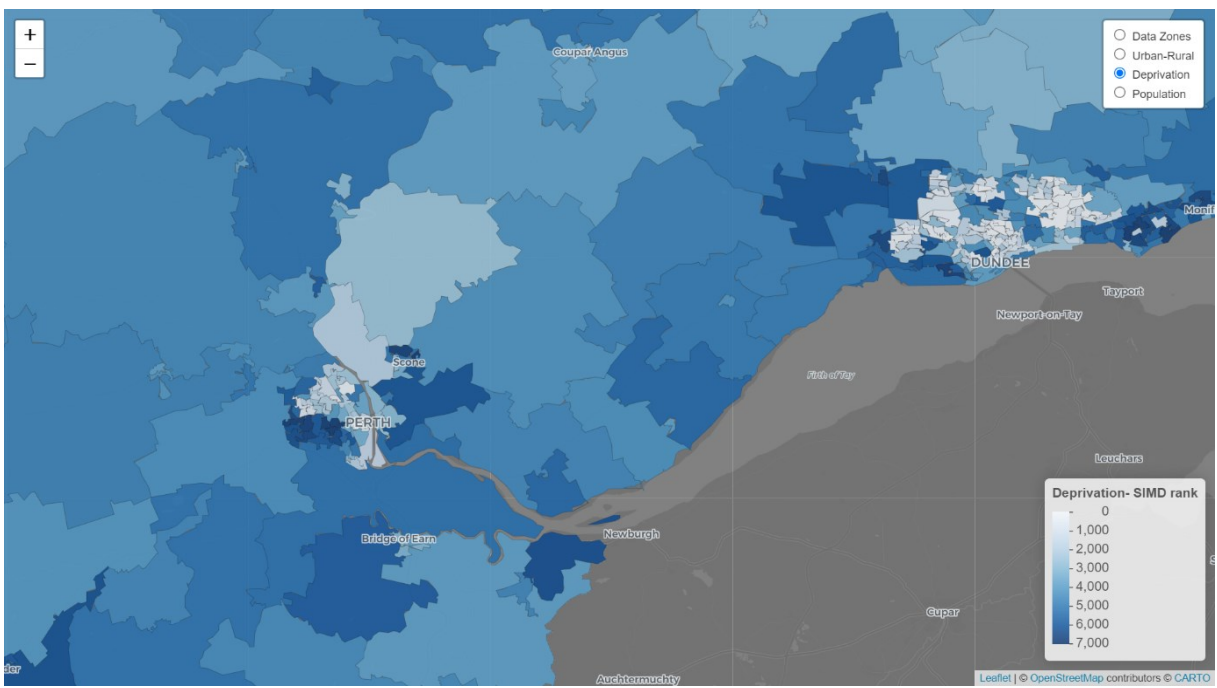

Supplementary Figure 4.6. Zoom into part of Tayside with cities of Perth and Dundee with marked deprivation at the Data Zone level

# Supplementary Note 5. Data processing- supplementary tables and figures

Supplementary Table 5.1. The time difference (in days) between SARS-Cov-2 cases that had mismatch of collection dates between Scottish SARS-CoV-2test data (Dataset I in the Table 3 in the Methods section) and Sequence and case IDs (Dataset II in the Table 3 in the Methods section)

| Time difference [day] | Count |
|---|---|
| 1 | 1647 |
| 2 | 222 |
| 3 | 73 |
| 4 | 50 |
| 5 | 51 |
| 6 | 20 |
| 7 | 23 |
| 8 | 19 |
| 9 | 12 |
| 10 | 13 |
| 11 | 5 |
| 12 | 3 |
| 13 | 15 |
| 14 | 5 |
| 15 | 4 |
| 16 | 3 |
| 17 | 3 |
| 18 | 3 |
| 19 | 7 |
| 20 | 7 |
| 21 | 2 |
| 22 | 1 |
| 25 | 1 |
| 26 | 1 |
| 29 | 1 |
| 30 | 3 |
| 31 | 4 |
| 38 | 1 |
| 39 | 1 |
| 42 | 3 |
| 50 | 1 |
| 51 | 1 |
| 53 | 1 |
| 59 | 1 |
| 60 | 1 |
| 65 | 1 |
| 70 | 1 |
| 71 | 8 |
| 72 | 1 |
| 78 | 1 |
| 84 | 3 |
| 88 | 1 |
| 90 | 1 |
| 93 | 1 |
| 103 | 1 |
| 111 | 1 |
| 176 | 1 |
| 259 | 1 |
| 365 | 1 |
| **SUM:** | **2231** |

| Area of residence | Area of workplace | Number of people | Connections between **Data Zone A** and **Data Zone B**: |
|---|---|---|---|
| **Data Zone A** | **Data Zone B** | 2 | 2 |
| **Data Zone B** | Data Zone C | 5 | 5 |
| **Data Zone B** | **Data Zone A** | 1 | 1 |
| Data Zone C | **Data Zone A** | 7 | 7 |
| Data Zone D | **Data Zone A** | 1 | 0 |
| Data Zone C | Data Zone D | 3 | 0 |
|  |  |  | **Total: 15 connections** |

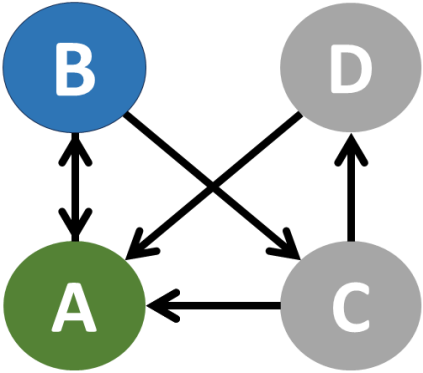

People exchanged between Data Zone C and Data Zones A & B are counted as connections because both A & B have connection to C. Data Zone D has only connection with A (but not B) so people going from/to Data Zone D are not included.

Supplementary Figure 5.1. Schematic representation of "Commuting connections" variable definition

**Supplementary Note 6. Dundee city**

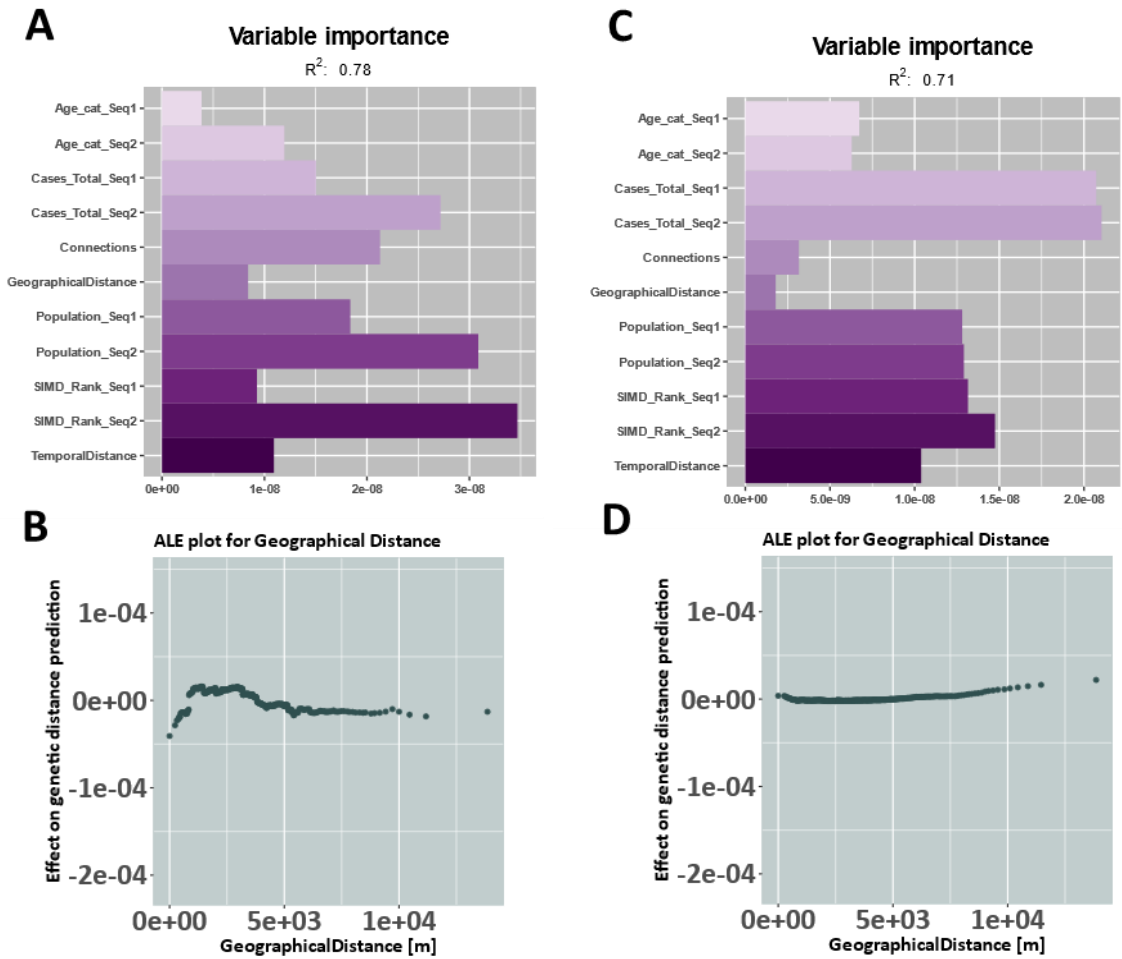

Supplementary Figure 6.1. Feature importance of analysed variables for a Random Forest model describing the correlation between genetic distance and other model variables for Dundee city in the period from 04th August 2020 to 30th July 2021; with additional ALE plots for the geographical distance.

## Supplementary Note 7. Grampian region results

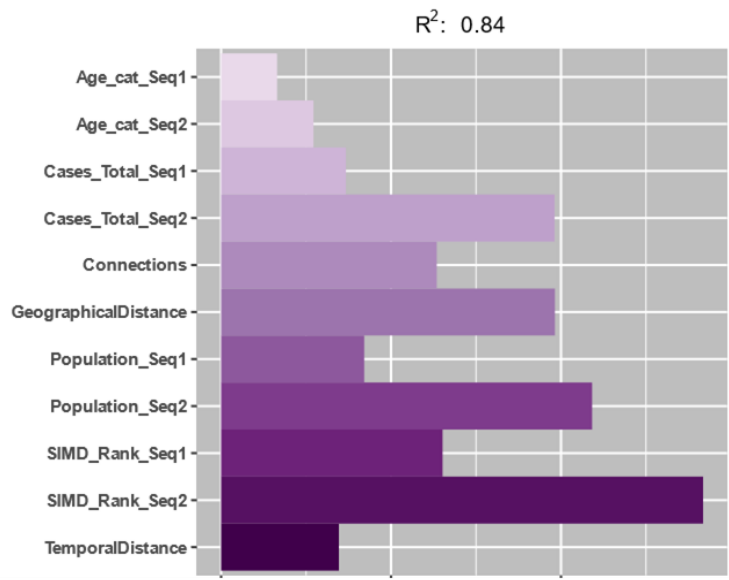

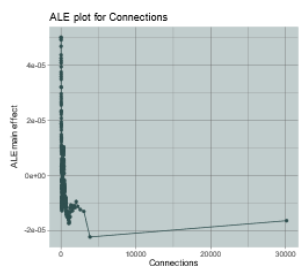 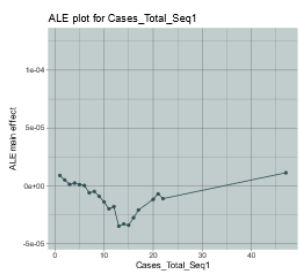 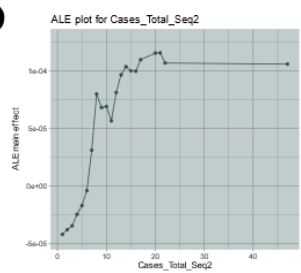

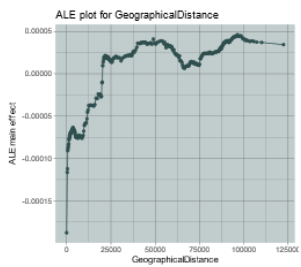 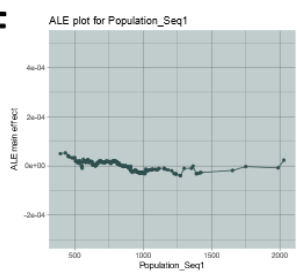 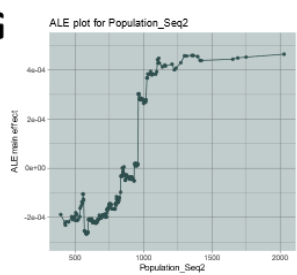

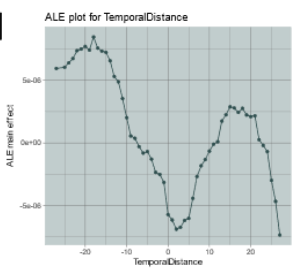 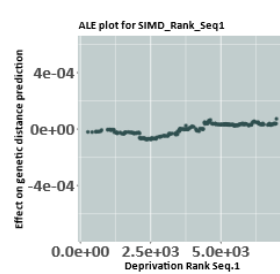 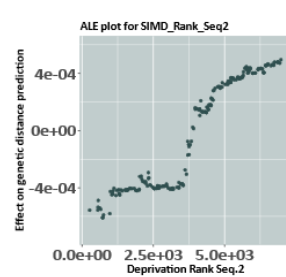

Supplementary Figure 7.1. Feature important (A) and ALE plots (B-J) of analysed variables for a Random Forest model describing the genetic distance between pairs of SARS-COV-2 sequences (with sequence 1 being genetically more similar to the wild type than sequence 2) for the Grampian region in the lockdown period from 02$^{nd}$ January 2021 to 30$^{th}$ April 2021; B) ALE plot for Commuting Connections variable; C) ALE plot for Contemporaneous Cases Sequence 1 variable; D) ALE plot for Contemporaneous Cases Sequence 2 variable; E) ALE plot for Geographical Distance variable; F) ALE plot for Population Size Sequence 1 variable; G) ALE plot for Population Size Sequence 2 variable; H) ALE plot for Temporal Distance variable; I) ALE plot for Deprivation (SIMD Rank) Sequence 1 variable; J) ALE plot for Deprivation (SIMD Rank) Sequence 2 variable;

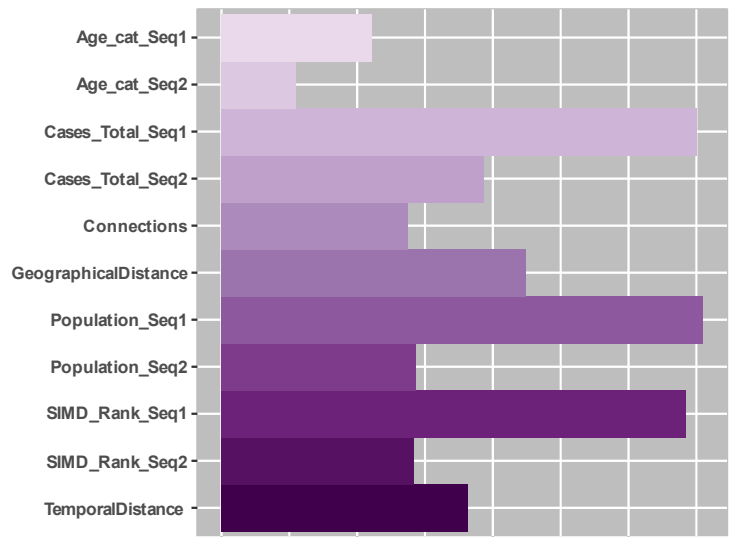
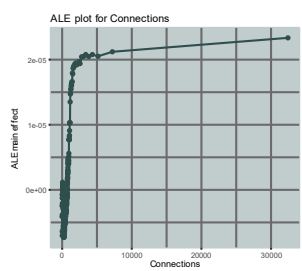 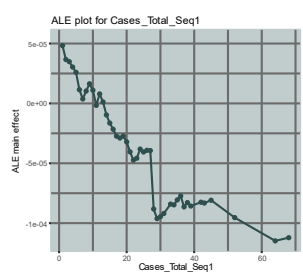 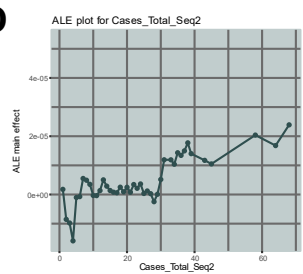
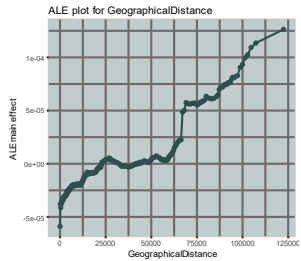 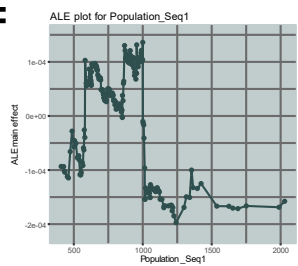 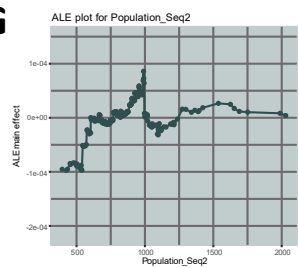
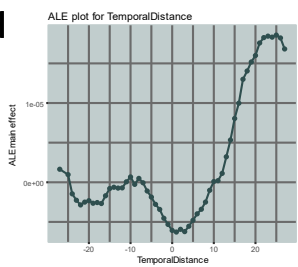 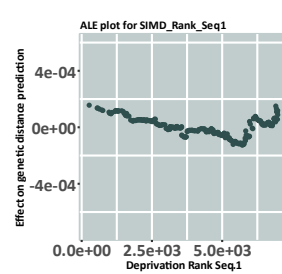 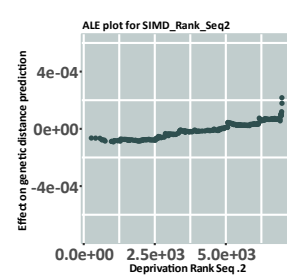

Supplementary Figure 7.2. Feature important (A) and ALE plots (B-J) of analysed variables for a Random Forest model describing the genetic distance between pairs of SARS-COV-2 sequences (with sequence 1 being genetically more similar to the wild type than sequence 2) for Grampian region out of lockdown periods from 04[th] August 2020-01[st] January 2021 & 01[st] May 2021-30[th] July 2021; B) ALE plot for Commuting Connections variable; C) ALE plot for Contemporaneous Cases Sequence 1 variable; D) ALE plot for Contemporaneous Cases Sequence 2 variable; E) ALE plot for Geographical Distance variable; F) ALE plot for Population Size Sequence 1 variable; G) ALE plot for Population Size Sequence 2 variable; H) ALE plot for Temporal Distance variable; I) ALE plot for Deprivation (SIMD Rank) Sequence 1 variable; J) ALE plot for Deprivation (SIMD Rank) Sequence 2 variable;

**Supplementary references**

# Supplementary Dataset 1

|  | beta | x | y | Runs | Major outbreaks | No/Minor outbreaks |
|---|---|---|---|---|---|---|
| Scenario 1 | 0.12 | 1 | 1 | 100 | 84 | 16 |
| Scenario 2 | 0.12 | 1 | 1.5 | 100 | 91 | 9 |
| Scenario 3 | 0.12 | 1 | 0.5 | 100 | 74 | 26 |
| Scenario 4 | 0.12 | 0 | 1 | 100 | 54 | 46 |
| Scenario 5 | 0.12 | 0 | 1.5 | 100 | 64 | 36 |
| Scenario 6 | 0.12 | 0 | 0.5 | 100 | 30 | 70 |
| Scenario 7 | 0.12 | 0.5 | 1 | 100 | 80 | 20 |
| Scenario 8 | 0.12 | 0.5 | 1.5 | 100 | 89 | 11 |
| Scenario 9 | 0.12 | 0.5 | 0.5 | 100 | 52 | 48 |
| Scenario 10 | 0.12 | 1.5 | 1 | 100 | 88 | 12 |
| Scenario 11 | 0.12 | 1.5 | 1.5 | 100 | 91 | 9 |
| Scenario 12 | 0.12 | 1.5 | 0.5 | 100 | 83 | 17 |

(INF/SUSC)

### Scenario 1 (1,1) — Group1/Group2

| | Cut-of | Case1 | Case2 |
|---|---|---|---|
| Ordered | | | |
| | 100 | 0.989122 | 0.992164 |
| | 50 | 0.990347 | 0.991271 |
| | 30 | 0.986494 | 0.993235 |
| | 10 | 0.982874 | 0.991359 |
| | 2 | 0.989122 | 0.992164 |
| Random | | | |
| | 100 | 0.990592 | 0.990692 |
| | 50 | 0.990814 | 0.990804 |
| | 30 | 0.989998 | 0.989719 |
| | 10 | 0.987196 | 0.987019 |
| | 2 | 0.989203 | 0.991632 |

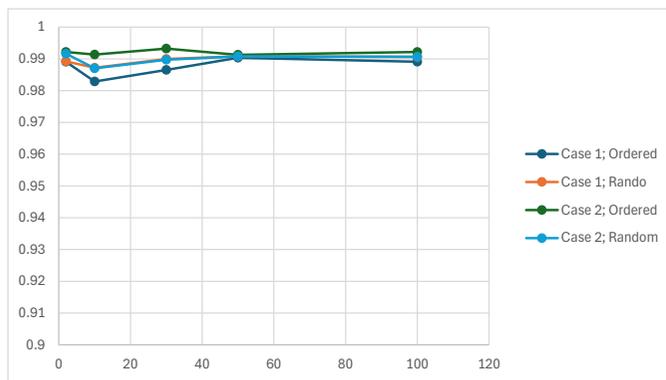

|  | Mean |  |  |  | Standard deviation |  |  |  |
|---|---|---|---|---|---|---|---|---|
|  | Case1/Group1 | Case1/Group2 | Case2/Group1 | Case2/Group2 | Case1/Group1 | Case1/Group2 | Case2/Group1 | Case2/Group2 |
|  | 6054510 | 6121090 | 6063850 | 6111750 | 715508 | 731172 | 712163 | 732750 |
|  | 5673780 | 5729080 | 5676440 | 5726420 | 594656 | 607017 | 590671 | 608335 |
|  | 2472670 | 2506530 | 2481150 | 2498050 | 284873 | 283795 | 279050 | 284581 |
|  | 122528 | 124663 | 123059 | 124132 | 12772.1 | 13341 | 12664.3 | 12527.5 |
|  | 5196.45 | 5250.37 | 5200.07 | 5246.75 | 327.009 | 328.488 | 316.814 | 327.061 |
|  | 6059030 | 6116570 | 6059330 | 6116270 | 712499 | 730764 | 712587 | 730636 |
|  | 5675120 | 5727740 | 5675090 | 5727760 | 591549 | 606480 | 591431 | 606578 |
|  | 2477080 | 2502110 | 2476730 | 2502460 | 281029 | 283255 | 281084 | 283296 |
|  | 122799 | 124392 | 122788 | 124403 | 12642.7 | 12804.2 | 12586.8 | 12873.4 |
|  | 5195.06 | 5251.76 | 5201.46 | 5245.36 | 323.118 | 325.392 | 320.763 | 330.124 |

### Scenario 2 (1,1.5) — Group1/Group2

| | Cut-of | Case1 | Case2 |
|---|---|---|---|
| Ordered | | | |
| | 100 | 1.06788 | 1.05889 |
| | 50 | 1.06845 | 1.05937 |
| | 30 | 1.07944 | 1.05987 |
| | 10 | 1.12528 | 1.06884 |
| | 2 | 1.20899 | 1.06381 |
| Random | | | |
| | 100 | 1.06337 | 1.06338 |
| | 50 | 1.06389 | 1.06391 |
| | 30 | 1.06959 | 1.06963 |
| | 10 | 1.09667 | 1.09669 |
| | 2 | 1.13316 | 1.1347 |

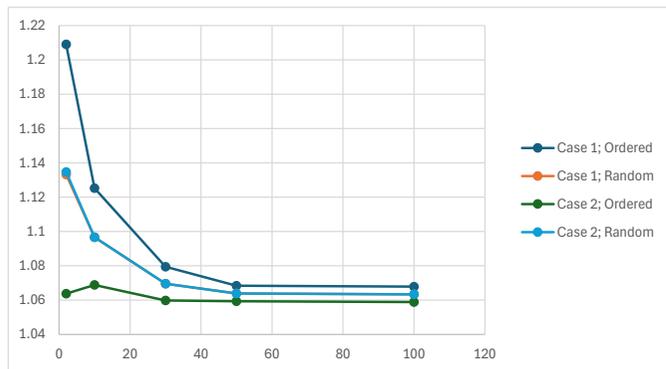

|  | Mean |  |  |  | Standard deviation |  |  |  |
|---|---|---|---|---|---|---|---|---|
|  | Case1/Group1 | Case1/Group2 | Case2/Group1 | Case2/Group2 | Case1/Group1 | Case1/Group2 | Case2/Group1 | Case2/Group2 |
|  | 9090420 | 8512620 | 9053260 | 8549790 | 737486 | 703522 | 729327 | 710378 |
|  | 8474320 | 7931400 | 8439330 | 7966390 | 615810 | 585089 | 607621 | 590089 |
|  | 3755880 | 3479460 | 3722820 | 3512520 | 327765 | 307863 | 326012 | 304736 |
|  | 177110 | 157392 | 172817 | 161686 | 17964.2 | 15613.1 | 16527.5 | 16003.1 |
|  | 6881.87 | 5692.26 | 6481.45 | 6092.68 | 296.601 | 248.358 | 265.019 | 261.837 |
|  | 9071820 | 8531230 | 9071870 | 8531180 | 732326 | 706019 | 732476 | 705788 |
|  | 8456790 | 7948930 | 8456860 | 7948860 | 610809 | 586259 | 610455 | 586645 |
|  | 3739320 | 3496020 | 3739380 | 3495960 | 325893 | 305270 | 325933 | 305253 |
|  | 174963 | 159540 | 174964 | 159538 | 17133.3 | 15675 | 17138.4 | 15670.6 |
|  | 6679.54 | 5894.59 | 6683.78 | 5890.35 | 275.537 | 260.362 | 285.835 | 248.732 |

## Scenario 3 (1, 0.5)

| | Cut-of | Group1/ Group2 | |
|---|---|---|---|
| | | Case1 | Case2 |
| Ordered | | | |
| | 100 | 0.807034 | 0.824485 |
| | 50 | 0.805642 | 0.824342 |
| | 30 | 0.792204 | 0.821048 |
| | 10 | 0.732844 | 0.813501 |
| | 2 | 0.652949 | 0.816061 |
| Random | | | |
| | 100 | 0.815715 | 0.81572 |
| | 50 | 0.814894 | 0.814993 |
| | 30 | 0.806595 | 0.806426 |
| | 10 | 0.772125 | 0.772386 |
| | 2 | 0.727802 | 0.733548 |

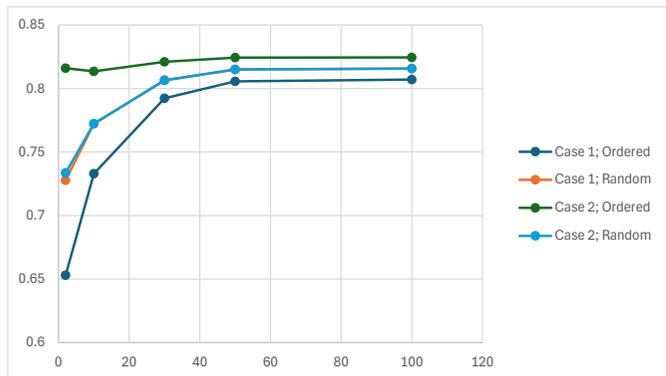

| | Mean | | | | Standard deviation | | | |
|---|---|---|---|---|---|---|---|---|
| Case1/Group1 | Case1/Group2 | Case2/Group1 | Case2/Group2 | Case1/Group1 | Case1/Group2 | Case2/Group1 | Case2/Group2 |
| 2621880 | 3248790 | 2652960 | 3217710 | 435114 | 532146 | 435955 | 530406 |
| 2469580 | 3065360 | 2501000 | 3033940 | 375576 | 455910 | 374620 | 455232 |
| 1107460 | 1397950 | 1129600 | 1375810 | 157627 | 188815 | 156869 | 186453 |
| 67853.1 | 92588.7 | 71971 | 88470.7 | 7039.24 | 9128.04 | 7052.88 | 8828.01 |
| 2860.36 | 4380.69 | 3253.82 | 3987.23 | 252.033 | 377.779 | 282.776 | 341.09 |
| 2637420 | 3233250 | 2637420 | 3233250 | 434923 | 530904 | 435036 | 530740 |
| 2485210 | 3049730 | 2485370 | 3049560 | 374460 | 455039 | 374457 | 455048 |
| 1118600 | 1386810 | 1118470 | 1386940 | 156577 | 187051 | 156544 | 187073 |
| 69905.4 | 90536.4 | 69918.7 | 90523 | 6998.19 | 8935.69 | 6989.26 | 8940.84 |
| 2860.36 | 4380.69 | 3253.82 | 3987.23 | 252.033 | 377.779 | 282.776 | 341.09 |

## Scenario 4 (0, 1)

| | Cut-of | Group1/ Group2 | |
|---|---|---|---|
| | | Case1 | Case2 |
| Ordered | | | |
| | 100 | 0.461961 | 0.473391 |
| | 50 | 0.461635 | 0.4739 |
| | 30 | 0.448597 | 0.474855 |
| | 10 | 0.38573 | 0.467355 |
| | 2 | 0.242399 | 0.468066 |
| Random | | | |
| | 100 | 0.467606 | 0.467702 |
| | 50 | 0.467907 | 0.467578 |
| | 30 | 0.461494 | 0.461722 |
| | 10 | 0.425489 | 0.425262 |
| | 2 | 0.345282 | 0.346394 |

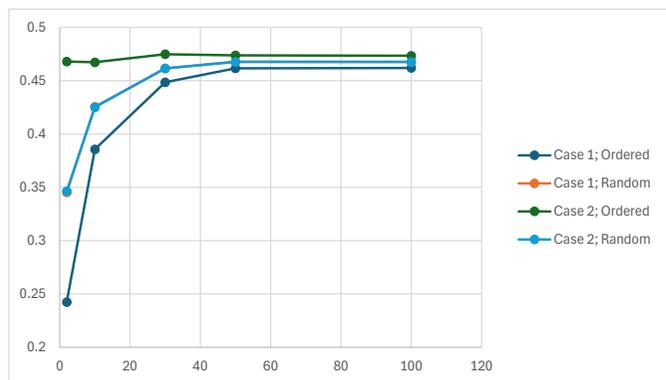

| | Mean | | | | Standard deviation | | | |
|---|---|---|---|---|---|---|---|---|
| Case1/Group1 | Case1/Group2 | Case2/Group1 | Case2/Group2 | Case1/Group1 | Case1/Group2 | Case2/Group1 | Case2/Group2 |
| 330889 | 716270 | 336445 | 710713 | 117499 | 252327 | 119070 | 250500 |
| 323407 | 700569 | 329237 | 694739 | 112938 | 243090 | 114715 | 240931 |
| 190991 | 425752 | 198571 | 418172 | 55309.3 | 125208 | 58157.2 | 121519 |
| 18021.5 | 46720.4 | 20620.4 | 44121.5 | 3730.08 | 9772.79 | 4195.96 | 9185.88 |
| 588.481 | 2427.74 | 961.667 | 2054.56 | 112.174 | 458.314 | 182.664 | 386.417 |
| 333644 | 713515 | 333690 | 713468 | 118029 | 251267 | 117994 | 251305 |
| 326400 | 697575 | 326244 | 697732 | 113596 | 241856 | 113525 | 241916 |
| 194748 | 421995 | 194814 | 421929 | 56475.5 | 123319 | 56569.5 | 123225 |
| 19324.6 | 45417.3 | 19317.3 | 45424.6 | 3927.3 | 9480.97 | 3955.34 | 9462.97 |
| 774.148 | 2242.07 | 776 | 2240.22 | 146.292 | 424.328 | 149.218 | 420.753 |

## Scenario 5 (0, 1.5)

| | Cut-of | Group1/ Group2 | |
|---|---|---|---|
| | | Case1 | Case2 |
| Ordered | | | |
| | 100 | 0.602736 | 0.616244 |
| | 50 | 0.602342 | 0.616428 |
| | 30 | 0.584859 | 0.618306 |
| | 10 | 0.520781 | 0.609158 |
| | 2 | 0.342265 | 0.609086 |
| Random | | | |
| | 100 | 0.609533 | 0.609389 |
| | 50 | 0.609288 | 0.60942 |
| | 30 | 0.601524 | 0.601291 |
| | 10 | 0.563616 | 0.563828 |
| | 2 | 0.465496 | 0.461737 |

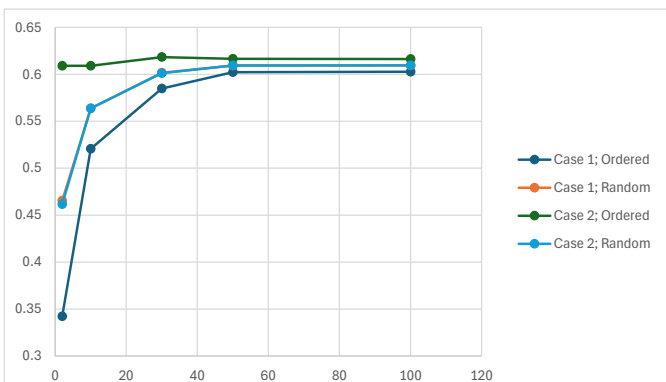

| | Mean | | | | Standard deviation | | | |
|---|---|---|---|---|---|---|---|---|
| Case1/Group1 | Case1/Group2 | Case2/Group1 | Case2/Group2 | Case1/Group1 | Case1/Group2 | Case2/Group1 | Case2/Group2 |
| 1128170 | 1871750 | 1143810 | 1856110 | 289175 | 459824 | 288715 | 459977 |
| 1077310 | 1788530 | 1092890 | 1772950 | 266623 | 420272 | 265904 | 420547 |
| 533611 | 912376 | 552468 | 893519 | 109002 | 171049 | 109522 | 169091 |
| 38738.4 | 74385.2 | 42823.7 | 70299.9 | 5837.89 | 10654.7 | 6458.85 | 9883.07 |
| 1312.42 | 3834.52 | 1948.27 | 3198.67 | 185.7 | 523.407 | 280.234 | 427.823 |
| 1136080 | 1863850 | 1135910 | 1864010 | 288574 | 459815 | 288744 | 459626 |
| 1085030 | 1780810 | 1085180 | 1780670 | 265932 | 420237 | 265981 | 420192 |
| 543105 | 902882 | 542974 | 903014 | 109093 | 169842 | 108927 | 169979 |
| 40776.2 | 72347.5 | 40786 | 72337.6 | 6139.75 | 10235.3 | 6094.5 | 10270.3 |
| 1634.86 | 3512.08 | 1625.83 | 3521.11 | 233.184 | 475.884 | 233.113 | 475.63 |

## Scenario 6 (0,0.5)

| | Cut-of | Group1/ Group2 Case1 | Case2 |
|---|---|---|---|
| Ordered | | | |
| | 100 | 0.26225 | 0.276463 |
| | 50 | 0.261835 | 0.27668 |
| | 30 | 0.254886 | 0.2778 |
| | 10 | 0.20996 | 0.273664 |
| | 2 | 0.117031 | 0.26943 |
| Random | | | |
| | 100 | 0.269155 | 0.269479 |
| | 50 | 0.269336 | 0.269091 |
| | 30 | 0.266155 | 0.266324 |
| | 10 | 0.24056 | 0.241431 |
| | 2 | 0.188957 | 0.187772 |

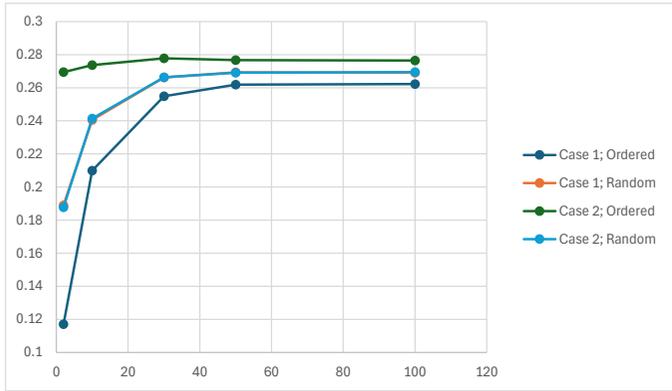

| | Mean | | | | Standard deviation | | | |
|---|---|---|---|---|---|---|---|---|
| Case1/Group1 | Case1/Group2 | Case2/Group1 | Case2/Group2 | Case1/Group1 | Case1/Group2 | Case2/Group1 | Case2/Group2 |
| 62847.1 | 239646 | 65515.5 | 236977 | 30970.1 | 117720 | 33330.1 | 115439 |
| 62109.4 | 237208 | 64867.6 | 234450 | 30212.1 | 115189 | 32669.5 | 112793 |
| 44386 | 174141 | 47508.8 | 171018 | 18039.1 | 71698 | 20068.7 | 69591 |
| 5626.53 | 26798.1 | 6966.87 | 25457.7 | 1617.84 | 8129.19 | 2147.2 | 7592.44 |
| 166.5 | 1422.7 | 337.3 | 1251.9 | 48.5498 | 397.555 | 96.8839 | 349.016 |
| 64150.9 | 238342 | 64211.7 | 238281 | 32088.6 | 116504 | 32006.4 | 116598 |
| 63511.2 | 235807 | 63465.7 | 235852 | 31376.5 | 113917 | 31289.4 | 114005 |
| 45935.9 | 172591 | 45959 | 172568 | 18958.1 | 70620.8 | 18968.4 | 70620.6 |
| 6287.53 | 26137.1 | 6305.87 | 26118.7 | 1883.57 | 7847.43 | 1865.39 | 7870.6 |
| 252.567 | 1336.63 | 251.233 | 1337.97 | 72.2399 | 373.908 | 73.1478 | 372.666 |

## Scenario 7 (0.5, 1)

| | Cut-of | Group1/ Group2 Case1 | Case2 |
|---|---|---|---|
| Ordered | | | |
| | 100 | 0.808789 | 0.825206 |
| | 50 | 0.807486 | 0.825061 |
| | 30 | 0.791769 | 0.822514 |
| | 10 | 0.740057 | 0.814217 |
| | 2 | 0.656183 | 0.81691 |
| Random | | | |
| | 100 | 0.816882 | 0.817039 |
| | 50 | 0.816124 | 0.816337 |
| | 30 | 0.806986 | 0.807036 |
| | 10 | 0.776417 | 0.77631 |
| | 2 | 0.734517 | 0.731142 |

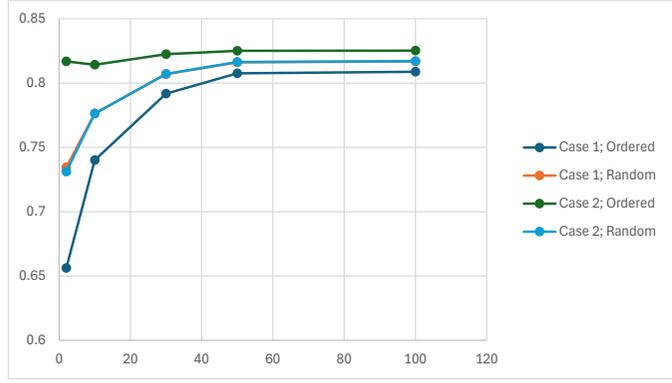

| | Mean | | | | Standard deviation | | | |
|---|---|---|---|---|---|---|---|---|
| Case1/Group1 | Case1/Group2 | Case2/Group1 | Case2/Group2 | Case1/Group1 | Case1/Group2 | Case2/Group1 | Case2/Group2 |
| 2579030 | 3188750 | 2607710 | 3160070 | 528301 | 635353 | 531254 | 632499 |
| 2431870 | 3011650 | 2460870 | 2982650 | 476700 | 570385 | 477477 | 569433 |
| 1094530 | 1382380 | 1117850 | 1359060 | 216327 | 256749 | 215889 | 255862 |
| 67186.3 | 90785.3 | 70897.4 | 87074.3 | 8925.94 | 11282.7 | 9228.89 | 10672.6 |
| 2837.45 | 4324.18 | 3219.98 | 3941.65 | 312.126 | 473.205 | 360.102 | 418.916 |
| 2593230 | 3174550 | 2593510 | 3174270 | 529356 | 633489 | 529268 | 633584 |
| 2446190 | 2997330 | 2446540 | 2996970 | 476436 | 569611 | 476696 | 569333 |
| 1106170 | 1370740 | 1106210 | 1370700 | 215789 | 25591 | 215646 | 256039 |
| 69044.6 | 88927.1 | 69039.2 | 88932.5 | 9043.78 | 10938.3 | 9020.72 | 10948.8 |
| 3032.74 | 4128.89 | 3024.69 | 4136.94 | 332.251 | 448.492 | 339.154 | 443.382 |

## Scenario 8 (0.5,1.5)

| | Cut-of | Group1/ Group2 Case1 | Case2 |
|---|---|---|---|
| Ordered | | | |
| | 100 | 0.930708 | 0.932933 |
| | 50 | 0.930488 | 0.933048 |
| | 30 | 0.923113 | 0.934444 |
| | 10 | 0.898269 | 0.931316 |
| | 2 | 0.855574 | 0.932258 |
| Random | | | |
| | 100 | 0.931809 | 0.93183 |
| | 50 | 0.931743 | 0.93179 |
| | 30 | 0.928888 | 0.928636 |
| | 10 | 0.914073 | 0.915227 |
| | 2 | 0.893721 | 0.892558 |

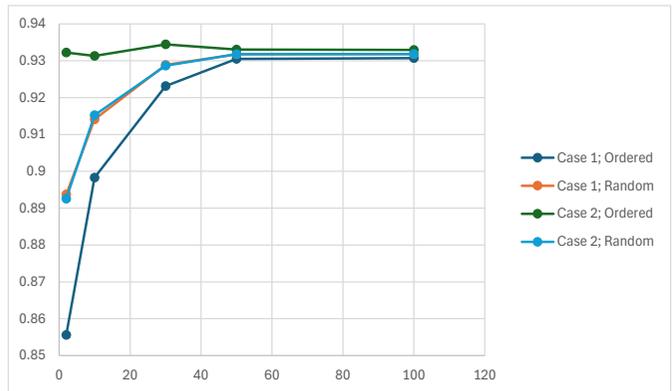

| | Mean | | | | Standard deviation | | | |
|---|---|---|---|---|---|---|---|---|
| Case1/Group1 | Case1/Group2 | Case2/Group1 | Case2/Group2 | Case1/Group1 | Case1/Group2 | Case2/Group1 | Case2/Group2 |
| 5124440 | 5505960 | 5130780 | 5499630 | 500902 | 545776 | 502398 | 544642 |
| 4786770 | 5144360 | 4793580 | 5137550 | 437465 | 483052 | 439347 | 480252 |
| 2059360 | 2230890 | 2072430 | 2217820 | 244414 | 271837 | 245528 | 267391 |
| 106493 | 118554 | 108522 | 116525 | 9739.32 | 11482 | 9842.06 | 10870.1 |
| 4503.53 | 5263.75 | 4712.43 | 5054.85 | 229.412 | 278.921 | 234.568 | 258.842 |
| 5127580 | 5502830 | 5127640 | 5502770 | 500378 | 544194 | 500543 | 544035 |
| 4790110 | 5141020 | 4790230 | 5140890 | 437120 | 480579 | 437243 | 480500 |
| 2066040 | 2224210 | 2065750 | 2224500 | 244250 | 268949 | 244231 | 268968 |
| 107472 | 117575 | 107543 | 117504 | 9746.71 | 11106.2 | 9707.06 | 11142.4 |
| 4609.56 | 5157.72 | 4606.39 | 5160.89 | 228.557 | 274.078 | 234.857 | 263.48 |

## Scenario 9

**(0.5, 0.5)**

Ordered

| | Cut-of | Group1/ Group2 Case1 | Case2 |
|---|---|---|---|
| | 100 | 0.570563 | 0.598273 |
| | 50 | 0.569316 | 0.597705 |
| | 30 | 0.552663 | 0.5928 |
| | 10 | 0.480685 | 0.581507 |
| | 2 | 0.380479 | 0.583487 |

Random

| | Cut-of | Case1 | Case2 |
|---|---|---|---|
| | 100 | 0.584346 | 0.584247 |
| | 50 | 0.583542 | 0.583224 |
| | 30 | 0.572287 | 0.572664 |
| | 10 | 0.528847 | 0.530025 |
| | 2 | 0.476595 | 0.473469 |

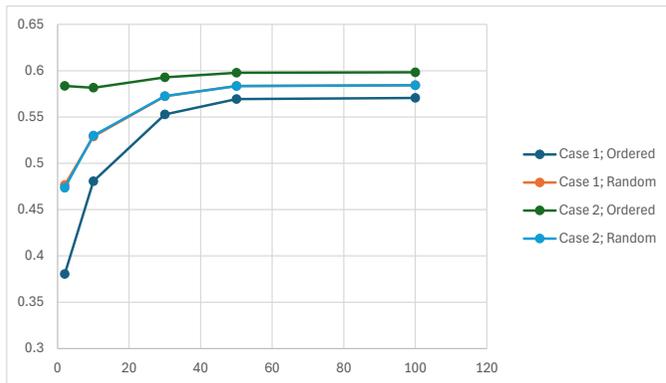

| | Mean | | | | Standard deviation | | | |
|---|---|---|---|---|---|---|---|---|
| Case1/Group1 | Case1/Group2 | Case2/Group1 | Case2/Group2 | Case1/Group1 | Case1/Group2 | Case2/Group1 | Case2/Group2 |
| 640260 | 1122160 | 659715 | 1102700 | 220254 | 379589 | 227710 | 371157 |
| 617022 | 1083790 | 636279 | 1064540 | 207749 | 358381 | 214820 | 350146 |
| 333703 | 603808 | 348918 | 588593 | 99044 | 176480 | 103896 | 170058 |
| 27573.8 | 57363.6 | 31230.8 | 53706.7 | 6076.5 | 11938.3 | 6810.75 | 11012 |
| 1077.08 | 2830.85 | 1440 | 2467.92 | 223.791 | 545.831 | 294.668 | 473.831 |
| 650022 | 1112390 | 649953 | 1112460 | 223608 | 375119 | 223558 | 375162 |
| 626758 | 1074060 | 626542 | 1074270 | 210878 | 354046 | 210891 | 354017 |
| 341239 | 596272 | 341382 | 596129 | 101175 | 173119 | 101195 | 173112 |
| 29380.9 | 55556.5 | 29423.7 | 55513.7 | 6397.5 | 11486.9 | 6439.49 | 11442.2 |
| 1261.35 | 2646.58 | 1255.73 | 2652.19 | 260.181 | 508.668 | 258.31 | 511.088 |

## Scenario 10

**(1.5, 1)**

Ordered

| | Cut-of | Group1/ Group2 Case1 | Case2 |
|---|---|---|---|
| | 100 | 1.06963 | 1.06483 |
| | 50 | 1.07006 | 1.06453 |
| | 30 | 1.07886 | 1.06301 |
| | 10 | 1.13027 | 1.06881 |
| | 2 | 1.20952 | 1.06763 |

Random

| | Cut-of | Case1 | Case2 |
|---|---|---|---|
| | 100 | 1.06736 | 1.06709 |
| | 50 | 1.06735 | 1.06723 |
| | 30 | 1.07086 | 1.07094 |
| | 10 | 1.09882 | 1.09936 |
| | 2 | 1.13758 | 1.13487 |

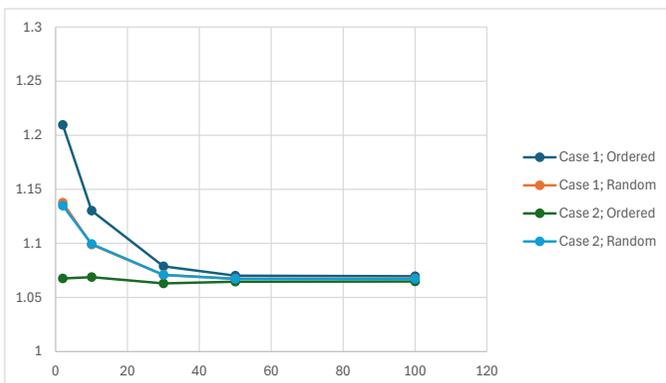

| | Mean | | | | Standard deviation | | | |
|---|---|---|---|---|---|---|---|---|
| Case1/Group1 | Case1/Group2 | Case2/Group1 | Case2/Group2 | Case1/Group1 | Case1/Group2 | Case2/Group1 | Case2/Group2 |
| 9181980 | 8584250 | 9162010 | 8604220 | 647254 | 620197 | 648861 | 619531 |
| 8522130 | 7964180 | 8500800 | 7985500 | 529103 | 501332 | 528461 | 501067 |
| 3727310 | 3454870 | 3700760 | 3481410 | 299517 | 277112 | 293469 | 278578 |
| 177188 | 156766 | 172531 | 161423 | 14150.9 | 12744.4 | 13151.7 | 12541.4 |
| 6916.42 | 5718.33 | 6524.01 | 6110.74 | 253.618 | 230.053 | 235.114 | 229.251 |
| 9172560 | 8593660 | 9171420 | 8594800 | 646812 | 618127 | 646429 | 618596 |
| 8511680 | 7946300 | 8511250 | 7975050 | 527194 | 499412 | 527062 | 499495 |
| 3713970 | 3468210 | 3714110 | 3468070 | 295585 | 276742 | 295473 | 276860 |
| 174839 | 159115 | 174880 | 159074 | 13506 | 12481 | 13509.5 | 12475.3 |
| 6723.97 | 5910.78 | 6716.47 | 5918.28 | 246.561 | 224.284 | 241.467 | 233.827 |

## Scenario 11

**(1.5, 1.5)**

Ordered

| | Cut-of | Group1/ Group2 Case1 | Case2 |
|---|---|---|---|
| | 100 | 1.12295 | 1.10839 |
| | 50 | 1.12404 | 1.10867 |
| | 30 | 1.14211 | 1.10983 |
| | 10 | 1.24984 | 1.12389 |
| | 2 | 1.43217 | 1.11602 |

Random

| | Cut-of | Case1 | Case2 |
|---|---|---|---|
| | 100 | 1.11565 | 1.11564 |
| | 50 | 1.11629 | 1.11637 |
| | 30 | 1.12586 | 1.12584 |
| | 10 | 1.18503 | 1.18507 |
| | 2 | 1.26191 | 1.26431 |

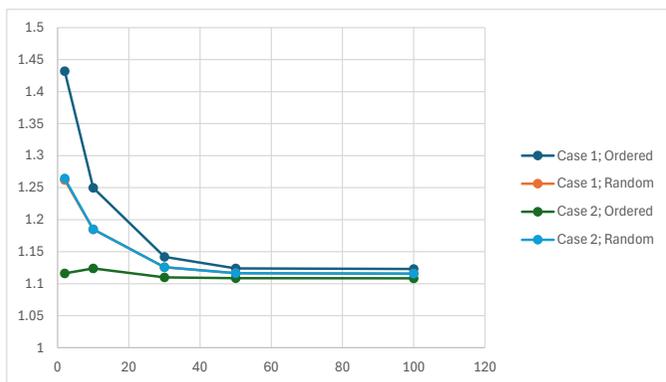

| | Mean | | | | Standard deviation | | | |
|---|---|---|---|---|---|---|---|---|
| Case1/Group1 | Case1/Group2 | Case2/Group1 | Case2/Group2 | Case1/Group1 | Case1/Group2 | Case2/Group1 | Case2/Group2 |
| 11826600 | 10531700 | 11753800 | 10604400 | 797030 | 716489 | 786093 | 726036 |
| 10952400 | 9743750 | 10881300 | 9814780 | 607300 | 545317 | 590444 | 557562 |
| 4842450 | 4239900 | 4777570 | 4304780 | 310616 | 277632 | 304552 | 277232 |
| 230492 | 184417 | 219556 | 195354 | 18627.4 | 15192.6 | 17180.2 | 15325.7 |
| 8347.02 | 5828.22 | 7476.23 | 6699.01 | 284.164 | 218.414 | 252.941 | 235.12 |
| 11790200 | 10568000 | 11790200 | 10568100 | 790563 | 719817 | 790247 | 720159 |
| 10916700 | 9779450 | 10917000 | 9779070 | 597737 | 549883 | 597431 | 550137 |
| 4810020 | 4272320 | 4809990 | 4272350 | 306466 | 276001 | 30628 | 276152 |
| 225022 | 189887 | 225026 | 189884 | 17782.3 | 15022.6 | 17706.9 | 15087.1 |
| 7908.3 | 6266.95 | 7914.96 | 6260.29 | 277.555 | 217.732 | 260.67 | 236.296 |

| Scenario 12 | | Group1/ Group2 | |
|---|---|---|---|
| (1.5, 0.5) | Cut-of | Case1 | Case2 |
| Ordered | | | |
| | 100 | 0.932572 | 0.937465 |
| | 50 | 0.932296 | 0.937404 |
| | 30 | 0.926192 | 0.937889 |
| | 10 | 0.897194 | 0.934281 |
| | 2 | 0.852338 | 0.935455 |
| Random | | | |
| | 100 | 0.934956 | 0.935075 |
| | 50 | 0.934754 | 0.93494 |
| | 30 | 0.932017 | 0.932028 |
| | 10 | 0.915569 | 0.915546 |
| | 2 | 0.892103 | 0.893868 |

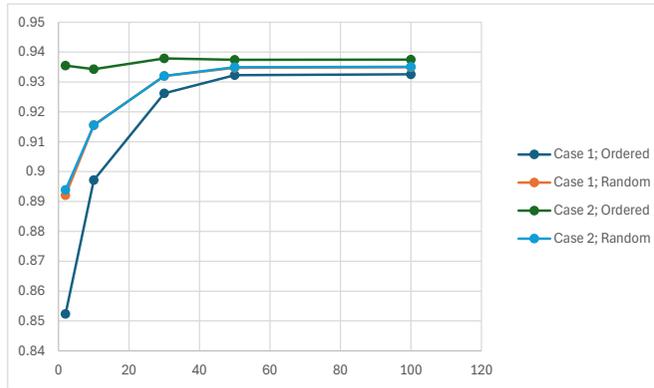

| | Mean | | | | Standard deviation | | | |
|---|---|---|---|---|---|---|---|---|
| Case1/Group1 | Case1/Group2 | Case2/Group1 | Case2/Group2 | Case1/Group1 | Case1/Group2 | Case2/Group1 | Case2/Group2 |
| 5097420 | 5465980 | 5111230 | 5452180 | 549466 | 592817 | 550003 | 591362 |
| 4760500 | 5106210 | 4773960 | 5092750 | 479301 | 511726 | 477496 | 511538 |
| 2064460 | 2228980 | 2077910 | 2215520 | 230003 | 245822 | 233655 | 239444 |
| 106146 | 118309 | 108415 | 116041 | 7714.95 | 9276.54 | 8156.6 | 8299.54 |
| 4478.84 | 5254.77 | 4704.51 | 5029.11 | 258.811 | 298.653 | 260.68 | 285.035 |
| | | | | | | | |
| 5104150 | 5459250 | 5104490 | 5458910 | 548911 | 591414 | 549007 | 591324 |
| 4766990 | 50.9972 | 4767480 | 5099230 | 477454 | 510947 | 477686 | 510766 |
| 2071180 | 2222260 | 2071190 | 2222240 | 231056 | 241960 | 231134 | 241933 |
| 107281 | 117174 | 107280 | 117176 | 7851.01 | 8703.46 | 7831.13 | 8723.57 |
| 4589.28 | 5144.34 | 4594.07 | 5139.54 | 266.583 | 285.651 | 253.841 | 299.01 |

|    | A    | B         | C         | D         | E         | F        | G         | H         |
|----|------|-----------|-----------|-----------|-----------|----------|-----------|-----------|
| 1  | Rank | Inf       | 0.002     | 0.0015    | 0.0013    | 0.001    | 0.0005    | 0.0001    |
| 2  | 10   | -0.00015  | -2.29E-05 | -5.05E-06 | 3.14E-05  | -2.70E-06| -1.59E-05 | -9.37E-07 |
| 3  | 25   | -0.000136 | -8.94E-06 | 7.14E-06  | 4.31E-05  | 9.66E-06 | -1.15E-05 | -8.46E-07 |
| 4  | 61   | -0.000135 | -7.38E-06 | 8.81E-06  | 4.55E-05  | 1.17E-05 | -1.13E-05 | -8.33E-07 |
| 5  | 115  | -0.000137 | -8.96E-06 | 7.40E-06  | 4.47E-05  | 1.04E-05 | -1.26E-05 | -9.51E-07 |
| 6  | 144  | -0.00014  | -1.21E-05 | 2.58E-06  | 4.35E-05  | 1.03E-05 | -1.29E-05 | -9.75E-07 |
| 7  | 145  | -4.85E-05 | 7.66E-05  |           |           |          |           | -4.02E-07 |
| 8  | 181  |           |           |           | 9.29E-05  | 4.52E-05 | -6.56E-06 |           |
| 9  | 197  | 1.46E-05  | 0.000146  | 0.000187  |           |          |           |           |
| 10 | 213  | -6.83E-05 | 5.61E-05  | 0.0001    | 6.37E-05  | 4.94E-05 | 2.58E-06  | 8.04E-07  |
| 11 | 233  | -6.44E-06 | 0.000123  | 0.000103  | 6.67E-05  | 5.24E-05 | 1.8E-06   |           |
| 12 | 234  |           | 0.000107  | 9.72E-05  | 5.91E-05  | 4.62E-05 | 3.13E-06  | 7.78E-07  |
| 13 | 239  | 2.07E-05  |           |           |           |          |           |           |
| 14 | 259  | -3.58E-05 | 8.81E-05  | 0.000111  | 7.25E-05  | 5.88E-05 |           |           |
| 15 | 261  | -3.49E-05 | 8.93E-05  | 0.000112  | 7.33E-05  | 6.01E-05 | 6.44E-06  | 5.59E-07  |
| 16 | 276  | -3.62E-05 | 8.86E-05  | 0.000111  | 7.25E-05  | 5.96E-05 | 6.67E-06  | 1.2E-06   |
| 17 | 284  | -3.88E-05 | 8.50E-05  | 0.000112  | 7.53E-05  | 6.22E-05 | 7.44E-06  | 1.17E-06  |
| 18 | 327  |           |           |           |           |          | 6.16E-06  |           |
| 19 | 334  | -3.32E-05 | 8.95E-05  | 0.000112  | 7.29E-05  |          |           | 1.53E-06  |
| 20 | 344  | -1.85E-05 | 0.000101  | 8.29E-05  | 5.80E-05  | 4.82E-05 | 7.65E-06  |           |
| 21 | 352  | -3.62E-05 | 8.39E-05  | 8.99E-05  | 7.14E-05  | 5.97E-05 | 9.16E-06  | 1.33E-06  |
| 22 | 357  |           |           | 7.99E-05  | 6.03E-05  | 5.09E-05 |           |           |
| 23 | 366  | -5.36E-05 | 6.78E-05  | 7.00E-05  | 4.10E-05  | 3.50E-05 | 6.67E-06  | 1.54E-06  |
| 24 | 375  |           |           |           | 3.79E-05  |          | 4.79E-06  | 1.54E-06  |
| 25 | 378  |           |           | 7.94E-05  |           |          |           |           |
| 26 | 380  | -4.36E-05 | 7.66E-05  | 7.96E-05  | 4.14E-05  | 3.91E-05 | 7.58E-06  | 2.31E-06  |
| 27 | 408  | -2.59E-05 | 9.80E-05  | 8.14E-05  | 4.81E-05  | 4.34E-05 | 8.11E-06  | 2.43E-06  |
| 28 | 416  | -2.73E-05 | 9.63E-05  | 8.03E-05  | 4.69E-05  | 4.25E-05 | 7.73E-06  | 2.4E-06   |
| 29 | 432  |           |           | 8.74E-05  |           |          |           |           |
| 30 | 434  | -4.37E-05 | 8.05E-05  |           | 2.53E-05  | 2.51E-05 | 5.18E-06  | 2.3E-06   |
| 31 | 487  |           |           | 9.26E-05  |           |          |           |           |
| 32 | 496  | -3.58E-05 | 8.74E-05  |           | 3.54E-05  | 3.50E-05 | 8.35E-06  | 2.31E-06  |
| 33 | 513  | -2.72E-05 | 9.74E-05  | 9.94E-05  |           | 4.23E-05 | 1.17E-05  | 2.19E-06  |
| 34 | 530  |           |           |           | 4.04E-05  |          |           |           |
| 35 | 540  | -2.3E-05  | 0.000102  | 0.000103  | 4.46E-05  | 4.55E-05 | 1.68E-05  | 1.96E-06  |
| 36 | 544  |           |           |           |           |          | 1.72E-05  |           |
| 37 | 551  | -3.56E-05 | 9.12E-05  | 0.000104  | 5.54E-05  | 5.52E-05 | 2.22E-05  | 2.05E-06  |
| 38 | 557  | -3.66E-05 | 9.03E-05  | 0.000103  | 5.49E-05  | 5.52E-05 | 2.15E-05  |           |
| 39 | 573  |           |           | 0.000118  |           |          |           |           |
| 40 | 582  | -2.74E-05 |           |           |           | 6.42E-05 |           | 2E-06     |
| 41 | 592  | -3.22E-05 | 8.63E-05  | 0.000116  | 5.58E-05  | 6.38E-05 | 2.07E-05  |           |
| 42 | 594  |           |           | 0.000124  |           |          |           |           |
| 43 | 663  | -2.15E-05 | 9.67E-05  |           |           | 6.77E-05 |           |           |
| 44 | 716  | -2.18E-05 | 9.64E-05  | 0.000127  | 5.77E-05  | 6.71E-05 | 2.57E-05  |           |
| 45 | 748  | -3.36E-05 | 8.32E-05  | 0.000117  | 5.08E-05  | 6.15E-05 | 2.35E-05  | 1.7E-06   |
| 46 | 760  |           |           | 0.000117  |           | 6.13E-05 |           |           |
| 47 | 772  | -3.33E-05 | 8.34E-05  |           | 5.14E-05  |          | 2.44E-05  |           |
| 48 | 782  | -3.54E-05 | 8.19E-05  | 1.15E-04  | 5.08E-05  | 6.27E-05 | 2.40E-05  | 2.24E-06  |
| 49 | 813  | -0.000107 | -4.81E-06 | 6.00E-05  | -4.30E-06 | 9.94E-06 | 7.31E-06  | 1.52E-06  |
| 50 | 818  | -0.000105 | -3.06E-06 | 6.05E-05  | -3.97E-06 | 1.06E-05 | 7.95E-06  | 1.62E-06  |
| 51 | 826  | -0.000103 | -1.42E-06 | 6.13E-05  | -3.73E-06 | 1.09E-05 | 8.42E-06  |           |
| 52 | 844  | -0.000105 | -2.84E-06 |           | -1.60E-06 |          | 9.13E-06  |           |
| 53 | 917  |           |           | 6.9E-05   |           | 1.42E-05 |           | 1.48E-06  |
| 54 | 925  |           |           |           | 4.01E-06  |          |           |           |
| 55 | 931  | -4.28E-05 | 6.21E-05  |           |           | 1.48E-05 | 9.27E-06  |           |
| 56 | 933  | -5.23E-05 | 5.19E-05  | 6.89E-05  | 1.83E-06  | 1.19E-05 | 5.24E-06  | 1.83E-06  |
| 57 | 937  |           |           |           | 2.96E-06  |          |           |           |
| 58 | 975  | -3.10E-05 |           |           |           |          |           |           |
| 59 | 976  |           | 8.71E-05  |           |           |          |           |           |
| 60 | 991  |           |           |           |           |          |           | 1.4E-06   |
| 61 | 997  | -4.16E-05 | 7.83E-05  | 6.35E-05  | -1.72E-07 | 8.58E-06 | 4.25E-06  | 1.4E-06   |
| 62 | 1043 |           |           | 6.47E-05  | 5.72E-07  | 9.65E-06 | 6.70E-06  |           |
| 63 | 1063 | -6.56E-05 | 4.43E-05  |           |           |          |           |           |
| 64 | 1093 | -6.94E-05 | 3.98E-05  | 6.72E-05  | 2.06E-06  | 1.15E-05 | 7.25E-06  | 1.63E-06  |
| 65 | 1107 | -7.24E-05 | 3.71E-05  |           |           |          |           | 1.89E-06  |
| 66 | 1114 | -7.22E-05 | 3.73E-05  | 6.67E-05  | 9.28E-07  | 7.4E-06  | 5.06E-06  | 1.41E-06  |
| 67 | 1165 | -8.76E-05 | 2.10E-05  | 5.16E-05  | -6.07E-06 | 1.82E-06 | 4.67E-06  | 1.34E-06  |
| 68 | 1174 | 4.17E-05  | 0.00014   |           | -6.18E-06 | 1.74E-06 | 4.62E-06  | 1.33E-06  |
| 69 | 1206 | -1.70E-05 | 8.55E-05  | 7.24E-05  |           |          |           | 1.4E-06   |
| 70 | 1211 | -5.62E-05 | 4.39E-05  | 6.63E-05  | -2.84E-06 | 6.09E-06 | 8.19E-06  | 1.42E-06  |
| 71 | 1240 |           |           | 5.02E-05  |           | 8.82E-06 |           |           |
| 72 | 1250 | -8.39E-05 | 5.48E-06  |           | -7.99E-06 | 4.20E-06 | 6.03E-06  | 1.58E-06  |
| 73 | 1260 |           |           | 4.67E-05  |           |          |           |           |
| 74 | 1262 | -8.42E-05 | 4.76E-06  |           | -8.68E-06 |          |           | 1.59E-06  |
| 75 | 1294 | -8.37E-05 | 5.24E-06  | 5.1E-05   |           | 4.08E-06 | 5.67E-07  | 1.29E-06  |
| 76 | 1306 | -8.73E-05 | 2.34E-06  | 3.22E-05  | -6.30E-06 | 3.54E-06 |           | 1.34E-06  |
| 77 | 1414 | -8.84E-05 | 8.59E-07  | 2.95E-05  | -8.45E-06 | 2.94E-06 | 3.17E-06  | 1.32E-06  |
| 78 | 1437 | -8.51E-05 |           | 6.44E-06  |           |          |           | 1.48E-06  |
| 79 | 1441 | -8.94E-05 | 2.25E-06  | 2.93E-05  | -1.08E-05 | 8.22E-07 | 2.48E-06  | 1.48E-06  |
| 80 | 1448 | -8.66E-05 | 6.16E-06  | 3.15E-05  | -9.36E-06 | 2.98E-06 | 1.04E-06  | 1.6E-06   |
| 81 | 1485 | -9.45E-05 | -2.32E-07 | 2.54E-05  | -9.91E-06 | 3.24E-06 | 2.28E-06  | 1.65E-06  |
| 82 | 1501 | -1.01E-04 | -5.59E-06 | 2.05E-05  | -1.32E-05 | 1.30E-06 | 1.3E-07   | 1.68E-06  |
| 83 | 1504 | -8.66E-05 | 6.99E-06  |           | 1.95E-05  | 2.09E-05 |           | 1.5E-06   |
| 84 | 1535 |           |           | 3.74E-05  |           |          |           |           |
| 85 | 1576 | -7.59E-05 | 1.75E-05  |           |           |          | 1.16E-06  |           |
| 86 | 1579 |           |           | 3.54E-05  | 1.90E-05  | 1.95E-05 |           |           |
| 87 | 1617 |           |           |           |           |          |           | 1.7E-06   |
| 88 | 1654 | -7.66E-05 | 1.62E-05  | 3.43E-05  | 1.87E-05  | 1.92E-05 | 8.54E-07  | 1.64E-06  |
| 89 | 1655 | -7.91E-05 | 1.34E-05  | 3.29E-05  | 1.88E-05  | 1.92E-05 | 1.66E-07  | 2.33E-06  |
| 90 | 1667 | -7.93E-05 | 1.30E-05  |           |           | 1.94E-05 |           |           |
| 91 | 1683 | -9.59E-05 | -6.98E-06 | 2.28E-05  | 3.06E-05  | 1.49E-06 | -2.01E-06 | 1.63E-06  |
| 92 | 1686 |           |           |           | 6.22E-06  |          |           | 1.51E-06  |
| 93 | 1708 | -0.000109 | -2.43E-05 |           |           | 3.6E-06  |           |           |
| 94 | 1720 |           |           | 1.73E-05  | 6.12E-06  | 4.12E-06 | -6.28E-07 | 2.27E-06  |
| 95 | 1725 | -0.000109 | -2.46E-05 |           |           |          |           |           |
| 96 | 1734 |           |           | 1.7E-05   |           |          |           |           |
| 97 | 1757 |           |           |           | 7.37E-06  | 5.54E-06 |           |           |
| 98 | 1758 | -1.11E-04 | -2.58E-05 | 1.52E-05  | 5.61E-06  | 4.33E-06 | -6.56E-07 | 2.27E-06  |
| 99 | 1759 | -6.88E-05 | 1.73E-05  |           |           |          |           |           |
| 100| 1781 |           |           | 2E-05     |           |          | 1.27E-06  |           |
| 101| 1787 | -8.73E-05 | -3.45E-07 |           | 9.17E-06  | 8.37E-06 | 2.08E-06  |           |
| 102| 1791 | -8.54E-05 | 1.19E-06  | 2.53E-05  | 1.04E-05  | 9.33E-06 |           | 2.07E-06  |

ALE_plot_S1_SL_All_2_12_08_24

| | A | B | C | D | E | F | G | H |
|---|---|---|---|---|---|---|---|---|
| 103 | 1816 | | | | 1.00E-05 | 8.47E-06 | | |
| 104 | 1851 | -8.94E-05 | -2.82E-06 | 2.37E-05 | | | 2.92E-07 | |
| 105 | 1909 | | | | 8.91E-06 | 7.77E-06 | 1E-07 | |
| 106 | 1942 | -8.96E-05 | -2.89E-06 | 1.77E-05 | 9.52E-06 | 8.21E-06 | 7.75E-07 | 2.18E-06 |
| 107 | 1956 | -8.99E-05 | -3.12E-06 | 1.77E-05 | | | | 2.19E-06 |
| 108 | 1992 | | | | 2.27E-05 | 1.88E-05 | 1.42E-06 | |
| 109 | 2018 | -7.3E-05 | 1.41E-05 | 3.36E-05 | | | | |
| 110 | 2042 | | | | 2.13E-05 | | | |
| 111 | 2047 | -2.52E-05 | 5.21E-05 | 5.34E-05 | | 1.72E-05 | | |
| 112 | 2049 | | | | 2.08E-05 | | 2.11E-06 | |
| 113 | 2076 | | | | | 1.82E-05 | | |
| 114 | 2080 | -2.34E-05 | 5.45E-05 | 5.42E-05 | | | 2.46E-06 | 1.7E-06 |
| 115 | 2085 | -2.11E-05 | 5.72E-05 | 5.51E-05 | 2.14E-05 | 1.82E-05 | 2.48E-06 | |
| 116 | 2096 | | | | 2.31E-05 | 1.91E-05 | | |
| 117 | 2105 | -1.98E-05 | 5.84E-05 | 5.55E-05 | | | | |
| 118 | 2116 | -2.17E-05 | 5.68E-05 | 5.33E-05 | 1.46E-05 | 1.09E-05 | 6.73E-07 | 1.77E-06 |
| 119 | 2127 | -2.6E-05 | 5.1E-05 | 4.88E-05 | 1.39E-05 | 1.09E-05 | 5.98E-07 | 1.8E-06 |
| 120 | 2128 | | | | 1.31E-05 | 9.92E-06 | 5.59E-07 | |
| 121 | 2148 | | | | | | | 1.87E-06 |
| 122 | 2156 | -1.81E-05 | 5.88E-05 | 5.42E-05 | | | | |
| 123 | 2175 | | | | 1.97E-05 | | | 1.95E-06 |
| 124 | 2185 | 0.000114 | | 0.000148 | | 1.01E-05 | | |
| 125 | 2226 | 0.000101 | 8.9E-05 | 0.000129 | 1.82E-05 | 1.03E-05 | -2.89E-06 | 1.88E-06 |
| 126 | 2261 | | | | 1.43E-05 | 7.24E-06 | -1.62E-06 | 1.83E-06 |
| 127 | 2272 | | | 0.000142 | | | | |
| 128 | 2351 | 0.000195 | 0.000183 | | | | | |
| 129 | 2357 | 0.00019 | 0.000179 | 0.00014 | 1.35E-05 | 7.77E-06 | -1.25E-06 | 1.55E-06 |
| 130 | 2365 | | | | | 7.98E-06 | -8.79E-06 | |
| 131 | 2369 | 0.000185 | 0.000173 | | | | | |
| 132 | 2376 | 1.82E-04 | 1.72E-04 | 1.35E-04 | 1.43E-05 | 9.1E-06 | | 1.61E-06 |
| 133 | 2394 | | | | | | -7.48E-06 | |
| 134 | 2415 | 1.73E-04 | 1.63E-04 | 1.23E-04 | 1.15E-05 | 7.39E-06 | | 1.85E-06 |
| 135 | 2433 | | | | | | -8.71E-06 | |
| 136 | 2448 | 9.71E-05 | 8.38E-05 | 7.17E-05 | 7.86E-06 | 6.97E-06 | | 1.84E-06 |
| 137 | 2451 | 9.75E-05 | 8.42E-05 | 7.20E-05 | 8.36E-06 | 7.63E-06 | -7.85E-06 | 1.94E-06 |
| 138 | 2539 | 7.03E-05 | 5.65E-05 | 5.32E-05 | 5.21E-06 | 6.62E-06 | -6.56E-06 | 1.66E-06 |
| 139 | 2570 | 7.63E-05 | 6.36E-05 | 5.75E-05 | 3.76E-06 | 5.67E-06 | -7.62E-06 | 1.63E-06 |
| 140 | 2585 | | | | | | | 1.53E-06 |
| 141 | 2678 | 8.01E-05 | 6.05E-05 | 5.61E-05 | 2.64E-06 | | -8.93E-06 | |
| 142 | 2696 | | | | | 6.1E-06 | | |
| 143 | 2717 | 8.14E-05 | 6.19E-05 | 5.82E-05 | 3.76E-06 | 5.88E-06 | -7.84E-06 | 1.44E-06 |
| 144 | 2769 | 6.24E-05 | 4.96E-05 | | | | | |
| 145 | 2803 | | | 5.32E-05 | 2.54E-06 | 4.96E-06 | -8.31E-06 | |
| 146 | 2822 | 6.71E-05 | 5.54E-05 | 6.03E-05 | | | | 1.43E-06 |
| 147 | 2831 | | | | 8.56E-06 | | | |
| 148 | 2856 | 5.84E-05 | 4.65E-05 | 5.10E-05 | 5.55E-06 | 5.37E-06 | -9.21E-06 | |
| 149 | 2929 | 5.47E-05 | 4.34E-05 | 4.61E-05 | 5.06E-06 | 4.94E-06 | -9.63E-06 | 1.46E-06 |
| 150 | 2942 | 5.47E-05 | 4.34E-05 | | | | | |
| 151 | 2964 | | | 4.62E-05 | 5.52E-06 | 5.19E-06 | -1.16E-05 | 1.44E-06 |
| 152 | 2980 | 5.34E-05 | 4.21E-05 | | 4.05E-06 | | -1.15E-05 | 1.39E-06 |
| 153 | 2982 | 5.38E-05 | 4.25E-05 | 4.64E-05 | 5.21E-06 | 6.25E-06 | | |
| 154 | 2984 | | | | | 8.67E-06 | | |
| 155 | 3004 | 3.58E-05 | 2.02E-05 | 3.54E-05 | 4.36E-06 | 8.14E-06 | -9.24E-06 | 1.41E-06 |
| 156 | 3011 | | | | | 8.6E-06 | | |
| 157 | 3061 | 5.82E-05 | 4.54E-05 | 4.68E-05 | 5.38E-06 | 9.76E-06 | -8.1E-06 | |
| 158 | 3085 | | | | | | | 1.73E-06 |
| 159 | 3120 | 5.61E-05 | 4.35E-05 | 4.4E-05 | | | | |
| 160 | 3143 | | | | 6.53E-06 | | | |
| 161 | 3159 | | | | | 1.07E-05 | -7.48E-06 | |
| 162 | 3199 | | | | | | | 1.39E-06 |
| 163 | 3240 | 4.76E-05 | 3.5E-05 | 4.04E-05 | 5.96E-06 | 9.08E-06 | -7.17E-06 | |
| 164 | 3258 | 8.75E-05 | 7.69E-05 | 5.53E-05 | 1.39E-05 | | | |
| 165 | 3341 | 0.000156 | 1.46E-04 | | | 1.13E-05 | | |
| 166 | 3366 | | | 8.28E-06 | | | | 1.38E-06 |
| 167 | 3380 | 0.000119 | 0.000108 | 8.13E-06 | 1.22E-05 | 1.15E-05 | -7.48E-06 | |
| 168 | 3410 | | | | | | | 1.42E-06 |
| 169 | 3428 | | | | 1.27E-05 | | | |
| 170 | 3437 | 0.00012 | 0.000107 | | | | | |
| 171 | 3470 | | | 3.91E-06 | | 1.07E-05 | -7.86E-06 | |
| 172 | 3505 | 1.09E-04 | 9.65E-05 | | 9.92E-06 | 8.96E-06 | | 1.26E-06 |
| 173 | 3516 | | | -1.48E-06 | | | -8.3E-06 | |
| 174 | 3576 | | | | 7.64E-06 | | | |
| 175 | 3578 | 0.000112 | 0.000101 | -4.36E-06 | | 5.83E-06 | -8.07E-06 | 1.07E-06 |
| 176 | 3601 | 0.000111 | 9.99E-05 | -4.81E-06 | 8.81E-06 | 6.47E-06 | -8.14E-06 | 1.07E-06 |
| 177 | 3615 | 0.000105 | 9.29E-05 | -5.92E-06 | | 5.85E-06 | | |
| 178 | 3633 | | | | | | | 1.2E-06 |
| 179 | 3671 | 0.000127 | 0.000116 | | 9.99E-06 | 9.07E-06 | -7.68E-06 | |
| 180 | 3682 | 9.67E-05 | 8.77E-05 | -2.62E-05 | 3.97E-06 | 6.35E-06 | -8.87E-06 | 1.13E-06 |
| 181 | 3701 | | | -2.54E-05 | | | | |
| 182 | 3745 | | | | 8.74E-06 | | | |
| 183 | 3759 | 0.000112 | 0.000102 | -1.37E-05 | | 1.95E-06 | -9.71E-06 | |
| 184 | 3766 | | | -7.45E-06 | | | | 1.11E-06 |
| 185 | 3776 | 1.11E-04 | 1.01E-04 | -1.42E-05 | -6.27E-06 | 1.52E-06 | -9.68E-06 | |
| 186 | 3881 | | 8.79E-06 | | | | | |
| 187 | 3886 | 0.000109 | 8.79E-06 | -1.68E-05 | -6.53E-06 | 1.48E-06 | -9.77E-06 | 1.09E-06 |
| 188 | 3906 | | | | | | -9.97E-06 | |
| 189 | 3910 | | | -3.60E-06 | -2.01E-07 | | | |
| 190 | 3930 | | 2.6E-05 | | | 4.48E-06 | | |
| 191 | 3943 | 0.000126 | | | 1.5E-05 | 2.19E-05 | 2.28E-05 | 9.4E-07 |
| 192 | 3946 | | | | | | | 9.55E-07 |
| 193 | 3997 | 1.21E-04 | 2.53E-05 | 6.47E-06 | -2.38E-05 | -2.32E-05 | -9.59E-06 | 1.43E-06 |
| 194 | 4014 | 9.69E-05 | -5.43E-07 | | | -2.19E-05 | -7.69E-06 | |
| 195 | 4029 | 9.89E-05 | 6.99E-07 | 2.71E-07 | -2.31E-05 | -1.87E-05 | -5.44E-06 | 1.46E-06 |
| 196 | 4043 | | | | | | | 1.45E-06 |
| 197 | 4047 | 0.000125 | 2.67E-05 | 1.04E-06 | -2.79E-05 | -2.03E-05 | 1.42E-06 | |
| 198 | 4071 | 0.00011 | 1.47E-05 | -4.26E-06 | -3.01E-05 | -2.26E-05 | 1.69E-06 | 1.73E-06 |
| 199 | 4088 | | | -3.26E-06 | | | | |
| 200 | 4090 | | | | -2.7E-05 | -2.03E-05 | | |
| 201 | 4129 | 9.35E-05 | -1.61E-06 | | | | 4.60E-06 | |
| 202 | 4133 | 9.04E-05 | -3.91E-06 | -1.04E-06 | -2.81E-05 | -2.14E-05 | 4.42E-06 | 1.68E-06 |
| 203 | 4177 | | | | | -2.12E-05 | | |
| 204 | 4190 | 0.0001 | 3.41E-06 | | | | 4.14E-06 | |

| | A | B | C | D | E | F | G | H |
|---|---|---|---|---|---|---|---|---|
| 205 | 4229 | 9.72E-05 | 1.88E-06 | -2.08E-05 | -3.40E-05 | -2.44E-05 | 4.02E-06 | 1.54E-06 |
| 206 | 4286 | | | -1.74E-05 | -3.16E-05 | -2.23E-05 | | |
| 207 | 4292 | | | | | | 5.00E-06 | |
| 208 | 4324 | 1.70E-04 | 6.82E-05 | | | | | |
| 209 | 4328 | | | -7.37E-06 | | | | |
| 210 | 4338 | 1.50E-04 | 5.34E-05 | -1.71E-05 | -3.40E-05 | -2.41E-05 | 4.48E-06 | 1.73E-06 |
| 211 | 4352 | | | -2.05E-05 | -3.64E-05 | | 3.64E-06 | 1.66E-06 |
| 212 | 4361 | 1.51E-04 | 5.51E-05 | -1.94E-05 | -3.23E-05 | -1.9E-05 | 7.92E-06 | 1.54E-06 |
| 213 | 4386 | | | | -3.11E-05 | -1.82E-05 | 9.98E-06 | |
| 214 | 4405 | 6.86E-05 | -2.97E-05 | -5.29E-05 | -3.11E-05 | -1.82E-05 | 9.81E-06 | |
| 215 | 4413 | 6.87E-05 | -2.94E-05 | | | | 1.01E-05 | 1.44E-06 |
| 216 | 4424 | 7.25E-05 | -2.58E-05 | -5.22E-05 | -3.11E-05 | -1.85E-05 | 9.92E-06 | 1.43E-06 |
| 217 | 4433 | 6.88E-05 | -2.98E-05 | | | | | |
| 218 | 4445 | | | | | | 1.05E-05 | |
| 219 | 4451 | 2.01E-05 | -8.19E-05 | -6.29E-05 | -3.66E-05 | -2.08E-05 | 9.62E-06 | 1.57E-06 |
| 220 | 4513 | 3.60E-05 | -6.21E-05 | | | | | |
| 221 | 4541 | 8.34E-05 | -1.98E-05 | 1.20E-05 | -2.37E-06 | -1.14E-05 | | 1.35E-06 |
| 222 | 4542 | 7.05E-05 | -3.40E-05 | -4.07E-07 | -8.62E-06 | -1.14E-05 | 1.12E-05 | |
| 223 | 4543 | 5.48E-05 | -4.71E-05 | -1.33E-05 | -1.42E-05 | -1.23E-05 | 1.12E-05 | |
| 224 | 4561 | 5.30E-05 | -4.91E-05 | | | -1.22E-05 | 1.13E-05 | 1.34E-06 |
| 225 | 4572 | 5.30E-05 | -4.92E-05 | -2.48E-05 | -1.85E-05 | -1.23E-05 | 1.13E-05 | |
| 226 | 4621 | 5.79E-05 | -4.47E-05 | -2.70E-05 | -2.09E-05 | -1.37E-05 | 8.40E-06 | 1.47E-06 |
| 227 | 4623 | | | -2.79E-05 | | | | |
| 228 | 4624 | 6.06E-05 | -4.46E-05 | -2.44E-05 | -1.83E-05 | -1.18E-05 | 8.41E-06 | 1.41E-06 |
| 229 | 4627 | 5.97E-05 | -4.50E-05 | | -1.77E-05 | | | |
| 230 | 4629 | 5.97E-05 | -4.51E-05 | -2.8E-05 | -1.76E-05 | -1.11E-05 | 8.57E-06 | 1.1E-06 |
| 231 | 4639 | 6.3E-05 | -4.24E-05 | | -1.37E-05 | | | |
| 232 | 4641 | 6.16E-05 | -4.39E-05 | -2.88E-05 | | -1.11E-05 | 7.62E-06 | 9.29E-07 |
| 233 | 4643 | 6.10E-05 | -4.46E-05 | -2.95E-05 | -1.45E-05 | -1.20E-05 | 6.97E-06 | 9.48E-07 |
| 234 | 4644 | 6.90E-05 | -3.78E-05 | | -1.14E-05 | -9.89E-06 | 8.54E-06 | |
| 235 | 4649 | | | -2.86E-05 | | | | 9.98E-07 |
| 236 | 4660 | 1.45E-04 | 4.07E-05 | | | | | |
| 237 | 4679 | 0.000125 | 2.23E-05 | -2.97E-05 | -5.86E-06 | -5.00E-06 | 9.80E-06 | 1E-06 |
| 238 | 4683 | | | | -5.92E-06 | | | |
| 239 | 4690 | 8.06E-05 | -1.75E-05 | -4.38E-05 | | -6.83E-06 | | |
| 240 | 4693 | | | | -9.96E-06 | | 6.39E-06 | 8.98E-07 |
| 241 | 4701 | | | | | -7.37E-06 | | 3.88E-07 |
| 242 | 4710 | | -2.14E-05 | -4.19E-05 | | | | |
| 243 | 4730 | 6.38E-05 | | | -9.48E-06 | | 6.81E-06 | |
| 244 | 4759 | 5.37E-05 | -2.92E-05 | -5.43E-05 | | -9.74E-06 | | |
| 245 | 4764 | | | | -1.66E-05 | -1.21E-05 | 6.15E-06 | -2.83E-06 |
| 246 | 4780 | 4.30E-05 | -3.72E-05 | -5.73E-05 | | | | |
| 247 | 4853 | | | | -2.02E-05 | -1.22E-05 | 4.99E-06 | -2.76E-06 |
| 248 | 4860 | | -3.60E-05 | -5.66E-05 | | | | |
| 249 | 4875 | 4.96E-05 | | | | | | -2.68E-06 |
| 250 | 4878 | | | | -1.77E-05 | | 4.87E-06 | |
| 251 | 4883 | 4.90E-05 | -3.36E-05 | -5.82E-05 | -1.84E-05 | -1.18E-05 | 4.99E-06 | -2.72E-06 |
| 252 | 4939 | 8.40E-06 | -7.77E-05 | -8.40E-05 | -2.64E-05 | -1.62E-05 | 4.58E-06 | -2.7E-06 |
| 253 | 4954 | 3.38E-05 | -4.64E-05 | -6.80E-05 | -1.75E-05 | | | |
| 254 | 5000 | 2.90E-05 | -5.02E-05 | -7E-05 | -1.88E-05 | -1.57E-05 | 4.88E-06 | -2.81E-06 |
| 255 | 5003 | 2.91E-05 | -5.01E-05 | -7.00E-05 | -1.84E-05 | -1.52E-05 | 8.74E-06 | |
| 256 | 5011 | 2.69E-05 | -5.39E-05 | -7.08E-05 | -1.82E-05 | -1.49E-05 | 1.00E-05 | |
| 257 | 5022 | | | -6.45E-05 | | -1.08E-05 | | |
| 258 | 5023 | 3.31E-05 | -4.84E-05 | -6.45E-05 | -1.38E-05 | -1.06E-05 | 1.24E-05 | -2.53E-06 |
| 259 | 5027 | | | -6.84E-05 | | | 1.34E-05 | |
| 260 | 5043 | | | | -1.43E-05 | | | |
| 261 | 5053 | 2.83E-05 | -5.23E-05 | -7.26E-05 | | -1.07E-05 | 1.2E-05 | -2.58E-06 |
| 262 | 5072 | 2.68E-05 | -5.25E-05 | | -1.53E-05 | -1.16E-05 | 1.26E-05 | |
| 263 | 5087 | 2.82E-05 | -5.12E-05 | -6.99E-05 | -1.30E-05 | -9.74E-06 | 1.31E-05 | |
| 264 | 5099 | | | -6.56E-05 | | | | |
| 265 | 5105 | 6.39E-05 | -2.67E-05 | | -1.38E-05 | -1.06E-05 | | |
| 266 | 5131 | 3.16E-05 | -5.15E-05 | -6.99E-05 | -2.08E-05 | -1.61E-05 | 8.14E-06 | -2.68E-06 |
| 267 | 5133 | 3.31E-05 | -4.97E-05 | -6.62E-05 | -2.05E-05 | -1.56E-05 | 8.87E-06 | -2.67E-06 |
| 268 | 5136 | | | | | -1.49E-05 | 8.76E-06 | |
| 269 | 5159 | 2.45E-05 | -5.53E-05 | -6.83E-05 | -2.09E-05 | | | |
| 270 | 5171 | | | | | -1.48E-05 | | -2.08E-06 |
| 271 | 5187 | 4.37E-05 | -4.18E-05 | | | | | |
| 272 | 5232 | | | -9.11E-05 | | | 9.46E-06 | |
| 273 | 5264 | | -0.000117 | | -2.34E-05 | | | |
| 274 | 5277 | -6.34E-05 | | | | -1.59E-05 | | |
| 275 | 5292 | -6.37E-05 | -0.000118 | -9.17E-05 | -2.37E-05 | -1.61E-05 | 9.38E-06 | -1.95E-06 |
| 276 | 5306 | -6.32E-05 | -0.000117 | -9.11E-05 | -2.30E-05 | -1.56E-05 | 9.74E-06 | |
| 277 | 5309 | -6.38E-05 | | | | -1.57E-05 | 9.73E-06 | -2.02E-06 |
| 278 | 5327 | | | -8.82E-05 | | | | |
| 279 | 5332 | | -0.000107 | | -2.01E-05 | | | |
| 280 | 5355 | -7.32E-05 | -0.000117 | -9.95E-05 | -2.18E-05 | -1.57E-05 | 9.76E-06 | -2.22E-06 |
| 281 | 5360 | -7.57E-05 | -0.00012 | -0.0001 | -2.22E-05 | -1.58E-05 | 9.63E-06 | |
| 282 | 5378 | -6.84E-05 | -0.000112 | | -1.7E-05 | | | |
| 283 | 5383 | | | -9.54E-05 | | -1.28E-05 | 1.11E-05 | |
| 284 | 5404 | | | | | | | -2.17E-06 |
| 285 | 5421 | -7.22E-05 | -0.000115 | | -1.76E-05 | -1.37E-05 | | |
| 286 | 5472 | | | -9.5E-05 | | | 1.30E-05 | |
| 287 | 5474 | -7.85E-05 | -0.000117 | -9.52E-05 | -1.94E-05 | -1.46E-05 | 1.27E-05 | |
| 288 | 5495 | -7.50E-05 | -0.000113 | -9.38E-05 | -1.91E-05 | -1.45E-05 | 1.31E-05 | -1.88E-06 |
| 289 | 5506 | -7.38E-05 | -1.12E-04 | -9.35E-05 | | -1.47E-05 | | |
| 290 | 5548 | -8.61E-05 | -0.000119 | | -1.46E-05 | -1.09E-05 | 1.4E-05 | |
| 291 | 5584 | -8.81E-05 | -1.22E-04 | -9.56E-05 | -1.66E-05 | -1.26E-05 | 1.21E-05 | -3.98E-06 |
| 292 | 5594 | -0.000115 | -0.000147 | | -1.82E-05 | -1.34E-05 | 9.66E-06 | -4.31E-06 |
| 293 | 5602 | | | -1.14E-04 | | | | |
| 294 | 5640 | -0.000114 | -0.000146 | -0.000114 | -1.82E-05 | -1.36E-05 | 1.03E-05 | -4.32E-06 |
| 295 | 5652 | -1.14E-04 | -1.46E-04 | -1.14E-04 | -1.78E-05 | -1.33E-05 | 1.05E-05 | -4.31E-06 |
| 296 | 5653 | -0.000116 | -0.000147 | | -2.03E-05 | -1.58E-05 | 8.61E-06 | -4.32E-06 |
| 297 | 5707 | | | | | | | -4.53E-06 |
| 298 | 5802 | -9.31E-05 | -1.24E-04 | -0.000103 | -1.63E-05 | -1.30E-05 | 1.15E-05 | |
| 299 | 5803 | -7.78E-05 | -0.00011 | -0.000101 | -2.03E-05 | -1.87E-05 | 1.08E-05 | -4.63E-06 |
| 300 | 5804 | -5.97E-05 | -8.78E-05 | -7.3E-05 | -1.13E-05 | | | |
| 301 | 5814 | | | | | -2.84E-05 | 7.51E-06 | |
| 302 | 5851 | -4.60E-06 | -5.6E-05 | -9.07E-05 | | | | -4.68E-06 |
| 303 | 5883 | | | | -2.08E-05 | | 6.64E-06 | |
| 304 | 5918 | -9.32E-06 | -5.98E-05 | -9.04E-05 | -2.06E-05 | -3.16E-05 | | |
| 305 | 5919 | | | | | | 7.11E-06 | |
| 306 | 5927 | -1.05E-05 | -6.15E-05 | -9.07E-05 | -2.09E-05 | | | |

| | A | B | C | D | E | F | G | H |
|---|---|---|---|---|---|---|---|---|
| 307 | 5969 | | | | | -3.85E-05 | -3.73E-06 | -5.12E-06 |
| 308 | 5985 | -3.18E-05 | -8.18E-05 | -0.000106 | | | | |
| 309 | 6021 | -3.12E-05 | -8.1E-05 | -0.000105 | -2.51E-05 | -3.83E-05 | -3.23E-06 | |
| 310 | 6028 | -5.39E-05 | -0.000104 | -0.000129 | | | | |
| 311 | 6037 | | | | | | | -5.05E-06 |
| 312 | 6057 | | | | -3.54E-05 | -4.71E-05 | -8.10E-06 | |
| 313 | 6072 | -6.83E-05 | -0.000117 | -0.000138 | -4.16E-05 | | | -4.83E-06 |
| 314 | 6083 | | | | | -5.25E-05 | -1.41E-05 | -6.28E-06 |
| 315 | 6094 | -7.54E-05 | -0.000123 | -0.000139 | -4.35E-05 | -5.35E-05 | -1.48E-05 | -6.27E-06 |
| 316 | 6139 | -0.000114 | -0.000174 | -0.000146 | -4.87E-05 | -5.92E-05 | -2.51E-05 | -5.55E-06 |
| 317 | 6152 | -0.000119 | -0.000179 | -0.000152 | -4.95E-05 | -5.86E-05 | -2.38E-05 | -5.78E-06 |
| 318 | 6156 | -0.000123 | -0.000183 | -0.000154 | -5.17E-05 | -6.10E-05 | -2.47E-05 | -5.69E-06 |
| 319 | 6178 | | | | | | | -5.53E-06 |
| 320 | 6199 | -0.000122 | -0.000183 | -0.000154 | -4.74E-05 | -5.48E-05 | -2.23E-05 | |
| 321 | 6250 | -8.81E-05 | -0.000151 | -0.000109 | -4.39E-05 | -6.15E-05 | -2.48E-05 | -6.16E-06 |
| 322 | 6273 | | | | | | | -6.96E-06 |
| 323 | 6274 | | | -0.00012 | -4.68E-05 | | -2.64E-05 | -6.93E-06 |
| 324 | 6278 | -9.69E-05 | -0.000159 | | | -6.25E-05 | | |
| 325 | 6321 | -9.07E-05 | -0.000153 | | | | | |
| 326 | 6334 | | | -0.000118 | | | | |
| 327 | 6407 | -0.000104 | -0.000165 | | -5.31E-05 | -6.74E-05 | -3.02E-05 | -7.53E-06 |
| 328 | 6484 | -0.000122 | -0.000183 | -0.000133 | -5.47E-05 | -6.84E-05 | -3.00E-05 | -7.42E-06 |
| 329 | 6523 | -0.000136 | -0.000199 | -0.000137 | -5.67E-05 | -6.89E-05 | -3.08E-05 | -7.4E-06 |
| 330 | 6566 | -0.000136 | -0.000199 | -0.000136 | -5.58E-05 | -6.79E-05 | -2.98E-05 | |
| 331 | 6606 | -0.00014 | -0.000202 | -0.000138 | | | | -7.44E-06 |
| 332 | 6613 | | | | -6.07E-05 | | | |
| 333 | 6629 | | | | | -8.71E-05 | -3.1E-05 | -8.32E-06 |
| 334 | 6749 | -0.000111 | -0.000181 | -0.00014 | -6.47E-05 | -8.89E-05 | -3.08E-05 | -8.67E-06 |
| 335 | 6772 | | | | | | | -8.76E-06 |
| 336 | 6790 | -9.84E-05 | -0.000168 | -0.000136 | -6.2E-05 | -8.54E-05 | -2.94E-05 | |
| 337 | 6811 | | | | | | | -8.58E-06 |
| 338 | 6823 | -0.000124 | -0.000188 | -0.000141 | -6.26E-05 | -8.64E-05 | -2.79E-05 | -8.58E-06 |
| 339 | 6876 | -0.000128 | -0.000197 | -0.000141 | -6.17E-05 | -8.57E-05 | -2.73E-05 | -8.68E-06 |
| 340 | 6904 | | | | | | | -8.88E-06 |
| 341 | 6940 | -0.000119 | -0.000187 | -0.00014 | -6.1E-05 | -8.46E-05 | -2.34E-05 | |

| | A | B | C | D | E | F | G | H |
|---|---|---|---|---|---|---|---|---|
| 1 | Rank | Inf | 0.002 | 0.0015 | 0.0013 | 0.001 | 0.0005 | 0.0001 |
| 2 | 10 | -6.99E-06 | 2.63E-05 | 1.59E-05 | -2.04E-04 | -0.000316 | -0.000178 | 4.91E-07 |
| 3 | 61 | 1.07E-05 | 4.63E-05 | 4.13E-05 | -1.82E-04 | -0.000299 | -0.000145 | 4.71E-07 |
| 4 | 115 | 1.73E-05 | 5.28E-05 | 4.75E-05 | -1.77E-04 | -0.000287 | -0.000126 | -2.91E-06 |
| 5 | 144 | 3.24E-05 | 6.75E-05 | 6.96E-05 | -1.57E-04 | -0.00027 | -0.00011 | -2.91E-06 |
| 6 | 145 | 3.55E-05 | 6.99E-05 | 7.16E-05 | -1.54E-04 | -0.000267 | -9.86E-05 | |
| 7 | 197 | | | | | | | -2.11E-06 |
| 8 | 213 | 1.05E-05 | 4.40E-05 | 4.08E-05 | -1.69E-04 | -0.000282 | -0.000117 | -1.65E-06 |
| 9 | 233 | 1.66E-05 | 5.03E-05 | 4.75E-05 | -1.65E-04 | -0.000279 | -0.000112 | |
| 10 | 234 | | | | | | -0.000124 | -1.71E-06 |
| 11 | 239 | 1.74E-05 | 5.12E-05 | 4.75E-05 | -1.66E-04 | -0.000279 | | |
| 12 | 259 | 1.23E-05 | 4.57E-05 | 4.41E-05 | -1.53E-04 | -0.000271 | -0.000112 | |
| 13 | 261 | 1.23E-05 | 4.60E-05 | 4.41E-05 | -1.53E-04 | -0.000271 | -0.000111 | -2.46E-06 |
| 14 | 276 | 1.32E-05 | 4.67E-05 | 4.50E-05 | -1.53E-04 | -0.000271 | -0.000109 | -2.52E-06 |
| 15 | 284 | 1.39E-05 | 4.69E-05 | 4.52E-05 | -1.52E-04 | -0.000269 | -0.000109 | -2.52E-06 |
| 16 | 334 | 1.56E-05 | 4.83E-05 | 4.79E-05 | -1.47E-04 | -0.000263 | -0.000107 | -1.83E-06 |
| 17 | 344 | 3.53E-05 | 6.78E-05 | | | | | -1.99E-06 |
| 18 | 352 | 2.88E-05 | 6.11E-05 | 3.13E-05 | -1.53E-04 | -0.000261 | -0.000104 | -2.04E-06 |
| 19 | 357 | | | | | -0.000265 | -0.000115 | |
| 20 | 366 | 2.83E-05 | 6.10E-05 | 3.06E-05 | -1.51E-04 | -0.00026 | -0.000118 | -2.66E-06 |
| 21 | 375 | | | | -0.000145 | -0.000251 | | -3.05E-06 |
| 22 | 378 | 2.36E-05 | 5.61E-05 | 2.62E-05 | | | -0.000125 | |
| 23 | 380 | | | | -0.000149 | -0.000256 | -0.000124 | |
| 24 | 390 | 2.96E-05 | 6.22E-05 | 3.42E-05 | | | | -1.12E-06 |
| 25 | 408 | 2.94E-05 | 6.21E-05 | 3.42E-05 | -1.47E-04 | -0.000253 | -0.000122 | -6.45E-07 |
| 26 | 416 | | | 3.40E-05 | -1.47E-04 | | -0.000122 | -8.14E-07 |
| 27 | 432 | 2.25E-05 | 5.57E-05 | | | -0.000276 | | |
| 28 | 434 | | | | | | -0.000137 | -2.08E-06 |
| 29 | 487 | 2.44E-05 | 5.81E-05 | 3.44E-05 | -1.47E-04 | -0.000271 | -0.000131 | |
| 30 | 496 | | | | -0.000147 | | | -1.58E-06 |
| 31 | 513 | 2.53E-05 | 5.91E-05 | 3.57E-05 | | -0.00027 | -0.000127 | |
| 32 | 526 | | | | -1.38E-04 | | | |
| 33 | 530 | 2.74E-05 | 6.11E-05 | 3.85E-05 | -1.38E-04 | -0.000263 | -0.000125 | |
| 34 | 540 | 2.17E-05 | 5.52E-05 | 2.99E-05 | -0.000153 | -0.000279 | -0.000128 | 1.64E-06 |
| 35 | 544 | | 5.52E-05 | | -0.000153 | | | |
| 36 | 551 | 2.07E-05 | | | | | -0.00013 | |
| 37 | 556 | | | 2.82E-05 | | -0.000286 | | 1.39E-06 |
| 38 | 557 | 2.11E-05 | 5.52E-05 | | -1.56E-04 | | -0.000129 | 1.61E-06 |
| 39 | 573 | 2.33E-05 | 5.75E-05 | 3.17E-05 | -1.53E-04 | -0.00028 | -0.000118 | |
| 40 | 582 | | | 3.27E-05 | | -0.000278 | -0.000117 | |
| 41 | 592 | 2.34E-05 | 5.77E-05 | 3.17E-05 | -1.58E-04 | -0.000283 | -0.00012 | |
| 42 | 594 | 3.01E-05 | 6.50E-05 | 4.47E-05 | -1.32E-04 | -0.000256 | -0.000114 | 2.88E-06 |
| 43 | 663 | | | | -1.43E-04 | | | |
| 44 | 716 | 2.52E-05 | 5.99E-05 | 3.99E-05 | -0.000144 | -0.000264 | -0.000117 | |
| 45 | 748 | | 5.98E-05 | 4.04E-05 | -1.43E-04 | -0.000264 | -0.000121 | 1.61E-06 |
| 46 | 760 | 2.51E-05 | | | | | | |
| 47 | 772 | 2.54E-05 | 6.00E-05 | 4.08E-05 | -1.42E-04 | -0.000263 | -0.00012 | |
| 48 | 782 | 2.57E-05 | 6.03E-05 | 4.1E-05 | -0.000141 | -0.000263 | -0.00012 | 1.59E-07 |
| 49 | 813 | 2.78E-05 | 6.29E-05 | 4.23E-05 | -1.36E-04 | -0.000258 | -0.000118 | -6.57E-07 |
| 50 | 818 | 2.86E-05 | | | | | | |
| 51 | 826 | 2.86E-05 | 6.33E-05 | 4.24E-05 | -1.37E-04 | -0.000258 | -0.00012 | |
| 52 | 844 | | 6.21E-05 | | | | | |
| 53 | 885 | 3.13E-05 | | 4.69E-05 | | -0.000246 | -0.000113 | -2.13E-06 |
| 54 | 917 | 3.11E-05 | 6.64E-05 | | -1.30E-04 | | | |
| 55 | 931 | 2.69E-05 | 6.03E-05 | 4.57E-05 | -0.00013 | -0.000239 | -0.000109 | -2.29E-06 |
| 56 | 933 | 2.20E-05 | 5.08E-05 | 3.89E-05 | -0.000138 | -0.000244 | -0.000109 | -2.35E-06 |
| 57 | 962 | | | | | | -0.00011 | |
| 58 | 991 | 1.98E-05 | 4.75E-05 | | | | | |
| 59 | 997 | | | 3.25E-05 | -0.000144 | -0.00025 | -0.000112 | -4.91E-06 |
| 60 | 1043 | 1.62E-05 | 4.36E-05 | 2.70E-05 | -1.52E-04 | -0.000258 | -0.000113 | |
| 61 | 1063 | 1.62E-05 | 4.36E-05 | | | | | -4.02E-06 |
| 62 | 1093 | | 4.1E-05 | 2.49E-05 | -1.55E-04 | -0.00026 | -0.000118 | -4.04E-06 |
| 63 | 1094 | 1.35E-05 | | | | | | |
| 64 | 1107 | | | 2.52E-05 | -1.54E-04 | -0.000258 | -0.000114 | |
| 65 | 1114 | | | | | | | -4.46E-06 |
| 66 | 1165 | 1.48E-05 | 4.25E-05 | 2.79E-05 | -1.49E-04 | -0.000253 | -0.00011 | -4.63E-06 |
| 67 | 1206 | 1.54E-05 | 4.26E-05 | | | | -0.00011 | |
| 68 | 1211 | 1.46E-05 | 4.18E-05 | 2.44E-05 | -0.000152 | -0.000251 | -0.00011 | -4.59E-06 |
| 69 | 1250 | 1.61E-05 | 4.30E-05 | 2.72E-05 | -1.49E-04 | -0.000244 | -1.05E-04 | |
| 70 | 1260 | | | | -1.51E-04 | | | -3.94E-06 |
| 71 | 1262 | 1.20E-05 | 3.93E-05 | 2.88E-05 | -1.48E-04 | -0.00024 | -9.83E-05 | |
| 72 | 1294 | 1.26E-05 | 3.98E-05 | 2.91E-05 | -1.48E-04 | -0.000237 | -9.74E-05 | -3.22E-06 |
| 73 | 1301 | 1.31E-05 | 4.03E-05 | 2.93E-05 | -0.000147 | -0.000235 | -8.02E-05 | |
| 74 | 1306 | 1.3E-05 | 4.02E-05 | 2.92E-05 | -1.47E-04 | -0.000236 | -8.04E-05 | -3.04E-06 |
| 75 | 1316 | 1.28E-05 | | | | -0.000237 | | |
| 76 | 1404 | | | 3.11E-05 | -1.41E-04 | | | |
| 77 | 1414 | 1.12E-06 | 2.68E-05 | 2.43E-05 | -1.55E-04 | -0.000253 | -0.000123 | -3.23E-06 |
| 78 | 1437 | 1.91E-06 | 2.81E-05 | 2.54E-05 | -1.49E-04 | -0.000245 | -9.89E-05 | -2.22E-06 |
| 79 | 1441 | -7.47E-06 | 1.21E-05 | -8.79E-06 | -2.04E-04 | -0.000269 | -0.000104 | |
| 80 | 1448 | -8.69E-06 | 1.21E-05 | -9.5E-06 | -0.000209 | -0.000271 | -0.000102 | -2.64E-06 |
| 81 | 1485 | -1.13E-05 | 9.55E-06 | -1.17E-05 | -2.12E-04 | -0.000274 | -1.06E-04 | -1.93E-06 |
| 82 | 1501 | | | | | -0.000284 | | -8.72E-07 |
| 83 | 1504 | -4.58E-06 | 1.24E-05 | -8.47E-06 | -0.000204 | | -9.97E-05 | |
| 84 | 1535 | 2.45E-06 | 2.00E-05 | 6.35E-06 | -8.81E-05 | -0.000177 | -8.31E-05 | |
| 85 | 1576 | | | | | -0.000172 | -7.63E-05 | -1.59E-07 |
| 86 | 1579 | 1.89E-06 | 2.00E-05 | | -7.54E-05 | | | -6.94E-08 |
| 87 | 1617 | | | 8.90E-06 | | | | |
| 88 | 1654 | -2.61E-06 | 1.36E-05 | -1.51E-06 | -8.59E-05 | -0.000183 | -8.09E-05 | 1.74E-07 |
| 89 | 1655 | 9.92E-08 | 1.63E-05 | 3.07E-06 | -8.09E-05 | -0.000174 | -7.92E-05 | 6.75E-07 |
| 90 | 1667 | 4.31E-07 | 1.66E-05 | 3.29E-06 | -7.94E-05 | -0.000172 | -7.77E-05 | 1.07E-06 |
| 91 | 1683 | | | | | | | -1.83E-06 |
| 92 | 1686 | -3.75E-06 | 1.13E-05 | -1.34E-06 | -7.28E-05 | -0.000157 | -7.37E-05 | -1.71E-06 |
| 93 | 1708 | -4.03E-06 | 1.09E-05 | -1.25E-06 | -7.32E-05 | -0.000156 | -7.44E-05 | -2.01E-06 |
| 94 | 1720 | | | | | -0.000159 | -7.57E-05 | |
| 95 | 1725 | 7.21E-06 | 1.96E-05 | 1.54E-05 | -2.94E-05 | -0.000106 | -5.11E-05 | |
| 96 | 1734 | | | | | -0.00017 | | -2.19E-06 |
| 97 | 1748 | 7.51E-06 | 2.00E-05 | 1.64E-05 | -7.47E-05 | | | |
| 98 | 1757 | | | | | -0.00016 | -4.91E-05 | |
| 99 | 1758 | 4.37E-06 | 1.65E-05 | 1.34E-05 | -8.23E-05 | -0.000161 | -5.10E-05 | -5.4E-06 |
| 100 | 1759 | 6.66E-06 | 1.81E-05 | 1.48E-05 | -7.96E-05 | | | |
| 101 | 1781 | | | | | | -3.89E-05 | |
| 102 | 1787 | 3.3E-06 | 1.59E-05 | 1.3E-05 | -8.32E-05 | -0.000156 | -3.89E-05 | -5.79E-06 |

| | A | B | C | D | E | F | G | H |
|---|---|---|---|---|---|---|---|---|
| 103 | 1791 | 3.02E-06 | 1.59E-05 | 1.23E-05 | -8.7E-05 | -0.000158 | -3.80E-05 | -5.79E-06 |
| 104 | 1851 | | | | | -0.000164 | | -5.57E-06 |
| 105 | 1909 | | | | -9.41E-05 | | -4.24E-05 | -5.43E-06 |
| 106 | 1928 | 2.65E-05 | 4.08E-05 | 2.77E-05 | | | | |
| 107 | 1942 | 9.78E-06 | 2.13E-05 | 6.10E-06 | -8.08E-05 | -0.000145 | -3.76E-05 | |
| 108 | 1956 | | 2.17E-05 | | | -0.000145 | -3.78E-05 | |
| 109 | 1992 | 9.52E-06 | | 7.65E-06 | -7.06E-05 | -0.000134 | -2.77E-05 | |
| 110 | 2018 | 9.32E-06 | 2.25E-05 | 7.72E-06 | -6.97E-05 | -0.000133 | -2.77E-05 | |
| 111 | 2027 | | 2.23E-05 | | | | | |
| 112 | 2042 | 1.13E-05 | | 1.28E-05 | -6.19E-05 | | -2.58E-05 | -7E-06 |
| 113 | 2047 | | | | | -0.000124 | | |
| 114 | 2049 | | 2.24E-05 | | | | | |
| 115 | 2076 | 1.02E-05 | | 1.29E-05 | -6.02E-05 | | -2.11E-05 | |
| 116 | 2080 | 1.08E-05 | 2.36E-05 | 1.45E-05 | -5.45E-05 | -1.03E-04 | -2.07E-05 | |
| 117 | 2085 | 1.07E-05 | 2.34E-05 | 1.43E-05 | -5.46E-05 | -0.000103 | -2.07E-05 | -6.07E-06 |
| 118 | 2096 | 1.13E-05 | | 1.53E-05 | -5.01E-05 | -9.77E-05 | -1.62E-05 | |
| 119 | 2105 | 1.16E-05 | 2.46E-05 | 1.54E-05 | -5.02E-05 | | -1.64E-05 | |
| 120 | 2116 | 1.12E-05 | 2.43E-05 | 1.58E-05 | -4.87E-05 | -9.63E-05 | -1.61E-05 | -5.95E-06 |
| 121 | 2127 | 1.02E-05 | 2.31E-05 | 1.50E-05 | -5.01E-05 | -9.74E-05 | -3.74E-05 | -6.10E-06 |
| 122 | 2148 | 1.20E-05 | 2.55E-05 | 1.99E-05 | -3.01E-05 | | | -5.03E-06 |
| 123 | 2156 | 1.18E-05 | 2.55E-05 | 1.98E-05 | -3.03E-05 | -7.82E-05 | -2.76E-05 | -4.7E-06 |
| 124 | 2175 | 1.17E-05 | 2.57E-05 | 1.96E-05 | -3.40E-05 | | -2.29E-05 | |
| 125 | 2185 | | | | | -7.29E-05 | | |
| 126 | 2226 | 1.8E-05 | 3.12E-05 | 2.73E-05 | -2.37E-05 | | -2.23E-05 | -7.14E-06 |
| 127 | 2244 | 1.85E-05 | 3.27E-05 | 3.13E-05 | 6.70E-06 | -3.71E-05 | 4.50E-06 | -4.97E-06 |
| 128 | 2272 | | | | | -4.43E-05 | | |
| 129 | 2357 | 1.84E-05 | 3.23E-05 | 3.05E-05 | 5.51E-06 | -4.5E-05 | 2.26E-06 | -5.55E-06 |
| 130 | 2365 | 1.95E-05 | 3.35E-05 | 3.12E-05 | 5.89E-06 | | | |
| 131 | 2369 | | | | | | 4.21E-06 | |
| 132 | 2376 | 2.04E-05 | 3.43E-05 | 3.26E-05 | 8.27E-06 | -4.59E-05 | 4.23E-06 | -6.14E-06 |
| 133 | 2394 | | 3.38E-05 | | 6.22E-06 | | 3.57E-06 | |
| 134 | 2415 | 2.34E-05 | 3.63E-05 | 3.85E-05 | 1.49E-05 | -3.23E-05 | 6.28E-06 | |
| 135 | 2446 | | | | | -9.32E-06 | | -6.15E-06 |
| 136 | 2448 | | 3.63E-05 | | | | 7.94E-06 | |
| 137 | 2449 | 2.70E-05 | 3.68E-05 | 4.57E-05 | 4.69E-05 | 3.50E-05 | | -5.94E-06 |
| 138 | 2451 | 2.60E-05 | 3.60E-05 | 4.47E-05 | 4.70E-05 | 3.29E-05 | 9.21E-06 | -5.95E-06 |
| 139 | 2539 | 1.26E-05 | 1.48E-05 | 2.18E-05 | -1.2E-05 | -4.65E-06 | 5.69E-06 | -7.38E-06 |
| 140 | 2570 | 1.34E-05 | 1.55E-05 | 2.55E-05 | -1.18E-06 | | | -7.53E-06 |
| 141 | 2585 | | | | | 5.22E-06 | 1.21E-06 | |
| 142 | 2587 | 1.37E-05 | 1.59E-05 | 2.57E-05 | -9.98E-07 | | | |
| 143 | 2594 | 1.30E-05 | 1.37E-05 | 2.56E-05 | 6.97E-06 | 1.59E-05 | 1.5E-05 | -7.27E-06 |
| 144 | 2696 | | 2.25E-05 | | | | | |
| 145 | 2704 | 1.73E-05 | | 3.35E-05 | 7.06E-06 | | | |
| 146 | 2717 | 1.56E-05 | 2.07E-05 | | | 1.31E-06 | 2.87E-06 | -7.12E-06 |
| 147 | 2769 | | | 3.21E-05 | -1.26E-06 | | | |
| 148 | 2803 | | | | | | -1.04E-06 | |
| 149 | 2831 | 7.75E-06 | 4.82E-06 | 3.18E-05 | 6.44E-06 | 1.87E-05 | 2.16E-05 | |
| 150 | 2856 | 4.88E-06 | 2.27E-06 | 2.88E-05 | 1.64E-07 | | 1.24E-05 | 1.95E-07 |
| 151 | 2929 | | | | | 1.20E-05 | | 1.52E-07 |
| 152 | 2964 | 2.13E-06 | 6.26E-07 | 2.30E-05 | 1.25E-06 | | 1.98E-05 | 7.63E-08 |
| 153 | 2975 | | | | | 2.11E-05 | 2.12E-05 | |
| 154 | 2980 | 3.72E-06 | 3.44E-06 | 2.23E-05 | -5.03E-06 | | | 2.85E-07 |
| 155 | 2982 | 3.66E-06 | 3.46E-06 | 2.21E-05 | -4.62E-06 | 2.08E-05 | 2.03E-05 | 7.51E-07 |
| 156 | 2984 | -9.19E-06 | -9.5E-06 | | | | | |
| 157 | 3004 | -1.10E-05 | -1.04E-05 | 1.58E-05 | -1.28E-05 | 9.04E-06 | 6.82E-06 | 1.79E-06 |
| 158 | 3011 | | | 1.57E-05 | -1.26E-05 | | | 1.62E-06 |
| 159 | 3014 | -1.23E-05 | -1.21E-05 | 1.63E-05 | -9.15E-06 | 1.36E-05 | 8.69E-06 | |
| 160 | 3061 | -1.03E-05 | -9.33E-06 | | | 8.44E-06 | 9.48E-06 | |
| 161 | 3085 | | | 1.97E-05 | -1.56E-05 | | | |
| 162 | 3120 | | -9E-06 | | | | | |
| 163 | 3143 | -7.56E-06 | | | | 6.01E-06 | 7.58E-06 | |
| 164 | 3159 | | | 2.21E-05 | | | | |
| 165 | 3199 | -3.54E-06 | -1.72E-06 | 3.49E-05 | 1.16E-05 | 4.21E-06 | 2.16E-05 | 1.49E-06 |
| 166 | 3214 | | | | 1.16E-05 | | | |
| 167 | 3240 | -1.64E-05 | -1.47E-05 | 1.98E-05 | 3.46E-06 | 3.97E-06 | -1.96E-05 | |
| 168 | 3258 | -1.44E-05 | -1.14E-05 | | | 4.44E-05 | -2.54E-07 | |
| 169 | 3315 | | | 2.21E-05 | 4.76E-06 | | | 1.50E-06 |
| 170 | 3366 | -7.97E-06 | -4.20E-06 | | | | -9.97E-07 | 1.58E-06 |
| 171 | 3370 | | | 1.90E-05 | | 5.23E-05 | | |
| 172 | 3380 | -8.37E-06 | -4.74E-06 | 1.86E-05 | 9.04E-06 | 5.23E-05 | 7.62E-07 | 1.55E-06 |
| 173 | 3410 | -8.92E-06 | -4.56E-06 | | 7.19E-06 | | 4.92E-06 | |
| 174 | 3437 | | | 1.69E-05 | | 5.18E-05 | | |
| 175 | 3470 | -9.05E-06 | -4.87E-06 | | 7.9E-06 | | 4.46E-06 | 1.6E-06 |
| 176 | 3505 | | | 1.93E-05 | 1.37E-05 | 6.28E-05 | | |
| 177 | 3516 | | -4.42E-06 | | | | 5.58E-06 | |
| 178 | 3564 | -1.17E-05 | | | | | | |
| 179 | 3576 | | | 9.22E-06 | | 5.11E-05 | | 4.5E-07 |
| 180 | 3601 | -1.15E-05 | -5.25E-06 | 9.49E-06 | 1.38E-05 | 5.21E-05 | 6.19E-06 | 5.88E-07 |
| 181 | 3615 | | -5.12E-06 | | 1.3E-05 | | 7.12E-06 | |
| 182 | 3633 | -6.58E-06 | | 1.66E-05 | | 7.12E-05 | | |
| 183 | 3671 | -1.26E-05 | -9.15E-06 | 8.6E-06 | 5.22E-06 | 7.34E-05 | -3.36E-06 | |
| 184 | 3682 | | | | | | -8.61E-06 | 5.14E-07 |
| 185 | 3701 | | -9.12E-06 | | 1.68E-06 | | | |
| 186 | 3704 | -1.34E-05 | | | | | | |
| 187 | 3759 | | -3.15E-05 | -2.34E-05 | -3.88E-06 | 6.41E-05 | -1.25E-05 | |
| 188 | 3766 | -3.51E-05 | | | | | | 1.79E-07 |
| 189 | 3776 | | | -2.32E-05 | | 7.08E-05 | -8.35E-06 | |
| 190 | 3790 | | -3.18E-05 | | | | | |
| 191 | 3840 | | | | 5.84E-06 | | | |
| 192 | 3862 | -1.86E-06 | | | | | | |
| 193 | 3886 | -1.78E-06 | -2.99E-05 | -2.02E-05 | 5.57E-06 | 7.20E-05 | -8.13E-06 | -2.12E-07 |
| 194 | 3894 | | | -1.93E-05 | | | | |
| 195 | 3906 | | -3.04E-05 | | | | | |
| 196 | 3910 | -2.3E-05 | | -5.15E-05 | 1.57E-05 | 9.15E-05 | 4.35E-05 | 6.92E-08 |
| 197 | 3943 | -2.43E-05 | -4.09E-05 | | | 9.97E-05 | | |
| 198 | 3946 | -2.43E-05 | -4.08E-05 | -6.32E-05 | 1.04E-05 | 9.99E-05 | 8.51E-05 | 1.14E-06 |
| 199 | 3997 | -4.99E-05 | | -0.000106 | -0.000132 | | -5.92E-05 | 1.04E-06 |
| 200 | 4014 | | -6.25E-05 | | | 1.80E-06 | | 1.35E-06 |
| 201 | 4029 | -4.77E-05 | -6.08E-05 | -0.000102 | -0.000126 | | -3.71E-05 | 1.33E-06 |
| 202 | 4043 | -4.74E-05 | -6.04E-05 | -0.000102 | -0.000121 | 8.20E-06 | -1.51E-05 | 1.42E-06 |
| 203 | 4047 | -7.56E-05 | -9.68E-05 | -0.000137 | -0.000155 | -2.32E-05 | -5.91E-05 | |
| 204 | 4071 | -7.73E-05 | -9.76E-05 | -0.000138 | | -2.71E-05 | -7.34E-05 | 1.91E-06 |

| | A | B | C | D | E | F | G | H |
|---|---|---|---|---|---|---|---|---|
| 205 | 4083 | | | | -9.22E-05 | | | |
| 206 | 4088 | -7.43E-05 | -9.32E-05 | -0.00013 | | | | |
| 207 | 4090 | | | | -9.04E-05 | -6.67E-06 | -4.65E-05 | |
| 208 | 4129 | -8.17E-05 | -0.000106 | -0.000142 | | | -8.99E-05 | |
| 209 | 4133 | -8.15E-05 | -1.06E-04 | -0.000142 | -0.000104 | -1.94E-05 | -9.00E-05 | 2.92E-08 |
| 210 | 4177 | -8.00E-05 | | | | | | |
| 211 | 4190 | | | | | -1.12E-05 | | |
| 212 | 4192 | | -9.31E-05 | -0.000119 | -6.66E-05 | | -6.89E-05 | |
| 213 | 4229 | -8.3E-05 | | | | | | -9.50E-07 |
| 214 | 4286 | | -9.75E-05 | -0.000124 | -8.07E-05 | -1E-05 | -6.92E-05 | |
| 215 | 4292 | -8.6E-05 | | | | | | |
| 216 | 4324 | | | | | -1.62E-05 | | 4.12E-07 |
| 217 | 4328 | | -0.000109 | -0.000137 | | | -7.5E-05 | |
| 218 | 4338 | -0.000127 | | | -0.000102 | -2.86E-05 | -8.82E-05 | -1.44E-07 |
| 219 | 4352 | | -0.000137 | | | | | |
| 220 | 4361 | -0.000124 | -0.000135 | -0.000138 | -9.66E-05 | -2.04E-05 | -7.37E-05 | 4.00E-08 |
| 221 | 4386 | | | | | | | 4E-08 |
| 222 | 4405 | | | | | -3.99E-05 | | 3.44E-08 |
| 223 | 4413 | -0.000125 | -0.000139 | -0.00014 | -9.24E-05 | | -6.56E-05 | |
| 224 | 4424 | | | | | -4.14E-05 | | -1.32E-07 |
| 225 | 4433 | | | | | | | -1.58E-07 |
| 226 | 4445 | | | -0.000142 | -0.000102 | | -7.14E-05 | |
| 227 | 4451 | -0.000133 | -0.000147 | | | | | -1.24E-07 |
| 228 | 4502 | | | | | -3.58E-05 | | |
| 229 | 4513 | -0.000119 | -0.000133 | -0.000125 | -8.14E-05 | | -2.17E-05 | |
| 230 | 4518 | | | | | -3.57E-05 | | |
| 231 | 4541 | | -0.000136 | -1.32E-04 | -8.20E-05 | | -2.4E-05 | |
| 232 | 4542 | -0.000118 | -0.000136 | -0.000132 | -8.12E-05 | -2.83E-05 | -2.41E-05 | 2.47E-06 |
| 233 | 4543 | -0.000119 | | | | -3.07E-05 | -3.5E-05 | 3.03E-06 |
| 234 | 4561 | -1.19E-04 | -0.000137 | -0.000133 | -8.3E-05 | | | |
| 235 | 4572 | -1.24E-04 | -1.40E-04 | -1.38E-04 | -8.89E-05 | -3.42E-05 | -4.88E-05 | |
| 236 | 4581 | -8.97E-05 | -0.000103 | -8.69E-05 | 4.75E-05 | 0.000112 | | 8.70E-07 |
| 237 | 4606 | | | | | 1.04E-04 | | |
| 238 | 4622 | | -9.69E-05 | -7.26E-05 | 5.41E-05 | | | |
| 239 | 4623 | -9.04E-05 | | | | | -3.16E-05 | |
| 240 | 4624 | -9.3E-05 | -9.86E-05 | -7.46E-05 | 5.54E-05 | 1.06E-04 | -2.69E-05 | 6.63E-07 |
| 241 | 4627 | | | | | | | 2.92E-06 |
| 242 | 4629 | -6.53E-05 | -6.85E-05 | -4.66E-05 | 4.06E-05 | 8.11E-05 | -4.3E-05 | 2.64E-06 |
| 243 | 4639 | -6.47E-05 | | | | 8.38E-05 | -3.20E-05 | |
| 244 | 4641 | -6.47E-05 | -6.7E-05 | -4.39E-05 | 4.58E-05 | | -3.27E-05 | |
| 245 | 4643 | | | | | 8.27E-05 | | 2.00E-06 |
| 246 | 4644 | | -6.35E-05 | -3.58E-05 | 6.36E-05 | | -2.08E-05 | |
| 247 | 4649 | -6.27E-05 | | | | | | |
| 248 | 4679 | -5.93E-05 | -5.98E-05 | -2.89E-05 | 7.59E-05 | 0.00012 | 2.19E-06 | 1.40E-06 |
| 249 | 4683 | -5.85E-05 | | | | 0.000121 | 7.23E-06 | |
| 250 | 4690 | | -6.1E-05 | -3.03E-05 | 7.11E-05 | | | 1.37E-06 |
| 251 | 4693 | -6.19E-05 | | | | | | 1.36E-06 |
| 252 | 4701 | | | | | 0.000119 | 1.68E-05 | |
| 253 | 4710 | -4.78E-05 | -4.83E-05 | -1.08E-05 | 0.000103 | 0.000162 | 5.18E-05 | 3.09E-06 |
| 254 | 4730 | -4.96E-05 | -5.07E-05 | -1.46E-05 | 1.01E-04 | 0.000162 | 4.95E-05 | |
| 255 | 4732 | | | -1.45E-05 | | | | |
| 256 | 4759 | -4.94E-05 | -5.02E-05 | | 1.01E-04 | 0.000161 | 4.91E-05 | |
| 257 | 4780 | | | -1.57E-05 | | | | 2.68E-06 |
| 258 | 4821 | -4.15E-05 | -4.39E-05 | | 0.000123 | | 5.14E-05 | |
| 259 | 4853 | | -4.35E-05 | -7.26E-06 | | | | |
| 260 | 4875 | -4.19E-05 | -4.36E-05 | -6.35E-06 | 1.25E-04 | 0.000178 | 5.39E-05 | |
| 261 | 4883 | -4.21E-05 | -4.38E-05 | -6.45E-06 | 0.000125 | 0.000178 | 5.37E-05 | 1.55E-06 |
| 262 | 4930 | | | | 1.24E-04 | 0.000177 | | |
| 263 | 4931 | -3.86E-05 | | | | | | |
| 264 | 4939 | -4.04E-05 | -4.28E-05 | -3.74E-06 | 1.49E-04 | 0.00021 | 6.52E-05 | 6.12E-07 |
| 265 | 4954 | -4.49E-05 | -4.65E-05 | -1.18E-05 | 1.52E-04 | 0.000218 | 7.24E-05 | |
| 266 | 5000 | | | | | | 6.06E-05 | 2.73E-07 |
| 267 | 5003 | -5.77E-05 | -5.53E-05 | -3.39E-05 | 1.35E-04 | 0.00021 | 6.06E-05 | 2.22E-07 |
| 268 | 5011 | -6.00E-05 | -5.82E-05 | -4.03E-05 | 0.000129 | 0.000204 | 5.40E-05 | 8.58E-06 |
| 269 | 5022 | -5.09E-05 | -4.88E-05 | -2.24E-05 | 1.50E-04 | 0.000233 | 8.21E-05 | 9.53E-06 |
| 270 | 5023 | -5.10E-05 | -4.88E-05 | -2.26E-05 | 1.51E-04 | 0.000234 | 8.19E-05 | |
| 271 | 5027 | | -5.15E-05 | -2.84E-05 | | 0.000228 | | |
| 272 | 5043 | | | | 0.000149 | | | |
| 273 | 5053 | -4.61E-05 | -4.79E-05 | -2.39E-05 | 0.000151 | 0.000258 | 9.26E-05 | 7.06E-06 |
| 274 | 5054 | | | | 1.53E-04 | | | |
| 275 | 5072 | -4.79E-05 | | | | | 8.96E-05 | 7.84E-06 |
| 276 | 5087 | | -4.27E-05 | -1.26E-05 | 1.61E-04 | 0.000267 | | |
| 277 | 5088 | -4.45E-05 | -4.12E-05 | -1.10E-05 | 1.65E-04 | 0.000297 | 0.000108 | 6.56E-06 |
| 278 | 5099 | | -3.98E-05 | -1.11E-05 | 1.64E-04 | 0.000293 | | |
| 279 | 5105 | -4.33E-05 | | | | | | |
| 280 | 5131 | | | | | | 9.29E-05 | 4.53E-06 |
| 281 | 5133 | -4.56E-05 | -4.23E-05 | -1.45E-05 | 0.000157 | 0.000286 | 9.54E-05 | 4.43E-06 |
| 282 | 5136 | -4.22E-05 | | | | 1.64E-04 | | 4.29E-06 |
| 283 | 5159 | -2.01E-05 | -2.85E-05 | 1.19E-05 | 0.000228 | 0.00035 | 0.000153 | 3.91E-06 |
| 284 | 5171 | -1.95E-05 | -2.82E-05 | 1.26E-05 | 2.29E-04 | 0.000352 | 0.000159 | |
| 285 | 5190 | -1.1E-05 | -2.03E-05 | 2.62E-05 | 2.48E-04 | 0.000385 | | |
| 286 | 5226 | -1.33E-05 | | | | | 0.000161 | |
| 287 | 5232 | | | -1.33E-06 | 2.08E-04 | 0.000314 | | |
| 288 | 5241 | -1.63E-05 | -2.66E-05 | | | | | |
| 289 | 5292 | -1.47E-05 | -2.52E-05 | 4.51E-07 | 2.10E-04 | 0.000317 | 0.000162 | 4.97E-06 |
| 290 | 5300 | | -3.14E-05 | | | | | |
| 291 | 5306 | -1.56E-05 | -3.16E-05 | -5.92E-07 | 0.000206 | 0.000312 | 0.000161 | 5.05E-06 |
| 292 | 5309 | | | -1.58E-06 | | | 0.00016 | 4.95E-06 |
| 293 | 5327 | | -3.1E-05 | | | | | |
| 294 | 5332 | -4.62E-06 | -3.1E-05 | 1.8E-05 | 0.000232 | 0.000332 | 0.000193 | |
| 295 | 5355 | -7.80E-06 | -3.44E-05 | 1.67E-05 | 2.32E-04 | 0.000331 | 0.000193 | 5.25E-06 |
| 296 | 5360 | -7.6E-06 | -3.44E-05 | 1.68E-05 | 0.000232 | 0.000333 | 0.000193 | 5.21E-06 |
| 297 | 5378 | | | 1.54E-05 | | | 0.000195 | |
| 298 | 5383 | -1.04E-05 | -3.71E-05 | | | 0.000334 | | |
| 299 | 5404 | | -3.67E-05 | 1.58E-05 | 2.34E-04 | | 0.000201 | 5.65E-06 |
| 300 | 5444 | -8.52E-06 | | | | | | |
| 301 | 5472 | | | | | 0.000312 | | |
| 302 | 5474 | -1.2E-05 | -4.89E-05 | 2.28E-06 | 2.31E-04 | 0.000312 | 0.000208 | |
| 303 | 5485 | | | | | | | 5.33E-06 |
| 304 | 5495 | -2.21E-05 | | -8.47E-06 | 0.000215 | 0.000302 | 0.000197 | 4.84E-06 |
| 305 | 5506 | -2.13E-05 | -5.48E-05 | | | 0.000304 | 0.000199 | |
| 306 | 5527 | | | -5.67E-06 | 2.18E-04 | | | |

|     | A    | B         | C         | D         | E         | F         | G         | H         |
| --- | ---- | --------- | --------- | --------- | --------- | --------- | --------- | --------- |
| 307 | 5548 |           | -5.07E-05 |           |           |           |           |           |
| 308 | 5581 | -1.85E-05 |           |           |           |           | 0.000198  |           |
| 309 | 5584 |           | -5.00E-05 | -5.88E-06 | 2.17E-04  | 0.000302  | 0.000195  |           |
| 310 | 5594 | -7.52E-06 | -3.99E-05 | 1.58E-05  | 2.67E-04  | 0.000374  | 0.000222  | -2.19E-06 |
| 311 | 5640 | -8.59E-06 |           |           |           |           |           |           |
| 312 | 5652 | -8.53E-06 | -4.02E-05 | 1.74E-05  | 2.65E-04  | 0.000372  | 0.00022   | -3.43E-06 |
| 313 | 5653 | -9.21E-07 | -3.33E-05 | 2.15E-05  | 0.000269  | 0.00038   | 0.000225  | -2.97E-06 |
| 314 | 5660 |           | -2.94E-05 |           |           |           |           |           |
| 315 | 5707 | 8.59E-06  |           |           |           |           |           |           |
| 316 | 5802 | 5.04E-07  | -3.08E-05 | 1.66E-05  | 2.67E-04  | 0.000371  | 0.000226  | 6.9E-07   |
| 317 | 5803 | 1.84E-06  | -2.98E-05 | 1.71E-05  | 0.000268  | 0.000373  | 0.000228  |           |
| 318 | 5804 | 1.19E-06  | -3.19E-05 |           |           |           |           | 6.51E-07  |
| 319 | 5834 |           | 9.15E-06  | 1.98E-05  | 0.00026   | 0.000367  |           |           |
| 320 | 5851 | 3.87E-05  |           |           |           |           | 0.00022   | -2.35E-06 |
| 321 | 5918 | 1.06E-05  | -1.17E-05 | -4.27E-06 | 2.49E-04  | 0.000366  | 0.000218  |           |
| 322 | 5919 |           | -1.19E-05 |           |           |           |           |           |
| 323 | 5927 |           |           | -4.43E-06 | 2.49E-04  | 0.000365  | 0.000217  |           |
| 324 | 5985 | 1.58E-05  |           |           |           |           |           |           |
| 325 | 6021 | 1.65E-05  | -1.22E-05 | 1.06E-06  | 2.57E-04  | 0.000376  | 0.000224  | 3.92E-06  |
| 326 | 6028 | 2.23E-05  |           |           |           |           |           |           |
| 327 | 6057 | 1.09E-05  | -2.18E-05 | -5.85E-06 | 2.51E-04  | 0.000367  | 0.000221  |           |
| 328 | 6072 | 1.34E-05  | -1.88E-05 | 9.6E-07   | 0.000268  | 0.000383  | 0.000236  | 1.34E-06  |
| 329 | 6083 |           |           |           |           |           |           | 9.71E-07  |
| 330 | 6094 |           | -2.30E-05 | -7.99E-06 |           |           |           | 2.35E-06  |
| 331 | 6103 | 5.29E-06  | -3.17E-05 | -1.39E-05 | 0.000259  | 0.000375  | 0.000231  | 1.94E-06  |
| 332 | 6139 | 3.2E-06   | -3.33E-05 | -1.53E-05 | 2.59E-04  | 0.000379  | 0.000234  |           |
| 333 | 6152 | -1.4E-06  | -3.69E-05 | -1.89E-05 | 0.000256  | 0.000376  | 0.000224  |           |
| 334 | 6156 |           | -3.36E-05 | -1.14E-05 | 2.68E-04  | 0.000395  | 0.000241  | 1.14E-05  |
| 335 | 6178 | 2.49E-06  |           |           |           |           |           |           |
| 336 | 6199 | -2.99E-06 | -3.78E-05 | -2.43E-05 | 2.54E-04  | 0.00039   | 0.000241  | 1.17E-05  |
| 337 | 6246 | 0.000211  | 7.07E-05  |           |           |           |           |           |
| 338 | 6250 |           |           |           |           |           |           | -1.3E-06  |
| 339 | 6273 |           |           | -2.15E-05 | 2.58E-04  | 0.000409  |           |           |
| 340 | 6274 |           |           |           |           |           | 0.000237  | -1.61E-06 |
| 341 | 6278 | 0.000197  | 8.05E-05  | -1.22E-05 |           |           |           |           |
| 342 | 6303 |           |           |           | 0.000266  |           |           |           |
| 343 | 6307 |           |           |           |           | 0.000409  |           | -1.71E-06 |
| 344 | 6321 | 0.000191  | 7.11E-05  | 2.67E-06  | 0.000278  | 0.000435  | 0.000251  |           |
| 345 | 6334 |           |           |           |           | 0.000434  | 0.000253  |           |
| 346 | 6407 | 0.000188  | 6.88E-05  | 2.71E-06  | 0.000277  | 0.000436  |           |           |
| 347 | 6484 | 0.000168  | 5.76E-05  | -9.29E-06 | 0.000261  | 0.000421  | 0.000211  | -3.52E-06 |
| 348 | 6503 |           |           |           |           |           | 0.000213  |           |
| 349 | 6523 | 0.000169  | 5.97E-05  | -4.07E-06 | 0.000264  |           |           | -3.93E-06 |
| 350 | 6565 |           |           |           |           | 0.000432  |           |           |
| 351 | 6566 | 0.000168  | 5.84E-05  | -6.02E-06 | 2.62E-04  | 0.000431  | 0.000214  | -1.56E-07 |
| 352 | 6606 |           |           |           |           |           | 0.000214  |           |
| 353 | 6608 | 0.000172  | 6.14E-05  | -1.39E-06 | 0.000264  |           |           |           |
| 354 | 6613 |           |           |           |           | 0.000432  |           | -3.92E-07 |
| 355 | 6731 |           |           |           |           |           | 0.00021   |           |
| 356 | 6741 | 0.000161  | 5.34E-05  | -9.54E-06 | 0.00026   | 0.000433  |           |           |
| 357 | 6772 |           |           |           |           |           |           | -3.45E-07 |
| 358 | 6790 | 0.000125  | 1.79E-05  | -6.17E-05 | 0.000247  | 0.000431  | 0.000206  |           |
| 359 | 6807 | 0.000127  | 1.88E-05  | -6.06E-05 | 0.00025   | 0.000433  | 0.000209  | 1.53E-06  |
| 360 | 6811 |           |           |           |           |           | 0.000212  |           |
| 361 | 6823 |           |           |           |           |           |           | -4.93E-06 |
| 362 | 6876 | 0.000115  | 1.14E-05  | -6.77E-05 | 0.000247  | 0.000432  | 0.000215  |           |
| 363 | 6940 | 0.000128  | 2.25E-05  | -5.86E-05 | 0.000232  | 0.000414  | 0.000191  | -3.08E-06 |

| | A | B | C | D | E | F | G | H |
|---|---|---|---|---|---|---|---|---|
| 1 | Rank | Inf | 0.002 | 0.0015 | 0.0013 | 0.001 | 0.0005 | 0.0001 |
| 2 | 10 | 3.32E-05 | 0.000139 | 0.000133 | 0.000155 | 0.000177 | 0.000106 | -4.53E-06 |
| 3 | 25 | | | | | | | -4.53E-06 |
| 4 | 61 | 3.46E-05 | 0.000138 | 0.000131 | 0.000152 | 0.000172 | 0.000106 | -3.92E-06 |
| 5 | 93 | 3.21E-05 | 0.000137 | 0.000131 | 0.000152 | 0.000172 | 0.000106 | -4.00E-06 |
| 6 | 144 | 3.05E-05 | 0.000134 | 0.000129 | 0.00015 | 0.000171 | 0.000106 | |
| 7 | 145 | 3.02E-05 | 0.000134 | 0.000128 | 0.000149 | 0.00017 | 0.000105 | -3.36E-06 |
| 8 | 181 | 3.05E-05 | 0.000131 | 0.000126 | 0.000147 | 0.000167 | 0.000103 | -3.44E-06 |
| 9 | 194 | 2.22E-05 | 0.00013 | 0.000125 | 0.000146 | 0.000166 | 0.000102 | -3.44E-06 |
| 10 | 197 | 2.32E-05 | 0.000131 | 0.000126 | 0.000147 | 0.000167 | 0.000103 | -3.62E-06 |
| 11 | 213 | 2.02E-05 | 0.000128 | 0.000125 | 0.000145 | 0.000166 | 9.91E-05 | -3.07E-06 |
| 12 | 233 | 2.15E-05 | 0.000128 | 0.000125 | 0.000145 | 0.000166 | 9.80E-05 | -3.12E-06 |
| 13 | 234 | 2.04E-05 | 0.000124 | 0.000116 | 0.000135 | 0.000157 | 9.40E-05 | -1.69E-06 |
| 14 | 245 | 1.26E-05 | 0.000123 | 0.000114 | 0.000133 | 0.000154 | 9.22E-05 | 8.03E-08 |
| 15 | 259 | 9.25E-06 | 0.00012 | 0.000112 | 0.000131 | 0.000153 | 9.15E-05 | -2.62E-07 |
| 16 | 261 | | | 0.000113 | 0.000132 | 0.000153 | 9.18E-05 | |
| 17 | 276 | 9.83E-06 | 0.00012 | 0.000113 | 0.000132 | 0.000154 | 9.21E-05 | -5.64E-08 |
| 18 | 284 | 1.87E-05 | 0.000119 | | | | 9.18E-05 | -1.83E-07 |
| 19 | 327 | | 0.000117 | 0.000113 | 0.000131 | 0.000152 | | |
| 20 | 334 | 1.73E-05 | | 0.000112 | 0.00013 | 0.000151 | 9.05E-05 | -2.11E-07 |
| 21 | 344 | 1.97E-05 | 0.000116 | 0.00011 | 0.000128 | 0.000148 | 8.92E-05 | 1.09E-07 |
| 22 | 351 | | 0.000115 | | | | | -2.76E-08 |
| 23 | 352 | | | | | 0.00015 | 9.01E-05 | |
| 24 | 357 | 2.11E-05 | 0.000116 | 0.000113 | 0.000129 | | | -1.34E-07 |
| 25 | 366 | 1.71E-05 | | 0.000109 | 0.000126 | 0.000146 | 8.75E-05 | |
| 26 | 375 | 1.76E-05 | 0.000106 | 9.72E-05 | 0.000114 | 0.000137 | 7.97E-05 | -7.24E-07 |
| 27 | 378 | 1.75E-05 | | 9.73E-05 | 0.000115 | 0.000137 | 8.00E-05 | -8.79E-07 |
| 28 | 380 | 1.77E-05 | 0.000105 | 9.63E-05 | 0.000113 | 0.000134 | 7.79E-05 | -9.38E-07 |
| 29 | 390 | 1.92E-05 | 0.000104 | 9.57E-05 | 0.000113 | 0.000133 | 7.65E-05 | 5.54E-10 |
| 30 | 408 | 2.24E-05 | 0.00011 | 0.000101 | 0.000118 | 0.000138 | 8.06E-05 | |
| 31 | 416 | 2.17E-05 | 0.000109 | 0.0001 | 0.000117 | | | -1.09E-06 |
| 32 | 432 | 8.34E-06 | | 0.000101 | 0.000117 | 0.000139 | 8.39E-05 | |
| 33 | 434 | 7.23E-06 | 0.000106 | 0.0001 | 0.000117 | 0.000139 | 8.36E-05 | -1.11E-06 |
| 34 | 435 | | 0.000107 | | | | 8.42E-05 | -7.62E-07 |
| 35 | 487 | 1.04E-05 | 0.000107 | 0.000104 | 0.000121 | 0.000141 | 8.51E-05 | |
| 36 | 496 | 1.06E-05 | 0.000107 | 0.000104 | 0.000121 | 0.000141 | 8.50E-05 | -5.09E-07 |
| 37 | 513 | 8.82E-06 | | 0.000103 | 0.00012 | 0.000141 | | |
| 38 | 526 | 9.05E-06 | 0.000109 | 0.000105 | 0.000121 | 0.000143 | 8.55E-05 | -5.94E-07 |
| 39 | 530 | | 0.000109 | | | | | -9.24E-07 |
| 40 | 540 | 5.17E-06 | | 0.000105 | 0.000122 | 0.000143 | 8.73E-05 | |
| 41 | 544 | | 0.000108 | | | | | -1.26E-06 |
| 42 | 551 | 5.14E-06 | | 0.000105 | 0.000122 | 0.000144 | 8.73E-05 | |
| 43 | 556 | 5.72E-06 | 0.00011 | 0.000108 | 0.000125 | 0.000146 | 9.00E-05 | -1.50E-06 |
| 44 | 557 | 5.60E-06 | 0.000109 | 0.000108 | 0.000125 | 0.000146 | 8.95E-05 | -1.48E-06 |
| 45 | 573 | 1.74E-06 | 0.000101 | 9.79E-05 | 0.000115 | 0.000136 | 8.17E-05 | -1.27E-06 |
| 46 | 582 | 3.59E-06 | 9.70E-05 | 9.03E-05 | 0.000104 | 0.000133 | 7.58E-05 | -1.22E-06 |
| 47 | 594 | 3.39E-06 | 9.48E-05 | 8.72E-05 | 0.0001 | 0.000126 | 7.36E-05 | -1.03E-06 |
| 48 | 663 | 5.92E-06 | 9.91E-05 | 9.23E-05 | 0.000105 | 0.000131 | 7.70E-05 | -9.41E-07 |
| 49 | 716 | 6.29E-06 | 9.90E-05 | 9.22E-05 | 0.000105 | 0.000131 | 7.71E-05 | -9.52E-07 |
| 50 | 760 | | | | | | 8.31E-05 | -8.42E-07 |
| 51 | 772 | -1.29E-06 | 9.78E-05 | 9.35E-05 | 0.000106 | 0.000131 | | |
| 52 | 782 | -1.46E-06 | 9.81E-05 | 9.33E-05 | 0.000106 | 0.000131 | 8.35E-05 | 1.36E-06 |
| 53 | 801 | | | | | | 8.34E-05 | 1.40E-06 |
| 54 | 813 | -6.00E-06 | 9.52E-05 | 9.27E-05 | 0.000106 | 0.000132 | | |
| 55 | 818 | -7.52E-06 | 9.38E-05 | | | 0.000131 | 8.16E-05 | |
| 56 | 826 | -9.36E-06 | 9.23E-05 | 9.31E-05 | 0.000106 | 0.00013 | 8.12E-05 | 1.02E-06 |
| 57 | 844 | -1.45E-05 | 8.41E-05 | 8.08E-05 | 9.74E-05 | 0.000118 | 6.96E-05 | 9.53E-07 |
| 58 | 885 | -1.35E-05 | 8.38E-05 | 7.99E-05 | 9.67E-05 | 0.000117 | 6.91E-05 | 1.23E-06 |
| 59 | 917 | -1.44E-05 | 7.88E-05 | 7.65E-05 | 9.58E-05 | 0.000116 | 6.87E-05 | 7.73E-07 |
| 60 | 925 | -1.40E-05 | | 7.55E-05 | 9.44E-05 | 0.000115 | 6.79E-05 | 7.01E-07 |
| 61 | 931 | -1.45E-05 | 8.02E-05 | 7.43E-05 | 9.33E-05 | 0.000113 | | 6.33E-07 |
| 62 | 933 | -1.45E-05 | 7.88E-05 | 7.33E-05 | 9.14E-05 | 0.00011 | 6.78E-05 | 6.27E-07 |
| 63 | 937 | | | | | | 7.17E-05 | |
| 64 | 991 | -1.74E-05 | 7.43E-05 | 6.85E-05 | 8.94E-05 | 0.000109 | | |
| 65 | 997 | -1.74E-05 | 7.42E-05 | 6.84E-05 | 8.93E-05 | 0.000109 | 7.13E-05 | 7.50E-07 |
| 66 | 1043 | | | | | | 7.39E-05 | 7.91E-07 |
| 67 | 1050 | -1.76E-05 | 7.46E-05 | 6.94E-05 | 9.04E-05 | 0.00011 | | 8.50E-07 |
| 68 | 1063 | -2.13E-05 | 7.34E-05 | 6.82E-05 | 8.94E-05 | 0.000109 | 7.35E-05 | 7.84E-07 |
| 69 | 1093 | -2.17E-05 | 7.25E-05 | 6.61E-05 | 8.76E-05 | 0.000107 | 7.26E-05 | 1.02E-06 |
| 70 | 1094 | -3.26E-05 | | 6.57E-05 | 8.74E-05 | | 7.29E-05 | |
| 71 | 1107 | | 7.98E-05 | 7.29E-05 | 9.49E-05 | 0.000113 | | |
| 72 | 1114 | -3.65E-05 | | | | | | 8.79E-07 |
| 73 | 1165 | -3.76E-05 | 7.55E-05 | 7.17E-05 | 9.14E-05 | 0.000109 | 7.16E-05 | 8.49E-07 |
| 74 | 1174 | -3.75E-05 | 7.52E-05 | 7.15E-05 | 9.11E-05 | 0.000109 | | 7.84E-07 |
| 75 | 1206 | | | | | | 7.07E-05 | 9.14E-07 |
| 76 | 1208 | -3.75E-05 | 7.48E-05 | 7.04E-05 | 9.03E-05 | 0.000108 | | |
| 77 | 1211 | | | | | | 7.04E-05 | 9.61E-07 |
| 78 | 1240 | -3.63E-05 | 7.09E-05 | 6.68E-05 | 8.63E-05 | 0.000105 | 6.77E-05 | 8.98E-07 |
| 79 | 1250 | -3.63E-05 | 7.10E-05 | 6.66E-05 | 8.59E-05 | 0.000105 | | |
| 80 | 1260 | -3.73E-05 | 6.70E-05 | 6.16E-05 | 8.20E-05 | 0.0001 | 6.56E-05 | 1.28E-06 |
| 81 | 1262 | -3.94E-05 | 6.79E-05 | 6.46E-05 | 8.48E-05 | 0.000104 | 6.93E-05 | |
| 82 | 1294 | -3.96E-05 | 6.59E-05 | 6.11E-05 | 8.13E-05 | 0.0001 | 6.76E-05 | 1.05E-06 |
| 83 | 1301 | -3.97E-05 | 6.55E-05 | | | | 6.69E-05 | 9.39E-07 |
| 84 | 1306 | -3.98E-05 | 6.50E-05 | 6.02E-05 | 8.06E-05 | 9.90E-05 | | 9.57E-07 |
| 85 | 1316 | | | 6.07E-05 | | 9.94E-05 | 6.67E-05 | |
| 86 | 1357 | -4.04E-05 | 6.15E-05 | 5.65E-05 | 7.73E-05 | 9.55E-05 | 6.49E-05 | 8.93E-07 |
| 87 | 1411 | -4.14E-05 | 6.03E-05 | 5.46E-05 | 7.57E-05 | 9.40E-05 | 6.39E-05 | 8.66E-07 |
| 88 | 1427 | -4.13E-05 | 6.12E-05 | 5.46E-05 | 7.63E-05 | 9.43E-05 | 6.40E-05 | 1.16E-06 |
| 89 | 1437 | -4.15E-05 | 6.11E-05 | 5.45E-05 | 7.62E-05 | 9.48E-05 | 6.42E-05 | 1.11E-06 |
| 90 | 1441 | | | 5.41E-05 | 7.57E-05 | 9.30E-05 | | 1.11E-06 |
| 91 | 1448 | | 5.89E-05 | | | | | |
| 92 | 1485 | -4.17E-05 | | | | 9.13E-05 | 6.27E-05 | 1.08E-06 |
| 93 | 1501 | | | 5.40E-05 | 7.61E-05 | | | |
| 94 | 1535 | | 5.48E-05 | | | | | |
| 95 | 1576 | -4.15E-05 | 5.53E-05 | 5.56E-05 | 7.80E-05 | 9.35E-05 | 6.36E-05 | 1.70E-06 |
| 96 | 1579 | | | | | 9.90E-05 | | |
| 97 | 1617 | | | 5.87E-05 | 8.19E-05 | | | |
| 98 | 1654 | -4.19E-05 | 5.43E-05 | 5.67E-05 | 7.90E-05 | 9.55E-05 | 6.34E-05 | 1.66E-06 |
| 99 | 1655 | -4.24E-05 | 5.52E-05 | | | | 6.36E-05 | |
| 100 | 1667 | -4.28E-05 | 5.31E-05 | 5.48E-05 | 7.58E-05 | 9.27E-05 | 6.37E-05 | 1.67E-06 |
| 101 | 1683 | | 5.19E-05 | 5.39E-05 | 7.49E-05 | 9.19E-05 | 6.39E-05 | |
| 102 | 1686 | -4.65E-05 | 2.30E-05 | 4.04E-05 | 5.72E-05 | 7.25E-05 | 5.06E-05 | 1.75E-06 |

| | A | B | C | D | E | F | G | H |
|---|---|---|---|---|---|---|---|---|
| 103 | 1708 | | 2.39E-05 | 4.27E-05 | 5.98E-05 | 7.64E-05 | | |
| 104 | 1720 | -4.56E-05 | | | | | 5.60E-05 | |
| 105 | 1725 | | | 3.94E-05 | 5.53E-05 | 7.17E-05 | | 1.63E-06 |
| 106 | 1748 | -4.48E-05 | 2.26E-05 | | | | 5.56E-05 | 1.65E-06 |
| 107 | 1757 | | | 3.86E-05 | 5.32E-05 | 6.97E-05 | | |
| 108 | 1758 | -4.61E-05 | 1.98E-05 | 3.79E-05 | 5.28E-05 | 6.90E-05 | 5.39E-05 | 1.56E-06 |
| 109 | 1759 | -4.61E-05 | 1.82E-05 | 3.67E-05 | 5.13E-05 | 6.70E-05 | 5.42E-05 | 1.20E-06 |
| 110 | 1781 | -4.54E-05 | 2.16E-05 | 4.02E-05 | 5.71E-05 | 7.23E-05 | 5.91E-05 | 1.18E-06 |
| 111 | 1787 | | | 4.07E-05 | 5.77E-05 | 7.27E-05 | | |
| 112 | 1791 | -3.90E-05 | 4.64E-05 | 4.36E-05 | 6.08E-05 | 7.62E-05 | 6.43E-05 | 1.18E-06 |
| 113 | 1827 | -4.02E-05 | 3.50E-05 | 3.82E-05 | 5.56E-05 | 7.15E-05 | 6.07E-05 | |
| 114 | 1851 | -4.12E-05 | 3.44E-05 | | | | 6.01E-05 | 1.32E-06 |
| 115 | 1882 | | | 3.88E-05 | 5.70E-05 | 7.20E-05 | | |
| 116 | 1909 | -3.73E-05 | 3.36E-05 | 3.84E-05 | 5.66E-05 | 7.12E-05 | 5.88E-05 | 1.32E-06 |
| 117 | 1942 | | 2.93E-05 | | | | | |
| 118 | 1956 | -3.99E-05 | 3.03E-05 | 4.22E-05 | 6.07E-05 | 7.40E-05 | 5.60E-05 | 1.20E-06 |
| 119 | 1992 | -4.12E-05 | 1.18E-05 | 3.48E-05 | 5.16E-05 | 6.45E-05 | 5.71E-05 | |
| 120 | 2015 | -4.13E-05 | 1.09E-05 | | | | 5.70E-05 | 1.29E-06 |
| 121 | 2018 | -3.77E-05 | 2.22E-05 | 5.26E-05 | 7.08E-05 | 7.85E-05 | 5.54E-05 | 1.21E-06 |
| 122 | 2027 | -5.05E-05 | 5.56E-06 | | | | 5.50E-05 | 1.19E-06 |
| 123 | 2033 | -5.04E-05 | 4.85E-06 | 4.18E-05 | 5.99E-05 | 7.05E-05 | | 9.31E-07 |
| 124 | 2047 | | | 4.18E-05 | 6.02E-05 | 7.09E-05 | 5.71E-05 | |
| 125 | 2049 | -5.11E-05 | 3.27E-06 | 4.11E-05 | 5.84E-05 | 6.87E-05 | 5.68E-05 | 9.19E-07 |
| 126 | 2076 | | 4.91E-08 | | | | | |
| 127 | 2080 | -5.32E-05 | -5.79E-07 | 2.73E-05 | 4.49E-05 | 5.36E-05 | 5.80E-05 | 1.01E-06 |
| 128 | 2085 | -5.30E-05 | -1.50E-06 | 2.60E-05 | 4.38E-05 | 5.26E-05 | 5.74E-05 | 1.01E-06 |
| 129 | 2105 | -5.07E-05 | -1.56E-06 | 2.68E-05 | 4.44E-05 | 5.31E-05 | 5.79E-05 | 1.09E-06 |
| 130 | 2115 | | -3.28E-06 | | | | | |
| 131 | 2116 | -5.11E-05 | -3.30E-06 | 2.36E-05 | 4.13E-05 | 4.94E-05 | 5.57E-05 | 1.09E-06 |
| 132 | 2127 | | | 2.47E-05 | 4.26E-05 | 5.11E-05 | | |
| 133 | 2128 | -5.13E-05 | -6.83E-06 | 2.03E-05 | 3.77E-05 | 4.24E-05 | 5.06E-05 | 1.37E-06 |
| 134 | 2148 | -4.10E-05 | 2.17E-05 | 2.46E-05 | 3.98E-05 | 4.23E-05 | 6.10E-05 | 1.36E-06 |
| 135 | 2156 | -5.15E-05 | 1.30E-05 | | | | 6.08E-05 | |
| 136 | 2175 | | | 2.62E-05 | 3.96E-05 | 4.17E-05 | | |
| 137 | 2226 | -5.27E-05 | -9.80E-06 | 2.46E-05 | 3.81E-05 | 3.96E-05 | 6.02E-05 | 1.38E-06 |
| 138 | 2244 | | | 2.19E-05 | 3.37E-05 | 3.48E-05 | | |
| 139 | 2261 | -5.36E-05 | -1.60E-05 | 2.07E-05 | 3.19E-05 | 3.24E-05 | 5.70E-05 | 2.01E-06 |
| 140 | 2272 | -5.37E-05 | | 1.92E-05 | 3.05E-05 | 3.10E-05 | 5.63E-05 | |
| 141 | 2351 | -5.29E-05 | -1.40E-05 | 2.04E-05 | | 3.39E-05 | 5.90E-05 | 1.99E-06 |
| 142 | 2357 | -5.35E-05 | -1.56E-05 | 1.83E-05 | 2.86E-05 | 3.15E-05 | 5.53E-05 | 1.94E-06 |
| 143 | 2365 | -5.49E-05 | -1.98E-05 | 1.50E-05 | 2.52E-05 | | 5.08E-05 | 2.11E-06 |
| 144 | 2369 | | | 1.49E-05 | 2.44E-05 | 2.91E-05 | | |
| 145 | 2376 | -5.40E-05 | -7.39E-06 | | | | | 2.24E-06 |
| 146 | 2384 | | | 1.94E-05 | 2.85E-05 | 3.14E-05 | 5.28E-05 | |
| 147 | 2388 | -5.41E-05 | -1.55E-05 | | | | | |
| 148 | 2415 | | | 1.83E-05 | 2.67E-05 | 2.87E-05 | 5.34E-05 | 2.50E-06 |
| 149 | 2433 | | -2.60E-05 | | | | | |
| 150 | 2446 | -5.73E-05 | -2.79E-05 | 1.73E-05 | 2.52E-05 | 2.69E-05 | 5.27E-05 | 2.46E-06 |
| 151 | 2448 | | -2.76E-05 | 1.76E-05 | 2.56E-05 | 2.74E-05 | | |
| 152 | 2449 | -5.69E-05 | -2.88E-05 | 1.62E-05 | 2.45E-05 | 2.58E-05 | 5.09E-05 | |
| 153 | 2451 | -5.52E-05 | -2.78E-05 | 1.86E-05 | 2.65E-05 | 2.72E-05 | 5.57E-05 | 2.22E-06 |
| 154 | 2494 | | | 1.85E-05 | 2.63E-05 | 2.70E-05 | | |
| 155 | 2552 | -5.41E-05 | -2.57E-05 | 2.01E-05 | 2.80E-05 | 2.85E-05 | 5.80E-05 | |
| 156 | 2585 | | -2.55E-05 | | | | 5.68E-05 | 2.22E-06 |
| 157 | 2587 | -5.40E-05 | | | | 2.73E-05 | | |
| 158 | 2594 | | | 1.81E-05 | 2.59E-05 | | | |
| 159 | 2665 | | -2.15E-05 | | | | 5.56E-05 | |
| 160 | 2677 | -5.58E-05 | | 1.52E-05 | 2.32E-05 | 2.55E-05 | | |
| 161 | 2678 | | -1.68E-05 | | | | | 2.09E-06 |
| 162 | 2696 | | | | | | 5.84E-05 | |
| 163 | 2704 | -5.57E-05 | -2.40E-05 | 1.10E-05 | 1.96E-05 | 2.17E-05 | 5.71E-05 | 2.02E-06 |
| 164 | 2769 | -5.55E-05 | | 8.10E-06 | 1.62E-05 | 1.86E-05 | 5.73E-05 | |
| 165 | 2808 | -5.54E-05 | -2.42E-05 | 8.05E-06 | 1.62E-05 | 1.86E-05 | 5.74E-05 | |
| 166 | 2822 | | | | | | | 2.17E-06 |
| 167 | 2831 | -5.57E-05 | -2.98E-05 | 5.43E-06 | 1.29E-05 | 1.59E-05 | 5.50E-05 | 2.27E-06 |
| 168 | 2856 | -4.31E-05 | 1.79E-05 | | | | 5.56E-05 | 2.19E-06 |
| 169 | 2929 | | | 6.92E-06 | 1.46E-05 | 2.10E-05 | | |
| 170 | 2942 | -4.54E-05 | -1.91E-05 | 6.87E-06 | 1.45E-05 | 2.09E-05 | 5.54E-05 | 1.62E-06 |
| 171 | 2950 | -4.63E-05 | -2.77E-05 | 7.53E-07 | 8.45E-06 | 1.37E-05 | 4.76E-05 | 1.63E-06 |
| 172 | 2964 | | | | | | 4.92E-05 | 1.59E-06 |
| 173 | 2980 | -4.78E-05 | | 1.26E-06 | 8.87E-06 | 1.38E-05 | | |
| 174 | 2982 | -4.92E-05 | -3.51E-05 | -3.74E-06 | 3.45E-06 | 7.73E-06 | 4.28E-05 | 9.33E-07 |
| 175 | 2984 | | -1.51E-05 | | | | | |
| 176 | 3004 | | | | | | 5.77E-05 | |
| 177 | 3011 | -4.55E-05 | | 2.55E-07 | 6.58E-06 | 1.23E-05 | | |
| 178 | 3014 | | | | | | | 8.68E-07 |
| 179 | 3040 | -4.47E-05 | -2.08E-05 | -2.90E-06 | 3.48E-06 | 8.80E-06 | 5.48E-05 | |
| 180 | 3061 | | -2.30E-05 | | | | 5.50E-05 | 6.39E-07 |
| 181 | 3071 | -4.66E-05 | | -6.31E-06 | 6.41E-07 | 5.16E-06 | | |
| 182 | 3085 | | -3.05E-05 | | | | 5.02E-05 | 1.63E-07 |
| 183 | 3120 | -4.73E-05 | | -8.18E-06 | -2.18E-06 | 2.05E-06 | | |
| 184 | 3143 | | -3.43E-05 | | | | 4.90E-05 | 2.47E-08 |
| 185 | 3159 | -4.74E-05 | | -9.87E-06 | -3.85E-06 | 3.26E-07 | | |
| 186 | 3214 | | -3.13E-05 | | | | 4.94E-05 | |
| 187 | 3225 | -4.73E-05 | | -1.16E-05 | -6.79E-06 | -3.34E-06 | | -1.19E-07 |
| 188 | 3232 | | -3.65E-05 | | | | 4.97E-05 | |
| 189 | 3240 | -4.71E-05 | | -1.07E-05 | -5.71E-06 | -2.18E-06 | | |
| 190 | 3258 | | 5.15E-05 | | | | 4.41E-05 | |
| 191 | 3278 | -2.40E-05 | 5.07E-05 | -1.39E-05 | -8.79E-06 | -5.53E-06 | 4.31E-05 | -1.19E-07 |
| 192 | 3315 | -7.37E-05 | | -1.45E-05 | -9.42E-06 | -6.83E-06 | | -4.94E-08 |
| 193 | 3320 | | 4.88E-06 | | | | 4.37E-05 | |
| 194 | 3351 | | | | | | | 3.66E-07 |
| 195 | 3366 | -7.18E-05 | 5.50E-06 | -1.28E-05 | -7.95E-06 | -5.67E-06 | 4.25E-05 | 3.65E-07 |
| 196 | 3370 | -7.07E-05 | | -1.07E-05 | -5.35E-06 | -3.11E-06 | | |
| 197 | 3380 | -7.07E-05 | 6.11E-06 | -1.09E-05 | -5.61E-06 | -3.46E-06 | 4.28E-05 | 3.29E-07 |
| 198 | 3384 | -6.79E-05 | | -1.12E-05 | -6.11E-06 | -3.98E-06 | | |
| 199 | 3386 | -6.88E-05 | -4.62E-06 | -1.13E-05 | -6.17E-06 | -4.09E-06 | 4.21E-05 | -6.13E-07 |
| 200 | 3428 | | | | | | 4.09E-05 | |
| 201 | 3437 | | -9.04E-06 | | | | | |
| 202 | 3470 | -6.88E-05 | -9.72E-06 | -1.10E-05 | -5.84E-06 | -4.28E-06 | 4.02E-05 | -6.61E-07 |
| 203 | 3505 | | -2.10E-05 | | | | 3.92E-05 | -6.06E-07 |
| 204 | 3516 | -6.98E-05 | | -1.10E-05 | -5.92E-06 | -6.25E-06 | | |

|     | A    | B         | C         | D         | E         | F         | G         | H         |
| --- | ---- | --------- | --------- | --------- | --------- | --------- | --------- | --------- |
| 205 | 3518 |           | -2.27E-05 |           |           |           | 3.91E-05  | -8.01E-07 |
| 206 | 3519 | -6.98E-05 |           | -1.08E-05 | -5.64E-06 | -6.02E-06 |           | -7.84E-07 |
| 207 | 3564 |           | -2.40E-05 |           |           |           | 4.06E-05  |           |
| 208 | 3578 | -7.08E-05 |           | -1.25E-05 | -7.39E-06 | -8.53E-06 |           | -9.79E-07 |
| 209 | 3593 |           | -3.42E-05 |           |           |           | 3.71E-05  |           |
| 210 | 3601 | -7.11E-05 |           | -1.25E-05 | -7.27E-06 | -1.02E-05 |           |           |
| 211 | 3633 | -7.22E-05 | -5.16E-05 | -2.24E-05 | -2.03E-05 | -2.45E-05 | 2.91E-05  | -1.59E-06 |
| 212 | 3671 |           | -5.00E-05 |           |           |           | 2.89E-05  |           |
| 213 | 3685 | -7.10E-05 |           | -2.20E-05 | -1.98E-05 | -2.38E-05 |           | -1.76E-06 |
| 214 | 3704 |           | -5.28E-05 |           |           |           | 2.47E-05  |           |
| 215 | 3745 | -6.98E-05 |           | -1.59E-05 | -1.46E-05 | -2.06E-05 |           |           |
| 216 | 3766 |           |           |           |           |           |           | -1.92E-06 |
| 217 | 3776 | -6.69E-05 | -1.14E-05 | -1.16E-05 | -1.12E-05 | -1.53E-05 | 2.20E-05  |           |
| 218 | 3790 |           | -7.03E-06 |           |           |           | 2.00E-05  | -2.10E-06 |
| 219 | 3827 | -6.74E-05 |           | -1.42E-05 | -1.29E-05 | -1.73E-05 |           |           |
| 220 | 3848 |           | -8.59E-06 |           |           |           |           |           |
| 221 | 3862 | -7.17E-05 |           | -5.70E-05 | -6.72E-05 | -7.33E-05 | 4.99E-06  | -2.47E-06 |
| 222 | 3886 | -7.17E-05 | -2.40E-05 | -5.71E-05 | -6.71E-05 | -7.37E-05 | 5.00E-06  | -2.47E-06 |
| 223 | 3894 | -7.35E-05 | -3.61E-05 |           |           |           |           | -2.47E-06 |
| 224 | 3895 |           |           |           |           |           | 4.58E-06  |           |
| 225 | 3901 | -7.45E-05 |           | -6.01E-05 | -7.05E-05 | -7.88E-05 |           | -2.70E-06 |
| 226 | 3906 |           |           |           |           |           |           | -2.41E-06 |
| 227 | 3910 |           | -3.32E-05 |           |           |           | 1.01E-05  |           |
| 228 | 3925 | -7.33E-05 |           | -5.23E-05 | -6.02E-05 | -6.51E-05 |           |           |
| 229 | 3939 | -7.35E-05 | -3.77E-05 | -5.34E-05 | -6.15E-05 | -6.68E-05 | 6.44E-06  | -2.47E-06 |
| 230 | 3943 |           | -3.80E-05 |           |           |           | 6.40E-06  |           |
| 231 | 3946 | -7.33E-05 |           | -4.81E-05 | -5.40E-05 | -5.90E-05 |           | -2.47E-06 |
| 232 | 4014 |           | -3.99E-05 |           |           |           | 8.44E-06  |           |
| 233 | 4029 |           |           | -4.95E-05 | -5.86E-05 | -6.28E-05 |           |           |
| 234 | 4043 | -7.06E-05 |           |           |           |           |           |           |
| 235 | 4047 | -7.10E-05 | -4.54E-05 | -5.28E-05 | -6.27E-05 | -6.72E-05 | 5.80E-06  | -2.62E-06 |
| 236 | 4052 | -7.14E-05 | -5.40E-05 | -5.76E-05 | -7.00E-05 | -7.39E-05 | 1.02E-06  | -2.49E-06 |
| 237 | 4071 | -7.18E-05 | -5.38E-05 | -5.78E-05 | -7.04E-05 | -7.43E-05 | 5.73E-07  | -2.52E-06 |
| 238 | 4083 |           |           |           |           |           |           | -2.14E-06 |
| 239 | 4088 | -7.07E-05 | -5.48E-05 | -5.80E-05 | -7.07E-05 | -7.49E-05 | 3.29E-07  | -2.17E-06 |
| 240 | 4090 | -7.09E-05 |           | -5.80E-05 | -7.08E-05 | -7.48E-05 |           | -2.14E-06 |
| 241 | 4129 |           |           |           |           |           |           | -2.36E-06 |
| 242 | 4133 | -6.84E-05 | -7.46E-05 | -7.28E-05 | -8.80E-05 | -8.87E-05 | -6.22E-06 | -2.62E-06 |
| 243 | 4155 |           |           |           |           |           |           | -2.60E-06 |
| 244 | 4177 | -6.86E-05 | -7.50E-05 | -7.15E-05 | -8.71E-05 | -8.78E-05 | -5.44E-06 |           |
| 245 | 4192 |           |           |           |           |           | -8.66E-06 | -3.07E-06 |
| 246 | 4200 |           | -7.97E-05 |           |           |           |           |           |
| 247 | 4229 | -7.11E-05 |           | -7.17E-05 | -8.71E-05 | -8.55E-05 |           |           |
| 248 | 4240 | -7.11E-05 | -7.95E-05 | -7.21E-05 | -8.77E-05 | -8.65E-05 | -9.45E-06 | -1.69E-06 |
| 249 | 4273 |           |           |           |           |           | -1.00E-05 | -1.76E-06 |
| 250 | 4286 | -6.81E-05 | -6.30E-05 | -6.25E-05 | -7.74E-05 | -7.42E-05 |           |           |
| 251 | 4302 | -6.81E-05 | -6.50E-05 | -6.30E-05 | -7.80E-05 | -7.46E-05 | -9.64E-06 | -1.75E-06 |
| 252 | 4324 |           |           |           |           |           | -1.04E-05 |           |
| 253 | 4340 | -6.72E-05 | -6.52E-05 | -6.38E-05 | -7.91E-05 | -7.60E-05 | -1.16E-05 | -2.79E-07 |
| 254 | 4352 | -6.70E-05 | -6.85E-05 | -6.44E-05 | -8.00E-05 | -7.69E-05 | -1.20E-05 | -4.16E-07 |
| 255 | 4360 | -6.88E-05 | -9.17E-05 | -7.03E-05 | -8.77E-05 | -8.58E-05 | -2.16E-05 | -5.75E-07 |
| 256 | 4361 | -7.01E-05 | -0.000103 | -7.06E-05 | -8.78E-05 | -8.63E-05 |           | -3.00E-07 |
| 257 | 4405 |           | -0.000116 | -7.50E-05 | -9.27E-05 | -9.05E-05 | -2.43E-05 |           |
| 258 | 4413 | -6.73E-05 |           |           |           |           |           |           |
| 259 | 4424 | -2.95E-05 | -5.70E-05 | -7.92E-05 | -9.73E-05 | -9.57E-05 | -3.06E-05 | -3.06E-07 |
| 260 | 4445 |           |           |           |           |           | -3.62E-05 |           |
| 261 | 4501 |           | -0.000115 |           |           |           |           |           |
| 262 | 4502 | -5.60E-05 |           | -8.88E-05 | -0.00011  | -0.000112 |           |           |
| 263 | 4513 |           | -0.000127 |           |           |           | -3.96E-05 |           |
| 264 | 4518 | -5.62E-05 | -0.000134 | -9.03E-05 | -0.000111 | -0.000114 | -4.14E-05 | 2.43E-06  |
| 265 | 4542 |           | -0.000133 |           |           |           | -4.32E-05 |           |
| 266 | 4543 | -4.95E-05 |           | -9.38E-05 | -0.000117 | -0.000121 |           |           |
| 267 | 4562 | -3.37E-06 | -8.08E-05 |           |           |           | -4.82E-05 | 3.28E-06  |
| 268 | 4572 |           |           | -9.60E-05 | -0.000119 | -0.000123 |           |           |
| 269 | 4623 | -2.65E-05 | -9.72E-05 |           |           |           | -4.28E-05 |           |
| 270 | 4624 | -2.04E-05 | -9.13E-05 | -0.000101 | -0.000125 | -0.000129 | -4.49E-05 |           |
| 271 | 4629 | -5.33E-05 |           | -0.000107 | -0.000133 | -0.000139 | -4.79E-05 | 3.00E-06  |
| 272 | 4639 |           | -0.000138 |           |           |           |           | 3.08E-06  |
| 273 | 4641 | -6.96E-05 |           | -0.000108 | -0.000134 | -0.00014  | -4.78E-05 |           |
| 274 | 4643 | -6.95E-05 | -0.000139 |           |           |           | -4.92E-05 | 3.31E-06  |
| 275 | 4644 |           |           | -0.000108 | -0.000134 | -0.000139 |           |           |
| 276 | 4648 |           | -0.000141 |           |           |           |           |           |
| 277 | 4649 | -7.04E-05 | -0.000146 | -0.000108 | -0.000135 | -0.000139 | -4.93E-05 | 2.89E-06  |
| 278 | 4679 | -8.13E-06 |           |           |           |           | -4.62E-05 |           |
| 279 | 4683 |           |           | -0.000113 | -0.00014  | -0.000144 |           |           |
| 280 | 4686 |           | -4.89E-05 |           |           |           |           |           |
| 281 | 4690 | 5.80E-06  |           |           |           |           | -4.83E-05 |           |
| 282 | 4693 |           |           | -0.000116 | -0.000144 | -0.000148 |           |           |
| 283 | 4701 | -7.36E-06 | -8.64E-05 |           |           |           | -5.44E-05 | 2.76E-06  |
| 284 | 4730 |           |           | -0.000111 | -0.000141 | -0.000145 |           |           |
| 285 | 4732 |           | -0.000112 |           |           |           |           |           |
| 286 | 4780 | -1.46E-05 |           |           |           |           | -5.76E-05 |           |
| 287 | 4821 |           | -0.00013  | -0.000112 | -0.000141 | -0.000143 |           | -1.69E-06 |
| 288 | 4826 | -2.06E-05 |           |           |           |           |           |           |
| 289 | 4860 |           |           |           |           |           | -6.61E-05 | -1.77E-06 |
| 290 | 4875 |           | -0.000138 | -0.000112 | -0.00014  | -0.000142 |           |           |
| 291 | 4883 | -2.26E-05 |           | -0.000113 | -0.000141 | -0.000144 | -7.09E-05 | -1.72E-06 |
| 292 | 4899 |           | -0.000132 |           |           |           |           |           |
| 293 | 4930 | -2.47E-05 |           |           |           |           | -7.53E-05 | -1.72E-06 |
| 294 | 4931 |           |           | -0.000115 | -0.000143 | -0.000145 |           |           |
| 295 | 4933 |           | -0.000132 |           |           |           |           |           |
| 296 | 4939 | -2.63E-05 |           | -0.000121 | -0.000151 | -0.000152 | -7.93E-05 | -1.94E-06 |
| 297 | 4954 |           | -0.000125 |           |           |           |           |           |
| 298 | 5003 | -2.69E-05 | -0.000127 | -0.000122 | -0.000153 | -0.000155 | -8.11E-05 | -1.78E-06 |
| 299 | 5022 | -2.36E-05 |           |           |           |           | -7.58E-05 |           |
| 300 | 5023 |           | -0.000163 | -0.000134 | -0.000164 | -0.000165 |           | 6.30E-07  |
| 301 | 5027 | -2.60E-05 |           | -0.000141 | -0.000171 | -0.000173 | -8.22E-05 | 5.15E-07  |
| 302 | 5043 | -2.64E-05 | -0.000175 | -0.000145 | -0.000175 | -0.000176 | -8.69E-05 | 2.17E-06  |
| 303 | 5053 |           | -0.000177 |           |           |           |           |           |
| 304 | 5072 | -1.94E-05 |           |           |           |           | -8.39E-05 |           |
| 305 | 5087 |           |           | -0.000143 | -0.00017  | -0.000173 |           |           |
| 306 | 5088 | -1.54E-05 | -0.00019  | -0.000142 | -0.00017  | -0.000172 | -8.33E-05 | 2.02E-06  |

| | A | B | C | D | E | F | G | H |
|---|---|---|---|---|---|---|---|---|
| 307 | 5089 | -1.69E-05 | -0.000202 | -0.000143 | -0.000171 | -0.000173 | -8.48E-05 | 1.85E-06 |
| 308 | 5092 | -2.06E-05 | | | | | -9.11E-05 | |
| 309 | 5099 | | -0.000229 | -0.000151 | -0.00018 | -0.000181 | | 8.52E-07 |
| 310 | 5105 | -1.83E-05 | | | | | -9.60E-05 | 9.64E-07 |
| 311 | 5131 | | | -0.00015 | -0.00018 | -0.00018 | | |
| 312 | 5133 | 2.21E-05 | -0.000189 | -0.000151 | -0.000181 | -0.000181 | -9.69E-05 | 4.66E-07 |
| 313 | 5136 | 1.99E-05 | -0.000192 | | | | -9.55E-05 | |
| 314 | 5137 | | | -0.000149 | -0.00018 | -0.00018 | | |
| 315 | 5143 | 1.50E-05 | -0.000196 | | | | | |
| 316 | 5159 | | -0.000196 | -0.00015 | -0.00018 | -0.000181 | -9.72E-05 | 1.31E-06 |
| 317 | 5178 | 1.07E-05 | -0.000206 | -0.000153 | -0.000184 | -0.000186 | -0.0001 | 1.22E-06 |
| 318 | 5222 | 4.10E-05 | -0.000163 | -0.000155 | -0.000186 | -0.000187 | -0.000103 | |
| 319 | 5241 | | | | | | | 1.39E-06 |
| 320 | 5264 | 3.62E-05 | | | | | -0.000104 | |
| 321 | 5292 | 3.55E-05 | -0.000178 | -0.000158 | -0.000191 | -0.000193 | -0.000108 | 1.12E-06 |
| 322 | 5306 | 3.54E-05 | -0.000181 | -0.000163 | -0.000193 | -0.000195 | -0.000109 | 2.61E-06 |
| 323 | 5309 | | -0.000181 | -0.000162 | -0.000193 | -0.000194 | -0.000109 | |
| 324 | 5332 | 3.59E-05 | -0.000196 | -0.000166 | -0.000197 | -0.000199 | -0.000116 | 2.65E-06 |
| 325 | 5355 | 3.56E-05 | | | | | -0.000122 | |
| 326 | 5360 | | -0.000195 | -0.000172 | -0.000201 | -0.000204 | | 2.35E-06 |
| 327 | 5404 | 4.19E-05 | -0.000194 | -0.000179 | -0.000208 | -0.000211 | -0.000125 | 1.88E-06 |
| 328 | 5421 | 0.000102 | -0.000137 | -0.000179 | -0.000208 | -0.000211 | -0.000125 | 2.12E-06 |
| 329 | 5430 | 7.53E-05 | | | | | | |
| 330 | 5444 | 6.45E-05 | -0.000166 | -0.00018 | -0.00021 | -0.000212 | -0.000124 | 1.80E-06 |
| 331 | 5457 | | | | | | -0.000124 | |
| 332 | 5472 | 6.16E-05 | -0.000168 | -0.00018 | -0.000211 | -0.000214 | -0.000125 | 1.66E-06 |
| 333 | 5485 | | | -0.000181 | -0.00021 | -0.000213 | | |
| 334 | 5495 | 5.88E-05 | -0.000171 | | | | -0.000123 | 1.59E-06 |
| 335 | 5506 | 5.75E-05 | -0.00017 | -0.000178 | -0.000206 | -0.000209 | -0.000124 | 1.50E-06 |
| 336 | 5527 | 0.000127 | -6.86E-05 | | | | | |
| 337 | 5542 | 0.000125 | -7.20E-05 | -0.000183 | -0.000211 | -0.000216 | -0.000128 | 1.17E-06 |
| 338 | 5548 | 0.000125 | | -0.000183 | -0.000211 | -0.000215 | | |
| 339 | 5581 | | -7.92E-05 | -0.000193 | -0.000221 | -0.000226 | -0.000133 | |
| 340 | 5584 | 0.000117 | -9.19E-05 | | | | -0.000135 | 8.69E-07 |
| 341 | 5594 | 0.000118 | -9.11E-05 | -0.000191 | -0.000219 | -0.000222 | -0.000134 | |
| 342 | 5602 | 0.000118 | -9.15E-05 | -0.000191 | -0.00022 | -0.000223 | -0.000134 | 2.81E-07 |
| 343 | 5627 | | | 0.000164 | 0.000141 | 0.000111 | | |
| 344 | 5639 | 0.000143 | | | | | | |
| 345 | 5640 | | -7.92E-05 | | | | -0.000127 | -8.81E-07 |
| 346 | 5653 | 0.000126 | -9.19E-05 | 6.31E-05 | 4.22E-05 | 2.04E-05 | -0.000129 | |
| 347 | 5660 | 0.000127 | | 6.76E-05 | 4.61E-05 | 2.65E-05 | | |
| 348 | 5677 | | -0.000106 | | | | | -8.04E-07 |
| 349 | 5707 | | | | | | -0.000126 | |
| 350 | 5740 | 0.000121 | | 5.78E-05 | 3.88E-05 | 1.53E-05 | | |
| 351 | 5763 | 0.000121 | -0.000117 | 5.78E-05 | 3.86E-05 | 1.51E-05 | -0.000127 | -9.91E-07 |
| 352 | 5797 | 0.000116 | | | | | | -1.25E-06 |
| 353 | 5800 | | -5.17E-05 | | | | -0.000123 | |
| 354 | 5802 | 0.00012 | -6.23E-05 | 6.13E-05 | 4.11E-05 | 1.78E-05 | -0.000124 | -3.02E-07 |
| 355 | 5803 | 0.000122 | | 6.70E-05 | 4.66E-05 | 2.59E-05 | | |
| 356 | 5814 | 0.000118 | -0.00012 | 4.16E-05 | 1.92E-05 | -9.52E-06 | -0.000149 | -4.96E-07 |
| 357 | 5824 | | -0.000129 | | | | | |
| 358 | 5825 | | | | | | -0.000145 | |
| 359 | 5834 | 0.00011 | | 4.92E-05 | 2.73E-05 | 1.33E-07 | | |
| 360 | 5851 | | -0.000124 | | | | -0.000145 | |
| 361 | 5865 | 0.000109 | | | | | | |
| 362 | 5866 | 0.000104 | -0.000149 | 3.65E-05 | 1.51E-05 | -1.42E-05 | -0.000149 | -8.10E-07 |
| 363 | 5883 | | -0.000151 | 3.28E-05 | 1.22E-05 | -1.85E-05 | -0.000151 | -4.77E-07 |
| 364 | 5888 | 0.000101 | | 2.41E-05 | 4.44E-06 | -2.82E-05 | | -5.86E-07 |
| 365 | 5917 | 0.000104 | -0.000148 | 4.16E-05 | 2.17E-05 | -8.81E-06 | -0.000145 | |
| 366 | 5918 | 0.000103 | -0.000155 | 3.78E-05 | 1.82E-05 | -1.29E-05 | -0.000146 | -8.02E-07 |
| 367 | 5927 | | -5.34E-05 | | | | | |
| 368 | 5985 | | | | | | -0.000145 | |
| 369 | 6021 | 0.000111 | -6.43E-05 | 2.26E-05 | -4.43E-07 | -3.38E-05 | -0.000145 | -9.71E-07 |
| 370 | 6028 | 0.000111 | -7.06E-05 | 2.09E-05 | -2.10E-06 | -3.63E-05 | -0.000147 | -7.48E-07 |
| 371 | 6037 | | 0.00026 | | | | | |
| 372 | 6041 | 0.000146 | 0.000165 | 2.07E-05 | -1.71E-06 | -3.84E-05 | -0.000146 | |
| 373 | 6057 | | | | | | | -1.01E-06 |
| 374 | 6063 | 0.000101 | 0.000114 | -3.98E-06 | -2.16E-05 | -6.21E-05 | -0.000161 | |
| 375 | 6072 | 9.19E-05 | 8.18E-05 | -2.76E-05 | -4.17E-05 | -9.34E-05 | -0.000167 | -1.08E-06 |
| 376 | 6083 | 9.02E-05 | | -3.62E-05 | -5.20E-05 | -0.00011 | -0.000171 | -1.18E-07 |
| 377 | 6103 | | 7.25E-05 | | | | | 5.33E-07 |
| 378 | 6139 | 9.26E-05 | 5.54E-05 | -4.24E-05 | -5.72E-05 | -0.000117 | -0.00017 | 5.72E-07 |
| 379 | 6152 | 8.97E-05 | | | | | | |
| 380 | 6156 | | 4.27E-05 | | | | | 1.78E-07 |
| 381 | 6165 | 9.03E-05 | 4.33E-05 | -4.43E-05 | -5.93E-05 | -0.00012 | -0.000171 | |
| 382 | 6185 | 9.05E-05 | | -4.79E-05 | -6.29E-05 | -0.000124 | -0.00017 | |
| 383 | 6199 | 9.26E-05 | 4.75E-05 | -4.83E-05 | -6.34E-05 | -0.000125 | -0.000171 | 2.02E-08 |
| 384 | 6250 | 0.000158 | 0.0002 | -4.27E-05 | -5.80E-05 | -0.00012 | -0.000168 | -5.18E-07 |
| 385 | 6273 | 0.000151 | 0.00017 | -3.91E-05 | -5.45E-05 | -0.000118 | -0.000172 | |
| 386 | 6274 | | | | | | | -5.49E-07 |
| 387 | 6278 | 0.000139 | 0.000138 | -5.38E-05 | -6.87E-05 | -0.000138 | -0.00017 | |
| 388 | 6307 | 0.000137 | | -5.48E-05 | -6.91E-05 | -0.000135 | -0.000166 | -3.19E-07 |
| 389 | 6334 | | 0.00015 | | | | | |
| 390 | 6493 | 0.000138 | 0.000138 | -6.04E-05 | -7.30E-05 | -0.000141 | -0.000168 | 2.60E-08 |
| 391 | 6503 | 0.00014 | | -5.78E-05 | -7.07E-05 | -0.000139 | -0.000164 | |
| 392 | 6529 | 0.00014 | 0.000139 | -5.82E-05 | -7.08E-05 | -0.00014 | -0.000164 | |
| 393 | 6565 | 0.00022 | | | | | | -2.13E-07 |
| 394 | 6566 | 0.000197 | 0.000128 | -6.52E-05 | -7.86E-05 | -0.000148 | -0.00017 | -1.86E-07 |
| 395 | 6581 | 0.000192 | 0.00012 | -6.84E-05 | -8.34E-05 | -0.000152 | -0.000172 | |
| 396 | 6608 | 0.000191 | | -7.33E-05 | -8.88E-05 | -0.000158 | -0.000175 | 3.19E-07 |
| 397 | 6613 | | 0.000115 | | | | | |
| 398 | 6629 | | | | | | | 4.96E-07 |
| 399 | 6741 | 0.000183 | 8.40E-05 | -9.10E-05 | -0.000105 | -0.000174 | -0.000188 | |
| 400 | 6772 | 0.000181 | 7.47E-05 | -8.67E-05 | -9.91E-05 | -0.000166 | -0.000182 | |
| 401 | 6773 | | | | | | | 7.35E-07 |
| 402 | 6790 | | | -8.71E-05 | -9.95E-05 | -0.000167 | -0.000183 | |
| 403 | 6807 | 0.00018 | 7.25E-05 | | | | | |
| 404 | 6823 | 0.00018 | 7.29E-05 | -9.94E-05 | -0.000113 | -0.000183 | -0.00019 | |
| 405 | 6899 | 0.000209 | 0.000103 | -0.000118 | -0.000127 | -0.000195 | -0.000195 | 4.39E-07 |
| 406 | 6940 | 0.000261 | 0.000125 | -0.000118 | -0.000125 | -0.000194 | -0.0002 | |
| 407 | 6946 | 0.000255 | 0.000122 | -0.000127 | -0.000134 | -0.000201 | -0.000204 | -9.67E-07 |

|    | A | B | C | D | E | F | G | H |
|----|---|---|---|---|---|---|---|---|
| 1 | Rank | Inf | 0.002 | 0.0015 | 0.0013 | 0.001 | 0.0005 | 0.0001 |
| 2 | 10 | -0.000233 | 0.000103 | 0.000113 | 6.60E-05 | 7.97E-05 | 1.76E-05 | -8.47E-06 |
| 3 | 25 | -0.000233 | 0.000103 | 0.000113 | 6.61E-05 | 7.97E-05 | 1.79E-05 | |
| 4 | 61 | -0.000231 | 0.000104 | 0.000115 | 6.78E-05 | 8.16E-05 | 2.02E-05 | -9.17E-06 |
| 5 | 93 | -0.000231 | | | | | | -1.03E-05 |
| 6 | 115 | -0.000234 | 8.90E-05 | 0.000113 | 6.57E-05 | 7.93E-05 | 1.82E-05 | -1.03E-05 |
| 7 | 144 | -0.000235 | 8.70E-05 | 0.000112 | 6.52E-05 | 7.86E-05 | 1.75E-05 | |
| 8 | 145 | -0.000232 | 8.70E-05 | 0.000111 | 6.44E-05 | 7.78E-05 | 1.70E-05 | -1.04E-05 |
| 9 | 181 | -0.000241 | 7.83E-05 | 0.000108 | 6.08E-05 | 7.47E-05 | 1.40E-05 | -1.20E-05 |
| 10 | 194 | -0.000242 | 7.52E-05 | 0.000103 | | | | -1.20E-05 |
| 11 | 197 | -0.000242 | 7.58E-05 | 0.000104 | 5.89E-05 | 7.26E-05 | 1.17E-05 | -1.19E-05 |
| 12 | 213 | -0.000244 | 7.59E-05 | 0.000103 | 5.87E-05 | 7.24E-05 | 1.25E-05 | -1.12E-05 |
| 13 | 233 | -0.000149 | 7.01E-05 | 9.38E-05 | 5.17E-05 | 6.59E-05 | 2.98E-06 | -1.12E-05 |
| 14 | 234 | -0.000144 | 6.94E-05 | 9.19E-05 | 5.00E-05 | 6.41E-05 | 7.94E-07 | -1.24E-05 |
| 15 | 239 | | | 9.68E-05 | 5.45E-05 | 6.79E-05 | | |
| 16 | 245 | -0.000196 | 6.86E-05 | | | | | -3.65E-06 |
| 17 | 259 | -0.000208 | 6.72E-05 | 9.20E-05 | 5.04E-05 | 6.37E-05 | 3.47E-06 | -2.14E-06 |
| 18 | 261 | -0.000208 | 6.70E-05 | 9.18E-05 | 5.02E-05 | 6.35E-05 | 3.30E-06 | -2.11E-06 |
| 19 | 276 | -0.000189 | 6.83E-05 | | | | | -2.14E-06 |
| 20 | 284 | -0.000217 | 6.55E-05 | 8.68E-05 | 4.49E-05 | 5.91E-05 | 2.05E-06 | -4.26E-06 |
| 21 | 327 | -0.000247 | 6.30E-05 | 9.17E-05 | 4.78E-05 | 5.84E-05 | | -5.10E-06 |
| 22 | 334 | -0.000247 | 6.28E-05 | 8.99E-05 | 4.64E-05 | 5.81E-05 | 5.28E-06 | -5.28E-06 |
| 23 | 344 | -0.000251 | 5.83E-05 | 8.33E-05 | 4.13E-05 | 5.21E-05 | 8.67E-07 | -4.90E-06 |
| 24 | 351 | | 5.80E-05 | 8.33E-05 | 4.12E-05 | 5.21E-05 | | |
| 25 | 352 | -0.000295 | | | | | | -5.05E-06 |
| 26 | 357 | -0.000317 | 5.52E-05 | 8.08E-05 | 4.01E-05 | 5.15E-05 | 4.69E-06 | |
| 27 | 366 | -0.000335 | 4.72E-05 | 6.80E-05 | 2.88E-05 | | | -5.79E-06 |
| 28 | 375 | -0.000346 | 4.14E-05 | 6.24E-05 | 2.43E-05 | 3.73E-05 | -3.92E-06 | -5.99E-06 |
| 29 | 378 | -0.000346 | 4.15E-05 | 6.28E-05 | 2.47E-05 | 3.77E-05 | | |
| 30 | 380 | -0.000347 | 4.04E-05 | 6.08E-05 | 2.29E-05 | 3.49E-05 | -7.24E-06 | -6.52E-06 |
| 31 | 390 | -0.000281 | 4.18E-05 | 6.23E-05 | 2.44E-05 | 3.66E-05 | -4.77E-06 | -6.52E-06 |
| 32 | 408 | -0.000268 | 4.44E-05 | 6.76E-05 | 2.83E-05 | 4.06E-05 | -2.58E-07 | -4.98E-06 |
| 33 | 416 | | | | 2.73E-05 | 4.00E-05 | | |
| 34 | 432 | -0.000282 | 4.12E-05 | 6.82E-05 | 2.75E-05 | 4.03E-05 | 1.13E-06 | -4.10E-06 |
| 35 | 434 | -0.000234 | 4.05E-05 | 6.72E-05 | 2.64E-05 | 3.93E-05 | -1.35E-06 | -4.09E-06 |
| 36 | 435 | -0.000196 | 4.08E-05 | 6.79E-05 | 2.69E-05 | 3.99E-05 | 2.41E-07 | -2.93E-06 |
| 37 | 487 | -0.000246 | 3.59E-05 | 7.16E-05 | 3.13E-05 | 4.42E-05 | 4.69E-06 | -2.68E-06 |
| 38 | 496 | -0.000308 | 3.46E-05 | 7.10E-05 | 3.09E-05 | 4.37E-05 | 3.84E-06 | -2.94E-06 |
| 39 | 513 | | | | | 4.42E-05 | | -3.31E-06 |
| 40 | 526 | -0.000325 | 3.53E-05 | 7.25E-05 | 3.24E-05 | 4.52E-05 | 7.01E-06 | |
| 41 | 530 | -0.000329 | | | | 4.46E-05 | 8.22E-06 | -2.70E-06 |
| 42 | 540 | | 3.33E-05 | 6.88E-05 | 2.95E-05 | | | |
| 43 | 544 | -0.000335 | | 6.88E-05 | 2.95E-05 | 3.91E-05 | 3.59E-06 | -2.60E-06 |
| 44 | 551 | -0.000316 | 3.33E-05 | 6.93E-05 | 2.99E-05 | 3.94E-05 | 3.92E-06 | -5.84E-07 |
| 45 | 556 | -0.000322 | 3.34E-05 | 7.01E-05 | 3.05E-05 | 3.97E-05 | 4.86E-06 | -6.54E-07 |
| 46 | 557 | -0.000322 | 3.36E-05 | 7.03E-05 | 3.06E-05 | 3.98E-05 | 5.11E-06 | -6.81E-07 |
| 47 | 573 | -0.000341 | 2.63E-05 | 5.94E-05 | 1.95E-05 | 3.07E-05 | 1.34E-06 | -9.89E-07 |
| 48 | 582 | -0.000343 | 2.47E-05 | 5.52E-05 | 1.46E-05 | 2.55E-05 | -5.93E-07 | -1.14E-06 |
| 49 | 592 | -0.00034 | 2.41E-05 | 5.58E-05 | 1.46E-05 | 2.59E-05 | | -7.55E-07 |
| 50 | 594 | -0.000341 | 2.37E-05 | 5.55E-05 | 1.44E-05 | 2.57E-05 | -3.18E-06 | -8.43E-07 |
| 51 | 663 | -0.000337 | 2.24E-05 | 5.71E-05 | 1.62E-05 | 2.75E-05 | -1.96E-06 | -6.80E-08 |
| 52 | 716 | -0.000338 | 2.24E-05 | 5.71E-05 | 1.62E-05 | 2.75E-05 | -1.85E-06 | -6.08E-08 |
| 53 | 748 | -0.000338 | | | | | 9.87E-06 | |
| 54 | 760 | | 2.39E-05 | 6.26E-05 | 2.22E-05 | 3.31E-05 | 8.79E-06 | -2.13E-08 |
| 55 | 772 | -0.000337 | 2.25E-05 | 6.12E-05 | 2.08E-05 | 3.15E-05 | 7.09E-06 | 1.39E-08 |
| 56 | 782 | -0.000338 | 2.22E-05 | 6.07E-05 | 2.04E-05 | 3.11E-05 | 6.52E-06 | 8.39E-07 |
| 57 | 801 | | 2.61E-05 | | | | | |
| 58 | 813 | -0.000337 | 2.59E-05 | 6.22E-05 | 2.16E-05 | 3.21E-05 | 7.73E-06 | 8.21E-07 |
| 59 | 818 | -0.000338 | 2.48E-05 | 6.13E-05 | 2.08E-05 | 3.14E-05 | 6.86E-06 | 6.66E-07 |
| 60 | 826 | -0.000321 | | 6.13E-05 | 2.08E-05 | 3.12E-05 | 6.35E-06 | |
| 61 | 844 | -0.000323 | 2.38E-05 | | | | 6.57E-06 | 1.22E-07 |
| 62 | 885 | -0.000347 | 2.23E-05 | 5.74E-05 | 1.75E-05 | 2.85E-05 | 6.45E-06 | 4.05E-07 |
| 63 | 917 | | | | | 2.76E-05 | | |
| 64 | 925 | -0.000357 | 2.30E-05 | 5.89E-05 | 1.88E-05 | | 8.15E-06 | -2.22E-07 |
| 65 | 931 | -0.000287 | 2.38E-05 | 6.05E-05 | 2.02E-05 | 2.97E-05 | | 1.92E-07 |
| 66 | 933 | | | | | | 7.78E-06 | |
| 67 | 991 | -0.000227 | 2.36E-05 | 6.12E-05 | 2.04E-05 | 3.10E-05 | | 2.14E-06 |
| 68 | 997 | -0.000232 | 2.34E-05 | 6.10E-05 | 2.02E-05 | 3.06E-05 | 5.83E-06 | 2.08E-06 |
| 69 | 1043 | -0.000281 | 2.03E-05 | 5.84E-05 | 1.78E-05 | 2.86E-05 | | |
| 70 | 1050 | | | | | 2.75E-05 | 5.00E-06 | 1.90E-06 |
| 71 | 1063 | -0.00028 | 2.02E-05 | 5.86E-05 | 1.78E-05 | | 4.79E-06 | 1.76E-06 |
| 72 | 1093 | -0.000272 | 2.01E-05 | 5.76E-05 | 1.69E-05 | 2.78E-05 | 3.79E-06 | 1.68E-06 |
| 73 | 1094 | -0.000267 | | | | | | |
| 74 | 1107 | | | | 1.85E-05 | 2.96E-05 | | |
| 75 | 1114 | | 2.11E-05 | 5.90E-05 | | | | |
| 76 | 1165 | -0.00029 | 1.96E-05 | 5.77E-05 | 1.76E-05 | 2.80E-05 | 3.46E-06 | 2.37E-06 |
| 77 | 1174 | -0.000289 | 2.06E-05 | 5.82E-05 | 1.81E-05 | 2.86E-05 | 3.51E-06 | |
| 78 | 1206 | | | | | | | 2.24E-07 |
| 79 | 1208 | -0.000291 | 2.08E-05 | 5.79E-05 | 1.80E-05 | 2.90E-05 | 3.93E-06 | 1.76E-06 |
| 80 | 1211 | | 2.05E-05 | 5.73E-05 | | | 3.71E-06 | |
| 81 | 1240 | -0.000263 | 1.96E-05 | 5.36E-05 | 1.48E-05 | 2.69E-05 | 1.32E-06 | 3.04E-06 |
| 82 | 1250 | -0.000304 | 1.71E-05 | 5.24E-05 | 1.36E-05 | 2.56E-05 | -5.17E-08 | 4.51E-06 |
| 83 | 1260 | -0.000314 | 1.52E-05 | 4.89E-05 | 1.06E-05 | 2.28E-05 | -2.27E-06 | 1.02E-06 |
| 84 | 1262 | -0.000279 | | | | 2.32E-05 | | -2.02E-07 |
| 85 | 1294 | | 1.20E-05 | 4.43E-05 | 6.51E-06 | | -7.16E-06 | |
| 86 | 1301 | -0.000286 | 1.19E-05 | 4.41E-05 | 6.29E-06 | 1.77E-05 | -7.23E-06 | -3.91E-07 |
| 87 | 1316 | -0.000251 | 1.34E-05 | 4.63E-05 | 8.90E-06 | 2.08E-05 | -4.64E-06 | 7.82E-07 |
| 88 | 1357 | | 1.00E-05 | 4.22E-05 | | 1.76E-05 | -5.63E-06 | -1.08E-06 |
| 89 | 1404 | -0.000252 | 1.09E-05 | 4.22E-05 | 1.38E-05 | | -5.36E-06 | -7.61E-07 |
| 90 | 1411 | -0.000252 | | | 1.37E-05 | 1.70E-05 | -5.48E-06 | -6.72E-07 |
| 91 | 1414 | | 1.17E-05 | 4.51E-05 | 1.51E-05 | 1.86E-05 | | |
| 92 | 1427 | -0.000249 | 1.12E-05 | 4.41E-05 | 1.46E-05 | 1.81E-05 | -4.97E-06 | -1.70E-06 |
| 93 | 1437 | -0.000249 | 1.13E-05 | 4.47E-05 | 1.52E-05 | 1.85E-05 | -4.06E-06 | -1.44E-06 |
| 94 | 1485 | -0.000252 | 1.02E-05 | 4.50E-05 | 1.52E-05 | 1.82E-05 | -4.22E-06 | -1.93E-06 |
| 95 | 1501 | | 2.49E-05 | 6.52E-05 | 3.19E-05 | 3.83E-05 | 2.28E-06 | |
| 96 | 1535 | -0.000246 | | | | | | -1.30E-06 |
| 97 | 1576 | -0.000248 | 2.36E-05 | 6.43E-05 | 3.13E-05 | 3.78E-05 | 1.62E-06 | -1.28E-06 |
| 98 | 1579 | | 2.51E-05 | 6.61E-05 | 3.29E-05 | | 3.22E-06 | |
| 99 | 1617 | -0.000226 | | | | | | |
| 100 | 1654 | -0.00028 | 1.17E-05 | 6.33E-05 | 3.01E-05 | 3.53E-05 | 1.82E-06 | -1.23E-06 |
| 101 | 1655 | -0.00027 | 1.20E-05 | 6.38E-05 | 3.09E-05 | 3.58E-05 | 1.85E-06 | -1.27E-06 |
| 102 | 1667 | -0.000279 | 1.04E-05 | 6.01E-05 | 2.89E-05 | 3.31E-05 | 8.35E-07 | -1.07E-06 |

|     | A    | B         | C         | D        | E         | F         | G         | H         |
|-----|------|-----------|-----------|----------|-----------|-----------|-----------|-----------|
| 103 | 1683 | -0.000284 | 6.10E-06  | 5.56E-05 | 2.35E-05  | 2.97E-05  | -2.36E-07 | -1.03E-07 |
| 104 | 1686 |           |           |          | 2.08E-05  | 2.73E-05  |           | -1.16E-06 |
| 105 | 1708 | -0.000285 | 5.17E-06  | 5.25E-05 | 1.96E-05  | 2.70E-05  | 4.19E-07  | 3.35E-07  |
| 106 | 1720 | -0.000285 |           |          |           | 2.63E-05  |           |           |
| 107 | 1725 | -0.000288 | 3.96E-06  | 4.91E-05 | 1.63E-05  |           | -7.42E-07 | 6.92E-09  |
| 108 | 1734 |           |           |          |           | 2.43E-05  |           |           |
| 109 | 1748 | -0.000327 | 3.34E-06  | 4.79E-05 | 1.53E-05  | 2.35E-05  | -1.14E-06 | -1.43E-06 |
| 110 | 1757 | -0.00033  |           | 4.51E-05 | 1.24E-05  |           |           | -1.83E-06 |
| 111 | 1758 |           | 1.56E-06  |          |           | 2.14E-05  | -3.96E-06 | -2.18E-06 |
| 112 | 1759 | -0.000332 | 6.11E-07  | 4.31E-05 | 1.08E-05  |           | -3.98E-06 | -1.51E-06 |
| 113 | 1781 | -0.000321 | 4.01E-07  | 4.43E-05 | 1.19E-05  | 2.22E-05  | -3.68E-06 | -1.18E-06 |
| 114 | 1787 | -0.000321 | 9.38E-07  | 4.46E-05 | 1.23E-05  | 2.25E-05  | -3.11E-06 |           |
| 115 | 1791 |           |           |          | 1.34E-05  | 2.36E-05  |           |           |
| 116 | 1816 | -0.000299 | 8.12E-06  | 5.84E-05 | 2.75E-05  | 3.73E-05  | 6.16E-06  |           |
| 117 | 1827 | -0.000303 | 7.22E-06  | 5.69E-05 |           | 3.65E-05  | 5.25E-06  | -2.50E-07 |
| 118 | 1851 |           |           |          | 2.66E-05  |           |           |           |
| 119 | 1882 |           | 6.47E-06  | 5.69E-05 |           |           | 4.26E-06  |           |
| 120 | 1901 | -0.000269 | 6.55E-06  | 5.73E-05 | 2.82E-05  | 3.73E-05  |           |           |
| 121 | 1909 | -0.000158 | 6.04E-06  | 5.42E-05 | 2.54E-05  | 3.49E-05  | 3.25E-06  | 6.51E-07  |
| 122 | 1942 | -0.000185 |           |          | 2.19E-05  |           |           |           |
| 123 | 1956 | -0.000181 | 1.54E-06  | 5.30E-05 | 2.20E-05  | 3.22E-05  | 3.90E-07  | 5.93E-07  |
| 124 | 1992 | -0.000256 |           |          |           | 2.96E-05  |           |           |
| 125 | 2015 | -0.000256 | -2.21E-06 | 4.69E-05 | 1.95E-05  | 2.93E-05  | 2.12E-07  | -1.63E-07 |
| 126 | 2018 |           | -3.92E-06 | 4.52E-05 | 1.76E-05  |           | -1.26E-06 | -2.82E-07 |
| 127 | 2027 | -0.000257 |           |          | 1.79E-05  | 2.84E-05  |           | -3.09E-07 |
| 128 | 2033 |           | -4.65E-06 | 4.30E-05 |           | 2.73E-05  | -1.58E-06 | -3.09E-07 |
| 129 | 2047 | -0.000252 | -3.88E-06 | 4.39E-05 | 2.02E-05  | 2.79E-05  | -9.28E-07 | 6.68E-07  |
| 130 | 2049 |           | -4.32E-06 | 4.37E-05 | 2.02E-05  |           | -1.18E-06 |           |
| 131 | 2080 | -0.000164 |           |          |           | 2.61E-05  |           | -1.36E-06 |
| 132 | 2085 | -0.000166 | -7.71E-06 | 3.85E-05 | 1.55E-05  | 2.68E-05  | -9.90E-06 | -1.28E-06 |
| 133 | 2105 | 1.32E-05  | -7.87E-06 | 3.81E-05 | 1.54E-05  | 2.67E-05  | -9.51E-06 | -6.57E-07 |
| 134 | 2116 | 4.21E-05  | -1.15E-05 | 3.08E-05 | 7.80E-06  | 2.27E-05  | -1.12E-05 | -8.74E-07 |
| 135 | 2127 | -3.99E-06 | -1.19E-05 | 3.07E-05 | 7.33E-06  | 2.23E-05  | -1.04E-05 |           |
| 136 | 2128 | -1.37E-05 | -1.62E-05 | 1.90E-05 | -2.79E-06 | 1.33E-05  | -1.70E-05 | -7.97E-07 |
| 137 | 2148 | 8.77E-05  |           |          |           | 1.06E-05  |           |           |
| 138 | 2156 |           | -1.71E-05 | 1.30E-05 | -8.40E-06 |           |           |           |
| 139 | 2175 | 8.02E-05  |           |          |           |           | -5.84E-06 |           |
| 140 | 2185 |           |           |          |           | 1.78E-05  |           |           |
| 141 | 2226 | 0.000103  | -2.31E-05 | 1.39E-05 | -8.24E-06 | 1.68E-05  | -7.22E-06 | -3.46E-07 |
| 142 | 2244 |           | -2.61E-05 | 7.82E-06 | -1.25E-05 |           | -5.64E-06 |           |
| 143 | 2261 | 7.11E-05  | -2.23E-05 | 1.36E-05 | -6.78E-06 | 1.89E-05  | -1.51E-06 | -5.18E-08 |
| 144 | 2272 |           |           |          |           |           |           | -1.65E-08 |
| 145 | 2351 | 3.80E-05  | -2.13E-05 | 1.62E-05 | -3.81E-06 | 2.13E-05  | 2.46E-06  |           |
| 146 | 2357 | 3.82E-05  | -2.38E-05 | 1.08E-05 | -8.06E-06 | 1.66E-05  | 3.82E-07  | -1.01E-07 |
| 147 | 2359 |           |           |          |           | 1.73E-05  |           |           |
| 148 | 2365 | 2.96E-05  |           |          | -1.04E-05 |           |           | -2.43E-06 |
| 149 | 2376 |           | -2.51E-05 | 8.10E-06 |           | 1.62E-05  | -1.70E-06 | -2.51E-06 |
| 150 | 2384 | 9.21E-05  | -2.43E-05 | 9.00E-06 | -1.03E-05 |           | -7.78E-07 |           |
| 151 | 2388 | 4.93E-05  |           |          |           | 1.55E-05  |           |           |
| 152 | 2394 |           | -2.23E-05 | 3.10E-05 | -7.59E-07 |           |           |           |
| 153 | 2415 | -9.02E-06 |           |          |           | 1.42E-05  | 8.37E-07  | 1.36E-07  |
| 154 | 2446 |           | -2.39E-05 | 2.55E-05 | -2.01E-06 |           | -2.35E-06 | -2.36E-06 |
| 155 | 2448 | -6.99E-06 |           |          |           | 1.12E-05  |           |           |
| 156 | 2449 |           | -2.32E-05 | 2.65E-05 | -1.41E-06 |           | -2.57E-06 | -2.10E-06 |
| 157 | 2451 | -8.46E-06 | -2.34E-05 | 2.58E-05 | -1.79E-06 | 1.09E-05  |           | -2.30E-06 |
| 158 | 2494 |           |           |          |           |           | -3.27E-06 |           |
| 159 | 2552 | -1.39E-05 | -2.68E-05 | 2.29E-05 | -4.57E-06 | 9.47E-06  | -3.83E-06 |           |
| 160 | 2570 | -7.56E-06 |           |          |           | 8.87E-06  |           |           |
| 161 | 2585 |           |           |          |           |           |           | -7.00E-07 |
| 162 | 2587 |           | -1.37E-05 | 3.71E-05 | 6.07E-06  |           | 8.35E-06  |           |
| 163 | 2594 | -1.50E-05 |           |          |           | 8.54E-06  |           |           |
| 164 | 2665 |           | -1.49E-05 | 3.43E-05 | 3.85E-06  | 8.23E-06  | 5.64E-06  |           |
| 165 | 2677 | -3.21E-06 | -1.44E-05 | 3.56E-05 | 4.57E-06  |           | 6.06E-06  | -1.14E-06 |
| 166 | 2678 | 4.81E-05  |           |          |           | 8.95E-06  |           |           |
| 167 | 2704 | 3.59E-05  | -1.60E-05 | 3.00E-05 | -1.20E-07 | 5.69E-06  | 3.33E-06  | -5.13E-07 |
| 168 | 2769 | 3.60E-05  |           |          |           | 6.38E-06  | 6.44E-06  | 1.40E-07  |
| 169 | 2808 |           | -1.69E-05 | 3.14E-05 | 1.10E-06  | 6.33E-06  | 6.52E-06  |           |
| 170 | 2822 | 3.29E-05  | -1.64E-05 | 3.23E-05 | 1.95E-06  |           | 7.18E-06  |           |
| 171 | 2831 |           |           |          |           | -3.15E-06 |           | -3.03E-07 |
| 172 | 2856 | 3.67E-05  | -1.36E-05 | 3.26E-05 | 3.24E-06  |           | 7.85E-06  | -8.86E-07 |
| 173 | 2929 | 3.70E-05  | -1.41E-05 | 3.30E-05 | 4.02E-06  | -5.21E-07 | 8.04E-06  | -8.15E-07 |
| 174 | 2942 | 3.61E-05  | -1.44E-05 | 3.18E-05 | 3.28E-06  | -7.60E-07 |           | -2.98E-07 |
| 175 | 2950 |           |           |          |           | -4.73E-06 | 6.41E-06  | -5.66E-07 |
| 176 | 2980 | 6.69E-05  | 6.93E-06  | 6.52E-05 | 2.68E-05  |           | 7.39E-06  | 1.99E-06  |
| 177 | 2984 |           | 9.11E-05  |          | 0.000111  | -2.20E-06 |           |           |
| 178 | 3004 | 0.000186  |           | 0.000137 |           |           |           |           |
| 179 | 3011 |           |           |          |           |           | 5.40E-06  |           |
| 180 | 3014 | 0.000207  |           |          |           | -3.52E-06 |           |           |
| 181 | 3040 | 0.000145  | 4.67E-05  | 6.72E-05 | 6.24E-05  | -5.36E-06 | 4.72E-06  | 2.61E-06  |
| 182 | 3071 |           |           |          |           |           | 1.14E-06  |           |
| 183 | 3085 | 0.000184  | 4.11E-05  | 5.50E-05 | 5.38E-05  | -9.11E-06 |           | 3.22E-06  |
| 184 | 3120 |           |           |          |           |           | 2.13E-06  |           |
| 185 | 3143 | 0.000165  |           |          | 4.81E-05  | -1.17E-05 |           |           |
| 186 | 3159 |           |           | 4.59E-05 |           |           |           | 3.13E-06  |
| 187 | 3199 |           | 3.51E-05  |          |           |           |           |           |
| 188 | 3214 | 0.000148  |           |          |           | -1.10E-05 | 3.50E-06  |           |
| 189 | 3225 |           | 2.43E-05  | 3.15E-05 | 4.12E-05  |           |           | 3.87E-06  |
| 190 | 3232 | 0.00014   |           |          |           | -1.44E-05 | 1.12E-06  |           |
| 191 | 3240 |           | 2.34E-05  | 3.04E-05 | 3.99E-05  |           |           |           |
| 192 | 3258 |           |           |          |           |           | 4.32E-07  |           |
| 193 | 3278 | 0.000144  | 1.95E-05  | 2.26E-05 | 3.46E-05  | -1.73E-05 | -3.30E-07 | 4.02E-06  |
| 194 | 3315 |           |           |          |           |           | 1.16E-07  |           |
| 195 | 3320 | 0.00013   | 1.73E-05  | 2.20E-05 | 3.40E-05  | -1.81E-05 |           |           |
| 196 | 3351 |           |           |          |           |           |           | 3.67E-06  |
| 197 | 3366 | 0.000146  | 1.25E-05  | 1.11E-05 | 2.63E-05  | -2.19E-05 | -1.35E-06 | 3.65E-06  |
| 198 | 3370 |           |           |          |           | -2.03E-05 | 1.60E-08  |           |
| 199 | 3380 | 0.00014   | 1.20E-05  | 4.19E-06 | 2.55E-05  | -2.08E-05 | -8.15E-07 | 2.43E-06  |
| 200 | 3384 | 0.000154  | 2.72E-05  | 3.67E-06 |           |           |           |           |
| 201 | 3386 |           |           |          |           | -1.99E-05 | 7.42E-07  |           |
| 202 | 3428 |           |           |          | 2.73E-05  |           |           |           |
| 203 | 3437 | 0.000156  | 1.99E-05  | 2.45E-06 |           |           |           | 2.65E-06  |
| 204 | 3470 | 0.000155  | 1.99E-05  | 1.99E-06 | 2.71E-05  | -2.06E-05 | 7.23E-08  | 2.69E-06  |

|     | A    | B        | C         | D         | E        | F         | G         | H         |
|-----|------|----------|-----------|-----------|----------|-----------|-----------|-----------|
| 205 | 3503 | 0.000208 | 2.32E-05  | 1.27E-05  | 3.40E-05 |           |           |           |
| 206 | 3505 |          |           |           |          |           | 1.71E-06  |           |
| 207 | 3512 |          |           |           |          |           |           | 3.00E-06  |
| 208 | 3516 | 0.000194 | 2.22E-05  | 1.07E-05  | 3.32E-05 | -1.57E-05 |           |           |
| 209 | 3518 |          | 2.15E-05  | 9.65E-06  | 3.24E-05 | -1.63E-05 | 8.32E-07  | 2.98E-06  |
| 210 | 3519 | 0.000193 |           |           |          | -1.63E-05 | 8.90E-07  | 2.98E-06  |
| 211 | 3561 |          | 2.02E-05  | 8.72E-06  | 2.95E-05 |           |           |           |
| 212 | 3564 | 0.000178 | 2.01E-05  | 8.30E-06  | 2.93E-05 | -2.03E-05 | 6.14E-07  | 3.50E-06  |
| 213 | 3576 | 0.000182 |           |           |          |           |           |           |
| 214 | 3578 |          |           |           |          |           | 2.60E-06  |           |
| 215 | 3593 |          |           |           |          | -2.31E-05 |           |           |
| 216 | 3601 |          | 1.83E-05  | 5.74E-06  | 2.76E-05 |           |           | 3.33E-06  |
| 217 | 3633 | 4.93E-05 |           |           |          |           |           |           |
| 218 | 3647 |          |           |           |          | -3.13E-05 | -4.96E-06 |           |
| 219 | 3671 | 0.000128 | 1.44E-05  | -5.02E-07 | 2.23E-05 |           |           | 3.31E-06  |
| 220 | 3682 | 0.000169 | 1.47E-05  | 7.18E-07  | 2.38E-05 | -3.02E-05 | -4.43E-06 |           |
| 221 | 3685 |          |           |           |          |           |           | 2.57E-06  |
| 222 | 3701 | 0.000101 |           |           | 2.62E-05 | -2.79E-05 | -4.51E-06 |           |
| 223 | 3704 |          | 1.50E-05  | 2.08E-06  |          |           |           |           |
| 224 | 3745 | 0.000159 |           |           |          |           | -4.75E-06 |           |
| 225 | 3759 |          |           |           |          | -2.67E-05 |           |           |
| 226 | 3766 |          |           |           |          |           |           | 2.25E-06  |
| 227 | 3776 |          | 1.12E-05  | -5.36E-06 | 2.66E-05 |           |           |           |
| 228 | 3790 | 0.000112 |           |           |          | -2.70E-05 | -4.73E-06 |           |
| 229 | 3827 | 0.000162 | 1.08E-05  | -8.27E-06 | 2.46E-05 | -2.81E-05 | -5.34E-06 | 1.02E-06  |
| 230 | 3848 |          |           |           |          |           | -7.10E-07 |           |
| 231 | 3862 |          |           |           |          | -2.94E-05 |           | -2.69E-06 |
| 232 | 3886 | 9.51E-05 | 8.67E-06  | -9.28E-06 | 2.37E-05 | -2.94E-05 | -1.39E-06 | -2.67E-06 |
| 233 | 3894 |          |           |           |          | -2.96E-05 |           | -2.86E-06 |
| 234 | 3895 | 9.42E-05 | 9.30E-06  | -7.72E-06 | 2.53E-05 |           |           |           |
| 235 | 3901 |          |           |           |          |           |           | -3.96E-06 |
| 236 | 3910 |          |           |           |          | -2.70E-05 | 2.00E-06  |           |
| 237 | 3939 | 9.56E-05 | 4.70E-06  | -8.65E-06 | 2.42E-05 |           |           | -3.07E-06 |
| 238 | 3946 |          |           |           |          |           | -2.83E-07 |           |
| 239 | 3953 |          |           |           |          | -2.77E-05 |           |           |
| 240 | 3997 | 6.97E-05 | -5.64E-07 | -1.59E-05 | 1.82E-05 |           | -5.91E-06 |           |
| 241 | 4014 |          |           |           | 2.22E-05 | -3.03E-05 |           | -4.40E-06 |
| 242 | 4029 |          | 1.27E-06  |           |          |           |           |           |
| 243 | 4043 | 8.42E-05 |           | -1.42E-05 |          |           |           |           |
| 244 | 4047 |          |           |           |          | -3.32E-05 | -6.99E-06 |           |
| 245 | 4052 |          |           |           |          |           |           | -5.18E-06 |
| 246 | 4071 |          | 4.11E-06  | -8.94E-06 | 2.58E-05 |           |           | -5.35E-06 |
| 247 | 4082 | 0.000105 |           |           |          |           |           |           |
| 248 | 4083 | 0.000108 | 3.14E-06  | -1.11E-05 | 2.53E-05 | -2.83E-05 | -9.29E-07 | -3.23E-06 |
| 249 | 4088 | 9.33E-05 | 2.88E-06  | -1.20E-05 | 2.48E-05 | -2.89E-05 | -1.33E-06 | -3.34E-06 |
| 250 | 4090 |          |           | -1.18E-05 |          |           |           |           |
| 251 | 4129 | 7.49E-05 |           |           |          |           |           | -3.64E-06 |
| 252 | 4133 |          | 3.23E-06  |           |          | -2.92E-05 | -1.30E-06 |           |
| 253 | 4155 |          |           | -1.32E-05 | 2.35E-05 |           |           |           |
| 254 | 4165 |          |           |           |          |           |           | -2.03E-06 |
| 255 | 4177 | 7.81E-05 |           |           |          | -2.95E-05 | -7.80E-07 |           |
| 256 | 4190 |          | 1.41E-06  | -1.36E-05 | 2.22E-05 |           |           | -2.00E-06 |
| 257 | 4229 | 8.72E-05 | 5.93E-06  | -1.34E-05 | 2.28E-05 | -2.99E-05 | -1.31E-06 | -1.89E-06 |
| 258 | 4240 |          | 1.30E-06  | -1.92E-05 | 2.07E-05 | -2.71E-05 | 3.53E-06  |           |
| 259 | 4273 |          |           |           |          |           |           | -1.87E-06 |
| 260 | 4286 | 8.33E-05 | 2.88E-06  | -1.53E-05 | 1.92E-05 | -9.97E-06 | 2.53E-06  | -1.59E-07 |
| 261 | 4292 | 8.73E-05 |           |           |          | -4.30E-06 | 3.08E-05  |           |
| 262 | 4302 |          | 8.23E-07  | -1.71E-05 | 1.85E-05 |           |           |           |
| 263 | 4320 |          |           |           |          | -5.42E-06 | 2.87E-05  |           |
| 264 | 4324 | 0.00011  | 1.25E-07  | -1.93E-05 | 1.92E-05 | -3.96E-06 | 3.02E-05  | 9.16E-07  |
| 265 | 4328 | 0.00011  | -2.17E-07 | -2.12E-05 | 1.84E-05 |           |           |           |
| 266 | 4340 | 0.00014  | -4.63E-06 | -2.94E-05 | 1.19E-05 | -1.10E-05 | 2.15E-05  | 1.72E-06  |
| 267 | 4352 | 0.000121 | -8.58E-06 | -3.88E-05 | 3.46E-06 | -1.79E-05 | 1.16E-05  | 1.57E-06  |
| 268 | 4360 |          |           |           |          | -2.01E-05 | 9.03E-06  | 1.42E-06  |
| 269 | 4361 | 0.000126 | -8.38E-06 | -3.52E-05 | 8.41E-06 |           |           |           |
| 270 | 4386 | 0.000127 | -6.87E-06 | -3.30E-05 | 9.97E-06 | -1.95E-05 | 1.04E-05  |           |
| 271 | 4405 |          |           |           |          |           | 1.02E-05  | 2.03E-06  |
| 272 | 4413 | 0.000132 | -4.64E-06 | -2.96E-05 | 1.25E-05 | -1.77E-05 |           |           |
| 273 | 4424 | 0.000237 | -5.40E-06 | -3.72E-05 | 7.28E-06 | -2.11E-05 | 7.83E-06  | 1.27E-06  |
| 274 | 4501 |          |           |           |          |           | 6.57E-06  |           |
| 275 | 4513 | 0.000169 | -6.08E-06 | -3.67E-05 | 6.84E-06 | -2.11E-05 | 6.03E-06  | 1.24E-06  |
| 276 | 4518 | 0.000167 | -8.17E-06 | -3.97E-05 | 4.37E-06 | -2.35E-05 | 3.48E-06  | 1.07E-06  |
| 277 | 4541 | 0.000227 | 2.32E-05  | -4.73E-05 | -1.29E-06 | -2.79E-05 |           |           |
| 278 | 4542 | 0.000218 | 1.90E-05  | -4.72E-05 | -1.30E-06 | -2.79E-05 | 3.10E-07  | 9.68E-07  |
| 279 | 4561 |          | 1.14E-05  |           |          | -2.75E-05 | 8.25E-07  |           |
| 280 | 4562 |          |           | -4.82E-05 | -2.00E-06 |           |           |           |
| 281 | 4572 | 0.000206 |           |           |          |           |           | 1.09E-06  |
| 282 | 4622 |          |           |           |          |           | 6.06E-06  |           |
| 283 | 4623 | 0.000402 | 1.17E-05  | -4.11E-05 | 1.97E-06 | -2.37E-05 |           |           |
| 284 | 4624 | 0.000405 | 9.64E-06  | -4.39E-05 | -2.25E-07 | -2.63E-05 | 4.46E-06  | 8.83E-07  |
| 285 | 4629 | 0.000362 |           |           |          | -3.77E-05 | 2.52E-07  | 2.31E-07  |
| 286 | 4639 |          | -1.10E-05 |           |          |           |           |           |
| 287 | 4641 |          |           | -4.94E-05 | -4.11E-06 |           |           |           |
| 288 | 4643 | 0.000316 | -1.09E-05 | -4.97E-05 | -4.30E-06 | -3.77E-05 | 1.11E-06  | 1.12E-06  |
| 289 | 4644 |          |           |           |          |           |           | 6.76E-07  |
| 290 | 4648 | 0.000301 | -1.16E-05 | -5.06E-05 |          | -3.81E-05 | -1.49E-06 |           |
| 291 | 4649 |          |           |           | -5.68E-06 |           |           | 7.24E-08  |
| 292 | 4679 |          |           |           |          |           | -6.79E-07 |           |
| 293 | 4683 | 0.000301 | -1.16E-05 |           |          | -3.85E-05 |           | -3.39E-07 |
| 294 | 4690 | 0.000303 |           | -5.21E-05 | -8.53E-06 | -3.83E-05 | -4.26E-07 |           |
| 295 | 4693 |          | -1.11E-05 |           |          |           |           |           |
| 296 | 4701 |          |           | -5.63E-05 | -1.17E-05 |           | -6.37E-06 | 6.41E-08  |
| 297 | 4710 | 0.000305 |           |           |          | -3.21E-05 |           |           |
| 298 | 4730 |          | -1.36E-05 |           |          |           |           |           |
| 299 | 4732 |          |           | -5.89E-05 | -1.71E-05 |           |           | 1.93E-07  |
| 300 | 4780 |          |           |           |          | -3.62E-05 |           |           |
| 301 | 4802 | 0.000305 | -1.43E-05 |           |          |           | -1.33E-06 |           |
| 302 | 4821 |          |           | -5.87E-05 | -1.56E-05 |           |           |           |
| 303 | 4826 |          |           |           |          |           |           | -9.13E-07 |
| 304 | 4865 | 0.000313 | -1.07E-05 | -5.62E-05 |          | -3.33E-05 | 3.34E-06  |           |
| 305 | 4875 | 0.000333 | -1.06E-05 |           | -1.68E-05 | -3.34E-05 | 3.23E-06  |           |
| 306 | 4878 |          |           | -5.78E-05 |          |           |           |           |

|  | A | B | C | D | E | F | G | H |
|---|---|---|---|---|---|---|---|---|
| 307 | 4883 |  |  |  | -1.88E-05 |  |  | -7.12E-07 |
| 308 | 4930 | 0.000299 | -1.30E-05 |  |  | -3.37E-05 | 1.43E-06 |  |
| 309 | 4931 |  |  | -5.59E-05 | -1.99E-05 |  |  | 1.30E-07 |
| 310 | 4939 | 0.000297 | -1.21E-05 | -5.79E-05 | -2.13E-05 | -3.39E-05 | 1.27E-06 |  |
| 311 | 4954 |  |  |  |  |  |  | -8.21E-08 |
| 312 | 5000 |  | -9.51E-06 |  |  | -3.08E-05 | 4.38E-06 |  |
| 313 | 5003 | 0.00028 |  | -5.77E-05 | -2.05E-05 |  |  |  |
| 314 | 5011 |  |  |  |  | -2.95E-05 | 6.29E-06 |  |
| 315 | 5022 | 0.000273 | -1.03E-05 | -6.27E-05 | -2.31E-05 |  |  |  |
| 316 | 5023 |  |  | -4.07E-05 | -8.46E-05 |  | -4.03E-05 | -3.01E-06 | 1.39E-06 |
| 317 | 5027 | 0.000267 |  |  | -3.60E-05 |  |  |  |
| 318 | 5043 | 0.000262 | -4.31E-05 | -8.49E-05 | -3.70E-05 | -3.98E-05 | -3.37E-06 | 3.39E-06 |
| 319 | 5053 | 0.000261 |  |  |  | -4.02E-05 | -3.58E-06 | 3.14E-06 |
| 320 | 5058 |  |  | -4.36E-05 | -8.60E-05 | -3.76E-05 |  |  |  |
| 321 | 5087 | 0.000339 |  |  |  |  |  |  |
| 322 | 5088 |  |  | -5.49E-05 | -8.97E-05 | -4.12E-05 | -4.19E-05 | -5.97E-06 | 2.06E-06 |
| 323 | 5089 | 0.000343 | -5.39E-05 | -8.86E-05 |  | -4.03E-05 | -5.72E-06 |  |
| 324 | 5092 |  |  |  |  | -4.58E-05 |  |  | 2.32E-06 |
| 325 | 5099 | 0.000331 |  |  |  |  |  |  |
| 326 | 5105 | 0.000341 | -5.68E-05 | -9.38E-05 | -4.58E-05 | -4.33E-05 | -8.12E-06 | 2.63E-06 |
| 327 | 5131 |  |  |  |  | -5.91E-05 |  |  |  |
| 328 | 5133 | 0.000314 | -5.87E-05 |  |  | -4.89E-05 | -7.28E-06 | 2.12E-06 |
| 329 | 5136 | 0.000312 |  | -0.000107 | -6.24E-05 | -5.02E-05 |  | 2.34E-06 |
| 330 | 5143 |  |  | -6.05E-05 | -0.000109 | -6.39E-05 |  | -8.82E-06 | 2.59E-06 |
| 331 | 5159 | 0.000316 | -6.10E-05 | -0.000108 |  | -5.12E-05 | -7.63E-06 |  |
| 332 | 5171 |  |  |  |  | -6.79E-05 |  |  |  |
| 333 | 5178 | 0.000297 |  |  |  | -5.43E-05 |  | 3.20E-06 |
| 334 | 5187 |  |  |  |  |  |  |  | 3.15E-06 |
| 335 | 5222 |  |  | -7.07E-05 | -0.000118 | -7.19E-05 |  | -1.44E-05 | 3.19E-06 |
| 336 | 5241 | 0.000291 |  |  |  | -5.76E-05 |  |  |
| 337 | 5264 |  |  | -6.96E-05 | -0.000115 | -6.95E-05 |  | -1.21E-05 | 2.89E-06 |
| 338 | 5277 |  |  |  |  |  | -5.59E-05 |  |  |
| 339 | 5288 | 0.000324 |  |  |  |  |  | -1.50E-05 |  |
| 340 | 5292 |  |  | -7.00E-05 | -0.000116 | -7.06E-05 |  |  | 2.33E-06 |
| 341 | 5306 | 0.000323 | -6.75E-05 | -0.000114 | -6.68E-05 | -5.68E-05 | -1.42E-05 | 2.38E-06 |
| 342 | 5309 | 0.000323 | -6.78E-05 | -0.000117 | -7.09E-05 | -5.67E-05 |  |  |
| 343 | 5332 |  |  |  |  |  |  | -1.94E-05 | 2.08E-06 |
| 344 | 5355 | 0.000382 | -5.98E-05 | -0.000112 | -6.52E-05 | -5.23E-05 |  |  |
| 345 | 5360 |  |  |  |  |  |  | -1.87E-05 | 2.03E-06 |
| 346 | 5383 | 0.000379 |  |  |  |  |  |  |
| 347 | 5404 |  |  | -5.91E-05 | -0.000113 | -6.64E-05 | -5.33E-05 | -1.94E-05 | 1.79E-06 |
| 348 | 5421 | 0.000358 |  |  |  |  |  |  |
| 349 | 5430 |  |  | -6.27E-05 | -0.000113 |  |  |  |  |
| 350 | 5444 | 0.000352 | -6.41E-05 |  | -6.61E-05 | -5.33E-05 | -1.86E-05 | 2.31E-06 |
| 351 | 5457 |  |  |  | -0.000112 |  |  |  |  |
| 352 | 5472 | 0.00035 |  |  | -6.74E-05 | -5.45E-05 | -1.91E-05 | 1.97E-06 |
| 353 | 5485 |  |  | -5.40E-05 |  |  |  |  | 2.71E-06 |
| 354 | 5495 |  |  |  | -0.00011 | -6.65E-05 | -5.36E-05 |  |  |
| 355 | 5506 | 0.000376 | -5.83E-05 | -0.000111 | -6.65E-05 | -5.36E-05 | -1.77E-05 | 3.02E-06 |
| 356 | 5527 | 0.000373 | -6.03E-05 |  |  |  |  |  |
| 357 | 5542 | 0.000373 | -6.16E-05 | -0.000114 | -7.07E-05 | -5.68E-05 | -2.04E-05 | 2.16E-06 |
| 358 | 5548 | 0.000373 |  |  |  | -5.67E-05 | -2.04E-05 | 2.17E-06 |
| 359 | 5581 |  |  | -6.20E-05 | -0.000114 | -7.03E-05 | -5.76E-05 | -2.44E-05 |  |
| 360 | 5584 | 0.000391 |  |  |  |  |  |  |
| 361 | 5594 |  |  | -6.12E-05 | -0.000113 | -6.96E-05 | -5.65E-05 | -2.30E-05 | 2.52E-06 |
| 362 | 5602 | 0.00036 | -6.14E-05 | -0.000112 | -6.86E-05 | -5.46E-05 | -1.97E-05 |  |
| 363 | 5640 | 0.000365 |  |  |  |  |  |  | 4.48E-06 |
| 364 | 5652 |  |  | -5.45E-05 | -0.000107 | -6.42E-05 | -4.91E-05 | -1.08E-05 |  |
| 365 | 5653 | 0.000362 | -5.65E-05 | -0.000109 |  |  |  |  |
| 366 | 5660 |  |  |  |  | -6.45E-05 | -4.99E-05 | -1.21E-05 |  |
| 367 | 5677 | 0.000359 |  |  |  |  |  |  | 4.27E-06 |
| 368 | 5707 |  |  | -6.03E-05 |  |  |  |  |  |
| 369 | 5740 |  |  |  | -0.000114 | -6.96E-05 |  |  |  |
| 370 | 5763 | 0.000336 | -6.63E-05 | -0.000114 | -6.97E-05 | -5.63E-05 | -1.89E-05 | 6.02E-06 |
| 371 | 5797 | 0.000337 | -6.51E-05 |  |  |  |  |  |
| 372 | 5800 |  |  |  | -0.000105 | -6.33E-05 | -4.64E-05 | -1.85E-05 |  |
| 373 | 5802 | 0.00033 | -6.54E-05 | -0.000107 | -6.48E-05 | -4.77E-05 | -1.78E-05 | 7.10E-06 |
| 374 | 5803 | 0.000332 | -6.37E-05 | -0.000105 | -6.12E-05 | -4.31E-05 | -1.21E-05 | 7.72E-06 |
| 375 | 5804 |  |  | -6.60E-05 | -0.000106 | -6.09E-05 | -4.16E-05 |  |  |
| 376 | 5814 | 0.000319 | -7.08E-05 | -0.000118 | -7.15E-05 | -5.25E-05 | -1.97E-05 | 5.63E-06 |
| 377 | 5824 | 0.000318 | -7.12E-05 | -0.000118 | -7.25E-05 | -5.34E-05 | -1.85E-05 |  |
| 378 | 5825 | 0.00032 |  |  |  |  |  |  |
| 379 | 5834 |  |  | -6.86E-05 | -0.000112 |  |  |  |  |
| 380 | 5851 | 0.000312 | -6.90E-05 | -0.000114 | -7.72E-05 | -5.43E-05 | -1.71E-05 | 6.24E-06 |
| 381 | 5866 | 0.000305 | -7.40E-05 | -0.000121 | -8.47E-05 | -6.02E-05 | -2.28E-05 |  |
| 382 | 5883 | 0.000344 | -7.40E-05 | -0.00012 | -8.39E-05 |  |  | 6.33E-06 |
| 383 | 5888 | 0.000274 |  |  |  | -6.46E-05 | -2.54E-05 | 5.85E-06 |
| 384 | 5917 |  |  | -7.83E-05 |  |  |  |  |  |
| 385 | 5918 | 0.000275 |  | -0.000119 | -8.35E-05 | -6.27E-05 | -2.19E-05 |  |
| 386 | 5927 |  |  |  |  |  |  |  | 6.02E-06 |
| 387 | 5985 |  |  | -7.30E-05 | -0.000111 | -7.35E-05 | -5.23E-05 | -1.36E-05 |  |
| 388 | 6021 | 0.000267 | -7.30E-05 | -0.000111 | -7.38E-05 | -5.30E-05 | -1.38E-05 | 7.31E-06 |
| 389 | 6028 | 0.000268 | -7.22E-05 | -0.00011 | -7.27E-05 | -5.22E-05 | -1.20E-05 |  |
| 390 | 6041 | 0.000271 |  |  |  |  |  |  |
| 391 | 6057 |  |  | -7.78E-05 | -0.000113 | -7.43E-05 | -5.33E-05 | -1.23E-05 | 6.76E-06 |
| 392 | 6072 | 0.000252 | -8.25E-05 | -0.000118 | -7.84E-05 | -5.70E-05 | -1.59E-05 |  |
| 393 | 6083 | 0.000249 | -8.34E-05 | -0.000121 | -8.12E-05 | -5.96E-05 | -1.66E-05 | 6.76E-06 |
| 394 | 6103 | 0.000244 | -8.32E-05 |  |  |  |  | 7.13E-06 |
| 395 | 6139 |  |  |  | -0.000122 | -8.01E-05 | -5.70E-05 | -1.46E-05 | 7.09E-06 |
| 396 | 6156 | 0.000233 | -8.71E-05 | -0.000125 | -8.17E-05 | -5.89E-05 | -1.72E-05 | 7.35E-06 |
| 397 | 6165 | 0.000233 | -8.71E-05 | -0.000124 | -8.16E-05 | -5.86E-05 | -1.67E-05 |  |
| 398 | 6185 |  |  | -8.16E-05 |  |  |  |  |  |
| 399 | 6199 | 0.000233 | -8.18E-05 | -0.000124 | -8.07E-05 | -5.75E-05 | -1.51E-05 | 7.49E-06 |
| 400 | 6246 |  |  | -7.76E-05 | -0.000114 | -7.26E-05 | -4.74E-05 | -2.63E-06 |  |
| 401 | 6249 | 0.00027 |  |  |  |  |  |  |
| 402 | 6250 | 0.000347 | -7.74E-05 | -0.000119 | -7.67E-05 | -5.19E-05 | -5.71E-06 |  |
| 403 | 6273 | 0.000431 | -7.95E-05 | -0.00012 | -7.76E-05 | -5.23E-05 | -7.30E-06 | 6.25E-06 |
| 404 | 6274 | 0.00013 | -8.81E-05 |  |  |  |  | 6.55E-06 |
| 405 | 6278 |  |  |  | -0.000128 | -8.45E-05 | -5.83E-05 | -7.72E-06 |  |
| 406 | 6307 |  |  |  |  |  |  |  | 9.12E-06 |
| 407 | 6321 |  |  | -7.75E-05 | -0.000104 | -6.17E-05 | -3.66E-05 |  |  |
| 408 | 6334 | 0.000137 |  |  |  |  | 9.22E-06 |  |

| | A | B | C | D | E | F | G | H |
|---|---|---|---|---|---|---|---|---|
| 409 | 6484 | 0.000132 | -8.28E-05 | -0.000112 | -6.81E-05 | -4.56E-05 | 3.25E-06 | |
| 410 | 6493 | | | | | | | 9.58E-06 |
| 411 | 6503 | | | | | | | 9.73E-06 |
| 412 | 6523 | | -6.01E-05 | -8.74E-05 | -4.30E-05 | -1.76E-05 | 2.12E-05 | |
| 413 | 6529 | 0.000137 | -6.18E-05 | -9.03E-05 | -4.52E-05 | -2.08E-05 | 1.88E-05 | |
| 414 | 6534 | | | | | | | 1.03E-05 |
| 415 | 6565 | 0.000188 | 5.01E-05 | -8.98E-05 | -4.47E-05 | -1.97E-05 | 1.93E-05 | |
| 416 | 6566 | 0.000175 | 2.49E-05 | -9.13E-05 | -4.63E-05 | -2.11E-05 | 1.85E-05 | 8.37E-06 |
| 417 | 6581 | | 2.81E-05 | -8.69E-05 | -4.22E-05 | -1.73E-05 | 2.55E-05 | |
| 418 | 6606 | 3.98E-05 | | | | | | |
| 419 | 6613 | 3.66E-05 | 2.41E-05 | -8.84E-05 | -4.33E-05 | -1.66E-05 | 2.56E-05 | |
| 420 | 6629 | | | | | | | 9.72E-06 |
| 421 | 6692 | | 3.27E-05 | | | | 3.35E-05 | |
| 422 | 6731 | | | -6.90E-05 | -2.50E-05 | 3.67E-06 | | |
| 423 | 6741 | 3.81E-05 | | | | | | |
| 424 | 6749 | | 3.65E-05 | -5.53E-05 | -1.28E-05 | 1.27E-05 | 3.59E-05 | 1.13E-05 |
| 425 | 6772 | 1.09E-05 | 1.36E-05 | -8.72E-05 | -4.60E-05 | -1.46E-05 | 2.72E-05 | |
| 426 | 6773 | 8.08E-06 | 1.23E-05 | -8.83E-05 | -4.74E-05 | -1.58E-05 | 2.63E-05 | |
| 427 | 6811 | -1.97E-07 | | | | | | 1.16E-05 |
| 428 | 6823 | | 1.34E-06 | -0.000115 | -7.21E-05 | -3.72E-05 | 1.35E-05 | |
| 429 | 6899 | -3.81E-06 | | | | | | 8.66E-06 |
| 430 | 6904 | | 9.10E-08 | -0.000116 | -7.84E-05 | -4.58E-05 | 9.12E-06 | 1.02E-05 |
| 431 | 6940 | -2.94E-06 | -2.82E-06 | -0.000117 | -7.90E-05 | -4.27E-05 | 8.16E-06 | |
| 432 | 6946 | 7.59E-06 | 8.10E-06 | -0.000103 | -6.86E-05 | -3.49E-05 | 1.80E-05 | 7.81E-06 |

| | A | B | C | D | E | F | G | H |
|---|---|---|---|---|---|---|---|---|
| 1 | Rank | Inf | 0.002 | 0.0015 | 0.0013 | 0.001 | 0.0005 | 0.0001 |
| 2 | 0 | -9.21295E-05 | -8.92224E-05 | -0.000102058 | -0.0001112 | -9.98877E-05 | -8.11603E-05 | -2.42757E-06 |
| 3 | 152.255362 | | | | | -7.09744E-05 | | |
| 4 | 175.591771 | | | | -7.11856E-05 | | | |
| 5 | 187.663943 | | | -6.58587E-05 | | | | |
| 6 | 219.278896 | -6.00398E-05 | -5.64464E-05 | | | | | |
| 7 | 237.849598 | | | | | | | -1.6497E-06 |
| 8 | 332.867183 | | | | | | | -1.49087E-06 |
| 9 | 343.619636 | | | | | | -4.74191E-05 | |
| 10 | 381.102026 | | | | | | | -1.58475E-06 |
| 11 | 389.681155 | | | | | -5.39389E-05 | | |
| 12 | 391.536477 | | | | -5.65986E-05 | | | |
| 13 | 393.501455 | | | -5.58367E-05 | | | | |
| 14 | 407.861694 | -5.87543E-05 | -5.61915E-05 | | | | | |
| 15 | 414.327217 | | | | | | | -1.62466E-06 |
| 16 | 420.694093 | | | | | | | -1.67093E-06 |
| 17 | 448.822556 | | | | | | -4.42583E-05 | |
| 18 | 451.222982 | | | | | | | -1.77052E-06 |
| 19 | 452.190024 | | | | | -5.40008E-05 | | |
| 20 | 457.081803 | | | | -5.61455E-05 | | | |
| 21 | 463.63976 | | | | | | | -1.59709E-06 |
| 22 | 477.053894 | | | -5.51163E-05 | | | | |
| 23 | 490.591071 | -5.88337E-05 | -5.612E-05 | | | | | |
| 24 | 526.018331 | | | | | | | -1.73107E-06 |
| 25 | 534.54586 | | | | | | -4.40005E-05 | -1.74594E-06 |
| 26 | 537.993891 | | | | | | | -1.7877E-06 |
| 27 | 565.074714 | | | | | -5.51544E-05 | | |
| 28 | 567.343602 | | | | -5.84508E-05 | | | |
| 29 | 573.407801 | | | -5.72889E-05 | | | | |
| 30 | 581.594396 | | -5.79752E-05 | | | | | |
| 31 | 583.535849 | -5.99439E-05 | | | | | | |
| 32 | 588.572276 | | | | | | | -1.80245E-06 |
| 33 | 623.044237 | | | | | | | -8.13585E-07 |
| 34 | 627.975385 | | | | | | -4.65477E-05 | |
| 35 | 631.878733 | | | | | | | -7.38441E-07 |
| 36 | 639.906845 | | | | | -5.61716E-05 | | |
| 37 | 642.736466 | | | | -5.9616E-05 | | | |
| 38 | 643.835602 | | | -5.79013E-05 | | | | -5.3018E-07 |
| 39 | 652.133588 | | -5.86783E-05 | | | | | |
| 40 | 656.019555 | -6.00369E-05 | | | | | | |
| 41 | 693.760288 | | | | | | | -1.05898E-06 |
| 42 | 697.483108 | | | | | | -4.36565E-05 | |
| 43 | 709.32005 | | | | | | | -1.00237E-06 |
| 44 | 713.582714 | | | | | | | -1.06386E-06 |
| 45 | 730.956255 | | | | | -5.23473E-05 | | |
| 46 | 734.825335 | | | | -5.54609E-05 | | | |
| 47 | 754.279456 | | | -5.52465E-05 | | | | |
| 48 | 763.722166 | | | | | | | -1.17798E-06 |
| 49 | 770.958384 | -5.90123E-05 | -5.73087E-05 | | | | | |
| 50 | 776.243552 | | | | | | | -1.13253E-06 |
| 51 | 791.674702 | | | | | | -4.16151E-05 | |
| 52 | 802.135639 | | | | | | | -1.39242E-06 |
| 53 | 811.110558 | | | | | -5.23163E-05 | | -1.39106E-06 |
| 54 | 816.655 | | | | -5.42496E-05 | | | |
| 55 | 823.85188 | | | -5.41145E-05 | | | | |
| 56 | 832.761115 | -5.3552E-05 | -5.21123E-05 | | | | | -1.4032E-06 |
| 57 | 833.300667 | | | | | | | -1.4032E-06 |
| 58 | 840.253602 | | | | | | -3.50703E-05 | |
| 59 | 858.747743 | | | | | | | -2.52959E-06 |
| 60 | 880.868078 | | | | | -3.95597E-05 | | |
| 61 | 887.100055 | | | | -4.31856E-05 | | | |
| 62 | 888.59151 | | | | | | | -2.90283E-06 |
| 63 | 905.812426 | | | -4.08551E-05 | | | | |
| 64 | 916.831142 | | -4.83665E-05 | | | | | |
| 65 | 916.993 | -5.05202E-05 | | | | | | |
| 66 | 927.860637 | | | | | | -2.82322E-05 | |
| 67 | 943.525274 | | | | | | | -2.98376E-06 |
| 68 | 978.665959 | | | | | | | -3.05695E-06 |
| 69 | 988.496144 | | | | | -3.78708E-05 | | |
| 70 | 999.499224 | | | | -4.09505E-05 | | | |
| 71 | 1008.74621 | | | -4.00938E-05 | | | | |
| 72 | 1024.79646 | | -4.92356E-05 | | | | | -2.9717E-06 |
| 73 | 1026.23481 | -5.07717E-05 | | | | | | |
| 74 | 1028.41519 | | | | | | -2.64025E-05 | |
| 75 | 1030.17133 | | | | | | | -2.93313E-06 |
| 76 | 1047.23161 | | | | | -3.69478E-05 | | |
| 77 | 1048.1803 | | | | | | | -2.74514E-06 |
| 78 | 1050.61055 | | | | -3.98407E-05 | | | |
| 79 | 1062.68209 | | | -3.70868E-05 | | | | |
| 80 | 1079.0512 | | | | | | -2.43787E-05 | |
| 81 | 1081.1413 | | -4.48743E-05 | | | | | |
| 82 | 1084.77644 | -4.7187E-05 | | | | | | |
| 83 | 1099.39022 | | | | | | | -2.37419E-06 |
| 84 | 1139.1237 | | | | | -3.58901E-05 | | |
| 85 | 1146.78509 | | | | -3.86177E-05 | | | |
| 86 | 1147.35016 | | | | | | | -2.48031E-06 |
| 87 | 1160.4683 | | | -3.49992E-05 | | | | |
| 88 | 1168.28424 | | | | | | -2.27655E-05 | |
| 89 | 1174.95313 | -4.55897E-05 | -4.34361E-05 | | | | | |
| 90 | 1182.03253 | | | | | | | -2.55713E-06 |
| 91 | 1198.55144 | | | | | | | -2.7386E-06 |
| 92 | 1201.42394 | | | | | -3.56949E-05 | | |
| 93 | 1208.71765 | | | | -3.78598E-05 | | | |
| 94 | 1224.14562 | | | -3.42142E-05 | | | | |
| 95 | 1226.5363 | | | | | | -2.22866E-05 | -2.93448E-06 |
| 96 | 1234.64904 | | -4.17798E-05 | | | | | |
| 97 | 1234.73972 | -4.29745E-05 | | | | | | |
| 98 | 1254.15057 | | | | | | | -3.20475E-06 |
| 99 | 1264.51709 | | | | | -3.45284E-05 | | |
| 100 | 1271.81744 | | | | -3.6905E-05 | | | |
| 101 | 1286.31471 | | | -3.33934E-05 | | | | |
| 102 | 1287.83711 | | | | | | -2.11309E-05 | |

|  | A | B | C | D | E | F | G | H |
|---|---|---|---|---|---|---|---|---|
| 103 | 1289.98423 |  |  |  |  |  |  | -3.65479E-06 |
| 104 | 1302.17574 |  | -4.24419E-05 |  |  |  |  |  |
| 105 | 1307.7701 | -4.36223E-05 |  |  |  |  |  |  |
| 106 | 1325.11142 |  |  |  |  |  |  | -3.79351E-06 |
| 107 | 1360.28803 |  |  |  |  |  |  | -3.76084E-06 |
| 108 | 1360.87041 |  |  |  |  | -3.15193E-05 |  |  |
| 109 | 1362.46513 |  |  |  | -3.40121E-05 |  |  |  |
| 110 | 1372.79098 |  |  |  |  |  | -1.99297E-05 |  |
| 111 | 1375.61864 |  |  | -3.02209E-05 |  |  |  |  |
| 112 | 1392.65829 |  | -4.02728E-05 |  |  |  |  |  |
| 113 | 1395.1362 | -4.11422E-05 |  |  |  |  |  |  |
| 114 | 1398.85035 |  |  |  |  |  |  | -3.76689E-06 |
| 115 | 1439.13583 |  |  |  |  |  |  | -3.76709E-06 |
| 116 | 1440.33404 |  |  |  |  | -2.96604E-05 |  |  |
| 117 | 1447.6892 |  |  |  | -3.24643E-05 |  |  |  |
| 118 | 1455.59763 |  |  |  |  |  | -1.61495E-05 |  |
| 119 | 1458.06992 |  |  | -2.69474E-05 |  |  |  |  |
| 120 | 1472.07881 |  | -3.8548E-05 |  |  |  |  |  |
| 121 | 1472.41373 | -3.92197E-05 |  |  |  |  |  |  |
| 122 | 1490.38165 |  |  |  |  |  |  | -3.7879E-06 |
| 123 | 1498.73003 |  |  |  |  | -3.03838E-05 |  |  |
| 124 | 1513.56451 |  |  |  | -3.38609E-05 |  |  |  |
| 125 | 1525.71906 |  |  |  |  |  | -1.67084E-05 | -3.67134E-06 |
| 126 | 1529.01978 |  |  | -2.89072E-05 |  |  |  |  |
| 127 | 1544.51934 |  | -3.89898E-05 |  |  |  |  |  |
| 128 | 1544.97694 | -4.01505E-05 |  |  |  |  |  |  |
| 129 | 1563.72767 |  |  |  |  |  |  | -3.79084E-06 |
| 130 | 1570.08817 |  |  |  |  | -2.8678E-05 |  |  |
| 131 | 1574.97489 |  |  |  | -3.26457E-05 |  |  |  |
| 132 | 1579.18529 |  |  |  |  |  | -1.59592E-05 |  |
| 133 | 1585.32751 |  |  | -2.91166E-05 |  |  |  |  |
| 134 | 1603.35512 |  |  |  |  |  |  | -4.04174E-06 |
| 135 | 1606.18998 |  | -3.88799E-05 |  |  |  |  |  |
| 136 | 1608.12319 | -4.00105E-05 |  |  |  |  |  |  |
| 137 | 1631.66407 |  |  |  |  | -2.85729E-05 |  |  |
| 138 | 1634.2147 |  |  |  |  |  |  | -3.91303E-06 |
| 139 | 1638.98268 |  |  |  | -3.32652E-05 |  |  |  |
| 140 | 1639.26835 |  |  |  |  |  | -1.66221E-05 |  |
| 141 | 1646.61244 |  |  | -2.99136E-05 |  |  |  |  |
| 142 | 1654.88745 |  | -3.94662E-05 |  |  |  |  | -4.06253E-06 |
| 143 | 1660.41199 | -4.10782E-05 |  |  |  |  |  |  |
| 144 | 1695.59155 |  |  |  |  | -2.92131E-05 |  |  |
| 145 | 1695.76837 |  |  |  |  |  |  | -3.92947E-06 |
| 146 | 1702.62092 |  |  |  |  |  | -1.67682E-05 |  |
| 147 | 1705.04635 |  |  |  | -3.24692E-05 |  |  |  |
| 148 | 1723.93886 |  |  | -2.70308E-05 |  |  |  |  |
| 149 | 1730.21754 |  |  |  |  |  |  | -3.83066E-06 |
| 150 | 1737.31863 |  | -3.88378E-05 |  |  |  |  |  |
| 151 | 1740.18432 | -4.04835E-05 |  |  |  |  |  |  |
| 152 | 1742.30866 |  |  |  |  |  |  | -3.81191E-06 |
| 153 | 1752.14559 |  |  |  |  | -2.69747E-05 |  |  |
| 154 | 1752.40433 |  |  |  |  |  | -1.47253E-05 |  |
| 155 | 1760.97446 |  |  |  | -3.11799E-05 |  |  |  |
| 156 | 1781.19325 |  |  | -2.59158E-05 |  |  |  |  |
| 157 | 1783.1723 |  |  |  |  |  |  | -3.81018E-06 |
| 158 | 1813.21923 |  | -3.81813E-05 |  |  |  |  |  |
| 159 | 1816.35972 | -3.97071E-05 |  |  |  |  |  |  |
| 160 | 1826.78581 |  |  |  |  |  | -1.40495E-05 |  |
| 161 | 1830.3291 |  |  |  |  | -2.6027E-05 |  |  |
| 162 | 1838.23694 |  |  |  |  |  |  | -3.84989E-06 |
| 163 | 1840.00007 |  |  |  | -3.09923E-05 |  |  |  |
| 164 | 1852.8563 |  |  | -2.58563E-05 |  |  |  |  |
| 165 | 1868.01643 |  |  |  |  |  |  | -3.95103E-06 |
| 166 | 1874.46864 | -4.05525E-05 | -3.91909E-05 |  |  |  |  |  |
| 167 | 1881.15949 |  |  |  |  |  | -1.40089E-05 |  |
| 168 | 1888.02749 |  |  |  |  | -2.56422E-05 |  |  |
| 169 | 1890.56542 |  |  |  |  |  |  | -3.98196E-06 |
| 170 | 1902.00774 |  |  |  | -2.96652E-05 |  |  |  |
| 171 | 1916.0575 |  |  | -2.49909E-05 |  |  |  |  |
| 172 | 1940.38654 |  | -3.67392E-05 |  |  |  |  |  |
| 173 | 1943.44812 | -3.82028E-05 |  |  |  |  |  | -3.87961E-06 |
| 174 | 1944.59745 |  |  |  |  |  | -1.34266E-05 |  |
| 175 | 1959.34135 |  |  |  |  | -2.46622E-05 |  |  |
| 176 | 1976.27516 |  |  |  | -2.96603E-05 |  |  |  |
| 177 | 1984.45178 |  |  |  |  |  |  | -3.69538E-06 |
| 178 | 2004.82822 |  |  | -2.43237E-05 |  |  |  |  |
| 179 | 2016.64368 |  |  |  |  |  | -1.33044E-05 |  |
| 180 | 2017.4526 | -3.91119E-05 | -3.76728E-05 |  |  |  |  |  |
| 181 | 2026.57066 |  |  |  |  |  |  | -3.60225E-06 |
| 182 | 2026.9249 |  |  |  |  | -2.53597E-05 |  |  |
| 183 | 2030.17666 |  |  |  | -3.05608E-05 |  |  |  |
| 184 | 2051.45168 |  |  | -2.40348E-05 |  |  |  |  |
| 185 | 2060.21238 |  |  |  |  |  | -1.17559E-05 |  |
| 186 | 2070.30794 |  | -3.73009E-05 |  |  |  |  |  |
| 187 | 2077.60418 |  |  |  |  |  |  | -3.62164E-06 |
| 188 | 2077.89724 | -3.81112E-05 |  |  |  |  |  |  |
| 189 | 2094.46282 |  |  |  |  | -2.51374E-05 |  |  |
| 190 | 2101.98878 |  |  |  | -3.0113E-05 |  |  |  |
| 191 | 2124.51211 |  |  | -2.45492E-05 |  |  |  |  |
| 192 | 2133.77095 |  |  |  |  |  | -1.06477E-05 |  |
| 193 | 2135.62081 |  |  |  |  |  |  | -3.68937E-06 |
| 194 | 2143.98096 |  | -3.74781E-05 |  |  |  |  |  |
| 195 | 2147.92807 | -3.84313E-05 |  |  |  |  |  |  |
| 196 | 2150.17877 |  |  |  |  |  |  | -3.67505E-06 |
| 197 | 2152.72882 |  |  |  |  | -2.53896E-05 |  |  |
| 198 | 2165.16081 |  |  |  | -2.86573E-05 |  |  |  |
| 199 | 2185.20576 |  |  | -2.14805E-05 |  |  |  |  |
| 200 | 2188.05474 |  |  |  |  |  | -9.58835E-06 |  |
| 201 | 2190.52997 |  |  |  |  |  |  | -3.89127E-06 |
| 202 | 2218.15202 |  | -3.67237E-05 |  |  |  |  |  |
| 203 | 2222.24133 | -3.79694E-05 |  |  |  |  |  |  |
| 204 | 2225.19969 |  |  |  |  | -2.36604E-05 |  |  |

| | A | B | C | D | E | F | G | H |
|---|---|---|---|---|---|---|---|---|
| 205 | 2239.07729 | | | | | | | -3.94986E-06 |
| 206 | 2239.8592 | | | | -2.79209E-05 | | | |
| 207 | 2264.37002 | | | -1.99548E-05 | | | | |
| 208 | 2264.49393 | | | | | | -7.90171E-06 | |
| 209 | 2288.45784 | | -3.12169E-05 | | | | | |
| 210 | 2289.21231 | -3.30123E-05 | | | | | | |
| 211 | 2292.01386 | | | | | -2.24392E-05 | | |
| 212 | 2294.66479 | | | | | | | -3.52379E-06 |
| 213 | 2305.25442 | | | | -2.81165E-05 | | | |
| 214 | 2321.3495 | | | | | | -9.48967E-06 | |
| 215 | 2326.60683 | | | -2.05135E-05 | | | | |
| 216 | 2333.00752 | | | | | | | -3.53641E-06 |
| 217 | 2355.00155 | | -3.49463E-05 | | | | | |
| 218 | 2357.30331 | -3.77054E-05 | | | | | | |
| 219 | 2357.63196 | | | | | -2.20418E-05 | | |
| 220 | 2368.23063 | | | | -2.73427E-05 | | | |
| 221 | 2384.62754 | | | | | | -7.28552E-06 | |
| 222 | 2386.45138 | | | -1.93756E-05 | | | | |
| 223 | 2386.81169 | | | | | | | -3.52358E-06 |
| 224 | 2410.359 | | -3.56192E-05 | | | | | |
| 225 | 2415.0575 | -3.88193E-05 | | | | -2.16539E-05 | | |
| 226 | 2436.32384 | | | | -2.70023E-05 | | | |
| 227 | 2445.30369 | | | | | | -6.39085E-06 | -3.44856E-06 |
| 228 | 2464.06966 | | | -1.92042E-05 | | | | |
| 229 | 2475.51507 | | | | | | | -3.46692E-06 |
| 230 | 2482.80291 | | -3.56563E-05 | | | | | |
| 231 | 2483.23 | | | | | -2.14702E-05 | | |
| 232 | 2484.50722 | -3.92796E-05 | | | | | | |
| 233 | 2491.23418 | | | | -2.69315E-05 | | | |
| 234 | 2494.72805 | | | | | | -5.90807E-06 | |
| 235 | 2521.45101 | | | | | | | -3.41626E-06 |
| 236 | 2530.38438 | | | -1.95654E-05 | | | | |
| 237 | 2553.79698 | | -3.55244E-05 | | | -2.13008E-05 | | |
| 238 | 2560.58438 | -3.84989E-05 | | | | | | |
| 239 | 2572.22828 | | | | | | -5.53581E-06 | |
| 240 | 2572.66698 | | | | -2.62618E-05 | | | |
| 241 | 2586.14666 | | | | | | | -3.56483E-06 |
| 242 | 2594.17584 | | | -1.76167E-05 | | | | |
| 243 | 2630.61179 | | | | | -1.98618E-05 | | |
| 244 | 2632.3209 | | -3.38677E-05 | | | | | |
| 245 | 2634.41475 | -3.74817E-05 | | | | | | |
| 246 | 2636.14996 | | | | | | | -3.51992E-06 |
| 247 | 2636.83014 | | | | | | -5.98544E-06 | |
| 248 | 2644.48884 | | | | -2.57025E-05 | | | |
| 249 | 2683.27445 | | | -1.66442E-05 | | | | |
| 250 | 2701.27085 | | | | | | | -3.91372E-06 |
| 251 | 2720.54539 | | | | | -1.86149E-05 | | |
| 252 | 2722.37808 | | -3.26471E-05 | | | | | |
| 253 | 2726.79743 | -3.60861E-05 | | | | | | |
| 254 | 2726.86193 | | | | | | -4.2997E-06 | |
| 255 | 2732.59925 | | | | -2.40772E-05 | | | |
| 256 | 2742.03012 | | | | | | | -3.84997E-06 |
| 257 | 2744.55641 | | | -1.61704E-05 | | | | |
| 258 | 2766.93145 | | | | | -1.79252E-05 | | |
| 259 | 2773.16254 | | | | | | -4.37048E-06 | |
| 260 | 2776.6656 | | -3.20506E-05 | | | | | |
| 261 | 2779.01158 | -3.48047E-05 | | | | | | |
| 262 | 2783.5409 | | | | -2.30318E-05 | | | |
| 263 | 2796.72402 | | | | | | | -3.74355E-06 |
| 264 | 2811.53247 | | | -1.55054E-05 | | | | |
| 265 | 2844.51584 | | | | | -1.79898E-05 | | |
| 266 | 2851.67122 | | | | | | -3.25862E-06 | |
| 267 | 2857.11112 | | -3.23082E-05 | | | | | |
| 268 | 2860.67817 | -3.52255E-05 | | | | | | |
| 269 | 2863.06908 | | | | -2.35917E-05 | | | |
| 270 | 2865.92248 | | | | | | | -3.61957E-06 |
| 271 | 2902.93064 | | | -1.58671E-05 | | | | |
| 272 | 2925.6432 | | | | | | | -3.53271E-06 |
| 273 | 2928.10242 | | | | | -1.79082E-05 | -3.26252E-06 | |
| 274 | 2945.00972 | | -3.25354E-05 | | | | | |
| 275 | 2948.21138 | -3.50825E-05 | | | -2.38854E-05 | | | |
| 276 | 2962.41541 | | | | | | | -3.55793E-06 |
| 277 | 2970.98023 | | | -1.59825E-05 | | | | |
| 278 | 2989.39374 | | | | | | -3.03599E-06 | |
| 279 | 2990.08879 | | | | | -1.72355E-05 | | |
| 280 | 3006.87354 | | -3.24804E-05 | | | | | |
| 281 | 3010.32839 | | | | -2.33521E-05 | | | |
| 282 | 3012.44152 | -3.45144E-05 | | | | | | |
| 283 | 3013.97614 | | | | | | | -3.58682E-06 |
| 284 | 3040.968 | | | -1.60393E-05 | | | | |
| 285 | 3059.60502 | | | | | | -3.32932E-06 | |
| 286 | 3063.51102 | | | | | -1.72541E-05 | | |
| 287 | 3079.10933 | | | | | | | -3.79827E-06 |
| 288 | 3080.47757 | | -3.27746E-05 | | | | | |
| 289 | 3081.48553 | | | | -2.409E-05 | | | |
| 290 | 3085.28223 | -3.47564E-05 | | | | | | |
| 291 | 3112.11597 | | | -1.72193E-05 | | | | |
| 292 | 3116.69753 | | | | | | -5.25791E-06 | -3.74936E-06 |
| 293 | 3125.75545 | | | | | -1.84093E-05 | | |
| 294 | 3142.96739 | | | | -2.45765E-05 | | | |
| 295 | 3143.21274 | | -3.38754E-05 | | | | | |
| 296 | 3145.24837 | -3.64035E-05 | | | | | | |
| 297 | 3155.24677 | | | | | | | -3.86E-06 |
| 298 | 3166.8055 | | | -1.7303E-05 | | | | |
| 299 | 3171.34794 | | | | | | -5.7501E-06 | |
| 300 | 3180.32304 | | | | | -1.70335E-05 | | |
| 301 | 3191.26351 | | -3.68944E-05 | | -2.50505E-05 | | | -3.90592E-06 |
| 302 | 3192.47933 | -3.84104E-05 | | | | | | |
| 303 | 3210.96526 | | | | | | | -3.92413E-06 |
| 304 | 3224.23541 | | | | | | -8.9102E-06 | |
| 305 | 3227.89273 | | | -2.04423E-05 | | | | |
| 306 | 3241.7268 | | | | | -2.08233E-05 | | |

| | A | B | C | D | E | F | G | H |
|---|---|---|---|---|---|---|---|---|
| 307 | 3247.2174 | | | | | | | -3.95168E-06 |
| 308 | 3260.32997 | | | | -2.61277E-05 | | | |
| 309 | 3265.23317 | | -3.8209E-05 | | | | | |
| 310 | 3267.64462 | -3.95345E-05 | | | | | | |
| 311 | 3292.35642 | | | | | | -8.66651E-06 | |
| 312 | 3296.82926 | | | -2.00501E-05 | | | | |
| 313 | 3305.36395 | | | | | | | -4.15541E-06 |
| 314 | 3318.89814 | | | | | -2.08719E-05 | | |
| 315 | 3343.96033 | | | | -2.65897E-05 | | | |
| 316 | 3358.09027 | | -3.86199E-05 | | | | | |
| 317 | 3366.54976 | -3.98143E-05 | | | | | | |
| 318 | 3375.85162 | | | | | | | -4.09441E-06 |
| 319 | 3382.2143 | | | | | | -8.7767E-06 | |
| 320 | 3388.34124 | | | -1.91022E-05 | | | | |
| 321 | 3390.69488 | | | | | -2.04498E-05 | | |
| 322 | 3408.16002 | | | | -2.67538E-05 | | | |
| 323 | 3415.18639 | | -3.9495E-05 | | | | | -4.1598E-06 |
| 324 | 3418.24382 | -4.07277E-05 | | | | | | |
| 325 | 3422.63511 | | | | | | -8.93385E-06 | |
| 326 | 3442.01013 | | | -1.94428E-05 | | | | |
| 327 | 3453.35235 | | | | | -2.06882E-05 | | |
| 328 | 3462.00006 | | | | | | | -4.20362E-06 |
| 329 | 3471.1175 | | | | -2.76989E-05 | | | |
| 330 | 3484.01666 | | -3.9435E-05 | | | | | |
| 331 | 3490.43258 | | | | | | -9.79378E-06 | |
| 332 | 3490.75547 | -4.05802E-05 | | | | | | |
| 333 | 3506.77358 | | | | | | | -4.34671E-06 |
| 334 | 3522.39548 | | | -2.15081E-05 | | | | |
| 335 | 3531.69967 | | | | | -2.14681E-05 | | |
| 336 | 3558.68395 | | | | -2.89541E-05 | | | |
| 337 | 3559.12266 | | | | | | | -4.60914E-06 |
| 338 | 3564.48129 | | | | | | -1.02586E-05 | |
| 339 | 3567.08806 | | -3.97091E-05 | | | | | |
| 340 | 3571.53241 | -4.11368E-05 | | | | | | |
| 341 | 3578.72467 | | | | | | | -5.03202E-06 |
| 342 | 3588.50678 | | | -2.24412E-05 | | | | |
| 343 | 3595.64423 | | | | | -2.24662E-05 | | |
| 344 | 3621.26564 | | | | -2.96456E-05 | | | -5.18392E-06 |
| 345 | 3623.14111 | | | | | | -1.14965E-05 | |
| 346 | 3633.76641 | | -4.14996E-05 | | | | | |
| 347 | 3634.07366 | -4.1974E-05 | | | | | | |
| 348 | 3663.92716 | | | | | | | -5.22695E-06 |
| 349 | 3667.83448 | | | -2.33181E-05 | | | | |
| 350 | 3675.45552 | | | | | -2.31298E-05 | | |
| 351 | 3695.12802 | | | | -3.02406E-05 | | | |
| 352 | 3696.54938 | | | | | | -1.22809E-05 | |
| 353 | 3703.20115 | | | | | | | -5.20362E-06 |
| 354 | 3717.14053 | | -4.21356E-05 | | | | | |
| 355 | 3723.79877 | -4.25999E-05 | | | | | | |
| 356 | 3755.05618 | | | -2.30218E-05 | | | | |
| 357 | 3757.39255 | | | | | | | -4.75193E-06 |
| 358 | 3758.46592 | | | | | -2.30464E-05 | | |
| 359 | 3778.2672 | | | | -2.99101E-05 | | -1.2053E-05 | -4.80708E-06 |
| 360 | 3786.22669 | | -4.1973E-05 | | | | | |
| 361 | 3791.32724 | -4.29391E-05 | | | | | | |
| 362 | 3796.95788 | | | | | | | -4.38691E-06 |
| 363 | 3823.14527 | | | -2.43837E-05 | | | | |
| 364 | 3826.08232 | | | | | -2.46201E-05 | | |
| 365 | 3831.84122 | | | | | | -1.19788E-05 | |
| 366 | 3838.03919 | | | | -2.95606E-05 | | | |
| 367 | 3845.39777 | | | | | | | -4.50634E-06 |
| 368 | 3866.10552 | | -4.15907E-05 | | | | | |
| 369 | 3870.91003 | -4.32929E-05 | | | | | | |
| 370 | 3898.01183 | | | -2.69375E-05 | | | | |
| 371 | 3898.47695 | | | | | -2.77571E-05 | | |
| 372 | 3900.1838 | | | | | | -1.58774E-05 | |
| 373 | 3915.84973 | | | | -3.04955E-05 | | | |
| 374 | 3929.83922 | | | | | | | -4.52432E-06 |
| 375 | 3936.8329 | | -4.28655E-05 | | | | | |
| 376 | 3951.97258 | -4.42553E-05 | | | | | | |
| 377 | 3962.4032 | | | | | | | -4.50697E-06 |
| 378 | 3967.8998 | | | | | | -1.64754E-05 | |
| 379 | 3974.07561 | | | | | -2.79611E-05 | | |
| 380 | 3974.82888 | | | -2.71122E-05 | | | | |
| 381 | 4000.47523 | | | | -3.00464E-05 | | | |
| 382 | 4011.42381 | | -4.21074E-05 | | | | | |
| 383 | 4015.13214 | | | | | | | -4.60276E-06 |
| 384 | 4017.08881 | -4.43575E-05 | | | | | | |
| 385 | 4020.69295 | | | | | | -1.62357E-05 | |
| 386 | 4032.15582 | | | | | -2.83352E-05 | | |
| 387 | 4035.16439 | | | -2.78424E-05 | | | | -4.6378E-06 |
| 388 | 4039.51558 | | | | -3.01091E-05 | | | |
| 389 | 4069.39891 | | | | | | -1.60918E-05 | |
| 390 | 4070.17165 | | -4.14249E-05 | | | | | |
| 391 | 4078.19706 | -4.37721E-05 | | | | | | |
| 392 | 4088.60975 | | | | | -2.81246E-05 | | |
| 393 | 4092.96796 | | | -2.76312E-05 | | | | |
| 394 | 4097.23454 | | | | | | | -4.7126E-06 |
| 395 | 4113.21873 | | | | -3.03219E-05 | | | |
| 396 | 4139.89987 | | | | | | -1.67527E-05 | |
| 397 | 4141.74884 | | | | | | | -4.72535E-06 |
| 398 | 4149.95594 | | -4.16084E-05 | | | | | |
| 399 | 4160.38641 | -4.4329E-05 | | | | | | |
| 400 | 4175.90196 | | | | | -2.99382E-05 | | |
| 401 | 4178.03503 | | | -2.91632E-05 | | | | |
| 402 | 4183.22258 | | | | | | | -4.72477E-06 |
| 403 | 4196.1654 | | | | -3.21988E-05 | | | |
| 404 | 4196.48468 | | | | | | | -4.73566E-06 |
| 405 | 4210.39097 | | | | | | -1.79088E-05 | |
| 406 | 4225.08047 | | -4.23612E-05 | | | | | -5.18426E-06 |
| 407 | 4226.09688 | -4.42301E-05 | | | | | | |
| 408 | 4236.528 | | | | | -2.97726E-05 | | |

| | A | B | C | D | E | F | G | H |
|---|---|---|---|---|---|---|---|---|
| 409 | 4250.03898 | | | -2.87176E-05 | | | | |
| 410 | 4269.30716 | | | | -3.06315E-05 | | | |
| 411 | 4274.26265 | | | | | | -1.63973E-05 | |
| 412 | 4311.50701 | | | | | | | -5.09398E-06 |
| 413 | 4314.7917 | | -4.12736E-05 | | | | | |
| 414 | 4325.6179 | -4.29208E-05 | | | | | | |
| 415 | 4341.16362 | | | | | -2.93905E-05 | | |
| 416 | 4347.78404 | | | -2.939E-05 | | | | |
| 417 | 4370.81594 | | | | | | | -5.15811E-06 |
| 418 | 4371.88457 | | | | | | -1.60928E-05 | |
| 419 | 4373.9914 | | | | -3.12221E-05 | | | |
| 420 | 4389.99538 | | | | | | | -5.11491E-06 |
| 421 | 4401.98723 | | -4.21499E-05 | | | | | |
| 422 | 4408.95958 | -4.37408E-05 | | | | | | |
| 423 | 4411.24274 | | | | | -2.9854E-05 | | |
| 424 | 4420.45108 | | | -2.9857E-05 | | | | |
| 425 | 4421.33308 | | | | | | | -5.12625E-06 |
| 426 | 4423.39627 | | | | | | -1.6091E-05 | |
| 427 | 4437.26521 | | | | -3.15595E-05 | | | |
| 428 | 4444.4637 | | | | | | | -5.18054E-06 |
| 429 | 4480.88314 | | -4.2205E-05 | | | | | |
| 430 | 4488.76809 | -4.37951E-05 | | | | -2.96629E-05 | | |
| 431 | 4500.37279 | | | | | | | -5.09151E-06 |
| 432 | 4512.62318 | | | | | | -1.62107E-05 | |
| 433 | 4516.99386 | | | -2.96905E-05 | | | | |
| 434 | 4532.12851 | | | | -3.11087E-05 | | | |
| 435 | 4562.72611 | | | | | | | -5.10373E-06 |
| 436 | 4584.17987 | | -4.1779E-05 | | | | | |
| 437 | 4586.49009 | | | | | -2.87881E-05 | | |
| 438 | 4588.06751 | | | | | | -1.54203E-05 | |
| 439 | 4588.14367 | -4.34689E-05 | | | | | | |
| 440 | 4596.19632 | | | -2.88069E-05 | | | | |
| 441 | 4600.52145 | | | | -2.99045E-05 | | | |
| 442 | 4611.28721 | | | | | | | -5.05787E-06 |
| 443 | 4660.09508 | | | | | | -1.53572E-05 | |
| 444 | 4660.18091 | | | | | | | -4.84491E-06 |
| 445 | 4661.51134 | | -4.1586E-05 | | | | | |
| 446 | 4662.51339 | | | | | -2.86832E-05 | | |
| 447 | 4682.29368 | -4.33431E-05 | | | | | | |
| 448 | 4699.63567 | | | -2.879E-05 | | | | |
| 449 | 4726.51321 | | | | -2.92687E-05 | | | |
| 450 | 4749.57353 | | | | | | | -4.73438E-06 |
| 451 | 4764.68677 | | | | | | -1.54603E-05 | |
| 452 | 4784.06555 | | | | | -2.84075E-05 | | |
| 453 | 4786.98414 | | -4.15335E-05 | | | | | |
| 454 | 4793.17708 | -4.34438E-05 | | | | | | |
| 455 | 4807.33261 | | | -2.87816E-05 | | | | |
| 456 | 4825.81025 | | | | -2.88869E-05 | | | |
| 457 | 4825.98188 | | | | | | | -4.90075E-06 |
| 458 | 4875.16404 | | | | | | -1.451E-05 | |
| 459 | 4897.08661 | | | | | -2.81381E-05 | | |
| 460 | 4898.98813 | -4.29343E-05 | -4.09933E-05 | | | | | |
| 461 | 4903.41512 | | | -2.79338E-05 | | | | |
| 462 | 4904.36107 | | | | | | | -4.93788E-06 |
| 463 | 4908.48798 | | | | -2.87296E-05 | | | |
| 464 | 4938.33388 | | | | | | -1.4731E-05 | |
| 465 | 4966.56702 | | | | | -2.91211E-05 | | |
| 466 | 4977.01114 | | | | | | | -4.95083E-06 |
| 467 | 4977.32996 | | -4.21087E-05 | | | | | |
| 468 | 4987.9442 | -4.43693E-05 | | | | | | |
| 469 | 4998.0222 | | | -2.88794E-05 | | | | |
| 470 | 5009.29672 | | | | -3.01448E-05 | | | |
| 471 | 5028.09675 | | | | | | | -4.9865E-06 |
| 472 | 5029.91193 | | | | | | -1.69604E-05 | |
| 473 | 5074.70286 | | | | | -2.97013E-05 | | |
| 474 | 5075.96073 | | | | | | | -4.95364E-06 |
| 475 | 5084.87622 | | -4.2902E-05 | | | | | |
| 476 | 5086.86851 | -4.36959E-05 | | | | | | |
| 477 | 5098.42555 | | | -2.73866E-05 | | | | |
| 478 | 5108.28465 | | | | -2.94005E-05 | | | |
| 479 | 5122.08045 | | | | | | -1.86697E-05 | |
| 480 | 5128.05682 | | | | | | | -5.1988E-06 |
| 481 | 5175.51344 | | -4.75536E-05 | | | -3.51719E-05 | | -5.20093E-06 |
| 482 | 5192.698 | -5.03953E-05 | | | | | | |
| 483 | 5208.32005 | | | -3.52935E-05 | | | | |
| 484 | 5212.934 | | | | | | | -5.09446E-06 |
| 485 | 5215.54798 | | | | -3.49826E-05 | | | |
| 486 | 5221.93822 | | | | | | -2.36852E-05 | |
| 487 | 5299.93087 | | | | | -3.51353E-05 | | |
| 488 | 5321.95718 | | -4.80717E-05 | | | | | -5.04143E-06 |
| 489 | 5325.93819 | -5.08488E-05 | | -3.61844E-05 | -3.54868E-05 | | -2.44011E-05 | |
| 490 | 5342.85289 | | | | | | | -5.04742E-06 |
| 491 | 5388.63545 | | | | | -3.67226E-05 | | |
| 492 | 5406.19068 | | | | | | | -4.58497E-06 |
| 493 | 5410.8434 | | -5.1059E-05 | | | | | |
| 494 | 5411.76404 | | | | | | -2.63657E-05 | |
| 495 | 5421.81267 | -5.23336E-05 | | | | | | |
| 496 | 5425.49531 | | | -3.78288E-05 | | | | |
| 497 | 5440.93314 | | | | -3.7413E-05 | | | |
| 498 | 5455.3442 | | | | | | | -4.50699E-06 |
| 499 | 5526.56514 | | | | | -3.75506E-05 | | |
| 500 | 5552.86451 | | | | | | -2.62054E-05 | |
| 501 | 5553.97894 | | | | | | | -4.48477E-06 |
| 502 | 5564.52325 | | -5.16997E-05 | | | | | |
| 503 | 5576.52714 | -5.30493E-05 | | | | | | |
| 504 | 5581.8654 | | | -3.79651E-05 | | | | |
| 505 | 5588.50967 | | | | -3.76281E-05 | | | |
| 506 | 5635.00005 | | | | | | | -4.45314E-06 |
| 507 | 5685.94891 | | | | | -3.66347E-05 | | |
| 508 | 5698.75314 | | | | | | -2.61034E-05 | |
| 509 | 5703.96547 | | | | | | | -4.60257E-06 |
| 510 | 5716.02849 | | -5.28012E-05 | | | | | |

| | A | B | C | D | E | F | G | H |
|---|---|---|---|---|---|---|---|---|
| 511 | 5727.59542 | -5.38714E-05 | | -3.90971E-05 | | | | |
| 512 | 5737.35357 | | | | -3.88485E-05 | | | |
| 513 | 5741.33179 | | | | | | | -4.60591E-06 |
| 514 | 5794.22645 | | | | | | -2.62402E-05 | |
| 515 | 5801.26808 | | | | | -3.8636E-05 | | |
| 516 | 5803.36496 | | | | | | | -4.55602E-06 |
| 517 | 5848.78279 | | -5.25218E-05 | | | | | |
| 518 | 5856.34359 | | | -4.0225E-05 | | | | |
| 519 | 5861.94406 | -5.36579E-05 | | | | | | |
| 520 | 5864.14894 | | | | -3.96812E-05 | | | |
| 521 | 5874.47491 | | | | | | | -4.4712E-06 |
| 522 | 5929.53316 | | | | | | -2.63187E-05 | |
| 523 | 5949.73264 | | | | | | | -4.37503E-06 |
| 524 | 5950.54215 | | | | | -3.93968E-05 | | |
| 525 | 6005.87796 | | -5.29144E-05 | | | | | |
| 526 | 6012.49298 | -5.39618E-05 | | -4.02982E-05 | -3.97698E-05 | | | |
| 527 | 6080.29499 | | | | | | | -4.54239E-06 |
| 528 | 6088.85419 | | | | | | -2.60291E-05 | |
| 529 | 6125.30271 | | | | | -3.98537E-05 | | |
| 530 | 6125.52628 | | | | | | | -4.54423E-06 |
| 531 | 6158.33035 | | -5.32599E-05 | | | | | |
| 532 | 6167.64861 | | | -4.05645E-05 | -4.02934E-05 | | | |
| 533 | 6177.46101 | -5.41542E-05 | | | | | | |
| 534 | 6212.14941 | | | | | | | -4.35484E-06 |
| 535 | 6231.59705 | | | | | | -2.54828E-05 | |
| 536 | 6284.86568 | | | | | -3.89963E-05 | | |
| 537 | 6302.69923 | | | | | | | -4.22088E-06 |
| 538 | 6333.66385 | | -5.16334E-05 | | | | | |
| 539 | 6335.53299 | | | | -4.01891E-05 | | | |
| 540 | 6335.64029 | | | -4.022E-05 | | | | |
| 541 | 6344.78532 | -5.33717E-05 | | | | | | |
| 542 | 6365.96406 | | | | | | | -4.25021E-06 |
| 543 | 6366.13953 | | | | | | -2.63783E-05 | |
| 544 | 6433.84164 | | | | | | | -4.19348E-06 |
| 545 | 6435.70092 | | | | | -4.04676E-05 | | |
| 546 | 6517.0986 | | -5.12435E-05 | | | | | |
| 547 | 6525.03079 | | | | -4.01569E-05 | | | |
| 548 | 6529.39554 | | | -4.03692E-05 | | | | |
| 549 | 6533.34401 | -5.23893E-05 | | | | | -2.60028E-05 | -4.20529E-06 |
| 550 | 6628.03711 | | | | | -4.0043E-05 | | |
| 551 | 6634.93191 | | | | | | | -4.1787E-06 |
| 552 | 6692.62415 | | | | | | -2.57836E-05 | |
| 553 | 6693.46538 | | -5.06475E-05 | | | | | |
| 554 | 6695.2796 | | | | -3.99251E-05 | | | |
| 555 | 6700.8309 | | | -4.00087E-05 | | | | |
| 556 | 6718.40714 | -5.05689E-05 | | | | | | |
| 557 | 6747.99528 | | | | | | | -3.9594E-06 |
| 558 | 6773.20424 | | | | | -3.98857E-05 | | |
| 559 | 6824.45848 | | | | | | | -3.92489E-06 |
| 560 | 6857.15498 | | | | | | -2.53532E-05 | |
| 561 | 6866.63696 | | -4.8734E-05 | | | | | |
| 562 | 6866.88007 | | | | -3.87167E-05 | | | |
| 563 | 6868.68861 | | | -3.86544E-05 | | | | |
| 564 | 6884.75567 | -4.91584E-05 | | | | | | |
| 565 | 6888.51328 | | | | | | | -3.86801E-06 |
| 566 | 6956.16582 | | | | | -3.87057E-05 | | |
| 567 | 7010.36639 | | | | | | -2.5412E-05 | |
| 568 | 7012.35891 | | | | | | | -3.63809E-06 |
| 569 | 7064.47279 | | -4.71301E-05 | | -3.76766E-05 | | | |
| 570 | 7072.56898 | | | -3.60804E-05 | | | | |
| 571 | 7083.67003 | -4.66786E-05 | | | | | | |
| 572 | 7085.85485 | | | | | | | -3.72325E-06 |
| 573 | 7145.13661 | | | | | | | -4.2194E-06 |
| 574 | 7181.74905 | | | | | -3.94717E-05 | | |
| 575 | 7211.08349 | | | | | | -2.58947E-05 | |
| 576 | 7241.64177 | | | | | | | -4.19038E-06 |
| 577 | 7279.21477 | | | | -3.77762E-05 | | | |
| 578 | 7282.17643 | | -4.37912E-05 | | | | | |
| 579 | 7300.32619 | | | -3.69549E-05 | | | | |
| 580 | 7312.06002 | -4.52084E-05 | | | | | | |
| 581 | 7375.78057 | | | | | | | -4.19721E-06 |
| 582 | 7384.70897 | | | | | -3.9052E-05 | | |
| 583 | 7411.16276 | | | | | | -2.59304E-05 | |
| 584 | 7440.23322 | | | | | | | -4.13719E-06 |
| 585 | 7500.31115 | | | | -3.69171E-05 | | | |
| 586 | 7503.60058 | | -4.33663E-05 | | | | | |
| 587 | 7508.0907 | | | -3.68732E-05 | | | | |
| 588 | 7533.79053 | -4.48891E-05 | | | | | | |
| 589 | 7596.21855 | | | | | | | -4.08195E-06 |
| 590 | 7627.05393 | | | | | -3.88733E-05 | | |
| 591 | 7635.6697 | | | | | | -2.54131E-05 | |
| 592 | 7747.18749 | | -4.11077E-05 | -3.635E-05 | -3.62995E-05 | | | -3.99096E-06 |
| 593 | 7763.98387 | -4.32666E-05 | | | | | | |
| 594 | 7854.10457 | | | | | | -2.5022E-05 | |
| 595 | 7856.08505 | | | | | -3.87071E-05 | | |
| 596 | 7867.326 | | | | | | | -3.8923E-06 |
| 597 | 7951.08606 | | | | -3.51457E-05 | | | |
| 598 | 7953.44653 | | -3.65311E-05 | | | | | |
| 599 | 7962.50365 | | | -3.35501E-05 | | | | |
| 600 | 7983.51028 | -3.83336E-05 | | | | | | |
| 601 | 7993.82554 | | | | | | | -3.87817E-06 |
| 602 | 8047.23485 | | | | | -3.84108E-05 | -2.57079E-05 | |
| 603 | 8081.49116 | | | | | | | -3.85523E-06 |
| 604 | 8151.86242 | | | | | -3.5903E-05 | | |
| 605 | 8163.86096 | | -3.5885E-05 | | | | | |
| 606 | 8164.31687 | | | -3.29069E-05 | | | | |
| 607 | 8193.64148 | -3.76926E-05 | | | | | | |
| 608 | 8214.37289 | | | | | | | -3.4764E-06 |
| 609 | 8228.35838 | | | | | | -2.60856E-05 | |
| 610 | 8251.33791 | | | | | -3.78867E-05 | | |
| 611 | 8397.02597 | | | | -3.48584E-05 | | | |
| 612 | 8406.33621 | | -3.28081E-05 | | | | | |

| | A | B | C | D | E | F | G | H |
|---|---|---|---|---|---|---|---|---|
| 613 | 8407.71729 | | | | | | | -3.43494E-06 |
| 614 | 8415.38986 | | | -3.05541E-05 | | | | |
| 615 | 8423.81305 | -3.45912E-05 | | | | | | |
| 616 | 8446.37969 | | | | | | -2.63538E-05 | |
| 617 | 8494.33431 | | | | | -3.78692E-05 | | |
| 618 | 8657.26269 | | | | -3.42114E-05 | | | -3.19582E-06 |
| 619 | 8660.10487 | | -2.99133E-05 | | | | | |
| 620 | 8661.36947 | | | -2.88902E-05 | | | | |
| 621 | 8670.45801 | -3.14046E-05 | | | | | | |
| 622 | 8700.14619 | | | | | | -2.77876E-05 | |
| 623 | 8759.18221 | | | | | -3.91173E-05 | | |
| 624 | 8762.78256 | | | | | | | -3.10121E-06 |
| 625 | 8898.89407 | | | | -3.49456E-05 | | | |
| 626 | 8907.6587 | | -3.03484E-05 | | | | | |
| 627 | 8920.1647 | | | -2.87827E-05 | | | | |
| 628 | 8936.48047 | -3.13282E-05 | | | | | | |
| 629 | 8952.8477 | | | | | | -2.71369E-05 | -2.91922E-06 |
| 630 | 9026.3371 | | | | | -3.88612E-05 | | |
| 631 | 9068.45405 | | | | | | | -2.74543E-06 |
| 632 | 9138.23741 | | | | -3.39391E-05 | | | |
| 633 | 9143.93442 | | -2.95246E-05 | | | | | |
| 634 | 9150.99606 | | | -2.80361E-05 | | | | |
| 635 | 9160.63297 | | | | | | -2.72422E-05 | |
| 636 | 9166.48291 | -3.05518E-05 | | | | | | |
| 637 | 9227.18549 | | | | | | | -2.83728E-06 |
| 638 | 9283.21611 | | | | | -3.90104E-05 | | |
| 639 | 9406.42149 | | | | -3.31101E-05 | | | |
| 640 | 9425.58897 | | -2.53121E-05 | -2.66076E-05 | | | | |
| 641 | 9433.46611 | | | | | | | -2.57522E-06 |
| 642 | 9453.13097 | | | | | | -2.73795E-05 | |
| 643 | 9468.85728 | -2.58381E-05 | | | | | | |
| 644 | 9567.3598 | | | | | -3.85525E-05 | | -2.58687E-06 |
| 645 | 9617.20145 | | | | | | | -2.49113E-06 |
| 646 | 9674.25754 | | | | -3.38661E-05 | | | |
| 647 | 9703.73108 | | -2.46804E-05 | | | | | |
| 648 | 9705.08013 | | | -2.68124E-05 | | | | |
| 649 | 9719.58612 | | | | | | -2.73986E-05 | |
| 650 | 9729.08819 | -2.46147E-05 | | | | | | |
| 651 | 9744.67408 | | | | | | | -2.74576E-06 |
| 652 | 9783.59677 | | | | | | | -2.75334E-06 |
| 653 | 9821.90602 | | | | | -3.7025E-05 | | |
| 654 | 9983.65615 | | | | -3.0789E-05 | | | |
| 655 | 9994.72041 | | -1.40011E-05 | | | | | -2.93204E-06 |
| 656 | 9997.2692 | | | -2.0619E-05 | | | | |
| 657 | 10023.9516 | | | | | | -2.57479E-05 | |
| 658 | 10030.1469 | -1.19022E-05 | | | | | | |
| 659 | 10103.4439 | | | | | | | -2.8851E-06 |
| 660 | 10136.7887 | | | | | -3.63492E-05 | | |
| 661 | 10245.1636 | | | | -2.55321E-05 | | | |
| 662 | 10254.717 | | -8.92581E-06 | | | | | -2.91491E-06 |
| 663 | 10260.5165 | | | -1.10118E-05 | | | -2.49377E-05 | |
| 664 | 10275.0635 | -9.68467E-06 | | | | | | |
| 665 | 10394.2471 | | | | | -1.7731E-05 | | |
| 666 | 10497.5824 | | | | | | | -2.89292E-06 |
| 667 | 10575.5894 | | 1.03659E-06 | | -1.32799E-05 | | | |
| 668 | 10585.532 | | | 4.61168E-06 | | | | |
| 669 | 10607.3644 | -2.3669E-06 | | | | | -2.4928E-05 | |
| 670 | 10773.1487 | | | | | -1.60143E-05 | | |
| 671 | 10814.3709 | | | | | | | -2.77043E-06 |
| 672 | 10968.4862 | | 7.19223E-07 | | | | | |
| 673 | 10985.4848 | | | | -1.21454E-05 | | | |
| 674 | 10985.8163 | | | 8.64881E-07 | | | | |
| 675 | 10985.9415 | | | | | | | -2.83394E-06 |
| 676 | 11005.657 | -3.59273E-06 | | | | | | |
| 677 | 11033.9105 | | | | | | -2.35397E-05 | |
| 678 | 11193.5727 | | | | | -1.34542E-05 | | |
| 679 | 11335.9418 | | | | | | | -2.89009E-06 |
| 680 | 11402.5397 | | 1.82833E-08 | | | | | |
| 681 | 11416.6068 | | | 2.00182E-06 | | | | |
| 682 | 11430.0526 | | | | -8.39426E-06 | | | |
| 683 | 11435.2849 | -3.7829E-06 | | | | | | |
| 684 | 11523.0482 | | | | | | -2.61511E-05 | |
| 685 | 11642.6448 | | | | | | | -3.01395E-06 |
| 686 | 11650.0459 | | | | | -1.39847E-05 | | |
| 687 | 11840.219 | | -7.20163E-07 | | | | | |
| 688 | 11881.0453 | | | -5.24918E-07 | | | | |
| 689 | 11888.8349 | -5.05344E-06 | | | | | | |
| 690 | 11892.5219 | | | | -9.88437E-06 | | | |
| 691 | 11967.4633 | | | | | | | -2.85622E-06 |
| 692 | 12056.0446 | | | | | | -2.84245E-05 | |
| 693 | 12161.6406 | | | | | -1.51348E-05 | | |
| 694 | 12371.5658 | | -1.32271E-07 | | | | | |
| 695 | 12410.7566 | | | | | | | -2.50581E-06 |
| 696 | 12420.0869 | | | -1.14179E-06 | | | | |
| 697 | 12422.2873 | -4.95588E-06 | | | | | | |
| 698 | 12439.7609 | | | | -8.15378E-06 | | | |
| 699 | 12606.0159 | | | | | | -2.67908E-05 | |
| 700 | 12677.939 | | | | | -1.04719E-05 | | |
| 701 | 12752.5168 | | | | | | | -2.09471E-06 |
| 702 | 12880.4911 | | 3.4381E-06 | | | | | |
| 703 | 12958.3443 | -1.47853E-06 | | | | | | |
| 704 | 12964.0361 | | | 3.47737E-06 | | | | |
| 705 | 12989.0053 | | | | -1.71561E-06 | | | |
| 706 | 13168.0805 | | | | | | -8.83286E-06 | |
| 707 | 13300.1321 | | | | | -7.14993E-06 | | |
| 708 | 13497.8807 | | 3.7589E-06 | | | | | -1.89541E-06 |
| 709 | 13581.1977 | -1.5166E-06 | | | | | | |
| 710 | 13585.1437 | | | 4.78969E-06 | | | | |
| 711 | 13622.9335 | | | | 9.79045E-08 | | | |
| 712 | 13835.3813 | | | | | | -5.43476E-06 | |
| 713 | 13924.0882 | | | | | -4.58225E-06 | | |
| 714 | 14144.9612 | | 7.34387E-06 | | | | | |

| | A | B | C | D | E | F | G | H |
|---|---|---|---|---|---|---|---|---|
| 715 | 14219.5653 | | | | | | | -1.51779E-06 |
| 716 | 14220.4441 | 1.03728E-06 | | | | | | |
| 717 | 14255.2079 | | | 6.8579E-06 | | | | |
| 718 | 14287.1646 | | | | 2.05022E-06 | | | |
| 719 | 14428.4595 | | | | | | 5.84436E-06 | |
| 720 | 14550.2367 | | | | | 2.63372E-06 | | |
| 721 | 14812.6715 | | 7.7261E-06 | | | | | |
| 722 | 14849.6868 | | | | | | | -3.02084E-08 |
| 723 | 14887.1219 | 1.70794E-06 | | | | | | |
| 724 | 14901.0814 | | | 1.14522E-05 | | | | |
| 725 | 14975.9889 | | | | 7.44076E-06 | | | |
| 726 | 15138.2358 | | | | | | 9.11696E-06 | |
| 727 | 15328.7949 | | | | | 4.9267E-06 | | |
| 728 | 15483.9834 | | 8.92467E-06 | | | | | |
| 729 | 15591.9606 | 2.66412E-06 | | | | | | |
| 730 | 15623.0987 | | | 1.26689E-05 | | | | |
| 731 | 15713.775 | | | | 9.67337E-06 | | | |
| 732 | 15754.7989 | | | | | | | 4.21293E-07 |
| 733 | 15823.8497 | | | | | | 9.30704E-06 | |
| 734 | 15979.6292 | | | | | 6.18766E-06 | | |
| 735 | 16095.393 | | 1.11296E-05 | | | | | |
| 736 | 16185.7674 | 5.23258E-06 | | | | | | |
| 737 | 16191.3977 | | | 1.56487E-05 | | | | |
| 738 | 16215.3904 | | | | | | | 6.35292E-07 |
| 739 | 16247.6993 | | | | 1.39887E-05 | | | |
| 740 | 16332.605 | | | | | | 1.33921E-05 | |
| 741 | 16464.099 | | | | | 1.0231E-05 | | |
| 742 | 16556.7503 | | 1.26985E-05 | | | | | |
| 743 | 16588.6296 | 5.70495E-06 | | | | | | |
| 744 | 16621.1454 | | | 1.76742E-05 | | | | |
| 745 | 16672.9054 | | | | 1.73436E-05 | | | |
| 746 | 16721.385 | | | | | | 1.64341E-05 | |
| 747 | 16818.568 | | | | | 1.2709E-05 | | |
| 748 | 16862.3775 | | 1.30036E-05 | | | | | |
| 749 | 16908.5447 | 6.12733E-06 | | | | | | |
| 750 | 16924.8545 | | | | | | | 6.92925E-07 |
| 751 | 16934.2346 | | | 1.88986E-05 | | | | |
| 752 | 16989.8112 | | | | 1.94969E-05 | | | |
| 753 | 17026.5182 | | | | | | 1.71557E-05 | |
| 754 | 17105.938 | | | | | 1.56016E-05 | | |
| 755 | 17128.3539 | | 1.4742E-05 | | | | | |
| 756 | 17159.7672 | 8.11032E-06 | | | | | | |
| 757 | 17193.3832 | | | 2.10199E-05 | | | | |
| 758 | 17238.1684 | | | | 2.23724E-05 | | | |
| 759 | 17266.988 | | | | | | 1.86127E-05 | |
| 760 | 17347.0408 | | | | | | | 9.78063E-07 |
| 761 | 17354.498 | | | | | 1.64887E-05 | | |
| 762 | 17366.0918 | | 1.63898E-05 | | | | | |
| 763 | 17394.105 | 8.97232E-06 | | | | | | |
| 764 | 17411.8176 | | | 2.0873E-05 | | | | |
| 765 | 17459.4792 | | | | 2.21621E-05 | | | |
| 766 | 17476.572 | | | | | | 1.91332E-05 | |
| 767 | 17557.3092 | | | | | 1.7248E-05 | | |
| 768 | 17576.8759 | | 1.72099E-05 | | | | | |
| 769 | 17606.9412 | 9.89266E-06 | | | | | | |
| 770 | 17630.4592 | | | 2.17278E-05 | | | | |
| 771 | 17679.1998 | | | | 2.29859E-05 | | 1.9024E-05 | |
| 772 | 17778.1686 | | | | | 1.80972E-05 | | |
| 773 | 17796.7632 | | 1.90383E-05 | | | | | |
| 774 | 17819.4497 | 1.19638E-05 | | | | | | |
| 775 | 17834.714 | | | 2.26305E-05 | | | | |
| 776 | 17860.4174 | | | | | | | 1.50939E-06 |
| 777 | 17907.2934 | | | | 2.32177E-05 | | 1.85846E-05 | |
| 778 | 17973.0226 | | 2.04689E-05 | | | | | |
| 779 | 17977.1839 | | | | | 1.83475E-05 | | |
| 780 | 18004.7523 | 1.38765E-05 | | | | | | |
| 781 | 18021.6114 | | | 2.54545E-05 | | | | |
| 782 | 18118.6395 | | | | 2.34977E-05 | | 1.86785E-05 | |
| 783 | 18195.293 | | 2.011E-05 | | | | | |
| 784 | 18208.8785 | 1.32763E-05 | | | | 1.77545E-05 | | |
| 785 | 18215.7398 | | | 2.58381E-05 | | | | |
| 786 | 18311.2117 | | | | | | 1.7697E-05 | |
| 787 | 18326.0748 | | | | 2.39088E-05 | | | |
| 788 | 18369.1633 | | | | | | | 1.42476E-06 |
| 789 | 18379.6074 | | 2.0423E-05 | | | | | |
| 790 | 18402.0323 | 1.38418E-05 | | | | | | |
| 791 | 18415.7386 | | | | | 1.91843E-05 | | |
| 792 | 18423.3065 | | | 2.58947E-05 | | | | |
| 793 | 18534.3778 | | | | | | 1.84853E-05 | |
| 794 | 18539.1203 | | | | 2.54136E-05 | | | |
| 795 | 18566.7651 | | 2.02971E-05 | | | | | |
| 796 | 18582.181 | 1.36503E-05 | | | | | | |
| 797 | 18597.3264 | | | 2.57077E-05 | | | | |
| 798 | 18614.4141 | | | | | 1.97292E-05 | | |
| 799 | 18716.8867 | | | | | | 1.75512E-05 | |
| 800 | 18738.1964 | | | | 2.40931E-05 | | | |
| 801 | 18763.9411 | | 1.96658E-05 | | | | | |
| 802 | 18787.6104 | 1.27945E-05 | | | | | | |
| 803 | 18798.3104 | | | 2.45502E-05 | | | | |
| 804 | 18802.4533 | | | | | | | 1.32495E-06 |
| 805 | 18805.9513 | | | | | 1.83152E-05 | | |
| 806 | 18889.4297 | | | | | | 1.80389E-05 | |
| 807 | 18914.9044 | | | | 2.3991E-05 | | | |
| 808 | 18926.0768 | | 1.98784E-05 | | | | | |
| 809 | 18959.6546 | 1.3402E-05 | | | | | | |
| 810 | 18984.7636 | | | 2.56623E-05 | | | | |
| 811 | 19006.2106 | | | | | 1.86781E-05 | | |
| 812 | 19096.1181 | | | | | | 1.87352E-05 | |
| 813 | 19108.9369 | | | | 2.43944E-05 | | | |
| 814 | 19109.9483 | | 1.64887E-05 | | | | | |
| 815 | 19132.5474 | 9.08679E-06 | | | | | | |
| 816 | 19148.2527 | | | | | | | 3.6986E-06 |

| | A | B | C | D | E | F | G | H |
|---|---|---|---|---|---|---|---|---|
| 817 | 19165.3897 | | | 2.38263E-05 | | | | |
| 818 | 19185.1795 | | | | | 1.89326E-05 | | |
| 819 | 19282.3444 | | | | | | 1.92347E-05 | |
| 820 | 19283.3929 | | 1.60788E-05 | | | | | |
| 821 | 19299.4109 | | | | 2.41674E-05 | | | |
| 822 | 19311.8178 | 8.99952E-06 | | | | | | |
| 823 | 19355.1443 | | | 2.2764E-05 | | | | |
| 824 | 19383.4012 | | | | | 1.78639E-05 | | |
| 825 | 19461.6965 | | 1.5712E-05 | | | | | |
| 826 | 19462.4108 | | | | | | 1.85112E-05 | |
| 827 | 19480.9784 | | | | 2.27475E-05 | | | |
| 828 | 19487.987 | 8.85961E-06 | | | | | | |
| 829 | 19488.461 | | | | | | | 3.66212E-06 |
| 830 | 19529.7332 | | | 2.25323E-05 | | | | |
| 831 | 19569.7306 | | | | | 1.83624E-05 | | |
| 832 | 19621.467 | | 1.44851E-05 | | | | | |
| 833 | 19636.5979 | | | | | | 1.79624E-05 | |
| 834 | 19643.4189 | 7.96994E-06 | | | | | | |
| 835 | 19645.5525 | | | | 2.23306E-05 | | | |
| 836 | 19692.4934 | | | 2.11931E-05 | | | | |
| 837 | 19732.3226 | | | | | 1.65316E-05 | | |
| 838 | 19791.6908 | | 1.49611E-05 | | | | | |
| 839 | 19791.8652 | | | | | | | 3.67459E-06 |
| 840 | 19816.3992 | | | | | | 1.78803E-05 | |
| 841 | 19822.7255 | 9.87206E-06 | | | | | | |
| 842 | 19826.0502 | | | | 2.28186E-05 | | | |
| 843 | 19859.1024 | | | 2.12141E-05 | | | | |
| 844 | 19904.7097 | | | | | 1.92163E-05 | | |
| 845 | 19956.7406 | | 1.59351E-05 | | | | | |
| 846 | 19962.997 | 1.01059E-05 | | | | | | |
| 847 | 19968.7608 | | | | | | 1.86453E-05 | |
| 848 | 19971.9652 | | | | 2.23801E-05 | | | |
| 849 | 19987.6766 | | | 2.03509E-05 | | | | |
| 850 | 20032.4952 | | | | | 1.8405E-05 | | |
| 851 | 20068.8951 | | 1.47199E-05 | | | | | |
| 852 | 20072.4797 | | | | | | | 4.77051E-06 |
| 853 | 20090.8993 | 8.87096E-06 | | | | | 1.93092E-05 | |
| 854 | 20108.6329 | | | | 2.24346E-05 | | | |
| 855 | 20129.7088 | | | 1.96828E-05 | | | | |
| 856 | 20179.5476 | | | | | 1.8795E-05 | | |
| 857 | 20240.3781 | | 1.67751E-05 | | | | | |
| 858 | 20267.2655 | | | | | | 1.97943E-05 | |
| 859 | 20271.2613 | 1.03455E-05 | | | | | | |
| 860 | 20278.6864 | | | | 2.27996E-05 | | | |
| 861 | 20305.931 | | | 1.95377E-05 | | | | |
| 862 | 20342.424 | | | | | | | 5.23624E-06 |
| 863 | 20343.9223 | | | | | 1.75875E-05 | | |
| 864 | 20396.3117 | | 1.43782E-05 | | | | | |
| 865 | 20428.7293 | 8.92182E-06 | | | | | 1.90204E-05 | |
| 866 | 20437.3092 | | | | 2.12362E-05 | | | |
| 867 | 20465.7533 | | | 1.87243E-05 | | | | |
| 868 | 20521.8499 | | | | | 1.81321E-05 | | |
| 869 | 20563.1066 | | 1.43637E-05 | | | | | |
| 870 | 20579.2837 | 9.38701E-06 | | | | | | |
| 871 | 20579.4804 | | | | | | 1.95683E-05 | |
| 872 | 20604.6978 | | | | 2.1541E-05 | | | |
| 873 | 20638.8193 | | | 1.85608E-05 | | | | |
| 874 | 20676.0447 | | | | | 1.72162E-05 | | |
| 875 | 20684.1496 | | 1.45596E-05 | | | | | 4.33377E-06 |
| 876 | 20691.6391 | | | | | | 2.05382E-05 | |
| 877 | 20698.8786 | 9.77142E-06 | | | | | | |
| 878 | 20707.6709 | | | | 2.26326E-05 | | | |
| 879 | 20738.1417 | | | 1.9212E-05 | | | | |
| 880 | 20794.3027 | | | | | 1.75641E-05 | | |
| 881 | 20840.3722 | | 1.47799E-05 | | | | | |
| 882 | 20851.5023 | | | | | | 2.01253E-05 | |
| 883 | 20866.4145 | 1.01095E-05 | | | | | | |
| 884 | 20891.4074 | | | | 2.24047E-05 | | | |
| 885 | 20910.2851 | | | 1.89178E-05 | | | | |
| 886 | 20955.4654 | | | | | | | 4.37062E-06 |
| 887 | 20975.7049 | | | | | 1.73111E-05 | | |
| 888 | 21010.024 | | 1.54107E-05 | | | | | |
| 889 | 21019.2072 | 1.06721E-05 | | | | | | |
| 890 | 21019.6244 | | | | | | 2.0235E-05 | |
| 891 | 21031.9366 | | | | 2.5896E-05 | | | |
| 892 | 21051.7348 | | | 2.10232E-05 | | | | |
| 893 | 21093.2354 | | | | | 2.09297E-05 | | |
| 894 | 21136.6178 | | 1.58223E-05 | | | | | |
| 895 | 21172.8719 | | | | | | 2.00187E-05 | |
| 896 | 21181.0409 | 1.1187E-05 | | | | | | |
| 897 | 21199.8422 | | | | 2.64096E-05 | | | |
| 898 | 21225.7364 | | | 2.20157E-05 | | | | |
| 899 | 21271.8858 | | | | | 2.32094E-05 | | |
| 900 | 21273.1746 | | | | | | | 4.22186E-06 |
| 901 | 21312.1826 | | 1.76473E-05 | | | | | |
| 902 | 21332.5304 | | | | | | 1.95907E-05 | |
| 903 | 21345.1422 | 1.13981E-05 | | | | | | |
| 904 | 21353.8507 | | | | 2.78358E-05 | | | |
| 905 | 21367.9096 | | | 2.21759E-05 | | | | |
| 906 | 21438.6167 | | | | | 2.42358E-05 | | |
| 907 | 21480.8352 | | 1.74127E-05 | | | | | |
| 908 | 21488.4563 | | | | | | | 4.02546E-06 |
| 909 | 21511.2894 | 1.20639E-05 | | | | | | |
| 910 | 21514.2319 | | | | | | 1.96843E-05 | |
| 911 | 21526.2485 | | | | 2.80405E-05 | | | |
| 912 | 21534.5411 | | | 2.28934E-05 | | | | |
| 913 | 21602.1308 | | | | | 2.50034E-05 | | |
| 914 | 21638.3501 | | 1.66019E-05 | | | | | |
| 915 | 21654.1753 | 1.17093E-05 | | | | | | |
| 916 | 21657.7216 | | | | | | 1.95676E-05 | |
| 917 | 21663.0558 | | | | 2.83668E-05 | | | |
| 918 | 21667.8033 | | | 2.26697E-05 | | | | |

| | A | B | C | D | E | F | G | H |
|---|---|---|---|---|---|---|---|---|
| 919 | 21718.2266 | | | | | 2.54475E-05 | | |
| 920 | 21733.4394 | | | | | | | 4.00671E-06 |
| 921 | 21762.7546 | | 1.66943E-05 | | | | | |
| 922 | 21776.0541 | 1.20034E-05 | | | | | 1.98214E-05 | |
| 923 | 21795.1363 | | | | 2.71417E-05 | | | |
| 924 | 21800.5654 | | | 2.1451E-05 | | | | |
| 925 | 21836.7303 | | | | | 2.46295E-05 | | |
| 926 | 21871.2364 | | 1.41346E-05 | | | | | |
| 927 | 21898.2575 | 1.26064E-05 | | | | | | |
| 928 | 21901.8148 | | | | | | 1.93577E-05 | |
| 929 | 21922.1411 | | | | 2.87271E-05 | | | |
| 930 | 21941.8837 | | | 2.31345E-05 | | | | |
| 931 | 21988.4319 | | | | | | | 4.32182E-06 |
| 932 | 22005.3232 | | | | | 2.47403E-05 | | |
| 933 | 22057.4249 | | 1.45413E-05 | | | | | |
| 934 | 22074.007 | 1.10231E-05 | | | | | | |
| 935 | 22092.3657 | | | | 2.8483E-05 | | | |
| 936 | 22097.8677 | | | | | | 1.95096E-05 | |
| 937 | 22102.8746 | | | 2.36757E-05 | | | | |
| 938 | 22163.8574 | | | | | 2.60588E-05 | | |
| 939 | 22191.9165 | | 1.56651E-05 | | | | | |
| 940 | 22218.6345 | 1.209E-05 | | | | | | |
| 941 | 22248.6456 | | | | 3.07198E-05 | | | |
| 942 | 22262.846 | | | 2.53608E-05 | | | | |
| 943 | 22265.1327 | | | | | | 2.07257E-05 | |
| 944 | 22282.7533 | | | | | | | 4.62423E-06 |
| 945 | 22315.4728 | | | | | 2.59599E-05 | | |
| 946 | 22365.7237 | | 1.60581E-05 | | | | | |
| 947 | 22386.4415 | 1.24999E-05 | | | | | | |
| 948 | 22414.1816 | | | | 3.03582E-05 | | | |
| 949 | 22428.6076 | | | 2.59137E-05 | | | | |
| 950 | 22465.8506 | | | | | | 2.32585E-05 | |
| 951 | 22511.6539 | | | | | 2.71833E-05 | | |
| 952 | 22566.6911 | | 1.69043E-05 | | | | | |
| 953 | 22586.9623 | 1.19058E-05 | | | | | | |
| 954 | 22605.3683 | | | | 2.98251E-05 | | | |
| 955 | 22612.2363 | | | 2.54142E-05 | | | | |
| 956 | 22642.4105 | | | | | | | 4.75574E-06 |
| 957 | 22663.2658 | | | | | | 2.31522E-05 | |
| 958 | 22681.8126 | | | | | 2.75288E-05 | | |
| 959 | 22723.963 | | 1.84022E-05 | | | | | |
| 960 | 22745.8199 | 1.25273E-05 | | | | | | |
| 961 | 22776.7467 | | | | 3.11932E-05 | | | |
| 962 | 22787.6934 | | | 2.66932E-05 | | | | |
| 963 | 22850.8897 | | | | | 2.87994E-05 | | |
| 964 | 22852.1883 | | | | | | 2.3153E-05 | |
| 965 | 22905.2639 | | 1.84346E-05 | | | | | |
| 966 | 22927.5353 | 1.26602E-05 | | | | | | |
| 967 | 22952.763 | | | | 3.00241E-05 | | | |
| 968 | 22958.644 | | | 2.64245E-05 | | | | |
| 969 | 22967.208 | | | | | | | 4.50272E-06 |
| 970 | 23035.7146 | | | | | 2.79535E-05 | | |
| 971 | 23049.4433 | | | | | | 2.27699E-05 | |
| 972 | 23083.5615 | | 1.83496E-05 | | | | | |
| 973 | 23110.1763 | 1.30389E-05 | | | | | | |
| 974 | 23141.5659 | | | | 3.0782E-05 | | | |
| 975 | 23148.8004 | | | 2.63839E-05 | | | | |
| 976 | 23190.1005 | | | | | | | 4.58546E-06 |
| 977 | 23229.9001 | | | | | 2.65818E-05 | | |
| 978 | 23250.6367 | | | | | | 2.29023E-05 | |
| 979 | 23261.5861 | | 1.68996E-05 | | | | | |
| 980 | 23312.863 | 1.18548E-05 | | | | | | |
| 981 | 23339.0149 | | | | 2.83724E-05 | | | |
| 982 | 23343.3111 | | | 2.55936E-05 | | | | |
| 983 | 23409.889 | | | | | | | 5.05336E-06 |
| 984 | 23418.025 | | | | | 2.58636E-05 | | |
| 985 | 23447.7108 | | | | | | 2.15194E-05 | |
| 986 | 23469.3458 | | 1.68616E-05 | | | | | |
| 987 | 23488.8514 | 1.16256E-05 | | | | | | |
| 988 | 23510.6259 | | | | 2.82458E-05 | | | |
| 989 | 23513.1456 | | | 2.50739E-05 | | | | |
| 990 | 23554.7746 | | | | | | | 5.67288E-06 |
| 991 | 23594.288 | | | | | 2.38376E-05 | | |
| 992 | 23632.004 | | | | | | 2.08433E-05 | |
| 993 | 23635.231 | | 1.7415E-05 | | | | | |
| 994 | 23676.3043 | 1.21617E-05 | | | | | | |
| 995 | 23701.2568 | | | | 2.65644E-05 | | | |
| 996 | 23702.5403 | | | 2.5355E-05 | | | | |
| 997 | 23783.2266 | | | | | 2.30502E-05 | | |
| 998 | 23822.9783 | | 1.78624E-05 | | | | | |
| 999 | 23824.2995 | | | | | | 2.10633E-05 | |
| 1000 | 23842.1006 | 1.2703E-05 | | | | | | 5.35262E-06 |
| 1001 | 23861.9314 | | | | 2.60597E-05 | | | |
| 1002 | 23864.3079 | | | 2.477E-05 | | | | |
| 1003 | 23956.6943 | | | | | 2.32924E-05 | | |
| 1004 | 24003.088 | | 1.70008E-05 | | | | | |
| 1005 | 24009.5952 | | | | | | 2.03426E-05 | |
| 1006 | 24030.7483 | 1.23782E-05 | | | | | | |
| 1007 | 24051.0152 | | | | 2.53429E-05 | | | |
| 1008 | 24058.7763 | | | 2.41309E-05 | | | | 5.09517E-06 |
| 1009 | 24140.6838 | | | | | 2.24981E-05 | | |
| 1010 | 24210.3334 | | 1.81671E-05 | | | | | |
| 1011 | 24223.4612 | | | | | | 2.06983E-05 | |
| 1012 | 24248.2522 | 1.26144E-05 | | | | | | |
| 1013 | 24270.6633 | | | | 2.68185E-05 | | | |
| 1014 | 24277.7602 | | | 2.52177E-05 | | | | |
| 1015 | 24332.0263 | | | | | 2.41052E-05 | | |
| 1016 | 24354.7425 | | 1.85903E-05 | | | | | |
| 1017 | 24377.5376 | | | | | | 2.09816E-05 | |
| 1018 | 24391.8171 | 1.26671E-05 | | | | | | |
| 1019 | 24431.7202 | | | 2.47312E-05 | 2.60359E-05 | | | |
| 1020 | 24516.5922 | | | | | | | 5.11237E-06 |

| | A | B | C | D | E | F | G | H |
|---|---|---|---|---|---|---|---|---|
| 1021 | 24531.0771 | | | | | 2.23521E-05 | | |
| 1022 | 24610.6865 | | 1.81322E-05 | | | | | |
| 1023 | 24639.2262 | | | | | | 2.01893E-05 | |
| 1024 | 24641.8537 | 1.23882E-05 | | | | | | |
| 1025 | 24666.1693 | | | 2.40115E-05 | 2.47341E-05 | | | |
| 1026 | 24751.6589 | | | | | 2.21948E-05 | | |
| 1027 | 24791.6528 | | 1.739E-05 | | | | | |
| 1028 | 24828.1401 | 1.13976E-05 | | | | | | |
| 1029 | 24828.346 | | | | | | 2.09754E-05 | |
| 1030 | 24839.6176 | | | 2.11664E-05 | 2.40844E-05 | | | |
| 1031 | 24901.751 | | | | | 2.19124E-05 | | |
| 1032 | 24917.3795 | | | | | | | 5.35597E-06 |
| 1033 | 24936.8295 | | 1.7879E-05 | | | | | |
| 1034 | 24986.9424 | 1.27135E-05 | | | | | | |
| 1035 | 25004.5698 | | | | | | 2.04791E-05 | |
| 1036 | 25012.0104 | | | | 2.35594E-05 | | | |
| 1037 | 25012.2189 | | | 2.07909E-05 | | | | |
| 1038 | 25088.1287 | | | | | 2.10407E-05 | | |
| 1039 | 25130.5751 | | 1.7471E-05 | | | | | |
| 1040 | 25147.9465 | 1.19624E-05 | | | | | 1.99078E-05 | |
| 1041 | 25154.4331 | | | | 2.21138E-05 | | | |
| 1042 | 25157.5552 | | | 1.9968E-05 | | | | |
| 1043 | 25199.5345 | | | | | 2.02385E-05 | | |
| 1044 | 25227.9684 | | | | | | | 5.74012E-06 |
| 1045 | 25243.482 | | 1.70856E-05 | | | | | |
| 1046 | 25274.2831 | 1.16299E-05 | | | | | | |
| 1047 | 25278.028 | | | | | | 1.93433E-05 | |
| 1048 | 25287.6901 | | | 1.94825E-05 | 2.07343E-05 | | | |
| 1049 | 25344.7311 | | | | | 1.99713E-05 | | |
| 1050 | 25383.6903 | | 1.67424E-05 | | | | | |
| 1051 | 25421.883 | 1.04792E-05 | | | | | | |
| 1052 | 25432.7938 | | | | | 1.79208E-05 | 1.73336E-05 | |
| 1053 | 25433.5839 | | | 1.79259E-05 | | | | |
| 1054 | 25489.9854 | | | | | 1.81443E-05 | | |
| 1055 | 25531.1128 | | 1.59231E-05 | | | | | |
| 1056 | 25543.2398 | | | | | | | 5.99152E-06 |
| 1057 | 25554.7976 | 1.04556E-05 | | | | | | |
| 1058 | 25564.9555 | | | 1.80228E-05 | 1.77736E-05 | | 1.6994E-05 | |
| 1059 | 25618.395 | | | | | 1.81106E-05 | | |
| 1060 | 25659.7946 | | 1.54465E-05 | | | | | |
| 1061 | 25682.4025 | 9.87345E-06 | | | | | | |
| 1062 | 25683.5731 | | | | | | 1.66848E-05 | |
| 1063 | 25687.4936 | | | | 1.74826E-05 | | | |
| 1064 | 25689.341 | | | 1.72261E-05 | | | | |
| 1065 | 25725.8214 | | | | | 1.76765E-05 | | |
| 1066 | 25773.8106 | | 1.57321E-05 | | | | | |
| 1067 | 25787.5428 | | | | | | | 5.86785E-06 |
| 1068 | 25788.4691 | 9.93492E-06 | | | | | | |
| 1069 | 25793.6652 | | | 1.66744E-05 | 1.57074E-05 | | | |
| 1070 | 25795.4331 | | | | | | 1.60496E-05 | |
| 1071 | 25841.1415 | | | | | 1.66611E-05 | | |
| 1072 | 25912.6984 | | 1.62137E-05 | | | | | |
| 1073 | 25930.5983 | 1.0093E-05 | | 1.67176E-05 | 1.57177E-05 | | 1.55887E-05 | |
| 1074 | 25956.6932 | | | | | 1.60618E-05 | | |
| 1075 | 25986.0965 | | 1.54897E-05 | | | | | |
| 1076 | 26005.3316 | | | | 1.53596E-05 | | | |
| 1077 | 26010.4178 | 9.70453E-06 | | | | | | |
| 1078 | 26015.635 | | | 1.63326E-05 | | | | |
| 1079 | 26017.7722 | | | | | | 1.50487E-05 | |
| 1080 | 26059.9703 | | | | | 1.64845E-05 | | |
| 1081 | 26071.4823 | | | | | | | 5.74163E-06 |
| 1082 | 26129.7452 | | 1.60972E-05 | | | | | |
| 1083 | 26147.5137 | | | | 1.5792E-05 | | | |
| 1084 | 26148.0191 | 1.00674E-05 | | | | | | |
| 1085 | 26149.2511 | | | 1.66622E-05 | | | | |
| 1086 | 26177.2491 | | | | | | 1.49245E-05 | |
| 1087 | 26203.2282 | | | | | 1.64501E-05 | | |
| 1088 | 26238.365 | | 1.66245E-05 | | | | | |
| 1089 | 26250.8353 | 1.01769E-05 | | 1.65777E-05 | 1.5907E-05 | | 1.46996E-05 | 5.69217E-06 |
| 1090 | 26282.1561 | | | | | 1.66587E-05 | | |
| 1091 | 26339.8726 | | 1.64359E-05 | | | | | |
| 1092 | 26367.7895 | | | | 1.53274E-05 | | | |
| 1093 | 26371.7295 | | | 1.61381E-05 | | | | |
| 1094 | 26373.8117 | 9.24423E-06 | | | | | | |
| 1095 | 26391.3579 | | | | | | 1.35152E-05 | |
| 1096 | 26421.9528 | | | | | 1.64707E-05 | | |
| 1097 | 26483.4161 | | 1.70756E-05 | | | | | |
| 1098 | 26487.4112 | | | | | | | 5.56633E-06 |
| 1099 | 26499.9061 | | | | 1.57377E-05 | | | |
| 1100 | 26507.4424 | 9.44955E-06 | | 1.64784E-05 | | | | |
| 1101 | 26528.0911 | | | | | | 1.26934E-05 | |
| 1102 | 26539.0841 | | | | | 1.64171E-05 | | |
| 1103 | 26583.3547 | | 1.75231E-05 | | | | | |
| 1104 | 26595.6834 | | | | 1.65581E-05 | | | |
| 1105 | 26604.1903 | | | 1.72061E-05 | | | | |
| 1106 | 26605.8716 | 1.04722E-05 | | | | | | |
| 1107 | 26625.5413 | | | | | | 1.29597E-05 | |
| 1108 | 26632.8186 | | | | | 1.61406E-05 | | |
| 1109 | 26680.2276 | | 1.52626E-05 | | | | | |
| 1110 | 26691.9644 | | | | 1.45415E-05 | | | |
| 1111 | 26696.9109 | | | 1.40616E-05 | | | | |
| 1112 | 26699.648 | 8.26303E-06 | | | | | | |
| 1113 | 26702.4724 | | | | | | | 5.49756E-06 |
| 1114 | 26729.2475 | | | | | | 1.10658E-05 | |
| 1115 | 26733.3886 | | | | | 1.47051E-05 | | |
| 1116 | 26811.7441 | | 1.5936E-05 | | | | | |
| 1117 | 26832.3156 | | | | 1.58871E-05 | | | |
| 1118 | 26834.2494 | | | 1.50475E-05 | | | | |
| 1119 | 26834.6104 | 9.56223E-06 | | | | | | |
| 1120 | 26866.6251 | | | | | 1.50789E-05 | | |
| 1121 | 26872.0063 | | | | | | 1.12003E-05 | |
| 1122 | 26937.4552 | | 1.60047E-05 | | | | | |

| | A | B | C | D | E | F | G | H |
|---|---|---|---|---|---|---|---|---|
| 1123 | 26953.4221 | | | | 1.55821E-05 | | | |
| 1124 | 26957.0047 | | | 1.49328E-05 | | | | |
| 1125 | 26959.9342 | 9.93232E-06 | | | | | | |
| 1126 | 26980.7319 | | | | | | | 6.01068E-06 |
| 1127 | 26988.1197 | | | | | 1.48663E-05 | | |
| 1128 | 27004.99 | | | | | | 1.03885E-05 | |
| 1129 | 27046.0482 | | 1.53006E-05 | | | | | |
| 1130 | 27054.7638 | | | | 1.52349E-05 | | | |
| 1131 | 27056.4545 | | | 1.49958E-05 | | | | |
| 1132 | 27063.7155 | 9.94139E-06 | | | | | | |
| 1133 | 27089.8374 | | | | | 1.52124E-05 | | |
| 1134 | 27131.692 | | | | | | 1.09465E-05 | |
| 1135 | 27159.9315 | | 1.55739E-05 | | | | | |
| 1136 | 27181.7153 | | | | 1.55389E-05 | | | |
| 1137 | 27181.9819 | | | 1.52753E-05 | | | | |
| 1138 | 27189.5101 | 9.78041E-06 | | | | | | |
| 1139 | 27224.2911 | | | | | 1.54616E-05 | | |
| 1140 | 27246.5912 | | | | | | | 5.92101E-06 |
| 1141 | 27260.1492 | | | | | | 1.13289E-05 | |
| 1142 | 27281.1871 | | 1.55511E-05 | | | | | |
| 1143 | 27310.8904 | | | | 1.59028E-05 | | | |
| 1144 | 27311.9944 | | | 1.57111E-05 | | | | |
| 1145 | 27313.7794 | 1.00642E-05 | | | | | | |
| 1146 | 27352.7617 | | | | | 1.50919E-05 | | |
| 1147 | 27398.7228 | | | | | | 1.13942E-05 | |
| 1148 | 27418.5403 | | 1.49005E-05 | | | | | |
| 1149 | 27434.8187 | | | 1.39182E-05 | 1.5514E-05 | | | |
| 1150 | 27437.0906 | 9.05221E-06 | | | | | | |
| 1151 | 27449.7403 | | | | | 1.54547E-05 | | |
| 1152 | 27452.051 | | | | | | | 5.98145E-06 |
| 1153 | 27494.6254 | | | | | | 1.1446E-05 | |
| 1154 | 27523.3327 | | 1.51762E-05 | | | | | |
| 1155 | 27534.4468 | | | | 1.58035E-05 | | | |
| 1156 | 27535.1137 | | | 1.40747E-05 | | | | |
| 1157 | 27540.3418 | 9.1886E-06 | | | | | | |
| 1158 | 27579.1514 | | | | | 1.66391E-05 | | |
| 1159 | 27635.7919 | | | | | | 1.25313E-05 | |
| 1160 | 27657.4282 | | 1.61783E-05 | | | | | |
| 1161 | 27666.745 | | | 1.53402E-05 | 1.69708E-05 | | | |
| 1162 | 27671.4206 | 9.86564E-06 | | | | | | |
| 1163 | 27711.7158 | | | | | 1.65605E-05 | | |
| 1164 | 27762.873 | | | | | | 1.19488E-05 | |
| 1165 | 27770.5498 | | 1.50824E-05 | | | | | |
| 1166 | 27791.1096 | | | | 1.58175E-05 | | | |
| 1167 | 27792.0685 | | | 1.50638E-05 | | | | |
| 1168 | 27794.8353 | 9.35201E-06 | | | | | | |
| 1169 | 27811.1759 | | | | | | | 6.09207E-06 |
| 1170 | 27828.9768 | | | | | 1.51443E-05 | | |
| 1171 | 27866.0281 | | | | | | 1.23421E-05 | |
| 1172 | 27867.8085 | | 1.53687E-05 | | | | | |
| 1173 | 27883.3267 | | | 1.47316E-05 | 1.4963E-05 | | | |
| 1174 | 27888.7592 | 9.3508E-06 | | | | | | |
| 1175 | 27912.2004 | | | | | 1.44959E-05 | | |
| 1176 | 27958.5402 | | | | | | 1.08256E-05 | |
| 1177 | 27959.4913 | | 1.38314E-05 | | | | | |
| 1178 | 27968.8936 | | | | 1.44278E-05 | | | |
| 1179 | 27969.2 | | | 1.35818E-05 | | | | |
| 1180 | 27986.3108 | 7.76299E-06 | | | | | | |
| 1181 | 28016.4201 | | | | | 1.29686E-05 | | |
| 1182 | 28045.7778 | | | | | | | 6.16573E-06 |
| 1183 | 28059.7022 | | | | | | 1.00402E-05 | |
| 1184 | 28065.4964 | | 1.29306E-05 | | | | | |
| 1185 | 28077.8426 | | | | 1.30345E-05 | | | |
| 1186 | 28079.916 | | | 1.27573E-05 | | | | |
| 1187 | 28100.4324 | 7.63458E-06 | | | | | | |
| 1188 | 28125.1329 | | | | | 1.21055E-05 | | |
| 1189 | 28172.6331 | | | | | | 9.76668E-06 | |
| 1190 | 28181.5541 | | 1.27061E-05 | | | | | |
| 1191 | 28192.9977 | | | 1.27358E-05 | 1.27578E-05 | | | |
| 1192 | 28196.1753 | 7.49057E-06 | | | | | | |
| 1193 | 28226.755 | | | | | 1.18317E-05 | | |
| 1194 | 28293.1309 | | | | | | 9.75237E-06 | |
| 1195 | 28304.4755 | | 1.27242E-05 | | | | | |
| 1196 | 28322.6291 | | | 1.29066E-05 | 1.31437E-05 | | | |
| 1197 | 28331.117 | 7.83225E-06 | | | | | | |
| 1198 | 28347.3833 | | | | | 1.21472E-05 | | |
| 1199 | 28377.911 | | | | | | | 6.24827E-06 |
| 1200 | 28391.1265 | | | | | | 9.06551E-06 | |
| 1201 | 28398.7624 | | 1.32314E-05 | 1.3333E-05 | 1.30384E-05 | | | |
| 1202 | 28402.9159 | 7.89803E-06 | | | | | | |
| 1203 | 28426.3151 | | | | | 1.20548E-05 | | |
| 1204 | 28457.5839 | | | | | | 9.09543E-06 | |
| 1205 | 28488.9925 | | 1.30385E-05 | | | | | |
| 1206 | 28498.4113 | | | | 1.31153E-05 | | | |
| 1207 | 28498.926 | 7.70829E-06 | | 1.33547E-05 | | | | |
| 1208 | 28509.9621 | | | | | 1.20601E-05 | | |
| 1209 | 28554.5508 | | | | | | 8.83276E-06 | |
| 1210 | 28581.6697 | | 1.29398E-05 | | | | | |
| 1211 | 28587.2658 | | | | 1.3494E-05 | | | |
| 1212 | 28589.1902 | | | 1.3698E-05 | | | | |
| 1213 | 28600.7395 | 9.04761E-06 | | | | 1.27897E-05 | 8.89713E-06 | 6.5066E-06 |
| 1214 | 28641.2328 | | 1.11421E-05 | | | | | |
| 1215 | 28647.836 | | | | 1.17E-05 | | | |
| 1216 | 28650.5269 | | | 1.1857E-05 | | | | |
| 1217 | 28662.7165 | 7.99832E-06 | | | | | | |
| 1218 | 28684.8293 | | | | | 1.19565E-05 | | |
| 1219 | 28706.5802 | | | | | | 7.55947E-06 | |
| 1220 | 28727.8049 | | 8.11402E-06 | | | | | |
| 1221 | 28731.5289 | | | | 8.87733E-06 | | | |
| 1222 | 28739.3252 | | | 9.20958E-06 | | | | |
| 1223 | 28756.7972 | 5.57853E-06 | | | | | | |
| 1224 | 28769.3869 | | | | | 8.70997E-06 | | |

| | A | B | C | D | E | F | G | H |
|---|---|---|---|---|---|---|---|---|
| 1225 | 28809.2193 | | | | | | 7.80338E-06 | |
| 1226 | 28845.1423 | | | | | | | 6.46234E-06 |
| 1227 | 28886.4171 | | 8.63059E-06 | | | | | |
| 1228 | 28887.3803 | | | | 9.1322E-06 | | | |
| 1229 | 28890.5941 | | | 9.78061E-06 | | | | |
| 1230 | 28911.3104 | 5.62632E-06 | | | | | | |
| 1231 | 28925.4117 | | | | | 8.20636E-06 | | |
| 1232 | 28958.8786 | | | | | | 7.3429E-06 | |
| 1233 | 29013.8151 | | | | 9.63362E-06 | | | |
| 1234 | 29014.4463 | | 9.77627E-06 | | | | | |
| 1235 | 29021.2935 | | | 1.04264E-05 | | | | |
| 1236 | 29045.2388 | 6.07336E-06 | | | | | | |
| 1237 | 29054.7465 | | | | | 8.72682E-06 | | |
| 1238 | 29061.5242 | | | | | | | 6.70436E-06 |
| 1239 | 29079.2255 | | | | | | 7.97911E-06 | |
| 1240 | 29105.9925 | | | | 9.22707E-06 | | | |
| 1241 | 29111.8488 | | 9.76452E-06 | | | | | |
| 1242 | 29113.5786 | | | 9.60793E-06 | | | | |
| 1243 | 29137.0622 | 7.26772E-06 | | | | | | |
| 1244 | 29138.9491 | | | | | 9.0941E-06 | | |
| 1245 | 29163.3642 | | | | | | 9.10694E-06 | |
| 1246 | 29212.156 | | | | 1.00481E-05 | | | |
| 1247 | 29217.851 | | 1.08467E-05 | 1.02793E-05 | | | | |
| 1248 | 29243.8716 | 9.16478E-06 | | | | 1.24E-05 | 1.36266E-05 | |
| 1249 | 29286.7376 | | | | 9.7683E-06 | | | 6.74391E-06 |
| 1250 | 29289.6572 | | 1.02454E-05 | 9.23183E-06 | | | | |
| 1251 | 29309.877 | | | | | 1.1343E-05 | | |
| 1252 | 29309.9792 | 9.56066E-06 | | | | | | |
| 1253 | 29325.6278 | | | | | | 1.27749E-05 | |
| 1254 | 29366.0167 | | | | 7.29122E-06 | | | |
| 1255 | 29373.4921 | | | 7.61839E-06 | | | | |
| 1256 | 29373.745 | | 8.63094E-06 | | | | | |
| 1257 | 29379.4041 | | | | | 1.09584E-05 | | |
| 1258 | 29384.7942 | 9.59855E-06 | | | | | 1.30748E-05 | |
| 1259 | 29438.8424 | | | | | | | 6.75301E-06 |
| 1260 | 29454.0607 | | | | 6.63412E-06 | | | |
| 1261 | 29464.7099 | | | 8.02873E-06 | | | | |
| 1262 | 29467.2745 | | 9.99913E-06 | | | | | |
| 1263 | 29483.2525 | | | | | 1.20113E-05 | | |
| 1264 | 29493.0459 | 1.21087E-05 | | | | | | |
| 1265 | 29496.9943 | | | | | | 1.44194E-05 | |
| 1266 | 29544.5261 | | | | 6.54396E-06 | | | |
| 1267 | 29554.3557 | | | 8.76841E-06 | | | | |
| 1268 | 29555.6084 | | 1.12683E-05 | | | | | |
| 1269 | 29568.2298 | | | | | 1.22883E-05 | | |
| 1270 | 29590.4463 | 1.22146E-05 | | | | | | |
| 1271 | 29603.4873 | | | | | | 1.47736E-05 | |
| 1272 | 29672.2302 | | | | 5.68019E-06 | | | |
| 1273 | 29694.8983 | | | 8.67291E-06 | | | | |
| 1274 | 29696.7162 | | 1.14537E-05 | | | | | |
| 1275 | 29702.6476 | | | | | 1.17077E-05 | | |
| 1276 | 29726.124 | 1.37898E-05 | | | | | | |
| 1277 | 29736.6782 | | | | | | 1.56592E-05 | 6.63539E-06 |
| 1278 | 29763.8329 | | | | 6.22194E-06 | | | |
| 1279 | 29766.4649 | | | 8.87893E-06 | | | | |
| 1280 | 29770.72 | | 1.13725E-05 | | | | | |
| 1281 | 29773.2033 | | | | | 1.16833E-05 | | |
| 1282 | 29784.7537 | 1.3692E-05 | | | | | | |
| 1283 | 29794.3061 | | | | | | 1.56332E-05 | |
| 1284 | 29859.5178 | | | | 6.0121E-06 | | | |
| 1285 | 29881.9614 | | | 9.01841E-06 | | | | |
| 1286 | 29891.7539 | | 1.21252E-05 | | | | | |
| 1287 | 29894.6958 | | | | | 1.14273E-05 | | |
| 1288 | 29908.478 | 1.46909E-05 | | | | | | |
| 1289 | 29923.1912 | | | | | | 1.67147E-05 | |
| 1290 | 29973.393 | | | | 7.3269E-06 | | | |
| 1291 | 29991.1571 | | | 9.71498E-06 | | | | |
| 1292 | 29999.0353 | | 1.37368E-05 | | | | | |
| 1293 | 29999.4824 | | | | | 1.20271E-05 | | |
| 1294 | 30026.6702 | 1.50115E-05 | | | | | | |
| 1295 | 30035.7595 | | | | | | | 6.84672E-06 |
| 1296 | 30037.4637 | | | | | | 1.63088E-05 | |
| 1297 | 30069.4769 | | | | 7.07486E-06 | | | |
| 1298 | 30102.4013 | | | 8.88539E-06 | | | | |
| 1299 | 30109.2757 | | | | | 1.21588E-05 | | |
| 1300 | 30110.1604 | | 1.40777E-05 | | | | | |
| 1301 | 30128.3978 | 1.52442E-05 | | | | | | |
| 1302 | 30147.4799 | | | | | | 1.60997E-05 | |
| 1303 | 30155.1623 | | | | 6.46901E-06 | | | |
| 1304 | 30177.1935 | | | 8.08793E-06 | | | | |
| 1305 | 30187.359 | | | | | 1.25617E-05 | | |
| 1306 | 30190.5232 | | 1.47642E-05 | | | | | |
| 1307 | 30211.9554 | 1.65777E-05 | | | | | | |
| 1308 | 30219.5925 | | | | | | 1.66304E-05 | |
| 1309 | 30260.3573 | | | | 7.48742E-06 | | | |
| 1310 | 30281.845 | | | 8.50712E-06 | | | | |
| 1311 | 30287.187 | | | | | 1.2489E-05 | | |
| 1312 | 30297.8553 | | 1.30781E-05 | | | | | |
| 1313 | 30324.8662 | 1.58118E-05 | | | | | | |
| 1314 | 30330.7326 | | | | | | 1.71201E-05 | |
| 1315 | 30349.4167 | | | | 7.60957E-06 | | | |
| 1316 | 30374.7749 | | | 8.75767E-06 | | 1.28279E-05 | | |
| 1317 | 30390.806 | | 1.35589E-05 | | | | | |
| 1318 | 30413.0584 | 1.57774E-05 | | | | | | |
| 1319 | 30463.6822 | | | | | | 1.66497E-05 | |
| 1320 | 30484.738 | | | | 7.31303E-06 | | | |
| 1321 | 30502.3002 | | | | | | | 6.87241E-06 |
| 1322 | 30505.7729 | | | 7.62357E-06 | | | | |
| 1323 | 30507.4853 | | | | | 1.2664E-05 | | |
| 1324 | 30514.416 | | 1.20831E-05 | | | | | |
| 1325 | 30539.1606 | 1.56602E-05 | | | | | | |
| 1326 | 30556.0506 | | | | | | 1.683E-05 | |

| | A | B | C | D | E | F | G | H |
|---|---|---|---|---|---|---|---|---|
| 1327 | 30575.5901 | | | | 8.12471E-06 | | | |
| 1328 | 30597.0711 | | | 4.50229E-06 | | 1.03156E-05 | | |
| 1329 | 30606.8575 | | 7.12598E-06 | | | | | |
| 1330 | 30651.3066 | 1.06149E-05 | | | | | | |
| 1331 | 30692.424 | | | | | | 1.6774E-05 | |
| 1332 | 30696.7522 | | | | 7.40823E-06 | | | |
| 1333 | 30711.7211 | | | 5.03261E-06 | | 1.08199E-05 | | |
| 1334 | 30721.5623 | | 7.43704E-06 | | | | | |
| 1335 | 30752.4947 | 1.13291E-05 | | | | | | |
| 1336 | 30758.1891 | | | | | | | 6.72761E-06 |
| 1337 | 30789.7548 | | | | 7.23629E-06 | | | |
| 1338 | 30806.5454 | | | | | | 1.65263E-05 | |
| 1339 | 30816.5797 | | | | | 1.0447E-05 | | |
| 1340 | 30817.9353 | | | 4.96577E-06 | | | | |
| 1341 | 30835.1027 | | 7.7159E-06 | | | | | |
| 1342 | 30870.8702 | 1.17449E-05 | | | | | | |
| 1343 | 30914.863 | | | | 5.50888E-06 | | | |
| 1344 | 30928.0612 | | | | | 8.5861E-06 | | |
| 1345 | 30930.4092 | | | | | | 1.58813E-05 | |
| 1346 | 30933.1928 | | | 3.58207E-06 | | | | |
| 1347 | 30948.659 | | 7.14348E-06 | | | | | |
| 1348 | 30966.89 | 1.11858E-05 | | | | | | |
| 1349 | 30991.4763 | | | | 5.28411E-06 | | | |
| 1350 | 31013.5639 | | | | | 8.8759E-06 | | |
| 1351 | 31016.3284 | | | | | | 1.56104E-05 | |
| 1352 | 31022.043 | | | 2.96153E-06 | | | | |
| 1353 | 31042.6746 | | 6.25696E-06 | | | | | |
| 1354 | 31069.8885 | 1.19729E-05 | | | | | | |
| 1355 | 31096.2997 | | | | 5.14781E-06 | | | |
| 1356 | 31113.7753 | | | | | 8.99551E-06 | | |
| 1357 | 31123.775 | | | 2.75565E-06 | | | | |
| 1358 | 31131.0169 | | | | | | | 6.70008E-06 |
| 1359 | 31132.1334 | | | | | | 1.50078E-05 | |
| 1360 | 31146.5428 | | 7.01845E-06 | | | | | |
| 1361 | 31172.6175 | 1.32716E-05 | | | | | | |
| 1362 | 31199.9322 | | | | 5.52898E-06 | | | |
| 1363 | 31207.1353 | | | | | 9.12117E-06 | | |
| 1364 | 31217.1483 | | | 2.9894E-06 | | | | |
| 1365 | 31225.9847 | | | | | | 1.46946E-05 | |
| 1366 | 31232.5979 | | 6.97032E-06 | | | | | |
| 1367 | 31265.4033 | 1.25517E-05 | | | | | | |
| 1368 | 31299.2933 | | | | 5.08663E-06 | | | |
| 1369 | 31317.3354 | | | | | 9.13608E-06 | | |
| 1370 | 31329.6841 | | | 2.8309E-06 | | | | |
| 1371 | 31349.1369 | | 7.26403E-06 | | | | | |
| 1372 | 31350.7533 | | | | | | 1.48598E-05 | |
| 1373 | 31382.3041 | 1.46149E-05 | | | | | | |
| 1374 | 31416.8432 | | | | 5.23652E-06 | | | |
| 1375 | 31439.9485 | | | | | 9.59209E-06 | | |
| 1376 | 31453.9165 | | | 3.63541E-06 | | | | |
| 1377 | 31459.561 | | | | | | | 6.7257E-06 |
| 1378 | 31474.5689 | | 8.04407E-06 | | | | | |
| 1379 | 31486.1303 | | | | | | 1.47536E-05 | |
| 1380 | 31501.8654 | 1.3649E-05 | | | | | | |
| 1381 | 31550.5067 | | | | 5.97696E-06 | | | |
| 1382 | 31567.5416 | | | | | 1.12872E-05 | | |
| 1383 | 31594.4958 | | | 4.3541E-06 | | | | |
| 1384 | 31605.3867 | | 8.5781E-06 | | | | | |
| 1385 | 31650.0346 | 1.48181E-05 | | | | | | |
| 1386 | 31657.5333 | | | | | | 1.50889E-05 | |
| 1387 | 31697.6112 | | | | 6.77428E-06 | | | |
| 1388 | 31717.8105 | | | | | 1.1272E-05 | | |
| 1389 | 31746.8627 | | | 3.57355E-06 | | | | |
| 1390 | 31763.6902 | | 8.50811E-06 | | | | | |
| 1391 | 31792.5227 | 1.50089E-05 | | | | | | |
| 1392 | 31807.9196 | | | | | | 1.54924E-05 | |
| 1393 | 31835.5616 | | | | 6.44676E-06 | | | |
| 1394 | 31858.3516 | | | | | 1.11847E-05 | | |
| 1395 | 31863.1686 | | | | | | | 6.79507E-06 |
| 1396 | 31867.0441 | | | 4.25968E-06 | | | | |
| 1397 | 31907.7879 | | 9.11625E-06 | | | | | |
| 1398 | 31932.9992 | 1.53113E-05 | | | | | | |
| 1399 | 31961.7108 | | | | | | 1.60224E-05 | |
| 1400 | 31977.6373 | | | | 7.32566E-06 | | | |
| 1401 | 31994.8383 | | | | | 1.17398E-05 | | |
| 1402 | 32008.3782 | | | 4.65783E-06 | | | | |
| 1403 | 32035.2513 | | 1.01153E-05 | | | | | |
| 1404 | 32086.1795 | | | | | | | 6.84635E-06 |
| 1405 | 32090.1089 | 1.67148E-05 | | | | | | |
| 1406 | 32140.1411 | | | | 8.11355E-06 | | | |
| 1407 | 32144.8919 | | | | | | 1.60129E-05 | |
| 1408 | 32157.0499 | | | | | 1.23515E-05 | | |
| 1409 | 32178.4865 | | | 4.41868E-06 | | | | |
| 1410 | 32224.035 | | 1.06023E-05 | | | | | |
| 1411 | 32249.1055 | 1.70778E-05 | | | | | | |
| 1412 | 32287.9237 | | | | 7.94534E-06 | | | |
| 1413 | 32290.6317 | | | | | | 1.58968E-05 | |
| 1414 | 32296.4676 | | | | | 1.25824E-05 | | |
| 1415 | 32311.1344 | | | 4.55019E-06 | | | | |
| 1416 | 32332.2108 | | 1.08495E-05 | | | | | |
| 1417 | 32375.469 | 1.823E-05 | | | | | | |
| 1418 | 32416.4928 | | | | 9.02657E-06 | | | |
| 1419 | 32439.2461 | | | | | 1.33935E-05 | | |
| 1420 | 32443.2835 | | | | | | 1.5776E-05 | |
| 1421 | 32451.6064 | | | 5.184E-06 | | | | |
| 1422 | 32489.8346 | | 1.07031E-05 | | | | | |
| 1423 | 32523.6132 | | | | | | | 6.89352E-06 |
| 1424 | 32531.9753 | 1.77409E-05 | | | | | | |
| 1425 | 32546.0435 | | | | 8.60248E-06 | | | |
| 1426 | 32562.5572 | | | | | 1.25437E-05 | | |
| 1427 | 32575.9624 | | | | | | 1.54146E-05 | |
| 1428 | 32587.0405 | | | 5.45753E-06 | | | | |

| | A | B | C | D | E | F | G | H |
|---|---|---|---|---|---|---|---|---|
| 1429 | 32625.4705 | | 1.12286E-05 | | | | | |
| 1430 | 32666.7265 | 1.77872E-05 | | | | | | |
| 1431 | 32690.2679 | | | | 8.25119E-06 | | | |
| 1432 | 32712.1414 | | | | | 1.1828E-05 | | |
| 1433 | 32729.0927 | | | | | | 1.55586E-05 | |
| 1434 | 32730.7979 | | | 5.9242E-06 | | | | |
| 1435 | 32776.5591 | | 1.09794E-05 | | | | | |
| 1436 | 32809.7155 | 1.73998E-05 | | | | | | |
| 1437 | 32819.5788 | | | | | | | 6.99668E-06 |
| 1438 | 32835.1885 | | | | 8.00211E-06 | | | |
| 1439 | 32839.4762 | | | | | 1.17907E-05 | | |
| 1440 | 32868.2571 | | | 6.57744E-06 | | | | |
| 1441 | 32889.0114 | | | | | | 1.66663E-05 | |
| 1442 | 32902.7039 | | 1.17731E-05 | | | | | |
| 1443 | 32923.7993 | 1.91619E-05 | | | | | | |
| 1444 | 32943.0604 | | | | 7.86639E-06 | | | |
| 1445 | 32963.8474 | | | | | 1.19206E-05 | | |
| 1446 | 32976.3897 | | | 7.01524E-06 | | | | |
| 1447 | 33000.6391 | | | | | | 1.70914E-05 | |
| 1448 | 33025.4924 | | 1.27922E-05 | | | | | |
| 1449 | 33076.3093 | 1.96828E-05 | | | | | | |
| 1450 | 33096.7444 | | | | 7.70974E-06 | | | |
| 1451 | 33126.2889 | | | | | 1.21548E-05 | | |
| 1452 | 33150.0136 | | | 7.41489E-06 | | | | |
| 1453 | 33181.6549 | | | | | | 1.64913E-05 | 7.1144E-06 |
| 1454 | 33183.2183 | | 1.30346E-05 | | | | | |
| 1455 | 33207.1819 | 2.06953E-05 | | | | | | |
| 1456 | 33236.3355 | | | | 7.70969E-06 | | | |
| 1457 | 33265.9877 | | | | | 1.17867E-05 | | |
| 1458 | 33292.0119 | | | 7.71858E-06 | | | | |
| 1459 | 33324.9452 | | | | | | 1.63786E-05 | |
| 1460 | 33334.006 | | 1.28514E-05 | | | | | |
| 1461 | 33388.0259 | 2.01553E-05 | | | | | | |
| 1462 | 33421.0588 | | | | 6.98306E-06 | | | |
| 1463 | 33445.0905 | | | | | 1.08523E-05 | | |
| 1464 | 33465.463 | | | 8.48976E-06 | | | | |
| 1465 | 33483.0332 | | | | | | | 7.23538E-06 |
| 1466 | 33493.7568 | | | | | | 1.62642E-05 | |
| 1467 | 33516.1379 | | 1.38972E-05 | | | | | |
| 1468 | 33564.6957 | 2.00053E-05 | | | | | | |
| 1469 | 33597.5258 | | | | 6.31786E-06 | | | |
| 1470 | 33622.3629 | | | | | 1.06878E-05 | | |
| 1471 | 33644.5643 | | | 7.71039E-06 | | | | |
| 1472 | 33701.7774 | | | | | | 1.5562E-05 | |
| 1473 | 33705.6173 | | 1.23269E-05 | | | | | |
| 1474 | 33755.2167 | 1.93075E-05 | | | | | | |
| 1475 | 33793.6332 | | | | 5.66943E-06 | | | |
| 1476 | 33841.2334 | | | | | 1.0261E-05 | | |
| 1477 | 33859.092 | | | 7.51632E-06 | | | | |
| 1478 | 33909.6989 | | 1.27992E-05 | | | | | |
| 1479 | 33926.0744 | | | | | | 1.58198E-05 | |
| 1480 | 33954.5407 | | | | | | | 7.38941E-06 |
| 1481 | 33962.8742 | 1.89718E-05 | | | | | | |
| 1482 | 33990.6969 | | | | 5.51628E-06 | | | |
| 1483 | 34028.3981 | | | | | 9.99842E-06 | | |
| 1484 | 34043.2662 | | | 7.05412E-06 | | | | |
| 1485 | 34077.6403 | | 1.18771E-05 | | | | | |
| 1486 | 34091.2877 | | | | | | 1.52735E-05 | |
| 1487 | 34113.3416 | 1.81629E-05 | | | | | | |
| 1488 | 34134.7487 | | | | 5.05525E-06 | | | |
| 1489 | 34184.9247 | | | | | 8.88893E-06 | | |
| 1490 | 34206.8288 | | | 6.40093E-06 | | | | |
| 1491 | 34282.9562 | | 1.20041E-05 | | | | | |
| 1492 | 34330.3812 | | | | | | 1.51236E-05 | |
| 1493 | 34335.7697 | 1.79935E-05 | | | | | | |
| 1494 | 34377.0015 | | | | | | | 7.57367E-06 |
| 1495 | 34382.7457 | | | | 4.17929E-06 | | | |
| 1496 | 34427.5739 | | | | | 8.27769E-06 | | |
| 1497 | 34439.1217 | | | 5.89361E-06 | | | | |
| 1498 | 34518.9091 | | 1.21108E-05 | | | | | |
| 1499 | 34562.9486 | 1.8529E-05 | | | | | | |
| 1500 | 34571.993 | | | | | | 1.54938E-05 | |
| 1501 | 34612.7432 | | | | 4.30634E-06 | | | |
| 1502 | 34687.5344 | | | | | 8.32235E-06 | | |
| 1503 | 34702.5738 | | | 5.33864E-06 | | | | |
| 1504 | 34762.5259 | | 1.15224E-05 | | | | | |
| 1505 | 34825.0832 | 1.76776E-05 | | | | | | |
| 1506 | 34845.7342 | | | | | | 1.25939E-05 | |
| 1507 | 34861.5448 | | | | 3.94626E-06 | | | |
| 1508 | 34909.4631 | | | | | 8.13563E-06 | | |
| 1509 | 34917.6609 | | | | | | | 7.17546E-06 |
| 1510 | 34918.7628 | | | 5.71768E-06 | | | | |
| 1511 | 34982.0192 | | 1.24289E-05 | | | | | |
| 1512 | 35059.9813 | 1.8795E-05 | | | | | | |
| 1513 | 35120.3016 | | | | | | 1.22212E-05 | |
| 1514 | 35128.4874 | | | | 4.15311E-06 | | | |
| 1515 | 35187.547 | | | | | 7.96226E-06 | | |
| 1516 | 35188.8387 | | | 4.52098E-06 | | | | |
| 1517 | 35256.5659 | | 1.25397E-05 | | | | | |
| 1518 | 35330.2849 | 1.84702E-05 | | | | | | |
| 1519 | 35393.8961 | | | | 5.30919E-06 | | | |
| 1520 | 35416.7605 | | | | | | 1.12665E-05 | |
| 1521 | 35459.6613 | | | 4.5565E-06 | | 8.23821E-06 | | |
| 1522 | 35532.4971 | | 1.22996E-05 | | | | | |
| 1523 | 35614.1974 | 1.73465E-05 | | | | | | |
| 1524 | 35643.4259 | | | | | | | 7.13155E-06 |
| 1525 | 35681.4879 | | | | 4.57603E-06 | | | |
| 1526 | 35710.924 | | | | | | 1.07129E-05 | |
| 1527 | 35737.9154 | | | 3.90145E-06 | | | | |
| 1528 | 35738.5076 | | | | | 8.34697E-06 | | |
| 1529 | 35808.597 | | 1.11015E-05 | | | | | |
| 1530 | 35865.44 | 1.66824E-05 | | | | | | |

| | A | B | C | D | E | F | G | H |
|---|---|---|---|---|---|---|---|---|
| 1531 | 35951.4148 | | | | 4.3914E-06 | | | |
| 1532 | 35977.8794 | | | | | | 1.08194E-05 | |
| 1533 | 36021.0766 | | | 3.11108E-06 | | | | |
| 1534 | 36027.5683 | | | | | 8.84835E-06 | | |
| 1535 | 36073.8649 | | 9.50355E-06 | | | | | |
| 1536 | 36142.9664 | 1.57941E-05 | | | | | | |
| 1537 | 36228.1362 | | | | 3.46776E-06 | | | |
| 1538 | 36242.1291 | | | | | | 9.87702E-06 | |
| 1539 | 36243.3408 | | | | | | | 7.29016E-06 |
| 1540 | 36291.1614 | | | 1.828E-06 | | | | |
| 1541 | 36302.2119 | | | | | 7.49961E-06 | | |
| 1542 | 36346.1887 | | 8.5071E-06 | | | | | |
| 1543 | 36408.7411 | 1.52542E-05 | | | | | | |
| 1544 | 36476.7202 | | | | 3.2797E-06 | | | |
| 1545 | 36487.4423 | | | | | | 1.03903E-05 | |
| 1546 | 36541.1856 | | | 1.46287E-06 | | | | |
| 1547 | 36572.5934 | | | | | 7.91252E-06 | | |
| 1548 | 36600.736 | | 8.69507E-06 | | | | | |
| 1549 | 36657.6292 | 1.51502E-05 | | | | | | |
| 1550 | 36657.7816 | | | | | | | 7.95966E-06 |
| 1551 | 36733.1126 | | | | 3.69366E-06 | | 8.96843E-06 | |
| 1552 | 36758.3576 | | | 1.35019E-06 | | | | |
| 1553 | 36815.8119 | | | | | 7.16846E-06 | | |
| 1554 | 36834.0487 | | 7.87385E-06 | | | | | |
| 1555 | 36898.6966 | 1.56998E-05 | | | | | | |
| 1556 | 36938.0128 | | | | | | 8.64478E-06 | |
| 1557 | 36939.3037 | | | | 4.26836E-06 | | | |
| 1558 | 37012.1574 | | | 1.51374E-06 | | | | |
| 1559 | 37074.9284 | | | | | 5.79063E-06 | | |
| 1560 | 37086.0628 | | 6.74886E-06 | | | | | |
| 1561 | 37137.3135 | 1.48798E-05 | | | | | | |
| 1562 | 37160.2158 | | | | | | 8.17703E-06 | |
| 1563 | 37175.15 | | | 2.08081E-06 | | | | |
| 1564 | 37191.9066 | | | | | | | 8.0429E-06 |
| 1565 | 37212.8797 | | | 7.72504E-07 | | | | |
| 1566 | 37261.0804 | | | | | 5.83687E-06 | | |
| 1567 | 37264.7722 | | 6.92179E-06 | | | | | |
| 1568 | 37324.2615 | 1.47199E-05 | | | | | | |
| 1569 | 37418.213 | | | | | | 8.72543E-06 | |
| 1570 | 37431.1293 | | | | 1.8707E-06 | | | |
| 1571 | 37475.9904 | | | | | | | 8.98133E-06 |
| 1572 | 37482.9826 | | | 2.27223E-07 | | | | |
| 1573 | 37542.1947 | | 6.2686E-06 | | | | | |
| 1574 | 37548.3595 | | | | | 5.7485E-06 | | |
| 1575 | 37584.3471 | 1.4124E-05 | | | | | | |
| 1576 | 37633.1766 | | | | | | 7.71556E-06 | |
| 1577 | 37658.5457 | | | | 1.29622E-06 | | | |
| 1578 | 37688.8847 | | | -8.96344E-07 | | | | |
| 1579 | 37748.0865 | | 5.76643E-06 | | | | | |
| 1580 | 37757.6934 | | | | | 4.91556E-06 | | |
| 1581 | 37810.8142 | 1.3526E-05 | | | | | | |
| 1582 | 37870.6656 | | | | | | 7.18224E-06 | 9.10463E-06 |
| 1583 | 37898.1294 | | | | 9.32493E-07 | | | |
| 1584 | 37931.6307 | | | -1.04829E-06 | | | | |
| 1585 | 37996.7343 | | 5.6986E-06 | | | | | |
| 1586 | 38020.4857 | | | | | 5.37991E-06 | | |
| 1587 | 38047.3682 | 1.42553E-05 | | | | | | |
| 1588 | 38170.0453 | | | | | | 7.63755E-06 | |
| 1589 | 38203.479 | | | | 1.90597E-06 | | | |
| 1590 | 38234.3232 | | | -8.94825E-07 | | | | |
| 1591 | 38272.5081 | | 6.3215E-06 | | | | | |
| 1592 | 38287.7661 | | | | | 5.91927E-06 | | |
| 1593 | 38314.7059 | 1.45275E-05 | | | | | | |
| 1594 | 38363.5232 | | | | | | | 9.73789E-06 |
| 1595 | 38406.2124 | | | | | | 7.96606E-06 | |
| 1596 | 38423.6784 | | | | 1.91258E-06 | | | |
| 1597 | 38471.0092 | | | -6.67049E-07 | | | | |
| 1598 | 38505.484 | | 6.74873E-06 | | | | | |
| 1599 | 38535.7362 | | | | | 6.18406E-06 | | |
| 1600 | 38563.7774 | 1.49876E-05 | | | | | | |
| 1601 | 38656.0295 | | | | | | 8.89971E-06 | |
| 1602 | 38666.5838 | | | | 3.31622E-06 | | | |
| 1603 | 38687.4356 | | | | | | | 1.01577E-05 |
| 1604 | 38695.4107 | | | 2.3114E-07 | | | | |
| 1605 | 38731.0478 | | 7.48332E-06 | | | | | |
| 1606 | 38756.6562 | | | | | 6.73017E-06 | | |
| 1607 | 38760.6282 | 1.57085E-05 | | | | | | |
| 1608 | 38853.2002 | | | | | | 8.70265E-06 | |
| 1609 | 38858.7522 | | | | 2.80736E-06 | | | |
| 1610 | 38902.1243 | | | 4.34548E-07 | | | | |
| 1611 | 38959.9581 | | 7.70011E-06 | | | | | |
| 1612 | 38990.0605 | | | | | 6.70402E-06 | | |
| 1613 | 39014.992 | 1.61717E-05 | | | | | | |
| 1614 | 39113.8118 | | | | | | | 1.11205E-05 |
| 1615 | 39140.397 | | | | | | 8.66212E-06 | |
| 1616 | 39153.6 | | | | 2.93342E-06 | | | |
| 1617 | 39186.908 | | | -6.29255E-09 | | | | |
| 1618 | 39226.6801 | | 7.22019E-06 | | | | | |
| 1619 | 39282.5463 | | | | | 7.53817E-06 | | |
| 1620 | 39283.7963 | 1.60329E-05 | | | | | | |
| 1621 | 39392.3468 | | | | | | 8.33025E-06 | |
| 1622 | 39411.1303 | | | | 3.28257E-06 | | | |
| 1623 | 39450.3094 | | | 5.42448E-07 | | | | |
| 1624 | 39484.5347 | | 7.10623E-06 | | | | | |
| 1625 | 39547.1556 | | | | | 7.54173E-06 | | |
| 1626 | 39550.2138 | 1.55567E-05 | | | | | | |
| 1627 | 39678.2371 | | | | | | 8.13766E-06 | |
| 1628 | 39738.0383 | | | | 3.66796E-06 | | | |
| 1629 | 39784.2098 | | | 1.68565E-06 | | | | |
| 1630 | 39816.4802 | | 8.64017E-06 | | | | | |
| 1631 | 39855.8878 | | | | | 7.75771E-06 | | |
| 1632 | 39866.8148 | 1.67985E-05 | | | | | | |

| | A | B | C | D | E | F | G | H |
|---|---|---|---|---|---|---|---|---|
| 1633 | 39985.3114 | | | | | | 7.79491E-06 | |
| 1634 | 40010.4468 | | | | 3.39376E-06 | | | |
| 1635 | 40049.2954 | | | 2.18412E-06 | | | | |
| 1636 | 40104.0435 | | 9.18563E-06 | | | | | |
| 1637 | 40163.0234 | | | | | 8.95485E-06 | | |
| 1638 | 40169.3181 | | | | | | | 1.14928E-05 |
| 1639 | 40175.6358 | 1.72557E-05 | | | | | | |
| 1640 | 40295.2193 | | | | | | 8.81055E-06 | |
| 1641 | 40300.3611 | | | | 6.17873E-06 | | | |
| 1642 | 40323.2092 | | | 2.46469E-06 | | | | |
| 1643 | 40333.6217 | | 9.23362E-06 | | | | | |
| 1644 | 40370.0585 | | | | | 1.02687E-05 | | |
| 1645 | 40391.3304 | 1.70098E-05 | | | | | | |
| 1646 | 40512.1827 | | | | | | 8.64223E-06 | |
| 1647 | 40532.952 | | | | 5.32973E-06 | | | |
| 1648 | 40561.1785 | | | 1.43835E-06 | | | | |
| 1649 | 40591.7346 | | 9.29229E-06 | | | | | |
| 1650 | 40675.1254 | | | | | 9.37964E-06 | | |
| 1651 | 40675.8196 | 1.75883E-05 | | | | | | |
| 1652 | 40789.2378 | | | | | | 8.71242E-06 | |
| 1653 | 40798.8998 | | | | 5.67946E-06 | | | |
| 1654 | 40832.6319 | | | 1.69699E-06 | | | | |
| 1655 | 40863.6813 | | 8.87307E-06 | | | | | |
| 1656 | 40926.5213 | | | | | 9.19192E-06 | | |
| 1657 | 40928.349 | 1.72148E-05 | | | | | | |
| 1658 | 41089.2527 | | | | | | 8.28178E-06 | |
| 1659 | 41098.2253 | | | | 5.47775E-06 | | | |
| 1660 | 41100.9246 | | | | | | | 1.16682E-05 |
| 1661 | 41147.0758 | | | 8.08063E-07 | | | | |
| 1662 | 41194.4701 | | 7.12161E-06 | | | | | |
| 1663 | 41243.2233 | | | | | 9.45307E-06 | | |
| 1664 | 41246.7693 | 1.63067E-05 | | | | | | |
| 1665 | 41356.812 | | | | | | 8.5922E-06 | |
| 1666 | 41398.9615 | | | | 5.73494E-06 | | | |
| 1667 | 41454.4496 | | | 8.44661E-07 | | | | |
| 1668 | 41474.5457 | | 6.43891E-06 | | | | | |
| 1669 | 41557.2938 | 1.52725E-05 | | | | 1.00984E-05 | | |
| 1670 | 41683.1213 | | | | | | 8.39035E-06 | |
| 1671 | 41694.2022 | | | | | | | 1.13022E-05 |
| 1672 | 41744.116 | | | | 6.01409E-06 | | | |
| 1673 | 41808.8549 | | | 5.36369E-07 | | | | |
| 1674 | 41832.0501 | | 6.83057E-06 | | | | | |
| 1675 | 41875.5312 | 1.53924E-05 | | | | | | |
| 1676 | 41880.7081 | | | | | 9.86213E-06 | | |
| 1677 | 41959.6729 | | | | | | 7.10269E-06 | |
| 1678 | 42055.3658 | | | | 4.95007E-06 | | | |
| 1679 | 42092.5829 | | | -6.20798E-08 | | | | |
| 1680 | 42107.8856 | | 6.28178E-06 | | | | | 1.13805E-05 |
| 1681 | 42162.4898 | 1.45127E-05 | | | | | | |
| 1682 | 42165.5459 | | | | | 9.37676E-06 | | |
| 1683 | 42281.0991 | | | | | | 6.83622E-06 | |
| 1684 | 42371.5095 | | | | 4.54908E-06 | | | |
| 1685 | 42418.6616 | | | 2.09245E-07 | | | | |
| 1686 | 42418.885 | | 7.38951E-06 | | | | | |
| 1687 | 42431.4125 | | | | | | | 1.14563E-05 |
| 1688 | 42464.657 | 1.5326E-05 | | | | | | |
| 1689 | 42472.2077 | | | | | 1.03226E-05 | | |
| 1690 | 42555.8249 | | | | | | 7.20736E-06 | |
| 1691 | 42640.9694 | | | | -9.62654E-07 | | | 1.21013E-05 |
| 1692 | 42689.1533 | | | -5.97929E-06 | | | | |
| 1693 | 42690.4198 | | 1.04328E-06 | | | | | |
| 1694 | 42697.4124 | | | | | | | 1.20225E-05 |
| 1695 | 42719.1027 | 2.63916E-06 | | | | | | |
| 1696 | 42731.1121 | | | | | 2.13177E-07 | | |
| 1697 | 42777.4992 | | | | | | -7.92464E-06 | |
| 1698 | 42837.6809 | | | | | | | 1.21536E-05 |
| 1699 | 42901.5278 | | | | -1.75522E-06 | | | |
| 1700 | 42936.5393 | | | -4.52338E-06 | | | | |
| 1701 | 42943.7055 | | 1.86373E-06 | | | | | |
| 1702 | 42989.6715 | 2.21591E-06 | | | | | | |
| 1703 | 43015.2098 | | | | | 7.04824E-07 | | |
| 1704 | 43090.478 | | | | | | -8.48827E-06 | |
| 1705 | 43092.4377 | | | | | | | 1.2025E-05 |
| 1706 | 43118.9235 | | | | | | | 1.19794E-05 |
| 1707 | 43138.1928 | | | | -2.77893E-06 | | | |
| 1708 | 43204.8052 | | 1.64171E-06 | -5.84091E-06 | | | | |
| 1709 | 43252.8193 | | | | | | | 1.15361E-05 |
| 1710 | 43256.7631 | 2.27566E-06 | | | | | | |
| 1711 | 43289.4804 | | | | | 3.34696E-06 | | |
| 1712 | 43390.2633 | | | | | | -6.95477E-06 | |
| 1713 | 43449.5396 | | | | -3.97411E-07 | | | |
| 1714 | 43518.0838 | | 4.8358E-06 | | | | | 1.16589E-05 |
| 1715 | 43520.184 | | | -2.83923E-06 | | | | |
| 1716 | 43547.299 | 5.61227E-06 | | | | | | |
| 1717 | 43560.4749 | | | | | 8.4532E-06 | | |
| 1718 | 43708.3748 | | | | | | -6.44577E-06 | |
| 1719 | 43768.5545 | | | | 1.54675E-06 | | | |
| 1720 | 43797.0695 | | | | | | | 1.12384E-05 |
| 1721 | 43797.8347 | | 6.63761E-06 | | | | | |
| 1722 | 43800.2169 | | | -1.16876E-06 | | | | |
| 1723 | 43879.6034 | 9.95225E-06 | | | | | | |
| 1724 | 43923.71 | | | | | 1.12887E-05 | | |
| 1725 | 44079.9592 | | | | | | -4.76244E-06 | |
| 1726 | 44093.447 | | | | 5.97209E-06 | | | |
| 1727 | 44132.6286 | | | | | | | 1.0651E-05 |
| 1728 | 44138.1412 | | 9.36576E-06 | | | | | |
| 1729 | 44156.0186 | | | 1.42475E-06 | | | | |
| 1730 | 44185.6529 | 1.03999E-05 | | | | | | |
| 1731 | 44253.8732 | | | | | 1.20675E-05 | | |
| 1732 | 44455.0821 | | | | 6.69435E-06 | | | |
| 1733 | 44490.5983 | | | | | | -4.51629E-06 | |
| 1734 | 44515.3928 | | 1.04744E-05 | | | | | |

| | A | B | C | D | E | F | G | H |
|---|---|---|---|---|---|---|---|---|
| 1735 | 44536.6625 | | | 2.52094E-06 | | | | |
| 1736 | 44566.7271 | 1.18886E-05 | | | | | | |
| 1737 | 44631.9653 | | | | | 1.21118E-05 | | |
| 1738 | 44833.3994 | | | | 8.02331E-06 | | | |
| 1739 | 44838.461 | | | | | | | 1.13704E-05 |
| 1740 | 44863.4147 | | 1.19532E-05 | | | | | |
| 1741 | 44883.688 | | | | | | -2.88396E-06 | |
| 1742 | 44909.556 | | | 4.01823E-06 | | | | |
| 1743 | 44918.6072 | 1.40497E-05 | | | | | | |
| 1744 | 45011.6074 | | | | | 1.45783E-05 | | |
| 1745 | 45236.1019 | | | | 1.08695E-05 | | | |
| 1746 | 45236.9437 | | 1.30873E-05 | | | | | |
| 1747 | 45313.5193 | 1.48258E-05 | | | | | | |
| 1748 | 45328.6405 | | | 5.00042E-06 | | | | |
| 1749 | 45338.3342 | | | | | | -2.77254E-06 | |
| 1750 | 45457.1545 | | | | | 1.64743E-05 | | |
| 1751 | 45634.0668 | | 1.37204E-05 | | | | | |
| 1752 | 45647.4329 | | | | 1.0601E-05 | | | |
| 1753 | 45694.1092 | 1.54229E-05 | | | | | | |
| 1754 | 45714.5842 | | | | | | -3.32438E-06 | |
| 1755 | 45730.0039 | | | 5.61717E-06 | | | | |
| 1756 | 45748.2734 | | | | | | | 1.10394E-05 |
| 1757 | 45889.165 | | | | | 1.65153E-05 | | |
| 1758 | 46059.6556 | | 1.39891E-05 | | | | | |
| 1759 | 46090.2738 | | | | 1.15937E-05 | | | |
| 1760 | 46148.0933 | 1.59998E-05 | | | | | | |
| 1761 | 46180.1298 | | | 6.04344E-06 | | | | |
| 1762 | 46196.1371 | | | | | | -3.21016E-06 | |
| 1763 | 46371.1197 | | | | | 1.77187E-05 | | |
| 1764 | 46618.1001 | | 1.52944E-05 | | | | | |
| 1765 | 46656.4256 | | | | 1.42414E-05 | | | |
| 1766 | 46690.1775 | 1.77738E-05 | | | | | | |
| 1767 | 46736.908 | | | 8.06939E-06 | | | | |
| 1768 | 46801.0068 | | | | | | -2.39906E-06 | |
| 1769 | 46811.6163 | | | | | | | 1.04327E-05 |
| 1770 | 46886.3288 | | | | | 1.90851E-05 | | |
| 1771 | 47167.5784 | | 1.83796E-05 | | | | | |
| 1772 | 47281.2864 | | | | 1.77777E-05 | | | |
| 1773 | 47295.6458 | 2.01284E-05 | | | | | | |
| 1774 | 47396.1622 | | | 1.00662E-05 | | | | |
| 1775 | 47484.8921 | | | | | | -1.1948E-06 | |
| 1776 | 47555.7855 | | | | | 2.0056E-05 | | |
| 1777 | 47822.1837 | | 1.87008E-05 | | | | | |
| 1778 | 47916.0787 | 2.06166E-05 | | | | | | |
| 1779 | 47925.4372 | | | | 1.72582E-05 | | | |
| 1780 | 48068.6593 | | | 1.00494E-05 | | | | |
| 1781 | 48311.2121 | | | | | 1.85802E-05 | -3.13858E-06 | |
| 1782 | 48598.3569 | | 2.00172E-05 | | | | | |
| 1783 | 48713.7005 | 2.18047E-05 | | | | | | |
| 1784 | 48744.949 | | | | 1.66806E-05 | | | |
| 1785 | 48867.6456 | | | | | | | 9.77193E-06 |
| 1786 | 48883.3658 | | | 1.05843E-05 | | | | |
| 1787 | 49144.7174 | | | | | 1.74584E-05 | -2.76448E-06 | |
| 1788 | 49417.9936 | | 2.15445E-05 | | | | | |
| 1789 | 49536.231 | 2.23706E-05 | | | | | | |
| 1790 | 49599.7946 | | | | 1.8152E-05 | | | |
| 1791 | 49630.3434 | | | 1.05767E-05 | | | | |
| 1792 | 49920.8625 | | | | | | -3.19854E-06 | |
| 1793 | 49958.401 | | | | | 1.75058E-05 | | |
| 1794 | 50178.0783 | | 2.16599E-05 | | | | | |
| 1795 | 50318.0588 | 2.36442E-05 | | | | | | |
| 1796 | 50390.2973 | | | | 1.75588E-05 | | | |
| 1797 | 50573.1928 | | | 1.19647E-05 | | | | |
| 1798 | 50888.5687 | | | | | | -3.45573E-06 | |
| 1799 | 50889.2515 | | | | | 1.72102E-05 | | |
| 1800 | 51031.9683 | | 2.23258E-05 | | | | | |
| 1801 | 51178.3877 | 2.28688E-05 | | | | | | |
| 1802 | 51316.1995 | | | | 1.85673E-05 | | | |
| 1803 | 51332.9569 | | | | | | | 8.63781E-06 |
| 1804 | 51414.645 | | | 9.04043E-06 | | | | |
| 1805 | 51855.326 | | | | | 1.57638E-05 | | |
| 1806 | 51956.353 | | | | | | -6.49451E-06 | |
| 1807 | 52045.6487 | | 2.21336E-05 | | | | | |
| 1808 | 52146.9372 | 2.31739E-05 | | | | | | |
| 1809 | 52430.9089 | | | | 1.90038E-05 | | | |
| 1810 | 52532.248 | | | 9.16469E-06 | | | | |
| 1811 | 52595.4794 | | | | | | | 3.20819E-06 |
| 1812 | 52931.0501 | | | | | 1.59662E-05 | | |
| 1813 | 52983.5568 | | 2.44321E-05 | | | | -6.28875E-06 | |
| 1814 | 53093.2194 | 2.56384E-05 | | | | | | |
| 1815 | 53406.2275 | | | | 2.03173E-05 | | | |
| 1816 | 53478.2146 | | | 1.33921E-05 | | | | |
| 1817 | 53601.0783 | | | | | | | 2.11891E-06 |
| 1818 | 53773.4937 | | 2.93742E-05 | | | | | |
| 1819 | 53799.5687 | | | | | 1.64097E-05 | | |
| 1820 | 53907.8462 | 2.98611E-05 | | | | | | |
| 1821 | 53936.1977 | | | | | | -6.53113E-06 | |
| 1822 | 54278.9743 | | | | 3.42446E-05 | | | |
| 1823 | 54352.0808 | | | 1.67368E-05 | | | | |
| 1824 | 54577.7952 | | 3.51819E-05 | | | | | |
| 1825 | 54608.6472 | | | | | 2.35053E-05 | | |
| 1826 | 54624.8641 | 3.48793E-05 | | | | | | |
| 1827 | 54706.7424 | | | | | | | 2.73947E-06 |
| 1828 | 54757.7642 | | | | | | -6.04147E-06 | |
| 1829 | 54840.0552 | | | | 3.38288E-05 | | | |
| 1830 | 54888.1824 | | | 1.75574E-05 | | | | |
| 1831 | 55057.0726 | | 3.78332E-05 | | | | | |
| 1832 | 55161.6491 | 3.75978E-05 | | | | | | |
| 1833 | 55167.2499 | | | | | 2.56558E-05 | | |
| 1834 | 55375.2879 | | | | | | -2.16824E-06 | |
| 1835 | 55394.2752 | | | | 3.54772E-05 | | | |
| 1836 | 55402.9537 | | | 1.97053E-05 | | | | |

| | A | B | C | D | E | F | G | H |
|---|---|---|---|---|---|---|---|---|
| 1837 | 55588.5929 | | 4.04228E-05 | | | | | |
| 1838 | 55631.7544 | 4.13257E-05 | | | | | | |
| 1839 | 55661.8986 | | | | | 2.97327E-05 | | |
| 1840 | 55881.4455 | | | | 3.63776E-05 | | | |
| 1841 | 55882.4714 | | | 2.15077E-05 | | | | |
| 1842 | 56019.542 | | | | | | -1.59751E-06 | |
| 1843 | 56102.0895 | | 4.4875E-05 | | | | | |
| 1844 | 56190.1643 | 4.51977E-05 | | | | | | |
| 1845 | 56227.2958 | | | | | 3.06026E-05 | | |
| 1846 | 56493.5873 | | | | 3.90721E-05 | | | |
| 1847 | 56506.4497 | | | 2.31843E-05 | | | | |
| 1848 | 56679.9188 | | 4.70741E-05 | | | | | |
| 1849 | 56745.7375 | | | | | | -6.44144E-07 | |
| 1850 | 56762.1857 | 4.7352E-05 | | | | | | |
| 1851 | 56827.5991 | | | | | 3.11968E-05 | | |
| 1852 | 56838.7143 | | | | | | | 7.82374E-06 |
| 1853 | 57200.9811 | | | | 3.87675E-05 | | | |
| 1854 | 57209.25 | | | 2.33987E-05 | | | | |
| 1855 | 57405.4158 | | 4.74458E-05 | | | | | |
| 1856 | 57537.7012 | 4.77924E-05 | | | | | | |
| 1857 | 57647.1354 | | | | | 3.15569E-05 | | |
| 1858 | 57665.3182 | | | | | | -1.42633E-06 | |
| 1859 | 58059.725 | | | | 3.82701E-05 | | | |
| 1860 | 58066.9697 | | | 2.30069E-05 | | | | |
| 1861 | 58395.8175 | | 4.83487E-05 | | | | | |
| 1862 | 58514.0452 | 4.84843E-05 | | | | | | |
| 1863 | 58719.8107 | | | | | 2.97515E-05 | | |
| 1864 | 58938.5957 | | | | | | -3.80702E-06 | |
| 1865 | 59308.9459 | | | | 3.93891E-05 | | | |
| 1866 | 59321.6956 | | | 2.73344E-05 | | | | |
| 1867 | 59468.6203 | | | | | | | 9.58417E-06 |
| 1868 | 59636.5067 | | 5.25117E-05 | | | | | |
| 1869 | 59786.0474 | 5.13626E-05 | | | | | | |
| 1870 | 59994.7816 | | | | | 2.69119E-05 | | |
| 1871 | 60225.5901 | | | | | | -8.39461E-06 | |
| 1872 | 60351.9692 | | | 2.38284E-05 | 3.49737E-05 | | | |
| 1873 | 60407.0059 | | 5.22414E-05 | | | | | |
| 1874 | 60609.1056 | 5.02636E-05 | | | | | | |
| 1875 | 60863.9136 | | | | | 2.66404E-05 | | |
| 1876 | 61105.0668 | | | | | | -7.58456E-06 | |
| 1877 | 61164.9706 | | | 2.64029E-05 | 3.67295E-05 | | | |
| 1878 | 61315.9591 | | 5.67345E-05 | | | | | |
| 1879 | 61424.5219 | 5.39529E-05 | | | | | | |
| 1880 | 61620.5279 | | | | | 2.66615E-05 | | |
| 1881 | 61876.1977 | | | | | | -7.2542E-06 | |
| 1882 | 61919.0477 | | | | 3.53062E-05 | | | |
| 1883 | 61924.7434 | | | 2.63852E-05 | | | | |
| 1884 | 62063.8666 | | 5.66006E-05 | | | | | |
| 1885 | 62153.8948 | 5.47255E-05 | | | | | | |
| 1886 | 62390.6645 | | | | | 2.51144E-05 | | |
| 1887 | 62753.6491 | | | 2.50498E-05 | | | | |
| 1888 | 62756.9404 | | | | 3.36963E-05 | | | |
| 1889 | 62768.0915 | | | | | | -7.97572E-06 | |
| 1890 | 62842.603 | | | | | | | 1.31465E-05 |
| 1891 | 62877.7595 | | 5.55332E-05 | | | | | |
| 1892 | 63034.8309 | 5.40676E-05 | | | | | | |
| 1893 | 63283.502 | | | | | 2.45775E-05 | | |
| 1894 | 63565.6588 | | | | 4.46875E-05 | | | |
| 1895 | 63574.9781 | | | | | | -3.30981E-06 | |
| 1896 | 63578.1139 | | | 3.86111E-05 | | | | |
| 1897 | 63790.9409 | | 6.87857E-05 | | | | | |
| 1898 | 63820.8208 | 7.90651E-05 | | | | | | |
| 1899 | 64125.0084 | | | | | 3.32387E-05 | | |
| 1900 | 64218.3399 | | | | | | | 1.32364E-05 |
| 1901 | 64390.3088 | | | | 4.51591E-05 | | -3.28494E-06 | |
| 1902 | 64432.6947 | | | 3.97078E-05 | | | | |
| 1903 | 64639.458 | | 7.25057E-05 | | | | | |
| 1904 | 64907.0493 | 8.31535E-05 | | | | | | |
| 1905 | 65355.4571 | | | | | 3.43429E-05 | | |
| 1906 | 65953.3581 | | | | 4.63507E-05 | | | |
| 1907 | 66009.0874 | | | 4.25705E-05 | | | | |
| 1908 | 66269.9555 | | 7.53696E-05 | | | | -3.49744E-06 | |
| 1909 | 66544.7361 | 8.70957E-05 | | | | | | |
| 1910 | 67372.2453 | | | | | 3.14035E-05 | | |
| 1911 | 68008.9199 | | | | 4.9302E-05 | | | |
| 1912 | 68055.0661 | | | 4.6509E-05 | | | | |
| 1913 | 68249.8734 | | 8.39978E-05 | | | | | |
| 1914 | 68387.2405 | 9.55925E-05 | | | | | | |
| 1915 | 68754.03 | | | | | | -1.95558E-06 | |
| 1916 | 68951.2259 | | | | | 3.11978E-05 | | |
| 1917 | 69153.611 | | | | 4.66844E-05 | | | |
| 1918 | 69167.5781 | | | 4.63984E-05 | | | | |
| 1919 | 69276.9736 | | 8.31199E-05 | | | | | |
| 1920 | 69410.2109 | 9.8414E-05 | | | | | | |
| 1921 | 69516.9916 | | | | | | | 1.38044E-05 |
| 1922 | 69833.1516 | | | | | | 1.37706E-06 | |
| 1923 | 70034.7956 | | | | | 3.15116E-05 | | |
| 1924 | 70331.9008 | | | | 4.79401E-05 | | | |
| 1925 | 70366.0973 | | | 4.96062E-05 | | | | |
| 1926 | 70418.0213 | | 9.01862E-05 | | | | | |
| 1927 | 70551.7381 | 0.000110074 | | | | | | |
| 1928 | 71306.0162 | | | | | | 2.989E-06 | |
| 1929 | 71740.8182 | | | | | 2.98291E-05 | | |
| 1930 | 72538.3196 | | | | 4.73551E-05 | | | |
| 1931 | 72558.3624 | | | 5.14102E-05 | | | | |
| 1932 | 72688.6842 | | 9.16234E-05 | | | | | |
| 1933 | 72852.5147 | 0.000116574 | | | | | | |
| 1934 | 75277.7249 | | | | | | 4.83E-06 | |
| 1935 | 75630.0918 | | | | | 3.04777E-05 | | |
| 1936 | 77000.6274 | | | | 4.83387E-05 | | | |
| 1937 | 77049.079 | | | 5.51585E-05 | | | | |
| 1938 | 77265.2395 | | 9.42553E-05 | | | | | |

|  | A | B | C | D | E | F | G | H |
|---|---|---|---|---|---|---|---|---|
| 1939 | 77606.2628 | 0.000121216 | | | | | | |
| 1940 | 82162.2246 | | | | | | 6.89563E-06 | |
| 1941 | 82843.5087 | | | | | 3.30141E-05 | | |
| 1942 | 84970.6994 | | | | 6.59077E-05 | | | |
| 1943 | 85065.3088 | | | 6.01767E-05 | | | | |
| 1944 | 85233.6618 | | 9.79713E-05 | | | | | |
| 1945 | 85326.1619 | 0.000138662 | | | | | | |
| 1946 | 95605.6146 | | | | | | | 1.53661E-05 |
| 1947 | 102287.496 | | 0.000113368 | 8.66097E-05 | 0.000101554 | 3.81682E-05 | 1.27154E-05 | |
| 1948 | 102745.27 | 0.000220321 | | | | | | |

|     | A        | B         | C         | D         | E         | F         | G         | H         |
|-----|----------|-----------|-----------|-----------|-----------|-----------|-----------|-----------|
| 1   | Rank     | Inf       | 0.002     | 0.0015    | 0.0013    | 0.001     | 0.0005    | 0.0001    |
| 2   | 0        | -2.75E-06 | -7.16E-06 | -1.53E-05 | -1.56E-05 | -1.48E-05 | -1.84E-05 | -8.81E-06 |
| 3   | 144.9339 |           |           |           |           |           |           | -7.27E-06 |
| 4   | 170.1765 |           |           |           |           |           | -1.75E-05 |           |
| 5   | 187.6639 |           |           | -1.52E-05 | -1.54E-05 | -1.43E-05 |           |           |
| 6   | 188.5724 |           | -7.45E-06 |           |           |           |           |           |
| 7   | 195.6518 | -6.14E-06 |           |           |           |           |           |           |
| 8   | 295.1507 |           |           |           |           |           |           | -6.60E-06 |
| 9   | 368.8204 |           |           |           |           |           | -1.85E-05 |           |
| 10  | 370.1827 |           |           | -1.75E-05 | -1.77E-05 | -1.67E-05 |           |           |
| 11  | 373.158  |           | -1.12E-05 |           |           |           |           |           |
| 12  | 373.9716 | -1.10E-05 |           |           |           |           |           |           |
| 13  | 377.8136 |           |           |           |           |           |           | -6.19E-06 |
| 14  | 438.6157 |           |           |           |           |           |           | -5.40E-06 |
| 15  | 477.6567 |           |           |           |           |           | -1.92E-05 |           |
| 16  | 478.5557 |           |           |           |           | -1.84E-05 |           |           |
| 17  | 478.6781 |           |           | -1.95E-05 | -1.96E-05 |           |           |           |
| 18  | 482.499  |           | -1.32E-05 |           |           |           |           |           |
| 19  | 482.7168 | -1.47E-05 |           |           |           |           |           |           |
| 20  | 502.6359 |           |           |           |           |           |           | -4.98E-06 |
| 21  | 558.2789 |           |           |           |           |           |           | -4.83E-06 |
| 22  | 581.9667 |           |           |           |           |           | -1.99E-05 |           |
| 23  | 583.9893 |           |           | -2.08E-05 | -2.08E-05 | -1.93E-05 |           |           |
| 24  | 586.5086 |           | -1.47E-05 |           |           |           |           |           |
| 25  | 587.3981 | -1.74E-05 |           |           |           |           |           |           |
| 26  | 597.711  |           |           |           |           |           |           | -4.69E-06 |
| 27  | 639.0041 |           |           |           |           |           |           | -4.53E-06 |
| 28  | 658.6859 |           |           |           |           | -1.99E-05 | -2.00E-05 |           |
| 29  | 659.1005 |           |           | -2.14E-05 | -2.16E-05 |           |           |           |
| 30  | 664.0816 | -2.17E-05 | -1.65E-05 |           |           |           |           |           |
| 31  | 681.8055 |           |           |           |           |           |           | -4.39E-06 |
| 32  | 728.2594 |           |           |           |           |           |           | -4.28E-06 |
| 33  | 737.7113 |           |           |           |           |           | -1.99E-05 |           |
| 34  | 744.7291 |           |           |           | -2.22E-05 | -2.05E-05 |           |           |
| 35  | 744.7865 |           |           | -2.21E-05 |           |           |           |           |
| 36  | 751.2108 |           | -1.75E-05 |           |           |           |           |           |
| 37  | 752.3151 | -2.17E-05 |           |           |           |           |           |           |
| 38  | 776.7122 |           |           |           |           |           |           | -4.22E-06 |
| 39  | 819.4231 |           |           |           |           |           | -2.02E-05 |           |
| 40  | 819.4403 |           |           |           |           |           |           | -4.16E-06 |
| 41  | 825.4923 |           |           |           |           | -2.12E-05 |           |           |
| 42  | 826.4174 |           |           | -2.27E-05 | -2.28E-05 |           |           |           |
| 43  | 834.2102 |           | -1.80E-05 |           |           |           |           |           |
| 44  | 834.6203 | -2.42E-05 |           |           |           |           |           |           |
| 45  | 858.7477 |           |           |           |           |           |           | -4.02E-06 |
| 46  | 890.8636 |           |           |           |           |           | -2.04E-05 |           |
| 47  | 895.3621 |           |           |           |           | -2.14E-05 |           |           |
| 48  | 895.6228 |           |           |           | -2.31E-05 |           |           |           |
| 49  | 895.7544 |           |           | -2.31E-05 |           |           |           |           |
| 50  | 902.3289 |           | -1.82E-05 |           |           |           |           |           |
| 51  | 902.3985 | -2.57E-05 |           |           |           |           |           |           |
| 52  | 904.7207 |           |           |           |           |           |           | -3.75E-06 |
| 53  | 948.4332 |           |           |           |           |           |           | -3.77E-06 |
| 54  | 959.6388 |           |           |           |           |           | -2.06E-05 |           |
| 55  | 963.3911 |           |           |           |           | -2.17E-05 |           |           |
| 56  | 964.0542 |           |           | -2.33E-05 | -2.33E-05 |           |           |           |
| 57  | 971.302  | -2.64E-05 |           |           |           |           |           |           |
| 58  | 972.2128 |           | -1.89E-05 |           |           |           |           |           |
| 59  | 990.3324 |           |           |           |           |           |           | -3.83E-06 |
| 60  | 1019.581 |           |           |           |           |           |           | -3.64E-06 |
| 61  | 1027.458 |           |           |           |           |           | -2.04E-05 |           |
| 62  | 1033.16  |           |           |           |           | -2.16E-05 |           |           |
| 63  | 1034.717 |           |           |           | -2.32E-05 |           |           |           |
| 64  | 1035.186 |           |           | -2.32E-05 |           |           |           |           |
| 65  | 1040.942 | -2.61E-05 |           |           |           |           |           |           |
| 66  | 1041.19  |           | -1.90E-05 |           |           |           |           |           |
| 67  | 1064.842 |           |           |           |           |           |           | -3.35E-06 |
| 68  | 1107.281 |           |           |           |           |           | -2.03E-05 |           |
| 69  | 1113.154 |           |           |           |           | -2.16E-05 |           | -3.08E-06 |
| 70  | 1114.646 |           |           |           | -2.31E-05 |           |           |           |
| 71  | 1114.676 |           |           | -2.32E-05 |           |           |           |           |
| 72  | 1120.718 | -2.82E-05 |           |           |           |           |           |           |
| 73  | 1121.724 |           | -1.95E-05 |           |           |           |           |           |
| 74  | 1159.708 |           |           |           |           |           |           | -2.99E-06 |
| 75  | 1177.232 |           |           |           |           |           | -2.06E-05 |           |
| 76  | 1186.275 |           |           |           |           | -2.21E-05 |           |           |
| 77  | 1186.833 |           |           |           | -2.35E-05 |           |           |           |
| 78  | 1187.417 |           |           | -2.36E-05 |           |           |           |           |
| 79  | 1194.654 | -2.88E-05 |           |           |           |           |           |           |
| 80  | 1195.562 |           | -2.00E-05 |           |           |           |           |           |
| 81  | 1196.27  |           |           |           |           |           |           | -3.07E-06 |
| 82  | 1236.609 |           |           |           |           |           |           | -3.06E-06 |
| 83  | 1244.262 |           |           |           |           |           | -2.07E-05 |           |
| 84  | 1248.906 |           |           |           |           | -2.22E-05 |           |           |
| 85  | 1253.18  |           |           |           | -2.37E-05 |           |           |           |
| 86  | 1253.421 |           |           | -2.38E-05 |           |           |           |           |
| 87  | 1258.927 | -2.99E-05 |           |           |           |           |           |           |
| 88  | 1259.441 |           | -2.05E-05 |           |           |           |           |           |
| 89  | 1270.858 |           |           |           |           |           |           | -3.01E-06 |
| 90  | 1307.084 |           |           |           |           |           | -2.05E-05 |           |
| 91  | 1316.431 |           |           |           |           | -2.20E-05 |           |           |
| 92  | 1320.016 |           |           |           | -2.36E-05 |           |           |           |
| 93  | 1320.104 |           |           | -2.38E-05 |           |           |           | -2.97E-06 |
| 94  | 1328.313 | -3.05E-05 |           |           |           |           |           |           |
| 95  | 1329.953 |           | -2.08E-05 |           |           |           |           |           |
| 96  | 1368.977 |           |           |           |           |           |           | -2.84E-06 |
| 97  | 1377.725 |           |           |           |           |           | -2.07E-05 |           |
| 98  | 1390.677 |           |           |           |           | -2.21E-05 |           |           |
| 99  | 1393.394 |           |           |           | -2.37E-05 |           |           |           |
| 100 | 1393.476 |           |           | -2.39E-05 |           |           |           |           |
| 101 | 1403.714 | -3.10E-05 |           |           |           |           |           |           |
| 102 | 1404.434 |           | -2.08E-05 |           |           |           |           |           |

| | A | B | C | D | E | F | G | H |
|---|---|---|---|---|---|---|---|---|
| 103 | 1409.511 | | | | | | | -2.66E-06 |
| 104 | 1451.473 | | | | | | -2.06E-05 | |
| 105 | 1459.325 | | | | | -2.20E-05 | | |
| 106 | 1460.879 | | | | -2.37E-05 | | | |
| 107 | 1461.199 | | | -2.39E-05 | | | | |
| 108 | 1461.408 | | | | | | | -2.54E-06 |
| 109 | 1467.206 | -3.18E-05 | | | | | | |
| 110 | 1467.366 | | -2.05E-05 | | | | | |
| 111 | 1500.337 | | | | | | | -2.54E-06 |
| 112 | 1509.143 | | | | | | -2.07E-05 | |
| 113 | 1520.119 | | | | | -2.22E-05 | | |
| 114 | 1525.794 | | | | -2.37E-05 | | | |
| 115 | 1526.476 | | | -2.39E-05 | | | | |
| 116 | 1533.978 | -3.21E-05 | | | | | | |
| 117 | 1533.983 | | -2.05E-05 | | | | | |
| 118 | 1541.509 | | | | | | | -2.42E-06 |
| 119 | 1568.599 | | | | | | -2.05E-05 | |
| 120 | 1574.852 | | | | | | | -2.30E-06 |
| 121 | 1578.242 | | | | | -2.21E-05 | | |
| 122 | 1579.756 | | | | -2.36E-05 | | | |
| 123 | 1579.807 | | | -2.38E-05 | | | | |
| 124 | 1588.806 | -3.22E-05 | | | | | | |
| 125 | 1589.377 | | -2.04E-05 | | | | | |
| 126 | 1614.524 | | | | | | | -2.26E-06 |
| 127 | 1626.078 | | | | | | -2.05E-05 | |
| 128 | 1632.452 | | | | | -2.22E-05 | | |
| 129 | 1635.599 | | | | -2.37E-05 | | | |
| 130 | 1636.508 | | | -2.40E-05 | | | | |
| 131 | 1645.394 | | | | | | | -2.21E-06 |
| 132 | 1645.89 | -3.24E-05 | -2.06E-05 | | | | | |
| 133 | 1682.752 | | | | | | -2.05E-05 | |
| 134 | 1695.352 | | | | | | | -2.19E-06 |
| 135 | 1699.534 | | | | | -2.20E-05 | | |
| 136 | 1702.203 | | | | -2.35E-05 | | | |
| 137 | 1702.621 | | | -2.38E-05 | | | | |
| 138 | 1712.97 | -3.25E-05 | | | | | | |
| 139 | 1714.055 | | -2.00E-05 | | | | | |
| 140 | 1736.577 | | | | | | | -1.89E-06 |
| 141 | 1747.607 | | | | | | -2.04E-05 | |
| 142 | 1758.16 | | | | | -2.21E-05 | | |
| 143 | 1761.288 | | | -2.39E-05 | -2.36E-05 | | | |
| 144 | 1769.155 | -3.27E-05 | | | | | | |
| 145 | 1770.694 | | -2.10E-05 | | | | | |
| 146 | 1774.609 | | | | | | | -1.92E-06 |
| 147 | 1805.783 | | | | | | -2.04E-05 | |
| 148 | 1819.161 | | | | | -2.21E-05 | | |
| 149 | 1820.127 | | | | | | | -1.79E-06 |
| 150 | 1822.617 | | | -2.39E-05 | -2.36E-05 | | | |
| 151 | 1831.041 | -3.31E-05 | -2.06E-05 | | | | | |
| 152 | 1853.479 | | | | | | | -1.75E-06 |
| 153 | 1858.864 | | | | | | -2.04E-05 | |
| 154 | 1874.86 | | | | | -2.21E-05 | | |
| 155 | 1876.884 | | | -2.39E-05 | -2.36E-05 | | | |
| 156 | 1883.847 | | | | | | | -1.61E-06 |
| 157 | 1884.917 | -3.38E-05 | | | | | | |
| 158 | 1885.322 | | -2.08E-05 | | | | | |
| 159 | 1917.808 | | | | | | -2.04E-05 | |
| 160 | 1926.996 | | | | | | | -1.65E-06 |
| 161 | 1932.727 | | | | | -2.21E-05 | | |
| 162 | 1936.535 | | | | -2.37E-05 | | | |
| 163 | 1937.048 | | | -2.39E-05 | | | | |
| 164 | 1948.037 | | -2.09E-05 | | | | | |
| 165 | 1948.994 | -3.37E-05 | | | | | | |
| 166 | 1969.516 | | | | | | | -1.45E-06 |
| 167 | 1978.244 | | | | | | -2.03E-05 | |
| 168 | 1991.703 | | | | | -2.19E-05 | | |
| 169 | 1997.058 | | | | -2.36E-05 | | | |
| 170 | 1997.794 | | | -2.38E-05 | | | | |
| 171 | 2016.121 | | | | | | | -1.55E-06 |
| 172 | 2017.453 | -3.48E-05 | -2.08E-05 | | | | | |
| 173 | 2040.002 | | | | | | -2.03E-05 | |
| 174 | 2050.903 | | | | | | | -1.58E-06 |
| 175 | 2050.952 | | | | | -2.20E-05 | | |
| 176 | 2054.362 | | | | -2.37E-05 | | | |
| 177 | 2055.214 | | | -2.40E-05 | | | | |
| 178 | 2071.765 | | -2.13E-05 | | | | | |
| 179 | 2072.674 | -3.52E-05 | | | | | | |
| 180 | 2097.124 | | | | | | | -1.54E-06 |
| 181 | 2100.487 | | | | | | -2.03E-05 | |
| 182 | 2117.602 | | | | | -2.21E-05 | | |
| 183 | 2119.925 | | | | -2.37E-05 | | | |
| 184 | 2119.926 | | | -2.41E-05 | | | | |
| 185 | 2132.673 | | -2.10E-05 | | | | | |
| 186 | 2132.801 | -3.60E-05 | | | | | | |
| 187 | 2141.497 | | | | | | | -1.59E-06 |
| 188 | 2156.42 | | | | | | -2.03E-05 | |
| 189 | 2170.295 | | | | | -2.21E-05 | | |
| 190 | 2171.201 | | | | | | | -1.43E-06 |
| 191 | 2172.662 | | | | -2.37E-05 | | | |
| 192 | 2173.14 | | | -2.41E-05 | | | | |
| 193 | 2187.156 | | -2.12E-05 | | | | | |
| 194 | 2188.055 | -3.60E-05 | | | | | | |
| 195 | 2206.532 | | | | | | | -1.29E-06 |
| 196 | 2207.789 | | | | | | -2.02E-05 | |
| 197 | 2226.375 | | | | | -2.21E-05 | | |
| 198 | 2229.637 | | | | -2.37E-05 | | | |
| 199 | 2230.852 | | | -2.41E-05 | | | | |
| 200 | 2246.387 | | | | | | | -1.26E-06 |
| 201 | 2248.713 | | -2.15E-05 | | | | | |
| 202 | 2249.391 | -3.61E-05 | | | | | | |
| 203 | 2271.862 | | | | | | -2.02E-05 | |
| 204 | 2287.295 | | | | | -2.20E-05 | | -1.28E-06 |

| | A | B | C | D | E | F | G | H |
|---|---|---|---|---|---|---|---|---|
| 205 | 2292.045 | | | -2.41E-05 | -2.37E-05 | | | |
| 206 | 2304.699 | -3.61E-05 | -2.16E-05 | | | | | |
| 207 | 2319.583 | | | | | | | -1.35E-06 |
| 208 | 2322.467 | | | | | | -2.01E-05 | |
| 209 | 2333.008 | | | | | -2.20E-05 | | |
| 210 | 2335.76 | | | | -2.37E-05 | | | |
| 211 | 2335.925 | | | -2.42E-05 | | | | |
| 212 | 2345.339 | | | | | | | -6.07E-07 |
| 213 | 2351.322 | | -2.17E-05 | | | | | |
| 214 | 2351.434 | -3.70E-05 | | | | | | |
| 215 | 2367.323 | | | | | | -2.02E-05 | |
| 216 | 2377.592 | | | | | | | -2.54E-07 |
| 217 | 2378.368 | | | | | -2.20E-05 | | |
| 218 | 2381.04 | | | | -2.36E-05 | | | |
| 219 | 2381.049 | | | -2.42E-05 | | | | |
| 220 | 2396.97 | | -2.18E-05 | | | | | |
| 221 | 2397.274 | -3.77E-05 | | | | | | |
| 222 | 2408.784 | | | | | | | -1.17E-07 |
| 223 | 2410.495 | | | | | | -2.02E-05 | |
| 224 | 2427.205 | | | | | -2.19E-05 | | |
| 225 | 2428.789 | | | -2.42E-05 | -2.36E-05 | | | |
| 226 | 2436.234 | | | | | | | -1.95E-07 |
| 227 | 2443.38 | | -2.18E-05 | | | | | |
| 228 | 2446.08 | -3.80E-05 | | | | | | |
| 229 | 2458.389 | | | | | | -2.02E-05 | |
| 230 | 2473.332 | | | | | | | -1.22E-07 |
| 231 | 2476.906 | | | | | -2.18E-05 | | |
| 232 | 2479.826 | | | | -2.38E-05 | | | |
| 233 | 2480.07 | | | -2.42E-05 | | | | |
| 234 | 2496.102 | | -2.18E-05 | | | | | |
| 235 | 2498.213 | -3.77E-05 | | | | | | |
| 236 | 2511.856 | | | | | | | -4.08E-08 |
| 237 | 2512.864 | | | | | | -2.01E-05 | |
| 238 | 2527.541 | | | | | -2.16E-05 | | |
| 239 | 2530.563 | | | | -2.36E-05 | | | |
| 240 | 2530.601 | | | -2.39E-05 | | | | |
| 241 | 2538.423 | | | | | | | 1.49E-07 |
| 242 | 2544.662 | | -2.14E-05 | | | | | |
| 243 | 2546.178 | -3.66E-05 | | | | | | |
| 244 | 2557.147 | | | | | | -1.99E-05 | |
| 245 | 2573.982 | | | | | -2.15E-05 | | 2.14E-07 |
| 246 | 2579.596 | | | | -2.34E-05 | | | |
| 247 | 2579.8 | | | -2.38E-05 | | | | |
| 248 | 2596.075 | | -2.11E-05 | | | | | |
| 249 | 2598.56 | -3.71E-05 | | | | | | |
| 250 | 2605.879 | | | | | | -1.98E-05 | |
| 251 | 2610.494 | | | | | | | 5.33E-07 |
| 252 | 2630.612 | | | | | -2.14E-05 | | |
| 253 | 2635.333 | | | | -2.33E-05 | | | |
| 254 | 2635.9 | | | -2.37E-05 | | | | |
| 255 | 2649.055 | | | | | | | 6.68E-07 |
| 256 | 2651.347 | | -2.10E-05 | | | | | |
| 257 | 2652.499 | -3.71E-05 | | | | | | |
| 258 | 2661.437 | | | | | | -1.97E-05 | |
| 259 | 2681.02 | | | | | -2.14E-05 | | |
| 260 | 2686.614 | | | | -2.33E-05 | | | |
| 261 | 2687.29 | | | -2.38E-05 | | | | |
| 262 | 2688.889 | | | | | | | 7.95E-07 |
| 263 | 2709.29 | | -2.15E-05 | | | | | |
| 264 | 2710.011 | -3.74E-05 | | | | | | |
| 265 | 2716.596 | | | | | | -1.96E-05 | |
| 266 | 2730.532 | | | | | | | 8.51E-07 |
| 267 | 2736.201 | | | | | -2.14E-05 | | |
| 268 | 2740.47 | | | | -2.34E-05 | | | |
| 269 | 2740.951 | | | -2.38E-05 | | | | |
| 270 | 2762.44 | | -2.13E-05 | | | | | |
| 271 | 2763.578 | -3.73E-05 | | | | | | |
| 272 | 2766.931 | | | | | | | 8.77E-07 |
| 273 | 2768.478 | | | | | | -1.95E-05 | |
| 274 | 2783.054 | | | | | -2.13E-05 | | |
| 275 | 2785.942 | | | | -2.33E-05 | | | |
| 276 | 2786.713 | | | -2.38E-05 | | | | |
| 277 | 2797.639 | | | | | | | 8.98E-07 |
| 278 | 2805.188 | | -2.11E-05 | | | | | |
| 279 | 2805.846 | -3.75E-05 | | | | | | |
| 280 | 2808.983 | | | | | | -1.96E-05 | |
| 281 | 2828.81 | | | | | -2.13E-05 | | |
| 282 | 2829.1 | | | | | | | 9.16E-07 |
| 283 | 2838.052 | | | | -2.32E-05 | | | |
| 284 | 2838.824 | | | -2.37E-05 | | | | |
| 285 | 2861.21 | | -2.11E-05 | | | | | |
| 286 | 2861.341 | -3.74E-05 | | | | | | |
| 287 | 2863.069 | | | | | | -1.95E-05 | |
| 288 | 2867.16 | | | | | | | 9.49E-07 |
| 289 | 2886.112 | | | | | -2.11E-05 | | |
| 290 | 2891.306 | | | | -2.30E-05 | | | |
| 291 | 2891.631 | | | -2.36E-05 | | | | |
| 292 | 2903.905 | | | | | | | 9.19E-07 |
| 293 | 2910.749 | | -2.10E-05 | | | | | |
| 294 | 2912.964 | -3.74E-05 | | | | | | |
| 295 | 2913.789 | | | | | | -1.95E-05 | |
| 296 | 2935.355 | | | | | -2.11E-05 | | |
| 297 | 2942.496 | | | | | | | 9.88E-07 |
| 298 | 2942.566 | | | | -2.29E-05 | | | |
| 299 | 2943.5 | | | -2.35E-05 | | | | |
| 300 | 2963.767 | -3.72E-05 | | | | | | |
| 301 | 2964.22 | | -2.06E-05 | | | | | |
| 302 | 2966.471 | | | | | | -1.93E-05 | |
| 303 | 2980.808 | | | | | | | 9.76E-07 |
| 304 | 2994.143 | | | | | -2.10E-05 | | |
| 305 | 2998.929 | | | | -2.29E-05 | | | |
| 306 | 2999.927 | | | -2.34E-05 | | | | |

|     | A        | B         | C         | D         | E         | F         | G         | H        |
|-----|----------|-----------|-----------|-----------|-----------|-----------|-----------|----------|
| 307 | 3027.496 | -3.67E-05 |           |           |           |           |           |          |
| 308 | 3027.585 |           |           |           |           |           |           | 9.64E-07 |
| 309 | 3027.618 |           | -2.05E-05 |           |           |           | -1.93E-05 |          |
| 310 | 3046.456 |           |           |           |           | -2.09E-05 |           |          |
| 311 | 3050.117 |           |           |           | -2.28E-05 |           |           |          |
| 312 | 3050.598 |           |           | -2.34E-05 |           |           |           |          |
| 313 | 3052.941 |           |           |           |           |           |           | 1.00E-06 |
| 314 | 3067.207 |           |           |           |           |           | -1.91E-05 |          |
| 315 | 3067.754 | -3.68E-05 |           |           |           |           |           |          |
| 316 | 3068.133 |           | -2.06E-05 |           |           |           |           |          |
| 317 | 3081.486 |           |           |           |           |           |           | 1.04E-06 |
| 318 | 3094.136 |           |           |           |           | -2.07E-05 |           |          |
| 319 | 3098.952 |           |           |           | -2.27E-05 |           |           |          |
| 320 | 3099.609 |           |           | -2.32E-05 |           |           |           |          |
| 321 | 3117.218 |           |           |           |           |           |           | 1.21E-06 |
| 322 | 3118.2   |           |           |           |           |           | -1.90E-05 |          |
| 323 | 3119.321 | -3.82E-05 |           |           |           |           |           |          |
| 324 | 3120.746 |           | -2.04E-05 |           |           |           |           |          |
| 325 | 3139.474 |           |           |           |           | -2.06E-05 |           |          |
| 326 | 3145.76  |           |           |           | -2.26E-05 |           |           |          |
| 327 | 3146.529 |           |           | -2.32E-05 |           |           |           |          |
| 328 | 3149.379 |           |           |           |           |           |           | 1.15E-06 |
| 329 | 3161.951 |           |           |           |           |           | -1.89E-05 |          |
| 330 | 3164.909 | -3.81E-05 |           |           |           |           |           |          |
| 331 | 3165.954 |           | -2.01E-05 |           |           |           |           |          |
| 332 | 3188.888 |           |           |           |           |           |           | 1.18E-06 |
| 333 | 3192.479 |           |           |           |           | -2.05E-05 |           |          |
| 334 | 3200.88  |           |           |           | -2.24E-05 |           |           |          |
| 335 | 3201.286 |           |           | -2.31E-05 |           |           |           |          |
| 336 | 3221.607 |           |           |           |           |           | -1.89E-05 |          |
| 337 | 3226.987 | -3.79E-05 |           |           |           |           |           |          |
| 338 | 3227.893 |           | -2.00E-05 |           |           |           |           |          |
| 339 | 3231.615 |           |           |           |           |           |           | 1.22E-06 |
| 340 | 3247.217 |           |           |           |           | -2.04E-05 |           |          |
| 341 | 3254.018 |           |           |           | -2.25E-05 |           |           |          |
| 342 | 3256.142 |           |           | -2.31E-05 |           |           |           |          |
| 343 | 3273.4   |           |           |           |           |           |           | 1.24E-06 |
| 344 | 3277.356 |           |           |           |           |           | -1.88E-05 |          |
| 345 | 3283.14  | -3.76E-05 |           |           |           |           |           |          |
| 346 | 3284.12  |           | -1.99E-05 |           |           |           |           |          |
| 347 | 3300.516 |           |           |           |           | -2.03E-05 |           |          |
| 348 | 3304.596 |           |           |           |           |           |           | 1.36E-06 |
| 349 | 3310.593 |           |           |           | -2.22E-05 |           |           |          |
| 350 | 3311.775 |           |           | -2.28E-05 |           |           |           |          |
| 351 | 3328.002 |           |           |           |           |           | -1.86E-05 |          |
| 352 | 3336.817 | -3.66E-05 |           |           |           |           |           |          |
| 353 | 3338.83  |           | -1.98E-05 |           |           |           |           |          |
| 354 | 3349.315 |           |           |           |           |           |           | 1.40E-06 |
| 355 | 3364.628 |           |           |           |           | -2.02E-05 |           |          |
| 356 | 3370.399 |           |           |           | -2.23E-05 |           |           |          |
| 357 | 3372.486 |           |           | -2.28E-05 |           |           |           |          |
| 358 | 3390.695 |           |           |           |           |           |           | 1.35E-06 |
| 359 | 3392.057 |           |           |           |           |           | -1.86E-05 |          |
| 360 | 3396.791 | -3.69E-05 |           |           |           |           |           |          |
| 361 | 3399.22  |           | -2.00E-05 |           |           |           |           |          |
| 362 | 3417.276 |           |           |           |           | -2.01E-05 |           |          |
| 363 | 3424.477 |           |           | -2.27E-05 | -2.22E-05 |           |           | 1.33E-06 |
| 364 | 3438.996 |           |           |           |           |           | -1.85E-05 |          |
| 365 | 3446.7   | -3.72E-05 |           |           |           |           |           |          |
| 366 | 3450.943 |           | -1.99E-05 |           |           |           |           |          |
| 367 | 3461.898 |           |           |           |           |           |           | 1.36E-06 |
| 368 | 3471.129 |           |           |           |           | -1.99E-05 |           |          |
| 369 | 3477.201 |           |           |           | -2.20E-05 |           |           |          |
| 370 | 3478.932 |           |           | -2.25E-05 |           |           |           |          |
| 371 | 3493.785 |           |           |           |           |           | -1.85E-05 |          |
| 372 | 3499.883 | -3.69E-05 |           |           |           |           |           |          |
| 373 | 3501.961 |           |           |           |           |           |           | 1.42E-06 |
| 374 | 3504.497 |           | -1.99E-05 |           |           |           |           |          |
| 375 | 3526.131 |           |           |           |           | -1.98E-05 |           |          |
| 376 | 3534.178 |           |           |           | -2.19E-05 |           |           |          |
| 377 | 3535.302 |           |           | -2.25E-05 |           |           |           |          |
| 378 | 3551.723 |           |           |           |           |           |           | 1.39E-06 |
| 379 | 3551.788 |           |           |           |           |           | -1.84E-05 |          |
| 380 | 3561.471 | -3.69E-05 |           |           |           |           |           |          |
| 381 | 3566.475 |           | -1.96E-05 |           |           |           |           |          |
| 382 | 3584.968 |           |           |           |           | -1.96E-05 |           |          |
| 383 | 3597.274 |           |           |           | -2.17E-05 |           |           |          |
| 384 | 3598.757 |           |           | -2.22E-05 |           |           |           |          |
| 385 | 3599.274 |           |           |           |           |           |           | 1.48E-06 |
| 386 | 3618.577 |           |           |           |           |           | -1.82E-05 |          |
| 387 | 3627.075 | -3.67E-05 |           |           |           |           |           |          |
| 388 | 3631.022 |           | -1.95E-05 |           |           |           |           |          |
| 389 | 3635.794 |           |           |           |           |           |           | 1.45E-06 |
| 390 | 3647.26  |           |           |           |           | -1.94E-05 |           |          |
| 391 | 3655.669 |           |           |           | -2.15E-05 |           |           |          |
| 392 | 3656.926 |           |           | -2.20E-05 |           |           |           |          |
| 393 | 3670.949 |           |           |           |           |           | -1.80E-05 |          |
| 394 | 3682.726 | -3.55E-05 |           |           |           |           |           |          |
| 395 | 3683.592 |           |           |           |           |           |           | 1.55E-06 |
| 396 | 3687.728 |           | -1.92E-05 |           |           |           |           |          |
| 397 | 3703.541 |           |           |           |           | -1.93E-05 |           |          |
| 398 | 3709.17  |           |           |           | -2.15E-05 |           |           |          |
| 399 | 3711.328 |           |           | -2.20E-05 |           |           |           |          |
| 400 | 3723.015 |           |           |           |           |           |           | 1.51E-06 |
| 401 | 3725.145 |           |           |           |           |           | -1.79E-05 |          |
| 402 | 3737.635 | -3.64E-05 |           |           |           |           |           |          |
| 403 | 3747.051 |           | -1.92E-05 |           |           |           |           |          |
| 404 | 3767.648 |           |           |           |           | -1.95E-05 |           |          |
| 405 | 3778.267 |           |           |           | -2.17E-05 |           |           |          |
| 406 | 3779.448 |           |           | -2.23E-05 |           |           |           |          |
| 407 | 3782.658 |           |           |           |           |           |           | 1.49E-06 |
| 408 | 3791.202 |           |           |           |           |           | -1.79E-05 |          |

| | A | B | C | D | E | F | G | H |
|---|---|---|---|---|---|---|---|---|
| 409 | 3800.647 | -3.66E-05 | | | | | | |
| 410 | 3809.543 | | -1.92E-05 | | | | | |
| 411 | 3826.401 | | | | | -1.96E-05 | | |
| 412 | 3826.692 | | | | | | | 1.47E-06 |
| 413 | 3830.962 | | | | -2.16E-05 | | | |
| 414 | 3831.642 | | | -2.22E-05 | | | | |
| 415 | 3844.31 | | | | | | -1.78E-05 | |
| 416 | 3861.275 | -3.65E-05 | | | | | | |
| 417 | 3867.188 | | -1.89E-05 | | | | | |
| 418 | 3870.218 | | | | | | | 1.51E-06 |
| 419 | 3889.095 | | | | | -1.95E-05 | | |
| 420 | 3894.266 | | | | -2.16E-05 | | | |
| 421 | 3897.039 | | | -2.21E-05 | | | | |
| 422 | 3908.278 | | | | | | -1.78E-05 | |
| 423 | 3913.24 | | | | | | | 1.57E-06 |
| 424 | 3919.029 | -3.65E-05 | | | | | | |
| 425 | 3929.478 | | -1.85E-05 | | | | | |
| 426 | 3942.627 | | | | | -1.94E-05 | | |
| 427 | 3949.624 | | | | -2.14E-05 | | | 1.48E-06 |
| 428 | 3951.331 | | | -2.19E-05 | | | | |
| 429 | 3961.712 | | | | | | -1.78E-05 | |
| 430 | 3980.257 | -3.68E-05 | | | | | | |
| 431 | 3990.911 | | -1.84E-05 | | | | | |
| 432 | 4003.874 | | | | | | | 1.34E-06 |
| 433 | 4007.089 | | | | | -1.93E-05 | | |
| 434 | 4017.448 | | | | -2.13E-05 | | | |
| 435 | 4018.483 | | | -2.18E-05 | | | | |
| 436 | 4031.369 | | | | | | -1.77E-05 | |
| 437 | 4041.03 | -3.60E-05 | | | | | | |
| 438 | 4050.492 | | | | | | | 1.42E-06 |
| 439 | 4052.088 | | -1.80E-05 | | | | | |
| 440 | 4066.836 | | | | | -1.93E-05 | | |
| 441 | 4074.009 | | | | -2.12E-05 | | | |
| 442 | 4078.215 | | | -2.18E-05 | | | | |
| 443 | 4088.023 | | | | | | -1.75E-05 | |
| 444 | 4103.338 | | | | | | | 1.37E-06 |
| 445 | 4108.812 | -3.68E-05 | | | | | | |
| 446 | 4117.369 | | -1.80E-05 | | | | | |
| 447 | 4133.382 | | | | | -1.92E-05 | | |
| 448 | 4141.904 | | | | -2.10E-05 | | | |
| 449 | 4143.538 | | | -2.17E-05 | | | | |
| 450 | 4146.157 | | | | | | | 1.38E-06 |
| 451 | 4151.91 | | | | | | -1.75E-05 | |
| 452 | 4164.784 | -3.72E-05 | | | | | | |
| 453 | 4176.78 | | -1.77E-05 | | | | | |
| 454 | 4181.137 | | | | | | | 1.39E-06 |
| 455 | 4188.02 | | | | | -1.91E-05 | | |
| 456 | 4196.165 | | | | -2.08E-05 | | | |
| 457 | 4196.542 | | | -2.15E-05 | | | | |
| 458 | 4208.726 | | | | | | -1.73E-05 | |
| 459 | 4226.628 | -3.71E-05 | | | | | | |
| 460 | 4231.254 | | | | | | | 1.48E-06 |
| 461 | 4238.304 | | -1.75E-05 | | | | | |
| 462 | 4251.622 | | | | | -1.90E-05 | | |
| 463 | 4263.137 | | | | -2.05E-05 | | | |
| 464 | 4264.842 | | | -2.12E-05 | | | | |
| 465 | 4269.307 | | | | | | -1.72E-05 | |
| 466 | 4273.204 | | | | | | | 1.45E-06 |
| 467 | 4289.023 | -3.68E-05 | | | | | | |
| 468 | 4305.747 | | -1.74E-05 | | | | | |
| 469 | 4320.002 | | | | | -1.89E-05 | | |
| 470 | 4328.785 | | | | -2.03E-05 | | | |
| 471 | 4329.476 | | | | | | | 1.37E-06 |
| 472 | 4330.205 | | | -2.10E-05 | | | | |
| 473 | 4339.136 | | | | | | -1.73E-05 | |
| 474 | 4367.242 | -3.68E-05 | | | | | | |
| 475 | 4381.716 | | -1.70E-05 | | | | | |
| 476 | 4392.348 | | | | | | | 1.27E-06 |
| 477 | 4395.346 | | | | | -1.88E-05 | | |
| 478 | 4403.649 | | | | -1.99E-05 | | | |
| 479 | 4406.539 | | | -2.07E-05 | | | | |
| 480 | 4412.712 | | | | | | -1.73E-05 | |
| 481 | 4429.531 | -3.68E-05 | | | | | | |
| 482 | 4438.112 | | | | | | | 1.23E-06 |
| 483 | 4445.176 | | -1.69E-05 | | | | | |
| 484 | 4457.171 | | | | | -1.87E-05 | | |
| 485 | 4467.145 | | | | -1.95E-05 | | | |
| 486 | 4468.949 | | | -2.04E-05 | | | | |
| 487 | 4473.719 | | | | | | -1.72E-05 | |
| 488 | 4486.423 | | | | | | | 1.24E-06 |
| 489 | 4497.969 | -3.60E-05 | | | | | | |
| 490 | 4519.173 | | -1.66E-05 | | | | | |
| 491 | 4531.715 | | | | | -1.84E-05 | | |
| 492 | 4544.124 | | | | -1.92E-05 | | | |
| 493 | 4546.321 | | | -2.01E-05 | | | | |
| 494 | 4547.489 | | | | | | | 1.27E-06 |
| 495 | 4550.611 | | | | | | -1.71E-05 | |
| 496 | 4578.757 | -3.58E-05 | | | | | | |
| 497 | 4591.966 | | -1.60E-05 | | | | | |
| 498 | 4600.521 | | | | | -1.82E-05 | | |
| 499 | 4603.546 | | | | | | | 1.29E-06 |
| 500 | 4610.118 | | | | -1.93E-05 | | | |
| 501 | 4610.462 | | | -2.02E-05 | | | | |
| 502 | 4614.123 | | | | | | -1.70E-05 | |
| 503 | 4639.85 | -3.59E-05 | | | | | | |
| 504 | 4656.185 | | | | | | | 1.27E-06 |
| 505 | 4658.058 | | -1.56E-05 | | | | | |
| 506 | 4667.853 | | | | | -1.82E-05 | | |
| 507 | 4679.493 | | | | -1.93E-05 | | | |
| 508 | 4681.691 | | | -2.02E-05 | | | | |
| 509 | 4684.109 | | | | | | -1.70E-05 | |
| 510 | 4709.881 | | | | | | | 1.25E-06 |

| | A | B | C | D | E | F | G | H |
|---|---|---|---|---|---|---|---|---|
| 511 | 4710.014 | -3.62E-05 | | | | | | |
| 512 | 4725.021 | | -1.57E-05 | | | | | |
| 513 | 4739.726 | | | | | -1.78E-05 | | |
| 514 | 4748.253 | | | | -1.91E-05 | | | |
| 515 | 4750.742 | | | -1.99E-05 | | | | |
| 516 | 4752.263 | | | | | | -1.67E-05 | |
| 517 | 4753.373 | | | | | | | 1.52E-06 |
| 518 | 4786.35 | -3.59E-05 | | | | | | |
| 519 | 4803.155 | | -1.50E-05 | | | | | |
| 520 | 4819.101 | | | | | -1.78E-05 | | |
| 521 | 4825.641 | | | | | | | 1.58E-06 |
| 522 | 4829.208 | | | | -1.91E-05 | | | |
| 523 | 4831.342 | | | -1.98E-05 | | | | |
| 524 | 4832.421 | | | | | | -1.66E-05 | |
| 525 | 4865.633 | -3.58E-05 | | | | | | |
| 526 | 4877.778 | | | | | | | 1.64E-06 |
| 527 | 4878.49 | | -1.42E-05 | | | | | |
| 528 | 4890.612 | | | | | -1.74E-05 | | |
| 529 | 4900.819 | | | | -1.87E-05 | | | |
| 530 | 4903.08 | | | -1.95E-05 | | | | |
| 531 | 4903.415 | | | | | | -1.62E-05 | |
| 532 | 4924.778 | | | | | | | 1.81E-06 |
| 533 | 4933.846 | -3.45E-05 | | | | | | |
| 534 | 4947.974 | | -1.21E-05 | | | | | |
| 535 | 4958.299 | | | | | -1.71E-05 | | |
| 536 | 4969.322 | | | | -1.82E-05 | | | |
| 537 | 4969.801 | | | | | | -1.60E-05 | |
| 538 | 4970.072 | | | -1.88E-05 | | | | |
| 539 | 4980.422 | | | | | | | 1.82E-06 |
| 540 | 5020.617 | -3.39E-05 | | | | | | |
| 541 | 5034.884 | | -1.09E-05 | | | | | |
| 542 | 5044.723 | | | | | -1.69E-05 | | |
| 543 | 5049.423 | | | | | | | 1.85E-06 |
| 544 | 5060.625 | | | -1.79E-05 | | | | |
| 545 | 5060.955 | | | | | | -1.58E-05 | |
| 546 | 5062.414 | | | -1.86E-05 | | | | |
| 547 | 5106.548 | -3.38E-05 | | | | | | |
| 548 | 5123.209 | | -1.06E-05 | | | | | |
| 549 | 5123.439 | | | | | | | 1.83E-06 |
| 550 | 5133.146 | | | | | -1.63E-05 | | |
| 551 | 5147.389 | | | | -1.76E-05 | | | |
| 552 | 5147.403 | | | | | | -1.56E-05 | |
| 553 | 5148.966 | | | -1.86E-05 | | | | |
| 554 | 5174.975 | | | | | | | 1.97E-06 |
| 555 | 5189.657 | -3.29E-05 | | | | | | |
| 556 | 5210.285 | | -9.68E-06 | | | | | |
| 557 | 5219.256 | | | | | -1.57E-05 | | |
| 558 | 5227.825 | | | | | | -1.51E-05 | |
| 559 | 5229.954 | | | | -1.71E-05 | | | |
| 560 | 5231.986 | | | -1.81E-05 | | | | |
| 561 | 5247.001 | | | | | | | 2.01E-06 |
| 562 | 5283.878 | -3.27E-05 | | | | | | |
| 563 | 5299.94 | | -6.88E-06 | | | | | |
| 564 | 5309.289 | | | | | -1.57E-05 | | |
| 565 | 5325.341 | | | | | | -1.49E-05 | |
| 566 | 5325.345 | | | | -1.67E-05 | | | |
| 567 | 5327.425 | | | -1.80E-05 | | | | |
| 568 | 5337.035 | | | | | | | 2.03E-06 |
| 569 | 5380.067 | -3.27E-05 | | | | | | |
| 570 | 5397.685 | | -8.55E-06 | | | | | |
| 571 | 5405.434 | | | | | -1.59E-05 | | |
| 572 | 5412.877 | | | | | | -1.48E-05 | |
| 573 | 5417.653 | | | | -1.69E-05 | | | |
| 574 | 5420.287 | | | -1.82E-05 | | | | 1.99E-06 |
| 575 | 5462.95 | -3.25E-05 | | | | | | |
| 576 | 5475.502 | | | | | | | 2.16E-06 |
| 577 | 5478.875 | | -9.10E-06 | | | | | |
| 578 | 5489.049 | | | | | -1.57E-05 | | |
| 579 | 5499.807 | | | | | | -1.44E-05 | |
| 580 | 5500.736 | | | | -1.67E-05 | | | |
| 581 | 5504.384 | | | -1.78E-05 | | | | |
| 582 | 5555.858 | | | | | | | 2.13E-06 |
| 583 | 5564.589 | -3.16E-05 | | | | | | |
| 584 | 5583.713 | | -9.18E-06 | | | | | |
| 585 | 5592.803 | | | | | -1.55E-05 | | |
| 586 | 5597.015 | | | | | | -1.41E-05 | |
| 587 | 5603.962 | | | | -1.60E-05 | | | |
| 588 | 5607.092 | | | -1.71E-05 | | | | |
| 589 | 5642.615 | | | | | | | 2.22E-06 |
| 590 | 5671.302 | -3.14E-05 | | | | | | |
| 591 | 5685.635 | | -9.59E-06 | | | | | |
| 592 | 5697.92 | | | | | -1.57E-05 | | |
| 593 | 5707.361 | | | | | | -1.42E-05 | |
| 594 | 5713.655 | | | | -1.64E-05 | | | |
| 595 | 5714.513 | | | -1.76E-05 | | | | |
| 596 | 5730.869 | | | | | | | 2.22E-06 |
| 597 | 5777.103 | -3.08E-05 | | | | | | |
| 598 | 5793.301 | | -9.63E-06 | | | | | |
| 599 | 5801.367 | | | | | -1.53E-05 | | |
| 600 | 5813.795 | | | | | | -1.36E-05 | |
| 601 | 5822.43 | | | | -1.62E-05 | | | |
| 602 | 5825.747 | | | -1.71E-05 | | | | |
| 603 | 5829.61 | | | | | | | 2.24E-06 |
| 604 | 5885.66 | -3.08E-05 | | | | | | |
| 605 | 5900.515 | | -9.58E-06 | | | | | |
| 606 | 5906.884 | | | | | -1.49E-05 | | |
| 607 | 5915.629 | | | | | | -1.34E-05 | |
| 608 | 5919.917 | | | | | | | 2.35E-06 |
| 609 | 5925.612 | | | | -1.60E-05 | | | |
| 610 | 5926.765 | | | -1.70E-05 | | | | |
| 611 | 5992.925 | -2.99E-05 | | | | | | |
| 612 | 6011.555 | | -8.83E-06 | | | | | |

| | A | B | C | D | E | F | G | H |
|---|---|---|---|---|---|---|---|---|
| 613 | 6017.807 | | | | | | | 2.38E-06 |
| 614 | 6019.711 | | | | | -1.47E-05 | | |
| 615 | 6022.236 | | | | | | -1.34E-05 | |
| 616 | 6034.861 | | | | -1.57E-05 | | | |
| 617 | 6038.58 | | | -1.68E-05 | | | | |
| 618 | 6107.193 | -3.02E-05 | | | | | | |
| 619 | 6120.266 | | | | | | | 2.34E-06 |
| 620 | 6122.307 | | -9.80E-06 | | | | | |
| 621 | 6132.987 | | | | | -1.48E-05 | | |
| 622 | 6139.119 | | | | | | -1.35E-05 | |
| 623 | 6145.884 | | | | -1.60E-05 | | | |
| 624 | 6148.031 | | | -1.70E-05 | | | | |
| 625 | 6218.028 | -2.96E-05 | | | | | | |
| 626 | 6218.548 | | | | | | | 2.34E-06 |
| 627 | 6240.472 | | -9.74E-06 | | | | | |
| 628 | 6249.876 | | | | | -1.47E-05 | | |
| 629 | 6254.219 | | | | | | -1.35E-05 | |
| 630 | 6268.239 | | | | -1.61E-05 | | | |
| 631 | 6271.899 | | | -1.70E-05 | | | | |
| 632 | 6332.827 | | | | | | | 2.42E-06 |
| 633 | 6337.472 | -3.00E-05 | | | | | | |
| 634 | 6352.259 | | -9.81E-06 | | | | | |
| 635 | 6360.004 | | | | | -1.47E-05 | | |
| 636 | 6366.14 | | | | | | -1.36E-05 | |
| 637 | 6378.411 | | | | -1.59E-05 | | | |
| 638 | 6382.066 | | | -1.69E-05 | | | | |
| 639 | 6425.012 | | | | | | | 2.49E-06 |
| 640 | 6452.584 | -2.94E-05 | | | | | | |
| 641 | 6473.052 | | -9.52E-06 | | | | | |
| 642 | 6486.615 | | | | | -1.51E-05 | | |
| 643 | 6490.249 | | | | | | -1.37E-05 | |
| 644 | 6506.464 | | | | -1.62E-05 | | | |
| 645 | 6507.802 | | | -1.71E-05 | | | | |
| 646 | 6529.396 | | | | | | | 2.41E-06 |
| 647 | 6571.355 | -2.89E-05 | | | | | | |
| 648 | 6593.441 | | -9.77E-06 | | | | | |
| 649 | 6606.497 | | | | | -1.51E-05 | | |
| 650 | 6607.026 | | | | | | -1.37E-05 | |
| 651 | 6624.824 | | | | -1.62E-05 | | | |
| 652 | 6626.952 | | | -1.71E-05 | | | | |
| 653 | 6630.854 | | | | | | | 2.46E-06 |
| 654 | 6708.399 | -2.90E-05 | | | | | | |
| 655 | 6729.474 | | -9.61E-06 | | | | | |
| 656 | 6742.084 | | | | | | -1.38E-05 | |
| 657 | 6742.213 | | | | | -1.51E-05 | | |
| 658 | 6757.198 | | | | | | | 2.42E-06 |
| 659 | 6758.272 | | | | -1.61E-05 | | | |
| 660 | 6760.426 | | | -1.71E-05 | | | | |
| 661 | 6848.417 | -2.84E-05 | | | | | | |
| 662 | 6859.761 | | -8.70E-06 | | | | | |
| 663 | 6868.566 | | | | | | -1.39E-05 | 2.37E-06 |
| 664 | 6871.86 | | | | | -1.52E-05 | | |
| 665 | 6893.442 | | | | -1.62E-05 | | | |
| 666 | 6896.551 | | | -1.73E-05 | | | | |
| 667 | 6988.048 | -2.72E-05 | | | | | | |
| 668 | 6999.035 | | -8.84E-06 | | | | | 2.25E-06 |
| 669 | 7005.92 | | | | | | -1.41E-05 | |
| 670 | 7010.747 | | | | | -1.55E-05 | | |
| 671 | 7027.185 | | | | -1.65E-05 | | | |
| 672 | 7030.703 | | | -1.74E-05 | | | | |
| 673 | 7105.198 | | | | | | | 2.26E-06 |
| 674 | 7128.762 | -2.69E-05 | | | | | | |
| 675 | 7140.272 | | -9.06E-06 | | | | | |
| 676 | 7148.722 | | | | | | -1.41E-05 | |
| 677 | 7153.315 | | | | | -1.56E-05 | | |
| 678 | 7179.053 | | | | -1.65E-05 | | | |
| 679 | 7183.299 | | | -1.75E-05 | | | | |
| 680 | 7241.16 | | | | | | | 2.25E-06 |
| 681 | 7287.181 | -2.68E-05 | | | | | | |
| 682 | 7295.387 | | -9.83E-06 | | | | | |
| 683 | 7304.289 | | | | | | -1.42E-05 | |
| 684 | 7308.163 | | | | | -1.58E-05 | | |
| 685 | 7319.56 | | | | -1.66E-05 | | | |
| 686 | 7324.027 | | | -1.75E-05 | | | | |
| 687 | 7337.201 | | | | | | | 2.20E-06 |
| 688 | 7422.313 | -2.61E-05 | | | | | | |
| 689 | 7424.662 | | -9.88E-06 | | | | | |
| 690 | 7435.192 | | | | | | -1.42E-05 | |
| 691 | 7443.157 | | | | | -1.55E-05 | | |
| 692 | 7463.784 | | | | | | | 2.29E-06 |
| 693 | 7464.023 | | | | -1.64E-05 | | | |
| 694 | 7470.589 | | | -1.73E-05 | | | | |
| 695 | 7566.723 | -2.52E-05 | | | | | | |
| 696 | 7568.346 | | -1.00E-05 | | | | | |
| 697 | 7574.219 | | | | | | -1.42E-05 | |
| 698 | 7581.885 | | | | | | | 2.28E-06 |
| 699 | 7583.451 | | | | | -1.54E-05 | | |
| 700 | 7604.468 | | | | -1.63E-05 | | | |
| 701 | 7608.904 | | | -1.72E-05 | | | | |
| 702 | 7707.45 | | | | | | | 2.19E-06 |
| 703 | 7720.402 | | -9.63E-06 | | | | | |
| 704 | 7723.487 | -2.48E-05 | | | | | | |
| 705 | 7727.676 | | | | | | -1.40E-05 | |
| 706 | 7739.77 | | | | | -1.52E-05 | | |
| 707 | 7759.805 | | | | -1.61E-05 | | | |
| 708 | 7763.323 | | | -1.69E-05 | | | | |
| 709 | 7831.321 | | | | | | | 2.23E-06 |
| 710 | 7874.433 | | -8.05E-06 | | | | | |
| 711 | 7876.981 | -2.35E-05 | | | | | | |
| 712 | 7881.325 | | | | | | -1.38E-05 | |
| 713 | 7891.499 | | | | | -1.49E-05 | | |
| 714 | 7920.895 | | | | -1.57E-05 | | | |

| | A | B | C | D | E | F | G | H |
|---|---|---|---|---|---|---|---|---|
| 715 | 7924.344 | | | -1.65E-05 | | | | |
| 716 | 7944.511 | | | | | | | 2.18E-06 |
| 717 | 8021.143 | | -8.34E-06 | | | | | |
| 718 | 8024.528 | | | | | | -1.35E-05 | |
| 719 | 8026.594 | -2.34E-05 | | | | | | |
| 720 | 8044.126 | | | | | -1.41E-05 | | 2.15E-06 |
| 721 | 8066.939 | | | | -1.50E-05 | | | |
| 722 | 8070.147 | | | -1.56E-05 | | | | |
| 723 | 8151.241 | | | | | | | 2.27E-06 |
| 724 | 8177.566 | | -7.73E-06 | | | | | |
| 725 | 8178.364 | | | | | | -1.34E-05 | |
| 726 | 8193.091 | -2.24E-05 | | | | | | |
| 727 | 8205.38 | | | | | -1.29E-05 | | |
| 728 | 8226.392 | | | | -1.34E-05 | | | |
| 729 | 8230.637 | | | -1.41E-05 | | | | |
| 730 | 8271.287 | | | | | | | 2.39E-06 |
| 731 | 8340.939 | | -6.16E-06 | | | | | |
| 732 | 8345.053 | | | | | | -1.28E-05 | |
| 733 | 8351.512 | -2.16E-05 | | | | | | |
| 734 | 8363.525 | | | | | -1.26E-05 | | |
| 735 | 8398.73 | | | | -1.37E-05 | | | |
| 736 | 8408.331 | | | -1.42E-05 | | | | |
| 737 | 8507.681 | | -5.39E-06 | | | | | 2.44E-06 |
| 738 | 8512.606 | | | | | | -1.24E-05 | |
| 739 | 8518.922 | | | | | | | 2.49E-06 |
| 740 | 8523.973 | -2.09E-05 | | | | | | |
| 741 | 8542.828 | | | | | -1.22E-05 | | |
| 742 | 8571.784 | | | | -1.31E-05 | | | |
| 743 | 8579.613 | | | -1.33E-05 | | | | |
| 744 | 8647.799 | | | | | | | 2.46E-06 |
| 745 | 8686.462 | | -4.54E-06 | | | | | |
| 746 | 8701.666 | | | | | | -1.15E-05 | |
| 747 | 8710.987 | -2.01E-05 | | | | | | |
| 748 | 8725.354 | | | | | -1.01E-05 | | |
| 749 | 8765.839 | | | | -1.13E-05 | | | |
| 750 | 8768.931 | | | -1.11E-05 | | | | |
| 751 | 8801.033 | | | | | | | 2.37E-06 |
| 752 | 8880.492 | | -4.04E-06 | | | | | |
| 753 | 8886.937 | | | | | | -1.11E-05 | |
| 754 | 8900.296 | -1.99E-05 | | | | | | |
| 755 | 8915.997 | | | | | -9.42E-06 | | |
| 756 | 8950.165 | | | | -1.03E-05 | | | |
| 757 | 8952.681 | | | -1.04E-05 | | | | |
| 758 | 8961.594 | | | | | | | 2.42E-06 |
| 759 | 9068.611 | | -3.16E-06 | | | | | |
| 760 | 9081.236 | | | | | | -1.08E-05 | |
| 761 | 9091.769 | -1.90E-05 | | | | | | |
| 762 | 9110.773 | | | | | -9.10E-06 | | |
| 763 | 9133.323 | | | | | | | 2.46E-06 |
| 764 | 9142.904 | | | | -9.97E-06 | | | |
| 765 | 9147.638 | | | -1.02E-05 | | | | |
| 766 | 9283.216 | | -2.68E-06 | | | | | |
| 767 | 9301.822 | | | | | | -1.05E-05 | |
| 768 | 9313.474 | -1.84E-05 | | | | | | |
| 769 | 9320.099 | | | | | | | 2.46E-06 |
| 770 | 9324.464 | | | | | -9.16E-06 | | |
| 771 | 9353.571 | | | | -1.03E-05 | | | |
| 772 | 9357.644 | | | -1.06E-05 | | | | |
| 773 | 9490.477 | | -2.68E-06 | | | | | |
| 774 | 9503.959 | | | | | | -1.06E-05 | |
| 775 | 9519.996 | | | | | | | 2.45E-06 |
| 776 | 9522.776 | -1.84E-05 | | | | | | |
| 777 | 9538.846 | | | | | -9.24E-06 | | |
| 778 | 9575.071 | | | | -1.03E-05 | | | |
| 779 | 9582.355 | | | -1.08E-05 | | | | |
| 780 | 9722.89 | | -2.23E-06 | | | | | |
| 781 | 9741.49 | | | | | | | 2.56E-06 |
| 782 | 9748.489 | | | | | | -1.02E-05 | |
| 783 | 9767.177 | -1.79E-05 | | | | | | |
| 784 | 9795.754 | | | | | -9.03E-06 | | |
| 785 | 9850.681 | | | | -9.98E-06 | | | |
| 786 | 9856.069 | | | -1.05E-05 | | | | |
| 787 | 9983.656 | | -2.74E-06 | | | | | |
| 788 | 9996.202 | | | | | | | 2.50E-06 |
| 789 | 10014.25 | | | | | | -9.92E-06 | |
| 790 | 10023.73 | -1.72E-05 | | | | | | |
| 791 | 10048.74 | | | | | -8.80E-06 | | |
| 792 | 10090.19 | | | | -9.37E-06 | | | |
| 793 | 10096.82 | | | -9.91E-06 | | | | |
| 794 | 10230.58 | | -2.45E-06 | | | | | |
| 795 | 10236.56 | | | | | | | 2.42E-06 |
| 796 | 10270.42 | | | | | | -9.47E-06 | |
| 797 | 10282.16 | -1.70E-05 | | | | | | |
| 798 | 10323.21 | | | | | -8.61E-06 | | |
| 799 | 10359.37 | | | | -9.44E-06 | | | |
| 800 | 10372.11 | | | -9.77E-06 | | | | |
| 801 | 10504.91 | | -2.14E-06 | | | | | |
| 802 | 10511.36 | | | | | | | 2.41E-06 |
| 803 | 10559.87 | | | | | | -8.79E-06 | |
| 804 | 10571.04 | -1.75E-05 | | | | | | |
| 805 | 10603.94 | | | | | -8.30E-06 | | |
| 806 | 10642.08 | | | | -8.99E-06 | | | |
| 807 | 10649.85 | | | -9.44E-06 | | | | |
| 808 | 10789.56 | | -1.94E-06 | | | | | |
| 809 | 10800.25 | | | | | | | 2.50E-06 |
| 810 | 10830.24 | | | | | | -8.42E-06 | |
| 811 | 10840.04 | -1.70E-05 | | | | | | |
| 812 | 10891.13 | | | | | -8.01E-06 | | |
| 813 | 10938.86 | | | | -8.86E-06 | | | |
| 814 | 10944.48 | | | -9.24E-06 | | | | |
| 815 | 11068.8 | | -1.69E-06 | | | | | |
| 816 | 11109.58 | | | | | | | 2.53E-06 |

| | A | B | C | D | E | F | G | H |
|---|---|---|---|---|---|---|---|---|
| 817 | 11120.98 | | | | | | -7.85E-06 | |
| 818 | 11127.13 | -1.73E-05 | | | | | | |
| 819 | 11164.87 | | | | | -7.18E-06 | | |
| 820 | 11215.8 | | | | -7.95E-06 | | | |
| 821 | 11227.37 | | | -8.55E-06 | | | | |
| 822 | 11369.19 | | -1.21E-06 | | | | | |
| 823 | 11425.71 | | | | | | -7.52E-06 | |
| 824 | 11427.48 | | | | | | | 2.53E-06 |
| 825 | 11428.51 | -1.72E-05 | | | | | | |
| 826 | 11467.94 | | | | | -7.15E-06 | | |
| 827 | 11530.38 | | | | -8.10E-06 | | | |
| 828 | 11540.5 | | | -8.84E-06 | | | | |
| 829 | 11666.25 | | -1.47E-06 | | | | | |
| 830 | 11735.6 | | | | | | -7.13E-06 | |
| 831 | 11743.28 | -1.71E-05 | | | | | | |
| 832 | 11748.06 | | | | | | | 2.52E-06 |
| 833 | 11787.47 | | | | | -6.79E-06 | | |
| 834 | 11834.55 | | | | -7.62E-06 | | | |
| 835 | 11845.31 | | | -8.33E-06 | | | | |
| 836 | 11974.04 | | -1.47E-06 | | | | | |
| 837 | 12054.31 | | | | | | -6.95E-06 | |
| 838 | 12054.87 | -1.68E-05 | | | | | | |
| 839 | 12125.35 | | | | | | | 2.55E-06 |
| 840 | 12125.48 | | | | | -6.51E-06 | | |
| 841 | 12172.26 | | | | -7.57E-06 | | | |
| 842 | 12181.65 | | | -8.17E-06 | | | | |
| 843 | 12315.98 | | -1.96E-06 | | | | | |
| 844 | 12396.18 | | | | | | -7.07E-06 | |
| 845 | 12399.82 | -1.66E-05 | | | | | | |
| 846 | 12446.06 | | | | | -6.81E-06 | | |
| 847 | 12507.44 | | | | -7.70E-06 | | | |
| 848 | 12511.94 | | | | | | | 2.64E-06 |
| 849 | 12520.77 | | | -8.20E-06 | | | | |
| 850 | 12638.05 | | -2.34E-06 | | | | | |
| 851 | 12723.59 | -1.65E-05 | | | | | -7.17E-06 | |
| 852 | 12795.17 | | | | | -6.84E-06 | | |
| 853 | 12838.14 | | | | -7.66E-06 | | | |
| 854 | 12841.81 | | | -8.06E-06 | | | | |
| 855 | 12880.65 | | | | | | | 2.71E-06 |
| 856 | 12981.63 | | -3.32E-06 | | | | | |
| 857 | 13103.44 | | | | | | -7.09E-06 | |
| 858 | 13109.34 | -1.65E-05 | | | | | | |
| 859 | 13175.77 | | | | | -6.67E-06 | | |
| 860 | 13239.89 | | | | -7.42E-06 | | | |
| 861 | 13249.4 | | | -7.94E-06 | | | | |
| 862 | 13304.35 | | | | | | | 2.80E-06 |
| 863 | 13346.94 | | -2.40E-06 | | | | | |
| 864 | 13473.03 | | | | | | -7.11E-06 | |
| 865 | 13484.79 | -1.66E-05 | | | | | | |
| 866 | 13550.15 | | | | | -6.63E-06 | | |
| 867 | 13601.06 | | | | -6.99E-06 | | | |
| 868 | 13612.12 | | | -7.52E-06 | | | | |
| 869 | 13705.07 | | | | | | | 2.86E-06 |
| 870 | 13742.76 | | -2.95E-06 | | | | | |
| 871 | 13880.58 | | | | | | -6.88E-06 | |
| 872 | 13891.78 | -1.78E-05 | | | | | | |
| 873 | 13972.19 | | | | | -6.47E-06 | | |
| 874 | 14041.84 | | | | -6.82E-06 | | | |
| 875 | 14048.39 | | | -7.43E-06 | | | | |
| 876 | 14152.41 | | | | | | | 2.63E-06 |
| 877 | 14153.14 | | -2.97E-06 | | | | | |
| 878 | 14279.54 | | | | | | -7.55E-06 | |
| 879 | 14296.46 | -1.78E-05 | | | | | | |
| 880 | 14362.88 | | | | | -7.00E-06 | | |
| 881 | 14439.08 | | | | -7.02E-06 | | | |
| 882 | 14449.82 | | | -7.65E-06 | | | | |
| 883 | 14533.87 | | -3.39E-06 | | | | | |
| 884 | 14557.24 | | | | | | | 2.56E-06 |
| 885 | 14679.13 | | | | | | -7.84E-06 | |
| 886 | 14696.9 | -1.79E-05 | | | | | | |
| 887 | 14771.99 | | | | | -7.50E-06 | | |
| 888 | 14822.47 | | | | -7.69E-06 | | | |
| 889 | 14831.78 | | | -8.29E-06 | | | | |
| 890 | 14896.21 | | | | | | | 2.44E-06 |
| 891 | 14921.4 | | -3.76E-06 | | | | | |
| 892 | 15027.75 | | | | | | -8.09E-06 | |
| 893 | 15041.84 | -1.83E-05 | | | | | | |
| 894 | 15103.13 | | | | | -7.90E-06 | | |
| 895 | 15152.54 | | | | -7.95E-06 | | | |
| 896 | 15165.97 | | | -8.57E-06 | | | | |
| 897 | 15244.73 | | | | | | | 2.38E-06 |
| 898 | 15296.07 | | -4.44E-06 | | | | | |
| 899 | 15419.95 | | | | | | -8.74E-06 | |
| 900 | 15457.9 | -1.91E-05 | | | | | | |
| 901 | 15492.89 | | | | | -8.45E-06 | | |
| 902 | 15532.61 | | | | -8.19E-06 | | | |
| 903 | 15537.06 | | | -8.81E-06 | | | | |
| 904 | 15590.74 | | | | | | | 2.28E-06 |
| 905 | 15653.4 | | -4.11E-06 | | | | | |
| 906 | 15783.52 | | | | | | -9.08E-06 | |
| 907 | 15813.24 | -1.96E-05 | | | | | | |
| 908 | 15863.65 | | | | | -8.73E-06 | | |
| 909 | 15925.89 | | | | -8.05E-06 | | | |
| 910 | 15930.08 | | | -8.71E-06 | | | | |
| 911 | 15931.06 | | | | | | | 2.17E-06 |
| 912 | 16006.48 | | -5.37E-06 | | | | | |
| 913 | 16118.45 | | | | | | -9.10E-06 | |
| 914 | 16156.48 | -2.08E-05 | | | | | | |
| 915 | 16205.5 | | | | | -8.98E-06 | | |
| 916 | 16259.51 | | | | -8.24E-06 | | | |
| 917 | 16269.48 | | | -9.00E-06 | | | | |
| 918 | 16323.46 | | | | | | | 2.18E-06 |

| | A | B | C | D | E | F | G | H |
|---|---|---|---|---|---|---|---|---|
| 919 | 16350.96 | | -5.37E-06 | | | | | |
| 920 | 16445.96 | | | | | | -9.28E-06 | |
| 921 | 16490.78 | -2.15E-05 | | | | | | |
| 922 | 16520.98 | | | | | -9.40E-06 | | |
| 923 | 16570.63 | | | | -8.94E-06 | | | |
| 924 | 16573.31 | | | -9.76E-06 | | | | |
| 925 | 16643.36 | | -5.39E-06 | | | | | |
| 926 | 16653.78 | | | | | | | 2.21E-06 |
| 927 | 16748.7 | | | | | | -9.99E-06 | |
| 928 | 16800.65 | -2.17E-05 | | | | | | |
| 929 | 16820.71 | | | | | -9.89E-06 | | |
| 930 | 16857.84 | | | | -9.39E-06 | | | |
| 931 | 16859.96 | | | -1.02E-05 | | | | |
| 932 | 16945.28 | | -5.82E-06 | | | | | |
| 933 | 17022.11 | | | | | | | 2.07E-06 |
| 934 | 17052.38 | | | | | | -1.05E-05 | |
| 935 | 17103.47 | -2.28E-05 | | | | | | |
| 936 | 17117.69 | | | | | -1.06E-05 | | |
| 937 | 17148.9 | | | | -1.02E-05 | | | |
| 938 | 17150.32 | | | -1.09E-05 | | | | |
| 939 | 17227.71 | | -6.77E-06 | | | | | |
| 940 | 17289.38 | | | | | | | 1.95E-06 |
| 941 | 17295.7 | | | | | | -1.12E-05 | |
| 942 | 17355.35 | -2.36E-05 | | | | | | |
| 943 | 17360.79 | | | | | -1.15E-05 | | |
| 944 | 17388.63 | | | | -1.10E-05 | | | |
| 945 | 17388.74 | | | -1.18E-05 | | | | |
| 946 | 17458.03 | | -7.44E-06 | | | | | |
| 947 | 17517.91 | | | | | | | 1.80E-06 |
| 948 | 17529.05 | | | | | | -1.15E-05 | |
| 949 | 17575 | -2.45E-05 | | | | | | |
| 950 | 17577.96 | | | | | -1.18E-05 | | |
| 951 | 17601.68 | | | -1.22E-05 | -1.15E-05 | | | |
| 952 | 17674.3 | | -7.58E-06 | | | | | |
| 953 | 17735.19 | | | | | | -1.18E-05 | |
| 954 | 17751.77 | | | | | | | 1.79E-06 |
| 955 | 17799.78 | -2.45E-05 | | | | | | |
| 956 | 17799.93 | | | | | -1.26E-05 | | |
| 957 | 17824.13 | | | | -1.22E-05 | | | |
| 958 | 17824.15 | | | -1.29E-05 | | | | |
| 959 | 17891.4 | | -8.63E-06 | | | | | |
| 960 | 17942.23 | | | | | | -1.23E-05 | |
| 961 | 17956.78 | | | | | | | 1.82E-06 |
| 962 | 18001.51 | | | | | -1.29E-05 | | |
| 963 | 18004 | -2.42E-05 | | | | | | |
| 964 | 18019.7 | | | | -1.24E-05 | | | |
| 965 | 18020.07 | | | -1.31E-05 | | | | |
| 966 | 18086.3 | | -8.79E-06 | | | | | |
| 967 | 18145.18 | | | | | | | 1.72E-06 |
| 968 | 18150.61 | | | | | | -1.27E-05 | |
| 969 | 18210.35 | | | | | -1.34E-05 | | |
| 970 | 18217.33 | -2.41E-05 | | | | | | |
| 971 | 18240.32 | | | | -1.29E-05 | | | |
| 972 | 18241 | | | -1.34E-05 | | | | |
| 973 | 18289.59 | | -8.12E-06 | | | | | |
| 974 | 18342.13 | | | | | | | 1.74E-06 |
| 975 | 18351.05 | | | | | | -1.31E-05 | |
| 976 | 18396.29 | | | | | -1.39E-05 | | |
| 977 | 18408.99 | -2.43E-05 | | | | | | |
| 978 | 18428.71 | | | -1.42E-05 | -1.35E-05 | | | |
| 979 | 18488.89 | | -8.35E-06 | | | | | |
| 980 | 18513.48 | | | | | | | 1.71E-06 |
| 981 | 18539.12 | | | | | | -1.35E-05 | |
| 982 | 18592.46 | | | | | -1.38E-05 | | |
| 983 | 18613.33 | -2.32E-05 | | | | | | |
| 984 | 18621.52 | | | -1.42E-05 | -1.36E-05 | | | |
| 985 | 18675.32 | | | | | | | 1.74E-06 |
| 986 | 18683.98 | | -8.66E-06 | | | | | |
| 987 | 18734.89 | | | | | | -1.38E-05 | |
| 988 | 18790.3 | | | | | -1.42E-05 | | |
| 989 | 18802.21 | -2.34E-05 | | | | | | |
| 990 | 18804.54 | | | -1.43E-05 | | | | |
| 991 | 18805.58 | | | | -1.39E-05 | | | |
| 992 | 18812.32 | | | | | | | 1.73E-06 |
| 993 | 18854.91 | | -9.27E-06 | | | | | |
| 994 | 18903 | | | | | | -1.41E-05 | |
| 995 | 18957.75 | | | | | | | 1.57E-06 |
| 996 | 18957.81 | | | | | -1.47E-05 | | |
| 997 | 18973.69 | -2.31E-05 | | | | | | |
| 998 | 18974.05 | | | -1.48E-05 | -1.45E-05 | | | |
| 999 | 19029.14 | | -9.53E-06 | | | | | |
| 1000 | 19064.99 | | | | | | -1.43E-05 | |
| 1001 | 19078.23 | | | | | | | 1.50E-06 |
| 1002 | 19115.69 | | | | | -1.49E-05 | | |
| 1003 | 19131.03 | | | -1.49E-05 | | | | |
| 1004 | 19132 | | | | -1.46E-05 | | | |
| 1005 | 19135.21 | -2.29E-05 | | | | | | |
| 1006 | 19189.14 | | -9.45E-06 | | | | | |
| 1007 | 19218.25 | | | | | | | 1.47E-06 |
| 1008 | 19234.49 | | | | | | -1.49E-05 | |
| 1009 | 19296.49 | | | | | -1.54E-05 | | |
| 1010 | 19315.26 | | | -1.54E-05 | -1.50E-05 | | | |
| 1011 | 19324.66 | -2.22E-05 | | | | | | |
| 1012 | 19352.89 | | | | | | | 1.45E-06 |
| 1013 | 19380.3 | | -1.03E-05 | | | | | |
| 1014 | 19413.57 | | | | | | -1.51E-05 | |
| 1015 | 19470.54 | | | | | -1.53E-05 | | |
| 1016 | 19495.48 | | | -1.54E-05 | | | | |
| 1017 | 19495.77 | | | | -1.49E-05 | | | |
| 1018 | 19497.02 | | | | | | | 1.36E-06 |
| 1019 | 19505.71 | -2.11E-05 | | | | | | |
| 1020 | 19545.28 | | -1.09E-05 | | | | | |

| | A | B | C | D | E | F | G | H |
|---|---|---|---|---|---|---|---|---|
| 1021 | 19580.47 | | | | | | -1.57E-05 | |
| 1022 | 19591.45 | | | | | | | 1.30E-06 |
| 1023 | 19625.2 | | | | | -1.57E-05 | | |
| 1024 | 19646.05 | | | -1.57E-05 | -1.52E-05 | | | |
| 1025 | 19659.82 | -2.04E-05 | | | | | | |
| 1026 | 19714.12 | | -1.04E-05 | | | | | |
| 1027 | 19719.78 | | | | | | | 1.18E-06 |
| 1028 | 19753.04 | | | | | | -1.56E-05 | |
| 1029 | 19805.03 | | | | | -1.58E-05 | | |
| 1030 | 19835.02 | | | -1.59E-05 | -1.55E-05 | | | |
| 1031 | 19847.94 | -2.02E-05 | | | | | | |
| 1032 | 19857.3 | | | | | | | 1.14E-06 |
| 1033 | 19885.86 | | -9.78E-06 | | | | | |
| 1034 | 19928.83 | | | | | | -1.58E-05 | |
| 1035 | 19968.42 | | | | | | | 1.14E-06 |
| 1036 | 19978.52 | | | | | -1.55E-05 | | |
| 1037 | 19995.84 | | | -1.58E-05 | -1.52E-05 | | | |
| 1038 | 20011.07 | -1.96E-05 | | | | | | |
| 1039 | 20042.06 | | -1.09E-05 | | | | | |
| 1040 | 20070.97 | | | | | | | 1.10E-06 |
| 1041 | 20080.21 | | | | | | -1.60E-05 | |
| 1042 | 20134.47 | | | | | -1.58E-05 | | |
| 1043 | 20163.84 | | | -1.60E-05 | | | | |
| 1044 | 20166.19 | | | | -1.54E-05 | | | |
| 1045 | 20180.41 | -1.92E-05 | | | | | | |
| 1046 | 20193.34 | | | | | | | 1.06E-06 |
| 1047 | 20211.53 | | -1.10E-05 | | | | | |
| 1048 | 20245.73 | | | | | | -1.57E-05 | |
| 1049 | 20294.38 | | | | | -1.58E-05 | | |
| 1050 | 20311.85 | | | | | | | 1.03E-06 |
| 1051 | 20315.77 | | | -1.60E-05 | -1.55E-05 | | | |
| 1052 | 20336.52 | -1.86E-05 | | | | | | |
| 1053 | 20363.08 | | -1.10E-05 | | | | | |
| 1054 | 20389.41 | | | | | | -1.60E-05 | |
| 1055 | 20402 | | | | | | | 1.01E-06 |
| 1056 | 20442.1 | | | | | -1.63E-05 | | |
| 1057 | 20463.02 | | | -1.65E-05 | | | | |
| 1058 | 20463.6 | | | | -1.60E-05 | | | |
| 1059 | 20492.33 | -1.82E-05 | | | | | | |
| 1060 | 20516.13 | | | | | | | 9.53E-07 |
| 1061 | 20517.16 | | -1.09E-05 | | | | | |
| 1062 | 20552.29 | | | | | | -1.59E-05 | |
| 1063 | 20603.73 | | | | | -1.65E-05 | | |
| 1064 | 20614.32 | | | | | | | 8.88E-07 |
| 1065 | 20618.96 | | | -1.67E-05 | | | | |
| 1066 | 20619.05 | | | | -1.62E-05 | | | |
| 1067 | 20639.68 | -1.75E-05 | | | | | | |
| 1068 | 20672.41 | | -1.04E-05 | | | | | |
| 1069 | 20695.88 | | | | | | -1.61E-05 | |
| 1070 | 20723.05 | | | | | | | 8.87E-07 |
| 1071 | 20747.75 | | | | | -1.65E-05 | | |
| 1072 | 20765.87 | | | -1.70E-05 | | | | |
| 1073 | 20766.48 | | | | -1.65E-05 | | | |
| 1074 | 20794.31 | -1.66E-05 | | | | | | |
| 1075 | 20819.38 | | -9.99E-06 | | | | | |
| 1076 | 20840.64 | | | | | | | 9.04E-07 |
| 1077 | 20857.81 | | | | | | -1.61E-05 | |
| 1078 | 20913.78 | | | | | -1.65E-05 | | |
| 1079 | 20938.53 | | | -1.68E-05 | | | | |
| 1080 | 20938.71 | | | | -1.64E-05 | | | |
| 1081 | 20964.49 | | | | | | | 8.48E-07 |
| 1082 | 20968.31 | -1.58E-05 | | | | | | |
| 1083 | 20988.14 | | -9.76E-06 | | | | | |
| 1084 | 21019.62 | | | | | | -1.59E-05 | |
| 1085 | 21079.96 | | | | | -1.64E-05 | | |
| 1086 | 21081.75 | | | | | | | 8.51E-07 |
| 1087 | 21099.52 | | | -1.66E-05 | -1.62E-05 | | | |
| 1088 | 21127.74 | -1.52E-05 | | | | | | |
| 1089 | 21148.57 | | -1.01E-05 | | | | | |
| 1090 | 21180.21 | | | | | | -1.59E-05 | |
| 1091 | 21183.52 | | | | | | | 8.10E-07 |
| 1092 | 21236.01 | | | | | -1.64E-05 | | |
| 1093 | 21258.02 | | | -1.66E-05 | -1.61E-05 | | | |
| 1094 | 21289.73 | -1.53E-05 | | | | | | |
| 1095 | 21299.5 | | | | | | | 8.05E-07 |
| 1096 | 21311.48 | | -1.00E-05 | | | | | |
| 1097 | 21340.18 | | | | | | -1.58E-05 | |
| 1098 | 21404.3 | | | | | -1.66E-05 | | |
| 1099 | 21422.82 | | | | | | | 7.85E-07 |
| 1100 | 21428.36 | | | -1.69E-05 | -1.64E-05 | | | |
| 1101 | 21449.69 | -1.52E-05 | | | | | | |
| 1102 | 21467.78 | | -9.98E-06 | | | | | |
| 1103 | 21498.68 | | | | | | -1.58E-05 | |
| 1104 | 21524.19 | | | | | | | 7.08E-07 |
| 1105 | 21558.19 | | | | | -1.62E-05 | | |
| 1106 | 21577.35 | | | -1.67E-05 | -1.63E-05 | | | |
| 1107 | 21609.16 | -1.43E-05 | | | | | | |
| 1108 | 21620.06 | | | | | | | 5.21E-07 |
| 1109 | 21626.18 | | -1.13E-05 | | | | | |
| 1110 | 21652.69 | | | | | | -1.64E-05 | |
| 1111 | 21708.53 | | | | | -1.68E-05 | | |
| 1112 | 21721.54 | | | | | | | 5.13E-07 |
| 1113 | 21730.22 | | | -1.68E-05 | | | | |
| 1114 | 21731.23 | | | | -1.65E-05 | | | |
| 1115 | 21754.14 | -1.40E-05 | | | | | | |
| 1116 | 21767.63 | | -1.21E-05 | | | | | |
| 1117 | 21799.56 | | | | | | -1.66E-05 | |
| 1118 | 21817.14 | | | | | | | 4.45E-07 |
| 1119 | 21848.57 | | | | | -1.69E-05 | | |
| 1120 | 21870.74 | | | -1.70E-05 | | | | |
| 1121 | 21872.52 | | | | -1.67E-05 | | | |
| 1122 | 21896.91 | -1.17E-05 | | | | | | |

| | A | B | C | D | E | F | G | H |
|---|---|---|---|---|---|---|---|---|
| 1123 | 21898.94 | | | | | | | 4.42E-07 |
| 1124 | 21908.65 | | -1.12E-05 | | | | | |
| 1125 | 21924.96 | | | | | | -1.65E-05 | |
| 1126 | 21984.8 | | | | | | | 4.60E-07 |
| 1127 | 21992.6 | | | | | -1.64E-05 | | |
| 1128 | 22018.59 | | | -1.65E-05 | | | | |
| 1129 | 22019.94 | | | | -1.62E-05 | | | |
| 1130 | 22057.45 | -1.08E-05 | | | | | | |
| 1131 | 22067.32 | | -1.12E-05 | | | | | |
| 1132 | 22090.8 | | | | | | -1.63E-05 | |
| 1133 | 22096.68 | | | | | | | 4.27E-07 |
| 1134 | 22149.19 | | | | | -1.62E-05 | | |
| 1135 | 22174.25 | | | -1.64E-05 | | | | |
| 1136 | 22174.32 | | | | -1.60E-05 | | | |
| 1137 | 22190.34 | | | | | | | 3.02E-07 |
| 1138 | 22205.59 | -1.04E-05 | | | | | | |
| 1139 | 22212.78 | | -1.15E-05 | | | | | |
| 1140 | 22235.86 | | | | | | -1.62E-05 | |
| 1141 | 22277.46 | | | | | | | 2.35E-07 |
| 1142 | 22290.66 | | | | | -1.58E-05 | | |
| 1143 | 22313.87 | | | -1.59E-05 | | | | |
| 1144 | 22314.43 | | | | -1.54E-05 | | | |
| 1145 | 22336.97 | -9.17E-06 | | | | | | |
| 1146 | 22345.12 | | -1.13E-05 | | | | | |
| 1147 | 22360.27 | | | | | | | 2.57E-07 |
| 1148 | 22366.32 | | | | | | -1.69E-05 | |
| 1149 | 22432.52 | | | | | -1.58E-05 | | |
| 1150 | 22451.76 | | | | | | | 1.68E-07 |
| 1151 | 22453.77 | | | -1.59E-05 | -1.56E-05 | | | |
| 1152 | 22498.13 | -9.18E-06 | | | | | | |
| 1153 | 22500.78 | | -1.07E-05 | | | | | |
| 1154 | 22524.03 | | | | | | -1.69E-05 | |
| 1155 | 22546.74 | | | | | | | 1.19E-07 |
| 1156 | 22589.06 | | | | | -1.44E-05 | | |
| 1157 | 22609.81 | | | -1.52E-05 | | | | |
| 1158 | 22609.91 | | | | -1.49E-05 | | | |
| 1159 | 22649.15 | | | | | | | -8.49E-08 |
| 1160 | 22651.85 | -7.62E-06 | | | | | | |
| 1161 | 22652.12 | | -1.01E-05 | | | | | |
| 1162 | 22671.52 | | | | | | -1.63E-05 | |
| 1163 | 22714.28 | | | | | -1.41E-05 | | |
| 1164 | 22726.68 | | | | | | | -8.44E-08 |
| 1165 | 22744.72 | | | -1.48E-05 | -1.46E-05 | | | |
| 1166 | 22793.41 | | -9.86E-06 | | | | | |
| 1167 | 22797.57 | -5.88E-06 | | | | | | |
| 1168 | 22811.48 | | | | | | -1.63E-05 | |
| 1169 | 22824.36 | | | | | | | -9.58E-08 |
| 1170 | 22864.08 | | | | | -1.38E-05 | | |
| 1171 | 22895.94 | | | -1.42E-05 | -1.40E-05 | | | |
| 1172 | 22942.4 | | | | | | | -1.28E-07 |
| 1173 | 22946.51 | | -8.81E-06 | | | | | |
| 1174 | 22948.54 | -4.42E-06 | | | | | | |
| 1175 | 22954.99 | | | | | | -1.56E-05 | |
| 1176 | 23009.81 | | | | | -1.35E-05 | | |
| 1177 | 23016.75 | | | | | | | -2.10E-07 |
| 1178 | 23032.81 | | | -1.39E-05 | -1.36E-05 | | | |
| 1179 | 23082.37 | | -7.93E-06 | | | | | |
| 1180 | 23094.17 | -2.42E-06 | | | | | | |
| 1181 | 23100.11 | | | | | | -1.55E-05 | |
| 1182 | 23121.08 | | | | | | | -2.76E-07 |
| 1183 | 23166.09 | | | | | -1.31E-05 | | |
| 1184 | 23198.14 | | | -1.34E-05 | -1.32E-05 | | | |
| 1185 | 23229.9 | | | | | | | -3.79E-07 |
| 1186 | 23234.16 | | -6.94E-06 | | | | | |
| 1187 | 23241.34 | -4.26E-07 | | | | | | |
| 1188 | 23246.71 | | | | | | -1.51E-05 | |
| 1189 | 23310.4 | | | | | -1.28E-05 | | |
| 1190 | 23314.61 | | | | | | | -4.50E-07 |
| 1191 | 23330.25 | | | -1.30E-05 | -1.28E-05 | | | |
| 1192 | 23381.35 | | -7.48E-06 | | | | | |
| 1193 | 23396.39 | | | | | | -1.45E-05 | |
| 1194 | 23397.44 | 4.33E-06 | | | | | | |
| 1195 | 23418.63 | | | | | | | -5.11E-07 |
| 1196 | 23467.97 | | | | | -1.17E-05 | | |
| 1197 | 23495.54 | | | -1.21E-05 | -1.19E-05 | | | |
| 1198 | 23530.64 | | | | | | | -5.78E-07 |
| 1199 | 23541.13 | | -6.29E-06 | | | | | |
| 1200 | 23547.76 | | | | | | -1.34E-05 | |
| 1201 | 23550.56 | 6.70E-06 | | | | | | |
| 1202 | 23622.62 | | | | | -1.09E-05 | | |
| 1203 | 23624.16 | | | | | | | -5.39E-07 |
| 1204 | 23648.28 | | | | -1.05E-05 | | | |
| 1205 | 23648.66 | | | -1.07E-05 | | | | |
| 1206 | 23696.59 | | -3.69E-06 | | | | | |
| 1207 | 23702.54 | | | | | | -1.24E-05 | |
| 1208 | 23711.14 | 1.12E-05 | | | | | | |
| 1209 | 23726.12 | | | | | | | -4.38E-07 |
| 1210 | 23775.6 | | | | | -9.60E-06 | | |
| 1211 | 23817.45 | | | | -9.16E-06 | | | |
| 1212 | 23817.64 | | | -9.20E-06 | | | | |
| 1213 | 23847.51 | | | | | | | -3.70E-07 |
| 1214 | 23853.85 | | -2.93E-06 | | | | | |
| 1215 | 23858.53 | | | | | | -1.17E-05 | |
| 1216 | 23870.02 | 1.31E-05 | | | | | | |
| 1217 | 23952.93 | | | | | -8.93E-06 | | |
| 1218 | 23973.9 | | | | | | | -5.22E-07 |
| 1219 | 23980.46 | | | -8.26E-06 | -8.38E-06 | | | |
| 1220 | 24037.17 | | -9.55E-07 | | | | | |
| 1221 | 24046.03 | | | | | | -1.05E-05 | |
| 1222 | 24054.66 | 1.56E-05 | | | | | | |
| 1223 | 24112.53 | | | | | | | -7.36E-07 |
| 1224 | 24129.19 | | | | | -7.10E-06 | | |

| | A | B | C | D | E | F | G | H |
|---|---|---|---|---|---|---|---|---|
| 1225 | 24166.04 | | | | -6.78E-06 | | | |
| 1226 | 24167.76 | | | -6.54E-06 | | | | |
| 1227 | 24221.45 | | 1.48E-06 | | | | | |
| 1228 | 24228.43 | | | | | | -9.66E-06 | |
| 1229 | 24255.92 | 2.00E-05 | | | | | | |
| 1230 | 24259.91 | | | | | | | -8.19E-07 |
| 1231 | 24323.26 | | | | | -5.36E-06 | | |
| 1232 | 24356.04 | | | | -4.57E-06 | | | |
| 1233 | 24358.05 | | | -3.93E-06 | | | | |
| 1234 | 24389 | | | | | | | -7.94E-07 |
| 1235 | 24416.86 | | 1.04E-05 | | | | | |
| 1236 | 24425.03 | | | | | | -8.23E-06 | |
| 1237 | 24452.35 | 2.96E-05 | | | | | | |
| 1238 | 24535.04 | | | | | -2.58E-06 | | |
| 1239 | 24581.36 | | | | -1.40E-06 | | | |
| 1240 | 24581.57 | | | -3.58E-07 | | | | |
| 1241 | 24609.18 | | | | | | | -7.24E-07 |
| 1242 | 24652.49 | | 1.55E-05 | | | | | |
| 1243 | 24664.03 | | | | | | -6.03E-06 | |
| 1244 | 24677.65 | 3.81E-05 | | | | | | |
| 1245 | 24781.62 | | | | | 1.10E-07 | | |
| 1246 | 24804.34 | | | | | | | -8.38E-07 |
| 1247 | 24824.54 | | | 1.66E-06 | 8.49E-07 | | | |
| 1248 | 24880.2 | | 1.60E-05 | | | | | |
| 1249 | 24890.69 | | | | | | -3.19E-06 | |
| 1250 | 24902.75 | 3.31E-05 | | | | | | |
| 1251 | 24981.02 | | | | | 4.33E-06 | | |
| 1252 | 25004.84 | | | | | | | -9.32E-07 |
| 1253 | 25027.8 | | | | 5.00E-06 | | | |
| 1254 | 25028.09 | | | 6.44E-06 | | | | |
| 1255 | 25086.73 | | 1.35E-05 | | | | | |
| 1256 | 25095.13 | | | | | | -1.46E-06 | |
| 1257 | 25104.01 | 3.06E-05 | | | | | | |
| 1258 | 25184.81 | | | | | 1.51E-05 | | |
| 1259 | 25216.49 | | | | | | | -1.02E-06 |
| 1260 | 25219.72 | | | | 1.60E-05 | | | |
| 1261 | 25219.96 | | | 1.69E-05 | | | | |
| 1262 | 25264.88 | | 1.59E-05 | | | | | |
| 1263 | 25272.57 | | | | | | 6.59E-06 | |
| 1264 | 25283.44 | 3.50E-05 | | | | | | |
| 1265 | 25373.98 | | | | | 1.31E-05 | | |
| 1266 | 25412.8 | | | 1.38E-05 | 1.41E-05 | | | |
| 1267 | 25428.06 | | | | | | | -8.29E-07 |
| 1268 | 25465.01 | | 1.35E-05 | | | | | |
| 1269 | 25479.11 | | | | | | 5.77E-06 | |
| 1270 | 25480.03 | 3.06E-05 | | | | | | |
| 1271 | 25552.13 | | | | | 1.16E-05 | | |
| 1272 | 25597.96 | | | | 1.26E-05 | | | |
| 1273 | 25599.43 | | | 1.28E-05 | | | | |
| 1274 | 25632.03 | | | | | | | -8.05E-07 |
| 1275 | 25642.48 | | 1.24E-05 | | | | | |
| 1276 | 25650.77 | 3.10E-05 | | | | | | |
| 1277 | 25653.41 | | | | | | 6.49E-06 | |
| 1278 | 25726.88 | | | | | 1.10E-05 | | |
| 1279 | 25767.96 | | | 1.15E-05 | 1.14E-05 | | | |
| 1280 | 25803.36 | | 1.16E-05 | | | | | |
| 1281 | 25806.76 | 2.80E-05 | | | | | | |
| 1282 | 25825.45 | | | | | | 6.44E-06 | |
| 1283 | 25843.23 | | | | | | | -9.02E-07 |
| 1284 | 25900.66 | | | | | 1.09E-05 | | |
| 1285 | 25937.41 | | | | 1.16E-05 | | | |
| 1286 | 25938.01 | | | 1.16E-05 | | | | |
| 1287 | 25971.26 | | 1.12E-05 | | | | | |
| 1288 | 25971.67 | 2.63E-05 | | | | | | |
| 1289 | 25984.7 | | | | | | 7.37E-06 | |
| 1290 | 26031.66 | | | | | | | -9.14E-07 |
| 1291 | 26069.91 | | | | | 1.12E-05 | | |
| 1292 | 26099.64 | | | | 1.13E-05 | | | |
| 1293 | 26099.93 | | | 1.17E-05 | | | | |
| 1294 | 26139.66 | 2.66E-05 | | | | | | |
| 1295 | 26142.13 | | 1.09E-05 | | | | | |
| 1296 | 26156.03 | | | | | | 7.64E-06 | |
| 1297 | 26226.59 | | | | | 1.13E-05 | | |
| 1298 | 26234.73 | | | | | | | -9.98E-07 |
| 1299 | 26249.43 | | | | 1.11E-05 | | | |
| 1300 | 26250.93 | | | 1.14E-05 | | | | |
| 1301 | 26280.53 | 2.69E-05 | | | | | | |
| 1302 | 26282.65 | | 1.00E-05 | | | | | |
| 1303 | 26298.2 | | | | | | 7.77E-06 | |
| 1304 | 26380.99 | | | | | 1.11E-05 | | |
| 1305 | 26414.94 | | | 1.18E-05 | 1.13E-05 | | | |
| 1306 | 26429.32 | | | | | | | -1.10E-06 |
| 1307 | 26442 | 2.78E-05 | | | | | | |
| 1308 | 26451.69 | | 1.07E-05 | | | | | |
| 1309 | 26468.54 | | | | | | 7.99E-06 | |
| 1310 | 26539 | | | | | 1.12E-05 | | |
| 1311 | 26571.67 | | | | 1.17E-05 | | | |
| 1312 | 26572.96 | | | 1.21E-05 | | | | |
| 1313 | 26594.21 | 2.62E-05 | | | | | | |
| 1314 | 26608.28 | | 1.03E-05 | | | | | |
| 1315 | 26622.75 | | | | | | 8.27E-06 | |
| 1316 | 26628.95 | | | | | | | -1.25E-06 |
| 1317 | 26686.2 | | | | | 1.15E-05 | | |
| 1318 | 26719.25 | | | | 1.20E-05 | | | |
| 1319 | 26719.61 | | | 1.25E-05 | | | | |
| 1320 | 26736.18 | 2.67E-05 | | | | | | |
| 1321 | 26751.33 | | 1.06E-05 | | | | | |
| 1322 | 26775.66 | | | | | | 9.09E-06 | |
| 1323 | 26809.18 | | | | | | | -1.30E-06 |
| 1324 | 26831.36 | | | | | 1.23E-05 | | |
| 1325 | 26857.55 | | | | 1.26E-05 | | | |
| 1326 | 26858.42 | | | 1.31E-05 | | | | |

| | A | B | C | D | E | F | G | H |
|---|---|---|---|---|---|---|---|---|
| 1327 | 26875.22 | 2.66E-05 | | | | | | |
| 1328 | 26894.78 | | 1.07E-05 | | | | | |
| 1329 | 26912.85 | | | | | | 9.86E-06 | |
| 1330 | 26977.88 | | | | | 1.24E-05 | | |
| 1331 | 27011.98 | | | | 1.29E-05 | | | |
| 1332 | 27012.28 | | | 1.32E-05 | | | | |
| 1333 | 27030.43 | 2.68E-05 | | | | | | |
| 1334 | 27030.63 | | | | | | | -1.30E-06 |
| 1335 | 27046.72 | | 1.11E-05 | | | | | |
| 1336 | 27065.1 | | | | | | 1.02E-05 | |
| 1337 | 27125.25 | | | | | 1.23E-05 | | |
| 1338 | 27155.6 | | | | 1.22E-05 | | | |
| 1339 | 27157.81 | | | 1.24E-05 | | | | |
| 1340 | 27169.42 | 2.65E-05 | | | | | | |
| 1341 | 27191.63 | | 9.87E-06 | | | | | |
| 1342 | 27210.4 | | | | | | 1.04E-05 | |
| 1343 | 27216.88 | | | | | | | -1.30E-06 |
| 1344 | 27266.25 | | | | | 1.24E-05 | | |
| 1345 | 27299.2 | | | | 1.26E-05 | | | |
| 1346 | 27299.9 | | | 1.29E-05 | | | | |
| 1347 | 27309.24 | 2.65E-05 | | | | | | |
| 1348 | 27331.61 | | 1.01E-05 | | | | | |
| 1349 | 27353.59 | | | | | | 1.10E-05 | |
| 1350 | 27414.49 | | | | | | | -1.25E-06 |
| 1351 | 27417.79 | | | | | 1.36E-05 | | |
| 1352 | 27444.52 | | | | 1.35E-05 | | | |
| 1353 | 27446.2 | | | 1.43E-05 | | | | |
| 1354 | 27456.74 | 2.78E-05 | | | | | | |
| 1355 | 27479.88 | | 1.09E-05 | | | | | |
| 1356 | 27507.17 | | | | | | 1.25E-05 | |
| 1357 | 27561.29 | | | | | 1.39E-05 | | |
| 1358 | 27594.06 | | | | 1.41E-05 | | | |
| 1359 | 27594.38 | | | 1.48E-05 | | | | |
| 1360 | 27603.39 | 2.82E-05 | | | | | | |
| 1361 | 27605.82 | | | | | | | -1.33E-06 |
| 1362 | 27627.43 | | 1.15E-05 | | | | | |
| 1363 | 27645.13 | | | | | | 1.32E-05 | |
| 1364 | 27697.21 | | | | | 1.47E-05 | | |
| 1365 | 27729.54 | | | | 1.48E-05 | | | |
| 1366 | 27730.66 | | | 1.54E-05 | | | | |
| 1367 | 27737.98 | 2.80E-05 | | | | | | |
| 1368 | 27769.48 | | 1.21E-05 | | | | | |
| 1369 | 27788.63 | | | | | | 1.40E-05 | |
| 1370 | 27799.63 | | | | | | | -1.34E-06 |
| 1371 | 27844.27 | | | | | 1.50E-05 | | |
| 1372 | 27876.1 | | | | 1.52E-05 | | | |
| 1373 | 27876.46 | | | 1.62E-05 | | | | |
| 1374 | 27883.06 | 2.95E-05 | | | | | | |
| 1375 | 27905.31 | | 1.39E-05 | | | | | |
| 1376 | 27922.82 | | | | | | 1.45E-05 | |
| 1377 | 27969.85 | | | | | 1.56E-05 | | |
| 1378 | 27984.68 | | | | | | | -1.32E-06 |
| 1379 | 28003.97 | | | | 1.60E-05 | | | |
| 1380 | 28004.12 | | | 1.74E-05 | | | | |
| 1381 | 28008.14 | 2.95E-05 | | | | | | |
| 1382 | 28037.92 | | 1.38E-05 | | | | | |
| 1383 | 28057.89 | | | | | | 1.56E-05 | |
| 1384 | 28115.13 | | | | | 1.63E-05 | | |
| 1385 | 28152.09 | | | | 1.66E-05 | | | |
| 1386 | 28152.44 | | | 1.80E-05 | | | | |
| 1387 | 28156.37 | 2.94E-05 | | | | | | |
| 1388 | 28186.77 | | 1.39E-05 | | | | | |
| 1389 | 28201.28 | | | | | | 1.71E-05 | |
| 1390 | 28208.33 | | | | | | | -1.21E-06 |
| 1391 | 28252.77 | | | | | 1.70E-05 | | |
| 1392 | 28284.39 | | | | 1.73E-05 | | | |
| 1393 | 28284.43 | | | 1.85E-05 | | | | |
| 1394 | 28285.92 | 3.02E-05 | | | | | | |
| 1395 | 28325.39 | | 1.47E-05 | | | | | |
| 1396 | 28341.31 | | | | | | 1.78E-05 | |
| 1397 | 28384.66 | | | | | 1.72E-05 | | |
| 1398 | 28391.59 | | | | | | | -1.26E-06 |
| 1399 | 28411.35 | | | | 1.78E-05 | | | |
| 1400 | 28411.51 | | | 1.92E-05 | | | | |
| 1401 | 28412.3 | 3.08E-05 | | | | | | |
| 1402 | 28445.04 | | 1.51E-05 | | | | | |
| 1403 | 28459.82 | | | | | | 1.85E-05 | |
| 1404 | 28510.11 | | | | | 1.83E-05 | | |
| 1405 | 28540.14 | 3.26E-05 | | 2.00E-05 | 1.84E-05 | | | |
| 1406 | 28573 | | 1.91E-05 | | | | | |
| 1407 | 28588.09 | | | | | | 1.98E-05 | |
| 1408 | 28606.63 | | | | | | | -1.23E-06 |
| 1409 | 28651.93 | | | | | 1.97E-05 | | |
| 1410 | 28679.22 | 3.29E-05 | | | | | | |
| 1411 | 28683.07 | | | 2.04E-05 | 1.93E-05 | | | |
| 1412 | 28711.17 | | 2.00E-05 | | | | | |
| 1413 | 28724.56 | | | | | | 2.00E-05 | |
| 1414 | 28774.49 | | | | | 2.06E-05 | | |
| 1415 | 28793.53 | 3.34E-05 | | | | | | |
| 1416 | 28801.25 | | | 2.23E-05 | 2.09E-05 | | | |
| 1417 | 28833.25 | | | | | | | -1.28E-06 |
| 1418 | 28842.41 | | 1.80E-05 | | | | | |
| 1419 | 28857.77 | | | | | | 2.05E-05 | |
| 1420 | 28900.68 | | | | | 2.16E-05 | | |
| 1421 | 28924.88 | 3.31E-05 | | | | | | |
| 1422 | 28932.69 | | | 2.51E-05 | 2.41E-05 | | | |
| 1423 | 28977.87 | | 1.91E-05 | | | | | |
| 1424 | 28989.99 | | | | | | 2.27E-05 | |
| 1425 | 29046.86 | | | | | 2.25E-05 | | |
| 1426 | 29058.1 | 3.49E-05 | | | | | | |
| 1427 | 29073.77 | | | | 2.38E-05 | | | |
| 1428 | 29073.86 | | | 2.47E-05 | | | | |

| | A | B | C | D | E | F | G | H |
|---|---|---|---|---|---|---|---|---|
| 1429 | 29077.49 | | | | | | | -8.89E-07 |
| 1430 | 29109.92 | | 1.69E-05 | | | | | |
| 1431 | 29129.42 | | | | | | 2.30E-05 | |
| 1432 | 29171.69 | | | | | 2.26E-05 | | |
| 1433 | 29184.78 | 3.57E-05 | | | | | | |
| 1434 | 29201.93 | | | 2.51E-05 | 2.40E-05 | | | |
| 1435 | 29245.77 | | 1.73E-05 | | | | | |
| 1436 | 29260.33 | | | | | | 2.36E-05 | |
| 1437 | 29288.48 | | | | | | | -8.47E-07 |
| 1438 | 29298.6 | | | | | 2.26E-05 | | |
| 1439 | 29304.97 | 3.65E-05 | | | | | | |
| 1440 | 29330.87 | | | 2.53E-05 | 2.43E-05 | | | |
| 1441 | 29365.55 | | 1.71E-05 | | | | | |
| 1442 | 29375.47 | | | | | | 2.40E-05 | |
| 1443 | 29425.6 | | | | | 2.27E-05 | | |
| 1444 | 29430 | 3.73E-05 | | | | | | |
| 1445 | 29453.02 | | | 2.51E-05 | 2.41E-05 | | | |
| 1446 | 29485.85 | | 1.73E-05 | | | | | |
| 1447 | 29498.43 | | | | | | 2.43E-05 | |
| 1448 | 29539.43 | | | | | | | -9.27E-07 |
| 1449 | 29540.26 | 3.80E-05 | | | | | | |
| 1450 | 29541.03 | | | | | 2.31E-05 | | |
| 1451 | 29563.31 | | | | 2.46E-05 | | | |
| 1452 | 29563.51 | | | 2.59E-05 | | | | |
| 1453 | 29596.24 | | 1.75E-05 | | | | | |
| 1454 | 29614.88 | | | | | | 2.48E-05 | |
| 1455 | 29666.8 | 3.68E-05 | | | | | | |
| 1456 | 29675.77 | | | | | 2.35E-05 | | |
| 1457 | 29708.07 | | | 2.60E-05 | 2.47E-05 | | | |
| 1458 | 29743.43 | | 1.75E-05 | | | | | |
| 1459 | 29751.33 | | | | | | 2.49E-05 | |
| 1460 | 29787.43 | 3.69E-05 | | | | | | |
| 1461 | 29795.67 | | | | | 2.37E-05 | | |
| 1462 | 29812.71 | | | | | | | -9.67E-07 |
| 1463 | 29822.2 | | | 2.60E-05 | 2.47E-05 | | | |
| 1464 | 29859.52 | | 1.67E-05 | | | | | |
| 1465 | 29874.31 | | | | | | 2.51E-05 | |
| 1466 | 29903.64 | 3.80E-05 | | | | | | |
| 1467 | 29911.11 | | | | | 2.40E-05 | | |
| 1468 | 29940.79 | | | 2.59E-05 | 2.47E-05 | | | |
| 1469 | 29976.92 | | 1.67E-05 | | | | | |
| 1470 | 29989.66 | | | | | | 2.52E-05 | |
| 1471 | 30016.26 | 3.80E-05 | | | | | | |
| 1472 | 30029.91 | | | | | 2.39E-05 | | |
| 1473 | 30057.69 | | | | 2.48E-05 | | | |
| 1474 | 30057.7 | | | 2.59E-05 | | | | |
| 1475 | 30102.57 | | 1.66E-05 | | | | | |
| 1476 | 30104.35 | | | | | | | -9.85E-07 |
| 1477 | 30110.39 | | | | | | 2.53E-05 | |
| 1478 | 30140.64 | 3.86E-05 | | | | | | |
| 1479 | 30149.7 | | | | | 2.41E-05 | | |
| 1480 | 30185.75 | | | 2.60E-05 | 2.49E-05 | | | |
| 1481 | 30211.05 | | 1.71E-05 | | | | | |
| 1482 | 30221.9 | | | | | | 2.56E-05 | |
| 1483 | 30251.79 | 3.90E-05 | | | | | | |
| 1484 | 30273.07 | | | | | 2.43E-05 | | |
| 1485 | 30300.98 | | | | 2.50E-05 | | | |
| 1486 | 30301.17 | | | 2.60E-05 | | | | |
| 1487 | 30336.19 | | 1.62E-05 | | | | | |
| 1488 | 30345.4 | | | | | | 2.57E-05 | |
| 1489 | 30364.41 | 3.93E-05 | | | | | | |
| 1490 | 30386.5 | | | | | 2.43E-05 | | |
| 1491 | 30404.58 | | | | | | | -1.02E-06 |
| 1492 | 30423.29 | | | | 2.49E-05 | | | |
| 1493 | 30423.93 | | | 2.58E-05 | | | | |
| 1494 | 30469.53 | | 1.65E-05 | | | | | |
| 1495 | 30484 | | | | | | 2.60E-05 | |
| 1496 | 30501.57 | 3.99E-05 | | | | | | |
| 1497 | 30526.73 | | | | | 2.45E-05 | | |
| 1498 | 30552.29 | | | | 2.51E-05 | | | |
| 1499 | 30552.69 | | | 2.61E-05 | | | | |
| 1500 | 30584.82 | | 1.64E-05 | | | | | |
| 1501 | 30593.71 | | | | | | 2.60E-05 | |
| 1502 | 30604.5 | 3.96E-05 | | | | | | |
| 1503 | 30652.4 | | | | | 2.46E-05 | | |
| 1504 | 30670.74 | | | 2.62E-05 | 2.52E-05 | | | |
| 1505 | 30710.6 | | 1.63E-05 | | | | | |
| 1506 | 30717.34 | | | | | | 2.61E-05 | |
| 1507 | 30723.82 | | | | | | | -9.79E-07 |
| 1508 | 30727.07 | 3.84E-05 | | | | | | |
| 1509 | 30758.22 | | | | | 2.49E-05 | | |
| 1510 | 30781.39 | | | | 2.53E-05 | | | |
| 1511 | 30782.23 | | | 2.61E-05 | | | | |
| 1512 | 30833.69 | | 1.57E-05 | | | | | |
| 1513 | 30846.66 | | | | | | 2.62E-05 | |
| 1514 | 30856.12 | 3.71E-05 | | | | | | |
| 1515 | 30888.52 | | | | | 2.47E-05 | | |
| 1516 | 30914.86 | | | | 2.51E-05 | | | |
| 1517 | 30915.92 | | | 2.59E-05 | | | | |
| 1518 | 30954.47 | | 1.58E-05 | | | | | |
| 1519 | 30966.76 | | | | | | 2.64E-05 | |
| 1520 | 30972.33 | 3.72E-05 | | | | | | |
| 1521 | 31018.5 | | | | | 2.46E-05 | | |
| 1522 | 31044.82 | | | | | | | -1.01E-06 |
| 1523 | 31046.89 | | | | 2.49E-05 | | | |
| 1524 | 31048.2 | | | 2.57E-05 | | | | |
| 1525 | 31091.5 | | 1.56E-05 | | | | | |
| 1526 | 31101.8 | | | | | | 2.66E-05 | |
| 1527 | 31110.66 | 3.55E-05 | | | | | | |
| 1528 | 31156.7 | | | | | 2.46E-05 | | |
| 1529 | 31184.78 | | | | 2.49E-05 | | | |
| 1530 | 31186.33 | | | 2.57E-05 | | | | |

| | A | B | C | D | E | F | G | H |
|---|---|---|---|---|---|---|---|---|
| 1531 | 31224.81 | | 1.41E-05 | | | | | |
| 1532 | 31235.12 | | | | | | 2.67E-05 | |
| 1533 | 31236.57 | 3.55E-05 | | | | | | |
| 1534 | 31281.77 | | | | | 2.44E-05 | | |
| 1535 | 31306.22 | | | | 2.47E-05 | | | |
| 1536 | 31310.85 | | | 2.54E-05 | | | | |
| 1537 | 31359.72 | | 1.36E-05 | | | | | |
| 1538 | 31368.28 | 3.49E-05 | | | | | 2.66E-05 | |
| 1539 | 31371.92 | | | | | | | -1.15E-06 |
| 1540 | 31405.5 | | | | | 2.43E-05 | | |
| 1541 | 31442.9 | | | | 2.47E-05 | | | |
| 1542 | 31444.16 | | | 2.54E-05 | | | | |
| 1543 | 31482.36 | | 1.37E-05 | | | | | |
| 1544 | 31488.08 | | | | | | 2.66E-05 | |
| 1545 | 31489.12 | 3.49E-05 | | | | | | |
| 1546 | 31547.3 | | | | | 2.43E-05 | | |
| 1547 | 31580.25 | | | | 2.47E-05 | | | |
| 1548 | 31582.68 | | | 2.54E-05 | | | | |
| 1549 | 31627.29 | | 1.36E-05 | | | | | |
| 1550 | 31634.52 | 3.48E-05 | | | | | | |
| 1551 | 31634.83 | | | | | | 2.66E-05 | |
| 1552 | 31693.5 | | | | | 2.42E-05 | | |
| 1553 | 31715.23 | | | | 2.45E-05 | | | |
| 1554 | 31716.76 | | | 2.53E-05 | | | | |
| 1555 | 31725.94 | | | | | | | -1.34E-06 |
| 1556 | 31776.72 | | 1.35E-05 | | | | | |
| 1557 | 31777.36 | 3.42E-05 | | | | | | |
| 1558 | 31779.8 | | | | | | 2.66E-05 | |
| 1559 | 31833.11 | | | | | 2.42E-05 | | |
| 1560 | 31863.83 | | | | 2.44E-05 | | | |
| 1561 | 31864.02 | | | 2.52E-05 | | | | |
| 1562 | 31920.01 | 3.32E-05 | 1.32E-05 | | | | | |
| 1563 | 31921.96 | | | | | | 2.65E-05 | |
| 1564 | 31980.49 | | | | | 2.42E-05 | | |
| 1565 | 32008.38 | | | | 2.45E-05 | | | |
| 1566 | 32010.63 | | | 2.52E-05 | | | | |
| 1567 | 32073.63 | 3.16E-05 | | | | | | |
| 1568 | 32073.73 | | 1.36E-05 | | | | | |
| 1569 | 32075.93 | | | | | | 2.68E-05 | |
| 1570 | 32082.55 | | | | | | | -1.40E-06 |
| 1571 | 32139.93 | | | | | 2.44E-05 | | |
| 1572 | 32176.06 | | | | 2.46E-05 | | | |
| 1573 | 32177.26 | | | 2.52E-05 | | | | |
| 1574 | 32226.67 | 3.12E-05 | | | | | | |
| 1575 | 32227.52 | | 1.34E-05 | | | | | |
| 1576 | 32230.2 | | | | | | 2.67E-05 | |
| 1577 | 32289.75 | | | | | 2.41E-05 | | |
| 1578 | 32315.83 | | | | 2.46E-05 | | | |
| 1579 | 32317.58 | | | 2.52E-05 | | | | |
| 1580 | 32363.84 | 3.10E-05 | | | | | | |
| 1581 | 32369.16 | | 1.34E-05 | | | | | |
| 1582 | 32370.81 | | | | | | 2.66E-05 | |
| 1583 | 32413.74 | | | | | | | -1.41E-06 |
| 1584 | 32423.04 | | | | | 2.41E-05 | | |
| 1585 | 32450.57 | | | | 2.45E-05 | | | |
| 1586 | 32451.66 | | | 2.52E-05 | | | | |
| 1587 | 32517.93 | 3.08E-05 | | | | | | |
| 1588 | 32522.91 | | 1.36E-05 | | | | | |
| 1589 | 32523.93 | | | | | | 2.68E-05 | |
| 1590 | 32589.17 | | | | | 2.40E-05 | | |
| 1591 | 32622.79 | | | | 2.46E-05 | | | |
| 1592 | 32623.66 | | | 2.53E-05 | | | | |
| 1593 | 32681.23 | 3.11E-05 | | | | | | |
| 1594 | 32683.25 | | 1.40E-05 | | | | | |
| 1595 | 32686.58 | | | | | | 2.69E-05 | |
| 1596 | 32749.24 | | | | | 2.40E-05 | | |
| 1597 | 32786 | | | | 2.46E-05 | | | |
| 1598 | 32787.34 | | | 2.52E-05 | | | | |
| 1599 | 32827.49 | | | | | | | -1.47E-06 |
| 1600 | 32844.22 | 3.03E-05 | | | | | | |
| 1601 | 32849.62 | | 1.35E-05 | | | | | |
| 1602 | 32851.16 | | | | | | 2.69E-05 | |
| 1603 | 32910.28 | | | | | 2.36E-05 | | |
| 1604 | 32926.89 | | | | 2.44E-05 | | | |
| 1605 | 32930.27 | | | 2.48E-05 | | | | |
| 1606 | 32988.77 | 3.08E-05 | | | | | | |
| 1607 | 32994.08 | | 1.39E-05 | | | | 2.69E-05 | |
| 1608 | 33055.95 | | | | | 2.35E-05 | | |
| 1609 | 33093.73 | | | | 2.45E-05 | | | |
| 1610 | 33094.13 | | | 2.50E-05 | | | | |
| 1611 | 33124.88 | | | | | | | -1.48E-06 |
| 1612 | 33153.74 | 3.07E-05 | | | | | | |
| 1613 | 33158.44 | | 1.43E-05 | | | | | |
| 1614 | 33158.7 | | | | | | 2.71E-05 | |
| 1615 | 33228.48 | | | | | 2.38E-05 | | |
| 1616 | 33257.13 | | | | 2.46E-05 | | | |
| 1617 | 33261.39 | | | 2.52E-05 | | | | |
| 1618 | 33319.71 | 3.07E-05 | | | | | | |
| 1619 | 33322.2 | | | | | | 2.69E-05 | |
| 1620 | 33323.86 | | 1.41E-05 | | | | | |
| 1621 | 33395.72 | | | | | 2.37E-05 | | |
| 1622 | 33431.01 | | | | 2.46E-05 | | | |
| 1623 | 33431.73 | | | 2.52E-05 | | | | |
| 1624 | 33478.74 | | | | | | | -1.41E-06 |
| 1625 | 33484.8 | 3.07E-05 | | | | | | |
| 1626 | 33488.86 | | 1.45E-05 | | | | | |
| 1627 | 33489.47 | | | | | | 2.68E-05 | |
| 1628 | 33560.46 | | | | | 2.38E-05 | | |
| 1629 | 33581.91 | | | | 2.51E-05 | | | |
| 1630 | 33583.65 | | | 2.55E-05 | | | | |
| 1631 | 33663.64 | 3.07E-05 | | | | | | |
| 1632 | 33664.17 | | | | | | 2.70E-05 | |

| | A | B | C | D | E | F | G | H |
|---|---|---|---|---|---|---|---|---|
| 1633 | 33665.55 | | 1.52E-05 | | | | | |
| 1634 | 33739.22 | | | | | 2.42E-05 | | |
| 1635 | 33769.31 | | | | 2.56E-05 | | | |
| 1636 | 33772.71 | | | 2.61E-05 | | | | |
| 1637 | 33857.26 | | | | | | 2.70E-05 | |
| 1638 | 33859.09 | | 1.55E-05 | | | | | |
| 1639 | 33862.15 | 3.12E-05 | | | | | | |
| 1640 | 33924.69 | | | | | | | -1.53E-06 |
| 1641 | 33929.97 | | | | | 2.40E-05 | | |
| 1642 | 33965.47 | | | | 2.54E-05 | | | |
| 1643 | 33967.17 | | | 2.59E-05 | | | | |
| 1644 | 34030.82 | | | | | | 2.67E-05 | |
| 1645 | 34033.64 | | 1.48E-05 | | | | | |
| 1646 | 34033.74 | 2.97E-05 | | | | | | |
| 1647 | 34090.52 | | | | | 2.37E-05 | | |
| 1648 | 34110.12 | | | | 2.50E-05 | | | |
| 1649 | 34112.57 | | | 2.55E-05 | | | | |
| 1650 | 34194.17 | | | | | | 2.66E-05 | |
| 1651 | 34203.11 | | 1.49E-05 | | | | | |
| 1652 | 34209.15 | 2.97E-05 | | | | | | |
| 1653 | 34274.38 | | | | | 2.36E-05 | | |
| 1654 | 34295.21 | | | | | | | -1.89E-06 |
| 1655 | 34304.85 | | | | 2.51E-05 | | | |
| 1656 | 34307.96 | | | 2.56E-05 | | | | |
| 1657 | 34405.71 | | | | | | 2.65E-05 | |
| 1658 | 34410.36 | | 1.51E-05 | | | | | |
| 1659 | 34420.95 | 2.96E-05 | | | | | | |
| 1660 | 34490.79 | | | | | 2.35E-05 | | |
| 1661 | 34514.8 | | | 2.65E-05 | 2.56E-05 | | | |
| 1662 | 34592.88 | | | | | | 2.66E-05 | |
| 1663 | 34598.32 | | 1.58E-05 | | | | | |
| 1664 | 34603.2 | 2.85E-05 | | | | | | |
| 1665 | 34669.63 | | | | | 2.35E-05 | | |
| 1666 | 34695.66 | | | | 2.57E-05 | | | |
| 1667 | 34697.38 | | | 2.61E-05 | | | | |
| 1668 | 34711.5 | | | | | | | -1.90E-06 |
| 1669 | 34789.07 | | | | | | 2.66E-05 | |
| 1670 | 34793.46 | | 1.58E-05 | | | | | |
| 1671 | 34798.34 | 2.86E-05 | | | | | | |
| 1672 | 34877.36 | | | | | 2.35E-05 | | |
| 1673 | 34905.21 | | | | 2.57E-05 | | | |
| 1674 | 34909.77 | | | 2.61E-05 | | | | |
| 1675 | 35019.93 | | | | | | 2.68E-05 | |
| 1676 | 35021.62 | | 1.68E-05 | | | | | |
| 1677 | 35033.95 | 2.86E-05 | | | | | | |
| 1678 | 35113.62 | | | | | 2.35E-05 | | |
| 1679 | 35141.84 | | | | 2.60E-05 | | | |
| 1680 | 35146.33 | | | 2.64E-05 | | | | |
| 1681 | 35232.17 | | | | | | | -2.02E-06 |
| 1682 | 35240.86 | | | | | | 2.66E-05 | |
| 1683 | 35244.8 | | 1.74E-05 | | | | | |
| 1684 | 35251.8 | 2.83E-05 | | | | | | |
| 1685 | 35338.53 | | | | | 2.36E-05 | | |
| 1686 | 35373.12 | | | | 2.60E-05 | | | |
| 1687 | 35374.29 | | | 2.67E-05 | | | | |
| 1688 | 35501.33 | | | | | | 2.66E-05 | |
| 1689 | 35506.87 | | 1.84E-05 | | | | | |
| 1690 | 35523.18 | 2.86E-05 | | | | | | |
| 1691 | 35601.17 | | | | | 2.34E-05 | | |
| 1692 | 35634.35 | | | | 2.56E-05 | | | |
| 1693 | 35635.51 | | | 2.60E-05 | | | | |
| 1694 | 35746.27 | | 1.87E-05 | | | | | |
| 1695 | 35750.02 | | | | | | 2.63E-05 | |
| 1696 | 35768.74 | 2.94E-05 | | | | | | |
| 1697 | 35855.91 | | | | | 2.33E-05 | | |
| 1698 | 35865.44 | | | | | | | -2.38E-06 |
| 1699 | 35894.35 | | | | 2.53E-05 | | | |
| 1700 | 35905.77 | | | 2.58E-05 | | | | |
| 1701 | 36043.81 | | 1.96E-05 | | | | | |
| 1702 | 36046.09 | | | | | | 2.62E-05 | |
| 1703 | 36064.34 | 2.96E-05 | | | | | | |
| 1704 | 36143.98 | | | | | 2.34E-05 | | |
| 1705 | 36174.86 | | | | 2.54E-05 | | | |
| 1706 | 36177.18 | | | 2.59E-05 | | | | |
| 1707 | 36324.42 | | 2.01E-05 | | | | | |
| 1708 | 36329.83 | | | | | | 2.59E-05 | |
| 1709 | 36342.27 | 3.13E-05 | | | | | | |
| 1710 | 36442.46 | | | | | 2.32E-05 | | |
| 1711 | 36487.44 | | | | 2.50E-05 | | | |
| 1712 | 36488.09 | | | 2.55E-05 | | | | |
| 1713 | 36620.8 | | | | | | | -2.39E-06 |
| 1714 | 36627.86 | | 2.07E-05 | | | | | |
| 1715 | 36646.07 | | | | | | 2.56E-05 | |
| 1716 | 36664.12 | 3.13E-05 | | | | | | |
| 1717 | 36736.01 | | | | | 2.31E-05 | | |
| 1718 | 36766.18 | | | | 2.51E-05 | | | |
| 1719 | 36775.06 | | | 2.57E-05 | | | | |
| 1720 | 36919.41 | | 1.99E-05 | | | | | |
| 1721 | 36942.59 | | | | | | 2.54E-05 | |
| 1722 | 36961.4 | 3.22E-05 | | | | | | |
| 1723 | 37047.21 | | | | | 2.38E-05 | | |
| 1724 | 37080.15 | | | | 2.60E-05 | | | |
| 1725 | 37085.99 | | | 2.63E-05 | | | | |
| 1726 | 37224.83 | | 1.99E-05 | | | | | |
| 1727 | 37261.08 | | | | | | 2.49E-05 | |
| 1728 | 37281.23 | 3.20E-05 | | | | | | |
| 1729 | 37352.94 | | | | | | | -2.31E-06 |
| 1730 | 37355.93 | | | | | 2.44E-05 | | |
| 1731 | 37391.07 | | | | 2.65E-05 | | | |
| 1732 | 37392.09 | | | 2.67E-05 | | | | |
| 1733 | 37536 | | 1.97E-05 | | | | | |
| 1734 | 37574.34 | | | | | | 2.46E-05 | |

| | A | B | C | D | E | F | G | H |
|---|---|---|---|---|---|---|---|---|
| 1735 | 37582.88 | 3.31E-05 | | | | | | |
| 1736 | 37666.19 | | | | | 2.41E-05 | | |
| 1737 | 37698.43 | | | | 2.58E-05 | | | |
| 1738 | 37698.56 | | | 2.60E-05 | | | | |
| 1739 | 37859.43 | | 1.81E-05 | | | | | |
| 1740 | 37912.98 | 3.20E-05 | | | | | 2.44E-05 | |
| 1741 | 37993.11 | | | | | 2.38E-05 | | |
| 1742 | 38020.62 | | | | 2.55E-05 | | | |
| 1743 | 38023.9 | | | 2.58E-05 | | | | |
| 1744 | 38087.38 | | | | | | | -2.48E-06 |
| 1745 | 38163.97 | | 1.81E-05 | | | | | |
| 1746 | 38222.58 | 3.27E-05 | | | | | | |
| 1747 | 38237.08 | | | | | | 2.46E-05 | |
| 1748 | 38294.3 | | | | | 2.39E-05 | | |
| 1749 | 38319.43 | | | | 2.59E-05 | | | |
| 1750 | 38324.35 | | | 2.63E-05 | | | | |
| 1751 | 38470.25 | | 1.87E-05 | | | | | |
| 1752 | 38506.51 | 3.42E-05 | | | | | | |
| 1753 | 38527.29 | | | | | | 2.42E-05 | |
| 1754 | 38597.76 | | | | | 2.37E-05 | | |
| 1755 | 38632.52 | | | | 2.58E-05 | | | |
| 1756 | 38634.77 | | | 2.61E-05 | | | | |
| 1757 | 38760.25 | | 1.79E-05 | | | | | |
| 1758 | 38804.81 | 3.50E-05 | | | | | | |
| 1759 | 38831.88 | | | | | | | -2.26E-06 |
| 1760 | 38861.24 | | | | | | 2.38E-05 | |
| 1761 | 38910.94 | | | | | 2.31E-05 | | |
| 1762 | 38928.92 | | | | 2.50E-05 | | | |
| 1763 | 38937.65 | | | 2.55E-05 | | | | |
| 1764 | 39074.46 | | 1.81E-05 | | | | | |
| 1765 | 39127.28 | 3.61E-05 | | | | | | |
| 1766 | 39185.09 | | | | | | 2.38E-05 | |
| 1767 | 39242.88 | | | | | 2.30E-05 | | |
| 1768 | 39268.49 | | | | 2.48E-05 | | | |
| 1769 | 39273.94 | | | 2.53E-05 | | | | |
| 1770 | 39370.26 | | 1.80E-05 | | | | | |
| 1771 | 39417.96 | 3.66E-05 | | | | | | |
| 1772 | 39492.99 | | | | | | 2.39E-05 | |
| 1773 | 39543.65 | | | | | 2.28E-05 | | -1.97E-06 |
| 1774 | 39565.83 | | | | 2.48E-05 | | | |
| 1775 | 39570.11 | | | 2.53E-05 | | | | |
| 1776 | 39700.58 | | 1.70E-05 | | | | | |
| 1777 | 39742.93 | 3.73E-05 | | | | | | |
| 1778 | 39845.91 | | | | | | 2.39E-05 | |
| 1779 | 39876.23 | | | | | 2.25E-05 | | |
| 1780 | 39899.87 | | | | 2.44E-05 | | | |
| 1781 | 39905.1 | | | 2.49E-05 | | | | |
| 1782 | 40012.8 | | 1.78E-05 | | | | | |
| 1783 | 40035.7 | 3.79E-05 | | | | | | |
| 1784 | 40174.34 | | | | | | 2.42E-05 | |
| 1785 | 40192.1 | | | | | 2.31E-05 | | |
| 1786 | 40221.88 | | | | 2.47E-05 | | | |
| 1787 | 40227.56 | | | 2.51E-05 | | | | |
| 1788 | 40227.78 | | | | | | | -1.63E-06 |
| 1789 | 40325.47 | | 1.74E-05 | | | | | |
| 1790 | 40343.04 | 3.77E-05 | | | | | | |
| 1791 | 40468.21 | | | | | | 2.45E-05 | |
| 1792 | 40493.75 | | | | | 2.36E-05 | | |
| 1793 | 40526.77 | | | | 2.51E-05 | | | |
| 1794 | 40530.14 | | | 2.57E-05 | | | | |
| 1795 | 40600.65 | | 1.83E-05 | | | | | |
| 1796 | 40630.9 | 3.76E-05 | | | | | | |
| 1797 | 40776.93 | | | | | | 2.44E-05 | |
| 1798 | 40786.81 | | | | | 2.36E-05 | | |
| 1799 | 40803.55 | | | | 2.48E-05 | | | |
| 1800 | 40812.14 | | | 2.56E-05 | | | | |
| 1801 | 40912.48 | | 1.80E-05 | | | | | |
| 1802 | 40931.61 | 3.88E-05 | | | | | | |
| 1803 | 40965.91 | | | | | | | -1.38E-06 |
| 1804 | 41101.73 | | | | | | 2.46E-05 | |
| 1805 | 41101.74 | | | | | 2.38E-05 | | |
| 1806 | 41131.81 | | | | 2.49E-05 | | | |
| 1807 | 41138.09 | | | 2.58E-05 | | | | |
| 1808 | 41208.76 | | 1.77E-05 | | | | | |
| 1809 | 41212.15 | 3.88E-05 | | | | | | |
| 1810 | 41408.8 | | | | | 2.41E-05 | | |
| 1811 | 41418.32 | | | | | | 2.45E-05 | |
| 1812 | 41440.98 | | | | 2.48E-05 | | | |
| 1813 | 41444.35 | | | 2.56E-05 | | | | |
| 1814 | 41532.3 | | 1.75E-05 | | | | | |
| 1815 | 41536.08 | 3.92E-05 | | | | | | |
| 1816 | 41753.06 | | | | | 2.38E-05 | | |
| 1817 | 41770.87 | | | | | | 2.45E-05 | |
| 1818 | 41790.33 | | | | 2.48E-05 | | | |
| 1819 | 41799.96 | | | 2.57E-05 | | | | |
| 1820 | 41811.13 | | | | | | | -1.29E-06 |
| 1821 | 41850.5 | 3.86E-05 | | | | | | |
| 1822 | 41856.6 | | 1.75E-05 | | | | | |
| 1823 | 42103.02 | | | | | 2.41E-05 | | |
| 1824 | 42122.57 | | | | | | 2.47E-05 | |
| 1825 | 42129.44 | | | | 2.51E-05 | | | |
| 1826 | 42138.42 | | | 2.61E-05 | | | | |
| 1827 | 42189.07 | 3.81E-05 | | | | | | |
| 1828 | 42199.65 | | 1.79E-05 | | | | | |
| 1829 | 42470.75 | | | | | 2.42E-05 | | |
| 1830 | 42528.87 | | | | 2.53E-05 | | | |
| 1831 | 42529.64 | | | | | | 2.49E-05 | |
| 1832 | 42532.19 | | | 2.63E-05 | | | | |
| 1833 | 42569.49 | 3.74E-05 | | | | | | |
| 1834 | 42593 | | 1.74E-05 | | | | | |
| 1835 | 42688.31 | | | | | | | -1.16E-06 |
| 1836 | 42887.59 | | | | | 2.45E-05 | | |

|      | A        | B        | C        | D        | E        | F        | G        | H         |
|------|----------|----------|----------|----------|----------|----------|----------|-----------|
| 1837 | 42925.52 |          |          |          | 2.62E-05 |          |          |           |
| 1838 | 42934.19 |          |          | 2.73E-05 |          |          |          |           |
| 1839 | 42940.54 |          |          |          |          |          | 2.50E-05 |           |
| 1840 | 42961.34 | 3.82E-05 |          |          |          |          |          |           |
| 1841 | 42988.61 |          | 1.76E-05 |          |          |          |          |           |
| 1842 | 43316.87 |          |          |          |          | 2.44E-05 |          |           |
| 1843 | 43376.79 |          |          |          | 2.60E-05 |          |          |           |
| 1844 | 43383.73 |          |          | 2.72E-05 |          |          |          |           |
| 1845 | 43392.37 | 3.76E-05 |          |          |          |          |          |           |
| 1846 | 43396.33 |          |          |          |          |          | 2.50E-05 |           |
| 1847 | 43425.18 |          | 1.74E-05 |          |          |          |          |           |
| 1848 | 43691.84 |          |          |          |          |          |          | -1.31E-06 |
| 1849 | 43799.23 |          |          |          |          | 2.50E-05 |          |           |
| 1850 | 43858.18 |          |          |          | 2.70E-05 |          |          |           |
| 1851 | 43863.82 |          |          | 2.83E-05 |          |          |          |           |
| 1852 | 43871.62 | 3.71E-05 |          |          |          |          |          |           |
| 1853 | 43904.05 |          |          |          |          |          | 2.56E-05 |           |
| 1854 | 43924.51 |          | 1.68E-05 |          |          |          |          |           |
| 1855 | 44300.53 |          |          |          |          | 2.55E-05 |          |           |
| 1856 | 44345.91 | 3.90E-05 |          |          | 2.81E-05 |          |          |           |
| 1857 | 44357.23 |          |          | 2.91E-05 |          |          |          |           |
| 1858 | 44395.85 |          |          |          |          |          | 2.57E-05 |           |
| 1859 | 44410.5  |          | 1.78E-05 |          |          |          |          |           |
| 1860 | 44759.66 |          |          |          |          |          |          | -1.57E-06 |
| 1861 | 44766.72 |          |          |          |          | 2.58E-05 |          |           |
| 1862 | 44812.84 | 3.89E-05 |          |          |          |          |          |           |
| 1863 | 44819.24 |          |          |          | 2.92E-05 |          |          |           |
| 1864 | 44828.07 |          |          | 3.07E-05 |          |          |          |           |
| 1865 | 44883.38 |          | 1.83E-05 |          |          |          |          |           |
| 1866 | 44887.86 |          |          |          |          |          | 2.59E-05 |           |
| 1867 | 45288.61 |          |          |          |          | 2.61E-05 |          |           |
| 1868 | 45319.85 | 3.93E-05 |          |          |          |          |          |           |
| 1869 | 45325.83 |          |          |          | 3.01E-05 |          |          |           |
| 1870 | 45334.45 |          |          | 3.11E-05 |          |          |          |           |
| 1871 | 45386.01 |          | 1.93E-05 |          |          |          |          |           |
| 1872 | 45408.6  |          |          |          |          |          | 2.56E-05 |           |
| 1873 | 45748.99 |          |          |          |          |          |          | -1.79E-06 |
| 1874 | 45800.11 |          |          |          |          | 2.69E-05 |          |           |
| 1875 | 45821.28 | 4.05E-05 |          |          |          |          |          |           |
| 1876 | 45831.16 |          |          |          | 3.17E-05 |          |          |           |
| 1877 | 45835.77 |          |          | 3.28E-05 |          |          |          |           |
| 1878 | 45909.35 |          | 2.03E-05 |          |          |          |          |           |
| 1879 | 45954.23 |          |          |          |          |          | 2.53E-05 |           |
| 1880 | 46388.7  | 3.89E-05 |          |          |          | 2.64E-05 |          |           |
| 1881 | 46425.22 |          |          |          | 3.02E-05 |          |          |           |
| 1882 | 46430.82 |          |          | 3.14E-05 |          |          |          |           |
| 1883 | 46500.53 |          | 2.02E-05 |          |          |          |          |           |
| 1884 | 46601.51 |          |          |          |          |          | 2.54E-05 |           |
| 1885 | 46971.02 |          |          |          |          |          |          | -1.73E-06 |
| 1886 | 47043.23 | 3.85E-05 |          |          |          |          |          |           |
| 1887 | 47056.81 |          |          |          |          | 2.61E-05 |          |           |
| 1888 | 47107.94 |          |          |          | 2.99E-05 |          |          |           |
| 1889 | 47111.35 |          |          | 3.10E-05 |          |          |          |           |
| 1890 | 47169.05 |          | 1.92E-05 |          |          |          |          |           |
| 1891 | 47295.66 |          |          |          |          |          | 2.55E-05 |           |
| 1892 | 47829.2  | 3.77E-05 |          |          |          |          |          |           |
| 1893 | 47831.67 |          |          |          |          | 2.73E-05 |          |           |
| 1894 | 47867.85 |          |          |          | 3.03E-05 |          |          |           |
| 1895 | 47876.03 |          |          | 3.11E-05 |          |          |          |           |
| 1896 | 47941.69 |          | 2.01E-05 |          |          |          |          |           |
| 1897 | 48147.33 |          |          |          |          |          | 2.56E-05 |           |
| 1898 | 48423.28 |          |          |          |          |          |          | -1.43E-06 |
| 1899 | 48747.59 |          |          |          |          | 2.81E-05 |          |           |
| 1900 | 48755.75 | 3.82E-05 |          |          |          |          |          |           |
| 1901 | 48767.53 |          |          |          | 3.03E-05 |          |          |           |
| 1902 | 48783.16 |          |          | 3.09E-05 |          |          |          |           |
| 1903 | 48878.92 |          | 1.94E-05 |          |          |          |          |           |
| 1904 | 49140.09 |          |          |          |          |          | 2.64E-05 |           |
| 1905 | 49838.05 |          |          |          |          | 2.94E-05 |          |           |
| 1906 | 49861.69 | 3.96E-05 |          |          |          |          |          |           |
| 1907 | 49865.18 |          |          |          | 3.15E-05 |          |          |           |
| 1908 | 49881.48 |          |          | 3.21E-05 |          |          |          |           |
| 1909 | 50012.58 |          | 2.06E-05 |          |          |          |          |           |
| 1910 | 50360.65 |          |          |          |          |          | 2.77E-05 |           |
| 1911 | 50456.52 |          |          |          |          |          |          | -1.47E-06 |
| 1912 | 51019.59 |          |          |          |          | 3.00E-05 |          |           |
| 1913 | 51039.8  |          |          |          | 3.19E-05 |          |          |           |
| 1914 | 51054.67 |          |          | 3.25E-05 |          |          |          |           |
| 1915 | 51057.56 | 3.97E-05 |          |          |          |          |          |           |
| 1916 | 51206.05 |          | 2.16E-05 |          |          |          |          |           |
| 1917 | 51669.57 |          |          |          |          |          | 2.82E-05 |           |
| 1918 | 52249.99 |          |          |          |          | 3.06E-05 |          |           |
| 1919 | 52261.93 |          |          |          | 3.29E-05 |          |          |           |
| 1920 | 52271.59 |          |          | 3.33E-05 |          |          |          |           |
| 1921 | 52308.78 | 3.98E-05 |          |          |          |          |          |           |
| 1922 | 52378.73 |          | 2.24E-05 |          |          |          |          |           |
| 1923 | 52917.59 |          |          |          |          |          | 2.96E-05 |           |
| 1924 | 53319.95 |          |          |          |          | 3.20E-05 |          |           |
| 1925 | 53330.59 |          |          |          | 3.41E-05 |          |          |           |
| 1926 | 53339.4  |          |          | 3.46E-05 |          |          |          |           |
| 1927 | 53380.36 | 4.00E-05 |          |          |          |          |          |           |
| 1928 | 53426.11 |          |          |          |          |          |          | -1.68E-06 |
| 1929 | 53450.35 |          | 2.35E-05 |          |          |          |          |           |
| 1930 | 53909.8  |          |          |          |          |          | 3.01E-05 |           |
| 1931 | 54234.18 |          |          |          |          | 3.24E-05 |          |           |
| 1932 | 54237.43 |          |          | 3.47E-05 | 3.44E-05 |          |          |           |
| 1933 | 54283.99 | 4.09E-05 |          |          |          |          |          |           |
| 1934 | 54332.97 |          | 2.48E-05 |          |          |          |          |           |
| 1935 | 54767.51 |          |          |          |          |          | 3.05E-05 |           |
| 1936 | 55023.1  |          |          |          |          | 3.32E-05 |          |           |
| 1937 | 55032.57 |          |          |          | 3.50E-05 |          |          |           |
| 1938 | 55033.57 |          |          | 3.54E-05 |          |          |          |           |

| | A | B | C | D | E | F | G | H |
|---|---|---|---|---|---|---|---|---|
| 1939 | 55076.72 | 4.18E-05 | | | | | | |
| 1940 | 55102.19 | | 2.60E-05 | | | | | |
| 1941 | 55499.18 | | | | | | 2.99E-05 | |
| 1942 | 55591.24 | | | | | | | -2.09E-06 |
| 1943 | 55747.11 | | | | | 3.27E-05 | | |
| 1944 | 55748.59 | | | 3.53E-05 | 3.47E-05 | | | |
| 1945 | 55782.45 | 4.15E-05 | | | | | | |
| 1946 | 55805.48 | | 2.60E-05 | | | | | |
| 1947 | 56234.81 | | | | | | 2.93E-05 | |
| 1948 | 56439.45 | | | | | 3.26E-05 | | |
| 1949 | 56441.39 | | | 3.51E-05 | 3.42E-05 | | | |
| 1950 | 56443.53 | 4.02E-05 | | | | | | |
| 1951 | 56496.33 | | 2.45E-05 | | | | | |
| 1952 | 56994.8 | | | | | | 2.92E-05 | |
| 1953 | 57007.59 | | | | | | | -1.79E-06 |
| 1954 | 57191.07 | 4.01E-05 | | | | | | |
| 1955 | 57206.73 | | | | | 3.22E-05 | | |
| 1956 | 57216.04 | | | 3.47E-05 | 3.38E-05 | | | |
| 1957 | 57266.28 | | 2.40E-05 | | | | | |
| 1958 | 57849.9 | | | | | | 2.85E-05 | |
| 1959 | 58079.22 | 4.02E-05 | | | | | | |
| 1960 | 58140.33 | | | | | 3.19E-05 | | |
| 1961 | 58152.54 | | | 3.42E-05 | 3.34E-05 | | | |
| 1962 | 58214.62 | | 2.30E-05 | | | | | |
| 1963 | 59062.56 | | | | | | 2.87E-05 | |
| 1964 | 59282.93 | 4.09E-05 | | | | | | |
| 1965 | 59444.57 | | | | | | | -1.91E-06 |
| 1966 | 59456.98 | | | | | 3.24E-05 | | |
| 1967 | 59462.65 | | | 3.48E-05 | 3.41E-05 | | | |
| 1968 | 59580.91 | | 2.36E-05 | | | | | |
| 1969 | 60723.25 | | | | | | 2.92E-05 | |
| 1970 | 60967.58 | 4.22E-05 | | | | | | |
| 1971 | 61070.05 | | | | | 3.32E-05 | | |
| 1972 | 61083.35 | | | | 3.52E-05 | | | |
| 1973 | 61085.87 | | | 3.59E-05 | | | | |
| 1974 | 61178 | | 2.60E-05 | | | | | |
| 1975 | 62310.59 | | | | | | 2.93E-05 | |
| 1976 | 62557.8 | | | | | | | -2.07E-06 |
| 1977 | 62622.93 | 4.45E-05 | | | | | | |
| 1978 | 62720.79 | | | | | 3.43E-05 | | |
| 1979 | 62736.51 | | | | 3.64E-05 | | | |
| 1980 | 62742.38 | | | 3.70E-05 | | | | |
| 1981 | 62874.25 | | 2.68E-05 | | | | | |
| 1982 | 64403.78 | | | | | | 3.02E-05 | |
| 1983 | 64658.51 | 4.89E-05 | | | | | | |
| 1984 | 64840.67 | | | | | 3.73E-05 | | |
| 1985 | 64857.06 | | | 4.14E-05 | 4.05E-05 | | | |
| 1986 | 64986.32 | | 3.14E-05 | | | | | |
| 1987 | 67372.25 | | | | | | 3.44E-05 | |
| 1988 | 67643.52 | 5.00E-05 | | | | | | |
| 1989 | 67932.67 | | | | | 4.13E-05 | | |
| 1990 | 67953.14 | | | 4.30E-05 | 4.31E-05 | | | |
| 1991 | 68002.68 | | | | | | | -3.14E-06 |
| 1992 | 68050.85 | | 3.26E-05 | | | | | |
| 1993 | 70163.55 | | | | | | 3.49E-05 | |
| 1994 | 70270.85 | 4.94E-05 | | | | | | |
| 1995 | 70408.23 | | | | | 3.93E-05 | | |
| 1996 | 70420.92 | | 3.17E-05 | | 4.09E-05 | | | |
| 1997 | 70421.17 | | | 4.15E-05 | | | | |
| 1998 | 75009.44 | | | | | | 3.41E-05 | |
| 1999 | 75371.93 | 5.56E-05 | | | | | | |
| 2000 | 75597.27 | | 3.65E-05 | | | | | |
| 2001 | 75650.4 | | | | | 4.05E-05 | | |
| 2002 | 75673.42 | | | | 4.30E-05 | | | |
| 2003 | 75685.64 | | | 4.33E-05 | | | | |
| 2004 | 100841 | | | | | | | -4.71E-06 |
| 2005 | 101835.9 | 6.86E-05 | 5.56E-05 | 6.22E-05 | 6.32E-05 | 6.44E-05 | 5.44E-05 | |

| | A | B | C | D | E | F | G | H | I | J | K | L | M | N | O | P |
|---|---|---|---|---|---|---|---|---|---|---|---|---|---|---|---|---|
| 1 | Rank | Inf | 0.002 | 0.0015 | 0.001 | 0.0005 | 0.0001 | 0.0013 | Inf Random | 0.0020 Ran | 0.0015 Ran | 0.0010 Ran | 0.0005 Ran | 0.0001 Ran | 0.0013 Random | |
| 2 | 10 | -0.000015 | -2.29E-05 | -5.05E-06 | -2.70E-06 | -1.59E-05 | -9.37E-07 | 3.14E-06 | 0.000146 | 0.000149 | 5.94E-05 | -6.25E-05 | -4.51E-05 | 1.85E-06 | -1.36E-05 | |
| 3 | 25 | -0.000136 | -8.94E-06 | 7.14E-05 | 9.66E-06 | -1.15E-05 | -8.46E-07 | 4.31E-06 | 0.00015 | 0.000153 | 6.16E-05 | -5.70E-05 | -3.99E-05 | 1.84E-06 | -8.79E-06 | |
| 4 | 61 | -0.000135 | -7.38E-06 | 8.81E-06 | 1.17E-05 | -1.13E-05 | -8.33E-07 | 4.55E-06 | 0.000153 | 0.000156 | 6.57E-05 | -5.45E-05 | -3.82E-05 | 2.05E-06 | -5.67E-06 | |
| 5 | 115 | -0.000137 | -8.96E-06 | 7.40E-05 | 1.04E-05 | -1.26E-05 | -9.51E-07 | 4.47E-06 | 0.000153 | 0.000156 | 6.66E-05 | -5.35E-05 | -3.79E-05 | 8.20E-07 | -4.13E-06 | |
| 6 | 144 | -0.00014 | -1.21E-05 | 2.58E-06 | 1.03E-05 | -1.29E-05 | -9.75E-07 | 4.35E-06 | 0.000159 | 0.000162 | 7.15E-05 | -4.16E-05 | -3.63E-05 | 7.85E-07 | 7.23E-06 | |
| 7 | 145 | -4.85E-05 | 7.66E-05 | | | | -4.02E-07 | | 0.000207 | 0.000217 | 0.000112 | -5.87E-06 | 4.13E-06 | | 3.76E-05 | |
| 8 | 181 | | | | 4.52E-05 | -6.56E-06 | | 9.29E-05 | | | 0.000111 | | | | 3.42E-05 | |
| 9 | 197 | 1.46E-05 | 0.000146 | 0.000187 | | | | | 0.000231 | 0.000241 | | | 7.98E-06 | 1.70E-06 | | |
| 10 | 213 | -6.83E-05 | 5.61E-05 | 0.0001 | 4.94E-05 | 2.58E-06 | 8.04E-07 | 6.37E-05 | 0.000214 | 0.000224 | 0.000102 | -9.08E-06 | 8.37E-06 | | 3.31E-05 | |
| 11 | 233 | -6.44E-06 | 0.000123 | 0.000103 | 5.24E-05 | 1.80E-06 | | 6.67E-05 | 0.000216 | 0.000227 | 0.000103 | -6.51E-06 | | 1.89E-06 | | |
| 12 | 234 | | 0.000107 | 9.72E-05 | 4.62E-05 | 3.13E-06 | 7.78E-07 | 5.91E-05 | 0.00021 | 0.00022 | 9.99E-05 | | 7.11E-06 | | 3.11E-05 | |
| 13 | 239 | 2.07E-05 | | | | | | | | | | | | | | |
| 14 | 245 | | | | | | | | | | | 1.22E-06 | | | 4.05E-05 | |
| 15 | 259 | -3.58E-05 | 8.81E-05 | 0.000111 | 5.88E-05 | | | 7.25E-05 | 0.000201 | 0.000211 | 9.71E-05 | 4.91E-06 | 1.47E-05 | 1.57E-06 | | |
| 16 | 261 | -3.49E-05 | 8.93E-05 | 0.000112 | 6.01E-05 | 6.44E-06 | 5.59E-06 | 7.33E-05 | 0.000202 | 0.000212 | 9.80E-05 | 5.53E-06 | 1.55E-05 | 1.27E-06 | 4.30E-05 | |
| 17 | 276 | -3.62E-05 | 8.86E-05 | 0.000111 | 5.96E-05 | 6.67E-06 | 1.20E-06 | 7.25E-05 | 0.000202 | 0.000212 | 9.80E-05 | 5.27E-06 | 1.56E-05 | 1.24E-06 | 4.30E-05 | |
| 18 | 284 | -3.88E-05 | 8.50E-05 | 0.000112 | 6.22E-05 | 7.44E-06 | 1.17E-06 | 7.53E-05 | 0.000201 | 0.000211 | 9.81E-05 | 5.64E-06 | 1.59E-05 | 1.18E-06 | 4.30E-05 | |
| 19 | 327 | | | | | 6.16E-06 | | | | | | | | | | |
| 20 | 334 | -3.32E-05 | 8.95E-05 | 0.000112 | | | 1.53E-06 | 7.29E-05 | 0.000202 | 0.000213 | 9.96E-05 | 6.88E-06 | 1.64E-05 | 1.99E-06 | 4.47E-05 | |
| 21 | 344 | -1.85E-05 | 0.000101 | 8.29E-05 | 4.82E-05 | 7.65E-06 | | 5.80E-05 | 0.000235 | 0.000238 | | 1.95E-06 | | 1.73E-06 | 3.67E-05 | |
| 22 | 351 | | | | | | | | | | | | 1.56E-05 | | | |
| 23 | 352 | -3.62E-05 | 8.39E-05 | 8.99E-05 | 5.97E-05 | 9.16E-06 | 1.33E-06 | 7.14E-05 | 0.000208 | 0.000214 | 9.17E-05 | 5.07E-06 | 1.58E-05 | | 4.01E-05 | |
| 24 | 357 | | | 7.99E-05 | 5.09E-05 | | | 6.03E-05 | | | | -9.01E-07 | | | | |
| 25 | 366 | -5.36E-05 | 6.78E-05 | 7.00E-05 | 3.50E-05 | 6.67E-06 | 1.54E-06 | 4.10E-05 | 0.000192 | 0.000204 | 7.82E-05 | -3.20E-05 | 9.39E-06 | 1.39E-06 | 3.34E-05 | |
| 26 | 375 | | | | | 4.79E-06 | 1.54E-06 | 3.79E-05 | 0.000209 | | | 6.07E-07 | 9.29E-06 | | 3.90E-05 | |
| 27 | 378 | | | 7.94E-05 | | | | | | | | | | 1.29E-06 | | |
| 28 | 380 | -4.36E-05 | 7.66E-05 | 7.96E-05 | 3.91E-05 | 7.58E-06 | 2.31E-06 | 4.14E-05 | 0.000199 | 0.000199 | 7.11E-05 | -6.20E-06 | 7.91E-06 | | 3.11E-05 | |
| 29 | 390 | | | | | | | | 0.000207 | 0.000207 | | 5.38E-06 | | | | |
| 30 | 408 | -2.59E-05 | 9.80E-05 | 8.14E-05 | 4.34E-05 | 8.11E-06 | 2.43E-06 | 4.81E-05 | 9.07E-05 | 0.000101 | 6.03E-05 | 5.65E-06 | 1.00E-05 | 1.66E-06 | 3.83E-05 | |
| 31 | 416 | -2.73E-05 | 9.63E-05 | 8.03E-05 | 4.25E-05 | 7.73E-06 | 2.40E-06 | 4.69E-05 | 8.95E-05 | 0.0001 | 6.01E-05 | 4.57E-06 | 8.55E-06 | 1.58E-06 | 3.72E-05 | |
| 32 | 432 | | | 8.74E-05 | | | | | | | 5.66E-05 | | | | | |
| 33 | 434 | -4.37E-05 | 8.05E-05 | | 2.51E-05 | 5.18E-06 | 2.30E-06 | 2.53E-05 | | 9.45E-05 | | -5.47E-07 | 2.01E-06 | | | |
| 34 | 487 | | | 9.26E-05 | | | | | 9.28E-05 | | 6.10E-05 | | | 1.82E-06 | 4.22E-05 | |
| 35 | 496 | -3.58E-05 | 8.74E-05 | | 3.50E-05 | 8.35E-06 | 2.31E-06 | 3.54E-05 | 9.26E-05 | 9.69E-05 | | 2.41E-06 | 8.77E-06 | | 4.20E-05 | |
| 36 | 513 | -2.72E-05 | 9.74E-05 | 9.94E-05 | 4.23E-05 | 1.17E-05 | 2.19E-06 | | | | 6.37E-05 | | 1.19E-05 | 1.97E-06 | | |
| 37 | 526 | | | | | | | | 9.67E-05 | 0.000101 | | 8.99E-06 | | | | |
| 38 | 530 | | | | | | | 4.04E-06 | | 0.0001 | 6.60E-05 | 8.88E-06 | | | 4.73E-05 | |
| 39 | 540 | -2.30E-05 | 0.000102 | 0.000103 | 4.55E-05 | 1.68E-05 | 1.96E-06 | 4.46E-05 | 9.22E-05 | 9.66E-05 | 6.48E-05 | 7.38E-06 | 1.36E-05 | 2.36E-06 | 4.63E-05 | |
| 40 | 544 | | | | | 1.72E-05 | | | | | | | | | | |
| 41 | 551 | -3.56E-05 | 9.12E-05 | 0.000104 | 5.52E-05 | 2.22E-05 | 2.05E-06 | 5.54E-05 | 8.52E-05 | 8.96E-05 | 6.18E-05 | 6.21E-06 | 1.66E-05 | 1.97E-06 | 4.42E-05 | |
| 42 | 557 | -3.66E-05 | 9.03E-05 | 0.000103 | 5.52E-05 | 2.15E-05 | | 5.49E-05 | 8.40E-05 | 8.75E-05 | 6.09E-05 | 3.49E-06 | 1.54E-05 | 2.28E-06 | 4.22E-05 | |
| 43 | 573 | | | 0.000118 | | | | | | | 6.57E-05 | 9.22E-06 | | | | |
| 44 | 582 | -2.74E-05 | | | 6.42E-05 | | 2.00E-06 | | 8.79E-05 | 9.14E-05 | | | 2.13E-05 | | 4.74E-05 | |
| 45 | 592 | -3.22E-05 | 8.63E-05 | 0.000116 | 6.38E-05 | 2.07E-05 | | 5.58E-05 | 8.56E-05 | 8.96E-05 | 6.56E-05 | 8.54E-06 | 1.92E-05 | 1.75E-06 | 4.62E-05 | |
| 46 | 594 | | | 0.000124 | | | | | 9.08E-05 | 9.61E-05 | 7.40E-05 | 1.42E-06 | 2.05E-05 | 2.02E-06 | 5.09E-05 | |
| 47 | 663 | -2.15E-05 | 9.67E-05 | | 6.77E-05 | | | | 9.11E-05 | 9.43E-05 | | | 2.32E-05 | | | |
| 48 | 716 | -2.18E-05 | 9.64E-05 | 0.000127 | 6.71E-05 | 2.57E-05 | | 5.77E-05 | 8.95E-05 | 9.27E-05 | 7.02E-05 | 1.08E-05 | 2.17E-05 | | 4.62E-05 | |
| 49 | 748 | -3.36E-05 | 8.32E-05 | 0.000117 | 6.15E-05 | 2.35E-05 | 1.70E-06 | 5.08E-05 | 8.51E-05 | 8.71E-05 | 6.80E-05 | 7.91E-06 | 1.59E-05 | 1.32E-06 | 4.36E-05 | |
| 50 | 760 | | | 0.000117 | 6.13E-05 | | | | | | 6.79E-05 | 7.60E-06 | | 1.34E-06 | | |
| 51 | 772 | -3.33E-05 | 8.34E-05 | | | 2.44E-05 | | 5.14E-05 | 8.57E-05 | 8.80E-05 | | | 1.84E-05 | | 4.47E-05 | |
| 52 | 782 | -3.54E-05 | 8.19E-05 | 0.000115 | 6.27E-05 | 2.40E-05 | 2.24E-06 | 5.08E-05 | 8.52E-05 | 8.71E-05 | 6.73E-05 | 7.86E-06 | 1.82E-05 | 1.35E-06 | 4.43E-05 | |
| 53 | 813 | -0.000107 | -4.81E-06 | 6.00E-05 | 9.94E-06 | 7.31E-06 | 1.52E-06 | -4.30E-06 | 8.18E-05 | 8.27E-05 | 6.47E-05 | 5.78E-06 | 8.71E-06 | 7.17E-07 | 4.16E-05 | |
| 54 | 818 | -0.000105 | -3.06E-06 | 6.05E-05 | 1.06E-05 | 7.95E-06 | 1.62E-06 | -3.97E-06 | 8.21E-05 | 8.30E-05 | | | 9.03E-06 | 5.16E-07 | 3.99E-05 | |
| 55 | 826 | -0.000103 | -1.42E-06 | 6.13E-05 | 1.09E-05 | 8.42E-06 | | -3.73E-06 | 8.18E-05 | 8.29E-05 | 6.46E-05 | 5.39E-06 | 1.07E-05 | 5.21E-07 | 4.01E-05 | |
| 56 | 844 | -0.000105 | -2.84E-06 | | | 9.13E-06 | | -1.60E-06 | 8.12E-05 | 8.11E-05 | | 5.13E-06 | 1.14E-05 | | | |
| 57 | 885 | | | | | | | | | | 6.73E-05 | | | | 4.38E-05 | |
| 58 | 917 | | | 6.90E-05 | 1.42E-05 | | 1.48E-06 | | 8.39E-05 | 8.45E-05 | | | | | | |
| 59 | 925 | | | | | | | 4.01E-06 | | | | | | | | |
| 60 | 931 | -4.28E-05 | 6.21E-05 | | 1.48E-05 | 9.27E-06 | | | 8.43E-05 | 8.39E-05 | 7.15E-05 | 1.52E-05 | 1.55E-05 | 3.37E-07 | 4.89E-05 | |
| 61 | 933 | -5.23E-05 | 5.19E-05 | 6.89E-05 | 1.19E-05 | 5.24E-06 | 1.83E-06 | 1.83E-06 | 7.12E-05 | 6.88E-05 | 6.35E-05 | 8.93E-06 | 5.56E-06 | 3.46E-07 | 4.25E-05 | |
| 62 | 937 | | | | | | | 2.96E-06 | | | | | | | | |
| 63 | 962 | | | | | | | | | | | | | | 4.26E-05 | |
| 64 | 975 | -3.10E-05 | | | | | | | | | | | | -2.39E-07 | | |
| 65 | 976 | | 8.71E-05 | | | | | | | | | | | | | |
| 66 | 991 | | | | | | 1.40E-06 | | 7.60E-05 | 7.09E-05 | | | 5.75E-06 | | | |
| 67 | 997 | -4.16E-05 | 7.83E-05 | 6.35E-05 | 8.58E-06 | 4.25E-06 | 1.40E-06 | -1.72E-07 | 7.39E-05 | 6.86E-05 | 5.49E-05 | -2.55E-06 | 3.63E-06 | -3.06E-07 | 2.97E-05 | |
| 68 | 1043 | | | 6.47E-05 | 9.65E-06 | 6.70E-06 | | 5.72E-07 | 6.20E-05 | 5.87E-05 | | | 6.09E-06 | | 2.96E-05 | |
| 69 | 1063 | -6.56E-05 | 4.43E-05 | | | | | | | | 5.17E-05 | -2.34E-06 | | | 2.95E-05 | |
| 70 | 1093 | -6.94E-05 | 3.98E-05 | 6.72E-05 | 1.15E-05 | 7.25E-06 | 1.63E-06 | 2.06E-06 | 6.15E-05 | 5.87E-05 | 5.04E-05 | -3.99E-06 | 5.88E-06 | 4.11E-09 | 2.79E-05 | |
| 71 | 1107 | -7.24E-05 | 3.71E-05 | | | | 1.89E-06 | | 6.31E-05 | 5.98E-05 | 5.04E-05 | -2.95E-06 | | | | |
| 72 | 1114 | -7.22E-05 | 3.73E-05 | 6.67E-05 | 7.40E-06 | 5.06E-06 | 1.41E-06 | 9.28E-07 | 6.31E-05 | 5.99E-05 | | | 4.62E-06 | 1.17E-07 | 2.90E-05 | |
| 73 | 1165 | -8.76E-05 | 2.10E-05 | 5.16E-05 | 1.82E-05 | 4.67E-06 | 1.34E-06 | -6.07E-06 | 5.91E-05 | 5.60E-05 | 4.64E-05 | -2.97E-06 | 3.45E-06 | 5.32E-08 | 2.85E-05 | |
| 74 | 1174 | 4.17E-05 | 0.00014 | | 1.74E-06 | 4.62E-06 | 1.33E-06 | -6.18E-06 | | | | -3.69E-06 | 2.31E-06 | 6.99E-08 | 2.80E-05 | |
| 75 | 1206 | -1.70E-05 | 8.55E-05 | 7.24E-05 | | | 1.40E-06 | | 6.92E-05 | 6.63E-05 | 4.92E-05 | | | | | |
| 76 | 1211 | -5.62E-05 | 4.39E-05 | 6.63E-05 | 6.09E-06 | 8.19E-06 | 1.42E-06 | -2.84E-06 | 6.71E-05 | 6.39E-05 | 4.87E-05 | 5.24E-07 | 7.68E-06 | 6.00E-08 | 3.06E-05 | |
| 77 | 1240 | | | 5.02E-05 | 8.82E-06 | | | | | | | -5.71E-07 | | 1.39E-06 | 2.73E-05 | |
| 78 | 1250 | -8.39E-05 | 5.48E-05 | | 4.20E-06 | 6.03E-06 | 1.58E-06 | -7.99E-06 | 5.64E-05 | 4.94E-05 | 4.46E-05 | 1.48E-06 | 1.26E-05 | | 2.83E-05 | |
| 79 | 1260 | | | 4.67E-05 | | | | | 5.60E-05 | 4.92E-05 | | | 1.26E-05 | | | |
| 80 | 1262 | -8.42E-05 | 4.76E-06 | | | | 1.59E-06 | -8.68E-06 | 5.60E-05 | 4.86E-05 | 4.51E-05 | 3.30E-06 | 1.30E-05 | 2.48E-07 | 2.92E-05 | |
| 81 | 1294 | -8.37E-05 | 5.24E-05 | 5.10E-05 | 4.08E-05 | 5.67E-07 | 1.29E-06 | | 5.64E-05 | 4.93E-05 | 4.54E-05 | 1.35E-05 | 1.35E-05 | 3.75E-07 | 2.99E-05 | |
| 82 | 1301 | | | | | | | | 5.66E-05 | 4.97E-05 | | 5.09E-06 | | | 3.02E-05 | |
| 83 | 1306 | -8.73E-05 | 2.34E-06 | 3.22E-05 | 3.54E-06 | | 1.34E-06 | -6.30E-06 | 5.65E-05 | 4.95E-05 | 4.30E-05 | 4.75E-06 | 1.32E-05 | 4.00E-07 | 3.00E-05 | |
| 84 | 1404 | | | | | | | | | | | 7.65E-06 | | | 3.21E-05 | |
| 85 | 1414 | -8.84E-05 | 8.59E-07 | 2.95E-05 | 2.94E-06 | 3.17E-06 | 1.32E-06 | -8.45E-06 | 4.71E-05 | 3.85E-05 | 3.39E-05 | 3.64E-06 | 1.01E-05 | 2.98E-07 | 2.88E-05 | |
| 86 | 1437 | -8.51E-05 | 6.44E-06 | | | | 1.48E-06 | | 4.86E-05 | 4.08E-05 | 3.53E-05 | 6.37E-06 | | 6.22E-07 | 3.33E-05 | |
| 87 | 1441 | -8.94E-05 | 2.25E-05 | 2.93E-05 | 8.22E-07 | 2.48E-06 | 1.48E-06 | -1.08E-05 | 4.50E-05 | 3.45E-05 | 3.26E-05 | 3.47E-06 | 7.22E-06 | 5.97E-07 | 2.92E-05 | |
| 88 | 1448 | -8.66E-05 | 6.16E-06 | 3.15E-05 | 2.98E-06 | 1.04E-06 | 1.60E-06 | -9.36E-06 | 4.69E-05 | 3.63E-05 | 3.43E-05 | 5.91E-06 | 1.04E-05 | 5.90E-07 | 3.18E-05 | |
| 89 | 1485 | -9.45E-05 | -2.32E-07 | 2.54E-05 | 3.24E-06 | 2.28E-06 | 1.65E-06 | -9.91E-06 | 4.29E-05 | 3.28E-05 | 3.14E-05 | 3.98E-06 | 1.13E-05 | | 2.98E-05 | |
| 90 | 1501 | -0.000101 | -5.59E-06 | 2.05E-05 | 1.30E-06 | 1.30E-07 | 1.68E-06 | -1.32E-05 | 3.85E-05 | 2.27E-05 | 2.64E-05 | -1.76E-06 | 1.96E-06 | 9.50E-07 | 2.06E-05 | |
| 91 | 1504 | -8.66E-05 | 6.99E-06 | | 2.09E-05 | | | 1.50E-06 | 4.99E-05 | 3.91E-05 | | | | | | |
| 92 | 1535 | | | 3.74E-05 | | | | | 5.45E-05 | 4.34E-05 | 3.61E-05 | 1.25E-05 | 1.45E-05 | | 4.59E-05 | |
| 93 | 1542 | | | | | | | | | | | | | | 8.71E-07 | 4.65E-05 | |
| 94 | 1576 | -7.59E-05 | 1.75E-05 | | | 1.16E-06 | | | | 4.31E-05 | | | 1.50E-05 | | | |
| 95 | 1579 | | | 3.54E-05 | 1.95E-05 | | | 1.90E-06 | 5.10E-05 | | 3.03E-05 | | | 9.10E-07 | | |
| 96 | 1617 | | | | | | 1.70E-06 | | | | | 1.06E-05 | | | | |
| 97 | 1654 | -7.66E-05 | 1.62E-05 | 3.43E-05 | 1.92E-05 | 8.54E-07 | 1.64E-06 | 1.87E-06 | 4.93E-05 | 4.00E-05 | 2.80E-05 | 1.05E-05 | 1.45E-05 | 1.01E-06 | 4.24E-05 | |
| 98 | 1655 | -7.91E-05 | 1.34E-05 | 3.29E-05 | 1.92E-05 | 1.66E-07 | 2.33E-06 | 1.88E-05 | 4.76E-05 | 3.92E-05 | 2.83E-05 | 1.04E-05 | 1.47E-05 | 1.29E-06 | 4.22E-05 | |
| 99 | 1667 | -7.93E-05 | 1.30E-05 | | 1.94E-05 | | | | | 4.64E-05 | 3.64E-05 | 2.82E-05 | 1.27E-05 | | 1.32E-06 | 4.28E-05 | |
| 100 | 1683 | -9.59E-05 | -6.98E-06 | 2.28E-05 | 1.49E-05 | -2.01E-06 | 1.63E-06 | 3.06E-06 | | | | | 9.29E-06 | 1.10E-07 | 3.27E-05 | |
| 101 | 1686 | | | | | | 1.51E-06 | 6.22E-06 | 3.82E-05 | 2.95E-05 | 2.62E-05 | 1.58E-06 | 1.32E-05 | 1.43E-07 | 3.81E-05 | |
| 102 | 1708 | -0.000109 | -2.43E-05 | | 3.60E-06 | | | | 4.05E-05 | 3.12E-05 | 2.73E-05 | 1.66E-05 | 1.31E-05 | 1.02E-07 | 3.92E-05 | |

| | A | B | C | D | E | F | G | H | I | J | K | L | M | N | O | P |
|---|---|---|---|---|---|---|---|---|---|---|---|---|---|---|---|---|
| 103 | 1720 | | | 1.73E-05 | 4.12E-06 | -6.28E-07 | 2.27E-06 | 6.12E-06 | 3.96E-05 | 3.01E-05 | 2.70E-05 | 1.62E-05 | | | | |
| 104 | 1725 | -0.000109 | -2.46E-05 | | | | | | 4.83E-05 | 4.01E-05 | 3.61E-05 | 3.16E-05 | 1.74E-05 | | 5.25E-05 | |
| 105 | 1734 | | | 1.70E-05 | | | | | | | | | | 2.46E-08 | 4.85E-05 | |
| 106 | 1757 | | | | 5.54E-06 | | | 7.37E-06 | 4.39E-05 | 3.61E-05 | 3.68E-05 | 2.71E-05 | 1.63E-05 | | | |
| 107 | 1758 | -0.000111 | -2.58E-05 | 1.52E-05 | 4.33E-06 | -6.56E-07 | 2.27E-06 | 5.61E-06 | 4.03E-05 | 3.28E-05 | 3.51E-05 | 2.45E-05 | 1.45E-05 | -3.92E-07 | 3.94E-05 | |
| 108 | 1759 | -6.88E-05 | 1.73E-05 | | | | | | 6.07E-05 | 5.60E-05 | 4.42E-05 | 3.94E-05 | | | | |
| 109 | 1781 | | | 2.00E-05 | | 1.27E-06 | | | | | | | 2.07E-05 | | 4.53E-05 | |
| 110 | 1787 | -8.73E-05 | -3.45E-07 | | 8.37E-06 | 2.08E-06 | | 9.17E-06 | 4.73E-05 | 4.60E-05 | 3.98E-05 | 3.88E-05 | 2.07E-05 | -7.20E-07 | 4.55E-05 | |
| 111 | 1791 | -8.54E-05 | 1.19E-06 | 2.53E-05 | 9.33E-06 | | 2.07E-06 | 1.04E-05 | 4.74E-05 | 4.62E-05 | 3.97E-05 | 3.83E-05 | 2.11E-05 | | 4.55E-05 | |
| 112 | 1816 | | | | 8.47E-06 | | | 1.00E-05 | | | | | | | | |
| 113 | 1851 | -8.94E-05 | -2.82E-06 | 2.37E-05 | | 2.92E-07 | | | | | | | | -6.71E-07 | | |
| 114 | 1901 | | | | 7.77E-06 | 1.00E-07 | | 8.91E-06 | 4.69E-05 | 4.60E-05 | 3.87E-05 | 3.55E-05 | 1.99E-05 | -6.62E-07 | 4.29E-05 | |
| 115 | 1909 | | | | | | | | | | | | | | | | |
| 116 | 1942 | -8.96E-05 | -2.89E-06 | 1.77E-05 | 8.21E-06 | 7.75E-07 | 2.18E-06 | 9.52E-06 | 4.88E-05 | 4.63E-05 | 3.79E-05 | 3.85E-05 | 2.13E-05 | | 4.57E-05 | |
| 117 | 1956 | -8.99E-05 | -3.12E-06 | 1.77E-05 | | | 2.19E-06 | | 4.85E-05 | 4.61E-05 | 3.78E-05 | 3.84E-05 | 2.14E-05 | -7.25E-07 | | |
| 118 | 1992 | | | | 1.88E-05 | 1.42E-06 | | 2.27E-06 | | | | | | | 4.92E-05 | |
| 119 | 2018 | -7.30E-05 | 1.41E-05 | 3.36E-05 | | | | | 4.88E-05 | 4.66E-05 | 3.90E-05 | 4.24E-05 | 2.38E-05 | | 4.90E-05 | |
| 120 | 2042 | | | | | | | 2.13E-05 | | 9.53E-05 | 9.83E-05 | 8.59E-05 | | | | |
| 121 | 2047 | -2.52E-05 | 5.21E-05 | 5.34E-05 | 1.72E-05 | | | 2.08E-05 | | 9.55E-05 | | 4.24E-05 | 2.38E-05 | | 5.00E-05 | |
| 122 | 2049 | | | | | 2.11E-06 | | | | | | | | | | |
| 123 | 2076 | | | | 1.82E-05 | | | | 9.58E-05 | | 8.58E-05 | | | | | |
| 124 | 2080 | -2.34E-05 | 5.45E-05 | 5.42E-05 | | 2.46E-06 | 1.70E-06 | | | | | 4.54E-05 | 2.54E-05 | -9.57E-07 | 5.18E-05 | |
| 125 | 2085 | -2.11E-05 | 5.72E-05 | 5.51E-05 | 1.82E-05 | 2.48E-06 | | 2.14E-05 | 9.72E-05 | 9.41E-05 | 8.58E-05 | 4.53E-05 | 2.56E-05 | | 5.17E-05 | |
| 126 | 2096 | | | | 1.91E-05 | | | 2.31E-05 | | | | | | | | |
| 127 | 2105 | -1.98E-05 | 5.84E-05 | 5.55E-05 | | | | | 9.73E-05 | 9.39E-05 | 8.58E-05 | 4.56E-05 | | | | |
| 128 | 2115 | | | | | | | | | | | | | 2.15E-05 | | 4.85E-05 | |
| 129 | 2116 | -2.17E-05 | 5.68E-05 | 5.33E-05 | 1.09E-05 | 6.73E-07 | 1.77E-06 | 1.46E-05 | 9.16E-05 | 8.58E-05 | 8.49E-05 | 4.48E-05 | 2.15E-05 | -9.71E-07 | 4.86E-05 | |
| 130 | 2127 | -2.60E-05 | 5.10E-05 | 4.88E-05 | 1.09E-05 | 5.98E-07 | 1.80E-06 | 1.39E-05 | 8.73E-05 | 8.22E-05 | 8.20E-05 | 4.34E-05 | 2.00E-05 | -1.02E-06 | 4.76E-05 | |
| 131 | 2128 | | | | 9.92E-06 | 5.59E-07 | | 1.31E-05 | | | | | | | 4.67E-05 | |
| 132 | 2148 | | | | | | 1.87E-06 | | | | | | | | | |
| 133 | 2156 | -1.81E-05 | 5.88E-05 | 5.42E-05 | | | | | 8.98E-05 | 8.41E-05 | 8.41E-05 | 4.56E-05 | 2.06E-05 | -7.46E-07 | 4.97E-05 | |
| 134 | 2175 | | | | | | 1.95E-06 | 1.97E-05 | | | | | | | | |
| 135 | 2185 | 0.000114 | | 0.000148 | 1.01E-05 | | | | 0.000145 | 0.000138 | 0.000166 | 4.73E-05 | | -1.31E-06 | 7.14E-05 | |
| 136 | 2226 | 0.000101 | 8.90E-05 | 0.000129 | 1.03E-05 | -2.89E-06 | 1.88E-06 | 1.82E-05 | 0.000139 | 0.000131 | 0.00016 | 4.65E-05 | 1.97E-05 | -1.57E-06 | | |
| 137 | 2244 | | | | | | | | | | | | 2.64E-05 | -1.19E-06 | 7.77E-05 | |
| 138 | 2261 | | | | 7.24E-06 | -1.62E-06 | 1.83E-06 | 1.43E-05 | 0.000131 | 0.000124 | 0.000154 | 2.99E-05 | | | | |
| 139 | 2272 | | | 0.000142 | | | | | | | | | | -1.52E-06 | | |
| 140 | 2351 | 0.000195 | 0.000183 | | | | | | | | | | | | 6.98E-05 | |
| 141 | 2357 | 0.00019 | 0.000179 | 0.00014 | 7.77E-06 | -1.25E-06 | 1.55E-06 | 1.35E-05 | 0.000128 | 0.000124 | 0.000153 | 3.03E-05 | 2.52E-05 | -1.57E-06 | 6.81E-05 | |
| 142 | 2365 | | | | 7.98E-06 | -8.79E-06 | | | 0.000123 | 0.00012 | 0.000143 | | | | | |
| 143 | 2369 | 0.000185 | 0.000173 | | | | | | | | | | | | | |
| 144 | 2376 | 0.000182 | 0.000172 | 0.000135 | 9.10E-06 | | 1.61E-06 | 1.43E-05 | 0.000122 | 0.00012 | 0.000143 | 3.13E-05 | 2.59E-05 | -1.83E-06 | 6.86E-05 | |
| 145 | 2384 | | | | | | | | | | | | | | | 6.85E-05 | |
| 146 | 2394 | | | | | -7.48E-06 | | | | | | | 2.71E-05 | | | |
| 147 | 2415 | 0.000173 | 0.000163 | 0.000123 | 7.39E-06 | | 1.85E-06 | 1.15E-05 | 0.000121 | 0.000119 | 0.000143 | 3.09E-05 | 2.69E-05 | -1.62E-06 | 6.87E-05 | |
| 148 | 2433 | | | | | -8.71E-06 | | | 0.000121 | | 0.000143 | | | | | |
| 149 | 2448 | 9.71E-05 | 8.38E-05 | 7.17E-05 | 6.97E-06 | | 1.84E-06 | 7.86E-06 | | | | 2.90E-05 | | -1.71E-06 | 6.54E-05 | |
| 150 | 2449 | | | | | | | | 0.000125 | 0.000118 | 0.000149 | 3.31E-05 | | | | |
| 151 | 2451 | 9.75E-05 | 8.42E-05 | 7.20E-05 | 7.63E-06 | -7.85E-06 | 1.94E-06 | 8.36E-06 | 0.000124 | 0.000117 | 0.000148 | 3.37E-05 | 3.25E-05 | -1.60E-06 | 6.68E-05 | |
| 152 | 2539 | 7.03E-05 | 5.65E-05 | 5.32E-05 | 6.62E-06 | -6.56E-06 | 1.66E-06 | 5.21E-06 | 9.39E-05 | 7.59E-05 | 0.000108 | -6.60E-06 | 2.63E-05 | -2.08E-06 | 3.13E-05 | |
| 153 | 2552 | | | | | | | | 9.69E-05 | | 0.000108 | | | | | |
| 154 | 2570 | 7.63E-05 | 6.36E-05 | 5.75E-05 | 5.67E-06 | -7.62E-06 | 1.63E-06 | 3.76E-06 | | | | -3.67E-06 | 2.62E-05 | -2.10E-06 | 3.29E-05 | |
| 155 | 2585 | | | | | | 1.53E-06 | | 9.81E-05 | 7.86E-05 | 0.000114 | | 2.68E-05 | | | |
| 156 | 2587 | | | | | | | | | | | -1.54E-06 | | | | |
| 157 | 2594 | | | | | | | | 9.58E-05 | 7.62E-05 | 0.000109 | | | -2.15E-06 | 3.57E-05 | |
| 158 | 2678 | 8.01E-05 | 6.05E-05 | 5.61E-05 | | -8.93E-06 | | 2.64E-06 | | | | -4.50E-06 | 2.56E-05 | | | |
| 159 | 2696 | | | | 6.10E-06 | | | | | | | | | | 3.52E-05 | |
| 160 | 2717 | 8.14E-05 | 6.19E-05 | 5.82E-05 | 5.88E-06 | -7.84E-06 | 1.44E-06 | 3.76E-06 | 9.53E-05 | 7.58E-05 | 0.000109 | -7.53E-06 | 2.56E-05 | -2.11E-06 | 3.03E-05 | |
| 161 | 2769 | 6.24E-05 | 4.96E-05 | | | | | | 9.25E-05 | 7.09E-05 | 0.000104 | | | | | |
| 162 | 2803 | | | 5.32E-05 | 4.96E-06 | -8.31E-06 | | 2.54E-06 | | | | | 2.50E-05 | | | | |
| 163 | 2808 | | | | | | | | | | | -2.08E-06 | | | 3.16E-05 | |
| 164 | 2822 | 6.71E-05 | 5.54E-05 | 6.03E-05 | | | 1.43E-06 | | | | | | | | | |
| 165 | 2831 | | | | | | | 8.56E-06 | 9.42E-05 | 7.25E-05 | 0.000107 | 7.99E-06 | 2.88E-05 | | 3.43E-05 | |
| 166 | 2856 | 5.84E-05 | 4.65E-05 | 5.10E-05 | 5.37E-06 | -9.21E-06 | | 5.55E-06 | 9.04E-05 | 6.80E-05 | 0.000102 | | | 1.46E-06 | | |
| 167 | 2929 | 5.47E-05 | 4.34E-05 | 4.61E-05 | 4.94E-06 | -9.63E-06 | 1.46E-06 | 5.06E-06 | 8.73E-05 | 6.44E-05 | 0.000101 | 1.37E-06 | 2.63E-05 | 1.51E-06 | 2.87E-05 | |
| 168 | 2942 | 5.47E-05 | 4.34E-05 | | | | | | | | | | | | | |
| 169 | 2964 | | | 4.62E-05 | 5.19E-06 | -1.16E-06 | 1.44E-06 | 5.52E-06 | 8.58E-05 | 6.29E-05 | 0.000101 | 7.41E-06 | 2.69E-05 | 1.51E-06 | 3.04E-05 | |
| 170 | 2980 | 5.34E-05 | 4.21E-05 | | | -1.15E-06 | 1.39E-06 | 4.05E-06 | | | | 5.71E-06 | | | 2.72E-05 | |
| 171 | 2982 | 5.38E-05 | 4.25E-05 | 4.64E-05 | 6.25E-06 | | | 5.21E-06 | 8.85E-05 | 6.57E-05 | 0.000101 | 8.04E-06 | 2.64E-05 | 2.06E-06 | 2.88E-05 | |
| 172 | 2984 | | | | 8.67E-06 | | | | | | | | 2.49E-06 | | 2.45E-05 | |
| 173 | 3004 | 3.58E-05 | 2.02E-05 | 3.54E-05 | 8.14E-06 | -9.24E-06 | 1.41E-06 | 4.36E-06 | 8.13E-05 | 5.82E-05 | 9.60E-05 | 1.26E-06 | 2.64E-05 | 2.33E-06 | 2.26E-05 | |
| 174 | 3011 | | | | 8.60E-06 | | | | 8.18E-05 | 5.90E-05 | 9.61E-05 | | 2.66E-05 | | | |
| 175 | 3014 | | | | | | | | | | | | 1.68E-06 | | 2.30E-05 | |
| 176 | 3061 | 5.82E-05 | 4.54E-05 | 4.68E-05 | 9.76E-06 | -8.10E-06 | | 5.38E-06 | 9.01E-05 | 6.87E-05 | 9.79E-05 | | 2.63E-05 | | | |
| 177 | 3085 | | | | | | 1.73E-06 | | | | | -2.41E-06 | | | 1.97E-05 | |
| 178 | 3120 | 5.61E-05 | 4.35E-05 | 4.40E-05 | | | | | 8.91E-05 | 6.76E-05 | | | | 2.47E-06 | | |
| 179 | 3143 | | | | | | | 6.53E-06 | | | 9.81E-05 | | 2.53E-05 | | | |
| 180 | 3159 | | | | 1.07E-05 | -7.48E-06 | | | | | | | | | | |
| 181 | 3199 | | | | | | 1.39E-06 | | | 6.88E-05 | | 4.94E-06 | | | 2.54E-05 | |
| 182 | 3214 | | | | | | | | 8.82E-05 | | 9.97E-05 | | | | | |
| 183 | 3240 | 4.76E-05 | 3.50E-05 | 4.04E-05 | 9.08E-06 | -7.17E-06 | | 5.96E-06 | 7.96E-05 | 6.28E-05 | 9.82E-05 | 8.32E-07 | 2.18E-05 | | 2.35E-05 | |
| 184 | 3258 | 8.75E-05 | 7.69E-05 | 5.53E-05 | | | | 1.39E-06 | | | | 8.02E-06 | | | 3.20E-05 | |
| 185 | 3315 | | | | | | | | | 0.00012 | 0.000108 | 0.00012 | | | 2.15E-06 | | |
| 186 | 3341 | 0.000156 | 0.000146 | | 1.13E-05 | | | | | | | | | | | | |
| 187 | 3366 | | | 8.28E-06 | | | 1.38E-06 | | | 9.43E-05 | | | | 2.29E-06 | | | |
| 188 | 3370 | | | | | | | | 9.44E-05 | | | | | 2.24E-05 | | | |
| 189 | 3380 | 0.000119 | 0.000108 | 8.13E-06 | 1.15E-05 | -7.48E-06 | | 1.22E-05 | 8.64E-05 | 8.48E-05 | 0.000109 | 7.28E-06 | 2.24E-05 | 2.31E-06 | 3.12E-05 | |
| 190 | 3410 | | | | | | 1.42E-06 | | | | 0.000106 | 9.69E-06 | | | 2.96E-05 | |
| 191 | 3428 | | | | | | | 1.27E-05 | | | | | | | | |
| 192 | 3437 | 0.00012 | 0.000107 | | | | | | 8.24E-05 | 8.09E-05 | | | | | | |
| 193 | 3470 | | | 3.91E-06 | 1.07E-05 | -7.86E-06 | | | | | 0.000104 | 5.56E-06 | 2.21E-05 | | 2.67E-05 | |
| 194 | 3505 | 0.000109 | 9.65E-05 | | 8.96E-06 | | 1.26E-06 | 9.92E-06 | 7.95E-05 | 7.73E-05 | | | | 7.28E-07 | | |
| 195 | 3516 | | | -1.48E-06 | | -8.30E-06 | | | | | 0.000102 | 5.24E-06 | 2.18E-05 | | 2.66E-05 | |
| 196 | 3564 | | | | | | | | | | | | | 5.20E-07 | | | |
| 197 | 3576 | | | | | | | 7.64E-06 | | | | | | | | |
| 198 | 3578 | 0.000112 | 0.000101 | -4.36E-06 | 5.83E-06 | -8.07E-06 | 1.07E-06 | | 7.08E-05 | 6.82E-05 | | | 1.89E-05 | | | |
| 199 | 3601 | 0.000111 | 9.99E-05 | -4.81E-06 | 6.47E-06 | -8.14E-06 | 1.07E-06 | 8.81E-06 | 7.06E-05 | 6.78E-05 | 0.0001 | 7.16E-06 | 2.05E-05 | 5.41E-07 | 2.56E-05 | |
| 200 | 3615 | 0.000105 | 9.29E-05 | -5.92E-06 | 5.85E-06 | | | | 7.03E-05 | 6.72E-05 | | | | | | |
| 201 | 3633 | | | | | | | 1.20E-06 | | | 0.000101 | 8.79E-06 | 2.11E-05 | | 2.61E-05 | |
| 202 | 3671 | 0.000127 | 0.000116 | | 9.07E-06 | -7.68E-06 | | 9.99E-06 | 6.66E-05 | 6.31E-05 | 9.25E-05 | 1.25E-05 | 2.49E-05 | | 3.00E-05 | |
| 203 | 3682 | 9.67E-05 | 8.77E-05 | -2.62E-05 | 6.35E-06 | -8.87E-06 | 1.13E-06 | 3.97E-06 | 6.17E-05 | 5.47E-05 | 8.88E-05 | 4.51E-05 | 1.46E-05 | 3.48E-07 | 2.43E-05 | |
| 204 | 3701 | | | -2.54E-05 | | | | | | | | | | | | | |

| | A | B | C | D | E | F | G | H | I | J | K | L | M | N | O | P |
|---|---|---|---|---|---|---|---|---|---|---|---|---|---|---|---|---|
| 205 | 3704 | | | | | | | | | 5.23E-05 | | | | | | |
| 206 | 3745 | | | | | | | 8.74E-06 | 7.45E-06 | | | | 1.57E-05 | | | |
| 207 | 3759 | 0.000112 | 0.000102 | -1.37E-05 | 1.95E-06 | -9.71E-06 | | | | | 7.75E-05 | 7.48E-07 | | | 6.49E-06 | |
| 208 | 3766 | | | | | | 1.11E-06 | -7.45E-06 | | 4.31E-06 | | | 1.27E-05 | | | |
| 209 | 3776 | 0.000111 | 0.000101 | -1.42E-05 | 1.52E-06 | -9.68E-06 | | -6.27E-06 | 6.89E-05 | | 7.75E-05 | 1.32E-06 | | | 7.27E-06 | |
| 210 | 3790 | | | | | | | | | 5.73E-05 | | | | | | |
| 211 | 3840 | | | | | | | 5.87E-05 | | | | | 9.24E-06 | | | |
| 212 | 3862 | | | | | | | | | | | | | 6.32E-07 | | |
| 213 | 3881 | | 8.79E-06 | | | | | | | | | | | | | |
| 214 | 3886 | 0.000109 | 8.79E-06 | -1.68E-05 | 1.48E-06 | -9.77E-06 | 1.09E-06 | -6.53E-06 | 5.86E-05 | 5.37E-05 | 7.65E-05 | 1.17E-06 | 9.30E-06 | 6.31E-07 | 7.16E-06 | |
| 215 | 3894 | | | | | | | | 6.31E-05 | | | | | | | |
| 216 | 3901 | | | | | | | | | | | 8.06E-05 | | | | |
| 217 | 3906 | | | | | -9.97E-06 | | | | | | | | | | |
| 218 | 3910 | | | -3.60E-06 | | | | -2.01E-07 | | | | | | 8.98E-07 | | |
| 219 | 3930 | | 2.60E-05 | | 4.48E-06 | | | | | | 3.38E-05 | | | | | |
| 220 | 3943 | 0.000126 | | | | | | 9.40E-07 | 5.10E-05 | | 6.93E-05 | 2.51E-05 | | | 1.94E-05 | |
| 221 | 3946 | | | 1.50E-05 | 2.28E-05 | | 9.55E-07 | 2.19E-05 | 5.10E-05 | 3.38E-05 | 6.93E-05 | 2.63E-05 | 3.35E-05 | 1.49E-06 | 1.95E-05 | |
| 222 | 3997 | 0.000121 | 2.53E-05 | 6.47E-06 | -2.32E-05 | -9.59E-06 | 1.43E-06 | -2.38E-05 | 4.47E-05 | 2.69E-05 | 3.44E-05 | -2.78E-05 | 4.30E-06 | 1.48E-06 | -3.17E-05 | |
| 223 | 4014 | 9.69E-05 | -5.43E-07 | | -2.19E-05 | -7.69E-06 | | | 3.84E-05 | 2.21E-05 | 3.10E-05 | | | 1.70E-06 | -3.26E-05 | |
| 224 | 4029 | 9.89E-05 | 6.99E-07 | 2.71E-07 | -1.87E-05 | -5.44E-06 | 1.46E-06 | -2.31E-05 | 4.03E-05 | 2.41E-05 | | -2.26E-05 | 1.41E-05 | 1.69E-06 | | |
| 225 | 4043 | | | | | | 1.45E-06 | | 4.22E-05 | 2.63E-05 | 3.93E-05 | -3.54E-06 | 2.03E-05 | 1.72E-06 | -1.57E-05 | |
| 226 | 4047 | 0.000125 | 2.67E-05 | 1.04E-05 | -2.03E-05 | 1.42E-06 | | -2.79E-05 | | | 1.06E-06 | -4.52E-06 | | -5.36E-06 | -5.92E-05 | |
| 227 | 4071 | 0.00011 | 1.47E-05 | -4.26E-06 | -2.26E-05 | 1.69E-06 | 1.73E-06 | -3.01E-05 | 1.33E-05 | | -2.10E-05 | -3.64E-05 | -6.27E-06 | 2.66E-06 | -6.59E-05 | |
| 228 | 4082 | | | | | | | | | -7.95E-07 | | | | | | |
| 229 | 4083 | | | | | | | | 1.91E-05 | | | -3.12E-05 | | | | |
| 230 | 4088 | | | -3.26E-06 | | | | | | | -1.96E-05 | | | | -5.49E-05 | |
| 231 | 4090 | | | | -2.03E-05 | | | -2.70E-05 | | | | | -2.30E-06 | | | |
| 232 | 4129 | 9.35E-05 | -1.61E-06 | | | 4.60E-06 | | | -1.26E-05 | -2.95E-05 | -3.41E-05 | -3.79E-05 | | | -6.07E-05 | |
| 233 | 4133 | 9.04E-05 | -3.91E-06 | -1.04E-05 | -2.14E-05 | 4.42E-06 | 1.68E-06 | -2.81E-05 | -1.26E-05 | -2.96E-05 | -3.42E-05 | -3.81E-05 | -6.55E-06 | 2.26E-06 | -6.09E-05 | |
| 234 | 4177 | | | | -2.12E-05 | | | | 6.11E-06 | -1.22E-05 | | -3.78E-05 | -6.02E-06 | | | |
| 235 | 4190 | 0.0001 | 3.41E-06 | | | 4.14E-06 | | | | | -2.10E-05 | | | | -5.75E-05 | |
| 236 | 4229 | 9.72E-05 | 1.88E-06 | -2.08E-05 | -2.44E-05 | 4.02E-06 | 1.54E-06 | -3.40E-05 | -5.71E-06 | -2.31E-05 | -2.10E-05 | -4.96E-05 | -1.31E-05 | 1.81E-06 | -5.88E-05 | |
| 237 | 4286 | | | -1.74E-05 | -2.23E-05 | | | | -3.16E-05 | -9.18E-06 | -2.50E-05 | -2.34E-05 | -4.99E-05 | -1.30E-05 | | -6.04E-05 | |
| 238 | 4292 | | | | | 5.00E-06 | | | | | | | | | | | |
| 239 | 4324 | 0.00017 | 6.82E-05 | | | | | | | | -2.58E-05 | | | -1.49E-05 | | | |
| 240 | 4328 | | | -7.37E-06 | | | | | | 3.69E-05 | | -1.53E-05 | -5.52E-05 | | | | |
| 241 | 4338 | 0.00015 | 5.34E-05 | -1.71E-05 | -2.41E-05 | 4.48E-06 | 1.73E-06 | -3.40E-05 | -6.94E-05 | -6.74E-05 | -7.98E-05 | -6.06E-05 | -2.04E-05 | 1.70E-06 | -7.36E-05 | |
| 242 | 4352 | | | -2.05E-05 | | 3.64E-06 | 1.66E-06 | -3.64E-05 | | | | | -2.29E-05 | | | | |
| 243 | 4361 | 0.000151 | 5.51E-05 | -1.94E-05 | -1.90E-05 | 7.92E-06 | 1.54E-06 | -3.23E-05 | -7.64E-05 | -7.20E-05 | -8.29E-05 | -5.20E-05 | -1.65E-05 | 1.83E-06 | -6.87E-05 | |
| 244 | 4386 | | | | -1.82E-05 | 9.98E-06 | | -3.11E-05 | | | | | | | | | |
| 245 | 4405 | 6.86E-05 | -2.97E-05 | -5.29E-05 | -1.82E-05 | 9.81E-06 | | -3.11E-05 | -0.000108 | -0.000103 | -0.000104 | -5.69E-05 | -2.33E-05 | 1.80E-06 | -7.55E-05 | |
| 246 | 4413 | 6.87E-05 | -2.94E-05 | | | 1.01E-05 | 1.44E-06 | | | | | | | | | | |
| 247 | 4419 | | | | | | | | | | -8.98E-05 | | | | | | |
| 248 | 4424 | 7.25E-05 | -2.58E-05 | -5.22E-05 | -1.85E-05 | 9.92E-06 | 1.43E-06 | -3.11E-05 | -9.80E-05 | -8.87E-05 | -0.000105 | -5.64E-05 | -2.33E-05 | 1.59E-06 | -7.60E-05 | |
| 249 | 4433 | 6.88E-05 | -2.98E-05 | | | | | | | | | | | 1.53E-06 | | | |
| 250 | 4445 | | | | | 1.05E-05 | | | | | | | | | | | |
| 251 | 4451 | 2.01E-05 | -8.19E-05 | -6.29E-05 | -2.08E-05 | 9.62E-06 | 1.57E-06 | -3.66E-05 | -0.000133 | -0.000116 | -0.000127 | -6.63E-05 | -2.77E-05 | 1.49E-06 | -0.000101 | |
| 252 | 4513 | 3.60E-05 | -6.21E-05 | | | | | | -0.000123 | -0.000108 | -0.000102 | -4.79E-05 | -1.91E-05 | 1.35E-06 | | | |
| 253 | 4518 | | | | | | | | | | | | | | | -7.31E-05 | |
| 254 | 4541 | 8.34E-05 | -1.98E-05 | 1.20E-05 | -1.14E-05 | | 1.35E-06 | -2.37E-06 | -0.000112 | -9.49E-05 | -9.29E-05 | | -2.10E-05 | | | | |
| 255 | 4542 | 7.05E-05 | -3.40E-05 | -4.07E-07 | -1.14E-05 | 1.12E-05 | | -8.62E-06 | -0.000116 | -9.87E-05 | -0.0001 | -4.74E-05 | -2.08E-05 | 2.01E-06 | -7.30E-05 | |
| 256 | 4543 | 5.48E-05 | -4.71E-05 | -1.33E-05 | -1.23E-05 | 1.12E-05 | | -1.42E-06 | -0.000127 | | -0.000113 | -5.39E-05 | -2.33E-05 | 1.61E-06 | -8.81E-05 | |
| 257 | 4561 | 5.30E-05 | -4.91E-05 | | | -1.22E-05 | 1.13E-05 | 1.34E-06 | | -0.00013 | -0.000106 | -0.000118 | -5.35E-05 | -2.35E-05 | | -8.98E-05 | |
| 258 | 4572 | 5.30E-05 | -4.92E-05 | -2.48E-05 | -1.23E-05 | 1.13E-05 | | -1.85E-06 | -0.00013 | -0.000106 | -0.000118 | -5.41E-05 | -2.35E-05 | | -9.10E-05 | |
| 259 | 4581 | | | | | | | | -0.000103 | | | -0.000101 | 5.20E-06 | | | -4.81E-05 | |
| 260 | 4621 | 5.79E-05 | -4.47E-05 | -2.70E-05 | -1.37E-05 | 8.40E-06 | 1.47E-06 | -2.09E-05 | -0.000125 | -9.71E-05 | | | -2.60E-05 | 1.05E-06 | | | |
| 261 | 4623 | | | -2.79E-05 | | | | | | | | -0.00011 | -1.83E-05 | | | -6.56E-05 | |
| 262 | 4624 | 6.06E-05 | -4.46E-05 | -2.44E-05 | -1.18E-05 | 8.41E-06 | 1.41E-06 | -1.83E-05 | -0.00013 | -0.000101 | -0.000111 | -1.79E-05 | -2.50E-05 | 1.05E-06 | -6.76E-05 | |
| 263 | 4627 | 5.97E-05 | -4.50E-05 | | | | | -1.77E-05 | | | | -0.000112 | -1.89E-05 | -2.46E-05 | 1.18E-06 | -6.91E-05 | |
| 264 | 4629 | 5.97E-05 | -4.51E-05 | -2.80E-05 | -1.11E-05 | 8.57E-06 | 1.10E-06 | -1.76E-05 | -0.000124 | -9.44E-05 | -0.000112 | -1.89E-05 | -2.46E-05 | 1.03E-06 | -6.90E-05 | |
| 265 | 4639 | 6.30E-05 | -4.24E-05 | | | | | -1.37E-06 | -0.000124 | | -0.000112 | -1.63E-05 | -2.32E-05 | | -6.56E-05 | |
| 266 | 4641 | 6.16E-05 | -4.39E-05 | -2.88E-05 | -1.11E-05 | 7.62E-06 | 9.29E-07 | | -0.000125 | -9.46E-05 | | | | 1.90E-07 | | | |
| 267 | 4643 | 6.10E-05 | -4.46E-05 | -2.95E-05 | -1.20E-05 | 6.97E-06 | 9.48E-07 | -1.45E-05 | -0.000126 | -9.61E-05 | -0.000113 | -1.73E-05 | -2.43E-05 | 2.00E-07 | -6.68E-05 | |
| 268 | 4644 | 6.90E-05 | -3.78E-05 | | -9.89E-06 | 8.54E-06 | | -1.14E-05 | | | | -4.19E-06 | | | -5.10E-05 | |
| 269 | 4649 | | | -2.86E-05 | | | 9.98E-07 | | -8.91E-05 | | | | | | | | |
| 270 | 4660 | 0.000145 | 4.07E-05 | | | | | | -8.56E-05 | | | | | -2.44E-08 | | | |
| 271 | 4679 | 0.000125 | 2.23E-05 | -2.97E-05 | -5.00E-06 | 9.80E-06 | 1.00E-06 | -5.86E-06 | -9.80E-05 | -9.64E-05 | -0.000104 | 9.19E-08 | -2.00E-05 | | -5.06E-05 | |
| 272 | 4683 | | | | | | | -5.92E-06 | | | -0.000103 | | -1.96E-05 | -1.80E-07 | | | |
| 273 | 4690 | 8.06E-05 | -1.75E-05 | -4.38E-05 | -6.83E-06 | | | | | -0.000119 | | | -2.59E-06 | | | -5.36E-05 | |
| 274 | 4693 | | | | | | 6.39E-06 | 8.98E-07 | -9.96E-06 | -0.00012 | | | -0.000106 | -1.23E-06 | -2.04E-05 | -5.08E-05 | |
| 275 | 4701 | | | | -7.37E-06 | | 3.88E-07 | | | | | | -0.000106 | -1.23E-06 | -2.04E-05 | -5.08E-05 | |
| 276 | 4710 | | -2.14E-05 | -4.19E-05 | | | | | -0.000118 | -0.000118 | | | | | -5.75E-08 | | |
| 277 | 4730 | 6.38E-05 | | | | 6.81E-06 | | -9.48E-06 | | -0.000125 | -0.000105 | 5.46E-06 | -1.81E-05 | | -4.86E-05 | |
| 278 | 4732 | | | | | | | | -0.000125 | | | | | | | | |
| 279 | 4759 | 5.37E-05 | -2.92E-05 | -5.43E-05 | -9.74E-06 | | | | | -0.000128 | -0.000107 | 3.79E-06 | -1.86E-05 | | -5.13E-05 | |
| 280 | 4764 | | | -1.21E-05 | 6.15E-06 | -2.83E-06 | -1.66E-05 | -0.000141 | | | | | -2.80E-06 | | | |
| 281 | 4780 | 4.30E-05 | -3.72E-05 | -5.73E-05 | | | | | | | -0.000107 | | -1.79E-05 | -2.30E-06 | | | |
| 282 | 4821 | | | | | | | | | | | | 9.06E-06 | | | -4.51E-05 | |
| 283 | 4853 | | | | -1.22E-05 | 4.99E-06 | -2.76E-06 | -2.02E-05 | -0.000137 | -0.000122 | | | | -1.62E-06 | | | |
| 284 | 4860 | | -3.60E-05 | -5.66E-05 | | | | | | | -0.000107 | | | | | | |
| 285 | 4875 | 4.96E-05 | | | | | | -2.68E-06 | | -0.000136 | -0.000121 | | 1.12E-05 | -1.46E-05 | | -4.26E-05 | |
| 286 | 4878 | | | | | 4.87E-06 | | -1.77E-05 | | | -0.000106 | | 1.73E-05 | | | | |
| 287 | 4883 | 4.90E-05 | -3.36E-05 | -5.82E-05 | -1.18E-05 | 4.99E-06 | -2.72E-06 | -1.84E-05 | -0.000138 | -0.000123 | -0.000108 | 1.09E-05 | -1.51E-05 | -1.63E-06 | -4.34E-05 | |
| 288 | 4930 | | | | | | | | -0.000136 | | | | | | | | |
| 289 | 4933 | | | | | | | | | | | -0.000107 | | | | | |
| 290 | 4939 | 8.40E-06 | -7.77E-05 | -8.40E-05 | -1.62E-05 | 4.58E-06 | -2.70E-06 | -2.64E-06 | -0.000146 | -0.000135 | -0.000116 | 1.25E-05 | -2.04E-05 | -1.69E-06 | -4.62E-05 | |
| 291 | 4954 | 3.38E-05 | -4.64E-05 | -6.80E-05 | | | | -1.75E-06 | -0.000144 | -0.000133 | -0.000118 | 2.36E-05 | -1.20E-05 | | -4.07E-05 | |
| 292 | 5000 | 2.90E-05 | -5.02E-05 | -7.00E-05 | -1.57E-05 | 4.88E-06 | -2.81E-06 | -1.88E-05 | -0.00015 | -0.000141 | -0.000134 | 1.25E-05 | -1.86E-05 | -1.75E-06 | -5.77E-05 | |
| 293 | 5003 | 2.91E-05 | -5.01E-05 | -7.00E-05 | -1.52E-05 | 8.74E-06 | | -1.84E-05 | -0.00015 | -0.000141 | -0.000134 | 1.96E-06 | -1.70E-05 | -1.53E-06 | -5.74E-05 | |
| 294 | 5011 | 2.69E-05 | -5.39E-05 | -7.08E-05 | -1.49E-05 | 1.00E-05 | | -1.82E-05 | -0.000153 | -0.000144 | -0.000135 | 1.81E-06 | -1.65E-05 | 2.59E-06 | -5.74E-05 | |
| 295 | 5022 | | | -6.45E-05 | -1.08E-05 | | | | -0.000147 | | -0.00013 | 1.09E-05 | -7.29E-06 | 3.15E-06 | -4.86E-05 | |
| 296 | 5023 | 3.31E-05 | -4.84E-05 | -6.45E-05 | -1.06E-05 | 1.24E-05 | -2.53E-06 | -1.38E-05 | -0.000147 | -0.000139 | -0.000129 | 1.12E-05 | -7.22E-06 | | -4.83E-05 | |
| 297 | 5027 | | | -6.84E-05 | | 1.34E-05 | | | | | | | 1.01E-05 | | 3.21E-06 | -4.93E-05 | |
| 298 | 5043 | | | | | | | -1.43E-05 | | | | | | | | | |
| 299 | 5053 | 2.83E-05 | -5.23E-05 | -7.26E-05 | -1.07E-05 | 1.20E-05 | -2.58E-06 | | -0.000143 | -0.000136 | -0.000129 | 1.97E-05 | -7.70E-06 | 2.12E-06 | -4.79E-05 | |
| 300 | 5072 | 2.68E-05 | -5.25E-05 | | -1.16E-05 | 1.26E-05 | | -1.53E-06 | | -0.000138 | | | 1.73E-05 | | 2.24E-06 | -4.99E-05 | |
| 301 | 5087 | 2.82E-05 | -5.12E-05 | -6.99E-05 | -9.74E-06 | 1.31E-05 | | -1.30E-06 | -0.000141 | | -0.000128 | 1.64E-05 | -1.33E-05 | | -5.01E-05 | |
| 302 | 5088 | | | | | | | | -0.000138 | -0.000134 | -0.000123 | | -1.23E-05 | | -4.75E-05 | |
| 303 | 5089 | | | | | | | | | | | | 1.94E-05 | | | | |
| 304 | 5099 | | | -6.56E-05 | | | | | | | | | | | | | |
| 305 | 5105 | 6.39E-05 | -2.67E-05 | | -1.06E-05 | | | -1.38E-05 | -0.000129 | -0.000126 | -0.000122 | | -1.20E-05 | | | |
| 306 | 5131 | 3.16E-05 | -5.15E-05 | -6.99E-05 | -1.61E-05 | 8.14E-06 | -2.68E-06 | -2.08E-05 | | | | 8.31E-06 | | 1.32E-06 | -5.42E-05 | |

| | A | B | C | D | E | F | G | H | I | J | K | L | M | N | O | P |
|---|---|---|---|---|---|---|---|---|---|---|---|---|---|---|---|---|
| 307 | 5133 | 3.31E-05 | -4.97E-05 | -6.62E-05 | -1.56E-05 | 8.87E-06 | -2.67E-06 | -2.05E-05 | -0.000144 | -0.000138 | -0.000125 | 9.99E-06 | -1.50E-05 | 1.30E-06 | -5.45E-05 | |
| 308 | 5136 | | | | -1.49E-05 | 8.76E-06 | | | | -0.000132 | | 2.23E-05 | -1.24E-05 | | -4.73E-05 | |
| 309 | 5159 | 2.45E-05 | -5.53E-05 | -6.83E-05 | | | | -2.09E-05 | -0.000128 | | -0.000119 | 4.65E-05 | | 1.19E-06 | -3.34E-05 | |
| 310 | 5171 | | | | -1.48E-05 | | -2.08E-06 | | -0.000127 | -0.000132 | -0.000119 | 5.37E-05 | -3.87E-06 | | -2.73E-05 | |
| 311 | 5187 | 4.37E-05 | -4.18E-05 | | | | | | | -0.000122 | | | | 1.55E-06 | | |
| 312 | 5190 | | | | | | | | -0.000118 | | | | | | | | |
| 313 | 5226 | | | | | | | | | | | -0.000117 | | | | -2.12E-05 | |
| 314 | 5232 | | | -9.11E-05 | | 9.46E-06 | | | -0.00017 | -0.000148 | | | 2.86E-05 | -8.38E-06 | | | |
| 315 | 5264 | | -0.000117 | | | | | -2.34E-05 | | | -0.000123 | | | | | | |
| 316 | 5277 | -6.34E-05 | | | -1.59E-05 | | | | | | | | | | | | |
| 317 | 5292 | -6.37E-05 | -0.000118 | -9.17E-05 | -1.61E-05 | 9.38E-06 | -1.95E-06 | -2.37E-05 | -0.00017 | -0.000147 | -0.000122 | 3.11E-05 | -7.46E-06 | 2.54E-06 | -2.35E-05 | |
| 318 | 5300 | | | | | | | | | | -0.000127 | | | | | | |
| 319 | 5306 | -6.32E-05 | -0.000117 | -9.11E-05 | -1.56E-05 | 9.74E-06 | | -2.30E-05 | -0.00017 | -0.000148 | -0.000128 | 2.84E-05 | -7.54E-06 | 2.48E-06 | -2.46E-05 | |
| 320 | 5309 | -6.38E-05 | | | -1.57E-05 | 9.73E-06 | -2.02E-06 | | | -0.000149 | | 9.37E-06 | | 2.40E-06 | -2.97E-05 | |
| 321 | 5327 | | | -8.82E-05 | | | | | -0.000168 | | | | | | | | |
| 322 | 5332 | | -0.000107 | | | | | -2.01E-05 | | -0.000142 | -0.000107 | 2.09E-05 | -2.39E-06 | | -1.87E-05 | | |
| 323 | 5355 | -7.32E-05 | -0.000117 | -9.95E-05 | -1.57E-05 | 9.76E-06 | -2.22E-06 | -2.18E-05 | -0.000177 | -0.00015 | -0.00011 | 1.61E-05 | -5.38E-06 | 2.54E-06 | -2.26E-05 | |
| 324 | 5360 | -7.57E-05 | -0.00012 | -0.0001 | -1.58E-05 | 9.63E-06 | | -2.22E-05 | -0.000177 | -0.000151 | -0.00011 | 1.65E-05 | -6.76E-06 | 2.69E-06 | -2.25E-05 | |
| 325 | 5378 | -6.84E-05 | -0.000112 | | | | | -1.70E-05 | | | -0.000108 | 1.66E-05 | -3.66E-06 | 2.52E-06 | -2.36E-05 | | |
| 326 | 5383 | | | -9.54E-05 | -1.28E-05 | 1.11E-05 | | | -0.000183 | -0.000155 | | | | | | | |
| 327 | 5404 | | | | | | -2.17E-06 | | | | -0.00012 | | | | | | |
| 328 | 5421 | -7.22E-05 | -0.000115 | | -1.37E-05 | | | -1.76E-05 | | -0.000156 | | 9.25E-06 | -4.26E-06 | | -2.78E-05 | | |
| 329 | 5444 | | | | | | | | -0.000183 | | | | | | | | |
| 330 | 5472 | | | -9.50E-05 | | 1.30E-05 | | | | | | | | | | | |
| 331 | 5474 | -7.85E-05 | -0.000117 | -9.52E-05 | -1.46E-05 | 1.27E-05 | | -1.94E-05 | -0.0002 | -0.000195 | -0.000145 | 7.06E-06 | -3.53E-06 | | -3.11E-05 | | |
| 332 | 5485 | | | | | | | | | | -0.000147 | | | 1.47E-06 | | | |
| 333 | 5495 | -7.50E-05 | -0.000113 | -9.38E-05 | -1.45E-05 | 1.31E-05 | -1.88E-06 | -1.91E-05 | -0.000205 | -0.000201 | -0.000147 | 6.03E-06 | -5.15E-06 | 1.37E-06 | -3.29E-05 | |
| 334 | 5506 | -7.38E-05 | -0.000112 | -9.35E-05 | -1.47E-05 | | | | -0.000205 | -0.000201 | | | | | | | |
| 335 | 5527 | | | | | | | | | | | -0.000145 | | -3.96E-06 | | | |
| 336 | 5548 | -8.61E-05 | -0.000119 | | -1.09E-05 | 1.40E-05 | | -1.46E-05 | | -0.000199 | | 1.00E-05 | | | -2.99E-05 | | |
| 337 | 5581 | | | | | | | | -0.000205 | | | | | | | | |
| 338 | 5584 | -8.81E-05 | -0.000122 | -9.56E-05 | -1.26E-05 | 1.21E-05 | -3.98E-06 | -1.66E-05 | -0.000205 | -0.0002 | -0.000146 | 8.50E-06 | -6.87E-06 | -3.20E-06 | -3.16E-05 | |
| 339 | 5594 | -0.000115 | -0.000147 | | -1.34E-05 | 9.66E-06 | -4.31E-06 | -1.82E-05 | -0.000205 | -0.0002 | -0.000145 | 2.07E-05 | -5.23E-06 | -3.63E-06 | -2.61E-05 | |
| 340 | 5602 | | | -0.000114 | | | | | | | | 1.30E-05 | -8.19E-06 | | -3.28E-05 | | |
| 341 | 5640 | -0.000114 | -0.000146 | -0.000114 | -1.36E-05 | 1.03E-05 | -4.32E-06 | -1.82E-05 | -0.000208 | -0.000202 | | | | -3.67E-06 | | | |
| 342 | 5652 | -0.000114 | -0.000146 | -0.000114 | -1.33E-05 | 1.05E-05 | -4.31E-06 | -1.78E-05 | -0.000208 | -0.000203 | -0.000143 | 1.19E-05 | -7.57E-06 | -3.93E-06 | -3.29E-05 | |
| 343 | 5653 | -0.000116 | -0.000147 | | -1.58E-05 | 8.61E-06 | -4.32E-06 | -2.03E-05 | -0.000206 | -0.000201 | -0.000139 | 1.84E-05 | -5.16E-06 | -2.87E-06 | -2.94E-05 | |
| 344 | 5677 | | | | | | | | | | | -0.000132 | | | | | |
| 345 | 5707 | | | | | | -4.53E-06 | | | | | | 2.17E-05 | | | -2.48E-05 | |
| 346 | 5763 | | | | | | | | | | -0.000185 | | | | | | |
| 347 | 5802 | -9.31E-05 | -0.000124 | -0.000103 | -1.30E-05 | 1.15E-05 | | -1.63E-05 | -0.000201 | -0.000185 | -0.000135 | 1.72E-05 | -4.11E-06 | -1.15E-06 | -2.85E-05 | |
| 348 | 5803 | -7.78E-05 | -0.00011 | -0.000101 | -1.87E-05 | 1.08E-05 | -4.63E-06 | -2.03E-05 | -0.000196 | -0.000181 | -0.000134 | 1.76E-05 | -4.62E-06 | -1.36E-06 | -2.82E-05 | |
| 349 | 5804 | -5.97E-05 | -8.78E-05 | -7.30E-05 | | | | -1.13E-05 | | -0.000163 | -0.000126 | -2.90E-06 | | -1.32E-06 | -3.24E-05 | |
| 350 | 5814 | | | | -2.84E-05 | 7.51E-06 | | | -0.000176 | | | | -2.19E-05 | -1.65E-06 | | | |
| 351 | 5851 | -4.60E-06 | -5.60E-05 | -9.07E-05 | | | -4.68E-06 | | -0.000112 | -0.000102 | -0.000133 | | | | | | |
| 352 | 5883 | | | | | 6.64E-06 | | -2.08E-05 | | | | -1.32E-05 | -2.58E-05 | | -4.38E-05 | | |
| 353 | 5918 | -9.32E-06 | -5.98E-05 | -9.04E-05 | -3.16E-05 | | | -2.06E-05 | -0.000119 | -0.000107 | -0.000137 | -1.24E-05 | -2.50E-05 | -2.76E-06 | -4.35E-05 | |
| 354 | 5919 | | | | | 7.11E-06 | | | | | | | | | | | |
| 355 | 5927 | -1.05E-05 | -6.15E-05 | -9.07E-05 | | | | -2.09E-05 | | -0.00013 | -0.000118 | -0.000137 | | | | | |
| 356 | 5969 | | | | -3.85E-05 | -3.73E-06 | -5.12E-06 | | | | | -2.28E-05 | -3.68E-05 | -2.67E-06 | -5.17E-05 | | |
| 357 | 5985 | -3.18E-05 | -8.18E-05 | -0.000106 | | | | | | | | | | | | | |
| 358 | 6021 | -3.12E-05 | -8.10E-05 | -0.000105 | -3.83E-05 | -3.23E-06 | | -2.51E-05 | -0.000156 | -0.000143 | -0.000135 | -1.61E-05 | -3.39E-05 | | -4.69E-05 | |
| 359 | 6028 | -5.39E-05 | -0.000104 | -0.000129 | | | | | | | | | | -1.92E-06 | | | |
| 360 | 6037 | | | | | | -5.05E-06 | | | | | | | | | | |
| 361 | 6057 | | | | -4.71E-05 | -8.10E-06 | | | -3.54E-05 | -0.000164 | -0.00015 | -0.000142 | -2.55E-05 | -3.81E-05 | | -5.52E-05 | |
| 362 | 6072 | -6.83E-05 | -0.000117 | -0.000138 | | | -4.83E-06 | -4.16E-06 | -0.000163 | -0.000147 | -0.000141 | -2.52E-05 | -4.91E-05 | -2.31E-06 | -5.69E-05 | |
| 363 | 6083 | | | | -5.25E-05 | -1.41E-05 | -6.28E-06 | | | | | | | -6.48E-06 | | | |
| 364 | 6094 | -7.54E-05 | -0.000123 | -0.000139 | -5.35E-05 | -1.48E-05 | -6.27E-06 | -4.35E-05 | -0.000178 | -0.000162 | -0.000151 | -3.75E-05 | -5.14E-05 | -6.45E-06 | -7.00E-05 | |
| 365 | 6103 | | | | | | | | -0.000187 | -0.000169 | | -3.32E-05 | -4.99E-05 | | -6.77E-05 | | |
| 366 | 6139 | -0.000114 | -0.000174 | -0.000146 | -5.92E-05 | -2.51E-05 | -5.55E-06 | -4.87E-05 | -0.000239 | -0.000237 | -0.000156 | -3.34E-05 | | -5.20E-05 | -6.95E-05 | |
| 367 | 6152 | -0.000119 | -0.000179 | -0.000152 | -5.86E-05 | -2.38E-05 | -5.78E-06 | -4.95E-05 | | -0.00024 | -0.000158 | -3.53E-05 | -5.13E-05 | -5.49E-06 | -7.25E-05 | |
| 368 | 6156 | -0.000123 | -0.000183 | -0.000154 | -6.10E-05 | -2.47E-05 | -5.69E-06 | -5.17E-05 | -0.000242 | -0.000241 | -0.000158 | -3.62E-05 | -5.21E-05 | -2.92E-06 | -7.36E-05 | |
| 369 | 6178 | | | | | | -5.53E-06 | | | | | | | | | | |
| 370 | 6185 | | | | | | | | | | | | | | -2.59E-06 | | |
| 371 | 6199 | -0.000122 | -0.000183 | -0.000154 | -5.48E-05 | -2.23E-05 | | -4.74E-05 | | -0.00024 | -0.000238 | -0.00016 | -3.44E-05 | -5.17E-05 | -2.60E-06 | -7.63E-05 | |
| 372 | 6250 | -8.81E-05 | -0.000151 | -0.000109 | -6.15E-05 | -2.48E-05 | -6.16E-06 | -4.39E-05 | | | | | | | -9.02E-06 | | |
| 373 | 6273 | | | | | | -6.96E-06 | | | -0.000214 | -0.000217 | -0.000141 | | -4.58E-05 | | -6.43E-05 | |
| 374 | 6274 | | | -0.00012 | | -2.64E-05 | -6.93E-06 | -4.68E-05 | | | | | -3.88E-05 | | -9.13E-06 | | |
| 375 | 6278 | -9.69E-05 | -0.000159 | | -6.25E-05 | | | | | | | | | | -9.14E-06 | | |
| 376 | 6303 | | | | | | | | -0.000218 | -0.000226 | | | | -4.73E-05 | | | |
| 377 | 6307 | | | | | | | | | | | -0.000138 | | | | -6.41E-05 | |
| 378 | 6321 | -9.07E-05 | -0.000153 | | | | | | -0.000215 | -0.000222 | -0.000122 | -2.27E-05 | | | | | |
| 379 | 6334 | | | -0.000118 | | | | | | | | -2.96E-05 | -4.63E-05 | | -6.50E-05 | | |
| 380 | 6407 | -0.000104 | -0.000165 | | -6.74E-05 | -3.02E-05 | -7.53E-06 | -5.31E-05 | -0.00022 | -0.000228 | -0.000128 | -3.41E-05 | | -9.38E-06 | -6.90E-05 | |
| 381 | 6484 | -0.000122 | -0.000183 | -0.000133 | -6.84E-05 | -3.00E-05 | -7.42E-06 | -5.47E-05 | -0.000242 | -0.000254 | -0.000175 | -3.98E-05 | -4.77E-05 | -9.42E-06 | -7.65E-05 | |
| 382 | 6523 | -0.000136 | -0.000199 | -0.000137 | -6.89E-05 | -3.08E-05 | -7.40E-06 | -5.67E-05 | -0.000241 | -0.000253 | -0.000173 | -3.71E-05 | -4.64E-05 | -9.51E-06 | -7.41E-05 | |
| 383 | 6566 | -0.000136 | -0.000199 | -0.000136 | -6.79E-05 | -2.98E-05 | | -5.58E-05 | -0.000241 | -0.000253 | -0.000172 | -3.66E-05 | -4.53E-05 | -5.49E-06 | -7.27E-05 | |
| 384 | 6606 | -0.00014 | -0.000202 | -0.000138 | | | -7.44E-06 | | | | | | | | | | |
| 385 | 6608 | | | | | | | | | -0.000235 | -0.000247 | -0.000169 | | -4.40E-05 | | -6.76E-05 | |
| 386 | 6613 | | | | | | | -6.07E-05 | | | | -3.03E-05 | | | | | |
| 387 | 6629 | | | | -8.71E-05 | -3.10E-05 | -8.32E-06 | | | | | | | -7.13E-06 | | | |
| 388 | 6741 | | | | | | | | -0.000248 | -0.000264 | -0.000177 | -2.80E-05 | | | -6.87E-05 | | |
| 389 | 6749 | -0.000111 | -0.000181 | -0.00014 | -8.89E-05 | -3.08E-05 | -8.67E-06 | -6.47E-05 | | | | | -4.52E-05 | -8.98E-06 | | | |
| 390 | 6772 | | | | | | -8.76E-06 | | | | | | | -9.22E-06 | | | |
| 391 | 6790 | -9.84E-05 | -0.000168 | -0.000136 | -8.54E-05 | -2.94E-05 | | -6.20E-05 | -0.000245 | -0.00026 | -0.000177 | -2.45E-05 | -4.22E-05 | | -6.68E-05 | |
| 392 | 6807 | | | | | | | | -0.000241 | -0.000257 | -0.000175 | -1.83E-05 | -3.97E-05 | -8.90E-06 | -5.96E-05 | | |
| 393 | 6811 | | | | | | -8.58E-06 | | | | | | | | | | |
| 394 | 6823 | -0.000124 | -0.000188 | -0.000141 | -8.64E-05 | -2.79E-05 | -8.58E-06 | -6.26E-05 | -0.000255 | -0.000271 | -0.000181 | -2.52E-05 | -4.37E-05 | -9.57E-06 | -6.81E-05 | |
| 395 | 6876 | -0.000128 | -0.000197 | -0.000141 | -8.57E-05 | -2.73E-05 | -8.68E-06 | -6.17E-05 | -0.000259 | -0.000274 | -0.000186 | -2.24E-05 | -4.21E-05 | | -6.58E-05 | | |
| 396 | 6904 | | | | | | -8.88E-06 | | | | | | | | | | |
| 397 | 6940 | -0.000119 | -0.000187 | -0.00014 | -8.46E-05 | -2.34E-05 | | -6.10E-05 | -0.000232 | -0.000251 | -0.000176 | -2.55E-05 | -4.30E-05 | -9.33E-06 | -6.74E-05 | |

| | A | B | C | D | E | F | G | H | I | J | K | L | M | N | O |
|---|---|---|---|---|---|---|---|---|---|---|---|---|---|---|---|
| 1 | Rank | Inf | 0.002 | 0.0015 | 0.001 | 0.0005 | 0.0001 | 0.0013 | 0.0020 Ran | 0.0015 Ran | 0.0010 Ran | 0.0005 Ran | 0.0001 Ran | 0.0013 Random | |
| 2 | 10 | -6.99E-06 | 2.63E-05 | 1.59E-05 | -0.000316 | -0.000178 | 4.91E-07 | -0.000204 | 0.00014 | 0.000156 | 4.87E-05 | -7.66E-05 | -5.78E-05 | 5.62E-06 | -2.44E-05 |
| 3 | 25 | | | | | | | | 0.000149 | 0.000164 | 5.34E-05 | -6.42E-05 | -4.82E-05 | | -1.16E-05 |
| 4 | 61 | 1.07E-05 | 4.63E-05 | 4.13E-05 | -0.000299 | -0.000145 | 4.71E-07 | -0.000182 | 0.000152 | 0.000168 | 5.85E-05 | -6.11E-05 | -4.63E-05 | 6.54E-06 | -8.38E-06 |
| 5 | 93 | | | | | | | | | | 5.96E-05 | | | | |
| 6 | 115 | 1.73E-05 | 5.28E-05 | 4.75E-05 | -0.000287 | -0.000126 | -2.91E-06 | -0.000177 | 0.00015 | 0.000166 | 5.95E-05 | -6.03E-05 | -4.64E-05 | 2.67E-06 | -7.57E-06 |
| 7 | 144 | 3.24E-05 | 6.75E-05 | 6.96E-05 | -0.00027 | -0.00011 | -2.91E-06 | -0.000157 | 0.00017 | 0.000185 | 6.68E-05 | -4.19E-05 | -4.28E-05 | 2.65E-06 | 9.20E-06 |
| 8 | 145 | 3.55E-05 | 6.99E-05 | 7.16E-05 | -0.000267 | -9.86E-05 | | -0.000154 | 0.000214 | 0.000233 | 0.000107 | -1.82E-05 | 1.14E-07 | | 3.38E-05 |
| 9 | 181 | | | | | | | | 0.000209 | 0.000228 | 0.000106 | | -4.77E-06 | | |
| 10 | 197 | | | | | | -2.11E-06 | | | | | | | 3.39E-06 | |
| 11 | 213 | 1.05E-05 | 4.40E-05 | 4.08E-05 | -0.000282 | -0.000117 | -1.65E-06 | -0.000169 | 0.000192 | 0.00021 | 8.79E-05 | -1.82E-05 | -3.40E-05 | | 3.62E-05 |
| 12 | 233 | 1.66E-05 | 5.03E-05 | 4.75E-05 | -0.000279 | -0.000112 | | -0.000165 | | | 8.84E-05 | -1.39E-05 | -1.99E-06 | 3.50E-06 | 3.85E-05 |
| 13 | 234 | | | | | -0.000124 | -1.71E-06 | | 0.000188 | 0.000205 | | | | | |
| 14 | 239 | 1.74E-05 | 5.12E-05 | 4.75E-05 | -0.000279 | | | -0.000166 | | | | -9.53E-06 | 3.94E-06 | | 4.15E-05 |
| 15 | 245 | | | | | | | | 0.000197 | 0.000213 | 0.000104 | | | | |
| 16 | 259 | 1.23E-05 | 4.57E-05 | 4.41E-05 | -0.000271 | -0.000112 | | -0.000153 | | | | -4.05E-06 | 8.41E-06 | | 4.73E-05 |
| 17 | 261 | 1.23E-05 | 4.60E-05 | 4.41E-05 | -0.000271 | -0.000111 | -2.46E-06 | -0.000153 | 0.000174 | 0.000189 | 9.37E-05 | -3.55E-06 | 8.86E-06 | 2.89E-06 | 4.80E-05 |
| 18 | 276 | 1.32E-05 | 4.67E-05 | 4.50E-05 | -0.000271 | -0.000109 | -2.52E-06 | -0.000153 | 0.000173 | 0.000189 | 9.26E-05 | -4.12E-06 | 8.36E-06 | | 4.70E-05 |
| 19 | 284 | 1.39E-05 | 4.69E-05 | 4.52E-05 | -0.000269 | -0.000109 | -2.52E-06 | -0.000152 | 0.000172 | 0.000187 | 9.23E-05 | -4.00E-06 | 8.19E-06 | 2.85E-06 | 4.72E-05 |
| 20 | 334 | 1.56E-05 | 4.83E-05 | 4.79E-05 | -0.000263 | -0.000107 | -1.83E-06 | -0.000147 | 0.000171 | 0.000186 | 9.30E-05 | -2.56E-06 | 8.72E-06 | 3.74E-06 | 4.91E-05 |
| 21 | 344 | 3.53E-05 | 6.78E-05 | | | -1.99E-06 | | | | | -9.50E-06 | 4.20E-06 | 3.68E-06 | 3.91E-05 | |
| 22 | 352 | 2.88E-05 | 6.11E-05 | 3.13E-05 | -0.000104 | | -2.04E-06 | -0.000153 | 0.000161 | 0.000176 | 7.81E-05 | -6.68E-06 | 7.54E-06 | 3.51E-06 | 4.20E-05 |
| 23 | 357 | | | | -0.000265 | -0.000115 | | | | | | 4.31E-06 | | | 3.76E-05 |
| 24 | 366 | 2.83E-05 | 6.10E-05 | 3.06E-05 | -0.00026 | -0.000118 | -2.66E-06 | -0.000151 | 0.000152 | 0.000168 | 6.81E-05 | -1.21E-05 | -4.10E-07 | 3.44E-06 | 3.31E-05 |
| 25 | 375 | | | | -0.000251 | | -3.05E-06 | -0.000145 | | | 7.52E-05 | | -1.40E-06 | 3.34E-06 | 4.03E-05 |
| 26 | 378 | 2.36E-05 | 5.61E-05 | 2.62E-05 | | -0.000125 | | | 0.000155 | 0.00017 | | | | | |
| 27 | 380 | | | | -0.000256 | -0.000124 | | -0.000149 | 0.000155 | 0.00017 | 6.60E-05 | -1.75E-05 | -3.62E-06 | | 3.21E-05 |
| 28 | 390 | 2.96E-05 | 6.22E-05 | 3.42E-05 | | | -1.12E-06 | | | | | -4.02E-06 | 6.44E-06 | 4.39E-06 | 4.38E-05 |
| 29 | 408 | 2.94E-05 | 6.21E-05 | 3.42E-05 | -0.000253 | -0.000122 | -6.45E-07 | -0.000147 | 4.85E-05 | 7.51E-05 | 5.35E-05 | -3.85E-06 | 6.19E-06 | 4.51E-06 | 4.34E-05 |
| 30 | 416 | | | 3.40E-05 | | -0.000122 | -8.14E-07 | -0.000147 | 4.78E-05 | 7.43E-05 | 5.33E-05 | -4.82E-06 | 5.31E-06 | 4.48E-06 | 4.29E-05 |
| 31 | 432 | 2.25E-05 | 5.57E-05 | | -0.000276 | | | | 4.97E-05 | 7.55E-05 | | -8.67E-06 | 9.41E-06 | | |
| 32 | 434 | | | | | -0.000137 | -2.08E-06 | | | | | | | 4.17E-06 | 3.69E-05 |
| 33 | 487 | 2.44E-05 | 5.81E-05 | 3.44E-05 | -0.000271 | -0.000131 | | -0.000147 | 5.65E-05 | 8.13E-05 | 5.98E-05 | -1.69E-06 | 1.74E-05 | | |
| 34 | 496 | | | | | | -1.58E-06 | -0.000147 | | | 5.98E-05 | | | 4.36E-06 | 4.06E-05 |
| 35 | 513 | 2.53E-05 | 5.91E-05 | 3.57E-05 | -0.00027 | -0.000127 | | | 5.95E-05 | 8.43E-05 | | 1.91E-06 | 2.13E-05 | | 4.34E-05 |
| 36 | 526 | | | | | | | -0.000138 | | | | | | | |
| 37 | 530 | 2.74E-05 | 6.11E-05 | 3.85E-05 | -0.000263 | -0.000125 | | -0.000138 | 6.01E-05 | 8.49E-05 | 6.36E-05 | 4.89E-06 | 2.39E-05 | | 4.60E-05 |
| 38 | 540 | 2.17E-05 | 5.52E-05 | 2.99E-05 | -0.000279 | -0.000128 | 1.64E-06 | -0.000153 | 5.66E-05 | 8.16E-05 | 6.28E-05 | 2.91E-06 | 2.33E-05 | 4.62E-06 | 4.40E-05 |
| 39 | 544 | | 5.52E-05 | | | | | -0.000153 | | | | | 2.98E-06 | | |
| 40 | 551 | 2.07E-05 | | | | -0.00013 | | | 5.06E-05 | 7.48E-05 | | | 2.72E-05 | | 4.31E-05 |
| 41 | 556 | | | 2.82E-05 | -0.000286 | | 1.39E-06 | | | | 5.96E-05 | 3.39E-06 | | 5.03E-06 | |
| 42 | 557 | 2.11E-05 | 5.52E-05 | | | -0.000129 | 1.61E-06 | -0.000156 | 4.90E-05 | 7.37E-05 | | | 2.41E-05 | 5.13E-06 | 4.09E-05 |
| 43 | 573 | 2.33E-05 | 5.75E-05 | 3.17E-05 | -0.00028 | -0.000118 | | -0.000153 | 5.65E-05 | 8.09E-05 | 6.68E-05 | 1.12E-05 | 3.49E-05 | | 4.76E-05 |
| 44 | 582 | | | 3.27E-05 | -0.000278 | -0.000117 | | | | | | | | | |
| 45 | 592 | 2.34E-05 | 5.77E-05 | 3.17E-05 | -0.000283 | -0.00012 | | -0.000158 | 5.44E-05 | 7.95E-05 | 6.68E-05 | 1.06E-05 | 3.15E-05 | 4.98E-06 | 4.69E-05 |
| 46 | 594 | 3.01E-05 | 6.50E-05 | 4.47E-05 | -0.000256 | -0.000114 | 2.88E-06 | -0.000132 | 6.32E-05 | 8.84E-05 | 7.60E-05 | 1.78E-05 | 3.23E-05 | | 5.27E-05 |
| 47 | 663 | | | | | | | -0.000143 | | | 7.78E-05 | 1.89E-05 | | | |
| 48 | 716 | 2.52E-05 | 5.99E-05 | 3.99E-05 | -0.000264 | -0.000117 | | -0.000144 | 5.41E-05 | 8.02E-05 | 7.64E-05 | 1.69E-05 | 2.90E-05 | 4.37E-06 | 4.75E-05 |
| 49 | 748 | | 5.98E-05 | 4.04E-05 | -0.000264 | -0.000121 | 1.61E-06 | -0.000143 | 5.03E-05 | | 7.41E-05 | 1.52E-05 | 2.53E-05 | 4.33E-06 | 4.57E-05 |
| 50 | 760 | 2.51E-05 | | | | | | | 5.03E-05 | 7.74E-05 | | | 2.51E-05 | | |
| 51 | 772 | 2.54E-05 | 6.00E-05 | 4.08E-05 | -0.000263 | -0.00012 | | -0.000142 | | 7.82E-05 | 7.46E-05 | 1.64E-05 | 2.85E-05 | 4.40E-06 | 4.71E-05 |
| 52 | 782 | 2.57E-05 | 6.03E-05 | 4.10E-05 | -0.000263 | -0.00012 | 1.59E-07 | -0.000141 | 5.08E-05 | 7.72E-05 | 7.22E-05 | 1.61E-05 | 2.80E-05 | | 4.64E-05 |
| 53 | 813 | 2.78E-05 | 6.29E-05 | 4.23E-05 | -0.000258 | -0.000118 | -6.57E-07 | -0.000136 | 4.53E-05 | 6.96E-05 | 6.99E-05 | 1.13E-05 | 1.50E-05 | 3.28E-06 | 4.22E-05 |
| 54 | 818 | 2.86E-05 | | | | | | | | | 6.93E-05 | 1.01E-05 | | 2.94E-06 | |
| 55 | 826 | 2.86E-05 | 6.33E-05 | 4.24E-05 | -0.000258 | -0.00012 | | -0.000137 | 4.54E-05 | 6.97E-05 | 6.95E-05 | 1.03E-05 | 1.77E-05 | | 4.18E-05 |
| 56 | 844 | | 6.21E-05 | | | | | | | | 6.94E-05 | 1.02E-05 | 1.80E-05 | | 4.17E-05 |
| 57 | 885 | 3.13E-05 | | 4.69E-05 | -0.000246 | -0.000113 | -2.13E-06 | | 5.29E-05 | 7.93E-05 | | | | | |
| 58 | 917 | 3.11E-05 | 6.64E-05 | | | | | -0.00013 | | | 7.29E-05 | 1.15E-05 | | 2.30E-06 | |
| 59 | 925 | | | | | | | | | | | | 1.61E-05 | | |
| 60 | 931 | 2.69E-05 | 6.03E-05 | 4.57E-05 | -0.000239 | -0.000109 | -2.29E-06 | -0.00013 | 5.71E-05 | 8.05E-05 | 7.53E-05 | 2.66E-06 | 2.01E-05 | | 5.64E-05 |
| 61 | 933 | 2.20E-05 | 5.08E-05 | 3.89E-05 | -0.000244 | -0.000109 | -2.35E-06 | -0.000138 | 3.41E-05 | 5.64E-05 | 6.07E-05 | 1.52E-05 | 7.42E-06 | 2.29E-06 | 4.36E-05 |
| 62 | 937 | | | | | | | | | | | | | 2.49E-08 | |
| 63 | 962 | | | | | -0.00011 | | | | | | | | | |
| 64 | 975 | | | | | | | | 4.04E-05 | 6.29E-05 | | | | | |
| 65 | 991 | 1.98E-05 | 4.75E-05 | | | | | | | | | | 7.57E-06 | | |
| 66 | 997 | | | 3.25E-05 | -0.00025 | -0.000112 | -4.91E-06 | -0.000144 | 3.64E-05 | 5.92E-05 | 5.01E-05 | 2.23E-06 | 5.96E-06 | -6.96E-07 | 3.16E-05 |
| 67 | 1043 | 1.62E-05 | 4.36E-05 | 2.70E-05 | -0.000258 | -0.000113 | | -0.000152 | 2.32E-05 | 4.70E-05 | 4.80E-05 | 5.26E-06 | 9.14E-06 | | 3.28E-05 |
| 68 | 1063 | 1.62E-05 | 4.36E-05 | | | -4.02E-06 | | | | | | | | | 3.26E-05 |
| 69 | 1093 | | 4.10E-05 | 2.49E-05 | -0.00026 | -0.000118 | -4.04E-06 | -0.000155 | 2.35E-05 | 4.78E-05 | 4.80E-05 | 5.31E-06 | 9.02E-06 | -4.52E-07 | 3.03E-05 |
| 70 | 1094 | 1.35E-05 | | | | | | | | 4.49E-05 | | | | | |
| 71 | 1107 | | | 2.52E-05 | -0.000258 | -0.000114 | | -0.000154 | 2.39E-05 | | 4.79E-05 | 5.85E-06 | | -2.51E-07 | |
| 72 | 1114 | | | | | | -4.46E-06 | | 2.40E-05 | 4.50E-05 | | | 6.45E-06 | -5.07E-07 | 3.03E-05 |
| 73 | 1165 | 1.48E-05 | 4.25E-05 | 2.79E-05 | -0.000253 | -0.00011 | -4.63E-06 | -0.000149 | 2.02E-05 | 4.19E-05 | 4.17E-05 | 5.41E-06 | 5.74E-06 | -5.49E-07 | 2.87E-05 |
| 74 | 1174 | | | | | | | | 8.00E-05 | 9.57E-05 | | | | -5.88E-07 | 2.86E-05 |
| 75 | 1206 | 1.54E-05 | 4.26E-05 | | | -0.00011 | | | | | 4.49E-05 | | 1.28E-05 | -5.88E-07 | |
| 76 | 1211 | 1.46E-05 | 4.18E-05 | 2.44E-05 | -0.000251 | -0.00011 | -4.59E-06 | -0.000152 | 6.09E-05 | 7.83E-05 | 4.46E-05 | 1.12E-05 | 1.39E-05 | -5.37E-07 | 3.19E-05 |
| 77 | 1240 | | | | | | | | 3.89E-05 | 5.09E-05 | | | 1.25E-05 | | 3.14E-05 |
| 78 | 1250 | 1.61E-05 | 4.30E-05 | 2.72E-05 | -0.000244 | -0.000105 | | -0.000149 | 3.93E-05 | 5.17E-05 | 3.81E-05 | 1.52E-05 | 1.94E-05 | | 3.28E-05 |
| 79 | 1260 | | | | | | -3.94E-06 | -0.000151 | | | | | 1.93E-05 | | |
| 80 | 1262 | 1.20E-05 | 3.93E-05 | 2.88E-05 | -0.00024 | -9.83E-05 | | -0.000148 | 3.77E-05 | 4.53E-05 | 3.85E-05 | 1.69E-05 | 1.95E-05 | -1.07E-07 | 3.33E-05 |
| 81 | 1294 | 1.26E-05 | 3.98E-05 | 2.91E-05 | -0.000237 | -9.74E-05 | -3.22E-06 | -0.000148 | 3.78E-05 | 4.57E-05 | 3.87E-05 | 1.86E-05 | 2.00E-05 | -2.04E-08 | 3.47E-05 |
| 82 | 1301 | 1.31E-05 | 4.03E-05 | 2.93E-05 | -0.000235 | -8.02E-05 | | -0.000147 | 3.80E-05 | 4.59E-05 | | 1.96E-05 | | | 3.52E-05 |
| 83 | 1306 | 1.30E-05 | 4.02E-05 | 2.92E-05 | -0.000236 | -8.04E-05 | -3.04E-06 | -0.000147 | 3.80E-05 | 4.57E-05 | 3.71E-05 | 1.93E-05 | 1.97E-05 | -8.35E-10 | 3.50E-05 |
| 84 | 1316 | 1.28E-05 | | | -0.000237 | | | | | | | | | | |
| 85 | 1404 | | | 3.11E-05 | | | | -0.000141 | | 4.85E-05 | | 2.08E-05 | | | 3.69E-05 |
| 86 | 1414 | 1.12E-06 | 2.68E-05 | 2.43E-05 | -0.000253 | -0.000123 | -3.23E-06 | -0.000155 | 2.63E-05 | 4.32E-05 | 2.85E-05 | 1.37E-05 | 1.59E-05 | 2.34E-08 | 2.87E-05 |
| 87 | 1427 | | | | | | | | | | | | 1.94E-05 | 1.70E-05 | |
| 88 | 1437 | 1.91E-06 | 2.81E-05 | 2.54E-05 | -0.000245 | -9.89E-05 | -2.22E-06 | -0.000149 | 2.89E-05 | 4.41E-05 | 3.01E-05 | 1.99E-05 | | 2.89E-07 | 3.25E-05 |
| 89 | 1441 | -7.47E-06 | 1.21E-05 | -8.79E-06 | -0.000269 | -0.000104 | | -0.000204 | 2.43E-05 | 3.81E-05 | 2.58E-05 | 1.53E-05 | 1.18E-05 | | 2.71E-05 |
| 90 | 1448 | -8.69E-06 | 1.21E-05 | -9.50E-06 | -0.000271 | -0.000102 | -2.64E-06 | -0.000148 | 2.59E-05 | 4.00E-05 | 2.77E-05 | 1.81E-05 | 1.53E-05 | 1.57E-07 | 2.87E-05 |
| 91 | 1485 | -1.13E-05 | 9.55E-06 | -1.17E-05 | -0.000274 | -0.000106 | -1.93E-06 | -0.000212 | 2.34E-05 | 3.73E-05 | 2.67E-05 | 1.65E-05 | 1.54E-05 | | 2.67E-05 |
| 92 | 1501 | | | | -0.000284 | | -8.72E-07 | | 2.01E-05 | 3.34E-05 | 2.59E-05 | 9.41E-06 | 5.06E-06 | 1.06E-06 | 2.23E-05 |
| 93 | 1504 | -4.58E-06 | 1.24E-05 | -8.47E-06 | | -9.97E-05 | | -0.000204 | 2.87E-05 | 4.01E-05 | | | | | |
| 94 | 1535 | 2.45E-06 | 2.00E-05 | 6.35E-06 | -0.000177 | -8.31E-05 | | -8.81E-05 | 3.27E-05 | 4.34E-05 | 3.15E-05 | 2.37E-05 | 1.99E-05 | 9.66E-07 | 4.03E-05 |
| 95 | 1542 | | | | | | | | | | | | | | 4.00E-05 |
| 96 | 1576 | | | | -0.000172 | -7.63E-05 | -1.59E-07 | | | | | | 2.07E-05 | | |
| 97 | 1579 | 1.89E-06 | 2.00E-05 | | | -6.94E-08 | -7.54E-05 | 2.94E-05 | 4.10E-05 | 2.84E-05 | | | 1.05E-06 | | |
| 98 | 1617 | | | 8.90E-06 | | | | | | | | 2.20E-05 | | | |
| 99 | 1654 | -2.61E-06 | 1.36E-05 | -1.51E-06 | -0.000183 | -8.09E-05 | 1.74E-07 | -8.59E-05 | 2.70E-05 | 3.83E-05 | 2.51E-05 | 2.14E-05 | 1.98E-05 | 1.02E-06 | 3.32E-05 |
| 100 | 1655 | 9.92E-08 | 1.63E-05 | 3.07E-06 | -0.000174 | -7.92E-05 | 6.75E-07 | -8.09E-05 | 2.53E-05 | 3.70E-05 | 2.45E-05 | 2.16E-05 | 2.03E-05 | 1.37E-06 | 3.30E-05 |
| 101 | 1667 | 4.31E-07 | 1.66E-05 | 3.29E-06 | -0.000172 | -7.77E-05 | 1.07E-06 | -7.94E-05 | 2.52E-05 | 3.69E-05 | 2.46E-05 | 2.30E-05 | 2.05E-05 | 1.50E-06 | 3.35E-05 |
| 102 | 1683 | | | | | | -1.83E-06 | | | | 1.74E-05 | 8.73E-06 | 1.29E-05 | 5.68E-07 | 1.86E-05 |

|  | A | B | C | D | E | F | G | H | I | J | K | L | M | N | O |
|---|---|---|---|---|---|---|---|---|---|---|---|---|---|---|---|
| 103 | 1686 | -3.75E-06 | 1.13E-05 | -1.34E-06 | -0.000157 | -7.37E-05 | -1.71E-06 | -7.28E-05 | 1.51E-05 | 2.44E-05 | 7.06E-06 | 1.34E-05 | 1.54E-05 | 5.93E-07 | 2.09E-05 |
| 104 | 1708 | -4.03E-06 | 1.09E-05 | -1.25E-05 | -0.000156 | -7.44E-05 | -2.01E-06 | -7.32E-05 | 1.64E-05 | 2.57E-05 | 7.67E-06 | 1.47E-05 |  |  | 2.15E-05 |
| 105 | 1720 |  |  |  | -0.000159 | -7.57E-05 |  |  | 1.55E-05 | 2.47E-05 | 7.62E-06 | 1.45E-05 | 1.55E-05 |  |  |
| 106 | 1725 | 7.21E-06 | 1.96E-05 | 1.54E-05 | -0.000106 | -5.11E-05 |  | -2.94E-05 | 3.08E-05 | 3.86E-05 |  | 3.75E-05 | 2.32E-05 | 6.23E-05 | 3.88E-05 |
| 107 | 1734 |  |  |  | -0.00017 |  | -2.19E-06 |  |  |  | 1.61E-05 |  |  |  | 3.38E-05 |
| 108 | 1748 | 7.51E-06 | 2.00E-05 | 1.64E-05 |  |  |  | -7.47E-05 |  |  |  |  |  |  |  |
| 109 | 1757 |  |  |  | -0.00016 | -4.91E-05 |  |  | 2.57E-05 | 3.62E-05 | 1.85E-05 | 2.92E-05 | 2.19E-05 |  |  |
| 110 | 1758 | 4.37E-06 | 1.65E-05 | 1.34E-05 | -0.000161 | -5.10E-05 | -5.40E-06 | -8.23E-05 | 2.29E-05 | 3.41E-05 | 1.76E-05 | 2.68E-05 | 2.03E-05 | 2.80E-07 | 2.30E-05 |
| 111 | 1759 | 6.66E-06 | 1.81E-05 | 1.48E-05 |  |  |  | -7.96E-05 | 4.31E-05 | 5.02E-05 |  |  |  |  |  |
| 112 | 1781 |  |  |  |  | -3.89E-05 |  |  |  |  | 2.23E-05 | 3.62E-05 | 2.67E-05 |  | 3.11E-05 |
| 113 | 1787 | 3.30E-06 | 1.59E-05 | 1.30E-05 | -0.000156 | -3.89E-05 | -5.79E-06 | -8.32E-05 | 3.46E-05 | 4.44E-05 | 2.23E-05 | 3.62E-05 | 2.67E-05 | 8.62E-08 |  |
| 114 | 1791 | 3.02E-06 | 1.59E-05 | 1.23E-05 | -0.000158 | -3.80E-05 | -5.79E-06 | -8.70E-05 | 3.48E-05 | 4.44E-05 | 2.23E-05 | 3.59E-05 | 2.69E-05 |  | 3.12E-05 |
| 115 | 1816 |  |  |  |  |  |  |  |  |  |  |  | 2.68E-05 |  |  |
| 116 | 1851 |  |  |  | -0.000164 |  | -5.57E-06 |  |  |  |  |  |  | 2.26E-07 |  |
| 117 | 1909 |  |  |  |  | -4.24E-05 | -5.43E-06 | -9.41E-05 | 3.51E-05 | 4.49E-05 | 2.41E-05 | 3.23E-05 | 2.46E-05 | 2.22E-07 | 2.76E-05 |
| 118 | 1928 | 2.65E-05 | 4.08E-05 | 2.77E-05 |  |  |  |  |  |  |  |  |  |  |  |
| 119 | 1942 | 9.78E-06 | 2.13E-05 | 6.10E-06 | -0.000145 | -3.76E-05 |  | -8.08E-05 | 3.24E-05 | 4.04E-05 | 2.14E-05 | 3.48E-05 | 2.62E-05 | 7.66E-08 | 2.92E-05 |
| 120 | 1956 |  | 2.17E-05 |  | -0.000145 | -3.78E-05 |  |  | 3.25E-05 | 4.05E-05 | 2.15E-05 | 3.49E-05 | 2.63E-05 |  |  |
| 121 | 1992 | 9.52E-06 |  | 7.65E-06 | -0.000134 | -2.77E-05 |  | -7.06E-05 |  |  |  |  |  |  | 3.24E-05 |
| 122 | 2018 | 9.32E-06 | 2.25E-05 | 7.72E-06 | -0.000133 | -2.77E-05 |  | -6.97E-05 | 3.20E-05 | 3.99E-05 | 2.04E-05 | 3.86E-05 | 2.88E-05 |  | 3.25E-05 |
| 123 | 2027 |  | 2.23E-05 |  |  |  |  |  |  |  |  |  |  |  |  |
| 124 | 2042 | 1.13E-05 |  | 1.28E-05 |  | -2.58E-05 | -7.00E-06 | -6.19E-05 |  | 8.37E-05 |  |  |  | -3.76E-07 |  |
| 125 | 2047 |  |  |  | -0.000124 |  |  |  | 4.70E-05 |  | 4.17E-05 | 3.87E-05 | 2.87E-05 |  |  |
| 126 | 2049 |  | 2.24E-05 |  |  |  |  |  |  |  |  |  |  |  | 3.23E-05 |
| 127 | 2076 | 1.02E-05 |  | 1.29E-05 |  | -2.11E-05 |  | -6.02E-05 | 5.07E-05 | 8.45E-05 |  |  |  |  |  |
| 128 | 2080 | 1.08E-05 | 2.36E-05 | 1.45E-05 | -0.000103 | -2.07E-05 |  | -5.45E-05 |  |  | 4.09E-05 | 4.15E-05 | 3.05E-05 |  | 3.30E-05 |
| 129 | 2085 | 1.07E-05 | 2.34E-05 | 1.43E-05 | -0.000103 | -2.07E-05 | -6.07E-06 | -5.46E-05 | 5.20E-05 | 8.47E-05 | 4.08E-05 | 4.14E-05 | 3.04E-05 | -3.92E-07 | 3.27E-05 |
| 130 | 2096 | 1.13E-05 |  | 1.53E-05 | -9.77E-05 | -1.62E-05 |  | -5.01E-05 | 5.26E-05 | 8.52E-05 | 4.10E-05 |  | 3.17E-05 |  | 3.40E-05 |
| 131 | 2105 | 1.16E-05 | 2.46E-05 | 1.54E-05 |  | -1.64E-05 |  | -5.02E-05 |  |  |  |  |  |  |  |
| 132 | 2115 |  |  |  |  |  |  |  |  |  |  | 3.49E-05 | 3.09E-05 |  | 3.27E-05 |
| 133 | 2116 | 1.12E-05 | 2.43E-05 | 1.58E-05 | -9.63E-05 | -1.61E-05 | -5.95E-06 | -4.87E-05 | 4.71E-05 | 7.98E-05 | 3.89E-05 | 3.49E-05 | 3.08E-05 | -4.28E-07 | 3.27E-05 |
| 134 | 2127 | 1.02E-05 | 2.31E-05 | 1.50E-05 | -9.74E-05 | -3.74E-05 | -6.10E-06 | -5.01E-05 | 4.32E-05 | 7.65E-05 | 3.36E-05 | 3.36E-05 | 2.88E-05 | -4.22E-07 | 3.18E-05 |
| 135 | 2128 |  |  |  |  |  |  |  | 4.23E-05 |  | 3.54E-05 | 3.27E-05 | 2.82E-05 | -6.09E-07 |  |
| 136 | 2148 | 1.20E-05 | 2.55E-05 | 1.99E-05 |  |  | -5.03E-06 | -3.01E-05 |  |  |  |  |  |  | 3.39E-05 |
| 137 | 2156 | 1.18E-05 | 2.55E-05 | 1.98E-05 | -7.82E-05 | -2.76E-05 | -4.70E-06 | -3.03E-05 | 4.38E-05 | 7.87E-05 | 3.85E-05 | 3.52E-05 | 2.78E-05 | -2.63E-07 | 3.43E-05 |
| 138 | 2175 | 1.17E-05 | 2.57E-05 | 1.96E-05 |  | -2.29E-05 |  | -3.40E-05 |  |  |  |  |  |  |  |
| 139 | 2185 |  |  |  | -7.29E-05 |  |  |  | 9.63E-05 | 0.000134 | 0.00011 | 3.57E-05 |  | -4.04E-07 |  |
| 140 | 2226 | 1.80E-05 | 3.12E-05 | 2.73E-05 |  | -2.23E-05 | -7.14E-06 | -2.37E-05 | 8.83E-05 | 0.000127 | 0.000102 | 3.54E-05 | 2.73E-05 | -1.08E-06 | 4.13E-05 |
| 141 | 2244 | 1.85E-05 | 3.27E-05 | 3.13E-05 | -3.71E-05 | 4.50E-05 | -4.97E-06 | 6.70E-05 |  |  |  | 3.07E-05 |  |  | 6.23E-05 |
| 142 | 2261 |  |  |  |  |  |  |  | 8.20E-05 | 0.00012 | 9.39E-05 |  |  | -1.51E-06 |  |
| 143 | 2272 |  |  |  | -4.43E-05 |  |  |  |  |  |  | 1.92E-05 |  |  |  |
| 144 | 2357 | 1.84E-05 | 3.23E-05 | 3.05E-05 | -4.50E-05 | 2.26E-05 | -5.55E-06 | 5.51E-05 | 8.25E-05 | 0.00012 | 9.36E-05 | 1.90E-05 | 2.92E-05 | -1.68E-06 | 5.89E-05 |
| 145 | 2365 | 1.95E-05 | 3.35E-05 | 3.12E-05 |  |  | 5.89E-06 |  |  | 0.000115 |  |  |  |  |  |
| 146 | 2369 |  |  |  |  | 4.21E-06 |  |  | 7.77E-05 |  | 8.87E-05 |  |  | -1.81E-06 |  |
| 147 | 2376 | 2.04E-05 | 3.43E-05 | 3.26E-05 | -4.59E-05 | 4.23E-05 | -6.14E-06 | 8.27E-05 | 7.70E-05 | 0.000114 | 8.85E-05 | 2.03E-05 | 2.99E-05 |  | 5.99E-05 |
| 148 | 2388 |  |  |  |  |  |  |  |  |  |  |  |  |  | 6.03E-05 |
| 149 | 2394 |  | 3.38E-05 |  |  | 3.57E-06 |  | 6.22E-06 |  |  |  |  | 3.04E-05 |  |  |
| 150 | 2415 | 2.34E-05 | 3.63E-05 | 3.85E-05 | -3.23E-05 | 6.28E-06 |  | 1.49E-05 | 7.66E-05 | 0.000114 | 8.74E-05 | 2.06E-05 | 3.02E-05 | -1.52E-06 | 5.97E-05 |
| 151 | 2433 |  |  |  |  |  |  |  | 7.67E-05 |  | 8.73E-05 |  |  |  |  |
| 152 | 2446 |  |  |  | -9.32E-06 |  | -6.15E-06 |  |  |  |  |  |  | -1.56E-06 |  |
| 153 | 2448 |  | 3.63E-05 |  |  | 7.94E-06 |  |  |  |  |  | 1.70E-05 |  |  | 5.11E-05 |
| 154 | 2449 | 2.70E-05 | 3.68E-05 | 4.57E-05 | 3.50E-05 |  | -5.94E-06 | 4.69E-06 | 7.84E-05 | 0.000107 | 9.37E-05 |  | 3.43E-05 |  |  |
| 155 | 2451 | 2.60E-05 | 3.60E-05 | 4.47E-05 | 3.29E-05 | 9.21E-06 | -5.95E-06 | 4.70E-05 | 7.78E-05 | 0.000106 | 9.24E-05 | 2.26E-05 | 3.39E-05 | -1.49E-06 | 5.49E-05 |
| 156 | 2539 | 1.26E-05 | 1.48E-05 | 2.18E-05 | -4.65E-06 | 5.69E-06 | -7.38E-06 | -1.20E-05 | 4.57E-05 | 5.40E-05 | 4.36E-05 | -1.81E-06 | 2.85E-05 | -2.11E-06 | 1.54E-05 |
| 157 | 2552 |  |  |  |  |  |  |  |  | 5.55E-05 | 4.36E-05 |  |  |  |  |
| 158 | 2570 | 1.34E-05 | 1.55E-05 | 2.55E-05 |  |  | -7.53E-06 | -1.18E-06 | 4.66E-05 |  |  | -1.36E-05 | 2.87E-05 | -2.21E-06 | 1.80E-05 |
| 159 | 2585 |  |  |  | 5.22E-06 | 1.21E-05 |  |  | 4.66E-05 | 5.74E-05 | 4.79E-05 |  | 2.93E-05 |  |  |
| 160 | 2587 | 1.37E-05 | 1.59E-05 | 2.57E-05 |  |  |  | -9.98E-07 |  |  |  | -1.20E-05 |  |  |  |
| 161 | 2594 | 1.30E-05 | 1.37E-05 | 2.56E-05 | 1.59E-05 | 1.50E-05 | -7.27E-06 | 6.97E-06 | 4.15E-05 | 5.24E-05 | 4.75E-05 |  | 2.94E-05 | -2.08E-06 | 2.07E-05 |
| 162 | 2678 |  |  |  |  |  |  |  |  |  |  | -1.62E-05 |  |  |  |
| 163 | 2696 |  | 2.25E-05 |  |  |  |  |  |  |  |  |  |  |  | 2.10E-05 |
| 164 | 2704 | 1.73E-05 |  | 3.35E-05 |  |  |  | 7.06E-06 |  |  |  |  |  |  |  |
| 165 | 2717 | 1.56E-05 | 2.07E-05 |  | 1.31E-06 | 2.87E-06 | -7.12E-06 |  | 3.98E-05 | 5.12E-05 | 4.57E-05 | -1.94E-05 | 2.75E-05 | -2.09E-06 | 1.63E-05 |
| 166 | 2769 |  |  | 3.21E-05 |  |  |  | -1.26E-06 | 3.74E-05 | 4.73E-05 |  |  |  |  |  |
| 167 | 2803 |  |  |  |  | -1.04E-06 |  |  |  |  | 4.10E-05 |  | 2.66E-05 |  |  |
| 168 | 2808 |  |  |  |  |  |  |  |  |  |  | -1.66E-05 |  |  | 1.54E-05 |
| 169 | 2831 | 7.75E-06 | 4.82E-06 | 3.18E-05 | 1.87E-05 | 2.16E-05 |  | 6.44E-06 | 4.01E-05 | 4.95E-05 | 4.42E-05 | -6.28E-06 | 3.04E-05 |  |  |
| 170 | 2856 | 4.88E-06 | 2.27E-06 | 2.88E-05 |  | 1.24E-05 | 1.95E-07 | 1.64E-05 | 3.76E-05 | 4.71E-05 | 4.01E-05 |  | 2.91E-05 | -1.02E-07 | 1.52E-05 |
| 171 | 2929 |  |  |  | 1.20E-05 |  | 1.52E-07 |  | 3.70E-05 | 4.61E-05 | 3.87E-05 | -1.02E-05 | 2.88E-05 | -1.28E-07 | 1.47E-05 |
| 172 | 2964 | 2.13E-05 | 6.26E-07 | 2.30E-05 |  | 1.98E-05 | 7.63E-08 | 1.25E-06 | 3.58E-05 | 4.42E-05 | 3.75E-05 | -4.63E-06 | 2.91E-05 | -1.32E-07 | 1.59E-05 |
| 173 | 2975 |  |  |  | 2.11E-05 | 2.12E-05 |  |  |  |  |  |  |  |  |  |
| 174 | 2980 | 3.72E-06 | 3.44E-06 | 2.23E-05 |  |  |  | 2.85E-07 | -5.03E-06 |  |  | -5.94E-06 |  |  | 1.44E-05 |
| 175 | 2982 | 3.66E-06 | 3.46E-06 | 2.21E-05 | 2.08E-05 | 2.03E-05 | 7.51E-07 | -4.62E-06 | 3.77E-05 | 4.66E-05 | 3.83E-05 | -4.21E-06 | 2.92E-05 | 4.81E-07 | 1.51E-05 |
| 176 | 2984 | -9.19E-06 | -9.50E-06 |  |  |  |  |  |  |  |  | -1.05E-05 |  |  | 1.06E-05 |
| 177 | 3004 | -1.10E-05 | -1.04E-05 | 1.58E-05 | 9.04E-06 | 6.82E-06 | 1.79E-06 | -1.28E-05 | 2.68E-05 | 3.67E-05 | 3.25E-05 | -1.13E-05 | 2.94E-05 | 7.03E-07 | 8.09E-06 |
| 178 | 3011 |  |  | 1.57E-05 |  |  | 1.62E-06 | -1.26E-05 | 2.76E-05 | 3.77E-05 | 3.26E-05 |  | 2.96E-05 |  |  |
| 179 | 3014 | -1.23E-05 | -1.21E-05 | 1.63E-05 | 1.36E-05 | 8.69E-06 |  | -9.15E-06 |  |  |  |  |  |  |  |
| 180 | 3061 | -1.03E-05 | -9.33E-06 |  |  | 8.44E-06 | 9.48E-06 |  |  | 3.71E-05 | 4.63E-05 | 3.59E-05 | -1.40E-05 | 2.98E-05 |  | 6.67E-06 |
| 181 | 3085 |  |  | 1.97E-05 |  |  |  | -1.56E-06 |  |  |  |  |  | 1.09E-06 |  |
| 182 | 3120 |  | -9.00E-06 |  |  |  |  |  |  |  |  | -1.53E-05 |  |  | 5.36E-06 |
| 183 | 3143 | -7.56E-06 |  |  | 6.01E-06 | 7.58E-06 |  |  | 3.84E-05 | 4.71E-05 | 3.66E-05 |  | 3.02E-05 |  |  |
| 184 | 3159 |  |  | 2.21E-05 |  |  |  |  |  |  |  |  |  |  |  |
| 185 | 3199 | -3.54E-06 | -1.72E-06 | 3.49E-05 | 4.21E-05 | 2.16E-05 | 1.49E-06 | 1.16E-05 |  |  |  | -1.21E-05 |  |  | 7.75E-06 |
| 186 | 3214 |  |  |  |  |  |  | 1.16E-05 |  | 4.70E-05 |  |  |  |  |  |
| 187 | 3225 |  |  |  |  |  |  |  | 4.00E-05 |  | 4.51E-05 |  |  |  |  |
| 188 | 3240 | -1.64E-05 | -1.47E-05 | 1.98E-05 | 3.97E-05 | -1.96E-05 |  | 3.46E-06 | 3.63E-05 | 4.20E-05 | 4.47E-05 | -1.62E-05 | 2.72E-05 |  | 6.89E-06 |
| 189 | 3258 | -1.44E-05 | -1.14E-05 |  | 4.44E-05 | -2.54E-07 |  |  |  |  |  |  |  |  |  |
| 190 | 3315 |  |  | 2.21E-05 |  |  | 1.50E-06 | 4.76E-06 |  | 8.19E-05 |  | -1.47E-05 |  | 7.59E-07 | 1.11E-05 |
| 191 | 3341 |  |  |  |  |  |  |  | 0.000111 |  | 6.93E-05 |  |  |  |  |
| 192 | 3366 | -7.97E-06 | -4.20E-06 |  |  | -9.97E-07 | 1.58E-06 |  |  |  |  |  |  | 7.03E-07 |  |
| 193 | 3370 |  |  | 1.90E-05 | 5.23E-05 |  |  |  | 9.01E-05 | 4.79E-05 |  |  | 2.64E-05 |  |  |
| 194 | 3380 | -8.37E-06 | -4.74E-06 | 1.86E-05 | 5.23E-05 | 7.62E-07 | 1.55E-06 | 9.04E-06 | 7.92E-05 | 3.98E-05 | 6.21E-05 | -1.35E-05 | 2.63E-05 | 6.86E-07 | 1.12E-05 |
| 195 | 3384 |  |  |  |  |  |  |  |  |  | 4.01E-05 |  |  |  |  |
| 196 | 3410 | -8.92E-06 | -4.56E-06 |  |  | 4.92E-06 |  | 7.19E-06 | 7.29E-05 |  | 6.18E-05 | -1.15E-05 |  |  | 1.04E-05 |
| 197 | 3437 |  |  | 1.69E-05 | 5.18E-05 |  |  |  |  |  |  |  |  |  |  |
| 198 | 3470 | -9.05E-06 | -4.87E-06 |  |  | 4.46E-06 | 1.60E-06 | 7.90E-06 | 6.93E-05 | 3.49E-05 | 5.93E-05 | -1.51E-05 | 2.69E-05 |  | 6.65E-06 |
| 199 | 3505 |  |  | 1.93E-05 | 6.28E-05 |  |  | 1.37E-05 |  |  |  |  |  | -1.03E-07 |  |
| 200 | 3516 |  | -4.42E-06 |  |  | 5.58E-06 |  |  |  |  | 3.40E-05 |  | 2.66E-05 |  |  |
| 201 | 3564 | -1.17E-05 |  |  |  |  |  |  | 6.79E-05 |  |  |  |  |  |  |
| 202 | 3576 |  |  | 9.22E-06 | 5.11E-05 |  | 4.50E-07 |  |  |  | 5.44E-05 | -1.98E-05 |  |  | 2.05E-06 |
| 203 | 3578 |  |  |  |  |  |  |  |  |  |  |  |  | -3.86E-07 |  |
| 204 | 3601 | -1.15E-05 | -5.25E-06 | 9.49E-06 | 5.21E-05 | 6.19E-06 | 5.88E-07 | 1.38E-05 | 6.73E-05 | 3.24E-05 | 5.46E-05 | -1.58E-05 | 2.72E-05 | -2.96E-07 | 2.53E-06 |

| | A | B | C | D | E | F | G | H | I | J | K | L | M | N | O |
|---|---|---|---|---|---|---|---|---|---|---|---|---|---|---|---|
| 205 | 3615 | | -5.12E-06 | | | 7.12E-06 | | 1.30E-05 | | 3.19E-05 | | | | -3.07E-07 | |
| 206 | 3633 | -6.58E-06 | | 1.66E-06 | 7.12E-06 | | | | 7.65E-05 | | 5.61E-05 | -1.40E-05 | 2.75E-05 | | 3.90E-06 |
| 207 | 3671 | -1.26E-05 | -9.15E-06 | 8.60E-06 | 7.34E-05 | -3.36E-06 | | 5.22E-06 | 6.33E-05 | 2.50E-05 | | | 3.01E-05 | | |
| 208 | 3682 | | | | | -8.61E-06 | 5.14E-07 | | 6.03E-05 | 2.26E-05 | 4.80E-05 | -2.30E-05 | 2.26E-05 | -3.87E-07 | 2.95E-08 |
| 209 | 3701 | | -9.12E-06 | | | | | 1.68E-06 | | | | | | | |
| 210 | 3704 | -1.34E-05 | | | | | | | | | | | 2.48E-05 | | |
| 211 | 3745 | | | | | | | | | 3.30E-05 | | | | | |
| 212 | 3759 | | -3.15E-05 | -2.34E-05 | 6.41E-05 | -1.25E-05 | | -3.88E-06 | 4.92E-05 | | 3.63E-05 | -2.52E-05 | | | -1.20E-05 |
| 213 | 3766 | -3.51E-05 | | | | | 1.79E-07 | | | | | | 1.66E-05 | | |
| 214 | 3776 | | | -2.32E-05 | 7.08E-05 | -8.35E-06 | | | 4.90E-05 | 3.02E-05 | 3.69E-05 | -2.38E-05 | | | -1.07E-05 |
| 215 | 3790 | | -3.18E-05 | | | | | | | | | | 1.74E-05 | | |
| 216 | 3840 | | | | | | | | 5.84E-06 | | | | | -3.37E-07 | |
| 217 | 3862 | -1.86E-06 | | | | | | | | | | | | | |
| 218 | 3881 | | | | | | | | | | 3.18E-05 | | | | | |
| 219 | 3886 | -1.78E-06 | -2.99E-05 | -2.02E-05 | 7.20E-05 | -8.13E-06 | -2.12E-07 | 5.57E-06 | 4.68E-05 | 3.18E-05 | 3.49E-05 | -2.39E-05 | 1.73E-05 | -3.35E-07 | -1.08E-05 |
| 220 | 3894 | | | -1.93E-05 | | | | | | 3.22E-05 | | | | | -1.01E-05 |
| 221 | 3906 | | -3.04E-05 | | | | | | 4.66E-05 | | 3.77E-05 | | | | |
| 222 | 3910 | -2.30E-05 | | -5.15E-05 | 9.15E-05 | 4.35E-05 | 6.92E-08 | 1.57E-05 | | | | | | | |
| 223 | 3943 | -2.43E-05 | -4.09E-05 | | 9.97E-05 | | | | 3.79E-05 | 2.06E-05 | 2.70E-05 | -6.67E-06 | 1.95E-05 | -1.48E-07 | -6.95E-06 |
| 224 | 3946 | -2.43E-05 | -4.08E-05 | -6.32E-05 | 9.99E-05 | 8.51E-06 | 1.14E-06 | 1.04E-05 | 3.78E-05 | 2.06E-05 | 2.70E-05 | -4.60E-06 | 2.84E-05 | -1.02E-07 | -7.05E-06 |
| 225 | 3997 | -4.99E-05 | | -0.000106 | | -5.92E-05 | 1.04E-06 | -0.000132 | 3.33E-05 | 1.59E-05 | 1.53E-05 | -5.02E-05 | -6.67E-08 | -5.75E-08 | -3.31E-05 |
| 226 | 4014 | | -6.25E-05 | | 1.80E-06 | | | 1.35E-06 | 3.04E-05 | 1.20E-05 | 1.41E-05 | -4.86E-05 | | 1.70E-07 | -3.27E-05 |
| 227 | 4029 | -4.77E-05 | -6.08E-05 | -0.000102 | | -3.71E-05 | 1.33E-06 | -0.000126 | | 1.37E-05 | | -4.56E-05 | 6.65E-06 | | |
| 228 | 4043 | -4.74E-05 | -6.04E-05 | -0.000102 | 8.20E-06 | -1.51E-05 | 1.42E-06 | -0.000121 | 3.37E-05 | 1.49E-05 | 1.61E-05 | -3.25E-05 | 1.15E-05 | 1.73E-06 | -2.34E-05 |
| 229 | 4047 | -7.56E-05 | -9.68E-05 | -0.000137 | -2.32E-05 | -5.91E-05 | | -0.000155 | 1.46E-05 | | -1.33E-05 | | -8.38E-06 | | -5.37E-05 |
| 230 | 4071 | -7.73E-05 | -9.76E-05 | -0.000138 | -2.71E-05 | -7.34E-05 | 1.91E-06 | | -1.07E-05 | -9.85E-06 | -1.95E-05 | -5.76E-05 | -9.81E-06 | 6.20E-06 | -6.72E-05 |
| 231 | 4083 | | | | | | | | -9.22E-05 | | -7.89E-06 | | | | |
| 232 | 4088 | -7.43E-05 | -9.32E-05 | -0.00013 | | | | | -6.47E-06 | | -1.97E-05 | -4.35E-05 | | | -5.78E-05 |
| 233 | 4090 | | | | | -6.67E-06 | -4.65E-05 | | -9.04E-05 | | | | -5.21E-06 | 4.53E-05 | |
| 234 | 4129 | -8.17E-05 | -0.000106 | -0.000142 | | -8.99E-05 | | | -2.36E-05 | -4.19E-05 | -3.46E-05 | -5.17E-05 | -1.33E-05 | | -6.87E-05 |
| 235 | 4133 | -8.15E-05 | -0.000106 | -0.000142 | -1.94E-05 | -9.00E-05 | 2.92E-08 | -0.000104 | -2.34E-05 | -4.17E-05 | -3.46E-05 | -5.18E-05 | -1.40E-05 | 7.93E-08 | -6.87E-05 |
| 236 | 4177 | -8.00E-05 | | | | | | | -3.77E-06 | -2.39E-05 | | -5.19E-05 | -1.37E-05 | | |
| 237 | 4190 | | | | | -1.12E-05 | | | | | | -1.93E-05 | | | -6.57E-05 |
| 238 | 4192 | | -9.31E-05 | -0.000119 | | -6.89E-05 | | -6.66E-05 | | | | | | | |
| 239 | 4229 | -8.30E-05 | | | | | | -9.50E-07 | -1.26E-05 | -3.48E-05 | -2.01E-05 | -6.01E-05 | -1.92E-05 | -2.51E-07 | -6.66E-05 |
| 240 | 4286 | | -9.75E-05 | -0.000124 | -1.00E-05 | -6.92E-05 | | -8.07E-05 | -1.48E-05 | -3.62E-05 | -2.10E-05 | -6.00E-05 | | | -6.76E-05 |
| 241 | 4292 | -8.60E-05 | | | | | | | | | | | | -3.36E-07 | |
| 242 | 4324 | | | | | -1.62E-05 | | 4.12E-07 | | | -3.77E-05 | | -2.04E-05 | | |
| 243 | 4328 | | -0.000109 | -0.000137 | | -7.50E-05 | | | 4.52E-05 | | -1.83E-05 | -6.45E-05 | | | |
| 244 | 4338 | -0.000127 | | | -2.86E-05 | -8.82E-05 | -1.44E-07 | -0.000102 | -5.91E-05 | -8.27E-05 | -6.41E-05 | -6.81E-05 | -2.52E-05 | -4.63E-07 | -7.77E-05 |
| 245 | 4352 | | -0.000137 | | | | | | | | | | -2.80E-05 | -7.14E-07 | |
| 246 | 4361 | -0.000124 | -0.000135 | -0.000138 | -2.04E-05 | -7.37E-05 | 4.00E-08 | -9.66E-05 | -6.14E-05 | -8.48E-05 | -6.72E-05 | -5.92E-05 | -2.02E-05 | -6.21E-07 | -7.43E-05 |
| 247 | 4386 | | | | | | 4.00E-08 | | | | | | | | |
| 248 | 4405 | | | | | -3.99E-05 | | 3.44E-06 | | | -0.000117 | | -6.52E-05 | -2.72E-05 | -6.27E-05 |
| 249 | 4413 | -0.000125 | -0.000139 | -0.00014 | | -6.56E-05 | | -9.24E-05 | -8.48E-05 | | -7.35E-05 | | | | -7.41E-05 |
| 250 | 4424 | | | | -4.14E-05 | | -1.32E-07 | | -8.15E-05 | -0.000111 | -7.51E-05 | -6.46E-05 | -2.70E-05 | -6.41E-07 | -7.50E-05 |
| 251 | 4433 | | | | | | -1.58E-07 | | | | -7.42E-05 | | | | -7.40E-05 |
| 252 | 4445 | | | -0.000142 | | -7.14E-05 | | -0.000102 | | | | | | | |
| 253 | 4451 | -0.000133 | -0.000147 | | | | -1.24E-07 | | -0.00011 | -0.000135 | -8.63E-05 | -7.32E-05 | -3.06E-05 | -7.28E-07 | -8.65E-05 |
| 254 | 4502 | | | | -3.58E-05 | | | | | | | | | | |
| 255 | 4513 | -0.000119 | -0.000133 | -0.000125 | | -2.17E-05 | | -8.14E-05 | -9.86E-05 | -0.000126 | | -5.41E-05 | -2.24E-05 | | |
| 256 | 4518 | | | | -3.57E-05 | | | | | | -7.16E-05 | | | | -6.73E-05 |
| 257 | 4541 | | -0.000136 | -0.000132 | | -2.40E-05 | | | -8.20E-05 | -8.72E-05 | -0.000113 | | -2.39E-05 | | |
| 258 | 4542 | -0.000118 | -0.000136 | -0.000132 | -2.83E-05 | -2.41E-05 | 2.47E-06 | -8.12E-05 | -9.22E-05 | -0.000118 | -7.62E-05 | -5.32E-05 | -2.38E-05 | 8.17E-07 | -6.62E-05 |
| 259 | 4543 | -0.000119 | | | -3.07E-05 | -3.50E-05 | 3.03E-06 | | -0.000103 | -0.000126 | -8.21E-05 | -5.74E-05 | -2.56E-05 | 9.18E-07 | -7.32E-05 |
| 260 | 4561 | -0.000119 | -0.000137 | -0.000133 | | | | -8.30E-05 | -0.000106 | -0.00013 | -8.60E-05 | -5.72E-05 | -2.56E-05 | | |
| 261 | 4572 | -0.000124 | -0.00014 | -0.000138 | -3.42E-05 | -4.88E-05 | | -8.89E-05 | -0.000106 | -0.00013 | -8.65E-05 | -5.76E-05 | -2.56E-05 | | -8.11E-05 |
| 262 | 4581 | -8.97E-05 | -0.000103 | -8.69E-05 | 0.000112 | | 8.70E-07 | 4.75E-05 | -8.37E-05 | -0.000109 | -7.02E-05 | -8.39E-06 | | | |
| 263 | 4606 | | | | 0.000104 | | | | | | | | | | -6.64E-05 |
| 264 | 4621 | | | | | | | | | | | | -2.80E-05 | 6.74E-07 | |
| 265 | 4622 | | -9.69E-05 | -7.26E-05 | | | | 5.41E-05 | | -0.00012 | | | | | -6.64E-05 |
| 266 | 4623 | -9.04E-05 | | | | -3.16E-05 | | | -8.97E-05 | | -7.14E-05 | -2.54E-05 | | | -6.64E-05 |
| 267 | 4624 | -9.30E-05 | -9.86E-05 | -7.46E-05 | 0.000106 | -2.69E-05 | 6.63E-07 | 5.54E-05 | -9.12E-05 | -0.000121 | -7.03E-05 | -2.48E-05 | -2.68E-05 | 6.74E-07 | -6.42E-05 |
| 268 | 4627 | | | | | | 2.92E-06 | | -9.32E-05 | | -7.09E-05 | -2.48E-05 | -2.61E-05 | 1.36E-06 | -6.53E-05 |
| 269 | 4629 | -6.53E-05 | -6.85E-05 | -4.66E-05 | 8.11E-05 | -4.30E-05 | 2.64E-06 | 4.06E-05 | -9.34E-05 | -0.000114 | -7.10E-05 | -2.49E-05 | -2.61E-05 | 1.14E-06 | -6.54E-05 |
| 270 | 4639 | -6.47E-05 | | | 8.38E-05 | -3.20E-05 | | | -9.39E-05 | | -7.03E-05 | -2.24E-05 | -2.46E-05 | | -6.38E-05 |
| 271 | 4641 | -6.47E-05 | -6.70E-05 | -4.39E-05 | | -3.27E-05 | | 4.58E-05 | -9.44E-05 | -0.000114 | | | -2.54E-05 | 4.31E-07 | |
| 272 | 4643 | | | | 8.27E-05 | | 2.00E-06 | | -9.54E-05 | -0.000114 | -7.13E-05 | -2.32E-05 | -2.62E-05 | 4.38E-07 | -6.46E-05 |
| 273 | 4644 | | -6.35E-05 | -3.58E-05 | | -2.08E-05 | | 6.36E-05 | | | | -1.18E-05 | | | -5.17E-05 |
| 274 | 4649 | -6.27E-05 | | | | | | | | | -0.000107 | | -2.41E-05 | | |
| 275 | 4660 | | | | | | | | -4.96E-05 | | -6.19E-05 | | | | |
| 276 | 4679 | -5.93E-05 | -5.98E-05 | -2.89E-05 | 0.00012 | 2.19E-06 | 1.40E-06 | 7.59E-05 | -6.53E-05 | -0.000116 | -6.96E-05 | -2.53E-06 | -2.03E-05 | 8.09E-08 | -4.91E-05 |
| 277 | 4683 | -5.85E-05 | | | 0.000121 | 7.23E-06 | | | | | -6.92E-05 | -1.92E-06 | | | -4.81E-05 |
| 278 | 4690 | | -6.10E-05 | -3.03E-05 | | | 1.37E-06 | 7.11E-05 | -8.90E-05 | -0.000134 | | | -2.56E-05 | 2.12E-08 | |
| 279 | 4693 | -6.19E-05 | | | | | 1.36E-06 | | | | -7.21E-05 | | | 1.88E-08 | |
| 280 | 4701 | | | | 0.000119 | 1.68E-05 | | | | | | | -4.30E-06 | | -4.95E-05 |
| 281 | 4710 | -4.78E-05 | -4.83E-05 | -1.08E-05 | 0.000162 | 5.18E-05 | 3.09E-06 | 0.000103 | -8.96E-05 | -0.000133 | | | -1.89E-05 | 6.90E-07 | |
| 282 | 4730 | -4.96E-05 | -5.07E-05 | -1.46E-05 | 0.000162 | 4.95E-05 | | 0.000101 | -9.89E-05 | -0.000144 | -7.02E-05 | 4.35E-07 | | | -4.74E-05 |
| 283 | 4732 | | | -1.45E-05 | | | | | | | | | -2.15E-05 | | |
| 284 | 4759 | -4.94E-05 | -5.02E-05 | | 0.000161 | 4.91E-05 | | 0.000101 | | -0.000146 | -7.06E-05 | -5.67E-07 | | | -4.87E-05 |
| 285 | 4764 | | | | | | | | -0.00011 | | | | -2.88E-05 | -1.60E-06 | |
| 286 | 4780 | | | -1.57E-05 | | | 2.68E-06 | | | | -7.12E-05 | -1.32E-06 | | -8.94E-07 | -5.17E-05 |
| 287 | 4821 | -4.15E-05 | -4.39E-05 | | | 5.14E-05 | | 0.000123 | | | | | -2.59E-05 | -8.23E-07 | |
| 288 | 4853 | | -4.35E-05 | -7.26E-06 | | | | | | -0.000109 | -0.000142 | -7.36E-05 | | | |
| 289 | 4875 | -4.19E-05 | -4.36E-05 | -6.35E-06 | 0.000178 | 5.39E-05 | | 0.000125 | -0.000108 | -0.000141 | -7.26E-05 | 4.72E-06 | | | -4.57E-05 |
| 290 | 4878 | | | | | | | | | | | | -2.48E-05 | | |
| 291 | 4883 | -4.21E-05 | -4.38E-05 | -6.45E-06 | 0.000178 | 5.37E-05 | 1.55E-06 | 0.000125 | -0.00011 | -0.000143 | -7.83E-05 | 4.42E-06 | -2.56E-05 | -8.42E-07 | -4.72E-05 |
| 292 | 4930 | | | | 0.000177 | | | 0.000124 | | | -8.18E-05 | | | -4.29E-07 | |
| 293 | 4931 | -3.86E-05 | | | | | | | | | | | | | |
| 294 | 4933 | | | | | | | | | | | | -2.32E-05 | | |
| 295 | 4939 | -4.04E-05 | -4.28E-05 | -3.74E-06 | 0.00021 | 6.52E-05 | 6.12E-07 | 0.000149 | -0.000122 | -0.000153 | -9.03E-05 | 8.35E-06 | -2.45E-05 | -5.56E-07 | -4.61E-05 |
| 296 | 4954 | -4.49E-05 | -4.65E-05 | -1.18E-05 | 0.000218 | 7.24E-05 | | 0.000152 | -0.00012 | -0.000149 | -8.85E-05 | 1.85E-05 | -1.96E-05 | -5.67E-07 | -4.03E-05 |
| 297 | 5000 | | | | | 6.06E-05 | 2.73E-06 | | -0.000125 | -0.000158 | -0.000102 | -2.99E-07 | -2.43E-05 | -5.69E-07 | -5.45E-05 |
| 298 | 5003 | -5.77E-05 | -5.53E-05 | -3.39E-06 | 0.00021 | 6.06E-05 | 2.22E-07 | 0.000135 | -0.000125 | -0.000158 | -0.000102 | | -2.27E-05 | -2.10E-07 | -5.40E-05 |
| 299 | 5011 | -6.00E-05 | -5.82E-05 | -4.03E-06 | 0.000204 | 5.40E-05 | 8.58E-06 | 0.000129 | -0.000129 | -0.000162 | -0.000102 | 1.63E-07 | -2.22E-05 | 8.08E-06 | -5.42E-05 |
| 300 | 5022 | -5.09E-05 | -4.88E-05 | -2.24E-05 | 0.000233 | 8.21E-05 | 9.53E-06 | 0.00015 | -0.000124 | -0.000157 | | 1.06E-05 | -1.46E-05 | 8.37E-06 | -4.58E-05 |
| 301 | 5023 | -5.10E-05 | -4.88E-05 | -2.26E-05 | 0.000234 | 8.19E-05 | | 0.000151 | -0.000122 | -0.000157 | -9.72E-05 | 1.09E-05 | -1.44E-05 | | -4.52E-05 |
| 302 | 5027 | | -5.15E-05 | -2.84E-05 | 0.000228 | | | | | | | | -1.48E-05 | | |
| 303 | 5043 | | | | | | | | 0.000149 | | -0.000148 | | | | | |
| 304 | 5053 | -4.61E-05 | -4.79E-05 | -2.39E-05 | 0.000258 | 9.26E-05 | 7.06E-06 | 0.000151 | -0.000122 | -0.000148 | -9.73E-05 | 1.66E-05 | -1.42E-05 | 6.59E-06 | -4.28E-05 |
| 305 | 5054 | | | | | | | 0.000153 | | | | | | | |
| 306 | 5072 | -4.79E-05 | | | | 8.96E-05 | 7.84E-06 | | -0.000124 | -0.000153 | -9.75E-05 | | -1.63E-05 | 6.85E-06 | |

| | A | B | C | D | E | F | G | H | I | J | K | L | M | N | O |
|---|---|---|---|---|---|---|---|---|---|---|---|---|---|---|---|
| 307 | 5087 | | -4.27E-05 | -1.26E-05 | 0.000267 | | | 0.000161 | | | | 1.75E-05 | -1.84E-05 | | -4.03E-05 |
| 308 | 5088 | -4.45E-05 | -4.12E-05 | -1.10E-05 | 0.000297 | 0.000108 | 6.56E-06 | 0.000165 | -0.000117 | -0.000146 | -9.01E-05 | 1.88E-05 | | 5.61E-06 | -3.87E-05 |
| 309 | 5099 | | -3.98E-05 | -1.11E-05 | 0.000293 | | | 0.000164 | | -0.000141 | | | -1.82E-05 | | |
| 310 | 5105 | -4.33E-05 | | | | | | | -0.000101 | | -8.69E-05 | 1.81E-05 | | | -3.88E-05 |
| 311 | 5131 | | | | | 9.29E-05 | 4.53E-06 | | | -0.000156 | | | -2.35E-05 | 5.37E-06 | |
| 312 | 5133 | -4.56E-05 | -4.23E-05 | -1.45E-05 | 0.000286 | 9.54E-05 | 4.43E-06 | 0.000157 | -0.000115 | -0.00016 | -9.20E-05 | 1.45E-05 | -2.28E-05 | 5.43E-06 | -4.22E-05 |
| 313 | 5136 | -4.22E-05 | | | | 0.000119 | 4.29E-06 | 0.000164 | -0.000107 | -0.000154 | | | | 5.59E-06 | |
| 314 | 5159 | -2.01E-05 | -2.85E-05 | 1.19E-05 | 0.00035 | 0.000153 | 3.91E-06 | 0.000228 | | -0.000148 | -8.76E-05 | 4.39E-05 | -1.69E-05 | 5.66E-06 | -2.72E-05 |
| 315 | 5171 | -1.95E-05 | -2.82E-05 | 1.26E-05 | 0.000352 | 0.000159 | | 0.000229 | -0.000106 | | -8.73E-05 | 4.93E-05 | | | -2.31E-05 |
| 316 | 5187 | | | | | | | | | -9.61E-05 | -0.000138 | | | | | |
| 317 | 5190 | -1.10E-05 | -2.03E-05 | 2.62E-05 | 0.000385 | | | 0.000248 | | | -8.10E-05 | | | | |
| 318 | 5226 | -1.33E-05 | | | | 0.000161 | | | -9.03E-05 | -0.000139 | | 5.29E-05 | -1.37E-05 | | -2.15E-05 |
| 319 | 5232 | | | -1.33E-06 | 0.000314 | | | 0.000208 | | | | | | | |
| 320 | 5241 | -1.63E-05 | -2.66E-05 | | | | | | | | -8.81E-05 | | | | |
| 321 | 5264 | | | | | | | | | | | | 3.96E-05 | | | -2.71E-05 |
| 322 | 5277 | | | | | | | | | -0.000155 | | | -1.45E-05 | | |
| 323 | 5292 | -1.47E-05 | -2.52E-05 | 4.51E-07 | 0.000317 | 0.000162 | 4.97E-06 | 0.00021 | -9.86E-05 | -0.000155 | -8.74E-05 | 4.18E-05 | -1.45E-05 | 5.97E-06 | -2.56E-05 |
| 324 | 5300 | | -3.14E-05 | | | | | | | -0.000149 | | | | | |
| 325 | 5306 | -1.56E-05 | -3.16E-05 | -5.92E-07 | 0.000312 | 0.000161 | 5.05E-06 | 0.000206 | -9.83E-05 | -0.000149 | -8.73E-05 | 3.92E-05 | -1.44E-05 | 6.11E-06 | -2.64E-05 |
| 326 | 5309 | | | -1.58E-06 | | 0.00016 | 4.95E-06 | | -9.96E-05 | -0.00015 | | | | 6.06E-06 | |
| 327 | 5327 | | -3.10E-05 | | | | | | | | | | -1.22E-05 | | |
| 328 | 5332 | -4.62E-06 | -3.10E-05 | 1.80E-05 | 0.000332 | 0.000193 | | 0.000232 | | -0.000139 | -6.01E-05 | 4.74E-05 | | | -9.92E-06 |
| 329 | 5355 | -7.80E-06 | -3.44E-05 | 1.67E-05 | 0.000331 | 0.000193 | 5.25E-06 | 0.000232 | -0.000105 | -0.000146 | -6.25E-05 | 4.38E-05 | -1.52E-05 | | -1.17E-05 |
| 330 | 5360 | -7.60E-06 | -3.44E-05 | 1.68E-05 | 0.000333 | 0.000193 | 5.21E-06 | 0.000232 | -0.000106 | -0.000146 | -6.24E-05 | 4.45E-05 | -1.63E-05 | 6.05E-06 | -1.14E-05 |
| 331 | 5378 | | | 1.54E-05 | | 0.000195 | | | -9.80E-05 | -0.000137 | | | | | |
| 332 | 5383 | -1.04E-05 | -3.71E-05 | | 0.000334 | | | | | | -7.28E-05 | | -1.36E-05 | | |
| 333 | 5404 | | -3.67E-05 | 1.58E-05 | | 0.000201 | 5.65E-06 | 0.000234 | | -0.000143 | | 4.28E-05 | | | -1.13E-05 |
| 334 | 5421 | | | | | | | | -0.000103 | | | | | | |
| 335 | 5444 | -8.52E-06 | | | | | | | | | | | | | |
| 336 | 5472 | | | | 0.000312 | | | | | | -7.58E-05 | | -1.53E-05 | 5.65E-06 | |
| 337 | 5474 | -1.20E-05 | -4.89E-05 | 2.28E-06 | 0.000312 | 0.000208 | | 0.000231 | -0.000156 | -0.000173 | -9.20E-05 | 3.97E-05 | -1.53E-05 | | -1.24E-05 |
| 338 | 5485 | | | | | | 5.33E-06 | | | -0.00018 | | | | | |
| 339 | 5495 | -2.21E-05 | | -8.47E-06 | 0.000302 | 0.000197 | 4.84E-06 | 0.000215 | -0.000162 | -0.000179 | -9.65E-05 | 3.75E-05 | -1.71E-05 | 5.64E-06 | -1.53E-05 |
| 340 | 5506 | -2.13E-05 | -5.48E-05 | | 0.000304 | 0.000199 | | | -0.000162 | | -9.61E-05 | 3.90E-05 | -1.73E-05 | 5.21E-06 | -1.44E-05 |
| 341 | 5527 | | | -5.67E-06 | | | | 0.000218 | | | | | | | |
| 342 | 5542 | | | | | | | | | | -0.000177 | | | | |
| 343 | 5548 | | -5.07E-05 | | | | | | -0.000159 | | | | | | |
| 344 | 5581 | -1.85E-05 | | | | 0.000198 | | | | | | | | | |
| 345 | 5584 | | -5.00E-05 | -5.88E-06 | 0.000302 | 0.000195 | | 0.000217 | -0.00016 | -0.000177 | -9.62E-05 | 3.72E-05 | -1.86E-05 | -5.64E-06 | -1.61E-05 |
| 346 | 5594 | -7.52E-06 | -3.99E-05 | 1.58E-05 | 0.000374 | 0.000222 | -2.19E-06 | 0.000267 | -0.000161 | -0.000178 | -9.62E-05 | 4.52E-05 | -1.88E-05 | -5.61E-06 | -1.27E-05 |
| 347 | 5640 | -8.59E-06 | | | | | | | | -0.000162 | -0.000179 | | | -1.95E-05 | -5.57E-06 | |
| 348 | 5652 | -8.53E-06 | -4.02E-05 | 1.74E-05 | 0.000372 | 0.00022 | -3.43E-06 | 0.000265 | -0.000162 | -0.00018 | -9.39E-05 | 4.30E-05 | -1.95E-05 | -6.05E-06 | -1.48E-05 |
| 349 | 5653 | -9.21E-07 | -3.33E-05 | 2.15E-05 | 0.00038 | 0.000225 | -2.97E-06 | 0.000269 | -0.000158 | -0.000175 | -9.02E-05 | 5.10E-05 | | -5.18E-06 | -8.98E-06 |
| 350 | 5660 | | -2.94E-05 | | | | | | | | | 5.01E-05 | | | -8.09E-06 |
| 351 | 5677 | | | | | | | | | | | | -1.71E-05 | | |
| 352 | 5707 | 8.59E-06 | | | | | | | | | -0.000144 | | | | |
| 353 | 5763 | | | | | | | | -0.000143 | | | | | | |
| 354 | 5802 | 5.04E-07 | -3.08E-05 | 1.66E-05 | 0.000371 | 0.000226 | 6.90E-07 | 0.000267 | -0.000143 | -0.00015 | -9.08E-05 | 4.77E-05 | -1.67E-05 | -3.75E-06 | -9.92E-06 |
| 355 | 5803 | 1.84E-06 | -2.98E-05 | 1.71E-05 | 0.000373 | 0.000228 | | 0.000268 | -0.000135 | -0.000143 | -8.98E-05 | 4.81E-05 | -1.71E-05 | -3.92E-06 | -1.01E-05 |
| 356 | 5804 | 1.19E-06 | -3.19E-05 | | | | 6.51E-07 | | -0.000125 | -0.00013 | | 1.94E-05 | -3.08E-05 | | -9.80E-06 |
| 357 | 5834 | | 9.15E-06 | 1.98E-05 | 0.000367 | | | 0.00026 | | 1.62E-05 | -8.55E-05 | | | -4.22E-05 | |
| 358 | 5851 | 3.87E-05 | | | | | 0.00022 | -2.35E-06 | | -7.73E-05 | | 8.53E-05 | -3.65E-05 | | -2.35E-05 |
| 359 | 5883 | | | | | | | | | | | | | -4.33E-06 | |
| 360 | 5918 | 1.06E-05 | -1.17E-05 | -4.27E-06 | 0.000366 | 0.000218 | | 0.000249 | -8.44E-05 | 2.58E-06 | -8.64E-05 | 9.64E-06 | -3.56E-05 | | -2.37E-05 |
| 361 | 5919 | | -1.19E-05 | | | | | | | 4.78E-07 | | | | | |
| 362 | 5927 | | | -4.43E-06 | 0.000365 | 0.000217 | | 0.000249 | -8.52E-05 | | -8.68E-05 | 9.20E-06 | -3.60E-05 | | -2.42E-05 |
| 363 | 5969 | | | | | | | | | | | | | -4.73E-06 | |
| 364 | 5985 | 1.58E-05 | | | | | | | | | | | | | |
| 365 | 6021 | 1.65E-05 | -1.22E-05 | 1.06E-06 | 0.000376 | 0.000224 | 3.92E-06 | 0.000257 | -0.000119 | -2.86E-05 | -9.37E-05 | 1.64E-05 | -3.62E-05 | -4.27E-05 | -1.89E-05 |
| 366 | 6028 | 2.23E-05 | | | | | | | | | | | | | |
| 367 | 6041 | | | | | | | | | | -3.12E-05 | -9.45E-05 | | | | |
| 368 | 6057 | 1.09E-05 | -2.18E-05 | -5.85E-06 | 0.000367 | 0.000221 | | 0.000251 | -0.000125 | | | | 6.80E-06 | -3.96E-05 | | -2.64E-05 |
| 369 | 6072 | 1.34E-05 | -1.88E-05 | 9.60E-07 | 0.000383 | 0.000236 | 1.34E-06 | 0.000268 | -0.000125 | -3.20E-05 | -9.34E-05 | 1.12E-05 | -4.56E-05 | -4.51E-06 | -2.48E-05 |
| 370 | 6083 | | | | | | 9.71E-07 | | | | -9.72E-05 | | | -7.81E-06 | |
| 371 | 6094 | | -2.30E-05 | -7.99E-06 | | | 2.35E-06 | | -0.000139 | -4.72E-05 | | -1.15E-06 | -4.75E-05 | -8.03E-06 | -3.51E-05 |
| 372 | 6103 | 5.29E-06 | -3.17E-05 | -1.39E-05 | 0.000375 | 0.000231 | 1.94E-06 | 0.000259 | -0.000148 | -5.54E-05 | -9.90E-05 | 3.11E-06 | -4.60E-05 | | -3.17E-05 |
| 373 | 6139 | 3.20E-06 | -3.33E-05 | -1.53E-05 | 0.000379 | 0.000234 | | 0.000259 | -0.000212 | -0.000133 | -0.000104 | 1.84E-06 | -4.70E-05 | -7.30E-06 | -3.31E-05 |
| 374 | 6152 | -1.40E-06 | -3.69E-05 | -1.89E-05 | 0.000376 | 0.000224 | | 0.000256 | | -0.000136 | -0.000106 | -1.55E-07 | -4.77E-05 | -8.19E-06 | -3.53E-05 |
| 375 | 6156 | | -3.36E-05 | -1.14E-05 | 0.000395 | 0.000241 | 1.14E-05 | 0.000268 | -0.000215 | -0.000138 | -0.000107 | -1.31E-06 | -4.86E-05 | -6.34E-06 | -3.63E-05 |
| 376 | 6178 | 2.49E-06 | | | | | | | | | | | | -6.20E-06 | |
| 377 | 6199 | -2.99E-06 | -3.78E-05 | -2.43E-05 | 0.00039 | 0.000241 | 1.17E-05 | 0.000254 | -0.000213 | -0.000136 | -0.000112 | -3.00E-06 | -4.77E-05 | -5.98E-06 | -4.11E-05 |
| 378 | 6246 | 0.000211 | 7.07E-05 | | | | | | -9.10E-05 | | | | | | |
| 379 | 6250 | | | | | | -1.30E-06 | | | | | | | -1.19E-05 | |
| 380 | 6273 | | | -2.15E-05 | 0.000409 | | | 0.000258 | -0.000102 | -0.00012 | -8.87E-05 | | -4.46E-05 | | |
| 381 | 6274 | | | | | 0.000237 | -1.61E-06 | | | | | -5.33E-06 | | -1.24E-05 | -3.79E-05 |
| 382 | 6278 | 0.000197 | 8.05E-05 | -1.22E-05 | | | | | | | | | | | |
| 383 | 6303 | | | | | | | 0.000266 | | | | | | -1.26E-05 | |
| 384 | 6307 | | | | 0.000409 | | -1.71E-06 | | -0.000105 | -0.000129 | -8.75E-05 | | -4.33E-05 | | |
| 385 | 6321 | 0.000191 | 7.11E-05 | 2.67E-06 | 0.000435 | 0.000251 | | 0.000278 | -9.95E-05 | -0.000122 | | 1.39E-05 | | | -2.32E-05 |
| 386 | 6334 | | | | 0.000434 | 0.000253 | | | | | -7.36E-05 | 7.95E-06 | -4.06E-05 | | -2.72E-05 |
| 387 | 6407 | 0.000188 | 6.88E-05 | 2.71E-06 | 0.000436 | | | 0.000277 | -0.000104 | -0.000127 | -7.56E-05 | | | -1.25E-05 | -3.04E-05 |
| 388 | 6484 | 0.000168 | 5.76E-05 | -9.29E-06 | 0.000421 | 0.000211 | -3.52E-06 | 0.000261 | -0.000127 | -0.000153 | -0.000127 | 6.57E-07 | -4.28E-05 | | -3.40E-05 |
| 389 | 6503 | | | | | 0.000213 | | | | | | | | -1.23E-05 | |
| 390 | 6523 | 0.000169 | 5.97E-05 | -4.07E-06 | | | -3.93E-06 | 0.000264 | -0.000124 | -0.000149 | -0.000121 | | -4.02E-05 | -1.25E-05 | |
| 391 | 6565 | | | | 0.000432 | | | | | | | | | | |
| 392 | 6566 | 0.000168 | 5.84E-05 | -6.02E-06 | 0.000431 | 0.000214 | -1.56E-07 | 0.000262 | -0.000124 | -0.000149 | -0.00012 | 4.79E-06 | -3.90E-05 | -9.72E-06 | -2.90E-05 |
| 393 | 6606 | | | | | 0.000214 | | | | | | | -4.00E-05 | | |
| 394 | 6608 | 0.000172 | 6.14E-05 | -1.39E-06 | | | | 0.000264 | -0.000112 | -0.000137 | -0.000118 | | | | |
| 395 | 6613 | | | | 0.000432 | | -3.92E-07 | | | | | | | | -2.44E-05 |
| 396 | 6629 | | | | | | | | | | | | -7.91E-06 | | -1.28E-05 |
| 397 | 6731 | | | | | 0.00021 | | | | | | | | | |
| 398 | 6741 | 0.000161 | 5.34E-05 | -9.54E-06 | 0.000433 | | | 0.00026 | -0.000128 | -0.000159 | -0.000124 | | | | -2.27E-05 |
| 399 | 6749 | | | | | | | | | | | -1.42E-05 | -4.26E-05 | -1.32E-05 | |
| 400 | 6772 | | | | | | -3.45E-07 | | | | | | | -1.32E-05 | |
| 401 | 6790 | 0.000125 | 1.79E-05 | -6.17E-06 | 0.000431 | 0.000206 | | 0.000247 | -0.000125 | -0.000156 | -0.000127 | -1.14E-05 | -4.06E-05 | | -2.33E-05 |
| 402 | 6807 | 0.000127 | 1.88E-05 | -6.06E-05 | 0.000433 | 0.000209 | 1.53E-06 | 0.00025 | -0.000122 | -0.000154 | -0.000125 | -7.15E-06 | -3.93E-05 | -1.24E-05 | -1.90E-05 |
| 403 | 6811 | | | | | 0.000212 | | | | | | | | | |
| 404 | 6823 | | | | | | -4.93E-06 | | -0.000134 | -0.000165 | -0.000131 | -1.26E-05 | -4.26E-05 | -1.31E-05 | -2.53E-05 |
| 405 | 6876 | 0.000115 | 1.14E-05 | -6.77E-05 | 0.000432 | 0.000215 | | 0.000247 | -0.000142 | -0.000173 | -0.000142 | -9.62E-05 | -4.15E-05 | -1.34E-05 | -2.42E-05 |
| 406 | 6940 | 0.000128 | 2.25E-05 | -5.86E-05 | 0.000414 | 0.000191 | -3.08E-06 | 0.000232 | -0.000123 | -0.000151 | -0.000136 | -1.07E-05 | -4.27E-05 | -8.11E-06 | -2.44E-05 |

|   | A | B | C | D | E | F | G | H | I | J | K | L | M | N | O |
|---|---|---|---|---|---|---|---|---|---|---|---|---|---|---|---|
| 1 | Rank | Inf | 0.002 | 0.0015 | 0.001 | 0.0005 | 0.0001 | 0.0013 | 0.0020 Ran | 0.0015 Ran | 0.0010 Ran | 0.0005 Ran | 0.0001 Ran | 0.0013 Random | |
| 2 | 10 | 3.32E-05 | 0.000139 | 0.000133 | 0.000177 | 0.000106 | -4.53E-06 | 0.000155 | 0.000122 | 0.000145 | 0.00016 | 0.000138 | 9.62E-05 | -5.42E-07 | 0.00013 |
| 3 | 25 | | | | | | -4.53E-06 | | 0.000122 | 0.000149 | 0.000159 | 0.000138 | 9.65E-05 | | 0.000129 |
| 4 | 61 | 3.46E-05 | 0.000138 | 0.000131 | 0.000172 | 0.000106 | -3.92E-06 | 0.000152 | 0.000122 | 0.000152 | 0.000159 | 0.000137 | 9.74E-05 | -4.94E-07 | 0.000129 |
| 5 | 93 | 3.21E-05 | 0.000137 | 0.000131 | 0.000172 | 0.000106 | -4.00E-06 | 0.000152 | | | | | 9.71E-05 | -7.56E-07 | |
| 6 | 115 | | | | | | | | 0.000116 | 0.000152 | 0.000156 | 0.000135 | | | 0.000127 |
| 7 | 144 | 3.05E-05 | 0.000134 | 0.000129 | 0.000171 | 0.000106 | | 0.00015 | 0.000115 | 0.00016 | 0.000156 | 0.000135 | 9.64E-05 | -6.46E-07 | 0.000127 |
| 8 | 145 | 3.02E-05 | 0.000134 | 0.000128 | 0.00017 | 0.000105 | -3.36E-06 | 0.000149 | 0.000115 | 0.000216 | 0.000155 | 0.000135 | 9.57E-05 | -6.42E-07 | 0.000126 |
| 9 | 181 | 3.05E-05 | 0.000131 | 0.000126 | 0.000167 | 0.000103 | -3.44E-06 | 0.000147 | 0.000109 | | 0.000151 | 0.00013 | 9.28E-05 | -9.90E-07 | 0.000121 |
| 10 | 194 | 2.22E-05 | 0.00013 | 0.000125 | 0.000166 | 0.000102 | -3.44E-06 | 0.000146 | 0.000108 | | 0.00015 | 0.000128 | 9.12E-05 | -9.73E-07 | 0.00012 |
| 11 | 197 | 2.32E-05 | 0.000131 | 0.000126 | 0.000167 | 0.000103 | -3.62E-06 | 0.000147 | 0.000108 | 0.000239 | 0.000152 | 0.00013 | 9.16E-05 | -1.37E-06 | 0.000122 |
| 12 | 213 | 2.02E-05 | 0.000128 | 0.000125 | 0.000166 | 9.91E-05 | -3.07E-06 | 0.000145 | 0.000106 | 0.000222 | 0.000148 | 0.000127 | 8.90E-05 | -7.50E-07 | 0.000119 |
| 13 | 233 | 2.15E-05 | 0.000128 | 0.000125 | 0.000166 | 9.80E-05 | -3.12E-06 | 0.000145 | 0.000181 | 0.000226 | 0.000144 | 0.000122 | 8.43E-05 | -7.18E-07 | 0.000113 |
| 14 | 234 | 2.04E-05 | 0.000124 | 0.000116 | 0.000157 | 9.40E-05 | -1.69E-06 | 0.000135 | 0.000178 | 0.00022 | 0.000137 | 0.000115 | 8.05E-05 | -8.24E-07 | 0.000107 |
| 15 | 245 | 1.26E-05 | 0.000123 | 0.000114 | 0.000154 | 9.22E-05 | 8.03E-08 | 0.000133 | 0.000112 | | 0.000135 | 0.000113 | 7.91E-05 | 4.15E-06 | 0.000106 |
| 16 | 259 | 9.25E-06 | 0.00012 | 0.000113 | 0.000153 | 9.15E-05 | -2.62E-07 | 0.000131 | 0.000102 | 0.00021 | 0.000134 | 0.000112 | 7.97E-05 | 4.28E-06 | 0.000105 |
| 17 | 261 | | | 0.000113 | 0.000153 | 9.18E-05 | | 0.000132 | 0.000102 | 0.000211 | 0.000134 | 0.000112 | 7.98E-05 | 4.35E-06 | 0.000105 |
| 18 | 276 | 9.83E-06 | 0.00012 | 0.000113 | 0.000154 | 9.21E-05 | -5.64E-08 | 0.000132 | 0.000111 | 0.000211 | 0.000134 | 0.000112 | 7.96E-05 | | 0.000105 |
| 19 | 284 | 1.87E-05 | 0.000119 | | | 9.18E-05 | -1.83E-07 | | 9.13E-05 | 0.00021 | 0.000132 | 0.00011 | 7.88E-05 | 3.17E-06 | 0.000102 |
| 20 | 327 | 1.72E-05 | 0.000117 | 0.000113 | 0.000152 | | | 0.000131 | | | 0.000131 | 0.000109 | 7.83E-05 | 2.31E-06 | 0.000102 |
| 21 | 334 | 1.73E-05 | | 0.000112 | 0.000151 | 9.05E-05 | -2.11E-07 | 0.00013 | 5.42E-05 | 0.000211 | | | | 2.28E-06 | |
| 22 | 344 | 1.97E-05 | 0.000116 | 0.00011 | 0.000148 | 8.92E-05 | 1.09E-07 | 0.000128 | 5.30E-05 | 0.000235 | 0.000129 | 0.000106 | 7.52E-05 | 2.51E-06 | 9.94E-05 |
| 23 | 351 | | 0.000115 | | | | -2.76E-08 | | 5.45E-05 | | | | | | |
| 24 | 352 | | | | 0.00015 | 9.01E-05 | | | | 0.000214 | | | 7.67E-05 | 2.27E-06 | |
| 25 | 357 | 2.11E-05 | 0.000116 | 0.000113 | | | -1.34E-07 | 0.000129 | 4.91E-05 | | 0.000131 | 0.000108 | | | 0.0001 |
| 26 | 366 | 1.71E-05 | | 0.000109 | 0.000146 | 8.75E-05 | | 0.000126 | 3.99E-05 | 0.000206 | 0.000127 | 0.000103 | 7.34E-05 | 1.72E-06 | 9.57E-05 |
| 27 | 375 | 1.76E-05 | 0.000106 | 9.72E-05 | 0.000137 | 7.97E-05 | -7.24E-07 | 0.000114 | 3.32E-05 | | 0.000119 | 9.61E-05 | 6.83E-05 | 1.52E-06 | 8.87E-05 |
| 28 | 378 | 1.75E-05 | | 9.73E-05 | 0.000137 | 8.00E-05 | -8.79E-07 | 0.000115 | 3.32E-05 | | 0.000119 | 9.63E-05 | 6.85E-05 | 1.37E-06 | 8.89E-05 |
| 29 | 380 | 1.77E-05 | 0.000105 | 9.63E-05 | 0.000134 | 7.79E-05 | -9.38E-07 | 0.000113 | 3.29E-05 | 0.000201 | 0.00012 | 9.04E-05 | 6.16E-05 | 3.56E-06 | 8.35E-05 |
| 30 | 390 | 1.92E-05 | 0.000104 | 9.57E-05 | 0.000133 | 7.65E-05 | 5.54E-10 | 0.000113 | 5.02E-05 | 0.000207 | 0.000115 | 9.08E-05 | 6.21E-05 | 8.60E-07 | 8.40E-05 |
| 31 | 408 | 2.24E-05 | 0.00011 | 0.000101 | 0.000138 | 8.06E-05 | | 0.000118 | 5.44E-05 | 0.000106 | 0.000117 | 9.30E-05 | 6.44E-05 | 5.13E-07 | 8.65E-05 |
| 32 | 416 | 2.17E-05 | 0.000109 | 0.0001 | | | -1.09E-06 | | 0.000105 | | | | 6.39E-05 | 5.57E-07 | |
| 33 | 432 | 8.34E-06 | | 0.000101 | 0.000139 | 8.39E-05 | | 0.000117 | 4.51E-05 | | 0.000118 | 9.38E-05 | 6.48E-05 | 7.49E-07 | 8.71E-05 |
| 34 | 434 | 7.23E-06 | 0.000106 | 0.0001 | 0.000139 | 8.36E-05 | -1.11E-06 | 0.000117 | 5.63E-05 | 0.0001 | 0.000116 | 9.30E-05 | 6.35E-05 | 6.71E-07 | 8.60E-05 |
| 35 | 435 | | 0.000107 | | | 8.42E-05 | -7.62E-07 | | 7.79E-05 | | 0.000117 | 9.34E-05 | 6.42E-05 | 1.19E-06 | 8.63E-05 |
| 36 | 487 | 1.04E-05 | 0.000107 | 0.000104 | 0.000141 | 8.51E-05 | | 0.000121 | 3.92E-05 | | 0.000118 | 9.50E-05 | 6.62E-05 | 1.35E-06 | 8.75E-05 |
| 37 | 496 | 1.06E-05 | 0.000107 | 0.000104 | 0.000141 | 8.50E-05 | -5.09E-07 | 0.000121 | 5.46E-06 | 0.000102 | 0.000118 | 9.49E-05 | 6.60E-05 | 1.22E-06 | 8.72E-05 |
| 38 | 513 | 8.82E-06 | | 0.000103 | 0.000141 | | | 0.00012 | -1.24E-06 | | 0.000118 | 9.51E-05 | | | 8.73E-05 |
| 39 | 526 | 9.05E-06 | 0.000109 | 0.000105 | 0.000143 | 8.55E-05 | -5.94E-07 | 0.000121 | -7.07E-07 | 0.000106 | 0.000119 | 9.60E-05 | 6.80E-05 | 1.07E-06 | 8.85E-05 |
| 40 | 530 | | 0.000109 | | | | -9.24E-07 | | | 0.000106 | | | 6.88E-05 | | |
| 41 | 540 | 5.17E-06 | | 0.000105 | 0.000143 | 8.73E-05 | | 0.000122 | -9.03E-07 | 0.0001 | 0.000118 | 9.45E-05 | | 8.70E-07 | 8.73E-05 |
| 42 | 544 | | 0.000108 | | | | -1.26E-06 | | | | | | 6.64E-05 | | |
| 43 | 551 | 5.14E-06 | | 0.000105 | 0.000144 | 8.73E-05 | | 0.000122 | 7.24E-06 | 9.23E-05 | 0.000118 | 9.49E-05 | 6.66E-05 | 1.18E-06 | 8.77E-05 |
| 44 | 556 | 5.72E-06 | 0.00011 | 0.000108 | 0.000146 | 9.00E-05 | -1.50E-06 | 0.000125 | 4.30E-06 | | 0.00012 | 9.65E-05 | 6.82E-05 | 1.10E-06 | 8.91E-05 |
| 45 | 557 | 5.60E-06 | 0.000109 | 0.000108 | 0.000146 | 8.95E-05 | -1.48E-06 | 0.000125 | 3.70E-06 | 9.07E-05 | 0.00012 | 9.66E-05 | 6.83E-05 | 9.47E-07 | 8.91E-05 |
| 46 | 573 | 1.74E-06 | 0.000101 | 9.79E-05 | 0.000136 | 8.17E-05 | -1.27E-06 | 0.000115 | -4.50E-06 | | 0.000112 | 8.83E-05 | 6.23E-05 | 2.05E-07 | 8.12E-05 |
| 47 | 582 | 3.59E-06 | 9.70E-05 | 9.03E-05 | 0.000133 | 7.58E-05 | -1.22E-06 | 0.000104 | -5.40E-06 | 9.44E-05 | | 0.000102 | 7.84E-05 | 5.94E-05 | 2.01E-07 | 7.30E-05 |
| 48 | 592 | | | | | | | | | 9.26E-05 | | | | | | |
| 49 | 594 | 3.39E-06 | 9.48E-05 | 8.72E-05 | 0.000126 | 7.36E-05 | -1.03E-06 | 0.0001 | -6.48E-06 | 9.91E-05 | 9.52E-05 | 7.15E-05 | 5.60E-05 | 4.07E-07 | 6.67E-05 |
| 50 | 663 | 5.92E-06 | 9.91E-05 | 9.23E-05 | 0.000131 | 7.70E-05 | -9.41E-07 | 0.000105 | -5.17E-06 | 9.79E-05 | 9.71E-05 | 7.36E-05 | 5.72E-05 | 8.37E-07 | 6.87E-05 |
| 51 | 716 | 6.29E-06 | 9.90E-05 | 9.22E-05 | 0.000131 | 7.71E-05 | -9.52E-07 | 0.000105 | -4.95E-06 | 9.60E-05 | 9.71E-05 | 7.35E-05 | 5.72E-05 | 8.40E-07 | 6.87E-05 |
| 52 | 748 | | | | | | | | | 9.13E-05 | | 8.63E-05 | | | |
| 53 | 760 | | | | | | 8.31E-05 | -8.42E-07 | | -5.73E-06 | | 0.000103 | | 6.08E-05 | | 7.47E-05 |
| 54 | 772 | -1.29E-06 | 9.78E-05 | 9.35E-05 | 0.000131 | | | 0.000106 | -6.03E-06 | 9.20E-05 | 0.000102 | 8.52E-05 | | 8.98E-07 | 7.29E-05 |
| 55 | 782 | -1.46E-06 | 9.81E-05 | 9.33E-05 | 0.000131 | 8.35E-05 | 1.36E-06 | 0.000106 | -6.29E-06 | 9.16E-05 | 0.000101 | 8.47E-05 | 5.98E-05 | 1.85E-06 | 7.24E-05 |
| 56 | 801 | | | | | 8.34E-05 | 1.40E-06 | | | | | 8.45E-05 | | 1.82E-06 | |
| 57 | 813 | -6.00E-06 | 9.52E-05 | 9.27E-05 | 0.000132 | | | 0.000106 | -1.40E-05 | 8.83E-05 | 9.98E-05 | | 5.97E-05 | | 7.18E-05 |
| 58 | 818 | -7.52E-06 | 9.38E-05 | | 0.000131 | 8.16E-05 | | | -1.42E-05 | 8.76E-05 | 9.91E-05 | 8.47E-05 | 5.91E-05 | 1.64E-06 | 7.13E-05 |
| 59 | 826 | -9.36E-06 | 9.23E-05 | 9.13E-05 | 0.00013 | 8.12E-05 | 1.02E-06 | 0.000106 | -1.25E-05 | 8.75E-05 | 9.87E-05 | 8.42E-05 | 5.87E-05 | | 7.08E-05 |
| 60 | 844 | -1.45E-05 | 8.41E-05 | 8.08E-05 | 0.000118 | 6.96E-05 | 9.53E-07 | 9.74E-05 | -1.76E-05 | 8.61E-05 | 9.14E-05 | 7.98E-05 | 5.12E-05 | 1.39E-06 | 6.39E-05 |
| 61 | 885 | -1.35E-05 | 8.38E-05 | 7.99E-05 | 0.000117 | 6.91E-05 | 1.23E-06 | 9.67E-05 | -2.56E-05 | | 8.95E-05 | 7.86E-05 | 5.05E-05 | 1.50E-06 | 6.27E-05 |
| 62 | 917 | -1.44E-05 | 7.88E-05 | 7.65E-05 | 0.000116 | 6.87E-05 | 7.73E-07 | 9.58E-05 | -2.29E-05 | 8.81E-05 | 8.81E-05 | | 5.01E-05 | 1.07E-06 | 6.23E-05 |
| 63 | 925 | -1.40E-05 | | 7.55E-05 | 0.000115 | 6.79E-05 | 7.01E-07 | 9.44E-05 | -3.00E-05 | | 8.77E-05 | 7.80E-05 | 5.02E-05 | 1.08E-06 | 6.16E-05 |
| 64 | 931 | -1.45E-05 | 8.02E-05 | 7.43E-05 | 0.000113 | | 6.33E-07 | 9.33E-05 | -2.38E-06 | 8.81E-05 | 8.79E-05 | 7.81E-05 | | 9.55E-07 | 6.18E-05 |
| 65 | 933 | -1.45E-05 | 7.88E-05 | 7.33E-05 | 0.00011 | 6.78E-05 | 6.27E-07 | 9.14E-05 | | 7.26E-05 | | 7.70E-05 | 5.00E-05 | | |
| 66 | 937 | | | | | 7.17E-05 | | | -4.36E-06 | | 8.61E-05 | | | 1.77E-06 | 6.06E-05 |
| 67 | 991 | -1.74E-05 | 7.43E-05 | 6.85E-05 | 0.000109 | | | 8.94E-05 | | 7.47E-05 | | | 4.99E-05 | | |
| 68 | 997 | -1.74E-05 | 7.42E-05 | 6.84E-05 | 0.000109 | 7.13E-05 | 7.50E-07 | 8.93E-05 | -6.46E-06 | 7.24E-05 | 8.41E-05 | 7.58E-05 | 4.99E-05 | 2.45E-06 | 5.85E-05 |
| 69 | 1043 | | | | | 7.39E-05 | 7.91E-07 | | -2.01E-05 | 6.09E-05 | 8.50E-05 | 7.69E-05 | 5.14E-05 | | 5.95E-05 |
| 70 | 1050 | -1.76E-05 | 7.46E-05 | 6.94E-05 | 0.00011 | | 8.50E-07 | 9.04E-05 | -2.14E-05 | | 8.24E-05 | 7.45E-05 | | 2.38E-06 | 5.69E-05 |
| 71 | 1063 | -2.13E-05 | 7.34E-05 | 6.82E-05 | 0.000109 | 7.35E-05 | 7.84E-07 | 8.94E-05 | | | | | 5.10E-05 | 2.17E-06 | |
| 72 | 1093 | -2.17E-05 | 7.25E-05 | 6.61E-05 | 0.000107 | 7.26E-05 | 1.02E-06 | 8.76E-05 | -1.76E-05 | 6.11E-05 | 8.36E-05 | 7.56E-05 | 5.04E-05 | 2.23E-06 | 5.84E-05 |
| 73 | 1094 | -3.26E-05 | | 6.57E-05 | | 7.29E-05 | | 8.74E-05 | -1.70E-05 | | 8.35E-05 | 7.57E-05 | | 2.12E-06 | 5.84E-05 |
| 74 | 1107 | | 7.98E-05 | 7.29E-05 | 0.000113 | | | 9.49E-05 | | 6.24E-05 | | | 5.49E-05 | | |
| 75 | 1114 | -3.65E-05 | | | | | 8.79E-05 | | | 6.24E-05 | | | | | |
| 76 | 1165 | -3.76E-05 | 7.55E-05 | 7.17E-05 | 0.000109 | 7.16E-05 | 8.49E-07 | 9.14E-05 | -2.98E-05 | 5.90E-05 | 8.20E-05 | 7.34E-05 | 5.39E-05 | 2.25E-06 | 5.72E-05 |
| 77 | 1174 | -3.75E-05 | 7.52E-05 | 7.15E-05 | 0.000109 | | 7.84E-07 | 9.11E-05 | | | | | 5.39E-05 | | |
| 78 | 1206 | | | | | 7.07E-05 | 9.14E-07 | | -3.24E-05 | 6.74E-05 | | 7.26E-05 | | 2.05E-06 | |
| 79 | 1208 | -3.75E-05 | 7.48E-05 | 7.04E-05 | 0.000108 | | | 9.03E-05 | -3.24E-05 | | 8.15E-05 | 7.30E-05 | 5.40E-05 | 2.39E-06 | 5.71E-05 |
| 80 | 1211 | | | | | 7.04E-05 | 9.61E-07 | | | 6.43E-05 | | | 5.36E-05 | | |
| 81 | 1240 | -3.63E-05 | 7.09E-05 | 6.68E-05 | 0.000105 | 6.77E-05 | 8.98E-07 | 8.63E-05 | -2.59E-05 | | 7.85E-05 | 7.02E-05 | 5.08E-05 | 2.54E-06 | 5.41E-05 |
| 82 | 1250 | -3.63E-05 | 7.10E-05 | 6.66E-05 | 0.000105 | | | 8.59E-05 | -3.93E-05 | 4.95E-05 | 7.79E-05 | 6.95E-05 | | 3.08E-06 | 5.36E-05 |
| 83 | 1260 | -3.73E-05 | 6.70E-05 | 6.16E-05 | 0.0001 | 6.56E-05 | 1.28E-06 | 8.20E-05 | -4.48E-05 | 4.90E-05 | 7.44E-05 | 6.65E-05 | 4.87E-05 | 2.39E-06 | 5.07E-05 |
| 84 | 1262 | -3.94E-05 | 6.79E-05 | 6.46E-05 | 0.000104 | 6.93E-05 | | 8.48E-05 | -4.38E-05 | 4.82E-05 | | 6.68E-05 | 4.92E-05 | 1.66E-06 | 5.09E-05 |
| 85 | 1294 | -3.96E-05 | 6.59E-05 | 6.11E-05 | 0.0001 | 6.76E-05 | 1.05E-06 | 8.13E-05 | -4.43E-05 | 4.88E-05 | 7.02E-05 | 6.42E-05 | 4.70E-05 | 1.27E-06 | 4.77E-05 |
| 86 | 1301 | -3.97E-05 | 6.55E-05 | | | 6.69E-05 | 9.39E-07 | | -4.46E-05 | 4.91E-05 | 6.97E-05 | 6.37E-05 | 4.67E-05 | 1.29E-06 | 4.71E-05 |
| 87 | 1306 | -3.98E-05 | 6.50E-05 | 6.02E-05 | 9.90E-05 | | 9.57E-07 | 8.06E-05 | | 4.91E-05 | | | 4.62E-05 | 1.28E-06 | |
| 88 | 1316 | | | 6.07E-05 | 9.94E-05 | 6.67E-05 | | | -3.45E-05 | | 7.16E-05 | 6.61E-05 | 4.65E-05 | | 4.97E-05 |
| 89 | 1357 | -4.04E-05 | 6.15E-05 | 5.65E-05 | 9.55E-05 | 6.49E-05 | 8.93E-07 | 7.73E-05 | -5.84E-05 | | 6.50E-05 | 6.07E-05 | 4.47E-05 | 9.29E-07 | 4.28E-05 |
| 90 | 1411 | -4.14E-05 | 6.03E-05 | 5.46E-05 | 9.40E-05 | 6.39E-05 | 8.66E-07 | 7.57E-05 | -5.98E-05 | | 6.40E-05 | 6.01E-05 | 4.38E-05 | 9.50E-07 | 4.16E-05 |
| 91 | 1414 | | | | | | | | -5.94E-05 | 3.86E-05 | 6.58E-05 | 6.16E-05 | 4.47E-05 | | 4.43E-05 |
| 92 | 1427 | -4.13E-05 | 6.12E-05 | 5.46E-05 | 9.43E-05 | 6.40E-05 | 1.16E-06 | 7.63E-05 | -5.89E-05 | | 6.54E-05 | 6.14E-05 | 4.45E-05 | 8.35E-07 | 4.39E-05 |
| 93 | 1437 | -4.15E-05 | 6.11E-05 | 5.45E-05 | 9.48E-05 | 6.42E-05 | 1.11E-06 | 7.62E-05 | -5.93E-05 | 4.16E-05 | 6.57E-05 | 6.16E-05 | 4.53E-05 | | 4.43E-05 |
| 94 | 1441 | | | 5.41E-05 | 9.30E-05 | | 1.11E-06 | 7.57E-05 | | 3.63E-05 | | | | | |
| 95 | 1448 | | 5.89E-05 | | | | | | | 3.80E-05 | | 5.96E-05 | | | |
| 96 | 1485 | -4.17E-05 | | | 9.13E-05 | 6.27E-05 | 1.08E-06 | | -6.01E-05 | | 3.35E-05 | 6.70E-05 | | 4.54E-05 | 5.75E-07 | 4.51E-05 |
| 97 | 1501 | | | 5.40E-05 | | | | 7.61E-05 | | | 2.68E-05 | 7.53E-05 | 7.27E-05 | | | 5.84E-05 |
| 98 | 1504 | | | | | | | | | 4.04E-05 | | | | | |
| 99 | 1535 | | 5.48E-05 | | | | | | -6.33E-05 | 4.50E-05 | | | 4.33E-05 | | |
| 100 | 1576 | -4.15E-05 | 5.53E-05 | 5.56E-05 | 9.35E-05 | 6.36E-05 | 1.70E-06 | 7.80E-05 | -6.35E-05 | 4.30E-05 | 7.48E-05 | 7.22E-05 | 4.32E-05 | 8.67E-07 | 5.79E-05 |
| 101 | 1579 | | | | 9.90E-05 | | | | | | | 7.41E-05 | | | 5.97E-05 |
| 102 | 1617 | | | 5.87E-05 | | | | 8.19E-05 | | | 7.68E-05 | | | | |

| | A | B | C | D | E | F | G | H | I | J | K | L | M | N | O |
|---|---|---|---|---|---|---|---|---|---|---|---|---|---|---|---|
| 103 | 1651 | | | | | | | | -5.91E-05 | | | | | | |
| 104 | 1654 | -4.19E-05 | 5.43E-05 | 5.67E-05 | 9.55E-05 | 6.34E-05 | 1.66E-06 | 7.90E-05 | -6.15E-05 | 4.10E-05 | 7.40E-05 | 7.21E-05 | 4.25E-05 | 8.63E-07 | 5.72E-05 |
| 105 | 1655 | -4.24E-05 | 5.52E-05 | | | 6.36E-05 | | | | 3.94E-05 | | 7.25E-05 | 4.26E-05 | | |
| 106 | 1667 | -4.28E-05 | 5.31E-05 | 5.48E-05 | 9.27E-05 | 6.37E-05 | 1.67E-06 | 7.58E-05 | -6.09E-05 | 3.84E-05 | 7.32E-05 | 7.03E-05 | 4.24E-05 | 8.86E-07 | 5.62E-05 |
| 107 | 1683 | | 5.19E-05 | 5.39E-05 | 9.19E-05 | 6.39E-05 | | 7.49E-05 | -6.43E-05 | | 7.03E-05 | 6.79E-05 | 4.25E-05 | 1.35E-06 | 5.27E-05 |
| 108 | 1686 | -4.65E-05 | 2.30E-05 | 4.04E-05 | 7.25E-05 | 5.06E-05 | 1.75E-06 | 5.72E-05 | -6.86E-05 | 3.09E-05 | 6.40E-05 | 6.16E-05 | 3.57E-05 | 1.10E-06 | 4.67E-05 |
| 109 | 1708 | | 2.39E-05 | 4.27E-05 | 7.64E-05 | | | 5.98E-05 | -6.92E-05 | 3.30E-05 | | 6.43E-05 | 3.80E-05 | 1.30E-06 | 4.69E-05 |
| 110 | 1720 | -4.56E-05 | | | | 5.60E-05 | | | | | 3.19E-05 | 6.41E-05 | 6.15E-05 | | | 4.67E-05 |
| 111 | 1725 | | | 3.94E-05 | 7.17E-05 | | 1.63E-06 | 5.53E-05 | -7.33E-05 | 4.33E-05 | | | 3.63E-05 | | |
| 112 | 1734 | | | | | | | | | | | | 5.64E-05 | | 1.19E-06 | |
| 113 | 1748 | -4.48E-05 | 2.26E-05 | | | 5.56E-05 | 1.65E-06 | | -9.88E-05 | | 6.18E-05 | 5.59E-05 | 3.61E-05 | | 4.39E-05 |
| 114 | 1757 | | 3.86E-05 | 6.97E-05 | | | | 5.32E-05 | | 3.74E-05 | 5.97E-05 | | | 1.05E-06 | 4.08E-05 |
| 115 | 1758 | -4.61E-05 | 1.98E-05 | 3.79E-05 | 6.90E-05 | 5.39E-05 | 1.56E-06 | 5.28E-05 | -9.96E-05 | 3.48E-05 | | 5.34E-05 | 3.34E-05 | 8.24E-07 | |
| 116 | 1759 | -4.61E-05 | 1.82E-05 | 3.67E-05 | 6.70E-05 | 5.42E-05 | 1.20E-06 | 5.13E-05 | -0.000101 | 5.79E-05 | 5.85E-05 | 5.17E-05 | 3.31E-05 | 7.90E-07 | 3.99E-05 |
| 117 | 1781 | -4.54E-05 | 2.16E-05 | 4.02E-05 | 7.23E-05 | 5.91E-05 | 1.18E-06 | 5.71E-05 | -9.90E-05 | | 6.01E-05 | 5.44E-05 | 3.61E-05 | | 4.25E-05 |
| 118 | 1787 | | | 4.07E-05 | 7.27E-05 | | | 5.77E-05 | -9.84E-05 | 4.86E-05 | 6.06E-05 | | 3.65E-05 | 7.75E-07 | 4.31E-05 |
| 119 | 1791 | -3.90E-05 | 4.64E-05 | 4.36E-05 | 7.62E-05 | 6.43E-05 | 1.18E-06 | 6.08E-05 | -9.58E-05 | 4.86E-05 | 6.19E-05 | 5.69E-05 | 3.81E-05 | | 4.43E-05 |
| 120 | 1816 | | | | | | | | | | | 7.18E-05 | 6.82E-05 | | | 5.59E-05 |
| 121 | 1827 | -4.02E-05 | 3.50E-05 | 3.82E-05 | 7.15E-05 | 6.07E-05 | | 5.56E-05 | -9.46E-05 | | | 6.66E-05 | 3.77E-05 | 1.14E-06 | |
| 122 | 1851 | -4.12E-05 | 3.44E-05 | | | 6.01E-05 | 1.32E-06 | | | | | 7.02E-05 | | | 5.43E-05 |
| 123 | 1882 | | | 3.88E-05 | 7.20E-05 | | | 5.70E-05 | -7.85E-05 | | | | | 3.70E-05 | 9.87E-07 | |
| 124 | 1901 | | | | | | | | | 4.83E-05 | 7.13E-05 | 6.63E-05 | | | 5.50E-05 |
| 125 | 1909 | -3.73E-05 | 3.36E-05 | 3.84E-05 | 7.12E-05 | 5.88E-05 | 1.32E-06 | 5.66E-05 | -3.53E-05 | | 6.98E-05 | 6.45E-05 | 3.54E-05 | 1.03E-06 | 5.36E-05 |
| 126 | 1928 | | | | | | | | | | | | 3.51E-05 | | |
| 127 | 1942 | | 2.93E-05 | | | | | | | 4.93E-05 | | | | | |
| 128 | 1956 | -3.99E-05 | 3.03E-05 | 4.22E-05 | 7.40E-05 | 5.60E-05 | 1.20E-06 | 6.07E-05 | -4.01E-05 | 4.91E-05 | 6.97E-05 | 6.45E-05 | 3.22E-05 | 9.76E-07 | 5.39E-05 |
| 129 | 1992 | -4.12E-05 | 1.18E-05 | 3.48E-05 | 6.45E-05 | 5.71E-05 | | 5.16E-05 | | | | 5.79E-05 | 3.24E-05 | 7.52E-07 | |
| 130 | 2015 | -4.13E-05 | 1.09E-05 | | | 5.70E-05 | 1.29E-06 | | -8.97E-05 | | 6.13E-05 | | 3.23E-05 | | 4.74E-05 |
| 131 | 2018 | -3.77E-05 | 2.22E-05 | 5.26E-05 | 7.85E-05 | 5.54E-05 | 1.21E-06 | 7.08E-05 | -8.93E-05 | 4.88E-05 | 6.52E-05 | 6.00E-05 | 3.09E-05 | 5.43E-07 | 5.06E-05 |
| 132 | 2027 | -5.05E-05 | 5.56E-06 | | | 5.50E-05 | 1.19E-06 | | -9.34E-05 | | 5.81E-05 | 5.42E-05 | | 4.66E-07 | 4.41E-05 |
| 133 | 2033 | -5.04E-05 | 4.85E-06 | 4.18E-05 | 7.05E-05 | | 9.31E-07 | 5.99E-05 | | | 5.74E-05 | 5.40E-05 | 3.00E-05 | 4.02E-07 | 4.36E-05 |
| 134 | 2042 | | | | | | | | | 0.000104 | | | | | |
| 135 | 2047 | | | 4.18E-05 | 7.09E-05 | 5.71E-05 | | 6.02E-05 | -9.17E-05 | | 5.76E-05 | 5.43E-05 | 3.04E-05 | | 4.38E-05 |
| 136 | 2049 | -5.11E-05 | 3.27E-06 | 4.11E-05 | 6.87E-05 | 5.68E-05 | 9.19E-07 | 5.84E-05 | -9.19E-05 | 0.000101 | | | 2.99E-05 | 6.16E-07 | |
| 137 | 2076 | | 4.91E-08 | | | | | | | | | 5.34E-05 | | | |
| 138 | 2080 | -5.32E-05 | -5.79E-07 | 2.73E-05 | 5.36E-05 | 5.80E-05 | 1.01E-06 | 4.49E-05 | -7.73E-05 | | 5.01E-05 | | 2.86E-05 | 9.24E-08 | 3.66E-05 |
| 139 | 2085 | -5.30E-05 | -1.50E-06 | 2.60E-05 | 5.26E-05 | 5.74E-05 | 1.01E-06 | 4.38E-05 | -7.90E-05 | 0.0001 | 5.01E-05 | 5.37E-05 | 2.84E-05 | 1.59E-07 | 3.67E-05 |
| 140 | 2096 | | | | | | | | | | | 4.76E-05 | | | |
| 141 | 2105 | -5.07E-05 | -1.56E-06 | 2.68E-05 | 5.31E-05 | 5.79E-05 | 1.09E-06 | 4.44E-05 | 1.81E-06 | 0.000101 | 5.03E-05 | | 2.85E-05 | 4.21E-07 | 3.66E-05 |
| 142 | 2115 | | -3.28E-06 | | | | | | -8.59E-06 | | | 4.51E-05 | | | 3.67E-05 |
| 143 | 2116 | -5.11E-05 | -3.30E-06 | 2.36E-05 | 4.94E-05 | 5.57E-05 | 1.09E-06 | 4.13E-05 | -5.06E-06 | 9.28E-05 | 4.72E-05 | 4.51E-05 | 2.44E-05 | 3.03E-07 | 3.66E-05 |
| 144 | 2127 | | | 2.47E-05 | 5.11E-05 | | | 4.26E-05 | -1.75E-05 | 8.94E-05 | | 4.48E-05 | 2.47E-05 | | |
| 145 | 2128 | -5.13E-05 | -6.83E-06 | 2.03E-05 | 4.24E-05 | 5.06E-05 | 1.37E-06 | 3.77E-05 | -2.09E-05 | | 3.94E-05 | 3.55E-05 | 1.87E-05 | 3.34E-07 | 2.90E-05 |
| 146 | 2148 | -4.10E-05 | 2.17E-05 | 2.46E-05 | 4.23E-05 | 6.10E-05 | 1.36E-06 | 3.98E-05 | 5.54E-05 | | | 3.66E-05 | 2.28E-05 | | |
| 147 | 2156 | -5.15E-05 | 1.30E-05 | | | 6.08E-05 | | | 1.27E-05 | 9.20E-05 | 3.90E-05 | | | 2.07E-07 | 2.90E-05 |
| 148 | 2175 | | | 2.62E-05 | 4.17E-05 | | | 3.96E-05 | | | | | 3.12E-05 | | |
| 149 | 2185 | | | | | | | | | 0.000149 | | 3.89E-05 | | | |
| 150 | 2226 | -5.27E-05 | -9.80E-06 | 2.46E-05 | 3.96E-05 | 6.02E-05 | 1.38E-06 | 3.81E-05 | 3.60E-05 | 0.00014 | 3.78E-05 | 3.81E-05 | 3.03E-05 | 1.74E-07 | 2.79E-05 |
| 151 | 2244 | | | 2.19E-05 | 3.48E-05 | | | 3.37E-05 | | | | 3.52E-05 | 3.16E-05 | | |
| 152 | 2261 | -5.36E-05 | -1.60E-05 | 2.07E-05 | 3.24E-05 | 5.70E-05 | 2.01E-06 | 3.19E-05 | 1.43E-05 | 0.000127 | 3.45E-05 | 3.52E-05 | 3.20E-05 | 5.97E-07 | 2.46E-05 |
| 153 | 2272 | -5.37E-05 | | 1.92E-05 | 3.10E-05 | 5.63E-05 | | 3.05E-05 | 2.82E-05 | | 3.51E-05 | | | 6.74E-07 | 2.47E-05 |
| 154 | 2351 | -5.29E-05 | -1.40E-05 | 2.04E-05 | 3.39E-05 | 5.90E-05 | 1.99E-06 | | | | 3.66E-05 | 3.62E-05 | 3.50E-05 | | 2.66E-05 |
| 155 | 2357 | -5.35E-05 | -1.56E-05 | 1.83E-05 | 3.15E-05 | 5.53E-05 | 1.94E-06 | 2.86E-05 | 2.62E-05 | 0.000126 | 3.32E-05 | 3.32E-05 | 3.26E-05 | 3.15E-07 | 2.27E-05 |
| 156 | 2365 | -5.49E-05 | -1.98E-05 | 1.50E-05 | | 5.08E-05 | 2.11E-06 | 2.52E-05 | 2.33E-05 | 0.000121 | 3.02E-05 | 3.09E-05 | | -1.49E-07 | 2.01E-05 |
| 157 | 2369 | | | 1.49E-05 | 2.91E-05 | | | 2.44E-05 | | | | | | | |
| 158 | 2376 | -5.40E-05 | -7.39E-06 | | | | 2.24E-06 | | | 0.00012 | 2.95E-05 | 3.02E-05 | 3.11E-05 | | 1.96E-05 |
| 159 | 2384 | | | 1.94E-05 | 3.14E-05 | 5.28E-05 | | 2.85E-05 | 3.86E-05 | | | | 3.24E-05 | | |
| 160 | 2388 | -5.41E-05 | -1.55E-05 | | | | | | 2.69E-05 | | 2.87E-05 | 2.94E-05 | | 5.09E-07 | 1.87E-05 |
| 161 | 2415 | | | 1.83E-05 | 2.87E-05 | 5.34E-05 | 2.50E-06 | 2.67E-05 | | 0.000119 | | 2.69E-05 | 3.36E-05 | | |
| 162 | 2433 | | -2.60E-05 | | | | | | -1.55E-05 | | 2.66E-05 | | | | 1.75E-05 |
| 163 | 2446 | -5.73E-05 | -2.79E-05 | 1.73E-05 | 2.69E-05 | 5.27E-05 | 2.46E-06 | 2.52E-05 | | | | 2.34E-05 | 3.17E-05 | -4.48E-07 | |
| 164 | 2448 | | -2.76E-05 | 1.76E-05 | 2.74E-05 | | | 2.56E-05 | -2.16E-05 | | 2.53E-05 | | 3.20E-05 | | 1.64E-05 |
| 165 | 2449 | -5.69E-05 | -2.88E-05 | 1.62E-05 | 2.58E-05 | 5.09E-05 | | 2.45E-05 | -2.13E-05 | 0.000117 | 2.42E-05 | 2.29E-05 | 3.03E-05 | -3.95E-07 | 1.47E-05 |
| 166 | 2451 | -5.52E-05 | -2.78E-05 | 1.86E-05 | 2.72E-05 | 5.57E-05 | 2.22E-06 | 2.65E-05 | -2.10E-05 | 0.000115 | 2.49E-05 | 2.35E-05 | | -4.66E-07 | 1.54E-05 |
| 167 | 2494 | | | 1.85E-05 | 2.70E-05 | | | 2.63E-05 | | | | | 3.05E-05 | | |
| 168 | 2539 | | | | | | | | | 7.36E-05 | | 2.37E-05 | | | |
| 169 | 2552 | -5.41E-05 | -2.57E-05 | 2.01E-05 | 2.85E-05 | 5.80E-05 | | 2.80E-05 | -2.38E-05 | | 2.43E-05 | | 3.04E-05 | | 1.48E-05 |
| 170 | 2570 | | | | | | | | | | | 2.40E-05 | | | |
| 171 | 2585 | | -2.55E-05 | | | 5.68E-05 | 2.22E-06 | | | 7.77E-05 | 2.52E-05 | | | | 1.57E-05 |
| 172 | 2587 | -5.40E-05 | | | 2.73E-05 | | | | -3.09E-05 | | | | 3.33E-05 | -1.93E-07 | |
| 173 | 2594 | | | 1.81E-05 | | | | 2.59E-05 | | 7.00E-05 | | 2.21E-05 | | | 1.39E-05 |
| 174 | 2665 | | -2.15E-05 | | | 5.56E-05 | | | -2.53E-05 | | 2.32E-05 | | 3.28E-05 | | |
| 175 | 2677 | -5.58E-05 | | 1.52E-05 | 2.55E-05 | | | 2.32E-05 | | | 2.36E-05 | 2.26E-05 | | | 1.43E-05 |
| 176 | 2678 | | -1.68E-05 | | | | 2.09E-06 | | -1.73E-05 | | | | 3.52E-05 | -6.07E-07 | |
| 177 | 2696 | | | | | 5.84E-05 | | | | | | | | | |
| 178 | 2704 | -5.57E-05 | -2.40E-05 | 1.10E-05 | 2.17E-05 | 5.71E-05 | 2.02E-06 | 1.96E-05 | -2.29E-05 | | 1.83E-05 | 1.78E-05 | 3.19E-05 | -3.55E-07 | 9.37E-06 |
| 179 | 2717 | | | | | | | | | 6.94E-05 | | | | | |
| 180 | 2769 | -5.55E-05 | | 8.10E-06 | 1.86E-05 | 5.73E-05 | | 1.62E-05 | -2.39E-05 | | 6.34E-05 | 1.80E-05 | | 3.24E-05 | -2.65E-07 | 9.27E-06 |
| 181 | 2808 | -5.54E-05 | -2.42E-05 | 8.05E-06 | 1.86E-05 | 5.74E-05 | | 1.62E-05 | | | | 1.80E-05 | 1.87E-05 | | | 9.30E-06 |
| 182 | 2822 | | | | | | 2.17E-06 | | -2.54E-05 | | | | | 3.05E-05 | | |
| 183 | 2831 | -5.57E-05 | -2.98E-05 | 5.43E-06 | 1.59E-05 | 5.50E-05 | 2.27E-06 | 1.29E-05 | | 6.50E-05 | 1.21E-05 | 1.36E-05 | | -2.32E-07 | 4.50E-06 |
| 184 | 2856 | -4.31E-05 | 1.79E-05 | | | 5.56E-05 | 2.19E-06 | | -1.86E-05 | 6.03E-05 | 1.61E-05 | 1.48E-05 | 3.08E-05 | -5.04E-07 | 8.30E-06 |
| 185 | 2929 | | | 6.92E-06 | 2.10E-05 | | | | 1.46E-05 | -1.91E-05 | 5.75E-05 | | 1.49E-05 | 3.09E-05 | | |
| 186 | 2942 | -4.54E-05 | -1.91E-05 | 6.87E-06 | 2.09E-05 | 5.54E-05 | 1.62E-06 | 1.45E-05 | -1.92E-05 | | | 1.34E-05 | | 3.07E-05 | -1.68E-07 | 5.42E-06 |
| 187 | 2950 | -4.63E-05 | -2.77E-05 | 7.53E-06 | 1.37E-05 | 4.76E-05 | 1.63E-06 | 8.45E-06 | -2.04E-05 | | -3.26E-07 | 1.03E-05 | 2.11E-05 | -1.98E-07 | -3.46E-06 |
| 188 | 2964 | | | | | 4.92E-05 | 1.59E-06 | | | | 5.62E-05 | | | | | |
| 189 | 2980 | -4.78E-05 | | 1.26E-05 | 1.38E-05 | | | 8.87E-06 | | | | 8.17E-07 | 1.07E-05 | | 6.69E-07 | -3.07E-06 |
| 190 | 2982 | -4.92E-05 | -3.51E-05 | -3.74E-06 | 7.73E-06 | 4.28E-05 | 9.33E-07 | 3.45E-06 | -2.22E-05 | 5.85E-05 | | 8.11E-06 | 1.83E-05 | 2.09E-07 | |
| 191 | 2984 | | -1.51E-05 | | | | | | | | | 7.11E-05 | | | | 4.76E-05 |
| 192 | 3004 | | | | | | 5.77E-05 | | | 6.29E-05 | 5.11E-05 | | | | | |
| 193 | 3011 | -4.55E-05 | | 2.55E-07 | 1.23E-05 | | | 6.58E-06 | | 5.19E-05 | | | 2.03E-05 | | | |
| 194 | 3014 | | | | | | | 8.68E-07 | | | | 4.55E-05 | 9.35E-06 | | -1.29E-08 | 3.32E-05 |
| 195 | 3040 | -4.47E-05 | -2.08E-05 | -2.90E-06 | 8.80E-06 | 5.48E-05 | | 3.48E-06 | 2.16E-05 | | 2.41E-05 | 7.30E-05 | 1.77E-05 | | 2.05E-05 |
| 196 | 3061 | | -2.30E-05 | | | 5.50E-05 | 6.39E-07 | | 2.12E-05 | 6.03E-05 | | | | -3.82E-07 | |
| 197 | 3071 | -4.66E-05 | | -6.31E-06 | 5.16E-06 | | 6.41E-07 | | | | | | 1.63E-05 | | |
| 198 | 3085 | | -3.05E-05 | | | 5.02E-05 | 1.63E-06 | | 2.99E-05 | | 1.61E-05 | 2.85E-06 | 1.62E-05 | -3.49E-07 | 1.35E-05 |
| 199 | 3120 | -4.73E-05 | | -8.18E-06 | 2.05E-06 | | -2.18E-06 | | 5.87E-05 | | | | | |
| 200 | 3143 | | -3.43E-05 | | | 4.90E-05 | 2.47E-08 | | 1.78E-05 | | 1.35E-05 | 8.67E-07 | | -4.63E-07 | 1.07E-05 |
| 201 | 3159 | -4.74E-05 | | -9.87E-06 | 3.26E-07 | | | -3.85E-05 | | | | | | |
| 202 | 3199 | | | | | | | | | 5.99E-05 | | | 1.64E-05 | | |
| 203 | 3214 | | -3.13E-05 | | | 4.94E-05 | | | | | | 1.45E-05 | 2.03E-06 | | | 1.22E-05 |
| 204 | 3225 | -4.73E-05 | | -1.16E-05 | -3.34E-06 | | -1.19E-07 | -6.79E-06 | 1.34E-05 | | | -1.74E-06 | 1.33E-05 | -6.00E-07 | |

| | A | B | C | D | E | F | G | H | I | J | K | L | M | N | O |
|---|---|---|---|---|---|---|---|---|---|---|---|---|---|---|---|
| 205 | 3232 | | -3.65E-05 | | | 4.97E-05 | | | | | 1.13E-05 | | | | 9.12E-06 |
| 206 | 3240 | -4.71E-05 | | -1.07E-05 | -2.18E-05 | | | -5.71E-06 | 9.26E-06 | 5.47E-05 | | | 1.35E-05 | -3.30E-07 | |
| 207 | 3258 | | 5.15E-05 | | | 4.41E-05 | | | | | | -4.28E-06 | | | -2.06E-06 |
| 208 | 3278 | -2.40E-05 | 5.07E-05 | -1.39E-05 | -5.53E-06 | 4.31E-05 | -1.19E-07 | -8.79E-06 | 1.61E-05 | | 5.91E-06 | -6.04E-06 | 9.59E-06 | -2.41E-06 | -3.65E-05 |
| 209 | 3315 | -7.37E-05 | | -1.45E-05 | -6.83E-06 | | -4.94E-08 | -9.42E-06 | -8.11E-06 | 0.000107 | | -6.66E-06 | 7.94E-06 | | |
| 210 | 3320 | | 4.88E-06 | | | 4.37E-05 | | | | | 4.52E-06 | | | | -4.64E-06 |
| 211 | 3351 | | | | | | 3.66E-07 | | -4.88E-06 | | | | | | -7.41E-08 | |
| 212 | 3366 | -7.18E-05 | 5.50E-06 | -1.28E-05 | -5.67E-06 | 4.25E-05 | 3.65E-07 | -7.95E-06 | | 9.09E-05 | 2.02E-06 | -5.99E-06 | 7.00E-06 | -6.81E-08 | -6.69E-06 |
| 213 | 3370 | -7.07E-05 | | -1.07E-05 | -3.11E-06 | | | -5.35E-06 | -4.15E-06 | | | -3.98E-06 | 8.80E-06 | | |
| 214 | 3380 | -7.07E-05 | 6.11E-06 | -1.09E-05 | -3.46E-06 | 4.28E-05 | 3.29E-07 | -5.61E-06 | -4.45E-06 | 7.95E-06 | 1.35E-07 | -4.14E-06 | 8.60E-06 | -3.27E-07 | -6.64E-06 |
| 215 | 3384 | -6.79E-05 | | -1.12E-05 | -3.98E-06 | | | -6.11E-06 | | | | -4.18E-06 | | -1.85E-07 | -6.72E-06 |
| 216 | 3386 | -6.88E-05 | -4.62E-06 | -1.13E-05 | -4.09E-06 | 4.21E-05 | -6.13E-07 | -6.17E-06 | -5.70E-06 | | 9.96E-08 | -4.22E-06 | 8.71E-06 | | -6.82E-06 |
| 217 | 3428 | | | | | 4.09E-05 | | | | | | | | | | |
| 218 | 3437 | | -9.04E-06 | | | | | | | | 7.59E-05 | | | | | |
| 219 | 3470 | -6.88E-05 | -9.72E-06 | -1.10E-05 | -4.28E-06 | 4.02E-05 | -6.61E-07 | -5.84E-06 | -6.18E-06 | | | -2.63E-06 | -5.00E-06 | 6.75E-06 | -3.10E-08 | -7.36E-06 |
| 220 | 3503 | | | | | | | | | | | 3.20E-06 | | | | 6.66E-07 |
| 221 | 3505 | | -2.10E-05 | | | 3.92E-05 | -6.06E-07 | | -1.90E-05 | 7.16E-05 | | | | 6.64E-06 | -6.99E-08 | |
| 222 | 3516 | -6.98E-05 | | -1.10E-05 | -6.25E-06 | | | -5.92E-06 | | | | | -4.53E-06 | | | 8.41E-08 |
| 223 | 3518 | | -2.27E-05 | | | 3.91E-05 | -8.01E-07 | | -2.03E-05 | | | 1.57E-06 | -5.23E-06 | 5.70E-06 | -1.38E-07 | |
| 224 | 3519 | -6.98E-05 | | -1.08E-05 | -6.02E-06 | | | -7.84E-07 | -5.64E-06 | | | 1.66E-06 | | | | 3.35E-07 |
| 225 | 3561 | | | | | | | | | -2.29E-05 | | | | 6.52E-06 | | |
| 226 | 3564 | | -2.40E-05 | | | 4.06E-05 | | | | | | -4.96E-06 | -7.42E-06 | | -1.59E-07 | -5.35E-06 |
| 227 | 3576 | | | | | | | | | | | | | 7.24E-06 | | |
| 228 | 3578 | -7.08E-05 | | -1.25E-05 | -8.53E-06 | | -9.79E-07 | -7.39E-06 | 2.31E-05 | 6.39E-05 | | -7.31E-06 | | | | |
| 229 | 3593 | | -3.42E-05 | | | 3.71E-05 | | | | | | -7.93E-06 | | | -3.62E-07 | -8.66E-06 |
| 230 | 3601 | -7.11E-05 | | -1.25E-05 | -1.02E-05 | | | -7.27E-06 | | 6.38E-05 | | | | | | |
| 231 | 3615 | | | | | | | | | | 6.33E-05 | | | | | |
| 232 | 3633 | -7.22E-05 | -5.16E-05 | -2.24E-05 | -2.45E-05 | 2.91E-05 | -1.59E-06 | -2.03E-05 | -4.01E-06 | | -1.85E-05 | -1.50E-05 | 2.54E-07 | -2.25E-06 | -1.98E-05 |
| 233 | 3671 | | -5.00E-05 | | | 2.89E-05 | | | 2.17E-05 | | 5.85E-05 | | -1.49E-05 | 3.84E-07 | | -1.95E-05 |
| 234 | 3682 | | | | | | | | | | 5.05E-05 | -1.69E-05 | | | | |
| 235 | 3685 | -7.10E-05 | | -2.20E-05 | -2.38E-05 | | -1.76E-06 | -1.98E-05 | 2.19E-05 | | | | -5.24E-07 | -2.19E-06 | | |
| 236 | 3701 | | | | | | | | | | | -1.64E-05 | -1.28E-05 | | | -1.72E-05 |
| 237 | 3704 | | -5.28E-05 | | | 2.47E-05 | | | 1.96E-06 | 4.68E-05 | | | -2.73E-06 | | | |
| 238 | 3745 | -6.98E-05 | | -1.59E-05 | -2.06E-05 | | | -1.46E-05 | | | | | | | | |
| 239 | 3759 | | | | | | | | | | | | -9.71E-06 | | | |
| 240 | 3766 | | | | | | -1.92E-06 | | | | 4.12E-05 | -8.36E-06 | | | -2.35E-06 | -1.33E-05 |
| 241 | 3776 | -6.69E-05 | -1.14E-05 | -1.16E-05 | -1.53E-05 | 2.20E-05 | | -1.12E-05 | -3.12E-05 | | | -8.05E-06 | -3.92E-06 | | -2.81E-06 | -1.75E-05 |
| 242 | 3790 | | -7.03E-06 | | | 2.00E-05 | -2.10E-06 | | | | 5.49E-06 | -1.53E-05 | | | | |
| 243 | 3827 | -6.74E-05 | | -1.42E-05 | -1.73E-05 | | | -1.29E-05 | -1.34E-05 | | -1.63E-05 | -9.81E-06 | -4.53E-06 | | -1.82E-05 |
| 244 | 3848 | | -8.59E-06 | | | | | | | | | | | | | |
| 245 | 3862 | -7.17E-05 | | -5.70E-05 | -7.33E-05 | 4.99E-06 | -2.47E-06 | -6.72E-05 | | | | -3.09E-05 | -1.02E-05 | -4.09E-06 | |
| 246 | 3886 | -7.17E-05 | -2.40E-05 | -5.71E-05 | -7.37E-05 | 5.00E-06 | -2.47E-06 | -6.71E-05 | -3.68E-05 | 5.35E-05 | -2.15E-05 | -3.10E-05 | -1.02E-05 | -4.08E-06 | -2.42E-05 |
| 247 | 3894 | -7.35E-05 | -3.61E-05 | | | | -2.47E-06 | | -3.74E-05 | | -2.25E-05 | -3.16E-05 | | -4.12E-06 | -2.56E-05 |
| 248 | 3895 | | | | | 4.58E-06 | | | | | | | | | | |
| 249 | 3901 | -7.45E-05 | | -6.01E-05 | -7.88E-05 | | -2.70E-06 | -7.05E-05 | | | | | -1.17E-05 | -4.35E-06 | | |
| 250 | 3906 | | | | | | -2.41E-06 | | | | | | -3.19E-05 | | | |
| 251 | 3910 | | -3.32E-05 | | | 1.01E-05 | | | -3.60E-05 | | -1.75E-05 | | | -3.98E-06 | -1.90E-05 |
| 252 | 3925 | -7.33E-05 | | -5.23E-05 | -6.51E-05 | | | -6.02E-05 | | | | | | | | |
| 253 | 3930 | | | | | | | | | | 3.69E-05 | | | | | |
| 254 | 3939 | -7.35E-05 | -3.77E-05 | -5.34E-05 | -6.68E-05 | 6.44E-06 | -2.47E-06 | -6.15E-05 | -3.54E-05 | | -2.46E-05 | -3.49E-05 | -1.29E-05 | -3.95E-06 | -2.49E-05 |
| 255 | 3943 | | -3.80E-05 | | | 6.40E-06 | | | | | | | | | | |
| 256 | 3946 | -7.33E-05 | | -4.81E-05 | -5.90E-05 | | -2.47E-06 | -5.40E-05 | | 3.69E-05 | | | -1.28E-05 | | | |
| 257 | 3997 | | | | | | | | | -4.06E-05 | 2.86E-05 | -2.65E-05 | -3.69E-05 | | | -2.71E-05 |
| 258 | 4014 | | -3.99E-05 | | | 8.44E-06 | | | | 2.30E-05 | | -3.50E-05 | -1.24E-05 | -4.00E-06 | |
| 259 | 4029 | | | -4.95E-05 | -6.28E-05 | | | -5.86E-05 | | 2.61E-05 | | | | | | |
| 260 | 4043 | -7.06E-05 | | | | | | | | -3.46E-05 | 3.04E-05 | -2.27E-05 | | | | -2.24E-05 |
| 261 | 4047 | -7.10E-05 | -4.54E-05 | -5.28E-05 | -6.72E-05 | 5.80E-06 | -2.62E-06 | -6.27E-05 | | 3.25E-06 | | -3.90E-05 | -1.53E-05 | -4.16E-06 | |
| 262 | 4052 | -7.14E-05 | -5.40E-05 | -5.76E-05 | -7.39E-05 | 1.02E-06 | -2.49E-06 | -7.00E-05 | -3.69E-05 | | -2.88E-05 | | | -4.09E-06 | -2.86E-05 |
| 263 | 4071 | -7.18E-05 | -5.38E-05 | -5.78E-05 | -7.43E-05 | 5.73E-07 | -2.52E-06 | -7.04E-05 | | | -2.77E-05 | -3.85E-05 | -1.57E-05 | -4.07E-06 | -2.80E-05 |
| 264 | 4082 | | | | | | | | | -2.03E-05 | 1.31E-06 | | | | -4.08E-06 | |
| 265 | 4083 | | | | | | -2.14E-06 | | | | | | -3.77E-05 | -1.45E-05 | | |
| 266 | 4088 | -7.07E-05 | -5.48E-05 | -5.80E-05 | -7.49E-05 | 3.29E-07 | -2.17E-06 | -7.07E-05 | -3.47E-05 | | -2.83E-05 | -3.80E-05 | -1.46E-05 | -4.16E-06 | -2.85E-05 |
| 267 | 4090 | -7.09E-05 | | -5.80E-05 | -7.48E-05 | | -2.14E-06 | -7.08E-05 | | | -2.83E-05 | -3.79E-05 | | -4.19E-06 | -2.84E-05 |
| 268 | 4129 | | | | | | -2.36E-06 | | -4.52E-05 | -2.62E-05 | | | -4.29E-06 | | | |
| 269 | 4133 | -6.84E-05 | -7.46E-05 | -7.28E-05 | -8.87E-05 | -6.22E-06 | -2.62E-06 | -8.80E-05 | | -2.62E-05 | | -4.25E-05 | -1.71E-05 | -4.41E-06 | |
| 270 | 4155 | | | | | | -2.60E-06 | | | | | -3.39E-05 | | | | -3.28E-05 |
| 271 | 4165 | | | | | | | | -4.53E-05 | | | | | | | |
| 272 | 4177 | -6.86E-05 | -7.50E-05 | -7.15E-05 | -8.78E-05 | -5.44E-06 | | -8.71E-05 | -4.46E-05 | -2.09E-05 | -3.33E-05 | -4.26E-05 | -1.69E-05 | -3.96E-06 | -3.43E-05 |
| 273 | 4190 | | | | | | | | | | | | -3.54E-05 | | -4.03E-06 | -3.44E-05 |
| 274 | 4192 | | | | | | -8.66E-06 | -3.07E-06 | | -4.66E-05 | | | | | | |
| 275 | 4200 | | -7.97E-05 | | | | | | | | | | | | | |
| 276 | 4229 | -7.11E-05 | | | -7.17E-05 | -8.55E-05 | | | -8.71E-05 | -4.46E-05 | -2.09E-05 | -3.33E-05 | -4.26E-05 | -1.69E-05 | -3.96E-06 | -3.43E-05 |
| 277 | 4240 | -7.11E-05 | -7.95E-05 | -7.21E-05 | -8.65E-05 | -9.45E-06 | -1.69E-06 | -8.77E-05 | -4.67E-05 | | -4.00E-05 | -4.33E-05 | -1.72E-05 | -1.14E-06 | -3.61E-05 |
| 278 | 4273 | | | | | -1.00E-05 | -1.76E-06 | | | | | | | | | |
| 279 | 4286 | -6.81E-05 | -6.30E-05 | -6.25E-05 | -7.42E-05 | | | -7.74E-05 | | -2.32E-05 | -3.63E-05 | -2.88E-05 | -3.76E-06 | | -2.39E-05 |
| 280 | 4292 | | | | | | | | | -3.91E-05 | | | | | -5.74E-07 | |
| 281 | 4302 | -6.81E-05 | -6.50E-05 | -6.30E-05 | -7.46E-05 | -9.64E-06 | -1.75E-06 | -7.80E-05 | | | -3.74E-05 | -2.99E-05 | -4.51E-06 | | -2.46E-05 |
| 282 | 4320 | | | | | | | | | -3.64E-05 | | | | | | |
| 283 | 4324 | | | | | -1.04E-05 | | | | | -2.44E-05 | -3.73E-05 | -2.79E-05 | -2.58E-06 | -3.51E-07 | -2.53E-05 |
| 284 | 4338 | | | | | | | | | -6.71E-05 | | | | | | |
| 285 | 4340 | -6.72E-05 | -6.52E-05 | -6.38E-05 | -7.60E-05 | -1.16E-05 | -2.79E-07 | -7.91E-05 | -2.16E-05 | | -4.35E-05 | -3.32E-05 | -7.27E-06 | 7.85E-07 | -3.02E-05 |
| 286 | 4352 | -6.70E-05 | -6.85E-05 | -6.44E-05 | -7.69E-05 | -1.20E-05 | -4.16E-07 | -8.00E-05 | -2.48E-05 | | -4.72E-05 | -3.61E-05 | -1.10E-05 | 7.08E-07 | -3.34E-05 |
| 287 | 4360 | -6.88E-05 | -9.17E-05 | -7.03E-05 | -8.58E-05 | -2.16E-05 | -5.75E-07 | -8.77E-05 | -2.54E-05 | | -5.97E-05 | -4.84E-05 | -1.61E-05 | 5.86E-07 | -4.66E-05 |
| 288 | 4361 | -7.01E-05 | -0.000103 | -7.06E-05 | -8.63E-05 | | -3.00E-07 | -8.78E-05 | | -7.00E-05 | -5.85E-05 | -4.78E-05 | -1.54E-05 | | -4.57E-05 |
| 289 | 4386 | | | | | | | | | -2.62E-05 | | | -1.49E-05 | | | |
| 290 | 4405 | | -0.000116 | -7.50E-05 | -9.05E-05 | -2.43E-05 | | -9.27E-05 | | -0.000101 | -5.92E-05 | -4.95E-05 | | | -4.78E-05 |
| 291 | 4413 | -6.73E-05 | | | | | | | | | | | | | 1.55E-06 | |
| 292 | 4419 | | | | | | | | | -1.77E-05 | -8.64E-05 | | | -1.40E-05 | | |
| 293 | 4424 | -2.95E-05 | -5.70E-05 | -7.92E-05 | -9.57E-05 | -3.06E-05 | -3.06E-07 | -9.73E-05 | 7.97E-06 | -8.51E-05 | -6.68E-05 | -5.54E-05 | -1.89E-05 | 1.17E-06 | -5.36E-05 |
| 294 | 4445 | | | | | | -3.62E-05 | | | | | | -6.02E-05 | | | |
| 295 | 4451 | | | | | | | | | | -0.00012 | | | | | |
| 296 | 4501 | | -0.000115 | | | | | | | | | -6.70E-05 | | | | -5.26E-05 |
| 297 | 4502 | -5.60E-05 | | -8.88E-05 | -0.000112 | | | -0.00011 | | | | | | | | |
| 298 | 4513 | | -0.000127 | | | -3.96E-05 | | | | -1.86E-05 | -0.000109 | -7.12E-05 | -6.63E-05 | -1.94E-05 | 1.69E-06 | -5.71E-05 |
| 299 | 4518 | -5.62E-05 | -0.000134 | -9.03E-05 | -0.000114 | -4.14E-05 | 2.43E-06 | -0.000111 | -1.98E-05 | | -7.25E-05 | -6.83E-05 | -2.16E-05 | 1.60E-06 | -5.82E-05 |
| 300 | 4541 | | | | | | | | | | -9.59E-05 | -7.55E-05 | -7.13E-05 | | 1.50E-06 | -6.10E-05 |
| 301 | 4542 | | -0.000133 | | | -4.32E-05 | | | -1.73E-05 | -1.00E-04 | | | -7.14E-05 | -2.43E-05 | | |
| 302 | 4543 | -4.95E-05 | | -9.38E-05 | -0.000121 | | | -0.000117 | | | -7.56E-05 | | | | -6.12E-05 |
| 303 | 4561 | | | | | | | | | -3.37E-06 | -0.000107 | | | | | |
| 304 | 4562 | -3.37E-06 | -8.08E-05 | | | -4.82E-05 | 3.28E-06 | | | | | | | -2.83E-05 | 1.86E-06 | |
| 305 | 4572 | | | -9.60E-05 | -0.000123 | | | -0.000119 | | | -0.000107 | -7.87E-05 | -7.68E-05 | | | -6.43E-05 |
| 306 | 4621 | | | | | | | | | | -0.000101 | | | | | |

| | A | B | C | D | E | F | G | H | I | J | K | L | M | N | O |
|---|---|---|---|---|---|---|---|---|---|---|---|---|---|---|---|
| 307 | 4623 | -2.65E-05 | -9.72E-05 | | | -4.28E-05 | | | 4.11E-05 | | | -7.55E-05 | -2.31E-05 | | |
| 308 | 4624 | -2.04E-05 | -9.13E-05 | -0.000101 | -0.000129 | -4.49E-05 | | -0.000125 | 3.88E-05 | -0.000105 | -8.24E-05 | -7.96E-05 | -2.56E-05 | 1.47E-06 | -6.79E-05 |
| 309 | 4629 | -5.33E-05 | | -0.000107 | -0.000139 | -4.79E-05 | 3.00E-06 | -0.000133 | | -9.83E-05 | | | -2.72E-05 | 1.12E-06 | |
| 310 | 4639 | | -0.000138 | | | | 3.08E-06 | | 3.25E-05 | | -8.93E-05 | -8.77E-05 | | | -7.57E-05 |
| 311 | 4641 | -6.96E-05 | | -0.000108 | -0.00014 | -4.78E-05 | | -0.000134 | | -9.79E-05 | | | | 1.31E-06 | |
| 312 | 4643 | -6.95E-05 | -0.000139 | | | -4.92E-05 | 3.31E-06 | | 1.51E-05 | -9.92E-05 | -8.99E-05 | -8.83E-05 | -2.79E-05 | | -7.63E-05 |
| 313 | 4644 | | | -0.000108 | -0.000139 | | | -0.000134 | | | | | | -3.49E-05 | 1.35E-06 | |
| 314 | 4648 | | -0.000141 | | | | | | 8.64E-06 | | -9.35E-05 | -9.20E-05 | | | -7.98E-05 |
| 315 | 4649 | -7.04E-05 | -0.000146 | -0.000108 | -0.000139 | -4.93E-05 | 2.89E-06 | -0.000135 | | -9.48E-05 | | -9.36E-05 | -3.56E-05 | 8.02E-07 | |
| 316 | 4679 | -8.13E-06 | | | | -4.62E-05 | | | | -0.000103 | -9.24E-05 | | | | -7.91E-05 |
| 317 | 4683 | | | -0.000113 | -0.000144 | | | | -0.00014 | 1.55E-05 | | | | | 5.04E-07 | |
| 318 | 4686 | | -4.89E-05 | | | | | | | | | | -3.58E-05 | | |
| 319 | 4690 | 5.80E-06 | | | | -4.83E-05 | | | | 1.86E-05 | -0.000122 | -9.56E-05 | -9.79E-05 | | | -8.25E-05 |
| 320 | 4693 | | | -0.000116 | -0.000148 | | | -0.000144 | | | | | | | | |
| 321 | 4701 | -7.36E-06 | -8.64E-05 | | | -5.44E-05 | 2.76E-06 | | 5.50E-06 | | -0.000105 | -0.000105 | -4.27E-05 | 5.77E-07 | -9.14E-05 |
| 322 | 4710 | | | | | | | | | -0.00012 | | | | | | |
| 323 | 4730 | | | -0.000111 | -0.000145 | | | -0.000141 | | -0.000129 | | | | | | |
| 324 | 4732 | | -0.000112 | | | | | | | | | | -0.000112 | -4.59E-05 | | |
| 325 | 4759 | | | | | | | | | -0.00013 | | | | | | |
| 326 | 4764 | | | | | | | | | | | -0.000106 | | | | -9.04E-05 |
| 327 | 4780 | -1.46E-05 | | | | -5.76E-05 | | | | | | | | | | |
| 328 | 4802 | | | | | | | | | 2.21E-06 | | | | | | |
| 329 | 4821 | | -0.00013 | -0.000112 | -0.000143 | | -1.69E-06 | -0.000141 | | | | | | | -1.09E-06 | |
| 330 | 4826 | -2.06E-05 | | | | | | | | | | | -0.000114 | -4.56E-05 | | |
| 331 | 4853 | | | | | | | | | | | -0.000125 | -0.000108 | | | -9.56E-05 |
| 332 | 4860 | | | | | -6.61E-05 | -1.77E-06 | | | | | | | | | |
| 333 | 4865 | | | | | | | | | -3.37E-06 | | | | | | |
| 334 | 4875 | | -0.000138 | -0.000112 | -0.000142 | | | -0.00014 | | -0.000124 | -0.000112 | -0.000114 | -4.67E-05 | -1.13E-06 | |
| 335 | 4878 | | | | | | | | | | | | | | | -9.61E-05 |
| 336 | 4883 | -2.26E-05 | | -0.000113 | -0.000144 | -7.09E-05 | -1.72E-06 | -0.000141 | -5.70E-06 | -0.000125 | | | | -1.13E-06 | |
| 337 | 4899 | | -0.000132 | | | | | | | | | -0.000113 | -4.69E-05 | | |
| 338 | 4930 | -2.47E-05 | | | | -7.53E-05 | -1.72E-06 | | -8.14E-06 | | -0.000111 | | | | | -9.55E-05 |
| 339 | 4931 | | | -0.000115 | -0.000145 | | | -0.000143 | | | | | | -3.02E-07 | |
| 340 | 4933 | | -0.000132 | | | | | | | | | | | -4.72E-05 | | |
| 341 | 4939 | -2.63E-05 | | -0.000121 | -0.000152 | -7.93E-05 | -1.94E-06 | -0.000151 | | -8.84E-06 | -0.000137 | -0.000114 | -0.000115 | | -4.03E-07 | -9.82E-05 |
| 342 | 4954 | | -0.000125 | | | | | | | -0.000136 | | -0.000113 | -4.60E-05 | | |
| 343 | 5000 | | | | | | | | | | -0.000143 | -0.000112 | | | | -9.56E-05 |
| 344 | 5003 | -2.69E-05 | -0.000127 | -0.000122 | -0.000155 | -8.11E-05 | -1.78E-06 | -0.000153 | -1.65E-05 | -0.000143 | -0.000112 | -0.000114 | -4.58E-05 | -4.17E-05 | |
| 345 | 5011 | | | | | | | | | -0.000146 | | | | | | -9.30E-05 |
| 346 | 5022 | -2.36E-05 | | | | -7.58E-05 | | | -1.69E-05 | | | | | | |
| 347 | 5023 | | -0.000163 | -0.000134 | -0.000165 | | 6.30E-07 | -0.000164 | | -0.000142 | -0.000137 | -0.000133 | -6.09E-05 | 9.88E-08 | -0.000114 |
| 348 | 5027 | -2.60E-05 | | -0.000141 | -0.000173 | -8.22E-05 | 5.15E-07 | -0.000171 | -1.82E-05 | | | | -6.78E-05 | 2.71E-07 | |
| 349 | 5043 | -2.64E-05 | -0.000175 | -0.000145 | -0.000176 | -8.69E-05 | 2.17E-06 | -0.000175 | -1.93E-05 | | -0.000139 | -0.000134 | -6.85E-05 | | -0.000115 |
| 350 | 5053 | | -0.000177 | | | | | | -2.02E-05 | -0.000139 | | | | 1.40E-06 | |
| 351 | 5058 | | | | | | | | | | | | -6.81E-05 | | | |
| 352 | 5072 | -1.94E-05 | | | | -8.39E-05 | | | | -0.000142 | -0.000137 | -0.000132 | | | -0.000113 |
| 353 | 5087 | | | -0.000143 | -0.000173 | | | -0.00017 | | | | | | | |
| 354 | 5088 | -1.54E-05 | -0.00019 | -0.000142 | -0.000172 | -8.33E-05 | 2.02E-06 | -0.00017 | -7.82E-06 | -0.000138 | -0.000138 | -0.000133 | -6.75E-05 | 7.68E-07 | -0.000114 |
| 355 | 5089 | -1.69E-05 | -0.000202 | -0.000143 | -0.000173 | -8.48E-05 | 1.85E-06 | -0.000171 | -7.80E-06 | | -0.000139 | -0.000133 | -6.88E-05 | | -0.000114 |
| 356 | 5092 | -2.06E-05 | | | | -9.11E-05 | | | | | | | | 9.41E-07 | |
| 357 | 5099 | | -0.000229 | -0.000151 | -0.000181 | | 8.52E-07 | -0.00018 | -1.35E-05 | | | -0.000139 | -7.28E-05 | | |
| 358 | 5105 | -1.83E-05 | | | | -9.60E-05 | 9.64E-07 | | 3.60E-06 | -0.000131 | -0.000145 | -0.000139 | -7.27E-05 | 6.09E-07 | -0.000119 |
| 359 | 5131 | | | -0.00015 | -0.00018 | | | -0.00018 | | | -0.000152 | | | | -0.00013 |
| 360 | 5133 | 2.21E-05 | -0.000189 | -0.000151 | -0.000181 | -9.69E-05 | 4.66E-07 | -0.000181 | 7.27E-07 | -0.000142 | | -0.000146 | -7.40E-05 | 1.02E-06 | |
| 361 | 5136 | 1.99E-05 | -0.000192 | | | -9.55E-05 | | | -1.20E-05 | -0.000136 | -0.000153 | | -7.30E-05 | 1.19E-07 | -0.000131 |
| 362 | 5137 | | | -0.000149 | -0.00018 | | | -0.00018 | | | | | -0.000146 | | |
| 363 | 5143 | 1.50E-05 | -0.000196 | | | | | | | | -0.000155 | | | | -0.000132 |
| 364 | 5159 | | -0.000196 | -0.00015 | -0.000181 | -9.72E-05 | 1.31E-06 | -0.00018 | -9.00E-06 | | -0.000155 | -0.000147 | -7.40E-05 | 3.94E-07 | -0.000133 |
| 365 | 5171 | | | | | | | | | -0.000137 | | | | | |
| 366 | 5178 | 1.07E-05 | -0.000206 | -0.000153 | -0.000186 | -0.0001 | 1.22E-06 | -0.000184 | -2.24E-05 | | | -0.000152 | -7.86E-05 | 5.66E-07 | |
| 367 | 5187 | | | | | | | | | -0.00013 | | | | | |
| 368 | 5222 | 4.10E-05 | -0.000163 | -0.000155 | -0.000187 | -0.000103 | | -0.000186 | -1.83E-05 | | -0.000164 | | -8.17E-05 | 6.26E-07 | -0.00014 |
| 369 | 5232 | | | | | | | | | -0.000157 | | -0.000156 | | | |
| 370 | 5241 | | | | | | 1.39E-06 | | | | -0.000166 | | | | -0.000142 |
| 371 | 5264 | 3.62E-05 | | | | -0.000104 | | | | | | | | | |
| 372 | 5277 | | | | | | | | -1.73E-05 | | | | | | |
| 373 | 5288 | | | | | | | | | | | | -8.35E-05 | 5.74E-07 | |
| 374 | 5292 | 3.55E-05 | -0.000178 | -0.000158 | -0.000193 | -0.000108 | 1.12E-06 | -0.000191 | | -0.000156 | -0.000169 | -0.000159 | | 4.29E-07 | -0.000144 |
| 375 | 5306 | 3.54E-05 | -0.000181 | -0.000163 | -0.000195 | -0.000109 | 2.61E-06 | -0.000193 | -1.58E-05 | -0.000156 | -0.000168 | -0.00016 | -8.23E-05 | 4.97E-07 | -0.000143 |
| 376 | 5309 | | -0.000181 | -0.000162 | -0.000194 | -0.000109 | | -0.000193 | -1.59E-05 | -0.000157 | -0.000169 | -0.00016 | -8.21E-05 | | -0.000145 |
| 377 | 5332 | 3.59E-05 | -0.000196 | -0.000166 | -0.000199 | -0.000116 | 2.65E-06 | -0.000197 | -1.84E-05 | -0.00015 | | -8.92E-05 | 4.18E-07 | | |
| 378 | 5355 | 3.56E-05 | | | | -0.000122 | | | | -0.000155 | -0.000166 | -0.000156 | | 4.88E-07 | -0.000141 |
| 379 | 5360 | | -0.000195 | -0.000172 | -0.000204 | | 2.35E-06 | -0.000201 | -3.13E-05 | -0.000156 | | | | | |
| 380 | 5383 | | | | | | | | | -0.000158 | -0.000167 | | -8.76E-05 | | -0.000142 |
| 381 | 5404 | 4.19E-05 | -0.000194 | -0.000179 | -0.000211 | -0.000125 | 1.88E-06 | -0.000208 | -2.45E-05 | | | -0.000158 | | 1.12E-07 | |
| 382 | 5421 | 0.000102 | -0.000137 | -0.000179 | -0.000211 | -0.000125 | 2.12E-06 | -0.000208 | | -0.000159 | -0.000164 | -0.000157 | -8.70E-05 | 1.52E-07 | -0.000141 |
| 383 | 5430 | 7.53E-05 | | | | | | | -4.51E-05 | | | | | | |
| 384 | 5444 | 6.45E-05 | -0.000166 | -0.00018 | -0.000212 | -0.000124 | 1.80E-06 | -0.00021 | -4.73E-05 | | -0.000164 | -0.000158 | -8.63E-05 | 1.50E-07 | -0.000142 |
| 385 | 5457 | | | | | -0.000124 | | | | | | | | | |
| 386 | 5472 | 6.16E-05 | -0.000168 | -0.00018 | -0.000214 | -0.000125 | 1.66E-06 | -0.000211 | -4.87E-05 | | -0.000165 | -0.00016 | -8.70E-05 | -4.64E-08 | -0.000144 |
| 387 | 5474 | | | | | | | | | -0.000193 | | | | | |
| 388 | 5485 | | | -0.000181 | -0.000213 | | | -0.00021 | | | | | -8.62E-05 | 4.95E-07 | |
| 389 | 5495 | 5.88E-05 | -0.000171 | | | -0.000123 | 1.59E-06 | | -3.64E-05 | -0.000197 | -0.000165 | -0.000158 | | 3.44E-07 | -0.000143 |
| 390 | 5506 | 5.75E-05 | -0.00017 | -0.000178 | -0.000209 | -0.000124 | 1.50E-06 | -0.000206 | -2.48E-05 | -0.000196 | -0.000159 | -0.00016 | -8.54E-05 | | -0.000143 |
| 391 | 5527 | 0.000127 | -6.86E-05 | | | | | | | | -0.000166 | -0.000161 | -8.52E-05 | 1.32E-07 | -0.000145 |
| 392 | 5542 | 0.000125 | -7.20E-05 | -0.000183 | -0.000216 | -0.000128 | 1.17E-06 | -0.000211 | -2.72E-05 | | -0.000168 | -0.000162 | -8.64E-05 | -2.47E-08 | -0.000146 |
| 393 | 5548 | 0.000125 | | -0.000183 | -0.000215 | | | -0.000211 | | | -0.000196 | -0.000167 | | -8.63E-05 | -2.06E-07 | -0.000146 |
| 394 | 5581 | | -7.92E-05 | -0.000193 | -0.000226 | -0.000133 | | -0.000221 | -6.58E-06 | | | -0.000165 | | | |
| 395 | 5584 | 0.000117 | -9.19E-05 | | | -0.000135 | 8.69E-07 | | | -0.000197 | -0.000181 | | -9.02E-05 | | |
| 396 | 5594 | 0.000118 | -9.11E-05 | -0.000191 | -0.000222 | -0.000134 | | -0.000219 | -1.81E-05 | -0.000198 | | -0.00016 | -8.99E-05 | -8.55E-07 | -0.000148 |
| 397 | 5602 | 0.000118 | -9.15E-05 | -0.000191 | -0.000223 | -0.000134 | 2.81E-07 | -0.00022 | -1.81E-05 | | -0.000181 | -0.00016 | -8.96E-05 | -9.05E-07 | -0.000148 |
| 398 | 5627 | | | 0.000164 | 0.000111 | | | 0.000141 | | | -8.81E-06 | 1.90E-05 | | | 3.15E-05 |
| 399 | 5639 | 0.000143 | | | | | | | | | | | | | |
| 400 | 5640 | | -7.92E-05 | | | -0.000127 | -8.81E-07 | | -4.60E-06 | -0.0002 | | | | | |
| 401 | 5652 | | | | | | | | | -0.0002 | -3.05E-05 | -7.22E-06 | -8.47E-05 | | 8.05E-06 |
| 402 | 5653 | 0.000126 | -9.19E-05 | 6.31E-05 | 2.04E-05 | -0.000129 | | 4.22E-05 | -9.40E-06 | -0.000197 | | | -8.76E-05 | -8.17E-07 | |
| 403 | 5660 | 0.000127 | | 6.76E-05 | 2.65E-05 | | | 4.61E-05 | | | | -3.74E-05 | -1.45E-05 | | 2.43E-06 |
| 404 | 5677 | | -0.000106 | | | | -8.04E-07 | | -1.25E-05 | | | | | | |
| 405 | 5707 | | | | | | -0.000126 | | | | | | | | |
| 406 | 5740 | 0.000121 | | 5.78E-05 | 1.53E-05 | | | 3.88E-05 | | | -4.67E-05 | -2.41E-05 | -8.73E-05 | | -6.59E-06 |
| 407 | 5763 | 0.000121 | -0.000117 | 5.78E-05 | 1.51E-05 | -0.000127 | -9.91E-07 | 3.86E-05 | -2.24E-05 | -0.000182 | -4.67E-05 | -2.42E-05 | -8.73E-05 | -9.12E-07 | -6.68E-06 |
| 408 | 5797 | 0.000116 | | | | | -1.25E-06 | | | | -4.67E-05 | | -8.66E-05 | -1.46E-06 | -6.73E-06 |

| | A | B | C | D | E | F | G | H | I | J | K | L | M | N | O |
|---|---|---|---|---|---|---|---|---|---|---|---|---|---|---|---|
| 409 | 5800 | | -5.17E-05 | | | -0.000123 | | | 1.16E-05 | | | -1.19E-05 | | | |
| 410 | 5802 | 0.00012 | -6.23E-05 | 6.13E-05 | 1.78E-05 | -0.000124 | -3.02E-07 | 4.11E-05 | 8.89E-06 | -0.000182 | -4.83E-05 | -1.44E-05 | -8.78E-05 | -7.13E-07 | -7.31E-06 |
| 411 | 5803 | 0.000122 | | 6.70E-05 | 2.59E-05 | | | 4.66E-05 | 9.22E-06 | -0.000178 | -4.53E-05 | -9.93E-06 | -8.40E-05 | -5.99E-07 | -3.34E-06 |
| 412 | 5804 | | | | | | | | | -0.000161 | | | | | |
| 413 | 5814 | 0.000118 | -0.00012 | 4.16E-05 | -9.52E-06 | -0.000149 | -4.96E-07 | 1.92E-05 | 5.83E-06 | | -6.07E-05 | -2.70E-05 | -9.85E-05 | -1.21E-06 | -1.72E-05 |
| 414 | 5824 | | -0.000129 | | | | | | 3.68E-07 | | | | -9.73E-05 | -1.45E-06 | |
| 415 | 5825 | | | | | -0.000145 | | | | | | -5.83E-05 | -2.38E-05 | | | -1.48E-05 |
| 416 | 5834 | 0.00011 | | 4.92E-05 | 1.33E-07 | | | 2.73E-05 | -1.04E-05 | | | | | | |
| 417 | 5851 | | -0.000124 | | | -0.000145 | | | -1.05E-05 | -9.97E-05 | -6.07E-05 | -2.35E-05 | -9.61E-05 | | -1.56E-05 |
| 418 | 5865 | 0.000109 | | | | | | | | | | | | | |
| 419 | 5866 | 0.000104 | -0.000149 | 3.65E-05 | -1.42E-05 | -0.000149 | -8.10E-07 | 1.51E-05 | -1.30E-05 | | -6.80E-05 | -3.27E-05 | -0.000101 | -2.24E-06 | -2.28E-05 |
| 420 | 5883 | | -0.000151 | 3.28E-05 | -1.85E-05 | -0.000151 | -4.77E-07 | 1.22E-05 | 5.87E-06 | | -7.00E-05 | -3.58E-05 | -0.000104 | -2.07E-06 | -2.46E-05 |
| 421 | 5888 | 0.000101 | | 2.41E-05 | -2.82E-05 | | -5.86E-07 | 4.44E-06 | | | | | -0.000108 | -2.17E-06 | |
| 422 | 5917 | 0.000104 | -0.000148 | 4.16E-05 | -8.81E-06 | -0.000145 | | 2.17E-05 | -1.90E-05 | | -4.29E-05 | -2.09E-06 | | | -1.83E-05 |
| 423 | 5918 | 0.000103 | -0.000155 | 3.78E-05 | -1.29E-05 | -0.000146 | -8.02E-07 | 1.82E-05 | -1.92E-05 | -0.000104 | -4.48E-05 | -4.03E-06 | -0.000101 | -2.75E-06 | -3.53E-06 |
| 424 | 5927 | | -5.34E-05 | | | | | | | -0.000118 | | | | | |
| 425 | 5985 | | | | | -0.000145 | | | | | | -7.15E-05 | -4.05E-05 | -9.54E-05 | | -2.53E-05 |
| 426 | 6021 | 0.000111 | -6.43E-05 | 2.26E-05 | -3.38E-05 | -0.000145 | -9.71E-07 | -4.43E-07 | -2.82E-05 | -0.000142 | | | -9.63E-05 | -2.77E-06 | |
| 427 | 6028 | 0.000111 | -7.06E-05 | 2.09E-05 | -3.63E-05 | -0.000147 | -7.48E-07 | -2.10E-06 | -2.80E-05 | | -7.29E-05 | -4.33E-05 | -9.65E-05 | -2.59E-06 | -2.61E-05 |
| 428 | 6037 | | | 0.00026 | | | | | | 7.34E-05 | | | | | |
| 429 | 6041 | 0.000146 | 0.000165 | 2.07E-05 | -3.84E-05 | -0.000146 | | -1.71E-06 | | | | -8.55E-05 | | -9.65E-05 | | -3.53E-05 |
| 430 | 6057 | | | | | | -1.01E-06 | | | 4.61E-05 | -0.00015 | | -6.53E-05 | | -3.03E-06 | |
| 431 | 6063 | 0.000101 | 0.000114 | -3.98E-05 | -6.21E-05 | -0.000161 | | -2.16E-05 | | | | | | | | |
| 432 | 6072 | 9.19E-05 | 8.18E-05 | -2.76E-05 | -9.34E-05 | -0.000167 | -1.08E-06 | -4.17E-05 | 3.81E-05 | -0.000148 | -0.000119 | -8.44E-05 | -0.000104 | -3.29E-06 | -7.02E-05 |
| 433 | 6083 | 9.02E-05 | | -3.62E-05 | -0.00011 | -0.000171 | -1.18E-07 | -5.20E-05 | 3.62E-05 | | -0.000128 | -9.44E-05 | -0.000106 | -2.88E-06 | -7.91E-05 |
| 434 | 6094 | | | | | | | | | -0.000165 | | | | | |
| 435 | 6103 | | 7.25E-05 | | | | 5.33E-07 | | | -0.000172 | -0.000138 | -9.67E-05 | -0.000109 | -2.18E-06 | -8.58E-05 |
| 436 | 6139 | 9.26E-05 | 5.54E-05 | -4.24E-05 | -0.000117 | -0.00017 | 5.72E-07 | -5.72E-05 | 4.34E-05 | -0.000233 | -0.000141 | -0.0001 | -0.00011 | -2.13E-06 | -8.85E-05 |
| 437 | 6152 | 8.97E-05 | | | | | | | | -0.000236 | | | | | |
| 438 | 6156 | | 4.27E-05 | | | | 1.78E-07 | | | 3.89E-05 | -0.000144 | -0.000104 | -0.000114 | -2.37E-06 | -9.11E-05 |
| 439 | 6165 | 9.03E-05 | 4.33E-05 | -4.43E-05 | -0.00012 | -0.000171 | | -5.93E-05 | 3.91E-05 | | | | | | |
| 440 | 6178 | | | | | | | | | | | -0.000144 | -0.000103 | -0.000113 | | -9.02E-05 |
| 441 | 6185 | 9.05E-05 | | -4.79E-05 | -0.000124 | -0.00017 | | -6.29E-05 | | | | | | | |
| 442 | 6199 | 9.26E-05 | 4.75E-05 | -4.83E-05 | -0.000125 | -0.000171 | 2.02E-08 | -6.34E-05 | 4.17E-05 | -0.000234 | -0.000145 | -0.000104 | -0.000113 | -2.40E-06 | -9.07E-05 |
| 443 | 6250 | 0.000158 | 0.0002 | -4.27E-05 | -0.00012 | -0.000168 | -5.18E-07 | -5.80E-05 | 0.000146 | | -0.000136 | -9.58E-05 | -0.000106 | -2.85E-06 | -8.23E-05 |
| 444 | 6273 | 0.000151 | 0.00017 | -3.91E-05 | -0.000118 | -0.000172 | | -5.45E-05 | 0.00017 | -0.000216 | -0.000136 | -9.55E-05 | -0.000107 | -3.19E-06 | -8.22E-05 |
| 445 | 6274 | | | | | | -5.49E-07 | | | | | -0.000149 | -0.000111 | -0.000106 | | -9.67E-05 |
| 446 | 6278 | 0.000139 | 0.000138 | -5.38E-05 | -0.000138 | -0.00017 | | -6.87E-05 | 1.05E-05 | | | | | | |
| 447 | 6303 | | | | | | | | | -0.000222 | -0.000149 | -0.000109 | -0.000103 | -2.28E-06 | -9.52E-05 |
| 448 | 6307 | 0.000137 | | -5.48E-05 | -0.000135 | -0.000166 | -3.19E-07 | -6.91E-05 | | | | | | | |
| 449 | 6321 | | | | | | | | | 1.58E-05 | -0.000218 | | | | | |
| 450 | 6334 | | 0.00015 | | | | | | | | | | | | |
| 451 | 6407 | | | | | | | | | -0.000225 | -0.00015 | -0.00011 | -0.000101 | | -9.54E-05 |
| 452 | 6484 | | | | | | | | | -0.000249 | | | | -5.90E-07 | |
| 453 | 6493 | 0.000138 | 0.000138 | -6.04E-05 | -0.000141 | -0.000168 | 2.60E-08 | -7.30E-05 | 1.56E-05 | | -0.000156 | | -0.000118 | -0.000105 | -6.84E-07 | -0.000102 |
| 454 | 6503 | 0.00014 | | -5.78E-05 | -0.000139 | -0.000164 | | -7.07E-05 | | | | | | | |
| 455 | 6523 | | | | | | | | 1.96E-05 | -0.000249 | | | | | |
| 456 | 6529 | 0.00014 | 0.000139 | -5.82E-05 | -0.00014 | -0.000164 | | -7.08E-05 | 1.94E-05 | | -0.000157 | -0.000117 | -0.000106 | | -0.000103 |
| 457 | 6534 | | | | | | | | | | -0.000158 | -0.000118 | -0.000107 | -3.69E-08 | -0.000104 |
| 458 | 6565 | 0.00022 | | | | | -2.13E-07 | | 0.00011 | | | | | | |
| 459 | 6566 | 0.000197 | 0.000128 | -6.52E-05 | -0.000148 | -0.00017 | -1.86E-07 | -7.86E-05 | 9.89E-05 | -0.000248 | -0.000162 | -0.000122 | -0.000109 | -3.37E-07 | -0.000108 |
| 460 | 6581 | 0.000192 | 0.00012 | -6.84E-05 | -0.000152 | -0.000172 | | -8.34E-05 | | | | | | -3.69E-07 | |
| 461 | 6606 | | | | | | | | 4.66E-05 | | | | | | |
| 462 | 6608 | 0.000191 | | -7.33E-05 | -0.000158 | -0.000175 | 3.19E-07 | -8.88E-05 | | -0.000241 | -0.000159 | -0.000123 | -0.000105 | | -0.000105 |
| 463 | 6613 | | 0.000115 | | | | | | 4.37E-05 | | | | | | |
| 464 | 6629 | | | | | | 4.96E-07 | | | | -0.000157 | -0.000121 | -0.000104 | 2.11E-07 | -0.000102 |
| 465 | 6741 | 0.000183 | 8.40E-05 | -9.10E-05 | -0.000174 | -0.000188 | | -0.000105 | 4.23E-05 | -0.000259 | | | | | |
| 466 | 6749 | | | | | | | | | | -0.000148 | -0.000112 | -0.000104 | | -9.13E-05 |
| 467 | 6772 | 0.000181 | 7.47E-05 | -8.67E-05 | -0.000166 | -0.000182 | | -9.91E-05 | 3.65E-05 | | -0.000172 | -0.000129 | -0.000109 | | -0.000115 |
| 468 | 6773 | | | | | | 7.35E-07 | | 3.51E-05 | | -0.000173 | -0.00013 | -0.000112 | 7.55E-07 | -0.000116 |
| 469 | 6790 | | | -8.71E-05 | -0.000167 | -0.000183 | | -9.95E-05 | | -0.000254 | | | | | |
| 470 | 6807 | | 0.00018 | 7.25E-05 | | | | | | -0.000251 | | | | | |
| 471 | 6823 | 0.00018 | 7.29E-05 | -9.94E-05 | -0.000183 | -0.00019 | | -0.000113 | 3.05E-05 | -0.000265 | -0.000198 | -0.000153 | -0.00012 | 5.39E-07 | -0.000138 |
| 472 | 6876 | | | | | | | | | -0.000269 | | | | | |
| 473 | 6899 | 0.000209 | 0.000103 | -0.000118 | -0.000195 | -0.000198 | 4.39E-07 | -0.000127 | 3.97E-05 | | -0.000215 | -0.000168 | -0.000129 | | -0.000152 |
| 474 | 6904 | | | | | | | | | | | | | 3.45E-07 | |
| 475 | 6940 | 0.000261 | 0.000125 | -0.000118 | -0.000194 | -0.0002 | | -0.000125 | 4.64E-05 | -0.000246 | -0.00021 | -0.000162 | -0.000127 | | -0.000148 |
| 476 | 6946 | 0.000255 | 0.000122 | -0.000127 | -0.000201 | -0.000204 | -9.67E-07 | -0.000134 | 4.68E-05 | | -0.00021 | -0.000163 | -0.000126 | -1.62E-06 | -0.000149 |

| | A | B | C | D | E | F | G | H | I | J | K | L | M | N | O |
|---|---|---|---|---|---|---|---|---|---|---|---|---|---|---|---|
| 1 | Rank | Inf | 0.002 | 0.0015 | 0.001 | 0.0005 | 0.0001 | 0.0013 | 0.0020 Ran | 0.0015 Ran | 0.0010 Ran | 0.0005 Ran | 0.0001 Ran | 0.0013 Random | |
| 2 | 10 | -0.000233 | 0.000103 | 0.000113 | 7.97E-05 | 1.76E-05 | -8.47E-06 | 6.60E-05 | 0.000169 | 0.000142 | 0.000194 | 0.000157 | 9.75E-05 | 1.31E-06 | 0.000188 |
| 3 | 25 | -0.000233 | 0.000103 | 0.000113 | 7.97E-05 | 1.79E-05 | | 6.61E-05 | 0.000169 | 0.00015 | 0.000194 | 0.000157 | 9.79E-05 | | 0.000187 |
| 4 | 61 | -0.000231 | 0.000104 | 0.000115 | 8.16E-05 | 2.02E-05 | -9.17E-06 | 6.78E-05 | 0.000168 | 0.000153 | 0.000194 | 0.000156 | 9.86E-05 | 1.40E-06 | 0.000187 |
| 5 | 93 | -0.000231 | | | | | -1.03E-05 | | | | | | | 1.34E-06 | |
| 6 | 115 | -0.000234 | 8.90E-05 | 0.000113 | 7.93E-05 | 1.82E-05 | -1.03E-05 | 6.57E-05 | 0.000162 | 0.000151 | 0.00019 | 0.000153 | 9.75E-05 | | 0.000183 |
| 7 | 144 | -0.000235 | 8.70E-05 | 0.000112 | 7.86E-05 | 1.75E-05 | | 6.52E-05 | 0.000159 | 0.000169 | 0.000189 | 0.000152 | 9.73E-05 | 1.48E-06 | 0.000183 |
| 8 | 145 | -0.000232 | 8.70E-05 | 0.000111 | 7.78E-05 | 1.70E-05 | -1.04E-05 | 6.44E-05 | 0.000159 | 0.000217 | 0.000189 | 0.000152 | 9.66E-05 | 1.57E-06 | 0.000182 |
| 9 | 181 | -0.000241 | 7.83E-05 | 0.000108 | 7.47E-05 | 1.40E-05 | -1.20E-05 | 6.08E-05 | 0.000152 | 0.000211 | 0.000185 | 0.000148 | 9.44E-05 | 1.28E-06 | 0.000178 |
| 10 | 194 | -0.000242 | 7.52E-05 | 0.000103 | | | -1.20E-05 | | 0.000151 | | 0.000184 | 0.000147 | 9.27E-05 | 1.19E-06 | 0.000177 |
| 11 | 197 | -0.000242 | 7.58E-05 | 0.000104 | 7.26E-05 | 1.17E-05 | -1.19E-05 | 5.89E-05 | 0.000151 | | 0.000186 | 0.000149 | 9.32E-05 | 7.55E-07 | 0.000178 |
| 12 | 213 | -0.000244 | 7.59E-05 | 0.000103 | 7.24E-05 | 1.25E-05 | -1.12E-05 | 5.87E-05 | 0.00015 | 0.000196 | 0.000184 | 0.000147 | 9.08E-05 | 1.43E-06 | 0.000177 |
| 13 | 233 | -0.000149 | 7.01E-05 | 9.38E-05 | 6.59E-05 | 2.98E-06 | -1.12E-05 | 5.17E-05 | 0.000236 | | | 0.000178 | 0.00014 | 8.54E-05 | 1.43E-06 | 0.00017 |
| 14 | 234 | -0.000144 | 6.94E-05 | 9.19E-05 | 6.41E-05 | 7.94E-07 | -1.24E-05 | 5.00E-05 | 0.000233 | 0.000192 | 0.000171 | 0.000134 | 8.22E-05 | 1.31E-06 | 0.000163 |
| 15 | 239 | | | 9.68E-05 | 6.79E-05 | | | 5.45E-05 | | | | | | | |
| 16 | 245 | -0.000196 | 6.86E-05 | | | | -3.65E-06 | | 0.000141 | 0.0002 | 0.000169 | 0.000132 | 8.15E-05 | 6.56E-06 | 0.00016 |
| 17 | 259 | -0.000208 | 6.72E-05 | 9.20E-05 | 6.37E-05 | 3.47E-06 | -2.14E-06 | 5.04E-05 | 0.000131 | | 0.000168 | 0.000131 | 8.22E-05 | 6.72E-06 | 0.00016 |
| 18 | 261 | -0.000208 | 6.70E-05 | 9.18E-05 | 6.35E-05 | 3.30E-06 | -2.11E-06 | 5.02E-05 | 0.000131 | 0.000178 | 0.000168 | 0.000132 | 8.23E-05 | 6.87E-06 | 0.00016 |
| 19 | 276 | -0.000189 | 6.83E-05 | | | | -2.14E-06 | | 0.000139 | 0.000176 | 0.000168 | 0.000131 | 8.22E-05 | | 0.00016 |
| 20 | 284 | -0.000217 | 6.55E-05 | 8.68E-05 | 5.91E-05 | 2.05E-06 | -4.26E-06 | 4.49E-05 | 0.000109 | 0.000175 | 0.000166 | 0.00013 | 8.18E-05 | 5.30E-06 | 0.000157 |
| 21 | 327 | -0.000247 | 6.30E-05 | 9.17E-05 | 5.84E-05 | | -5.10E-06 | 4.78E-05 | | | 0.000165 | 0.000129 | 8.17E-05 | 3.94E-06 | 0.000156 |
| 22 | 334 | -0.000247 | 6.28E-05 | 8.99E-05 | 5.81E-05 | 5.28E-06 | -5.28E-06 | 4.64E-05 | 7.51E-05 | 0.000171 | | | | 3.90E-06 | |
| 23 | 344 | -0.000251 | 5.83E-05 | 8.33E-05 | 5.21E-05 | 8.67E-07 | -4.90E-06 | 4.13E-05 | 7.35E-05 | | 0.000163 | 0.000125 | 7.92E-05 | 4.27E-06 | 0.000155 |
| 24 | 351 | | 5.80E-05 | 8.33E-05 | 5.21E-05 | | | 4.12E-05 | 7.68E-05 | | | | | | |
| 25 | 352 | -0.000295 | | | | | -5.05E-06 | | | 0.00016 | | | 8.07E-05 | 4.08E-06 | |
| 26 | 357 | -0.000317 | 5.52E-05 | 8.08E-05 | 5.15E-05 | 4.69E-06 | | 4.01E-05 | 7.10E-05 | | 0.000165 | 0.000126 | | | 0.000156 |
| 27 | 366 | -0.000335 | 4.72E-05 | 6.80E-05 | | | -5.79E-06 | 2.88E-05 | 6.11E-05 | 0.00015 | 0.000159 | 0.000121 | 7.71E-05 | 3.43E-06 | 0.00015 |
| 28 | 375 | -0.000346 | 4.14E-05 | 6.24E-05 | 3.73E-05 | -3.92E-06 | -5.99E-06 | 2.43E-05 | 5.53E-05 | | 0.000152 | 0.000115 | 7.22E-05 | 3.26E-06 | 0.000143 |
| 29 | 378 | -0.000346 | 4.15E-05 | 6.28E-05 | 3.77E-05 | | | 2.47E-05 | 5.52E-05 | 0.000151 | 0.000152 | 0.000115 | 7.23E-05 | 3.12E-06 | 0.000143 |
| 30 | 380 | -0.000347 | 4.04E-05 | 6.08E-05 | 3.49E-05 | -7.24E-06 | -6.52E-06 | 2.29E-05 | 5.51E-05 | 0.000152 | 0.000148 | 0.000112 | 6.72E-05 | 2.57E-06 | 0.000138 |
| 31 | 390 | -0.000281 | 4.18E-05 | 6.23E-05 | 3.66E-05 | -4.77E-06 | -6.52E-06 | 2.44E-05 | 7.04E-05 | | 0.000148 | 0.000112 | 6.75E-05 | 3.16E-06 | 0.000139 |
| 32 | 408 | -0.000268 | 4.44E-05 | 6.76E-05 | 4.06E-05 | -2.58E-07 | -4.98E-06 | 2.83E-05 | 7.33E-05 | 5.88E-05 | 0.00015 | 0.000114 | 6.88E-05 | 2.81E-06 | 0.000141 |
| 33 | 416 | | | | 4.00E-05 | | | 2.73E-05 | | 5.80E-05 | | | 6.85E-05 | 2.86E-06 | |
| 34 | 432 | -0.000282 | 4.12E-05 | 6.82E-05 | 4.03E-05 | 1.13E-06 | -4.10E-06 | 2.75E-05 | 6.06E-05 | 5.92E-05 | 0.000151 | 0.000114 | 6.94E-05 | | 0.000142 |
| 35 | 434 | -0.000234 | 4.05E-05 | 6.72E-05 | 3.93E-05 | -1.35E-06 | -4.09E-06 | 2.64E-05 | 6.99E-05 | | 0.00015 | 0.000113 | 6.79E-05 | 2.79E-06 | 0.000141 |
| 36 | 435 | -0.000196 | 4.08E-05 | 6.79E-05 | 3.99E-05 | 2.41E-07 | -2.93E-06 | 2.69E-05 | 9.03E-05 | | 0.00015 | 0.000114 | 6.87E-05 | 3.62E-06 | 0.000141 |
| 37 | 487 | -0.000246 | 3.59E-05 | 7.16E-05 | 4.42E-05 | 4.69E-06 | -2.68E-06 | 3.13E-05 | 5.28E-05 | 6.46E-05 | 0.000151 | 0.000116 | 7.10E-05 | 3.84E-06 | 0.000142 |
| 38 | 496 | -0.000308 | 3.46E-05 | 7.10E-05 | 4.37E-05 | 3.84E-06 | -2.94E-06 | 3.09E-05 | 2.21E-05 | | 0.000151 | 0.000115 | 7.08E-05 | 3.69E-06 | 0.000142 |
| 39 | 513 | | | | 4.42E-05 | | -3.31E-06 | | | 6.77E-05 | 0.000151 | 0.000116 | | | 0.000142 |
| 40 | 526 | -0.000325 | 3.53E-05 | 7.25E-05 | 4.52E-05 | 7.01E-06 | | 3.24E-05 | 1.60E-05 | | 0.000152 | 0.000117 | 7.26E-05 | 3.63E-06 | 0.000144 |
| 41 | 530 | -0.000329 | | | 4.46E-05 | 8.22E-06 | -2.70E-06 | | | | 6.78E-05 | | 7.32E-05 | | |
| 42 | 540 | | 3.33E-05 | 6.88E-05 | | | | 2.95E-05 | 1.26E-05 | 6.50E-05 | 0.000151 | 0.000115 | | 3.52E-06 | 0.000142 |
| 43 | 544 | -0.000335 | | 6.88E-05 | 3.91E-05 | 3.59E-06 | -2.60E-06 | 2.95E-05 | | | | | 7.03E-05 | | |
| 44 | 551 | -0.000316 | 3.33E-05 | 6.93E-05 | 3.94E-05 | 3.92E-06 | -5.84E-07 | 2.99E-05 | 2.12E-05 | 5.83E-05 | 0.000151 | 0.000115 | 7.05E-05 | 3.49E-06 | 0.000142 |
| 45 | 556 | -0.000322 | 3.34E-05 | 7.01E-05 | 3.97E-05 | 4.86E-06 | -6.54E-07 | 3.05E-05 | 1.69E-05 | | 0.000153 | 0.000118 | 7.25E-05 | 3.39E-06 | 0.000144 |
| 46 | 557 | -0.000322 | 3.36E-05 | 7.03E-05 | 3.98E-05 | 5.11E-06 | -6.81E-07 | 3.06E-05 | 1.60E-05 | 5.68E-05 | 0.000153 | 0.000118 | 7.25E-05 | 3.33E-06 | 0.000144 |
| 47 | 573 | -0.000341 | 2.63E-05 | 5.94E-05 | 3.07E-05 | 1.34E-06 | -9.89E-07 | 1.95E-05 | 6.01E-06 | 6.37E-05 | 0.000144 | 0.000109 | 6.69E-05 | 2.68E-06 | 0.000136 |
| 48 | 582 | -0.000343 | 2.47E-05 | 5.52E-05 | 2.55E-05 | -5.93E-07 | -1.14E-06 | 1.46E-05 | 4.36E-06 | | 0.000135 | 0.000101 | 6.30E-05 | 2.64E-06 | 0.000127 |
| 49 | 592 | -0.00034 | 2.41E-05 | 5.58E-05 | 2.59E-05 | | -7.55E-07 | 1.46E-05 | | 6.23E-05 | | | | | |
| 50 | 594 | -0.000341 | 2.37E-05 | 5.55E-05 | 2.57E-05 | -8.43E-07 | -3.18E-06 | 1.44E-05 | 3.40E-06 | 7.06E-05 | 0.000129 | 9.39E-05 | 5.96E-05 | 2.83E-06 | 0.00012 |
| 51 | 663 | -0.000337 | 2.24E-05 | 5.71E-05 | 2.75E-05 | -1.96E-06 | -6.80E-08 | 1.62E-05 | 4.53E-06 | | 0.000131 | 9.64E-05 | 6.10E-05 | 3.33E-06 | 0.000122 |
| 52 | 716 | -0.000338 | 2.24E-05 | 5.71E-05 | 2.75E-05 | -1.85E-06 | -6.08E-08 | 1.62E-05 | 4.85E-06 | 6.33E-05 | 0.000131 | 9.62E-05 | 6.10E-05 | 3.32E-06 | 0.000122 |
| 53 | 748 | -0.000338 | | | | 9.87E-06 | | | | | | 0.000108 | | | |
| 54 | 760 | | 2.39E-05 | 6.26E-05 | 3.31E-05 | 8.79E-06 | -2.13E-08 | 2.22E-05 | 2.97E-06 | 6.03E-05 | 0.000136 | | 6.53E-05 | | 0.000128 |
| 55 | 772 | -0.000337 | 2.25E-05 | 6.12E-05 | 3.15E-05 | 7.09E-06 | 1.39E-08 | 2.08E-05 | 2.70E-06 | 6.13E-05 | 0.000134 | 0.000107 | 6.40E-05 | 3.39E-06 | 0.000126 |
| 56 | 782 | -0.000338 | 2.22E-05 | 6.07E-05 | 3.11E-05 | 6.52E-06 | 8.39E-07 | 2.04E-05 | 2.22E-06 | 6.04E-05 | 0.000134 | 0.000106 | 6.36E-05 | 4.38E-06 | 0.000125 |
| 57 | 801 | | 2.61E-05 | | | | | | | | | 0.000106 | | | |
| 58 | 813 | -0.000337 | 2.59E-05 | 6.22E-05 | 3.21E-05 | 7.73E-06 | 8.21E-07 | 2.16E-05 | -1.09E-05 | 5.36E-05 | 0.000133 | | 6.36E-05 | 4.31E-06 | 0.000124 |
| 59 | 818 | -0.000338 | 2.48E-05 | 6.13E-05 | 3.14E-05 | 6.86E-06 | 6.66E-07 | 2.08E-05 | -1.17E-05 | | 0.000132 | 0.000106 | 6.31E-05 | | 0.000124 |
| 60 | 826 | -0.000321 | | 6.13E-05 | 3.12E-05 | 6.35E-06 | | 2.08E-05 | -9.24E-06 | 5.39E-05 | 0.000131 | 0.000106 | 6.27E-05 | 4.08E-06 | 0.000123 |
| 61 | 844 | -0.000323 | 2.38E-05 | | | 6.57E-06 | 1.22E-07 | | -1.35E-05 | | | 0.000122 | 9.84E-05 | 5.55E-05 | 3.58E-06 | 0.000115 |
| 62 | 885 | -0.000347 | 2.23E-05 | 5.74E-05 | 2.85E-05 | 6.45E-06 | 4.05E-07 | 1.75E-05 | -2.55E-05 | 6.55E-05 | 0.00012 | 9.77E-05 | 5.45E-05 | 3.69E-06 | 0.000113 |
| 63 | 917 | | | | 2.76E-05 | | | | -2.22E-05 | | | 9.70E-05 | 5.40E-05 | 2.92E-06 | |
| 64 | 925 | -0.000357 | 2.30E-05 | 5.89E-05 | | 8.15E-06 | -2.22E-07 | 1.88E-05 | -3.01E-05 | | 0.000118 | | 5.40E-05 | 2.99E-06 | 0.000113 |
| 65 | 931 | -0.000287 | 2.38E-05 | 6.05E-05 | 2.97E-05 | | 1.92E-07 | 2.02E-05 | -9.76E-08 | 6.90E-05 | 0.000119 | 9.96E-05 | | 2.97E-06 | 0.000113 |
| 66 | 933 | | | | | 7.78E-06 | | | 9.25E-07 | 4.59E-05 | | 9.81E-05 | 5.39E-05 | | 0.000112 |
| 67 | 937 | | | | | | | | | | 0.000116 | | | 4.39E-06 | |
| 68 | 975 | | | | | | | | | 5.19E-05 | | | | | |
| 69 | 991 | -0.000227 | 2.36E-05 | 6.12E-05 | 3.10E-05 | | 2.14E-06 | 2.04E-05 | | | | | 5.37E-05 | | |
| 70 | 997 | -0.000232 | 2.34E-05 | 6.10E-05 | 3.06E-05 | 5.83E-06 | 2.08E-06 | 2.02E-05 | 3.18E-07 | 4.86E-05 | 0.000114 | 9.68E-05 | 5.37E-05 | 4.97E-06 | 0.000109 |
| 71 | 1043 | -0.000281 | 2.03E-05 | 5.84E-05 | 2.86E-05 | | | 1.78E-05 | -1.32E-05 | 3.87E-05 | 0.000115 | 9.78E-05 | 5.52E-05 | 4.78E-06 | 0.00011 |
| 72 | 1050 | | | | 2.75E-05 | 5.00E-06 | 1.90E-06 | | -1.44E-05 | | 0.000112 | 9.52E-05 | | 4.81E-06 | 0.000107 |
| 73 | 1063 | -0.00028 | 2.02E-05 | 5.86E-05 | | 4.79E-06 | 1.76E-06 | 1.78E-05 | -2.17E-05 | | | 9.41E-05 | 5.44E-05 | 4.35E-06 | |
| 74 | 1093 | -0.000272 | 2.01E-05 | 5.76E-05 | 2.78E-05 | 3.79E-06 | 1.68E-06 | 1.69E-05 | -1.76E-05 | 3.88E-05 | 0.000113 | 9.29E-05 | 5.36E-05 | 4.42E-06 | 0.000108 |
| 75 | 1094 | -0.000267 | | | | | | | -1.78E-05 | 3.63E-05 | 0.000113 | 9.27E-05 | | | 0.000108 |
| 76 | 1107 | | | | 2.96E-05 | | | 1.85E-05 | | | | | 5.93E-05 | 4.41E-06 | |
| 77 | 1114 | | 2.11E-05 | 5.90E-05 | | | | | | 3.62E-05 | | | | | |
| 78 | 1165 | -0.00029 | 1.96E-05 | 5.77E-05 | 2.80E-05 | 3.46E-06 | 2.37E-06 | 1.76E-05 | -3.26E-05 | | 0.000112 | 9.09E-05 | 5.79E-05 | 4.52E-06 | 0.000106 |
| 79 | 1174 | -0.000289 | 2.06E-05 | 5.82E-05 | 2.86E-05 | 3.51E-06 | | 1.81E-05 | | 8.83E-05 | | | 5.78E-05 | 4.44E-06 | |
| 80 | 1206 | | | | | | 2.24E-07 | | -3.47E-05 | | 0.000111 | 8.99E-05 | | 4.24E-06 | 0.000105 |
| 81 | 1208 | -0.000291 | 2.08E-05 | 5.79E-05 | 2.90E-05 | 3.93E-06 | 1.76E-06 | 1.80E-05 | -3.45E-05 | | 0.000111 | 9.05E-05 | 5.79E-05 | | 0.000105 |
| 82 | 1211 | | 2.05E-05 | 5.73E-05 | | 3.71E-06 | | | | 6.97E-05 | | | 5.74E-05 | 4.06E-06 | |
| 83 | 1240 | -0.000263 | 1.96E-05 | 5.36E-05 | 2.69E-05 | 1.32E-06 | 3.04E-06 | 1.48E-05 | -2.77E-05 | 4.43E-05 | 0.000108 | 8.83E-05 | 5.44E-05 | 4.28E-06 | 0.000102 |
| 84 | 1250 | -0.000304 | 1.71E-05 | 5.24E-05 | 2.56E-05 | -5.17E-08 | 4.51E-06 | 1.36E-05 | -4.53E-05 | 4.46E-05 | 0.000107 | 8.74E-05 | 5.35E-05 | 4.79E-06 | 0.000102 |
| 85 | 1260 | -0.000314 | 1.52E-05 | 4.89E-05 | 2.28E-05 | -2.27E-06 | 1.02E-06 | 1.06E-05 | -5.19E-05 | | 0.000103 | 8.46E-05 | | 4.43E-06 | 9.85E-05 |
| 86 | 1262 | -0.000279 | | | 2.32E-05 | | -2.02E-07 | | -5.13E-05 | 3.91E-05 | 0.000103 | 8.52E-05 | 5.30E-05 | 3.61E-06 | 9.91E-05 |
| 87 | 1294 | | 1.20E-05 | 4.43E-05 | | -7.16E-06 | | 6.51E-06 | -5.15E-05 | | 3.96E-05 | 9.98E-05 | 8.24E-05 | 5.10E-05 | 3.10E-06 | 9.60E-05 |
| 88 | 1301 | -0.000286 | 1.19E-05 | 4.41E-05 | 1.77E-05 | -7.23E-06 | -3.91E-07 | 6.29E-06 | -5.16E-05 | 3.98E-05 | 9.91E-05 | 8.19E-05 | | 3.09E-06 | 9.52E-05 |
| 89 | 1306 | | | | | | | | | 3.97E-05 | | | 5.07E-05 | 3.07E-06 | |
| 90 | 1316 | -0.000251 | 1.34E-05 | 4.63E-05 | 2.08E-05 | -4.64E-06 | 7.82E-07 | 8.90E-06 | -3.98E-05 | | 9.87E-05 | 8.44E-05 | 5.10E-05 | | 9.75E-05 |
| 91 | 1357 | | 1.00E-05 | 4.22E-05 | 1.76E-05 | -5.63E-06 | -1.08E-06 | | -7.11E-05 | | 9.00E-05 | 7.73E-05 | 4.90E-05 | 2.74E-06 | 8.85E-05 |
| 92 | 1404 | -0.000252 | 1.09E-05 | 4.22E-05 | | -5.36E-06 | -7.61E-07 | 1.38E-05 | | 4.31E-05 | | | | | |
| 93 | 1411 | -0.000252 | | | 1.70E-05 | -5.48E-06 | -6.72E-07 | 1.37E-05 | -7.18E-05 | | 8.89E-05 | 7.66E-05 | 4.84E-05 | 2.68E-06 | 8.76E-05 |
| 94 | 1414 | | 1.17E-05 | 4.51E-05 | 1.86E-05 | | | 1.51E-05 | -7.12E-05 | 3.84E-05 | 9.06E-05 | 7.82E-05 | 4.91E-05 | | 9.11E-05 |
| 95 | 1427 | -0.000249 | 1.12E-05 | 4.41E-05 | 1.81E-05 | -4.97E-06 | -1.70E-06 | 1.46E-05 | -7.08E-05 | | 8.98E-05 | 7.82E-05 | 4.87E-05 | 2.53E-06 | 9.08E-05 |
| 96 | 1437 | -0.000249 | 1.13E-05 | 4.47E-05 | 1.85E-05 | -4.06E-06 | -1.44E-06 | 1.52E-05 | -7.12E-05 | 4.06E-05 | 9.03E-05 | 7.88E-05 | 4.96E-05 | 2.42E-06 | 9.14E-05 |
| 97 | 1441 | | | | | | | | | 3.61E-05 | | 7.77E-05 | | | |
| 98 | 1448 | | | | | | | | | 3.78E-05 | | | | | |
| 99 | 1485 | -0.000252 | 1.02E-05 | 4.50E-05 | 1.82E-05 | -4.22E-06 | -1.93E-06 | 1.52E-05 | -7.19E-05 | | 3.51E-05 | 9.06E-05 | | 4.99E-05 | 2.09E-06 | 9.14E-05 |
| 100 | 1501 | | 2.49E-05 | 6.52E-05 | 3.83E-05 | 2.28E-06 | | 3.19E-05 | | 3.17E-05 | 0.000104 | 9.16E-05 | | | 0.000107 |
| 101 | 1504 | | | | | | | | | 3.84E-05 | | | | | |
| 102 | 1535 | -0.000246 | | | | | -1.30E-06 | | -7.84E-05 | 4.18E-05 | | | 4.77E-05 | 2.18E-06 | |

| | A | B | C | D | E | F | G | H | I | J | K | L | M | N | O |
|---|---|---|---|---|---|---|---|---|---|---|---|---|---|---|---|
| 103 | 1576 | -0.000248 | 2.36E-05 | 6.43E-05 | 3.78E-05 | 1.62E-06 | -1.28E-06 | 3.13E-05 | -7.87E-05 | | 0.000103 | 9.10E-05 | 4.77E-05 | 2.21E-06 | 0.000107 |
| 104 | 1579 | | 2.51E-05 | 6.61E-05 | | 3.22E-06 | | 3.29E-05 | | 3.87E-05 | 0.000105 | 9.30E-05 | | | 0.000109 |
| 105 | 1617 | -0.000226 | | | | | | | | | | | 4.80E-05 | | |
| 106 | 1651 | | | | | | | | -7.59E-05 | | | | | | |
| 107 | 1654 | -0.00028 | 1.17E-05 | 6.33E-05 | 3.53E-05 | 1.82E-06 | -1.23E-06 | 3.01E-05 | -7.87E-05 | 3.62E-05 | 0.000102 | 9.08E-05 | 4.71E-05 | 2.22E-06 | 0.000106 |
| 108 | 1655 | -0.00027 | 1.20E-05 | 6.38E-05 | 3.58E-05 | 1.85E-06 | -1.27E-06 | 3.09E-05 | | 3.50E-05 | | 9.11E-05 | | | |
| 109 | 1667 | -0.000279 | 1.04E-05 | 6.01E-05 | 3.31E-05 | 8.35E-07 | -1.07E-06 | 2.89E-05 | -7.74E-05 | 3.50E-05 | 0.000101 | 8.82E-05 | 4.69E-05 | 2.27E-06 | 0.000104 |
| 110 | 1683 | -0.000284 | 6.10E-06 | 5.56E-05 | 2.97E-05 | -2.36E-07 | -1.03E-07 | 2.35E-05 | -7.97E-05 | | 9.74E-05 | 8.56E-05 | 4.65E-05 | 2.61E-06 | 0.000101 |
| 111 | 1686 | | | | | 2.73E-05 | | -1.16E-06 | 2.08E-05 | -8.40E-05 | 2.60E-05 | | 7.86E-05 | 3.97E-05 | 2.39E-06 | |
| 112 | 1708 | -0.000285 | 5.17E-06 | 5.25E-05 | 2.70E-05 | 4.19E-07 | 3.35E-07 | 1.96E-05 | -8.57E-05 | 2.73E-05 | 9.66E-05 | | 4.22E-05 | 2.90E-06 | 9.98E-05 |
| 113 | 1720 | -0.000285 | | | 2.63E-05 | | | | | 2.66E-05 | 9.64E-05 | 7.87E-05 | | | 9.95E-05 |
| 114 | 1725 | -0.000288 | 3.96E-06 | 4.91E-05 | | -7.42E-07 | 6.92E-09 | 1.63E-05 | -8.92E-05 | 4.01E-05 | | | 4.04E-05 | | |
| 115 | 1734 | | | | 2.43E-05 | | | | | | | 9.36E-05 | 7.63E-05 | | 2.68E-06 | 9.69E-05 |
| 116 | 1748 | -0.000327 | 3.34E-06 | 4.79E-05 | 2.35E-05 | -1.14E-06 | -1.43E-06 | 1.53E-05 | -0.000115 | | | 7.59E-05 | 4.01E-05 | | |
| 117 | 1757 | -0.00033 | | 4.51E-05 | | | -1.83E-06 | 1.24E-05 | | 3.74E-05 | 9.19E-05 | | | 2.52E-06 | 9.46E-05 |
| 118 | 1758 | | 1.56E-05 | | 2.14E-05 | -3.96E-06 | -2.18E-06 | | -0.000116 | 3.52E-05 | | 7.24E-05 | 3.74E-05 | 2.32E-06 | |
| 119 | 1759 | -0.000332 | 6.11E-07 | 4.31E-05 | | -3.98E-06 | -1.51E-06 | 1.08E-05 | -0.000117 | 5.21E-05 | 9.06E-05 | 7.09E-05 | 3.72E-05 | 2.21E-06 | 9.33E-05 |
| 120 | 1781 | -0.000321 | 4.01E-07 | 4.43E-05 | 2.22E-05 | -3.68E-06 | -1.18E-06 | 1.19E-05 | -0.000115 | | 9.12E-05 | 7.33E-05 | 4.00E-05 | | 9.56E-05 |
| 121 | 1787 | -0.000321 | 9.38E-07 | 4.46E-05 | 2.25E-05 | -3.11E-06 | | 1.23E-05 | -0.000114 | 4.53E-05 | 9.15E-05 | 7.35E-05 | 4.01E-05 | 2.08E-06 | |
| 122 | 1791 | | | | 2.36E-05 | | | 1.34E-05 | -0.000112 | 4.52E-05 | 9.28E-05 | | 4.14E-05 | | 9.89E-05 |
| 123 | 1816 | -0.000299 | 8.12E-06 | 5.84E-05 | 3.73E-05 | 6.16E-06 | | 2.75E-05 | | | 0.000103 | 8.15E-05 | | | 0.000111 |
| 124 | 1827 | -0.000303 | 7.22E-06 | 5.69E-05 | 3.65E-05 | 5.25E-06 | -2.50E-07 | | -0.000112 | | | 7.93E-05 | 4.15E-05 | 2.49E-06 | 0.000109 |
| 125 | 1851 | | | | | | | 2.66E-05 | | | 0.0001 | | 4.06E-05 | | |
| 126 | 1882 | | 6.47E-06 | 5.69E-05 | | 4.26E-06 | | | -9.84E-05 | | | | | 2.22E-06 | |
| 127 | 1901 | -0.000269 | 6.55E-06 | 5.73E-05 | 3.73E-05 | | | 2.82E-05 | | | 0.000101 | 7.75E-05 | | | 0.000108 |
| 128 | 1909 | -0.000158 | 6.04E-06 | 5.42E-05 | 3.49E-05 | 3.25E-06 | 6.51E-06 | 2.54E-05 | -5.96E-05 | 4.55E-05 | 9.89E-05 | 7.56E-05 | 3.86E-05 | 2.26E-06 | 0.000106 |
| 129 | 1942 | -0.000185 | | | | | | 2.19E-05 | | 4.21E-05 | | | | | |
| 130 | 1956 | -0.000181 | 1.54E-06 | 5.30E-05 | 3.22E-05 | 3.90E-07 | 5.93E-07 | 2.20E-05 | -6.00E-05 | 4.21E-05 | 0.000101 | 7.65E-05 | 3.58E-05 | 2.20E-06 | 0.000107 |
| 131 | 1992 | -0.000256 | | | 2.96E-05 | | | | | | | 7.07E-05 | 3.61E-05 | 1.92E-06 | 9.91E-05 |
| 132 | 2015 | -0.000256 | -2.21E-06 | 4.69E-05 | 2.93E-05 | 2.12E-06 | -1.63E-07 | 1.95E-05 | -0.000105 | | 9.07E-05 | 7.05E-05 | 3.60E-05 | | |
| 133 | 2018 | | -3.92E-06 | 4.52E-05 | | -1.26E-06 | -2.82E-07 | 1.76E-05 | -0.000105 | 4.21E-05 | 9.59E-05 | 7.43E-05 | 3.43E-05 | 1.62E-06 | 0.000104 |
| 134 | 2027 | -0.000257 | | | 2.84E-05 | | -3.09E-07 | 1.79E-05 | -0.000109 | | | 8.63E-05 | 3.39E-05 | 1.53E-06 | 9.36E-05 |
| 135 | 2033 | | -4.65E-06 | 4.30E-05 | 2.73E-05 | -1.58E-06 | -3.09E-07 | | -0.000109 | | 8.53E-05 | 6.91E-05 | | 1.45E-06 | 9.30E-05 |
| 136 | 2042 | | | | | | | | | 8.80E-05 | | | | | |
| 137 | 2047 | -0.000252 | -3.88E-06 | 4.39E-05 | 2.79E-05 | -9.28E-07 | 6.68E-07 | 2.02E-05 | -0.000108 | | 8.54E-05 | 6.92E-05 | 3.58E-05 | 1.75E-06 | 9.30E-05 |
| 138 | 2049 | | -4.32E-06 | 4.37E-05 | | -1.18E-06 | | 2.02E-05 | | | | 6.79E-05 | 3.50E-05 | | |
| 139 | 2076 | | | | | | | | | 8.88E-05 | 8.38E-05 | | | | 9.11E-05 |
| 140 | 2080 | -0.000164 | | | 2.61E-05 | | -1.36E-06 | | -6.01E-05 | | | 5.80E-05 | 3.39E-05 | 1.28E-06 | |
| 141 | 2085 | -0.000166 | -7.71E-06 | 3.85E-05 | 2.68E-05 | -9.90E-06 | -1.28E-06 | 1.55E-05 | -6.16E-05 | 8.88E-05 | 8.38E-05 | 5.78E-05 | 3.36E-05 | 1.33E-06 | 9.11E-05 |
| 142 | 2096 | | | | | | | | | | 7.82E-05 | | | | 8.53E-05 |
| 143 | 2105 | 1.32E-05 | -7.87E-06 | 3.81E-05 | 2.67E-05 | -9.51E-06 | -6.57E-07 | 1.54E-05 | -9.72E-07 | 8.94E-05 | | 5.78E-05 | 3.39E-05 | 1.72E-06 | 8.54E-05 |
| 144 | 2115 | | | | | | | | | | 7.59E-05 | | | | |
| 145 | 2116 | 4.21E-05 | -1.15E-05 | 3.08E-05 | 2.27E-05 | -1.12E-05 | -8.74E-07 | 7.80E-06 | -8.84E-06 | 8.53E-05 | 7.58E-05 | 5.41E-05 | 3.04E-05 | 1.73E-06 | 8.12E-05 |
| 146 | 2127 | -3.99E-06 | -1.19E-05 | 3.07E-05 | 2.23E-05 | -1.04E-05 | | 7.33E-06 | -2.01E-05 | 8.27E-05 | 7.52E-05 | 5.48E-05 | | | 8.19E-05 |
| 147 | 2128 | -1.37E-05 | -1.62E-05 | 1.90E-05 | 1.33E-05 | -1.70E-05 | -7.97E-06 | -2.79E-06 | -2.34E-05 | | 6.62E-05 | 4.42E-05 | 2.56E-05 | 1.87E-06 | 7.14E-05 |
| 148 | 2148 | 8.77E-05 | | | 1.06E-05 | | | | 6.94E-05 | | 6.80E-05 | 4.74E-05 | | | 7.55E-05 |
| 149 | 2156 | | -1.71E-05 | 1.30E-05 | | | | -8.40E-06 | 2.04E-05 | 8.46E-05 | | | 2.82E-05 | 1.70E-06 | |
| 150 | 2175 | 8.02E-05 | | | | -5.84E-06 | | | | | | 5.69E-05 | | | |
| 151 | 2185 | | | | 1.78E-05 | | | | | | 0.000141 | 7.06E-05 | | | 7.72E-05 |
| 152 | 2226 | 0.000103 | -2.31E-05 | 1.39E-05 | 1.68E-05 | -7.22E-06 | -3.46E-07 | -8.24E-06 | 4.51E-06 | 0.000131 | 6.95E-05 | 5.48E-05 | 2.77E-05 | 1.68E-06 | 7.63E-05 |
| 153 | 2244 | | -2.61E-05 | 7.82E-06 | | -5.64E-06 | | -1.25E-05 | | | | 5.10E-05 | | | 7.28E-05 |
| 154 | 2261 | 7.11E-05 | -2.23E-05 | 1.36E-05 | 1.89E-05 | -1.51E-06 | -5.18E-08 | -6.78E-06 | 2.60E-05 | 0.000124 | 6.55E-05 | 5.04E-05 | 2.76E-05 | 2.02E-06 | 7.22E-05 |
| 155 | 2272 | | | | | -1.65E-08 | | | 3.30E-05 | | | | | 2.06E-06 | |
| 156 | 2351 | 3.80E-05 | -2.13E-05 | 1.62E-05 | 2.13E-05 | 2.46E-06 | | -3.81E-06 | 3.01E-05 | | 6.63E-05 | 5.21E-05 | 2.98E-05 | | 7.37E-05 |
| 157 | 2357 | 3.82E-05 | -2.38E-05 | 1.08E-05 | 1.66E-05 | 3.82E-07 | -1.01E-07 | -8.06E-06 | 2.86E-05 | 0.000125 | 6.36E-05 | 4.96E-05 | 2.80E-05 | 1.86E-06 | 7.13E-05 |
| 158 | 2359 | | | | 1.73E-05 | | | | | | | | | | |
| 159 | 2365 | 2.96E-05 | | | | | -2.43E-06 | -1.04E-05 | 2.55E-05 | 0.000118 | 6.08E-05 | | 2.46E-05 | 1.27E-06 | 6.91E-05 |
| 160 | 2376 | | -2.51E-05 | 8.10E-06 | 1.62E-05 | -1.70E-06 | -2.51E-06 | | 3.15E-05 | 0.000118 | 6.02E-05 | 4.86E-05 | | | 6.85E-05 |
| 161 | 2384 | 9.21E-05 | -2.43E-05 | 9.00E-06 | | -7.78E-07 | | -1.03E-05 | | | | 4.92E-05 | 2.66E-05 | | |
| 162 | 2388 | 4.93E-05 | | | 1.55E-05 | | | | 1.45E-05 | | | 5.91E-05 | | 1.93E-06 | 6.65E-05 |
| 163 | 2394 | | -2.23E-05 | 3.10E-05 | | | | -7.59E-07 | | | | 5.67E-05 | 2.96E-05 | | |
| 164 | 2415 | -9.02E-06 | | | 1.42E-05 | 8.37E-07 | 1.36E-06 | | -4.37E-05 | 0.000117 | 5.67E-05 | 4.86E-05 | | | 6.44E-05 |
| 165 | 2446 | | -2.39E-05 | 2.55E-05 | | -2.35E-06 | -2.36E-06 | -2.01E-06 | -4.74E-05 | | 5.45E-05 | 4.55E-05 | 2.85E-05 | 5.90E-07 | 6.17E-05 |
| 166 | 2448 | -6.99E-06 | | | 1.12E-05 | | | | | | | | 2.90E-05 | | |
| 167 | 2449 | | -2.32E-05 | 2.65E-05 | | -2.57E-06 | -2.10E-06 | -1.41E-06 | -4.72E-05 | 0.00011 | 5.40E-05 | 4.49E-05 | 2.75E-05 | 6.83E-07 | 6.10E-05 |
| 168 | 2451 | -8.46E-06 | -2.34E-05 | 2.58E-05 | 1.09E-05 | | -2.30E-06 | -1.79E-06 | -4.68E-05 | 0.000109 | 5.46E-05 | 4.56E-05 | 2.81E-05 | 6.14E-07 | 6.16E-05 |
| 169 | 2494 | | | | | -3.27E-06 | | | | | | 4.42E-05 | | | |
| 170 | 2539 | | | | | | | | | | 6.34E-05 | 5.43E-05 | | | 6.13E-05 |
| 171 | 2552 | -1.39E-05 | -2.68E-05 | 2.29E-05 | 9.47E-06 | -3.83E-06 | | -4.57E-06 | -5.01E-05 | 6.53E-05 | | 4.38E-05 | 2.79E-05 | | |
| 172 | 2570 | -7.56E-06 | | | 8.87E-06 | | | | | | | 5.54E-05 | | | 6.16E-05 |
| 173 | 2585 | | | | | | -7.00E-07 | | -5.64E-05 | 6.70E-05 | | | 2.84E-05 | 9.75E-07 | |
| 174 | 2587 | | -1.37E-05 | 3.71E-05 | | 8.35E-06 | | 6.07E-06 | | | | 4.54E-05 | | | |
| 175 | 2594 | -1.50E-05 | | | 8.54E-06 | | | | | 6.10E-05 | 5.44E-05 | | | | 6.05E-05 |
| 176 | 2665 | | -1.49E-05 | 3.43E-05 | 8.23E-06 | 5.64E-06 | | 3.85E-06 | -4.82E-05 | | | 4.47E-05 | 2.51E-05 | | |
| 177 | 2677 | -3.21E-06 | -1.44E-05 | 3.56E-05 | | 6.06E-06 | -1.14E-06 | 4.57E-06 | -4.79E-05 | | 5.49E-05 | | | | 6.12E-05 |
| 178 | 2678 | 4.81E-05 | | | 8.95E-06 | | | | | | | | 2.77E-05 | 7.13E-07 | |
| 179 | 2696 | | | | | | | | | | | 4.59E-05 | | | |
| 180 | 2704 | 3.59E-05 | -1.60E-05 | 3.00E-05 | 5.69E-06 | 3.33E-06 | -5.13E-07 | -1.20E-07 | -5.13E-05 | | 4.97E-05 | 4.49E-05 | 2.41E-05 | 8.50E-07 | 5.55E-05 |
| 181 | 2717 | | | | | | | | | 5.82E-05 | | | | | |
| 182 | 2769 | 3.60E-05 | | | 6.38E-06 | 6.44E-06 | 1.40E-07 | | -5.19E-05 | 5.40E-05 | 4.91E-05 | | 2.46E-05 | 9.95E-07 | |
| 183 | 2808 | | -1.69E-05 | 3.14E-05 | 6.33E-06 | 6.52E-06 | | 1.10E-06 | -5.21E-05 | | 4.91E-05 | 4.56E-05 | 2.47E-05 | | 5.70E-05 |
| 184 | 2822 | 3.29E-05 | -1.64E-05 | 3.23E-05 | | 7.18E-06 | | 1.95E-06 | | | | 4.53E-05 | | | |
| 185 | 2831 | | | | -3.15E-06 | | -3.03E-07 | | -5.43E-05 | 5.62E-05 | 4.44E-05 | | 2.14E-05 | 1.03E-06 | 5.25E-05 |
| 186 | 2856 | 3.67E-05 | -1.36E-05 | 3.26E-05 | | 7.85E-06 | -8.86E-07 | 3.24E-06 | | 5.33E-05 | 4.89E-05 | 4.51E-05 | | 7.67E-07 | 5.61E-05 |
| 187 | 2929 | 3.70E-05 | -1.41E-05 | 3.30E-05 | -5.21E-07 | 8.04E-06 | -8.15E-07 | 4.02E-06 | -5.32E-05 | 5.18E-05 | 4.87E-05 | 4.51E-05 | 2.36E-05 | 7.88E-07 | 5.59E-05 |
| 188 | 2942 | 3.61E-05 | -1.44E-05 | 3.18E-05 | -7.60E-07 | | -2.98E-07 | 3.28E-06 | -5.32E-05 | | | 4.46E-05 | 2.33E-05 | | |
| 189 | 2950 | | | | | -4.73E-06 | 6.41E-06 | -5.66E-07 | | -5.47E-05 | | 4.00E-05 | 1.49E-05 | 9.69E-07 | 4.85E-05 |
| 190 | 2964 | | | | | | | | | | 5.03E-05 | | 4.04E-05 | | |
| 191 | 2980 | 6.69E-05 | 6.93E-06 | 6.52E-05 | | 7.39E-06 | 1.99E-06 | 2.68E-05 | | | 4.19E-05 | | 1.60E-05 | 1.75E-06 | 5.01E-05 |
| 192 | 2982 | | | | | | | | -5.60E-05 | 5.22E-05 | | 3.70E-05 | | 1.04E-06 | 4.69E-05 |
| 193 | 2984 | | 9.11E-05 | | -2.20E-06 | | | 0.000111 | | | 0.000115 | | | | |
| 194 | 3004 | 0.000186 | | 0.000137 | | | | | 3.60E-05 | 4.30E-05 | | | 2.52E-05 | | |
| 195 | 3011 | | | | | 5.40E-06 | | | | 4.40E-05 | | 4.02E-05 | | | |
| 196 | 3014 | 0.000207 | | | -3.52E-06 | | | | | | 8.31E-05 | | | 7.20E-07 | 4.45E-05 |
| 197 | 3040 | 0.000145 | 4.67E-05 | 6.72E-05 | -5.36E-06 | 4.72E-06 | 2.61E-06 | 6.24E-05 | -1.42E-05 | | 5.76E-05 | 3.61E-05 | 2.18E-05 | | 3.14E-05 |
| 198 | 3061 | | | | | | | | -1.47E-05 | 5.30E-05 | | | 2.14E-05 | 3.11E-07 | |
| 199 | 3071 | | | | | 1.14E-06 | | | | | 4.81E-05 | 3.19E-05 | | | 2.44E-05 |
| 200 | 3085 | 0.000184 | 4.11E-05 | 5.50E-05 | -9.11E-06 | | 3.22E-06 | 5.38E-05 | -4.99E-06 | | | | 1.97E-05 | 2.88E-07 | |
| 201 | 3120 | | | | | 2.13E-06 | | | | | | 3.06E-05 | | | |
| 202 | 3143 | 0.000165 | | | -1.17E-05 | | | 4.81E-05 | -1.86E-05 | 5.50E-05 | | 4.45E-05 | 1.86E-05 | 1.83E-07 | 2.12E-05 |
| 203 | 3159 | | | 4.59E-05 | | | 3.13E-06 | | | | | | | | |
| 204 | 3199 | | 3.51E-05 | | | | | | | | 4.48E-05 | 2.99E-05 | | | 2.21E-05 |

|  | A | B | C | D | E | F | G | H | I | J | K | L | M | N | O |
|---|---|---|---|---|---|---|---|---|---|---|---|---|---|---|---|
| 205 | 3214 | 0.000148 |  |  | -1.10E-05 | 3.50E-06 |  |  |  | 5.51E-05 |  |  |  |  |  |
| 206 | 3225 |  | 2.43E-05 | 3.15E-05 |  |  | 3.87E-06 | 4.12E-05 | -2.31E-05 |  | 3.98E-05 | 2.74E-05 | 1.81E-05 | 1.11E-07 | 1.75E-05 |
| 207 | 3232 | 0.00014 |  |  | -1.44E-05 | 1.12E-06 |  |  | -2.44E-05 |  |  |  |  |  |  |
| 208 | 3240 |  | 2.34E-05 | 3.04E-05 |  |  |  | 3.99E-05 |  | 4.84E-05 |  |  | 1.84E-05 |  |  |
| 209 | 3258 |  |  |  |  | 4.32E-07 |  |  |  |  | 3.33E-05 | 2.48E-05 |  | 1.50E-07 | 1.07E-05 |
| 210 | 3278 | 0.000144 | 1.95E-05 | 2.26E-05 | -1.73E-05 | -3.30E-07 | 4.02E-06 | 3.46E-05 | -1.53E-05 |  | 2.95E-05 | 2.26E-05 | 1.45E-05 | 1.66E-07 | 8.27E-06 |
| 211 | 3315 |  |  |  |  | 1.16E-07 |  |  |  | 8.56E-05 | 2.77E-05 | 2.22E-05 |  |  | 6.98E-06 |
| 212 | 3320 | 0.00013 | 1.73E-05 | 2.20E-05 | -1.81E-05 |  |  | 3.40E-05 | -3.74E-05 |  |  |  | 1.24E-05 |  |  |
| 213 | 3351 |  |  |  |  |  | 3.67E-06 |  |  |  |  |  |  | 1.93E-07 |  |
| 214 | 3366 | 0.000146 | 1.25E-05 | 1.11E-05 | -2.19E-05 | -1.35E-05 | 3.65E-06 | 2.63E-05 | -3.12E-05 |  | 2.75E-05 | 2.33E-05 | 1.04E-05 | 2.12E-07 | 6.78E-06 |
| 215 | 3370 |  |  |  | -2.03E-05 | 1.60E-08 |  |  |  | 4.54E-05 | 2.55E-05 | 2.51E-05 |  |  | 8.61E-06 |
| 216 | 3380 | 0.00014 | 1.20E-05 | 4.19E-06 | -2.08E-05 | -8.15E-07 | 2.43E-06 | 2.55E-05 | -3.12E-05 | 3.79E-05 | 2.33E-05 | 2.47E-05 | 1.06E-05 | -3.26E-07 | 8.24E-06 |
| 217 | 3384 | 0.000154 | 2.72E-05 | 3.67E-06 |  |  |  |  |  | 3.87E-05 | 2.31E-05 |  |  |  | 7.99E-06 |
| 218 | 3386 |  |  |  | -1.99E-05 | 7.42E-07 |  |  | -3.22E-05 |  | 2.31E-05 | 2.48E-05 | 1.07E-05 | -6.49E-07 | 7.95E-06 |
| 219 | 3428 |  |  |  |  |  |  |  | 2.73E-05 |  |  |  |  |  |  |
| 220 | 3437 | 0.000156 | 1.99E-05 | 2.45E-06 |  |  | 2.65E-06 |  | -3.34E-05 |  |  |  | 8.75E-06 |  |  |
| 221 | 3470 | 0.000155 | 1.99E-05 | 1.99E-05 | -2.06E-05 | 7.23E-08 | 2.69E-06 | 2.71E-05 | -3.38E-05 | 3.37E-05 | 1.99E-05 | 2.39E-05 | 7.80E-06 | -4.94E-07 | 7.09E-06 |
| 222 | 3503 | 0.000208 | 2.32E-05 | 1.27E-05 |  |  |  | 3.40E-05 |  |  |  |  | 7.75E-06 |  |  |
| 223 | 3505 |  |  |  |  | 1.71E-06 |  |  | -4.74E-05 |  |  |  | 7.75E-06 | -4.85E-07 |  |
| 224 | 3512 |  |  |  |  |  | 3.00E-06 |  |  |  |  | 2.51E-05 |  |  |  |
| 225 | 3516 | 0.000194 | 2.22E-05 | 1.07E-05 | -1.57E-05 |  |  | 3.32E-05 |  | 3.27E-05 | 2.13E-05 |  |  |  | 8.27E-06 |
| 226 | 3518 |  | 2.15E-05 | 9.65E-06 | -1.63E-05 | 8.32E-07 | 2.98E-06 | 3.24E-05 | -4.87E-05 |  | 2.08E-05 | 2.40E-05 | 6.83E-06 | -6.37E-07 | 7.77E-06 |
| 227 | 3519 | 0.000193 |  |  | -1.63E-05 | 8.90E-07 | 2.98E-06 |  |  |  |  |  | 7.10E-06 |  |  |
| 228 | 3548 |  |  |  |  |  |  |  | -4.64E-05 |  |  |  |  |  |  |
| 229 | 3561 |  | 2.02E-05 | 8.72E-06 |  |  |  | 2.95E-05 |  |  |  | 2.10E-05 |  |  |  |
| 230 | 3564 | 0.000178 | 2.01E-05 | 8.30E-06 | -2.03E-05 | 6.14E-07 | 3.50E-06 | 2.93E-05 |  |  | 1.82E-05 |  | 4.69E-06 | -3.62E-07 | 5.39E-06 |
| 231 | 3576 | 0.000182 |  |  |  |  |  |  | -3.22E-05 |  |  |  |  |  |  |
| 232 | 3578 |  |  |  |  | 2.60E-06 |  |  |  |  | 1.87E-05 | 2.18E-05 |  |  | 5.74E-06 |
| 233 | 3593 |  |  |  | -2.31E-05 |  |  |  |  |  |  |  |  | -4.38E-07 |  |
| 234 | 3601 |  | 1.83E-05 | 5.74E-06 |  |  | 3.33E-06 | 2.76E-05 | -5.45E-05 | 3.09E-05 |  |  | 3.81E-06 |  |  |
| 235 | 3615 |  |  |  |  |  |  |  |  | 3.00E-05 |  |  |  |  |  |
| 236 | 3633 | 4.93E-05 |  |  |  |  |  |  |  |  |  | 8.90E-06 | 1.25E-05 |  | -3.09E-06 | -3.94E-06 |
| 237 | 3647 |  |  |  | -3.13E-05 | -4.96E-06 |  |  | -6.64E-05 |  |  |  | -8.89E-06 |  |  |
| 238 | 3671 | 0.000128 | 1.44E-05 | -5.02E-07 |  |  | 3.31E-06 | 2.23E-05 |  | 2.63E-05 | 8.93E-06 | 1.24E-05 |  |  | -3.73E-06 |
| 239 | 3682 | 0.000169 | 1.47E-05 | 7.18E-07 | -3.02E-05 | -4.43E-06 |  | 2.38E-05 | -5.80E-06 | 2.39E-05 |  |  | -8.16E-06 |  |  |
| 240 | 3685 |  |  |  |  |  | 2.57E-06 |  |  |  |  |  | 1.15E-05 |  | -3.08E-06 |  |
| 241 | 3701 | 0.000101 |  |  | -2.79E-05 | -4.51E-06 |  | 2.62E-05 |  |  | 1.11E-05 |  |  |  | -1.63E-06 |
| 242 | 3704 |  | 1.50E-05 | 2.08E-06 |  |  |  |  | -2.73E-05 |  |  |  | -1.06E-05 |  |  |
| 243 | 3745 | 0.000159 |  |  |  | -4.75E-06 |  |  |  | 3.09E-05 |  | 1.27E-05 |  |  | 5.51E-06 |
| 244 | 3759 |  |  |  | -2.67E-05 |  |  |  |  |  | 2.26E-05 |  |  |  |  |
| 245 | 3766 |  |  |  |  |  | 2.25E-06 |  |  |  |  |  |  | -3.32E-06 |  |
| 246 | 3776 |  | 1.12E-05 | -5.36E-06 |  |  |  | 2.66E-05 | -7.35E-05 | 2.88E-05 | 2.23E-05 | 1.42E-05 | -1.17E-05 |  | 8.29E-06 |
| 247 | 3790 | 0.000112 |  |  | -2.70E-05 | -4.73E-06 |  |  | -7.59E-05 |  |  |  |  | -3.87E-06 |  |
| 248 | 3827 | 0.000162 | 1.08E-05 | -8.27E-06 | -2.81E-05 | -5.34E-06 | 1.02E-06 | 2.46E-05 | -5.95E-05 |  | 2.02E-05 | 1.26E-05 | -1.22E-05 |  | 5.93E-06 |
| 249 | 3848 |  |  |  |  | -7.10E-07 |  |  |  |  |  |  | -1.20E-05 |  |  |
| 250 | 3862 |  |  |  | -2.94E-05 |  | -2.69E-06 |  | -6.74E-05 |  | 3.03E-06 | -7.62E-06 |  | -4.91E-06 | -1.36E-05 |
| 251 | 3881 |  |  |  |  |  |  |  |  | 3.22E-05 |  |  |  |  |  |
| 252 | 3886 | 9.51E-05 | 8.67E-06 | -9.28E-06 | -2.94E-05 | -1.39E-06 | -2.67E-06 | 2.37E-05 | -7.41E-05 | 3.22E-05 | 2.98E-06 | -7.61E-06 | -1.28E-05 | -4.90E-06 | -1.36E-05 |
| 253 | 3894 |  |  |  | -2.96E-05 |  | -2.86E-06 |  | -7.44E-05 | 3.43E-05 | 1.13E-07 |  |  | -4.93E-06 | -1.58E-05 |
| 254 | 3895 | 9.42E-05 | 9.30E-06 | -7.72E-06 |  |  |  | 2.53E-05 |  |  |  |  |  |  |  |
| 255 | 3901 |  |  |  |  |  | -3.96E-06 |  |  |  |  | -9.68E-06 | -1.52E-05 | -5.13E-06 | -1.81E-05 |
| 256 | 3906 |  |  |  |  |  |  |  | -7.53E-05 |  | -6.42E-07 |  |  |  |  |
| 257 | 3910 |  |  |  | -2.70E-05 | 2.00E-06 |  |  |  |  |  |  |  | -4.76E-06 |  |
| 258 | 3939 | 9.56E-05 | 4.70E-06 | -8.65E-06 |  |  | -3.07E-06 | 2.42E-05 | -7.49E-05 |  | -2.39E-06 | -7.94E-06 | -1.63E-05 | -4.71E-06 | -1.64E-05 |
| 259 | 3943 |  |  |  |  |  |  |  |  | 1.97E-05 |  |  | -1.64E-05 |  |  |
| 260 | 3946 |  |  |  |  | -2.83E-07 |  |  |  |  | 1.97E-05 |  | -6.81E-06 |  |  |
| 261 | 3953 |  |  |  | -2.77E-05 |  |  |  |  |  |  |  |  |  | -1.31E-05 |
| 262 | 3997 | 6.97E-05 | -5.64E-07 | -1.59E-05 |  | -5.91E-06 |  | 1.82E-05 | -8.07E-05 | 1.55E-05 | -4.89E-06 |  |  | -4.93E-06 |  |
| 263 | 4014 |  |  |  | -3.03E-05 |  | -4.40E-06 | 2.22E-05 |  | 1.17E-05 | -3.58E-06 | -9.90E-06 | -1.57E-05 |  | -1.54E-05 |
| 264 | 4029 |  | 1.27E-06 |  |  |  |  |  |  | 1.32E-05 |  |  |  |  |  |
| 265 | 4043 | 8.42E-05 |  | -1.42E-05 |  |  |  |  | -7.54E-05 | 1.45E-05 |  |  |  |  |  |
| 266 | 4047 |  |  |  | -3.32E-05 | -6.99E-06 |  |  |  |  | -8.58E-06 | -1.40E-05 | -1.88E-05 | -5.27E-06 | -2.02E-05 |
| 267 | 4052 |  |  |  |  |  | -5.18E-06 |  | -7.74E-05 |  |  |  |  | -5.23E-06 |  |
| 268 | 4071 |  | 4.11E-06 | -8.94E-06 |  |  | -5.35E-06 | 2.58E-05 | -7.71E-05 | -5.56E-06 | -5.25E-06 | -1.23E-05 | -1.97E-05 | -5.27E-06 | -1.72E-05 |
| 269 | 4082 | 0.000105 |  |  |  |  |  |  |  |  |  |  |  | -5.28E-06 |  |
| 270 | 4083 | 0.000108 | 3.14E-06 | -1.11E-05 | -2.83E-05 | -9.29E-07 | -3.23E-06 | 2.53E-05 |  | -3.81E-06 | -5.31E-06 | -1.20E-05 | -1.82E-05 |  | -1.78E-05 |
| 271 | 4088 | 9.33E-05 | 2.88E-06 | -1.20E-05 | -2.89E-05 | -1.33E-06 | -3.34E-06 | 2.48E-05 | -8.42E-05 |  | -5.77E-06 | -1.23E-05 | -1.84E-05 | -5.34E-06 | -1.82E-05 |
| 272 | 4090 |  |  | -1.18E-05 |  |  |  |  | -8.51E-05 |  | -5.86E-06 |  |  |  |  |
| 273 | 4129 | 7.49E-05 |  |  |  |  | -3.64E-06 |  |  | -4.02E-06 |  |  | -2.10E-05 | -5.34E-06 |  |
| 274 | 4133 |  | 3.23E-06 |  | -2.92E-05 | -1.30E-06 |  |  | -0.0001 | -4.00E-05 | -1.15E-05 | -1.56E-05 |  | -5.50E-06 | -2.21E-05 |
| 275 | 4155 |  |  | -1.32E-05 |  |  |  | 2.35E-05 |  |  |  |  |  |  |  |
| 276 | 4165 |  |  |  |  |  | -2.03E-06 |  |  |  |  |  |  |  |  |
| 277 | 4177 | 7.81E-05 |  |  | -2.95E-05 | -7.80E-07 |  |  |  | -2.39E-05 | -1.13E-05 | -1.53E-05 | -1.80E-05 | -4.69E-06 | -2.15E-05 |
| 278 | 4190 |  | 1.41E-06 | -1.36E-05 |  |  | -2.00E-06 | 2.22E-05 | -0.000105 |  |  |  |  |  |  |
| 279 | 4192 |  |  |  |  |  |  |  |  |  |  |  | -1.75E-05 |  |  |
| 280 | 4229 | 8.72E-05 | 5.93E-06 | -1.34E-05 | -2.99E-05 | -1.31E-06 | -1.89E-06 | 2.28E-05 | -0.000101 | -3.47E-05 | -8.76E-06 | -1.54E-05 | -1.82E-05 | -4.76E-06 | -2.16E-05 |
| 281 | 4240 |  | 1.30E-06 | -1.92E-05 | -2.71E-05 | 3.53E-06 |  | 2.07E-05 | -0.000104 |  | -1.43E-05 | -1.63E-05 |  | -2.20E-05 | -2.45E-05 |
| 282 | 4273 |  |  |  |  |  | -1.87E-06 |  |  |  |  |  | -1.62E-05 |  |  |
| 283 | 4286 | 8.33E-05 | 2.88E-06 | -1.53E-05 | -9.97E-06 | 2.53E-06 | -1.59E-07 | 1.92E-05 | -9.94E-05 | -3.58E-05 | -9.68E-06 | 2.78E-07 |  | -1.13E-06 | -1.49E-05 |
| 284 | 4292 | 8.73E-05 |  |  | -4.30E-06 | 3.08E-06 |  |  |  |  |  |  |  |  |  |
| 285 | 4302 |  | 8.23E-07 | -1.71E-05 |  |  |  |  | 1.85E-05 | -9.93E-05 |  | -1.04E-05 | -8.11E-07 | -1.49E-05 |  | -1.56E-05 |
| 286 | 4320 |  |  |  | -5.42E-06 | 2.87E-05 |  |  |  |  |  |  |  | -1.55E-05 | -9.94E-07 |  |
| 287 | 4324 | 0.00011 | 1.25E-07 | -1.93E-05 | -3.96E-06 | 3.02E-05 | 9.16E-07 | 1.92E-05 |  | -3.65E-05 | -1.20E-05 | 6.78E-07 |  |  | -1.49E-05 |
| 288 | 4328 | 0.00011 | -2.17E-07 | -2.12E-05 |  |  |  | 1.84E-05 | -9.84E-05 |  |  |  |  |  |  |
| 289 | 4338 |  |  |  |  |  |  |  |  | -8.05E-05 |  |  |  |  |  |
| 290 | 4340 | 0.00014 | -4.63E-06 | -2.94E-05 | -1.10E-05 | 2.15E-05 | 1.72E-06 | 1.19E-05 | -9.05E-05 |  | -1.78E-05 | -4.51E-06 | -2.23E-05 | 2.89E-07 | -2.01E-05 |
| 291 | 4352 | 0.000121 | -8.58E-06 | -3.88E-05 | -1.19E-05 | 1.16E-05 | 1.57E-06 | 3.46E-06 | -9.32E-05 |  | -2.23E-05 | -7.40E-06 | -2.63E-05 | 1.24E-07 | -2.39E-05 |
| 292 | 4360 |  |  |  | -2.01E-05 | 9.03E-06 | 1.42E-06 |  | -9.40E-05 |  | -3.31E-05 | -2.08E-05 | -3.17E-05 | 1.88E-08 | -3.48E-05 |
| 293 | 4361 | 0.000126 | -8.38E-06 | -3.52E-05 |  |  |  | 8.41E-06 | -9.41E-05 | -8.35E-05 | -3.25E-05 | -2.02E-05 |  | 1.31E-07 | -3.41E-05 |
| 294 | 4386 | 0.000127 | -6.87E-06 | -3.30E-05 | -1.95E-05 |  | 9.97E-06 |  |  |  |  | -1.95E-05 | -3.10E-05 |  |  |
| 295 | 4405 |  |  |  |  |  | 1.02E-05 | 2.03E-06 |  |  | -0.000113 | -3.32E-05 |  |  |  | -3.52E-05 |
| 296 | 4413 | 0.000132 | -4.64E-06 | -2.96E-05 | -1.77E-05 |  |  | 1.25E-05 | -9.71E-05 |  |  |  | -2.95E-05 |  |  |
| 297 | 4419 |  |  |  |  |  |  |  |  |  |  | -3.60E-05 | -1.97E-05 |  |  | -3.72E-05 |
| 298 | 4424 | 0.000237 | -5.40E-06 | -3.72E-05 | -2.11E-05 | 7.83E-06 | 1.27E-06 | 7.28E-06 | -5.10E-05 | -0.000106 |  | -2.26E-05 | -3.61E-05 | -2.33E-07 |  |
| 299 | 4445 |  |  |  |  |  |  |  |  |  |  | -3.83E-05 |  |  |  | -3.96E-05 |
| 300 | 4451 |  |  |  |  |  |  |  |  |  | -0.00013 |  |  |  |  |  |
| 301 | 4501 |  |  |  |  |  | 6.57E-06 |  |  |  |  |  |  |  |  |  |
| 302 | 4502 |  |  |  |  |  |  |  |  | -6.76E-05 |  |  |  |  | 2.90E-07 |  |
| 303 | 4513 | 0.000169 | -6.08E-06 | -3.67E-05 | -2.11E-05 | 6.03E-06 | 1.24E-06 | 6.84E-06 | -7.07E-05 | -0.00012 | -4.26E-05 | -2.55E-05 | -3.61E-05 |  | -4.36E-05 |
| 304 | 4518 | 0.000167 | -8.17E-06 | -3.97E-05 | -2.35E-05 | 3.48E-06 | 1.07E-06 | 4.37E-06 | -7.21E-05 |  | -4.38E-05 | -2.74E-05 | -3.77E-05 | 1.11E-07 | -4.51E-05 |
| 305 | 4541 | 0.000227 | 2.32E-05 | -4.73E-06 | -2.79E-05 |  |  | -1.29E-06 |  | -0.000107 |  |  |  |  |  |  |
| 306 | 4542 | 0.000218 | 1.90E-05 | -4.72E-06 | -2.79E-05 | 3.10E-07 | 9.68E-07 | -1.30E-06 | -4.96E-05 | -0.000112 | -4.65E-05 | -3.05E-05 | -4.04E-05 | -1.16E-07 | -4.78E-05 |

| | A | B | C | D | E | F | G | H | I | J | K | L | M | N | O |
|---|---|---|---|---|---|---|---|---|---|---|---|---|---|---|---|
| 307 | 4543 | | | | | | | | | -0.000122 | | | | | |
| 308 | 4561 | | 1.14E-05 | | -2.75E-05 | 8.25E-07 | | | -5.19E-05 | -0.000125 | | | -4.02E-05 | | |
| 309 | 4562 | | | -4.82E-05 | | | | -2.00E-06 | | | -4.86E-05 | -3.28E-05 | | | -5.00E-05 |
| 310 | 4572 | 0.000206 | | | | | 1.09E-06 | | -5.81E-05 | -0.000125 | | | | | |
| 311 | 4581 | | | | | | | | | -0.0001 | | | | | |
| 312 | 4622 | | | | | | 6.06E-06 | | | -0.000116 | | | -3.63E-05 | -2.19E-07 | |
| 313 | 4623 | 0.000402 | 1.17E-05 | -4.11E-05 | -2.37E-05 | | | 1.97E-06 | | | -4.32E-05 | -2.75E-05 | | | -4.47E-05 |
| 314 | 4624 | 0.000405 | 9.64E-06 | -4.39E-05 | -2.63E-05 | 4.46E-06 | 8.83E-07 | -2.25E-07 | -2.82E-05 | -0.000116 | -4.70E-05 | -3.21E-05 | -3.88E-05 | | -4.80E-05 |
| 315 | 4629 | 0.000362 | | | -3.77E-05 | 2.52E-07 | 2.31E-07 | | | -0.000107 | | -4.63E-05 | -4.21E-05 | -2.94E-07 | |
| 316 | 4639 | | -1.10E-05 | | | | | | -3.58E-05 | | -5.32E-05 | | | -3.91E-05 | -5.37E-05 |
| 317 | 4641 | | | -4.94E-05 | | | | | -4.11E-06 | -0.000108 | | | -4.16E-05 | | |
| 318 | 4643 | 0.000316 | -1.09E-05 | -4.97E-05 | -3.77E-05 | 1.11E-06 | 1.12E-06 | -4.30E-06 | -5.29E-05 | -0.000109 | -5.36E-05 | -4.79E-05 | -4.23E-05 | -8.57E-08 | -5.40E-05 |
| 319 | 4644 | | | | | | 6.76E-07 | | | | | | -5.59E-05 | | | |
| 320 | 4648 | 0.000301 | -1.16E-05 | -5.06E-05 | -3.81E-05 | -1.49E-06 | | | | -6.17E-05 | | -5.75E-05 | | | | -5.84E-05 |
| 321 | 4649 | | | | | | | 7.24E-08 | -5.68E-06 | | -0.000104 | -5.90E-05 | -5.78E-05 | -4.45E-05 | -8.66E-07 | -5.99E-05 |
| 322 | 4679 | | | | | | -6.79E-07 | | | -5.11E-05 | -0.00011 | | | | | |
| 323 | 4683 | 0.000301 | -1.16E-05 | | -3.85E-05 | | -3.39E-07 | | | | | | | | | |
| 324 | 4686 | | | | | | | | | | | | | -4.50E-05 | | |
| 325 | 4690 | 0.000303 | | -5.21E-05 | -3.83E-05 | -4.26E-07 | | | -8.53E-06 | -4.83E-05 | -0.00013 | -6.39E-05 | -6.13E-05 | | | -6.40E-05 |
| 326 | 4693 | | -1.11E-05 | | | | | | | | | | | -4.68E-05 | | |
| 327 | 4701 | | | -5.63E-05 | | | -6.37E-06 | 6.41E-08 | -1.17E-05 | -6.19E-05 | | -7.36E-05 | -6.79E-05 | | -1.46E-06 | -7.23E-05 |
| 328 | 4710 | 0.000305 | | | -3.21E-05 | | | | | | -0.000129 | | | | | |
| 329 | 4730 | | -1.36E-05 | | | | | | | | -0.00014 | | | | | |
| 330 | 4732 | | | -5.89E-05 | | | | 1.93E-07 | -1.71E-05 | | | | -7.30E-05 | -4.78E-05 | -1.33E-06 | |
| 331 | 4759 | | | | | | | | | | -0.000142 | -7.88E-05 | | | | -7.64E-05 |
| 332 | 4780 | | | | -3.62E-05 | | | | | -6.65E-05 | | | | | | |
| 333 | 4802 | 0.000305 | -1.43E-05 | | | -1.33E-06 | | | | | | | | | | |
| 334 | 4821 | | | -5.87E-05 | | | | | -1.56E-05 | | | | | -4.40E-05 | | |
| 335 | 4826 | | | | | | | -9.13E-07 | | | | -8.51E-05 | -7.57E-05 | | -2.13E-06 | -8.29E-05 |
| 336 | 4853 | | | | | | | | | | -0.000138 | | | | | |
| 337 | 4860 | | | | | | | | | -7.42E-05 | | | | | | |
| 338 | 4865 | 0.000313 | -1.07E-05 | -5.62E-05 | -3.33E-05 | 3.34E-06 | | | | | | | | | | |
| 339 | 4875 | 0.000333 | -1.06E-05 | | -3.34E-05 | 3.23E-06 | | | -1.68E-05 | -7.31E-05 | -0.000137 | -8.84E-05 | -7.51E-05 | -4.48E-05 | | -8.36E-05 |
| 340 | 4878 | | | -5.78E-05 | | | | | | | | | | | | |
| 341 | 4883 | | | | | | | -7.12E-07 | -1.88E-05 | | -0.000139 | | | -4.63E-05 | -2.09E-06 | |
| 342 | 4899 | | | | | | | | | | | -8.86E-05 | -7.43E-05 | | | -8.42E-05 |
| 343 | 4930 | 0.000299 | -1.30E-05 | | -3.37E-05 | 1.43E-06 | | | | -7.48E-05 | | | | | -2.07E-06 | |
| 344 | 4931 | | | -5.59E-05 | | | | 1.30E-07 | -1.99E-05 | | | | | -4.80E-05 | | |
| 345 | 4933 | | | | | | | | | | | -9.05E-05 | -7.38E-05 | | | |
| 346 | 4939 | 0.000297 | -1.21E-05 | -5.79E-05 | -3.39E-05 | 1.27E-06 | | | -2.13E-05 | -7.52E-05 | -0.000149 | | | -5.07E-05 | -2.07E-06 | -8.90E-05 |
| 347 | 4954 | | | | | | | -8.21E-08 | | | -0.000146 | -8.86E-05 | -6.80E-05 | | | -8.66E-05 |
| 348 | 5000 | | | -9.51E-06 | | -3.08E-05 | 4.38E-06 | | | -7.28E-05 | -0.000154 | | | | | |
| 349 | 5003 | 0.00028 | | | -5.77E-05 | | | | -2.05E-05 | | -0.000154 | -9.07E-05 | -6.84E-05 | -4.99E-05 | -2.05E-06 | -8.79E-05 |
| 350 | 5011 | | | | | -2.95E-05 | 6.29E-06 | | | | -0.000157 | | | | | |
| 351 | 5022 | 0.000273 | -1.03E-05 | -6.27E-05 | | | | | -2.31E-05 | -6.84E-05 | -0.000152 | | | -5.08E-05 | -2.20E-06 | |
| 352 | 5023 | | -4.07E-05 | -8.46E-05 | -4.03E-05 | -3.01E-06 | 1.39E-06 | | | | -0.000151 | -0.000108 | -8.31E-05 | | | -0.000102 |
| 353 | 5027 | 0.000267 | | | | | | | -3.60E-06 | -6.92E-05 | | | | -5.90E-05 | -1.95E-06 | |
| 354 | 5043 | 0.000262 | -4.31E-05 | -8.49E-05 | -3.98E-05 | -3.37E-06 | 3.39E-06 | -3.70E-05 | -7.02E-05 | -0.00014 | -0.00011 | -8.53E-05 | -6.07E-05 | -7.98E-07 | -0.000104 |
| 355 | 5053 | 0.000261 | | | -4.02E-05 | -3.58E-06 | 3.14E-06 | | -7.13E-05 | -0.00014 | | | -6.06E-05 | | | |
| 356 | 5058 | | | -4.36E-05 | -8.60E-05 | | | | -3.76E-06 | | | | -8.52E-05 | | | |
| 357 | 5072 | | | | | | | | | | -0.000145 | -0.000107 | | | | -0.000102 |
| 358 | 5087 | 0.000339 | | | | | | | | | | | | | | |
| 359 | 5088 | | | -5.49E-05 | -8.97E-05 | -4.19E-05 | -5.97E-06 | 2.06E-06 | -4.12E-05 | -6.49E-05 | -0.000139 | -0.000109 | -8.44E-05 | -6.20E-05 | -1.53E-06 | -0.000104 |
| 360 | 5089 | 0.000343 | -5.39E-05 | -8.86E-05 | -4.03E-05 | -5.72E-06 | | | | -6.44E-05 | | -0.00011 | -8.42E-05 | -6.23E-05 | -1.57E-06 | -0.000104 |
| 361 | 5092 | | | | | | | 2.32E-06 | -4.58E-05 | -7.28E-05 | | | | | | |
| 362 | 5099 | 0.000331 | | | | | | | | | -0.000133 | -0.000117 | -9.24E-05 | -6.74E-05 | -2.31E-06 | |
| 363 | 5105 | 0.000341 | -5.68E-05 | -9.38E-05 | -4.33E-05 | -8.12E-06 | 2.63E-06 | -4.58E-05 | -7.44E-06 | | -0.000117 | -9.21E-05 | -6.74E-05 | -2.25E-06 | -0.00011 |
| 364 | 5131 | | | | | | | | -5.91E-05 | | -0.00015 | | | | | |
| 365 | 5133 | 0.000314 | -5.87E-05 | | -4.89E-05 | -7.28E-06 | 2.12E-06 | | -1.19E-05 | -0.000153 | -0.000122 | -9.80E-05 | -6.80E-05 | -2.56E-06 | -0.000115 |
| 366 | 5136 | 0.000312 | | -0.000107 | -5.02E-05 | | 2.34E-06 | -6.24E-05 | -2.47E-05 | -0.000146 | | -9.89E-05 | -6.70E-05 | | |
| 367 | 5143 | | -6.05E-05 | -0.000109 | | -8.82E-06 | 2.59E-06 | -6.39E-05 | | | -0.000124 | | | | -2.30E-06 | -0.000117 |
| 368 | 5159 | 0.000316 | -6.10E-05 | -0.000108 | -5.12E-05 | -7.63E-06 | | | -2.29E-05 | -0.000139 | -0.000124 | -0.000101 | -6.90E-05 | | -0.000118 |
| 369 | 5171 | | | | | | | | -6.79E-05 | | | | | | | |
| 370 | 5178 | 0.000297 | | | -5.43E-05 | | 3.20E-06 | | -3.25E-05 | | -0.000129 | -0.000107 | -7.49E-05 | -2.03E-06 | -0.000122 |
| 371 | 5187 | | | | | | | 3.15E-06 | | | -0.00013 | | | | | |
| 372 | 5222 | | -7.07E-05 | -0.000118 | | -1.44E-05 | 3.19E-06 | -7.19E-05 | -2.90E-05 | | | | | -7.77E-05 | -2.02E-06 | |
| 373 | 5226 | | | | | | | | | | -0.000131 | | -0.00011 | | | |
| 374 | 5241 | 0.000291 | | | -5.76E-05 | | | | | | | -0.000133 | | | -1.92E-06 | -0.000126 |
| 375 | 5264 | | -6.96E-05 | -0.000115 | | -1.21E-05 | 2.89E-06 | -6.95E-05 | -3.22E-05 | | | | | -7.77E-05 | | |
| 376 | 5277 | | | | -5.59E-05 | | | | | | -0.000148 | | -0.000111 | | | |
| 377 | 5288 | 0.000324 | | | | -1.50E-05 | | | | | | | | | | |
| 378 | 5292 | | -7.00E-05 | -0.000116 | | | | 2.33E-06 | -7.06E-05 | | -0.000149 | -0.000136 | | | -2.68E-06 | -0.000129 |
| 379 | 5300 | | | | | | | | | -1.24E-05 | -0.000144 | | | -7.73E-05 | | |
| 380 | 5306 | 0.000323 | -6.75E-05 | -0.000114 | -5.68E-05 | -1.42E-05 | 2.38E-06 | -6.68E-05 | -1.26E-05 | -0.000144 | -0.000135 | -0.000111 | -7.73E-05 | -2.64E-06 | -0.000128 |
| 381 | 5309 | 0.000323 | -6.78E-05 | -0.000117 | -5.67E-05 | | | -7.09E-05 | | -0.000145 | -0.000136 | -0.000111 | -7.73E-05 | | -0.000129 |
| 382 | 5332 | | | | | -1.94E-05 | 2.08E-06 | | -9.08E-05 | -0.000133 | -0.000143 | -0.000117 | -8.41E-05 | -2.68E-06 | -0.000137 |
| 383 | 5355 | 0.000382 | -5.98E-05 | -0.000112 | -5.23E-05 | | | -6.52E-05 | | -0.000138 | | | | | | |
| 384 | 5360 | | | | | -1.87E-05 | 2.03E-06 | | -2.49E-05 | -0.000138 | | | -8.40E-05 | | | |
| 385 | 5378 | | | | | | | | | -0.000128 | | | | | | |
| 386 | 5379 | | | | | | | | | | -0.000144 | -0.000116 | | | | -0.000137 |
| 387 | 5383 | 0.000379 | | | | | | | | | | | | | -3.03E-06 | |
| 388 | 5404 | | -5.91E-05 | -0.000113 | -5.33E-05 | -1.94E-05 | 1.79E-06 | -6.64E-05 | -1.51E-05 | -0.000135 | -0.000146 | | -8.55E-05 | | |
| 389 | 5421 | 0.000358 | | | | | | | 2.67E-06 | | | -0.000117 | | -2.98E-06 | -0.000136 |
| 390 | 5430 | | -6.27E-05 | -0.000113 | | | | | | | | | | | | |
| 391 | 5444 | 0.000352 | -6.41E-05 | | -5.33E-05 | -1.86E-05 | 2.31E-06 | -6.61E-05 | -6.42E-06 | | -0.000146 | -0.000118 | -8.41E-05 | -3.08E-06 | -0.000135 |
| 392 | 5457 | | | -0.000112 | | | | | | | | | | | | |
| 393 | 5472 | 0.00035 | | | -5.45E-05 | -1.91E-05 | 1.97E-06 | -6.74E-05 | -7.97E-06 | | -0.000147 | -0.000119 | -8.49E-05 | -3.25E-06 | -0.000137 |
| 394 | 5474 | | | | | | | | | -0.000166 | | | | | | |
| 395 | 5485 | | -5.40E-05 | | | | | 2.71E-06 | | | -0.000174 | -0.000148 | | -8.43E-05 | | |
| 396 | 5495 | | | -0.00011 | -5.36E-05 | | | | -6.65E-05 | 1.08E-05 | -0.000173 | | -0.000118 | | -3.37E-06 | -0.000137 |
| 397 | 5506 | 0.000376 | -5.83E-05 | -0.000111 | -5.36E-05 | -1.77E-05 | 3.02E-06 | -6.65E-05 | 2.31E-05 | | -0.000147 | -0.000119 | -8.35E-05 | | -0.000137 |
| 398 | 5527 | 0.000373 | -6.03E-05 | | | | | | | 5.83E-05 | | -0.000147 | -0.00012 | -8.35E-05 | -3.67E-06 | -0.000138 |
| 399 | 5542 | 0.000373 | -6.16E-05 | -0.000114 | -5.68E-05 | -2.04E-05 | 2.16E-06 | -7.07E-05 | 5.38E-05 | -0.000171 | | -0.000148 | -0.000121 | -8.48E-05 | -3.79E-06 | -0.000139 |
| 400 | 5548 | 0.000373 | | | -5.67E-05 | -2.04E-05 | 2.17E-06 | | | | | -0.000148 | -0.000121 | -8.47E-05 | -3.89E-06 | -0.000139 |
| 401 | 5581 | | -6.20E-05 | -0.000114 | -5.76E-05 | -2.44E-05 | | -7.03E-05 | 7.53E-05 | | | | | | | |
| 402 | 5584 | 0.000391 | | | | | | | | | -0.000172 | -0.000161 | -0.000132 | -8.83E-05 | -3.28E-06 | -0.000152 |
| 403 | 5594 | | -6.12E-05 | -0.000113 | -5.65E-05 | -2.30E-05 | 2.52E-06 | -6.96E-05 | 6.59E-05 | -0.000173 | -0.000161 | -0.000131 | -8.78E-05 | | -0.000151 |
| 404 | 5602 | 0.00036 | -6.14E-05 | -0.000112 | -5.46E-05 | -1.97E-05 | | -6.86E-05 | 6.59E-05 | | -0.000161 | | -8.76E-05 | -4.09E-06 | -0.000151 |
| 405 | 5627 | | | | | | | | | | | | -8.49E-05 | | | |
| 406 | 5639 | | | | | | | | | 7.29E-05 | | | | | | |
| 407 | 5640 | 0.000365 | | | | | | 4.48E-06 | | | -0.000174 | | | | | |
| 408 | 5652 | | -5.45E-05 | -0.000107 | -4.91E-05 | -1.08E-05 | | -6.42E-05 | | -0.000174 | -0.000153 | -0.000105 | -8.32E-05 | | -0.000143 |

| | A | B | C | D | E | F | G | H | I | J | K | L | M | N | O |
|---|---|---|---|---|---|---|---|---|---|---|---|---|---|---|---|
| 409 | 5653 | 0.000362 | -5.65E-05 | -0.000109 | | | | | 4.79E-05 | -0.00017 | -0.000162 | -0.000112 | -8.51E-05 | -4.12E-06 | -0.000151 |
| 410 | 5660 | | | | -4.99E-05 | -1.21E-05 | | -6.45E-05 | 4.63E-05 | | | | | | |
| 411 | 5677 | 0.000359 | | | | | 4.27E-06 | | | | | | | | |
| 412 | 5707 | | -6.03E-05 | | | | | | | -0.000145 | -0.000165 | | -8.59E-05 | | -0.000154 |
| 413 | 5740 | | | -0.000114 | | | | -6.96E-05 | | | | -0.00012 | | -5.45E-06 | |
| 414 | 5763 | 0.000336 | -6.63E-05 | -0.000114 | -5.63E-05 | -1.89E-05 | 6.02E-06 | -6.97E-05 | 3.55E-05 | | -0.000172 | -0.00012 | -8.77E-05 | -5.41E-06 | -0.000161 |
| 415 | 5797 | 0.000337 | -6.51E-05 | | | | | | | | -0.000172 | -0.00012 | | -6.20E-06 | -0.000162 |
| 416 | 5800 | | | -0.000105 | -4.64E-05 | -1.85E-05 | | -6.33E-05 | 7.27E-05 | | | | | | |
| 417 | 5802 | 0.00033 | -6.54E-05 | -0.000107 | -4.77E-05 | -1.78E-05 | 7.10E-06 | -6.48E-05 | 6.79E-05 | -0.000149 | -0.000175 | -0.000122 | -8.84E-05 | -5.32E-06 | -0.000163 |
| 418 | 5803 | 0.000332 | -6.37E-05 | -0.000105 | -4.31E-05 | -1.21E-05 | 7.72E-06 | -6.12E-05 | 6.83E-05 | -0.000141 | -0.000171 | -0.000118 | -8.47E-05 | | -0.00016 |
| 419 | 5804 | | -6.60E-05 | -0.000106 | -4.16E-05 | | | -6.09E-05 | | -0.000126 | | | | | |
| 420 | 5814 | 0.000319 | -7.08E-05 | -0.000118 | -5.25E-05 | -1.97E-05 | 5.63E-06 | -7.15E-05 | 6.51E-05 | | -0.000191 | -0.000136 | -0.000101 | -5.94E-06 | -0.000178 |
| 421 | 5824 | 0.000318 | -7.12E-05 | -0.000118 | -5.34E-05 | -1.85E-05 | | -7.25E-05 | | | -0.000191 | -0.000137 | -9.96E-05 | -6.10E-06 | -0.000179 |
| 422 | 5825 | 0.00032 | | | | | | | 6.79E-05 | | | | | | |
| 423 | 5834 | | -6.86E-05 | -0.000112 | | | | | | 2.48E-05 | | | | | |
| 424 | 5851 | 0.000312 | -6.90E-05 | -0.000114 | -5.43E-05 | -1.71E-05 | 6.24E-06 | -7.72E-05 | 6.82E-05 | | -0.000192 | -0.000134 | -9.86E-05 | | -0.000178 |
| 425 | 5866 | 0.000305 | -7.40E-05 | -0.000121 | -6.02E-05 | -2.28E-05 | | -8.47E-05 | 6.56E-05 | | -0.000199 | -0.000143 | -0.000104 | -6.70E-06 | -0.000187 |
| 426 | 5883 | 0.000344 | -7.40E-05 | -0.00012 | | | 6.33E-06 | -8.39E-05 | 8.10E-05 | | -0.000201 | -0.000146 | -0.000107 | -6.50E-06 | -0.000189 |
| 427 | 5888 | 0.000274 | | | -6.46E-05 | -2.54E-05 | 5.85E-06 | | | | -0.000211 | -0.000157 | -0.000111 | -6.54E-06 | -0.0002 |
| 428 | 5917 | | -7.83E-05 | | | | | | 5.78E-05 | | | | | | |
| 429 | 5918 | 0.000275 | | -0.000119 | -6.27E-05 | -2.19E-05 | | -8.35E-05 | 5.75E-05 | 1.10E-05 | -0.000213 | -0.000158 | -0.000105 | -7.00E-06 | -0.0002 |
| 430 | 5919 | | | | | | | | | 1.10E-05 | | | | | |
| 431 | 5927 | | | | | | 6.02E-06 | | | | | | | | |
| 432 | 5985 | | -7.30E-05 | -0.000111 | -5.23E-05 | -1.36E-05 | | -7.35E-05 | | | -0.000245 | -0.000196 | -0.0001 | | -0.000237 |
| 433 | 6021 | 0.000267 | -7.30E-05 | -0.000111 | -5.30E-05 | -1.38E-05 | 7.31E-06 | -7.38E-05 | 4.87E-05 | -1.70E-05 | -0.000247 | -0.000199 | -0.000101 | -6.90E-06 | -0.00024 |
| 434 | 6028 | 0.000268 | -7.22E-05 | -0.00011 | -5.22E-05 | -1.20E-05 | | -7.27E-05 | 4.85E-05 | | -0.000248 | -0.0002 | -0.000102 | -6.73E-06 | -0.00024 |
| 435 | 6037 | | | | | | | | 0.000167 | | | | | | |
| 436 | 6041 | 0.000271 | | | | | | | | -1.97E-05 | -0.000271 | -0.000212 | -0.000102 | | -0.000255 |
| 437 | 6057 | | -7.78E-05 | -0.000113 | -5.33E-05 | -1.23E-05 | 6.76E-06 | -7.43E-05 | 0.000145 | | | | | -7.00E-06 | |
| 438 | 6063 | | | | | | | | | | -0.000289 | -0.000233 | -0.00011 | | -0.000276 |
| 439 | 6072 | 0.000252 | -8.25E-05 | -0.000118 | -5.70E-05 | -1.59E-05 | | -7.84E-05 | 0.000138 | -2.11E-05 | | | | -7.19E-06 | |
| 440 | 6083 | 0.000249 | -8.34E-05 | -0.000121 | -5.96E-05 | -1.66E-05 | 6.76E-06 | -8.12E-05 | 0.000136 | | -0.000299 | -0.000245 | -0.00011 | -7.00E-06 | -0.000288 |
| 441 | 6094 | | | | | | | | | -3.53E-05 | | | | | |
| 442 | 6103 | 0.000244 | -8.32E-05 | | | | 7.13E-06 | | | -4.17E-05 | -0.000311 | -0.000255 | -0.000112 | -6.58E-06 | -0.000298 |
| 443 | 6139 | | | -0.000122 | -5.70E-05 | -1.46E-05 | 7.09E-06 | -8.01E-05 | 0.000141 | -0.000113 | -0.000319 | -0.00026 | -0.000114 | -6.54E-06 | -0.000302 |
| 444 | 6152 | | | | | | | | 0.000134 | -0.000116 | | | | | |
| 445 | 6156 | 0.000233 | -8.71E-05 | -0.000125 | -5.89E-05 | -1.72E-05 | 7.35E-06 | -8.17E-05 | | -0.000117 | -0.000322 | -0.000264 | -0.000118 | -6.80E-06 | -0.000304 |
| 446 | 6165 | 0.000233 | -8.71E-05 | -0.000124 | -5.86E-05 | -1.67E-05 | | -8.16E-05 | 0.000133 | | | | | | |
| 447 | 6178 | | | | | | | | | | | -0.000322 | -0.000262 | -0.000117 | | -0.000303 |
| 448 | 6185 | | -8.16E-05 | | | | | | | | | | | | |
| 449 | 6199 | 0.000233 | -8.18E-05 | -0.000124 | -5.75E-05 | -1.51E-05 | 7.49E-06 | -8.07E-05 | 0.000137 | -0.000116 | -0.000323 | -0.000263 | -0.000117 | -6.76E-06 | -0.000304 |
| 450 | 6246 | | -7.76E-05 | -0.000114 | -4.74E-05 | -2.63E-06 | | -7.26E-05 | | | | | | | |
| 451 | 6249 | 0.00027 | | | | | | | | 0.00015 | | | | | |
| 452 | 6250 | 0.000347 | -7.74E-05 | -0.000119 | -5.19E-05 | -5.71E-06 | | -7.67E-05 | 0.00021 | | -0.000314 | -0.000254 | -0.000109 | -7.21E-06 | -0.000294 |
| 453 | 6273 | 0.000431 | -7.95E-05 | -0.00012 | -5.23E-05 | -7.30E-06 | 6.25E-06 | -7.76E-05 | 0.000239 | -9.89E-05 | -0.000313 | -0.000252 | -0.00011 | -7.44E-06 | -0.000293 |
| 454 | 6274 | 0.00013 | -8.81E-05 | | | | 6.55E-06 | | | | -0.000325 | -0.000266 | -0.000108 | | -0.000306 |
| 455 | 6278 | | | -0.000128 | -5.83E-05 | -7.72E-06 | | -8.45E-05 | 5.10E-05 | | | | | -7.15E-06 | |
| 456 | 6303 | | | | | | | | | | -0.000325 | -0.000264 | -0.000105 | | -0.000305 |
| 457 | 6307 | | | | | | 9.12E-06 | | | -0.000107 | | | | | |
| 458 | 6321 | | -7.75E-05 | -0.000104 | -3.66E-05 | | | -6.17E-05 | 5.53E-05 | -9.73E-05 | | | | | |
| 459 | 6334 | 0.000137 | | | | 9.22E-06 | | | | | | | | | |
| 460 | 6407 | | | | | | | | | | -0.000102 | -0.000327 | -0.000264 | -0.000102 | -5.91E-06 | -0.000306 |
| 461 | 6484 | 0.000132 | -8.28E-05 | -0.000112 | -4.56E-05 | 3.25E-06 | | -6.81E-05 | 5.34E-05 | -0.000134 | | | | | |
| 462 | 6493 | | | | | | 9.58E-06 | | | | | -0.000334 | -0.000273 | -0.000107 | -6.05E-06 | -0.000312 |
| 463 | 6503 | | | | | | 9.73E-06 | | | | | | | | | |
| 464 | 6523 | | -6.01E-05 | -8.74E-05 | -1.76E-05 | 2.12E-05 | | -4.30E-05 | 6.11E-05 | -0.000131 | | | | | |
| 465 | 6529 | 0.000137 | -6.18E-05 | -9.03E-05 | -2.08E-05 | 1.88E-05 | | -4.52E-05 | 6.08E-05 | | -0.000335 | -0.000272 | -0.000108 | | -0.000313 |
| 466 | 6534 | | | | | | 1.03E-05 | | | | -0.000336 | -0.000274 | -0.00011 | -5.43E-06 | -0.000314 |
| 467 | 6565 | 0.000188 | 5.01E-05 | -8.98E-05 | -1.97E-05 | 1.93E-05 | | -4.47E-05 | 0.000149 | | | | | | |
| 468 | 6566 | 0.000175 | 2.49E-05 | -9.13E-05 | -2.11E-05 | 1.85E-05 | 8.37E-06 | -4.63E-05 | 0.000136 | -0.000131 | -0.000341 | -0.000278 | -0.000112 | -5.83E-06 | -0.000319 |
| 469 | 6581 | | 2.81E-05 | -8.69E-05 | -1.73E-05 | 2.55E-05 | | -4.22E-05 | 0.000141 | | | | | -5.93E-06 | |
| 470 | 6606 | 3.98E-05 | | | | | | | | | | | | | |
| 471 | 6608 | | | | | | | | | -0.000119 | -0.000337 | -0.000282 | -0.00011 | | -0.00032 |
| 472 | 6613 | 3.66E-05 | 2.41E-05 | -8.84E-05 | -1.66E-05 | 2.56E-05 | | -4.33E-05 | 9.50E-05 | | | | | | |
| 473 | 6629 | | | | | | 9.72E-06 | | | | -0.000334 | -0.000279 | -0.000108 | -5.38E-06 | -0.000317 |
| 474 | 6692 | | 3.27E-05 | | 3.35E-05 | | | | | | | | | | |
| 475 | 6731 | | | -6.90E-05 | 3.67E-06 | | | -2.50E-05 | | | | | | | |
| 476 | 6741 | 3.81E-05 | | | | | | | 9.19E-05 | -0.000142 | | | | | |
| 477 | 6749 | | 3.65E-05 | -5.53E-05 | 1.27E-05 | 3.59E-05 | 1.13E-05 | -1.28E-05 | | | -0.000328 | -0.000269 | -0.000109 | -2.55E-06 | -0.000299 |
| 478 | 6772 | 1.09E-05 | 1.36E-05 | -8.72E-05 | -1.46E-05 | 2.72E-05 | | -4.60E-05 | 8.55E-05 | | -0.000354 | -0.000288 | -0.000113 | | -0.000329 |
| 479 | 6773 | 8.08E-06 | 1.23E-05 | -8.83E-05 | -1.58E-05 | 2.63E-05 | | -4.74E-05 | 8.49E-05 | | | | | | |
| 480 | 6790 | | | | | | | | | -0.00014 | -0.000354 | -0.000289 | -0.000114 | | -0.00033 |
| 481 | 6807 | | | | | | | | | -0.000138 | | | | | |
| 482 | 6811 | -1.97E-07 | | | | | 1.16E-05 | | | | | | | | |
| 483 | 6823 | | 1.34E-06 | -0.000115 | -3.72E-05 | 1.35E-05 | | -7.21E-05 | 8.08E-05 | -0.000149 | -0.00038 | -0.00031 | -0.000122 | -2.66E-06 | -0.000352 |
| 484 | 6876 | | | | | | | | | -0.000157 | | | | | |
| 485 | 6899 | -3.81E-06 | | | | | 8.66E-06 | | 9.11E-05 | | -0.000399 | -0.000326 | -0.000131 | -3.90E-06 | -0.000369 |
| 486 | 6904 | | 9.10E-08 | -0.000116 | -4.58E-05 | 9.12E-06 | 1.02E-05 | -7.84E-05 | | | | | | | |
| 487 | 6940 | -2.94E-06 | -2.82E-06 | -0.000117 | -4.27E-05 | 8.16E-06 | | -7.90E-05 | 9.94E-05 | -0.000136 | -0.000395 | -0.000323 | -0.00013 | | -0.000365 |
| 488 | 6946 | 7.59E-06 | 8.10E-06 | -0.000103 | -3.49E-05 | 1.80E-05 | 7.81E-06 | -6.86E-05 | 9.97E-05 | | -0.000396 | -0.00032 | -0.000129 | -5.48E-06 | -0.000363 |

| | A | B | C | D | E | F | G | H | I | J | K | L | M | N | O |
|---|---|---|---|---|---|---|---|---|---|---|---|---|---|---|---|
| 1 | Rank | Inf | 0.002 | 0.0015 | 0.001 | 0.0005 | 0.0001 | 0.0013 | 0.0020 Ran | 0.0015 Ran | 0.0010 Ran | 0.0005 Ran | 0.0001 Ran | 0.0013 Random | |
| 2 | 0 | -9.21E-05 | -8.92E-05 | -0.000102 | -9.99E-05 | -8.12E-05 | -2.43E-06 | -0.000111 | -8.40E-05 | -6.83E-05 | -0.000102 | -0.000105 | -6.05E-05 | -3.56E-06 | -0.000117 |
| 3 | 152.2554 | | | | -7.10E-05 | | | | | | | -4.83E-05 | | | |
| 4 | 175.5918 | | | | | | -7.12E-05 | | | | | | | | -4.99E-05 |
| 5 | 187.6639 | | | -6.59E-05 | | | | | | | -4.46E-05 | | | | |
| 6 | 219.2789 | -6.00E-05 | -5.64E-05 | | | | | | -2.86E-05 | -1.68E-05 | | | | | |
| 7 | 237.8496 | | | | | | -1.65E-06 | | | | | | | -1.81E-06 | |
| 8 | 332.8672 | | | | | | -1.49E-06 | | | | | | | -1.76E-06 | |
| 9 | 343.6196 | | | | | -4.74E-05 | | | | | | | -2.15E-05 | | |
| 10 | 381.102 | | | | | | -1.58E-06 | | | | | | | -2.08E-06 | |
| 11 | 389.6812 | | | | -5.39E-05 | | | | | | | -4.43E-05 | | | |
| 12 | 391.5365 | | | | | | | -5.66E-05 | | | | | | | -4.77E-05 |
| 13 | 393.5015 | | | -5.58E-05 | | | | | | | -4.18E-05 | | | | |
| 14 | 407.8617 | -5.88E-05 | -5.62E-05 | | | | | | -3.33E-05 | -2.00E-05 | | | | | |
| 15 | 414.3272 | | | | | | -1.62E-06 | | | | | | | -2.25E-06 | |
| 16 | 420.6941 | | | | | | -1.67E-06 | | | | | | | -2.43E-06 | |
| 17 | 448.8226 | | | | | -4.43E-05 | | | | | | | -1.96E-05 | | |
| 18 | 451.223 | | | | | | -1.77E-06 | | | | | | | -2.15E-06 | |
| 19 | 452.19 | | | | -5.40E-05 | | | | | | | -4.69E-05 | | | |
| 20 | 457.0818 | | | | | | | -5.61E-05 | | | | | | | -5.05E-05 |
| 21 | 463.6398 | | | | | | -1.60E-06 | | | | | | | -2.07E-06 | |
| 22 | 477.0539 | | | -5.51E-05 | | | | | | | -4.18E-05 | | | | |
| 23 | 490.5911 | -5.88E-05 | -5.61E-05 | | | | | | -3.38E-05 | -2.16E-05 | | | | | |
| 24 | 526.0183 | | | | | | -1.73E-06 | | | | | | | -1.93E-06 | |
| 25 | 534.5459 | | | | | -4.40E-05 | -1.75E-06 | | | | | | -1.91E-05 | -2.05E-06 | |
| 26 | 537.9939 | | | | | | -1.79E-06 | | | | | | | -1.98E-06 | |
| 27 | 565.0747 | | | | -5.52E-05 | | | | | | | -5.02E-05 | | | |
| 28 | 567.3436 | | | | | | | -5.85E-05 | | | | | | | -5.37E-05 |
| 29 | 573.4078 | | | -5.73E-05 | | | | | | | -4.59E-05 | | | | |
| 30 | 581.5944 | | -5.80E-05 | | | | | | | -2.48E-05 | | | | | |
| 31 | 583.5358 | -5.99E-05 | | | | | | -3.60E-05 | | | | | | | |
| 32 | 588.5723 | | | | | | -1.80E-06 | | | | | | | -2.38E-06 | |
| 33 | 623.0442 | | | | | | -8.14E-07 | | | | | | | -8.29E-07 | |
| 34 | 627.9754 | | | | | -4.65E-05 | | | | | | | -2.24E-05 | | |
| 35 | 631.8787 | | | | | | -7.38E-07 | | | | | | | -4.76E-07 | |
| 36 | 639.9068 | | | | -5.62E-05 | | | | | | | -5.19E-05 | | | |
| 37 | 642.7365 | | | | | | | -5.96E-05 | | | | | | | -5.48E-05 |
| 38 | 643.8356 | | | -5.79E-05 | | | -5.30E-07 | | | | -4.79E-05 | | | -4.65E-07 | |
| 39 | 652.1336 | | -5.87E-05 | | | | | | | -2.76E-05 | | | | | |
| 40 | 656.0196 | -6.00E-05 | | | | | | -3.96E-05 | | | | | | | |
| 41 | 693.7603 | | | | | | -1.06E-06 | | | | | | | -1.08E-06 | |
| 42 | 697.4831 | | | | | -4.37E-05 | | | | | | | -2.02E-05 | | |
| 43 | 709.3201 | | | | | | -1.00E-06 | | | | | | | -8.41E-07 | |
| 44 | 713.5827 | | | | | | -1.06E-06 | | | | | | | -9.82E-07 | |
| 45 | 730.9563 | | | | -5.23E-05 | | | | | | | -4.76E-05 | | | |
| 46 | 734.8253 | | | | | | | -5.55E-05 | | | | | | | -5.01E-05 |
| 47 | 754.2795 | | | -5.52E-05 | | | | | | | -4.70E-05 | | | | |
| 48 | 763.7222 | | | | | | -1.18E-06 | | | | | | | -1.23E-06 | |
| 49 | 770.9584 | -5.90E-05 | -5.73E-05 | | | | | | -4.09E-05 | -2.94E-05 | | | | | |
| 50 | 776.2436 | | | | | | -1.13E-06 | | | | | | | -1.23E-06 | |
| 51 | 791.6747 | | | | | -4.16E-05 | | | | | | | -1.82E-05 | | |
| 52 | 802.1356 | | | | | | -1.39E-06 | | | | | | | -1.78E-06 | |
| 53 | 811.1106 | | | | -5.23E-05 | | -1.39E-06 | | | | | -4.40E-05 | | -1.78E-06 | |
| 54 | 816.655 | | | | | | | -5.42E-05 | | | | | | | -4.54E-05 |
| 55 | 823.8519 | | | -5.41E-05 | | | | | | | -4.38E-05 | | | | |
| 56 | 832.7611 | -5.36E-05 | -5.21E-05 | | | | -1.40E-06 | | -3.35E-05 | -2.16E-05 | | | | -1.67E-06 | |
| 57 | 833.3007 | | | | | | -1.40E-06 | | | | | | | -1.74E-06 | |
| 58 | 840.2536 | | | | | -3.51E-05 | | | | | | | -1.52E-05 | | |
| 59 | 858.7477 | | | | | | -2.53E-06 | | | | | | | -1.97E-06 | |
| 60 | 880.8681 | | | | -3.96E-05 | | | | | | | -3.11E-05 | | | |
| 61 | 887.1001 | | | | | | | -4.32E-05 | | | | | | | -3.05E-05 |
| 62 | 888.5915 | | | | | | -2.90E-06 | | | | | | | -2.43E-06 | |
| 63 | 905.8124 | | | -4.09E-05 | | | | | | | -2.16E-05 | | | | |
| 64 | 916.8311 | | -4.84E-05 | | | | | | | -1.76E-05 | | | | | |
| 65 | 916.993 | -5.05E-05 | | | | | | -3.02E-05 | | | | | | | |
| 66 | 927.8606 | | | | | -2.82E-05 | | | | | | | -5.68E-07 | | |
| 67 | 943.5253 | | | | | | -2.98E-06 | | | | | | | -2.63E-06 | |
| 68 | 978.666 | | | | | | -3.06E-06 | | | | | | | -2.76E-06 | |
| 69 | 988.4961 | | | | -3.79E-05 | | | | | | | -2.95E-05 | | | |
| 70 | 999.4992 | | | | | | | -4.10E-05 | | | | | | | -2.88E-05 |
| 71 | 1008.746 | | | -4.01E-05 | | | | | | | -2.25E-05 | | | | |
| 72 | 1024.796 | | -4.92E-05 | | | | -2.97E-06 | | | -2.06E-05 | | | | -2.50E-06 | |
| 73 | 1026.235 | -5.08E-05 | | | | | | -3.18E-05 | | | | | | | |
| 74 | 1028.415 | | | | | -2.64E-05 | | | | | | | -1.57E-06 | | |
| 75 | 1030.171 | | | | | | -2.93E-06 | | | | | | | -2.41E-06 | |
| 76 | 1047.232 | | | | -3.69E-05 | | | | | | | -3.01E-05 | | | |
| 77 | 1048.18 | | | | | | -2.75E-06 | | | | | | | -2.29E-06 | |
| 78 | 1050.611 | | | | | | -3.98E-05 | | | | | | | | -2.89E-05 |
| 79 | 1062.682 | | | -3.71E-05 | | | | | | | -2.01E-05 | | | | |
| 80 | 1079.051 | | | | | -2.44E-05 | | | | | | | -6.21E-07 | | |
| 81 | 1081.141 | | -4.49E-05 | | | | | | | -1.68E-05 | | | | | |
| 82 | 1084.776 | -4.72E-05 | | | | | | -2.83E-05 | | | | | | | |
| 83 | 1099.39 | | | | | | -2.37E-06 | | | | | | | -1.95E-06 | |
| 84 | 1139.124 | | | | -3.59E-05 | | | | | | | -2.92E-05 | | | |
| 85 | 1146.785 | | | | | | | -3.86E-05 | | | | | | | -2.82E-05 |
| 86 | 1147.35 | | | | | | -2.48E-06 | | | | | | | -2.24E-06 | |
| 87 | 1160.468 | | | -3.50E-05 | | | | | | | -1.88E-05 | | | | |
| 88 | 1168.284 | | | | | -2.28E-05 | | | | | | | 1.74E-06 | | |
| 89 | 1174.953 | -4.56E-05 | -4.34E-05 | | | | | | -2.73E-05 | -1.66E-05 | | | | | |
| 90 | 1182.033 | | | | | | -2.56E-06 | | | | | | | -2.20E-06 | |
| 91 | 1198.551 | | | | | | -2.74E-06 | | | | | | | -2.23E-06 | |
| 92 | 1201.424 | | | | -3.57E-05 | | | | | | | -2.97E-05 | | | |
| 93 | 1208.718 | | | | | | | -3.79E-05 | | | | | | | -2.88E-05 |
| 94 | 1224.146 | | | -3.42E-05 | | | | | | | -1.91E-05 | | | | |
| 95 | 1226.536 | | | | | | -2.23E-05 | -2.93E-06 | | | | | 4.14E-07 | -2.47E-06 | |
| 96 | 1234.649 | | -4.18E-05 | | | | | | | -1.71E-05 | | | | | |
| 97 | 1234.74 | -4.30E-05 | | | | | | -2.71E-05 | | | | | | | |
| 98 | 1254.151 | | | | | | -3.20E-06 | | | | | | | -3.06E-06 | |
| 99 | 1264.517 | | | | | -3.45E-05 | | | | | | | -2.84E-05 | | |
| 100 | 1271.817 | | | | | | | -3.69E-05 | | | | | | | -2.73E-05 |
| 101 | 1286.315 | | | -3.34E-05 | | | | | | | -1.93E-05 | | | | |
| 102 | 1287.837 | | | | | -2.11E-05 | | | | | | | 1.82E-06 | | |

| | A | B | C | D | E | F | G | H | I | J | K | L | M | N | O |
|---|---|---|---|---|---|---|---|---|---|---|---|---|---|---|---|
| 103 | 1289.984 | | | | | | -3.65E-06 | | | | | | | -3.67E-06 | |
| 104 | 1302.176 | | -4.24E-05 | | | | | | | -1.89E-05 | | | | | |
| 105 | 1307.77 | -4.36E-05 | | | | | | | -2.86E-05 | | | | | | |
| 106 | 1325.111 | | | | | | -3.79E-06 | | | | | | | -3.58E-06 | |
| 107 | 1360.288 | | | | | | -3.76E-06 | | | | | | | -3.57E-06 | |
| 108 | 1360.87 | | | | -3.15E-05 | | | | | | | -2.70E-05 | | | |
| 109 | 1362.465 | | | | | | | -3.40E-05 | | | | | | | -2.59E-05 |
| 110 | 1372.791 | | | | | -1.99E-05 | | | | | | | 2.44E-06 | | |
| 111 | 1375.619 | | | -3.02E-05 | | | | | | | -1.72E-05 | | | | |
| 112 | 1392.658 | | -4.03E-05 | | | | | | | -1.74E-05 | | | | | |
| 113 | 1395.136 | -4.11E-05 | | | | | | | -2.64E-05 | | | | | | |
| 114 | 1398.85 | | | | | | -3.77E-06 | | | | | | | -3.57E-06 | |
| 115 | 1439.136 | | | | | | -3.77E-06 | | | | | | | -3.61E-06 | |
| 116 | 1440.334 | | | | -2.97E-05 | | | | | | | -2.50E-05 | | | |
| 117 | 1447.689 | | | | | | | -3.25E-05 | | | | | | | -2.38E-05 |
| 118 | 1455.598 | | | | | -1.61E-05 | | | | | | | 9.21E-06 | | |
| 119 | 1458.07 | | | -2.69E-05 | | | | | | | -1.40E-05 | | | | |
| 120 | 1472.079 | | -3.85E-05 | | | | | | | -1.48E-05 | | | | | |
| 121 | 1472.414 | -3.92E-05 | | | | | | | -2.41E-05 | | | | | | |
| 122 | 1490.382 | | | | | | -3.79E-06 | | | | | | | -3.55E-06 | |
| 123 | 1498.73 | | | | -3.04E-05 | | | | | | | -2.51E-05 | | | |
| 124 | 1513.565 | | | | | | | -3.39E-05 | | | | | | | -2.57E-05 |
| 125 | 1525.719 | | | | | -1.67E-05 | -3.67E-06 | | | | | | 6.77E-06 | -3.52E-06 | |
| 126 | 1529.02 | | | -2.89E-05 | | | | | | | -1.57E-05 | | | | |
| 127 | 1544.519 | | -3.90E-05 | | | | | | | -1.60E-05 | | | | | |
| 128 | 1544.977 | -4.02E-05 | | | | | | | -2.54E-05 | | | | | | |
| 129 | 1563.728 | | | | | | -3.79E-06 | | | | | | | -3.53E-06 | |
| 130 | 1570.088 | | | | -2.87E-05 | | | | | | | -2.35E-05 | | | |
| 131 | 1574.975 | | | | | | | -3.26E-05 | | | | | | | -2.41E-05 |
| 132 | 1579.185 | | | | | -1.60E-05 | | | | | | | 7.85E-06 | | |
| 133 | 1585.328 | | | -2.91E-05 | | | | | | | -1.60E-05 | | | | |
| 134 | 1603.355 | | | | | | -4.04E-06 | | | | | | | -3.76E-06 | |
| 135 | 1606.19 | | -3.89E-05 | | | | | | | -1.68E-05 | | | | | |
| 136 | 1608.123 | -4.00E-05 | | | | | | | -2.57E-05 | | | | | | |
| 137 | 1631.664 | | | | -2.86E-05 | | | | | | | -2.25E-05 | | | |
| 138 | 1634.215 | | | | | | -3.91E-06 | | | | | | | -3.68E-06 | |
| 139 | 1638.983 | | | | | | | -3.33E-05 | | | | | | | -2.45E-05 |
| 140 | 1639.268 | | | | | -1.66E-05 | | | | | | | 6.18E-06 | | |
| 141 | 1646.612 | | | -2.99E-05 | | | | | | | -1.85E-05 | | | | |
| 142 | 1654.887 | | -3.95E-05 | | | | -4.06E-06 | | | -1.79E-05 | | | | -3.74E-06 | |
| 143 | 1660.412 | -4.11E-05 | | | | | | | -2.63E-05 | | | | | | |
| 144 | 1695.592 | | | | -2.92E-05 | | | | | | | -2.41E-05 | | | |
| 145 | 1695.768 | | | | | | -3.93E-06 | | | | | | | -3.54E-06 | |
| 146 | 1702.621 | | | | | -1.68E-05 | | | | | | | 6.10E-06 | | |
| 147 | 1705.046 | | | | | | | -3.25E-05 | | | | | | | -2.46E-05 |
| 148 | 1723.939 | | | -2.70E-05 | | | | | | | -1.59E-05 | | | | |
| 149 | 1730.218 | | | | | | -3.83E-06 | | | | | | | -3.43E-06 | |
| 150 | 1737.319 | | -3.88E-05 | | | | | | | -1.86E-05 | | | | | |
| 151 | 1740.184 | -4.05E-05 | | | | | | | -2.65E-05 | | | | | | |
| 152 | 1742.309 | | | | | | -3.81E-06 | | | | | | | -3.43E-06 | |
| 153 | 1752.146 | | | | -2.70E-05 | | | | | | | -2.30E-05 | | | |
| 154 | 1752.404 | | | | | -1.47E-05 | | | | | | | 7.11E-06 | | |
| 155 | 1760.974 | | | | | | | -3.12E-05 | | | | | | | -2.51E-05 |
| 156 | 1781.193 | | | -2.59E-05 | | | | | | | -1.45E-05 | | | | |
| 157 | 1783.172 | | | | | | -3.81E-06 | | | | | | | -3.38E-06 | |
| 158 | 1813.219 | | -3.82E-05 | | | | | | | -1.76E-05 | | | | | |
| 159 | 1816.36 | -3.97E-05 | | | | | | | -2.56E-05 | | | | | | |
| 160 | 1826.786 | | | | | -1.40E-05 | | | | | | | 8.89E-06 | | |
| 161 | 1830.329 | | | | -2.60E-05 | | | | | | | -2.19E-05 | | | |
| 162 | 1838.237 | | | | | | -3.85E-06 | | | | | | | -3.40E-06 | |
| 163 | 1840 | | | | | | | -3.10E-05 | | | | | | | -2.51E-05 |
| 164 | 1852.856 | | | -2.59E-05 | | | | | | | -1.50E-05 | | | | |
| 165 | 1868.016 | | | | | | -3.95E-06 | | | | | | | -3.22E-06 | |
| 166 | 1874.469 | -4.06E-05 | -3.92E-05 | | | | | | -2.66E-05 | -1.81E-05 | | | | | |
| 167 | 1881.159 | | | | | -1.40E-05 | | | | | | | 8.04E-06 | | |
| 168 | 1888.027 | | | | -2.56E-05 | | | | | | | -2.16E-05 | | | |
| 169 | 1890.565 | | | | | | -3.98E-06 | | | | | | | -3.13E-06 | |
| 170 | 1902.008 | | | | | | | -2.97E-05 | | | | | | | -2.37E-05 |
| 171 | 1916.058 | | | -2.50E-05 | | | | | | | -1.34E-05 | | | | |
| 172 | 1940.387 | | -3.67E-05 | | | | | | | -1.64E-05 | | | | | |
| 173 | 1943.448 | -3.82E-05 | | | | | -3.88E-06 | | -2.51E-05 | | | | | -3.03E-06 | |
| 174 | 1944.597 | | | | | -1.34E-05 | | | | | | | 8.76E-06 | | |
| 175 | 1959.341 | | | | -2.47E-05 | | | | | | | -2.19E-05 | | | |
| 176 | 1976.275 | | | | | | | -2.97E-05 | | | | | | | -2.45E-05 |
| 177 | 1984.452 | | | | | | -3.70E-06 | | | | | | | -2.94E-06 | |
| 178 | 2004.828 | | | -2.43E-05 | | | | | | | -1.27E-05 | | | | |
| 179 | 2016.644 | | | | | -1.33E-05 | | | | | | | 9.84E-06 | | |
| 180 | 2017.453 | -3.91E-05 | -3.77E-05 | | | | | | -2.58E-05 | -1.61E-05 | | | | | |
| 181 | 2026.571 | | | | | | -3.60E-06 | | | | | | | -2.85E-06 | |
| 182 | 2026.925 | | | | -2.54E-05 | | | | | | | -2.16E-05 | | | |
| 183 | 2030.177 | | | | | | | -3.06E-05 | | | | | | | -2.40E-05 |
| 184 | 2051.452 | | | -2.40E-05 | | | | | | | -1.14E-05 | | | | |
| 185 | 2060.212 | | | | | -1.18E-05 | | | | | | | 1.22E-05 | | |
| 186 | 2070.308 | | -3.73E-05 | | | | | | | -1.50E-05 | | | | | |
| 187 | 2077.604 | | | | | | -3.62E-06 | | | | | | | -2.71E-06 | |
| 188 | 2077.897 | -3.81E-05 | | | | | | | -2.46E-05 | | | | | | |
| 189 | 2094.463 | | | | -2.51E-05 | | | | | | | -2.07E-05 | | | |
| 190 | 2101.989 | | | | | | | -3.01E-05 | | | | | | | -2.22E-05 |
| 191 | 2124.512 | | | -2.45E-05 | | | | | | | -1.13E-05 | | | | |
| 192 | 2133.771 | | | | | -1.06E-05 | | | | | | | 1.25E-05 | | |
| 193 | 2135.621 | | | | | | -3.69E-06 | | | | | | | -2.93E-06 | |
| 194 | 2143.981 | | -3.75E-05 | | | | | | | -1.74E-05 | | | | | |
| 195 | 2147.928 | -3.84E-05 | | | | | | | -2.66E-05 | | | | | | |
| 196 | 2150.179 | | | | | | -3.68E-06 | | | | | | | -2.97E-06 | |
| 197 | 2152.729 | | | | -2.54E-05 | | | | | | | -2.18E-05 | | | |
| 198 | 2165.161 | | | | | | | -2.87E-05 | | | | | | | -2.25E-05 |
| 199 | 2185.206 | | | -2.15E-05 | | | | | | | -1.03E-05 | | | | |
| 200 | 2188.055 | | | | | -9.59E-06 | | | | | | | 1.34E-05 | | |
| 201 | 2190.53 | | | | | | -3.89E-06 | | | | | | | -3.33E-06 | |
| 202 | 2218.152 | | -3.67E-05 | | | | | | | -1.68E-05 | | | | | |
| 203 | 2222.241 | -3.80E-05 | | | | | | | -2.62E-05 | | | | | | |
| 204 | 2225.2 | | | | -2.37E-05 | | | | | | | -2.06E-05 | | | |

| | A | B | C | D | E | F | G | H | I | J | K | L | M | N | O |
|---|---|---|---|---|---|---|---|---|---|---|---|---|---|---|---|
| 205 | 2239.077 | | | | | | -3.95E-06 | | | | | | | -3.42E-06 | |
| 206 | 2239.859 | | | | | | | -2.79E-05 | | | | | | | -2.16E-05 |
| 207 | 2264.37 | | | -2.00E-05 | | | | | | | -8.93E-06 | | | | |
| 208 | 2264.494 | | | | | -7.90E-06 | | | | | | | 1.51E-05 | | |
| 209 | 2288.458 | | -3.12E-05 | | | | | | | -1.09E-05 | | | | | |
| 210 | 2289.212 | -3.30E-05 | | | | | | | -2.02E-05 | | | | | | |
| 211 | 2292.014 | | | | -2.24E-05 | | | | | | | -1.85E-05 | | | |
| 212 | 2294.665 | | | | | | -3.52E-06 | | | | | | | -3.33E-06 | |
| 213 | 2305.254 | | | | | | | -2.81E-05 | | | | | | | -2.05E-05 |
| 214 | 2321.349 | | | | | -9.49E-06 | | | | | | | 1.38E-05 | | |
| 215 | 2326.607 | | | -2.05E-05 | | | | | | | -9.59E-06 | | | | |
| 216 | 2333.008 | | | | | | -3.54E-06 | | | | | | | -3.45E-06 | |
| 217 | 2355.002 | | -3.49E-05 | | | | | | | -1.72E-05 | | | | | |
| 218 | 2357.303 | -3.77E-05 | | | | | | | -2.60E-05 | | | | | | |
| 219 | 2357.632 | | | | -2.20E-05 | | | | | | | -1.90E-05 | | | |
| 220 | 2368.231 | | | | | | | -2.73E-05 | | | | | | | -2.05E-05 |
| 221 | 2384.628 | | | | | -7.29E-06 | | | | | | | 1.51E-05 | | |
| 222 | 2386.451 | | | -1.94E-05 | | | | | | | -8.77E-06 | | | | |
| 223 | 2386.812 | | | | | | -3.52E-06 | | | | | | | -3.33E-06 | |
| 224 | 2410.359 | | -3.56E-05 | | | | | | | -1.80E-05 | | | | | |
| 225 | 2415.057 | -3.88E-05 | | | -2.17E-05 | | | | -2.64E-05 | | | -1.87E-05 | | | |
| 226 | 2436.324 | | | | | | | -2.70E-05 | | | | | | | -1.99E-05 |
| 227 | 2445.304 | | | | | -6.39E-06 | -3.45E-06 | | | | | | | 1.59E-05 | -3.23E-06 | |
| 228 | 2464.07 | | | -1.92E-05 | | | | | | | -7.99E-06 | | | | |
| 229 | 2475.515 | | | | | | -3.47E-06 | | | | | | | | -3.24E-06 | |
| 230 | 2482.803 | | -3.57E-05 | | | | | | | -1.77E-05 | | | | | |
| 231 | 2483.23 | | | | -2.15E-05 | | | | | | | -1.85E-05 | | | | |
| 232 | 2484.507 | -3.93E-05 | | | | | | | -2.63E-05 | | | | | | |
| 233 | 2491.234 | | | | | | | -2.69E-05 | | | | | | | | -2.02E-05 |
| 234 | 2494.728 | | | | | -5.91E-06 | | | | | | | | 1.66E-05 | | |
| 235 | 2521.451 | | | | | | -3.42E-06 | | | | | | | | -3.27E-06 | |
| 236 | 2530.384 | | | -1.96E-05 | | | | | | | -8.37E-06 | | | | |
| 237 | 2553.797 | | -3.55E-05 | | -2.13E-05 | | | | | -1.78E-05 | | -1.87E-05 | | | |
| 238 | 2560.584 | -3.85E-05 | | | | | | | -2.60E-05 | | | | | | |
| 239 | 2572.228 | | | | | -5.54E-06 | | | | | | | | 1.67E-05 | | |
| 240 | 2572.667 | | | | | | | -2.63E-05 | | | | | | | | -1.97E-05 |
| 241 | 2586.147 | | | | | | -3.56E-06 | | | | | | | | -3.26E-06 | |
| 242 | 2594.176 | | | -1.76E-05 | | | | | | | -7.80E-06 | | | | |
| 243 | 2630.612 | | | | -1.99E-05 | | | | | | | -1.78E-05 | | | | |
| 244 | 2632.321 | | -3.39E-05 | | | | | | | -1.76E-05 | | | | | |
| 245 | 2634.415 | -3.75E-05 | | | | | | | -2.65E-05 | | | | | | |
| 246 | 2636.15 | | | | | | -3.52E-06 | | | | | | | | -3.21E-06 | |
| 247 | 2636.83 | | | | | -5.99E-06 | | | | | | | | 1.60E-05 | | |
| 248 | 2644.489 | | | | | | | -2.57E-05 | | | | | | | | -1.99E-05 |
| 249 | 2683.274 | | | -1.66E-05 | | | | | | | -6.68E-06 | | | | |
| 250 | 2701.271 | | | | | | -3.91E-06 | | | | | | | | -3.25E-06 | |
| 251 | 2720.545 | | | | -1.86E-05 | | | | | | | -1.60E-05 | | | | |
| 252 | 2722.378 | | -3.26E-05 | | | | | | | -1.67E-05 | | | | | |
| 253 | 2726.797 | -3.61E-05 | | | | | | | -2.53E-05 | | | | | | |
| 254 | 2726.862 | | | | | -4.30E-06 | | | | | | | | 1.75E-05 | | |
| 255 | 2732.599 | | | | | | | -2.41E-05 | | | | | | | | -1.86E-05 |
| 256 | 2742.03 | | | | | | -3.85E-06 | | | | | | | | -3.29E-06 | |
| 257 | 2744.556 | | | -1.62E-05 | | | | | | | -6.86E-06 | | | | |
| 258 | 2766.931 | | | | -1.79E-05 | | | | | | | -1.54E-05 | | | | |
| 259 | 2773.163 | | | | | -4.37E-06 | | | | | | | | 1.80E-05 | | |
| 260 | 2776.666 | | -3.21E-05 | | | | | | | -1.62E-05 | | | | | |
| 261 | 2779.012 | -3.48E-05 | | | | | | | -2.44E-05 | | | | | | |
| 262 | 2783.541 | | | | | | | -2.30E-05 | | | | | | | | -1.70E-05 |
| 263 | 2796.724 | | | | | | -3.74E-06 | | | | | | | | -3.47E-06 | |
| 264 | 2811.532 | | | -1.55E-05 | | | | | | | -5.85E-06 | | | | |
| 265 | 2844.516 | | | | -1.80E-05 | | | | | | | -1.57E-05 | | | | |
| 266 | 2851.671 | | | | | -3.26E-06 | | | | | | | | 1.84E-05 | | |
| 267 | 2857.111 | | -3.23E-05 | | | | | | | -1.74E-05 | | | | | |
| 268 | 2860.678 | -3.52E-05 | | | | | | | -2.59E-05 | | | | | | |
| 269 | 2863.069 | | | | | | | -2.36E-05 | | | | | | | | -1.87E-05 |
| 270 | 2865.922 | | | | | | -3.62E-06 | | | | | | | | -3.30E-06 | |
| 271 | 2902.931 | | | -1.59E-05 | | | | | | | -6.68E-06 | | | | |
| 272 | 2925.643 | | | | | | -3.53E-06 | | | | | | | | -3.25E-06 | |
| 273 | 2928.102 | | | | -1.79E-05 | -3.26E-06 | | | | | | -1.56E-05 | 1.87E-05 | | |
| 274 | 2945.01 | | -3.25E-05 | | | | | | | -1.78E-05 | | | | | |
| 275 | 2948.211 | -3.51E-05 | | | | | | -2.39E-05 | -2.65E-05 | | | | | | -1.89E-05 |
| 276 | 2962.415 | | | | | | -3.56E-06 | | | | | | | | -3.35E-06 | |
| 277 | 2970.98 | | | -1.60E-05 | | | | | | | -6.55E-06 | | | | |
| 278 | 2989.394 | | | | | -3.04E-06 | | | | | | | | 1.95E-05 | | |
| 279 | 2990.089 | | | | -1.72E-05 | | | | | | | -1.52E-05 | | | | |
| 280 | 3006.874 | | -3.25E-05 | | | | | | | -1.83E-05 | | | | | |
| 281 | 3010.328 | | | | | | | -2.34E-05 | | | | | | | | -1.85E-05 |
| 282 | 3012.442 | -3.45E-05 | | | | | | | -2.68E-05 | | | | | | |
| 283 | 3013.976 | | | | | | -3.59E-06 | | | | | | | | -3.49E-06 | |
| 284 | 3040.968 | | | -1.60E-05 | | | | | | | -7.23E-06 | | | | |
| 285 | 3059.605 | | | | | -3.33E-06 | | | | | | | | 1.92E-05 | | |
| 286 | 3063.511 | | | | -1.73E-05 | | | | | | | -1.61E-05 | | | | |
| 287 | 3079.109 | | | | | | -3.80E-06 | | | | | | | | -3.96E-06 | |
| 288 | 3080.478 | | -3.28E-05 | | | | | | | -1.88E-05 | | | | | |
| 289 | 3081.486 | | | | | | | -2.41E-05 | | | | | | | | -1.96E-05 |
| 290 | 3085.282 | -3.48E-05 | | | | | | | -2.77E-05 | | | | | | |
| 291 | 3112.116 | | | -1.72E-05 | | | | | | | -8.63E-06 | | | | |
| 292 | 3116.698 | | | | | -5.26E-06 | -3.75E-06 | | | | | | | 1.52E-05 | -3.92E-06 | |
| 293 | 3125.755 | | | | -1.84E-05 | | | | | | | -1.86E-05 | | | | |
| 294 | 3142.967 | | | | | | | -2.46E-05 | | | | | | | | -2.22E-05 |
| 295 | 3143.213 | | -3.39E-05 | | | | | | | -2.18E-05 | | | | | |
| 296 | 3145.248 | -3.64E-05 | | | | | | | -3.05E-05 | | | | | | |
| 297 | 3155.247 | | | | | | -3.86E-06 | | | | | | | | -4.13E-06 | |
| 298 | 3166.805 | | | -1.73E-05 | | | | | | | -9.24E-06 | | | | |
| 299 | 3171.348 | | | | | -5.75E-06 | | | | | | | | 1.44E-05 | | |
| 300 | 3180.323 | | | | -1.70E-05 | | | | | | | -1.71E-05 | | | | |
| 301 | 3191.264 | | -3.69E-05 | | | | -3.91E-06 | -2.51E-05 | | -2.31E-05 | | | | -4.16E-06 | -2.31E-05 |
| 302 | 3192.479 | -3.84E-05 | | | | | | | -3.12E-05 | | | | | | |
| 303 | 3210.965 | | | | | | -3.92E-06 | | | | | | | | -4.12E-06 | |
| 304 | 3224.235 | | | | | -8.91E-06 | | | | | | | | 8.15E-06 | | |
| 305 | 3227.893 | | | -2.04E-05 | | | | | | | -1.15E-05 | | | | |
| 306 | 3241.727 | | | | -2.08E-05 | | | | | | | -2.16E-05 | | | | |

| | A | B | C | D | E | F | G | H | I | J | K | L | M | N | O |
|---|---|---|---|---|---|---|---|---|---|---|---|---|---|---|---|
| 307 | 3247.217 | | | | | | -3.95E-06 | | | | | | | -4.18E-06 | |
| 308 | 3260.33 | | | | | | | -2.61E-05 | | | | | | | -2.48E-05 |
| 309 | 3265.233 | | -3.82E-05 | | | | | | | -2.44E-05 | | | | | |
| 310 | 3267.645 | -3.95E-05 | | | | | | | -3.23E-05 | | | | | | |
| 311 | 3292.356 | | | | | -8.67E-06 | | | | | | | 7.24E-06 | | |
| 312 | 3296.829 | | | -2.01E-05 | | | | | | | -1.16E-05 | | | | |
| 313 | 3305.364 | | | | | | -4.16E-06 | | | | | | | -4.32E-06 | |
| 314 | 3318.898 | | | | -2.09E-05 | | | | | | | -2.25E-05 | | | |
| 315 | 3343.96 | | | | | | | -2.66E-05 | | | | | | | -2.56E-05 |
| 316 | 3358.09 | | -3.86E-05 | | | | | | | -2.52E-05 | | | | | |
| 317 | 3366.55 | -3.98E-05 | | | | | | | -3.31E-05 | | | | | | |
| 318 | 3375.852 | | | | | | -4.09E-06 | | | | | | | -4.21E-06 | |
| 319 | 3382.214 | | | | | -8.78E-06 | | | | | | | 7.53E-06 | | |
| 320 | 3388.341 | | | -1.91E-05 | | | | | | | -1.16E-05 | | | | |
| 321 | 3390.695 | | | | -2.04E-05 | | | | | | | -2.23E-05 | | | |
| 322 | 3408.16 | | | | | | | -2.68E-05 | | | | | | | -2.59E-05 |
| 323 | 3415.186 | | -3.95E-05 | | | | -4.16E-06 | | | -2.56E-05 | | | | -4.20E-06 | |
| 324 | 3418.244 | -4.07E-05 | | | | | | | -3.39E-05 | | | | | | |
| 325 | 3422.635 | | | | | -8.93E-06 | | | | | | | 7.70E-06 | | |
| 326 | 3442.01 | | | -1.94E-05 | | | | | | | -1.18E-05 | | | | |
| 327 | 3453.352 | | | | -2.07E-05 | | | | | | | -2.29E-05 | | | |
| 328 | 3462 | | | | | | -4.20E-06 | | | | | | | -4.35E-06 | |
| 329 | 3471.118 | | | | | | | | -2.77E-05 | | | | | | -2.74E-05 |
| 330 | 3484.017 | | -3.94E-05 | | | | | | | | -2.60E-05 | | | | |
| 331 | 3490.433 | | | | | | -9.79E-06 | | | | | | | 6.48E-06 | | |
| 332 | 3490.755 | -4.06E-05 | | | | | | | -3.43E-05 | | | | | | |
| 333 | 3506.774 | | | | | | | -4.35E-06 | | | | | | -4.49E-06 | |
| 334 | 3522.395 | | | -2.15E-05 | | | | | | | -1.47E-05 | | | | |
| 335 | 3531.7 | | | | -2.15E-05 | | | | | | | -2.44E-05 | | | |
| 336 | 3558.684 | | | | | | | -2.90E-05 | | | | | | | -2.87E-05 |
| 337 | 3559.123 | | | | | | -4.61E-06 | | | | | | | -4.58E-06 | |
| 338 | 3564.481 | | | | | -1.03E-05 | | | | | | | 5.41E-06 | | | |
| 339 | 3567.088 | | -3.97E-05 | | | | | | | -2.84E-05 | | | | | |
| 340 | 3571.532 | -4.11E-05 | | | | | | | -3.57E-05 | | | | | | |
| 341 | 3578.725 | | | | | | | -5.03E-06 | | | | | | -4.60E-06 | |
| 342 | 3588.507 | | | -2.24E-05 | | | | | | | -1.56E-05 | | | | |
| 343 | 3595.644 | | | | -2.25E-05 | | | | | | | -2.44E-05 | | | |
| 344 | 3621.266 | | | | | | | -5.18E-06 | -2.96E-05 | | | | | | -5.59E-06 | -2.87E-05 |
| 345 | 3623.141 | | | | | -1.15E-05 | | | | | | | 3.27E-06 | | | |
| 346 | 3633.766 | | -4.15E-05 | | | | | | | -3.03E-05 | | | | | |
| 347 | 3634.074 | -4.20E-05 | | | | | | | -3.66E-05 | | | | | | |
| 348 | 3663.927 | | | | | | | -5.23E-06 | | | | | | -5.56E-06 | |
| 349 | 3667.834 | | | -2.33E-05 | | | | | | | -1.67E-05 | | | | |
| 350 | 3675.456 | | | | -2.31E-05 | | | | | | | -2.53E-05 | | | |
| 351 | 3695.128 | | | | | | | -3.02E-05 | | | | | | | -2.97E-05 |
| 352 | 3696.549 | | | | | -1.23E-05 | | | | | | | 2.40E-06 | | | |
| 353 | 3703.201 | | | | | | | -5.20E-06 | | | | | | -5.58E-06 | |
| 354 | 3717.141 | | -4.21E-05 | | | | | | | -3.13E-05 | | | | | |
| 355 | 3723.799 | -4.26E-05 | | | | | | | -3.77E-05 | | | | | | |
| 356 | 3755.056 | | | -2.30E-05 | | | | | | | -1.70E-05 | | | | |
| 357 | 3757.393 | | | | | | | -4.75E-06 | | | | | | -5.33E-06 | |
| 358 | 3758.466 | | | | -2.30E-05 | | | | | | | -2.54E-05 | | | |
| 359 | 3778.267 | | | | | | -1.21E-05 | -4.81E-06 | -2.99E-05 | | | | | 3.02E-06 | -5.38E-06 | -2.94E-05 |
| 360 | 3786.227 | | -4.20E-05 | | | | | | | | -3.13E-05 | | | | |
| 361 | 3791.327 | -4.29E-05 | | | | | | | -3.83E-05 | | | | | | |
| 362 | 3796.958 | | | | | | | -4.39E-06 | | | | | | -5.18E-06 | |
| 363 | 3823.145 | | | -2.44E-05 | | | | | | | -1.76E-05 | | | | |
| 364 | 3826.082 | | | | -2.46E-05 | | | | | | | -2.64E-05 | | | |
| 365 | 3831.841 | | | | | -1.20E-05 | | | | | | | 2.65E-06 | | | |
| 366 | 3838.039 | | | | | | | | -2.96E-05 | | | | | | | -2.86E-05 |
| 367 | 3845.398 | | | | | | | -4.51E-06 | | | | | | -5.31E-06 | |
| 368 | 3866.106 | | | -4.16E-05 | | | | | | | -3.07E-05 | | | | |
| 369 | 3870.91 | -4.33E-05 | | | | | | | -3.69E-05 | | | | | | |
| 370 | 3898.012 | | | | -2.69E-05 | | | | | | | -2.07E-05 | | | | |
| 371 | 3898.477 | | | | | -2.78E-05 | | | | | | | -3.02E-05 | | | |
| 372 | 3900.184 | | | | | | -1.59E-05 | | | | | | | -1.61E-06 | | |
| 373 | 3915.85 | | | | | | | -3.05E-05 | | | | | | | -2.98E-05 |
| 374 | 3929.839 | | | | | | | -4.52E-06 | | | | | | -5.16E-06 | |
| 375 | 3936.833 | | -4.29E-05 | | | | | | | -3.22E-05 | | | | | |
| 376 | 3951.973 | -4.43E-05 | | | | | | | -3.86E-05 | | | | | | |
| 377 | 3962.403 | | | | | | | -4.51E-06 | | | | | | -5.17E-06 | |
| 378 | 3967.9 | | | | | | -1.65E-05 | | | | | | | -2.77E-06 | | |
| 379 | 3974.076 | | | | -2.80E-05 | | | | | | | -3.17E-05 | | | | |
| 380 | 3974.829 | | | -2.71E-05 | | | | | | | -2.09E-05 | | | | |
| 381 | 4000.475 | | | | | | | | -3.00E-05 | | | | | | | -2.96E-05 |
| 382 | 4011.424 | | -4.21E-05 | | | | | | | -3.19E-05 | | | | | |
| 383 | 4015.132 | | | | | | | -4.60E-06 | | | | | | -5.29E-06 | |
| 384 | 4017.089 | -4.44E-05 | | | | | | | -3.83E-05 | | | | | | |
| 385 | 4020.693 | | | | | -1.62E-05 | | | | | | | -3.00E-06 | | | |
| 386 | 4032.156 | | | | -2.83E-05 | | | | | | | -3.26E-05 | | | | |
| 387 | 4035.164 | | | -2.78E-05 | | | | -4.64E-06 | | | -2.22E-05 | | | | -5.37E-06 | |
| 388 | 4039.516 | | | | | | | | -3.01E-05 | | | | | | | -3.03E-05 |
| 389 | 4069.399 | | | | | | -1.61E-05 | | | | | | | -2.63E-06 | | |
| 390 | 4070.172 | | | -4.14E-05 | | | | | | | -3.23E-05 | | | | |
| 391 | 4078.197 | -4.38E-05 | | | | | | | -3.87E-05 | | | | | | |
| 392 | 4088.61 | | | | | -2.81E-05 | | | | | | | -3.27E-05 | | | |
| 393 | 4092.968 | | | -2.76E-05 | | | | | | | -2.25E-05 | | | | |
| 394 | 4097.235 | | | | | | | -4.71E-06 | | | | | | -5.44E-06 | |
| 395 | 4113.219 | | | | | | | | -3.03E-05 | | | | | | | -3.10E-05 |
| 396 | 4139.9 | | | | | | -1.68E-05 | | | | | | | -4.27E-06 | | |
| 397 | 4141.749 | | | | | | | -4.73E-06 | | | | | | -5.40E-06 | |
| 398 | 4149.956 | | -4.16E-05 | | | | | | | -3.31E-05 | | | | | |
| 399 | 4160.386 | -4.43E-05 | | | | | | | -3.94E-05 | | | | | | |
| 400 | 4175.902 | | | | -2.99E-05 | | | | | | | -3.37E-05 | | | | |
| 401 | 4178.035 | | | | -2.92E-05 | | | | | | | -2.36E-05 | | | | |
| 402 | 4183.223 | | | | | | | -4.72E-06 | | | | | | -5.61E-06 | |
| 403 | 4196.165 | | | | | | | | -3.22E-05 | | | | | | | -3.25E-05 |
| 404 | 4196.485 | | | | | | | -4.74E-06 | | | | | | -5.55E-06 | |
| 405 | 4210.391 | | | | | | -1.79E-05 | | | | | | | -5.11E-06 | | |
| 406 | 4225.08 | | | -4.24E-05 | | | | -5.18E-06 | | | -3.42E-05 | | | | -6.36E-06 | |
| 407 | 4226.097 | -4.42E-05 | | | | | | | -3.95E-05 | | | | | | |
| 408 | 4236.528 | | | | | -2.98E-05 | | | | | | | -3.42E-05 | | | |

| | A | B | C | D | E | F | G | H | I | J | K | L | M | N | O |
|---|---|---|---|---|---|---|---|---|---|---|---|---|---|---|---|
| 409 | 4250.039 | | | -2.87E-05 | | | | | | | -2.22E-05 | | | | |
| 410 | 4269.307 | | | | | | | -3.06E-05 | | | | | | | -3.15E-05 |
| 411 | 4274.263 | | | | | -1.64E-05 | | | | | | | -3.66E-06 | | |
| 412 | 4311.507 | | | | | | -5.09E-06 | | | | | | | -6.17E-06 | |
| 413 | 4314.792 | | -4.13E-05 | | | | | | | -3.34E-05 | | | | | |
| 414 | 4325.618 | -4.29E-05 | | | | | | | -3.89E-05 | | | | | | |
| 415 | 4341.164 | | | | -2.94E-05 | | | | | | | -3.39E-05 | | | |
| 416 | 4347.784 | | | -2.94E-05 | | | | | | | -2.33E-05 | | | | |
| 417 | 4370.816 | | | | | | -5.16E-06 | | | | | | | -6.18E-06 | |
| 418 | 4371.885 | | | | | -1.61E-05 | | | | | | | -3.57E-06 | | |
| 419 | 4373.991 | | | | | | | | -3.12E-05 | | | | | | -3.25E-05 |
| 420 | 4389.995 | | | | | | -5.11E-06 | | | | | | | -6.19E-06 | |
| 421 | 4401.987 | | | -4.21E-05 | | | | | | -3.39E-05 | | | | | |
| 422 | 4408.96 | -4.37E-05 | | | | | | | -3.88E-05 | | | | | | |
| 423 | 4411.243 | | | | -2.99E-05 | | | | | | | -3.39E-05 | | | |
| 424 | 4420.451 | | | -2.99E-05 | | | | | | | -2.32E-05 | | | | |
| 425 | 4421.333 | | | | | | -5.13E-06 | | | | | | | -6.14E-06 | |
| 426 | 4423.396 | | | | | -1.61E-05 | | | | | | | -4.28E-06 | | |
| 427 | 4437.265 | | | | | | | | -3.16E-05 | | | | | | -3.29E-05 |
| 428 | 4444.464 | | | | | | -5.18E-06 | | | | | | | -6.21E-06 | |
| 429 | 4480.883 | | -4.22E-05 | | | | | | | -3.42E-05 | | | | | |
| 430 | 4488.768 | -4.38E-05 | | | -2.97E-05 | | | | -3.83E-05 | | | -3.39E-05 | | | |
| 431 | 4500.373 | | | | | | -5.09E-06 | | | | | | | -6.14E-06 | |
| 432 | 4512.623 | | | | | -1.62E-05 | | | | | | | -4.18E-06 | | |
| 433 | 4516.994 | | | -2.97E-05 | | | | | | | -2.22E-05 | | | | |
| 434 | 4532.129 | | | | | | | | -3.11E-05 | | | | | | -3.23E-05 |
| 435 | 4562.726 | | | | | | -5.10E-06 | | | | | | | -5.97E-06 | |
| 436 | 4584.18 | | -4.18E-05 | | | | | | | -3.32E-05 | | | | | |
| 437 | 4586.49 | | | | -2.88E-05 | | | | | | | -3.34E-05 | | | |
| 438 | 4588.068 | | | | | -1.54E-05 | | | | | | | -3.22E-06 | | |
| 439 | 4588.144 | -4.35E-05 | | | | | | | -3.84E-05 | | | | | | |
| 440 | 4596.196 | | | -2.88E-05 | | | | | | | -2.15E-05 | | | | |
| 441 | 4600.521 | | | | | | | -2.99E-05 | | | | | | | -3.05E-05 |
| 442 | 4611.287 | | | | | | -5.06E-06 | | | | | | | -5.85E-06 | |
| 443 | 4660.095 | | | | | -1.54E-05 | | | | | | | -3.13E-06 | | |
| 444 | 4660.181 | | | | | | | -4.84E-06 | | | | | | -5.87E-06 | |
| 445 | 4661.511 | | -4.16E-05 | | | | | | | -3.32E-05 | | | | | |
| 446 | 4662.513 | | | | -2.87E-05 | | | | | | | -3.32E-05 | | | |
| 447 | 4682.294 | -4.33E-05 | | | | | | | -3.83E-05 | | | | | | |
| 448 | 4699.636 | | | -2.88E-05 | | | | | | | -2.11E-05 | | | | |
| 449 | 4726.513 | | | | | | | | -2.93E-05 | | | | | | -2.96E-05 |
| 450 | 4749.574 | | | | | | | -4.73E-06 | | | | | | -5.79E-06 | |
| 451 | 4764.687 | | | | | -1.55E-05 | | | | | | | -3.67E-06 | | |
| 452 | 4784.066 | | | | -2.84E-05 | | | | | | | -3.34E-05 | | | |
| 453 | 4786.984 | | -4.15E-05 | | | | | | | -3.28E-05 | | | | | |
| 454 | 4793.177 | -4.34E-05 | | | | | | | -3.86E-05 | | | | | | |
| 455 | 4807.333 | | | -2.88E-05 | | | | | | | -2.13E-05 | | | | |
| 456 | 4825.81 | | | | | | | -2.89E-05 | | | | | | | -2.91E-05 |
| 457 | 4825.982 | | | | | | | -4.90E-06 | | | | | | -5.98E-06 | |
| 458 | 4875.164 | | | | | -1.45E-05 | | | | | | | -2.70E-06 | | |
| 459 | 4897.087 | | | | -2.81E-05 | | | | | | | -3.29E-05 | | | |
| 460 | 4898.988 | -4.29E-05 | -4.10E-05 | | | | | | -3.78E-05 | -3.18E-05 | | | | | |
| 461 | 4903.415 | | | -2.79E-05 | | | | | | | -2.02E-05 | | | | |
| 462 | 4904.361 | | | | | | | -4.94E-06 | | | | | | -6.08E-06 | |
| 463 | 4908.488 | | | | | | | | -2.87E-05 | | | | | | -2.88E-05 |
| 464 | 4938.334 | | | | | -1.47E-05 | | | | | | | -2.30E-06 | | |
| 465 | 4966.567 | | | | -2.91E-05 | | | | | | | -2.83E-05 | | | |
| 466 | 4977.011 | | | | | | | -4.95E-06 | | | | | | -5.82E-06 | |
| 467 | 4977.33 | | | -4.21E-05 | | | | | | -3.16E-05 | | | | | |
| 468 | 4987.944 | -4.44E-05 | | | | | | | -4.00E-05 | | | | | | |
| 469 | 4998.022 | | | -2.89E-05 | | | | | | | -2.26E-05 | | | | |
| 470 | 5009.297 | | | | | | | -3.01E-05 | | | | | | | -3.21E-05 |
| 471 | 5028.097 | | | | | | -4.99E-06 | | | | | | | -6.25E-06 | |
| 472 | 5029.912 | | | | | -1.70E-05 | | | | | | | -1.88E-05 | | |
| 473 | 5074.703 | | | | -2.97E-05 | | | | | | | -3.87E-05 | | | |
| 474 | 5075.961 | | | | | | | -4.95E-06 | | | | | | -6.26E-06 | |
| 475 | 5084.876 | | -4.29E-05 | | | | | | | -3.65E-05 | | | | | |
| 476 | 5086.869 | -4.37E-05 | | | | | | | -3.96E-05 | | | | | | |
| 477 | 5098.426 | | | -2.74E-05 | | | | | | | -2.14E-05 | | | | |
| 478 | 5108.285 | | | | | | | -2.94E-05 | | | | | | | -3.15E-05 |
| 479 | 5122.08 | | | | | -1.87E-05 | | | | | | | -1.73E-05 | | |
| 480 | 5128.057 | | | | | | | -5.20E-06 | | | | | | -6.47E-06 | |
| 481 | 5175.513 | | -4.76E-05 | | -3.52E-05 | | -5.20E-06 | | | -4.06E-05 | | -4.24E-05 | | -6.53E-06 | |
| 482 | 5192.698 | -5.04E-05 | | | | | | | -4.39E-05 | | | | | | |
| 483 | 5208.32 | | | -3.53E-05 | | | | | | | -2.77E-05 | | | | |
| 484 | 5212.934 | | | | | | -5.09E-06 | | | | | | | -6.44E-06 | |
| 485 | 5215.548 | | | | | | | -3.50E-05 | | | | | | | -3.59E-05 |
| 486 | 5221.938 | | | | | -2.37E-05 | | | | | | | -2.40E-05 | | |
| 487 | 5299.931 | | | | -3.51E-05 | | | | | | | -4.29E-05 | | | |
| 488 | 5321.957 | | -4.81E-05 | | | | -5.04E-06 | | | -4.11E-05 | | | | -6.36E-06 | |
| 489 | 5325.938 | -5.08E-05 | | -3.62E-05 | | -2.44E-05 | | | -3.55E-05 | -4.44E-05 | | -2.79E-05 | | -2.54E-05 | -3.69E-05 |
| 490 | 5342.853 | | | | | | -5.05E-06 | | | | | | | -6.48E-06 | |
| 491 | 5388.635 | | | | -3.67E-05 | | | | | | | -4.42E-05 | | | |
| 492 | 5406.191 | | | | | | | -4.58E-06 | | | | | | -5.49E-06 | |
| 493 | 5410.843 | | -5.11E-05 | | | | | | | -4.33E-05 | | | | | |
| 494 | 5411.764 | | | | | -2.64E-05 | | | | | | | -2.77E-05 | | |
| 495 | 5421.813 | -5.23E-05 | | | | | | | -4.64E-05 | | | | | | |
| 496 | 5425.495 | | | -3.78E-05 | | | | | | | -2.97E-05 | | | | |
| 497 | 5440.933 | | | | | | | -3.74E-05 | | | | | | | -3.85E-05 |
| 498 | 5455.344 | | | | | | | -4.51E-06 | | | | | | -5.39E-06 | |
| 499 | 5526.565 | | | | -3.76E-05 | | | | | | | -4.46E-05 | | | |
| 500 | 5552.865 | | | | | -2.62E-05 | | | | | | | -2.80E-05 | | |
| 501 | 5553.979 | | | | | | | -4.48E-06 | | | | | | -4.98E-06 | |
| 502 | 5564.523 | | -5.17E-05 | | | | | | | -4.28E-05 | | | | | |
| 503 | 5576.527 | -5.30E-05 | | | | | | | -4.59E-05 | | | | | | |
| 504 | 5581.865 | | | -3.80E-05 | | | | | | | -2.91E-05 | | | | |
| 505 | 5588.51 | | | | | | | -3.76E-05 | | | | | | | -3.88E-05 |
| 506 | 5635 | | | | | | | -4.45E-06 | | | | | | -5.01E-06 | |
| 507 | 5685.949 | | | | -3.66E-05 | | | | | | | -4.38E-05 | | | |
| 508 | 5698.753 | | | | | -2.61E-05 | | | | | | | -2.89E-05 | | |
| 509 | 5703.965 | | | | | | | -4.60E-06 | | | | | | -5.35E-06 | |
| 510 | 5716.028 | | -5.28E-05 | | | | | | | -4.37E-05 | | | | | |

| | A | B | C | D | E | F | G | H | I | J | K | L | M | N | O |
|---|---|---|---|---|---|---|---|---|---|---|---|---|---|---|---|
| 511 | 5727.595 | -5.39E-05 | | -3.91E-05 | | | | | -4.62E-05 | | -2.97E-05 | | | | |
| 512 | 5737.354 | | | | | | | -3.88E-05 | | | | | | | -4.01E-05 |
| 513 | 5741.332 | | | | | | -4.61E-06 | | | | | | | -5.28E-06 | |
| 514 | 5794.226 | | | | | -2.62E-05 | | | | | | | -2.92E-05 | | |
| 515 | 5801.268 | | | | -3.86E-05 | | | | | | | -4.69E-05 | | | |
| 516 | 5803.365 | | | | | | -4.56E-06 | | | | | | | -5.19E-06 | |
| 517 | 5848.783 | | -5.25E-05 | | | | | | | -4.28E-05 | | | | | |
| 518 | 5856.344 | | | -4.02E-05 | | | | | | | -3.05E-05 | | | | |
| 519 | 5861.944 | -5.37E-05 | | | | | | | -4.55E-05 | | | | | | |
| 520 | 5864.149 | | | | | | | -3.97E-05 | | | | | | | -4.06E-05 |
| 521 | 5874.475 | | | | | | -4.47E-06 | | | | | | | -5.11E-06 | |
| 522 | 5929.533 | | | | | -2.63E-05 | | | | | | | -2.91E-05 | | |
| 523 | 5949.733 | | | | | | -4.38E-06 | | | | | | | -4.98E-06 | |
| 524 | 5950.542 | | | -3.94E-05 | | | | | | | -4.72E-05 | | | | |
| 525 | 6005.878 | | -5.29E-05 | | | | | | | -4.22E-05 | | | | | |
| 526 | 6012.493 | -5.40E-05 | | -4.03E-05 | | | | -3.98E-05 | -4.50E-05 | | -2.97E-05 | | | | -4.01E-05 |
| 527 | 6080.295 | | | | | | -4.54E-06 | | | | | | | -5.07E-06 | |
| 528 | 6088.854 | | | | | -2.60E-05 | | | | | | | -2.87E-05 | | |
| 529 | 6125.303 | | | | -3.99E-05 | | | | | | | -4.77E-05 | | | |
| 530 | 6125.526 | | | | | | -4.54E-06 | | | | | | | -5.06E-06 | |
| 531 | 6158.33 | | -5.33E-05 | | | | | | | -4.19E-05 | | | | | |
| 532 | 6167.649 | | | -4.06E-05 | | | | -4.03E-05 | | | -2.97E-05 | | | | -4.13E-05 |
| 533 | 6177.461 | -5.42E-05 | | | | | | | -4.53E-05 | | | | | | |
| 534 | 6212.149 | | | | | | -4.35E-06 | | | | | | | -4.89E-06 | |
| 535 | 6231.597 | | | | | -2.55E-05 | | | | | | | -2.86E-05 | | |
| 536 | 6284.866 | | | | -3.90E-05 | | | | | | | -4.65E-05 | | | |
| 537 | 6302.699 | | | | | | -4.22E-06 | | | | | | | -4.90E-06 | |
| 538 | 6333.664 | | -5.16E-05 | | | | | | | -4.05E-05 | | | | | |
| 539 | 6335.533 | | | | | | | -4.02E-05 | | | | | | | -4.11E-05 |
| 540 | 6335.64 | | | -4.02E-05 | | | | | | | -2.92E-05 | | | | |
| 541 | 6344.785 | -5.34E-05 | | | | | | | -4.41E-05 | | | | | | |
| 542 | 6365.964 | | | | | | -4.25E-06 | | | | | | | -4.98E-06 | |
| 543 | 6366.14 | | | | | -2.64E-05 | | | | | | | -2.95E-05 | | |
| 544 | 6433.842 | | | | | | -4.19E-06 | | | | | | | -4.97E-06 | |
| 545 | 6435.701 | | | | -4.05E-05 | | | | | | | -4.74E-05 | | | |
| 546 | 6517.099 | | -5.12E-05 | | | | | | | -3.99E-05 | | | | | |
| 547 | 6525.031 | | | | | | | -4.02E-05 | | | | | | | -4.13E-05 |
| 548 | 6529.396 | | | -4.04E-05 | | | | | | | -2.92E-05 | | | | |
| 549 | 6533.344 | -5.24E-05 | | | | -2.60E-05 | -4.21E-06 | | -4.28E-05 | | | | -2.92E-05 | -4.87E-06 | |
| 550 | 6628.037 | | | | -4.00E-05 | | | | | | | -4.70E-05 | | | |
| 551 | 6634.932 | | | | | | -4.18E-06 | | | | | | | -4.57E-06 | |
| 552 | 6692.624 | | | | | -2.58E-05 | | | | | | | -2.90E-05 | | |
| 553 | 6693.465 | | -5.06E-05 | | | | | | | -3.86E-05 | | | | | |
| 554 | 6695.28 | | | | | | | -3.99E-05 | | | | | | | -4.05E-05 |
| 555 | 6700.831 | | | -4.00E-05 | | | | | | | -2.81E-05 | | | | |
| 556 | 6718.407 | -5.06E-05 | | | | | | | -4.17E-05 | | | | | | |
| 557 | 6747.995 | | | | | | -3.96E-06 | | | | | | | -4.26E-06 | |
| 558 | 6773.204 | | | | -3.99E-05 | | | | | | | -4.69E-05 | | | |
| 559 | 6824.458 | | | | | | -3.92E-06 | | | | | | | -4.03E-06 | |
| 560 | 6857.155 | | | | | -2.54E-05 | | | | | | | -2.82E-05 | | |
| 561 | 6866.637 | | -4.87E-05 | | | | | | | -3.45E-05 | | | | | |
| 562 | 6866.88 | | | | | | | -3.87E-05 | | | | | | | -3.90E-05 |
| 563 | 6868.689 | | | -3.87E-05 | | | | | | | -2.64E-05 | | | | |
| 564 | 6884.756 | -4.92E-05 | | | | | | | -3.76E-05 | | | | | | |
| 565 | 6888.513 | | | | | | | -3.87E-06 | | | | | | -3.86E-06 | |
| 566 | 6956.166 | | | | -3.87E-05 | | | | | | | -4.54E-05 | | | |
| 567 | 7010.366 | | | | | -2.54E-05 | | | | | | | -2.84E-05 | | |
| 568 | 7012.359 | | | | | | -3.64E-06 | | | | | | | -3.83E-06 | |
| 569 | 7064.473 | | -4.71E-05 | | | | | -3.77E-05 | | -3.42E-05 | | | | | -3.81E-05 |
| 570 | 7072.569 | | | -3.61E-05 | | | | | | | -2.29E-05 | | | | |
| 571 | 7083.67 | -4.67E-05 | | | | | | | -3.66E-05 | | | | | | |
| 572 | 7085.855 | | | | | | -3.72E-06 | | | | | | | -3.91E-06 | |
| 573 | 7145.137 | | | | | | -4.22E-06 | | | | | | | -4.28E-06 | |
| 574 | 7181.749 | | | | -3.95E-05 | | | | | | | -4.51E-05 | | | |
| 575 | 7211.083 | | | | | -2.59E-05 | | | | | | | -2.88E-05 | | |
| 576 | 7241.642 | | | | | | -4.19E-06 | | | | | | | -4.24E-06 | |
| 577 | 7279.215 | | | | | | | -3.78E-05 | | | | | | | -3.81E-05 |
| 578 | 7282.176 | | -4.38E-05 | | | | | | | -3.13E-05 | | | | | |
| 579 | 7300.326 | | | -3.70E-05 | | | | | | | -2.26E-05 | | | | |
| 580 | 7312.06 | -4.52E-05 | | | | | | | -3.52E-05 | | | | | | |
| 581 | 7375.781 | | | | | | -4.20E-06 | | | | | | | -4.21E-06 | |
| 582 | 7384.709 | | | | -3.91E-05 | | | | | | | -4.47E-05 | | | |
| 583 | 7411.163 | | | | | -2.59E-05 | | | | | | | -2.87E-05 | | |
| 584 | 7440.233 | | | | | | -4.14E-06 | | | | | | | -4.13E-06 | |
| 585 | 7500.311 | | | | | | | -3.69E-05 | | | | | | | -3.70E-05 |
| 586 | 7503.601 | | -4.34E-05 | | | | | | | -2.97E-05 | | | | | |
| 587 | 7508.091 | | | -3.69E-05 | | | | | | | -2.15E-05 | | | | |
| 588 | 7533.791 | -4.49E-05 | | | | | | | -3.27E-05 | | | | | | |
| 589 | 7596.219 | | | | | | -4.08E-06 | | | | | | | -3.92E-06 | |
| 590 | 7627.054 | | | | -3.89E-05 | | | | | | | -4.44E-05 | | | |
| 591 | 7635.67 | | | | | -2.54E-05 | | | | | | | -2.81E-05 | | |
| 592 | 7747.187 | | -4.11E-05 | -3.63E-05 | | | -3.99E-06 | -3.63E-05 | | -2.55E-05 | -2.10E-05 | | | -3.78E-06 | -3.62E-05 |
| 593 | 7763.984 | -4.33E-05 | | | | | | | -2.77E-05 | | | | | | |
| 594 | 7854.105 | | | | | -2.50E-05 | | | | | | | -2.78E-05 | | |
| 595 | 7856.085 | | | | -3.87E-05 | | | | | | | -4.40E-05 | | | |
| 596 | 7867.326 | | | | | | -3.89E-06 | | | | | | | -3.71E-06 | |
| 597 | 7951.086 | | | | | | | -3.51E-05 | | | | | | | -3.46E-05 |
| 598 | 7953.447 | | -3.65E-05 | | | | | | | -1.91E-05 | | | | | |
| 599 | 7962.504 | | | -3.36E-05 | | | | | | | -1.80E-05 | | | | |
| 600 | 7983.51 | -3.83E-05 | | | | | | | -2.18E-05 | | | | | | |
| 601 | 7993.826 | | | | | | | -3.88E-06 | | | | | | -3.67E-06 | |
| 602 | 8047.235 | | | | -3.84E-05 | -2.57E-05 | | | | | | -4.28E-05 | -2.70E-05 | | |
| 603 | 8081.491 | | | | | | -3.86E-06 | | | | | | | -3.69E-06 | |
| 604 | 8151.862 | | | | | | | -3.59E-05 | | | | | | | -3.54E-05 |
| 605 | 8163.861 | | -3.59E-05 | | | | | | | -1.79E-05 | | | | | |
| 606 | 8164.317 | | | -3.29E-05 | | | | | | | -1.76E-05 | | | | |
| 607 | 8193.641 | -3.77E-05 | | | | | | | -1.98E-05 | | | | | | |
| 608 | 8214.373 | | | | | | | -3.48E-06 | | | | | | -3.40E-06 | |
| 609 | 8228.358 | | | | | -2.61E-05 | | | | | | | -2.68E-05 | | |
| 610 | 8251.338 | | | | -3.79E-05 | | | | | | | -4.23E-05 | | | |
| 611 | 8397.026 | | | | | | | -3.49E-05 | | | | | | | -3.49E-05 |
| 612 | 8406.336 | | -3.28E-05 | | | | | | | -1.49E-05 | | | | | |

| | A | B | C | D | E | F | G | H | I | J | K | L | M | N | O |
|---|---|---|---|---|---|---|---|---|---|---|---|---|---|---|---|
| 613 | 8407.717 | | | | | | -3.43E-06 | | | | | | | -3.35E-06 | |
| 614 | 8415.39 | | | -3.06E-05 | | | | | | | -1.53E-05 | | | | |
| 615 | 8423.813 | -3.46E-05 | | | | | | | -1.77E-05 | | | | | | |
| 616 | 8446.38 | | | | | -2.64E-05 | | | | | | | -2.69E-05 | | |
| 617 | 8494.334 | | | | -3.79E-05 | | | | | | | -4.20E-05 | | | |
| 618 | 8657.263 | | | | | | -3.20E-06 | -3.42E-05 | | | | | | -3.15E-06 | -3.40E-05 |
| 619 | 8660.105 | | -2.99E-05 | | | | | | | -1.41E-05 | | | | | |
| 620 | 8661.369 | | | -2.89E-05 | | | | | | | -1.33E-05 | | | | |
| 621 | 8670.458 | -3.14E-05 | | | | | | | -1.71E-05 | | | | | | |
| 622 | 8700.146 | | | | | -2.78E-05 | | | | | | | -2.82E-05 | | |
| 623 | 8759.182 | | | | -3.91E-05 | | | | | | | -4.25E-05 | | | |
| 624 | 8762.783 | | | | | | | -3.10E-06 | | | | | | -3.16E-06 | |
| 625 | 8898.894 | | | | | | | | -3.49E-05 | | | | | | -3.48E-05 |
| 626 | 8907.659 | | -3.03E-05 | | | | | | | -1.22E-05 | | | | | |
| 627 | 8920.165 | | | -2.88E-05 | | | | | | | -1.25E-05 | | | | |
| 628 | 8936.48 | -3.13E-05 | | | | | | | -1.52E-05 | | | | | | |
| 629 | 8952.848 | | | | | -2.71E-05 | -2.92E-06 | | | | | | -2.87E-05 | -3.01E-06 | |
| 630 | 9026.337 | | | | -3.89E-05 | | | | | | | -4.22E-05 | | | |
| 631 | 9068.454 | | | | | | -2.75E-06 | | | | | | | -2.94E-06 | |
| 632 | 9138.237 | | | | | | | | -3.39E-05 | | | | | | -3.28E-05 |
| 633 | 9143.934 | | -2.95E-05 | | | | | | | -9.68E-06 | | | | | |
| 634 | 9150.996 | | | -2.80E-05 | | | | | | | -1.07E-05 | | | | |
| 635 | 9160.633 | | | | | -2.72E-05 | | | | | | | -3.00E-05 | | |
| 636 | 9166.483 | -3.06E-05 | | | | | | | -1.19E-05 | | | | | | |
| 637 | 9227.185 | | | | | | | -2.84E-06 | | | | | | -3.14E-06 | |
| 638 | 9283.216 | | | | -3.90E-05 | | | | | | | -4.17E-05 | | | |
| 639 | 9406.421 | | | | | | | | -3.31E-05 | | | | | | -3.25E-05 |
| 640 | 9425.589 | | -2.53E-05 | -2.66E-05 | | | | | | | 4.54E-07 | -5.87E-06 | | | |
| 641 | 9433.466 | | | | | | -2.58E-06 | | | | | | | -2.86E-06 | |
| 642 | 9453.131 | | | | | -2.74E-05 | | | | | | | -3.06E-05 | | |
| 643 | 9468.857 | -2.58E-05 | | | | | | | -1.20E-06 | | | | | | |
| 644 | 9567.36 | | | | -3.86E-05 | | -2.59E-06 | | | | | | -4.15E-05 | | -2.94E-06 | |
| 645 | 9617.201 | | | | | | -2.49E-06 | | | | | | | -2.87E-06 | |
| 646 | 9674.258 | | | | | | | | -3.39E-05 | | | | | | -3.31E-05 |
| 647 | 9703.731 | | -2.47E-05 | | | | | | | | 3.27E-06 | | | | |
| 648 | 9705.08 | | | -2.68E-05 | | | | | | | | -5.42E-06 | | | |
| 649 | 9719.586 | | | | | -2.74E-05 | | | | | | | -3.09E-05 | | |
| 650 | 9729.088 | -2.46E-05 | | | | | | | | 4.43E-06 | | | | | |
| 651 | 9744.674 | | | | | | -2.75E-06 | | | | | | | -2.96E-06 | |
| 652 | 9783.597 | | | | | | -2.75E-06 | | | | | | | -2.95E-06 | |
| 653 | 9821.906 | | | | -3.70E-05 | | | | | | | -4.12E-05 | | | |
| 654 | 9983.656 | | | | | | | | -3.08E-05 | | | | | | -3.06E-05 |
| 655 | 9994.72 | | -1.40E-05 | | | | -2.93E-06 | | | | 1.73E-05 | | | -2.91E-06 | |
| 656 | 9997.269 | | | -2.06E-05 | | | | | | | | 6.43E-07 | | | |
| 657 | 10023.95 | | | | | -2.57E-05 | | | | | | | -2.85E-05 | | |
| 658 | 10030.15 | -1.19E-05 | | | | | | | | 1.58E-05 | | | | | |
| 659 | 10103.44 | | | | | | -2.89E-06 | | | | | | | -2.88E-06 | |
| 660 | 10136.79 | | | | -3.63E-05 | | | | | | | -3.92E-05 | | | |
| 661 | 10245.16 | | | | | | | | -2.55E-05 | | | | | | -2.40E-05 |
| 662 | 10254.72 | | -8.93E-06 | | | | -2.91E-06 | | | | 2.36E-05 | | | -2.81E-06 | |
| 663 | 10260.52 | | | -1.10E-05 | | -2.49E-05 | | | | | | 1.44E-05 | | -2.83E-05 | | |
| 664 | 10275.06 | -9.68E-06 | | | | | | | | 1.75E-05 | | | | | |
| 665 | 10394.25 | | | | -1.77E-05 | | | | | | | | -1.72E-05 | | | |
| 666 | 10497.58 | | | | | | -2.89E-06 | | | | | | | -3.01E-06 | |
| 667 | 10575.59 | | 1.04E-06 | | | | | | -1.33E-05 | | 2.45E-05 | | | | | -1.33E-05 |
| 668 | 10585.53 | | | 4.61E-06 | | | | | | | | 2.19E-05 | | | | |
| 669 | 10607.36 | -2.37E-06 | | | | -2.49E-05 | | | | 1.55E-05 | | | | -2.95E-05 | | |
| 670 | 10773.15 | | | | -1.60E-05 | | | | | | | | -1.85E-05 | | | |
| 671 | 10814.37 | | | | | | -2.77E-06 | | | | | | | -3.00E-06 | |
| 672 | 10968.49 | | 7.19E-07 | | | | | | | | 2.02E-05 | | | | | |
| 673 | 10985.48 | | | | | | | | -1.21E-05 | | | | | | | -1.30E-05 |
| 674 | 10985.82 | | | 8.65E-07 | | | | | | | | 1.51E-05 | | | | |
| 675 | 10985.94 | | | | | | -2.83E-06 | | | | | | | | -3.21E-06 | |
| 676 | 11005.66 | -3.59E-06 | | | | | | | | 1.10E-05 | | | | | | |
| 677 | 11033.91 | | | | | | -2.35E-05 | | | | | | | -2.86E-05 | | |
| 678 | 11193.57 | | | | -1.35E-05 | | | | | | | | -1.59E-05 | | | |
| 679 | 11335.94 | | | | | | -2.89E-06 | | | | | | | | -3.19E-06 | |
| 680 | 11402.54 | | 1.83E-08 | | | | | | | | 1.98E-05 | | | | | |
| 681 | 11416.61 | | | 2.00E-06 | | | | | | | | 1.62E-05 | | | | |
| 682 | 11430.05 | | | | | | | | -8.39E-06 | | | | | | | -7.24E-06 |
| 683 | 11435.28 | -3.78E-06 | | | | | | | | 1.07E-05 | | | | | | |
| 684 | 11523.05 | | | | | -2.62E-05 | | | | | | | | -3.18E-05 | | |
| 685 | 11642.64 | | | | | | | -3.01E-06 | | | | | | | -3.26E-06 | |
| 686 | 11650.05 | | | | -1.40E-05 | | | | | | | | -1.65E-05 | | | |
| 687 | 11840.22 | | -7.20E-07 | | | | | | | | 1.91E-05 | | | | | |
| 688 | 11881.05 | | | | -5.25E-07 | | | | | | | 1.14E-05 | | | | |
| 689 | 11888.83 | -5.05E-06 | | | | | | | | 7.32E-06 | | | | | | |
| 690 | 11892.52 | | | | | | | | -9.88E-06 | | | | | | | -9.70E-06 |
| 691 | 11967.46 | | | | | | -2.86E-06 | | | | | | | -2.91E-06 | |
| 692 | 12056.04 | | | | | -2.84E-05 | | | | | | | | -3.25E-05 | | |
| 693 | 12161.64 | | | | -1.51E-05 | | | | | | | | -1.71E-05 | | | |
| 694 | 12371.57 | | -1.32E-07 | | | | | | | | 1.72E-05 | | | | | |
| 695 | 12410.76 | | | | | | | -2.51E-06 | | | | | | | -2.89E-06 | |
| 696 | 12420.09 | | | -1.14E-06 | | | | | | | | 9.53E-06 | | | | |
| 697 | 12422.29 | -4.96E-06 | | | | | | | | 6.32E-06 | | | | | | |
| 698 | 12439.76 | | | | | | | | -8.15E-06 | | | | | | | -6.88E-06 |
| 699 | 12606.02 | | | | | | -2.68E-05 | | | | | | | -3.39E-05 | | |
| 700 | 12677.94 | | | | -1.05E-05 | | | | | | | | -1.23E-05 | | | |
| 701 | 12752.52 | | | | | | | -2.09E-06 | | | | | | | -2.52E-06 | |
| 702 | 12880.49 | | | 3.44E-06 | | | | | | | | 1.99E-05 | | | | |
| 703 | 12958.34 | -1.48E-06 | | | | | | | | 8.97E-06 | | | | | | |
| 704 | 12964.04 | | | | 3.48E-06 | | | | | | | 1.40E-05 | | | | |
| 705 | 12989.01 | | | | | | | | -1.72E-06 | | | | | | | -8.70E-07 |
| 706 | 13168.08 | | | | | -8.83E-06 | | | | | | | | -1.00E-05 | | |
| 707 | 13300.13 | | | | -7.15E-06 | | | | | | | | -8.56E-06 | | | |
| 708 | 13497.88 | | | 3.76E-06 | | | | -1.90E-06 | | | | 2.06E-05 | | | -2.26E-06 | |
| 709 | 13581.2 | -1.52E-06 | | | | | | | | 9.57E-06 | | | | | | |
| 710 | 13585.14 | | | 4.79E-06 | | | | | | | | 1.54E-05 | | | | |
| 711 | 13622.93 | | | | | | | | 9.79E-08 | | | | | | | 2.77E-06 |
| 712 | 13835.38 | | | | | -5.43E-06 | | | | | | | | -6.40E-06 | | |
| 713 | 13924.09 | | | | -4.58E-06 | | | | | | | | -4.98E-06 | | | |
| 714 | 14144.96 | | 7.34E-06 | | | | | | | | 2.36E-05 | | | | | |

| | A | B | C | D | E | F | G | H | I | J | K | L | M | N | O |
|---|---|---|---|---|---|---|---|---|---|---|---|---|---|---|---|
| 715 | 14219.57 | | | | | | -1.52E-06 | | | | | | | -1.44E-06 | |
| 716 | 14220.44 | 1.04E-06 | | | | | | | 1.15E-05 | | | | | | |
| 717 | 14255.21 | | | 6.86E-06 | | | | | | | | 1.72E-05 | | | |
| 718 | 14287.16 | | | | | | 2.05E-06 | | | | | | | | 4.51E-06 |
| 719 | 14428.46 | | | | | 5.84E-06 | | | | | | | 2.85E-06 | | |
| 720 | 14550.24 | | | | 2.63E-06 | | | | | | | -8.31E-07 | | | |
| 721 | 14812.67 | | 7.73E-06 | | | | | | | 2.66E-05 | | | | | |
| 722 | 14849.69 | | | | | | -3.02E-08 | | | | | | | 7.16E-08 | |
| 723 | 14887.12 | 1.71E-06 | | | | | | | 1.49E-05 | | | | | | |
| 724 | 14901.08 | | | 1.15E-05 | | | | | | | 1.93E-05 | | | | |
| 725 | 14975.99 | | | | | | | 7.44E-06 | | | | | | | 7.31E-06 |
| 726 | 15138.24 | | | | | 9.12E-06 | | | | | | | 3.37E-06 | | |
| 727 | 15328.79 | | | | 4.93E-06 | | | | | | | -1.80E-06 | | | |
| 728 | 15483.98 | | 8.92E-06 | | | | | | | 2.85E-05 | | | | | |
| 729 | 15591.96 | 2.66E-06 | | | | | | | 1.74E-05 | | | | | | |
| 730 | 15623.1 | | | 1.27E-05 | | | | | | | 1.96E-05 | | | | |
| 731 | 15713.78 | | | | | | | 9.67E-06 | | | | | | | 7.78E-06 |
| 732 | 15754.8 | | | | | | 4.21E-07 | | | | | | | 2.39E-07 | |
| 733 | 15823.85 | | | | | 9.31E-06 | | | | | | | 1.01E-06 | | |
| 734 | 15979.63 | | | | 6.19E-06 | | | | | | | -7.33E-07 | | | |
| 735 | 16095.39 | | 1.11E-05 | | | | | | | 2.97E-05 | | | | | |
| 736 | 16185.77 | 5.23E-06 | | | | | | | 1.95E-05 | | | | | | |
| 737 | 16191.4 | | | 1.56E-05 | | | | | | | 2.26E-05 | | | | |
| 738 | 16215.39 | | | | | | 6.35E-07 | | | | | | | 6.69E-07 | |
| 739 | 16247.7 | | | | | | | 1.40E-05 | | | | | | | 1.18E-05 |
| 740 | 16332.6 | | | | | 1.34E-05 | | | | | | | 5.49E-06 | | |
| 741 | 16464.1 | | | | 1.02E-05 | | | | | | | 2.78E-06 | | | |
| 742 | 16556.75 | | 1.27E-05 | | | | | | | 3.14E-05 | | | | | |
| 743 | 16588.63 | 5.70E-06 | | | | | | | 2.00E-05 | | | | | | |
| 744 | 16621.15 | | | 1.77E-05 | | | | | | | 2.39E-05 | | | | |
| 745 | 16672.91 | | | | | | | 1.73E-05 | | | | | | | 1.35E-05 |
| 746 | 16721.39 | | | | | 1.64E-05 | | | | | | | 8.29E-06 | | |
| 747 | 16818.57 | | | | 1.27E-05 | | | | | | | 4.14E-06 | | | |
| 748 | 16862.38 | | 1.30E-05 | | | | | | | 3.08E-05 | | | | | |
| 749 | 16908.54 | 6.13E-06 | | | | | | | 2.04E-05 | | | | | | |
| 750 | 16924.85 | | | | | | 6.93E-07 | | | | | | | 1.47E-06 | |
| 751 | 16934.23 | | | 1.89E-05 | | | | | | | 2.76E-05 | | | | |
| 752 | 16989.81 | | | | | | | 1.95E-05 | | | | | | | 1.50E-05 |
| 753 | 17026.52 | | | | | 1.72E-05 | | | | | | | 1.01E-05 | | |
| 754 | 17105.94 | | | | 1.56E-05 | | | | | | | 5.81E-06 | | | |
| 755 | 17128.35 | | 1.47E-05 | | | | | | | 3.33E-05 | | | | | |
| 756 | 17159.77 | 8.11E-06 | | | | | | | 2.28E-05 | | | | | | |
| 757 | 17193.38 | | | 2.10E-05 | | | | | | | 3.06E-05 | | | | |
| 758 | 17238.17 | | | | | | | 2.24E-05 | | | | | | | 1.73E-05 |
| 759 | 17266.99 | | | | | 1.86E-05 | | | | | | | 1.19E-05 | | |
| 760 | 17347.04 | | | | | | 9.78E-07 | | | | | | | 2.02E-06 | |
| 761 | 17354.5 | | | | 1.65E-05 | | | | | | | 7.07E-06 | | | |
| 762 | 17366.09 | | 1.64E-05 | | | | | | | 3.26E-05 | | | | | |
| 763 | 17394.1 | 8.97E-06 | | | | | | | 2.19E-05 | | | | | | |
| 764 | 17411.82 | | | 2.09E-05 | | | | | | | 3.00E-05 | | | | |
| 765 | 17459.48 | | | | | | | 2.22E-05 | | | | | | | 1.85E-05 |
| 766 | 17476.57 | | | | | 1.91E-05 | | | | | | | 1.27E-05 | | |
| 767 | 17557.31 | | | | 1.72E-05 | | | | | | | 7.13E-06 | | | |
| 768 | 17576.88 | | 1.72E-05 | | | | | | | 3.30E-05 | | | | | |
| 769 | 17606.94 | 9.89E-06 | | | | | | | 2.28E-05 | | | | | | |
| 770 | 17630.46 | | | 2.17E-05 | | | | | | | 3.06E-05 | | | | |
| 771 | 17679.2 | | | | | 1.90E-05 | 2.30E-05 | | | | | | | 1.29E-05 | | 1.88E-05 |
| 772 | 17778.17 | | | | 1.81E-05 | | | | | | | 8.18E-06 | | | |
| 773 | 17796.76 | | 1.90E-05 | | | | | | | 3.38E-05 | | | | | |
| 774 | 17819.45 | 1.20E-05 | | | | | | | 2.51E-05 | | | | | | |
| 775 | 17834.71 | | | 2.26E-05 | | | | | | | 3.17E-05 | | | | |
| 776 | 17860.42 | | | | | | 1.51E-06 | | | | | | | 2.60E-06 | |
| 777 | 17907.29 | | | | | 1.86E-05 | 2.32E-05 | | | | | | | 1.27E-05 | | 2.13E-05 |
| 778 | 17973.02 | | 2.05E-05 | | | | | | | 3.92E-05 | | | | | |
| 779 | 17977.18 | | | | 1.83E-05 | | | | | | | 9.08E-06 | | | |
| 780 | 18004.75 | 1.39E-05 | | | | | | | 3.15E-05 | | | | | | |
| 781 | 18021.61 | | | 2.55E-05 | | | | | | | 3.90E-05 | | | | |
| 782 | 18118.64 | | | | | 1.87E-05 | 2.35E-05 | | | | | | | 1.40E-05 | | 2.10E-05 |
| 783 | 18195.29 | | 2.01E-05 | | | | | | | 3.84E-05 | | | | | |
| 784 | 18208.88 | 1.33E-05 | | | 1.78E-05 | | | | | 3.12E-05 | | | 8.38E-06 | | | |
| 785 | 18215.74 | | | 2.58E-05 | | | | | | | 3.94E-05 | | | | | |
| 786 | 18311.21 | | | | | 1.77E-05 | | | | | | | 1.39E-05 | | | |
| 787 | 18326.07 | | | | | | 2.39E-05 | | | | | | | | | 2.13E-05 |
| 788 | 18369.16 | | | | | | 1.42E-06 | | | | | | | 2.44E-06 | |
| 789 | 18379.61 | | 2.04E-05 | | | | | | | 4.08E-05 | | | | | |
| 790 | 18402.03 | 1.38E-05 | | | | | | | 3.31E-05 | | | | | | |
| 791 | 18415.74 | | | | 1.92E-05 | | | | | | | 1.03E-05 | | | | |
| 792 | 18423.31 | | | 2.59E-05 | | | | | | | 4.06E-05 | | | | | |
| 793 | 18534.38 | | | | | 1.85E-05 | | | | | | | 1.45E-05 | | | |
| 794 | 18539.12 | | | | | | 2.54E-05 | | | | | | | | | 2.23E-05 |
| 795 | 18566.77 | | 2.03E-05 | | | | | | | 4.10E-05 | | | | | |
| 796 | 18582.18 | 1.37E-05 | | | | | | | 3.33E-05 | | | | | | |
| 797 | 18597.33 | | | 2.57E-05 | | | | | | | 3.89E-05 | | | | | |
| 798 | 18614.41 | | | | 1.97E-05 | | | | | | | 1.01E-05 | | | |
| 799 | 18716.89 | | | | | 1.76E-05 | | | | | | | 1.42E-05 | | | |
| 800 | 18738.2 | | | | | | 2.41E-05 | | | | | | | | | 2.15E-05 |
| 801 | 18763.94 | | 1.97E-05 | | | | | | | 3.99E-05 | | | | | |
| 802 | 18787.61 | 1.28E-05 | | | | | | | 3.26E-05 | | | | | | |
| 803 | 18798.31 | | | 2.46E-05 | | | | | | | 3.84E-05 | | | | | |
| 804 | 18802.45 | | | | | | 1.32E-06 | | | | | | | 2.16E-06 | |
| 805 | 18805.95 | | | | 1.83E-05 | | | | | | | 9.61E-06 | | | |
| 806 | 18889.43 | | | | | 1.80E-05 | | | | | | | 1.47E-05 | | | |
| 807 | 18914.9 | | | | | | 2.40E-05 | | | | | | | | | 2.18E-05 |
| 808 | 18926.08 | | 1.99E-05 | | | | | | | 4.00E-05 | | | | | |
| 809 | 18959.65 | 1.34E-05 | | | | | | | 3.29E-05 | | | | | | |
| 810 | 18984.76 | | | 2.57E-05 | | | | | | | 3.96E-05 | | | | | |
| 811 | 19006.21 | | | | 1.87E-05 | | | | | | | 1.06E-05 | | | |
| 812 | 19096.12 | | | | | 1.87E-05 | | | | | | | 1.43E-05 | | | |
| 813 | 19108.94 | | | | | | 2.44E-05 | | | | | | | | | 2.21E-05 |
| 814 | 19109.95 | | 1.65E-05 | | | | | | | 3.53E-05 | | | | | |
| 815 | 19132.55 | 9.09E-06 | | | | | | | 2.89E-05 | | | | | | |
| 816 | 19148.25 | | | | | | 3.70E-06 | | | | | | | 5.64E-06 | |

| | A | B | C | D | E | F | G | H | I | J | K | L | M | N | O |
|---|---|---|---|---|---|---|---|---|---|---|---|---|---|---|---|
| 817 | 19165.39 | | | 2.38E-05 | | | | | | | 3.67E-05 | | | | |
| 818 | 19185.18 | | | | 1.89E-05 | | | | | | | 1.21E-05 | | | |
| 819 | 19282.34 | | | | | 1.92E-05 | | | | | | | 1.49E-05 | | |
| 820 | 19283.39 | | 1.61E-05 | | | | | | | 3.51E-05 | | | | | |
| 821 | 19299.41 | | | | | | | 2.42E-05 | | | | | | | 2.24E-05 |
| 822 | 19311.82 | 9.00E-06 | | | | | | | 2.90E-05 | | | | | | |
| 823 | 19355.14 | | | 2.28E-05 | | | | | | | 3.72E-05 | | | | |
| 824 | 19383.4 | | | | 1.79E-05 | | | | | | | | 1.24E-05 | | |
| 825 | 19461.7 | | 1.57E-05 | | | | | | | 3.46E-05 | | | | | |
| 826 | 19462.41 | | | | | 1.85E-05 | | | | | | | 1.46E-05 | | |
| 827 | 19480.98 | | | | | | | 2.27E-05 | | | | | | | 2.15E-05 |
| 828 | 19487.99 | 8.86E-06 | | | | | | | 2.86E-05 | | | | | | |
| 829 | 19488.46 | | | | | | 3.66E-06 | | | | | | | 5.22E-06 | |
| 830 | 19529.73 | | | 2.25E-05 | | | | | | | 3.62E-05 | | | | |
| 831 | 19569.73 | | | | 1.84E-05 | | | | | | | 1.22E-05 | | | |
| 832 | 19621.47 | | 1.45E-05 | | | | | | | 3.15E-05 | | | | | |
| 833 | 19636.6 | | | | | 1.80E-05 | | | | | | | 1.40E-05 | | |
| 834 | 19643.42 | 7.97E-06 | | | | | | 2.55E-05 | | | | | | | |
| 835 | 19645.55 | | | | | | | 2.23E-05 | | | | | | | 2.08E-05 |
| 836 | 19692.49 | | | 2.12E-05 | | | | | | | 3.18E-05 | | | | |
| 837 | 19732.32 | | | | 1.65E-05 | | | | | | | 1.11E-05 | | | |
| 838 | 19791.69 | | 1.50E-05 | | | | | | | 3.02E-05 | | | | | |
| 839 | 19791.87 | | | | | | 3.67E-06 | | | | | | | 4.84E-06 | |
| 840 | 19816.4 | | | | | 1.79E-05 | | | | | | | 1.34E-05 | | |
| 841 | 19822.73 | 9.87E-06 | | | | | | | 2.54E-05 | | | | | | |
| 842 | 19826.05 | | | | | | | 2.28E-05 | | | | | | | 2.27E-05 |
| 843 | 19859.1 | | | 2.12E-05 | | | | | | | 3.26E-05 | | | | |
| 844 | 19904.71 | | | | 1.92E-05 | | | | | | | 1.58E-05 | | | |
| 845 | 19956.74 | | 1.59E-05 | | | | | | | 2.74E-05 | | | | | |
| 846 | 19963 | 1.01E-05 | | | | | | 2.23E-05 | | | | | | | |
| 847 | 19968.76 | | | | | 1.86E-05 | | | | | | | 1.41E-05 | | |
| 848 | 19971.97 | | | | | | | 2.24E-05 | | | | | | | 2.33E-05 |
| 849 | 19987.68 | | | 2.04E-05 | | | | | | | 3.15E-05 | | | | |
| 850 | 20032.5 | | | | 1.84E-05 | | | | | | | 1.44E-05 | | | |
| 851 | 20068.9 | | 1.47E-05 | | | | | | | 2.39E-05 | | | | | |
| 852 | 20072.48 | | | | | | 4.77E-06 | | | | | | | 6.16E-06 | |
| 853 | 20090.9 | 8.87E-06 | | | | 1.93E-05 | | | 1.95E-05 | | | | 1.43E-05 | | |
| 854 | 20108.63 | | | | | | | 2.24E-05 | | | | | | | 2.29E-05 |
| 855 | 20129.71 | | | 1.97E-05 | | | | | | | 3.00E-05 | | | | |
| 856 | 20179.55 | | | | 1.88E-05 | | | | | | | 1.44E-05 | | | |
| 857 | 20240.38 | | 1.68E-05 | | | | | | | 2.26E-05 | | | | | |
| 858 | 20267.27 | | | | | 1.98E-05 | | | | | | | 1.53E-05 | | |
| 859 | 20271.26 | 1.03E-05 | | | | | | | 1.83E-05 | | | | | | |
| 860 | 20278.69 | | | | | | | 2.28E-05 | | | | | | | 2.33E-05 |
| 861 | 20305.93 | | | 1.95E-05 | | | | | | | 2.88E-05 | | | | |
| 862 | 20342.42 | | | | | | 5.24E-06 | | | | | | | 6.50E-06 | |
| 863 | 20343.92 | | | | 1.76E-05 | | | | | | | 1.36E-05 | | | |
| 864 | 20396.31 | | 1.44E-05 | | | | | | | 2.01E-05 | | | | | |
| 865 | 20428.73 | 8.92E-06 | | | | 1.90E-05 | | | 1.63E-05 | | | | 1.49E-05 | | |
| 866 | 20437.31 | | | | | | | 2.12E-05 | | | | | | | 2.21E-05 |
| 867 | 20465.75 | | | 1.87E-05 | | | | | | | 2.84E-05 | | | | |
| 868 | 20521.85 | | | | 1.81E-05 | | | | | | | 1.45E-05 | | | |
| 869 | 20563.11 | | 1.44E-05 | | | | | | | 1.95E-05 | | | | | |
| 870 | 20579.28 | 9.39E-06 | | | | | | | 1.64E-05 | | | | | | |
| 871 | 20579.48 | | | | | 1.96E-05 | | | | | | | 1.54E-05 | | |
| 872 | 20604.7 | | | | | | | 2.15E-05 | | | | | | | 2.28E-05 |
| 873 | 20638.82 | | | 1.86E-05 | | | | | | | 2.74E-05 | | | | |
| 874 | 20676.04 | | | | 1.72E-05 | | | | | | | 1.42E-05 | | | |
| 875 | 20684.15 | | 1.46E-05 | | | | 4.33E-06 | | 1.98E-05 | | | | | 5.81E-06 | |
| 876 | 20691.64 | | | | | 2.05E-05 | | | | | | | 1.58E-05 | | |
| 877 | 20698.88 | 9.77E-06 | | | | | | | 1.69E-05 | | | | | | |
| 878 | 20707.67 | | | | | | | 2.26E-05 | | | | | | | 2.36E-05 |
| 879 | 20738.14 | | | 1.92E-05 | | | | | | | 2.74E-05 | | | | |
| 880 | 20794.3 | | | | 1.76E-05 | | | | | | | 1.49E-05 | | | |
| 881 | 20840.37 | | 1.48E-05 | | | | | | | 1.93E-05 | | | | | |
| 882 | 20851.5 | | | | | 2.01E-05 | | | | | | | 1.51E-05 | | |
| 883 | 20866.41 | 1.01E-05 | | | | | | | 1.69E-05 | | | | | | |
| 884 | 20891.41 | | | | | | | 2.24E-05 | | | | | | | 2.31E-05 |
| 885 | 20910.29 | | | 1.89E-05 | | | | | | | 2.66E-05 | | | | |
| 886 | 20955.47 | | | | | | 4.37E-06 | | | | | | | 5.51E-06 | |
| 887 | 20975.7 | | | | 1.73E-05 | | | | | | | 1.09E-05 | | | |
| 888 | 21010.02 | | 1.54E-05 | | | | | | | 1.63E-05 | | | | | |
| 889 | 21019.21 | 1.07E-05 | | | | | | | 2.66E-06 | | | | | | |
| 890 | 21019.62 | | | | | 2.02E-05 | | | | | | | 1.44E-05 | | |
| 891 | 21031.94 | | | | | | | 2.59E-05 | | | | | | | 2.51E-05 |
| 892 | 21051.73 | | | 2.10E-05 | | | | | | | 2.48E-05 | | | | |
| 893 | 21093.24 | | | | 2.09E-05 | | | | | | | 1.97E-05 | | | |
| 894 | 21136.62 | | 1.58E-05 | | | | | | | 1.61E-05 | | | | | |
| 895 | 21172.87 | | | | | 2.00E-05 | | | | | | | 1.42E-05 | | |
| 896 | 21181.04 | 1.12E-05 | | | | | | | 2.10E-06 | | | | | | |
| 897 | 21199.84 | | | | | | | 2.64E-05 | | | | | | | 2.51E-05 |
| 898 | 21225.74 | | | 2.20E-05 | | | | | | | 2.46E-05 | | | | |
| 899 | 21271.89 | | | | 2.32E-05 | | | | | | | 2.31E-05 | | | |
| 900 | 21273.17 | | | | | | 4.22E-06 | | | | | | | 5.04E-06 | |
| 901 | 21312.18 | | 1.76E-05 | | | | | | | 1.70E-05 | | | | | |
| 902 | 21332.53 | | | | | 1.96E-05 | | | | | | | 1.37E-05 | | |
| 903 | 21345.14 | 1.14E-05 | | | | | | | 1.76E-06 | | | | | | |
| 904 | 21353.85 | | | | | | | 2.78E-05 | | | | | | | 2.65E-05 |
| 905 | 21367.91 | | | 2.22E-05 | | | | | | | 2.46E-05 | | | | |
| 906 | 21438.62 | | | | | 2.42E-05 | | | | | | | 2.79E-05 | | |
| 907 | 21480.84 | | 1.74E-05 | | | | | | | 1.68E-05 | | | | | |
| 908 | 21488.46 | | | | | | 4.03E-06 | | | | | | | 4.96E-06 | |
| 909 | 21511.29 | 1.21E-05 | | | | | | | 2.62E-06 | | | | | | |
| 910 | 21514.23 | | | | | 1.97E-05 | | | | | | | 1.42E-05 | | |
| 911 | 21526.25 | | | | | | | 2.80E-05 | | | | | | | 2.75E-05 |
| 912 | 21534.54 | | | 2.29E-05 | | | | | | | 2.53E-05 | | | | |
| 913 | 21602.13 | | | | 2.50E-05 | | | | | | | 2.86E-05 | | | |
| 914 | 21638.35 | | 1.66E-05 | | | | | | | 1.54E-05 | | | | | |
| 915 | 21654.18 | 1.17E-05 | | | | | | | 1.45E-06 | | | | | | |
| 916 | 21657.72 | | | | | 1.96E-05 | | | | | | | 1.41E-05 | | |
| 917 | 21663.06 | | | | | | | 2.84E-05 | | | | | | | 2.76E-05 |
| 918 | 21667.8 | | | 2.27E-05 | | | | | | | 2.51E-05 | | | | |

| | A | B | C | D | E | F | G | H | I | J | K | L | M | N | O |
|---|---|---|---|---|---|---|---|---|---|---|---|---|---|---|---|
| 919 | 21718.23 | | | | 2.54E-05 | | | | | | | 3.03E-05 | | | |
| 920 | 21733.44 | | | | | | 4.01E-06 | | | | | | | 4.91E-06 | |
| 921 | 21762.75 | | 1.67E-05 | | | | | | | 1.55E-05 | | | | | |
| 922 | 21776.05 | 1.20E-05 | | | | 1.98E-05 | | | 2.13E-06 | | | | 1.42E-05 | | |
| 923 | 21795.14 | | | | | | | 2.71E-05 | | | | | | | 2.62E-05 |
| 924 | 21800.57 | | | 2.15E-05 | | | | | | | 2.32E-05 | | | | |
| 925 | 21836.73 | | | | 2.46E-05 | | | | | | | | 3.02E-05 | | |
| 926 | 21871.24 | | 1.41E-05 | | | | | | | 1.21E-05 | | | | | |
| 927 | 21898.26 | 1.26E-05 | | | | | | | 9.02E-07 | | | | | | |
| 928 | 21901.81 | | | | | 1.94E-05 | | | | | | | 1.51E-05 | | |
| 929 | 21922.14 | | | | | | | 2.87E-05 | | | | | | | 2.93E-05 |
| 930 | 21941.88 | | | 2.31E-05 | | | | | | | | 2.38E-05 | | | |
| 931 | 21988.43 | | | | | | 4.32E-06 | | | | | | | 5.52E-06 | |
| 932 | 22005.32 | | | | 2.47E-05 | | | | | | | 3.03E-05 | | | |
| 933 | 22057.42 | | 1.45E-05 | | | | | | | 1.09E-05 | | | | | |
| 934 | 22074.01 | 1.10E-05 | | | | | | | -1.41E-06 | | | | | | |
| 935 | 22092.37 | | | | | | | 2.85E-05 | | | | | | | 2.88E-05 |
| 936 | 22097.87 | | | | | 1.95E-05 | | | | | | | 1.61E-05 | | |
| 937 | 22102.87 | | | 2.37E-05 | | | | | | | 2.37E-05 | | | | |
| 938 | 22163.86 | | | | 2.61E-05 | | | | | | | | 3.13E-05 | | |
| 939 | 22191.92 | | 1.57E-05 | | | | | | | 1.24E-05 | | | | | |
| 940 | 22218.63 | 1.21E-05 | | | | | | | 3.47E-07 | | | | | | |
| 941 | 22248.65 | | | | | | | 3.07E-05 | | | | | | | 3.04E-05 |
| 942 | 22262.85 | | | 2.54E-05 | | | | | | | | 2.42E-05 | | | |
| 943 | 22265.13 | | | | | 2.07E-05 | | | | | | | 1.77E-05 | | |
| 944 | 22282.75 | | | | | | 4.62E-06 | | | | | | | 5.65E-06 | |
| 945 | 22315.47 | | | | 2.60E-05 | | | | | | | 3.12E-05 | | | |
| 946 | 22365.72 | | 1.61E-05 | | | | | | | 1.24E-05 | | | | | |
| 947 | 22386.44 | 1.25E-05 | | | | | | | 1.42E-06 | | | | | | |
| 948 | 22414.18 | | | | | | | 3.04E-05 | | | | | | | 2.99E-05 |
| 949 | 22428.61 | | | 2.59E-05 | | | | | | | 2.49E-05 | | | | |
| 950 | 22465.85 | | | | | 2.33E-05 | | | | | | | 1.85E-05 | | |
| 951 | 22511.65 | | | | 2.72E-05 | | | | | | | 3.15E-05 | | | |
| 952 | 22566.69 | | 1.69E-05 | | | | | | | 1.31E-05 | | | | | |
| 953 | 22586.96 | 1.19E-05 | | | | | | | 1.07E-06 | | | | | | |
| 954 | 22605.37 | | | | | | | 2.98E-05 | | | | | | | 2.87E-05 |
| 955 | 22612.24 | | | 2.54E-05 | | | | | | | | 2.44E-05 | | | |
| 956 | 22642.41 | | | | | | 4.76E-06 | | | | | | | 6.17E-06 | |
| 957 | 22663.27 | | | | | 2.32E-05 | | | | | | | 1.59E-05 | | |
| 958 | 22681.81 | | | | 2.75E-05 | | | | | | | 2.95E-05 | | | |
| 959 | 22723.96 | | 1.84E-05 | | | | | | | 1.21E-05 | | | | | |
| 960 | 22745.82 | 1.25E-05 | | | | | | | 9.82E-07 | | | | | | |
| 961 | 22776.75 | | | | | | | 3.12E-05 | | | | | | | 2.94E-05 |
| 962 | 22787.69 | | | 2.67E-05 | | | | | | | 2.48E-05 | | | | |
| 963 | 22850.89 | | | | 2.88E-05 | | | | | | | 3.00E-05 | | | |
| 964 | 22852.19 | | | | | 2.32E-05 | | | | | | | 1.56E-05 | | |
| 965 | 22905.26 | | 1.84E-05 | | | | | | | 1.15E-05 | | | | | |
| 966 | 22927.54 | 1.27E-05 | | | | | | | 3.01E-07 | | | | | | |
| 967 | 22952.76 | | | | | | | 3.00E-05 | | | | | | | 2.78E-05 |
| 968 | 22958.64 | | | 2.64E-05 | | | | | | | | 2.42E-05 | | | |
| 969 | 22967.21 | | | | | | 4.50E-06 | | | | | | | 5.88E-06 | |
| 970 | 23035.71 | | | | 2.80E-05 | | | | | | | 2.92E-05 | | | |
| 971 | 23049.44 | | | | | 2.28E-05 | | | | | | | 1.55E-05 | | |
| 972 | 23083.56 | | 1.83E-05 | | | | | | | 1.16E-05 | | | | | |
| 973 | 23110.18 | 1.30E-05 | | | | | | | 1.92E-07 | | | | | | |
| 974 | 23141.57 | | | | | | | 3.08E-05 | | | | | | | 2.85E-05 |
| 975 | 23148.8 | | | 2.64E-05 | | | | | | | 2.40E-05 | | | | |
| 976 | 23190.1 | | | | | | 4.59E-06 | | | | | | | 6.29E-06 | |
| 977 | 23229.9 | | | | 2.66E-05 | | | | | | | 2.74E-05 | | | |
| 978 | 23250.64 | | | | | 2.29E-05 | | | | | | | 1.55E-05 | | |
| 979 | 23261.59 | | 1.69E-05 | | | | | | | 7.65E-06 | | | | | |
| 980 | 23312.86 | 1.19E-05 | | | | | | | -2.84E-06 | | | | | | |
| 981 | 23339.01 | | | | | | | 2.84E-05 | | | | | | | 2.63E-05 |
| 982 | 23343.31 | | | 2.56E-05 | | | | | | | 2.26E-05 | | | | |
| 983 | 23409.89 | | | | | | 5.05E-06 | | | | | | | 6.18E-06 | |
| 984 | 23418.03 | | | | 2.59E-05 | | | | | | | 2.62E-05 | | | |
| 985 | 23447.71 | | | | | 2.15E-05 | | | | | | | 1.37E-05 | | |
| 986 | 23469.35 | | 1.69E-05 | | | | | | | 7.45E-06 | | | | | |
| 987 | 23488.85 | 1.16E-05 | | | | | | | -2.60E-06 | | | | | | |
| 988 | 23510.63 | | | | | | | 2.82E-05 | | | | | | | 2.64E-05 |
| 989 | 23513.15 | | | 2.51E-05 | | | | | | | 2.27E-05 | | | | |
| 990 | 23554.77 | | | | | | 5.67E-06 | | | | | | | 6.67E-06 | |
| 991 | 23594.29 | | | | 2.38E-05 | | | | | | | 2.44E-05 | | | |
| 992 | 23632 | | | | | 2.08E-05 | | | | | | | 1.36E-05 | | |
| 993 | 23635.23 | | 1.74E-05 | | | | | | | 6.29E-06 | | | | | |
| 994 | 23676.3 | 1.22E-05 | | | | | | | -4.52E-06 | | | | | | |
| 995 | 23701.26 | | | | | | | 2.66E-05 | | | | | | | 2.58E-05 |
| 996 | 23702.54 | | | 2.54E-05 | | | | | | | 2.19E-05 | | | | |
| 997 | 23783.23 | | | | 2.31E-05 | | | | | | | 2.78E-05 | | | |
| 998 | 23822.98 | | 1.79E-05 | | | | | | | 7.19E-06 | | | | | |
| 999 | 23824.3 | | | | | 2.11E-05 | | | | | | | 1.55E-05 | | |
| 1000 | 23842.1 | 1.27E-05 | | | | | 5.35E-06 | | -3.40E-06 | | | | | 5.86E-06 | |
| 1001 | 23861.93 | | | | | | | 2.61E-05 | | | | | | | 2.65E-05 |
| 1002 | 23864.31 | | | 2.48E-05 | | | | | | | 2.18E-05 | | | | |
| 1003 | 23956.69 | | | | 2.33E-05 | | | | | | | 2.83E-05 | | | |
| 1004 | 24003.09 | | 1.70E-05 | | | | | | | 6.31E-06 | | | | | |
| 1005 | 24009.6 | | | | | 2.03E-05 | | | | | | | 1.60E-05 | | |
| 1006 | 24030.75 | 1.24E-05 | | | | | | | -3.71E-06 | | | | | | |
| 1007 | 24051.02 | | | | | | | 2.53E-05 | | | | | | | 2.83E-05 |
| 1008 | 24058.78 | | | 2.41E-05 | | | 5.10E-06 | | | | 2.19E-05 | | | 5.71E-06 | |
| 1009 | 24140.68 | | | | 2.25E-05 | | | | | | | 2.97E-05 | | | |
| 1010 | 24210.33 | | 1.82E-05 | | | | | | | 6.83E-06 | | | | | |
| 1011 | 24223.46 | | | | | 2.07E-05 | | | | | | | 1.66E-05 | | |
| 1012 | 24248.25 | 1.26E-05 | | | | | | | -3.37E-06 | | | | | | |
| 1013 | 24270.66 | | | | | | | 2.68E-05 | | | | | | | 2.90E-05 |
| 1014 | 24277.76 | | | 2.52E-05 | | | | | | | 2.22E-05 | | | | |
| 1015 | 24332.03 | | | | 2.41E-05 | | | | | | | 3.00E-05 | | | |
| 1016 | 24354.74 | | 1.86E-05 | | | | | | | 6.58E-06 | | | | | |
| 1017 | 24377.54 | | | | | 2.10E-05 | | | | | | | 1.67E-05 | | |
| 1018 | 24391.82 | 1.27E-05 | | | | | | | -3.47E-06 | | | | | | |
| 1019 | 24431.72 | | | 2.47E-05 | | | | 2.60E-05 | | | 2.11E-05 | | | | 2.84E-05 |
| 1020 | 24516.59 | | | | | | 5.11E-06 | | | | | | | 5.87E-06 | |

| | A | B | C | D | E | F | G | H | I | J | K | L | M | N | O |
|---|---|---|---|---|---|---|---|---|---|---|---|---|---|---|---|
| 1021 | 24531.08 | | | | 2.24E-05 | | | | | | | 2.87E-05 | | | |
| 1022 | 24610.69 | | 1.81E-05 | | | | | | | 5.02E-06 | | | | | |
| 1023 | 24639.23 | | | | | 2.02E-05 | | | | | | | 1.54E-05 | | |
| 1024 | 24641.85 | 1.24E-05 | | | | | | | -3.20E-06 | | | | | | |
| 1025 | 24666.17 | | | 2.40E-05 | | | | 2.47E-05 | | | 2.16E-05 | | | | 2.75E-05 |
| 1026 | 24751.66 | | | | 2.22E-05 | | | | | | | 2.88E-05 | | | |
| 1027 | 24791.65 | | 1.74E-05 | | | | | | | 2.68E-07 | | | | | |
| 1028 | 24828.14 | 1.14E-05 | | | | | | | -7.67E-06 | | | | | | |
| 1029 | 24828.35 | | | | 2.10E-05 | | | | | | | | 1.54E-05 | | |
| 1030 | 24839.62 | | | 2.12E-05 | | | | 2.41E-05 | | | 1.48E-05 | | | | 2.65E-05 |
| 1031 | 24901.75 | | | | 2.19E-05 | | | | | | | 2.79E-05 | | | |
| 1032 | 24917.38 | | | | | | 5.36E-06 | | | | | | | 5.94E-06 | |
| 1033 | 24936.83 | | 1.79E-05 | | | | | | | 1.61E-06 | | | | | |
| 1034 | 24986.94 | 1.27E-05 | | | | | | | -5.51E-06 | | | | | | |
| 1035 | 25004.57 | | | | | 2.05E-05 | | | | | | | 1.58E-05 | | |
| 1036 | 25012.01 | | | | | | | 2.36E-05 | | | | | | | 2.56E-05 |
| 1037 | 25012.22 | | | 2.08E-05 | | | | | | | 1.41E-05 | | | | |
| 1038 | 25088.13 | | | | 2.10E-05 | | | | | | | 2.61E-05 | | | |
| 1039 | 25130.58 | | 1.75E-05 | | | | | | | -6.61E-07 | | | | | |
| 1040 | 25147.95 | 1.20E-05 | | | | 1.99E-05 | | | -7.50E-06 | | | | 1.51E-05 | | |
| 1041 | 25154.43 | | | | | | | 2.21E-05 | | | | | | | 2.46E-05 |
| 1042 | 25157.56 | | | 2.00E-05 | | | | | | | 1.28E-05 | | | | |
| 1043 | 25199.53 | | | | 2.02E-05 | | | | | | | 2.56E-05 | | | |
| 1044 | 25227.97 | | | | | | 5.74E-06 | | | | | | | 6.18E-06 | |
| 1045 | 25243.48 | | 1.71E-05 | | | | | | | -1.71E-06 | | | | | |
| 1046 | 25274.28 | 1.16E-05 | | | | | | | -8.69E-06 | | | | | | |
| 1047 | 25278.03 | | | | | 1.93E-05 | | | | | | | 1.41E-05 | | |
| 1048 | 25287.69 | | | 1.95E-05 | | | | 2.07E-05 | | | 1.19E-05 | | | | 2.32E-05 |
| 1049 | 25344.73 | | | | 2.00E-05 | | | | | | | 2.52E-05 | | | |
| 1050 | 25383.69 | | 1.67E-05 | | | | | | | -4.66E-06 | | | | | |
| 1051 | 25421.88 | 1.05E-05 | | | | | | | -1.22E-05 | | | | | | |
| 1052 | 25432.79 | | | | | 1.73E-05 | | 1.79E-05 | | | | | 1.22E-05 | | 2.16E-05 |
| 1053 | 25433.58 | | | 1.79E-05 | | | | | | | 8.61E-06 | | | | |
| 1054 | 25489.99 | | | | 1.81E-05 | | | | | | | 2.39E-05 | | | |
| 1055 | 25531.11 | | 1.59E-05 | | | | | | | -7.01E-06 | | | | | |
| 1056 | 25543.24 | | | | | | 5.99E-06 | | | | | | | 6.29E-06 | |
| 1057 | 25554.8 | 1.05E-05 | | | | | | | -1.29E-05 | | | | | | |
| 1058 | 25564.96 | | | 1.80E-05 | | 1.70E-05 | | 1.78E-05 | | | 7.85E-06 | | 1.17E-05 | | 2.12E-05 |
| 1059 | 25618.4 | | | | 1.81E-05 | | | | | | | 2.37E-05 | | | |
| 1060 | 25659.79 | | 1.54E-05 | | | | | | | -8.62E-06 | | | | | |
| 1061 | 25682.4 | 9.87E-06 | | | | | | | -1.37E-05 | | | | | | |
| 1062 | 25683.57 | | | | | 1.67E-05 | | | | | | | 1.10E-05 | | |
| 1063 | 25687.49 | | | | | | | 1.75E-05 | | | | | | | 2.08E-05 |
| 1064 | 25689.34 | | | 1.72E-05 | | | | | | | 7.77E-06 | | | | |
| 1065 | 25725.82 | | | | 1.77E-05 | | | | | | | 2.33E-05 | | | |
| 1066 | 25773.81 | | 1.57E-05 | | | | | | | -9.09E-06 | | | | | |
| 1067 | 25787.54 | | | | | | 5.87E-06 | | | | | | | 6.11E-06 | |
| 1068 | 25788.47 | 9.93E-06 | | | | | | | -1.42E-05 | | | | | | |
| 1069 | 25793.67 | | | 1.67E-05 | | | | 1.57E-05 | | | 5.19E-06 | | | | 1.96E-05 |
| 1070 | 25795.43 | | | | | 1.60E-05 | | | | | | | 1.04E-05 | | |
| 1071 | 25841.14 | | | | 1.67E-05 | | | | | | | 2.17E-05 | | | |
| 1072 | 25912.7 | | 1.62E-05 | | | | | | | -8.89E-06 | | | | | |
| 1073 | 25930.6 | 1.01E-05 | | 1.67E-05 | | 1.56E-05 | | 1.57E-05 | -1.62E-05 | | 4.16E-06 | | 1.04E-05 | | 1.99E-05 |
| 1074 | 25956.69 | | | | 1.61E-05 | | | | | | | 2.17E-05 | | | |
| 1075 | 25986.1 | | 1.55E-05 | | | | | | | -9.49E-06 | | | | | |
| 1076 | 26005.33 | | | | | | | 1.54E-05 | | | | | | | 1.94E-05 |
| 1077 | 26010.42 | 9.70E-06 | | | | | | | -1.66E-05 | | | | | | |
| 1078 | 26015.64 | | | 1.63E-05 | | | | | | | 3.63E-06 | | | | |
| 1079 | 26017.77 | | | | | 1.50E-05 | | | | | | | 9.43E-06 | | |
| 1080 | 26059.97 | | | | 1.65E-05 | | | | | | | 2.18E-05 | | | |
| 1081 | 26071.48 | | | | | | 5.74E-06 | | | | | | | 5.77E-06 | |
| 1082 | 26129.75 | | 1.61E-05 | | | | | | | -7.47E-06 | | | | | |
| 1083 | 26147.51 | | | | | | | 1.58E-05 | | | | | | | 1.87E-05 |
| 1084 | 26148.02 | 1.01E-05 | | | | | | | -1.56E-05 | | | | | | |
| 1085 | 26149.25 | | | 1.67E-05 | | | | | | | 3.98E-06 | | | | |
| 1086 | 26177.25 | | | | | 1.49E-05 | | | | | | | 9.38E-06 | | |
| 1087 | 26203.23 | | | | 1.65E-05 | | | | | | | 2.03E-05 | | | |
| 1088 | 26238.37 | | 1.66E-05 | | | | | | | -7.58E-06 | | | | | |
| 1089 | 26250.84 | 1.02E-05 | | 1.66E-05 | | 1.47E-05 | 5.69E-06 | 1.59E-05 | -1.59E-05 | | 3.70E-06 | | 9.50E-06 | 5.68E-06 | 1.87E-05 |
| 1090 | 26282.16 | | | | 1.67E-05 | | | | | | | 2.04E-05 | | | |
| 1091 | 26339.87 | | 1.64E-05 | | | | | | | -7.66E-06 | | | | | |
| 1092 | 26367.79 | | | | | | | 1.53E-05 | | | | | | | 1.86E-05 |
| 1093 | 26371.73 | | | 1.61E-05 | | | | | | | 3.67E-06 | | | | |
| 1094 | 26373.81 | 9.24E-06 | | | | | | | -1.65E-05 | | | | | | |
| 1095 | 26391.36 | | | | | 1.35E-05 | | | | | | | 8.80E-06 | | |
| 1096 | 26421.95 | | | | 1.65E-05 | | | | | | | 2.01E-05 | | | |
| 1097 | 26483.42 | | 1.71E-05 | | | | | | | -7.26E-06 | | | | | |
| 1098 | 26487.41 | | | | | | 5.57E-06 | | | | | | | 5.59E-06 | |
| 1099 | 26499.91 | | | | | | | 1.57E-05 | | | | | | | 1.86E-05 |
| 1100 | 26507.44 | 9.45E-06 | | 1.65E-05 | | | | | -1.62E-05 | | 3.90E-06 | | | | |
| 1101 | 26528.09 | | | | | 1.27E-05 | | | | | | | 7.36E-06 | | |
| 1102 | 26539.08 | | | | 1.64E-05 | | | | | | | 1.95E-05 | | | |
| 1103 | 26583.35 | | 1.75E-05 | | | | | | | -7.10E-06 | | | | | |
| 1104 | 26595.68 | | | | | | | 1.66E-05 | | | | | | | 1.93E-05 |
| 1105 | 26604.19 | | | 1.72E-05 | | | | | | | 4.67E-06 | | | | |
| 1106 | 26605.87 | 1.05E-05 | | | | | | | -1.57E-05 | | | | | | |
| 1107 | 26625.54 | | | | | 1.30E-05 | | | | | | | 6.98E-06 | | |
| 1108 | 26632.82 | | | | 1.61E-05 | | | | | | | 1.93E-05 | | | |
| 1109 | 26680.23 | | 1.53E-05 | | | | | | | -9.22E-06 | | | | | |
| 1110 | 26691.96 | | | | | | | 1.45E-05 | | | | | | | 1.70E-05 |
| 1111 | 26696.91 | | | 1.41E-05 | | | | | | | 2.02E-06 | | | | |
| 1112 | 26699.65 | 8.26E-06 | | | | | | | -1.71E-05 | | | | | | |
| 1113 | 26702.47 | | | | | | 5.50E-06 | | | | | | | 5.55E-06 | |
| 1114 | 26729.25 | | | | | 1.11E-05 | | | | | | | 4.45E-06 | | |
| 1115 | 26733.39 | | | | 1.47E-05 | | | | | | | 1.76E-05 | | | |
| 1116 | 26811.74 | | 1.59E-05 | | | | | | | -7.73E-06 | | | | | |
| 1117 | 26832.32 | | | | | | | 1.59E-05 | | | | | | | 1.72E-05 |
| 1118 | 26834.25 | | | 1.50E-05 | | | | | | | 2.49E-06 | | | | |
| 1119 | 26834.61 | 9.56E-06 | | | | | | | -1.56E-05 | | | | | | |
| 1120 | 26866.63 | | | | 1.51E-05 | | | | | | | 1.71E-05 | | | |
| 1121 | 26872.01 | | | | | 1.12E-05 | | | | | | | 3.90E-06 | | |
| 1122 | 26937.46 | | 1.60E-05 | | | | | | | -8.09E-06 | | | | | |

| | A | B | C | D | E | F | G | H | I | J | K | L | M | N | O |
|---|---|---|---|---|---|---|---|---|---|---|---|---|---|---|---|
| 1123 | 26953.42 | | | | | | | 1.56E-05 | | | | | | | 1.69E-05 |
| 1124 | 26957 | | | 1.49E-05 | | | | | | | 2.31E-06 | | | | |
| 1125 | 26959.93 | 9.93E-06 | | | | | | | -1.56E-05 | | | | | | |
| 1126 | 26980.73 | | | | | | 6.01E-06 | | | | | | | 6.77E-06 | |
| 1127 | 26988.12 | | | | 1.49E-05 | | | | | | | 1.63E-05 | | | |
| 1128 | 27004.99 | | | | | 1.04E-05 | | | | | | | 3.83E-06 | | |
| 1129 | 27046.05 | | 1.53E-05 | | | | | | -1.01E-05 | | | | | | |
| 1130 | 27054.76 | | | | | | | 1.52E-05 | | | | | | | 1.65E-05 |
| 1131 | 27056.45 | | | 1.50E-05 | | | | | | | 2.19E-06 | | | | |
| 1132 | 27063.72 | 9.94E-06 | | | | | | | -1.59E-05 | | | | | | |
| 1133 | 27089.84 | | | | 1.52E-05 | | | | | | | 1.79E-05 | | | |
| 1134 | 27131.69 | | | | | 1.09E-05 | | | | | | | 3.31E-06 | | |
| 1135 | 27159.93 | | 1.56E-05 | | | | | | -1.01E-05 | | | | | | |
| 1136 | 27181.72 | | | | | | | 1.55E-05 | | | | | | | 1.60E-05 |
| 1137 | 27181.98 | | | 1.53E-05 | | | | | | | 1.56E-06 | | | | |
| 1138 | 27189.51 | 9.78E-06 | | | | | | | -1.64E-05 | | | | | | |
| 1139 | 27224.29 | | | | 1.55E-05 | | | | | | | 1.77E-05 | | | |
| 1140 | 27246.59 | | | | | | 5.92E-06 | | | | | | | 6.67E-06 | |
| 1141 | 27260.15 | | | | | 1.13E-05 | | | | | | | 3.50E-06 | | |
| 1142 | 27281.19 | | 1.56E-05 | | | | | | -1.10E-05 | | | | | | |
| 1143 | 27310.89 | | | | | | | 1.59E-05 | | | | | | | 1.60E-05 |
| 1144 | 27311.99 | | | 1.57E-05 | | | | | | | 1.17E-06 | | | | |
| 1145 | 27313.78 | 1.01E-05 | | | | | | | -1.71E-05 | | | | | | |
| 1146 | 27352.76 | | | | 1.51E-05 | | | | | | | 1.80E-05 | | | |
| 1147 | 27398.72 | | | | | 1.14E-05 | | | | | | | 2.91E-06 | | |
| 1148 | 27418.54 | | 1.49E-05 | | | | | | -1.23E-05 | | | | | | |
| 1149 | 27434.82 | | | 1.39E-05 | | | | 1.55E-05 | | | 2.20E-07 | | | | 1.47E-05 |
| 1150 | 27437.09 | 9.05E-06 | | | | | | | -1.82E-05 | | | | | | |
| 1151 | 27449.74 | | | | 1.55E-05 | | | | | | | 1.86E-05 | | | |
| 1152 | 27452.05 | | | | | | 5.98E-06 | | | | | | | 6.64E-06 | |
| 1153 | 27494.63 | | | | | 1.14E-05 | | | | | | | 2.97E-06 | | |
| 1154 | 27523.33 | | 1.52E-05 | | | | | | -1.24E-05 | | | | | | |
| 1155 | 27534.45 | | | | | | | 1.58E-05 | | | | | | | 1.50E-05 |
| 1156 | 27535.11 | | | 1.41E-05 | | | | | | | -1.98E-07 | | | | |
| 1157 | 27540.34 | 9.19E-06 | | | | | | | -1.82E-05 | | | | | | |
| 1158 | 27579.15 | | | | 1.66E-05 | | | | | | | 1.99E-05 | | | |
| 1159 | 27635.79 | | | | | 1.25E-05 | | | | | | | 4.10E-06 | | |
| 1160 | 27657.43 | | 1.62E-05 | | | | | | -1.09E-05 | | | | | | |
| 1161 | 27666.74 | | | 1.53E-05 | | | | 1.70E-05 | | | 6.26E-07 | | | | 1.61E-05 |
| 1162 | 27671.42 | 9.87E-06 | | | | | | | -1.78E-05 | | | | | | |
| 1163 | 27711.72 | | | | 1.66E-05 | | | | | | | 1.98E-05 | | | |
| 1164 | 27762.87 | | | | | 1.19E-05 | | | | | | | 4.01E-06 | | |
| 1165 | 27770.55 | | 1.51E-05 | | | | | | -1.32E-05 | | | | | | |
| 1166 | 27791.11 | | | | | | | 1.58E-05 | | | | | | | 1.49E-05 |
| 1167 | 27792.07 | | | 1.51E-05 | | | | | | | -1.38E-06 | | | | |
| 1168 | 27794.84 | 9.35E-06 | | | | | | | -1.91E-05 | | | | | | |
| 1169 | 27811.18 | | | | | | 6.09E-06 | | | | | | | 6.72E-06 | |
| 1170 | 27828.98 | | | | 1.51E-05 | | | | | | | 1.91E-05 | | | |
| 1171 | 27866.03 | | | | | 1.23E-05 | | | | | | | 3.82E-06 | | |
| 1172 | 27867.81 | | 1.54E-05 | | | | | | -1.39E-05 | | | | | | |
| 1173 | 27883.33 | | | 1.47E-05 | | | | 1.50E-05 | | | -2.56E-06 | | | | 1.40E-05 |
| 1174 | 27888.76 | 9.35E-06 | | | | | | | -2.03E-05 | | | | | | |
| 1175 | 27912.2 | | | | 1.45E-05 | | | | | | | 1.85E-05 | | | |
| 1176 | 27958.54 | | | | | 1.08E-05 | | | | | | | 2.65E-06 | | |
| 1177 | 27959.49 | | 1.38E-05 | | | | | | -1.56E-05 | | | | | | |
| 1178 | 27968.89 | | | | | | | 1.44E-05 | | | | | | | 1.33E-05 |
| 1179 | 27969.2 | | | 1.36E-05 | | | | | | | -3.24E-06 | | | | |
| 1180 | 27986.31 | 7.76E-06 | | | | | | | -2.13E-05 | | | | | | |
| 1181 | 28016.42 | | | | 1.30E-05 | | | | | | | 1.75E-05 | | | |
| 1182 | 28045.78 | | | | | | 6.17E-06 | | | | | | | 7.08E-06 | |
| 1183 | 28059.7 | | | | | 1.00E-05 | | | | | | | 1.74E-06 | | |
| 1184 | 28065.5 | | 1.29E-05 | | | | | | -1.68E-05 | | | | | | |
| 1185 | 28077.84 | | | | | | | 1.30E-05 | | | | | | | 1.24E-05 |
| 1186 | 28079.92 | | | 1.28E-05 | | | | | | | -3.92E-06 | | | | |
| 1187 | 28100.43 | 7.63E-06 | | | | | | | -2.20E-05 | | | | | | |
| 1188 | 28125.13 | | | | 1.21E-05 | | | | | | | 1.71E-05 | | | |
| 1189 | 28172.63 | | | | | 9.77E-06 | | | | | | | 1.39E-06 | | |
| 1190 | 28181.55 | | 1.27E-05 | | | | | | -1.76E-05 | | | | | | |
| 1191 | 28193 | | | 1.27E-05 | | | | 1.28E-05 | | | -4.35E-06 | | | | 1.24E-05 |
| 1192 | 28196.18 | 7.49E-06 | | | | | | | -2.25E-05 | | | | | | |
| 1193 | 28226.76 | | | | 1.18E-05 | | | | | | | 1.64E-05 | | | |
| 1194 | 28293.13 | | | | | 9.75E-06 | | | | | | | 9.93E-07 | | |
| 1195 | 28304.48 | | 1.27E-05 | | | | | | -1.81E-05 | | | | | | |
| 1196 | 28322.63 | | | 1.29E-05 | | | | 1.31E-05 | | | -4.10E-06 | | | | 1.26E-05 |
| 1197 | 28331.12 | 7.83E-06 | | | | | | | -2.26E-05 | | | | | | |
| 1198 | 28347.38 | | | | 1.21E-05 | | | | | | | 1.68E-05 | | | |
| 1199 | 28377.91 | | | | | | 6.25E-06 | | | | | | | 6.94E-06 | |
| 1200 | 28391.13 | | | | | 9.07E-06 | | | | | | | 2.78E-07 | | |
| 1201 | 28398.76 | | 1.32E-05 | 1.33E-05 | | | | 1.30E-05 | -1.70E-05 | -3.24E-06 | | | | | 1.28E-05 |
| 1202 | 28402.92 | 7.90E-06 | | | | | | | -2.20E-05 | | | | | | |
| 1203 | 28426.32 | | | | 1.21E-05 | | | | | | | 1.64E-05 | | | |
| 1204 | 28457.58 | | | | | 9.10E-06 | | | | | | | -3.22E-07 | | |
| 1205 | 28488.99 | | 1.30E-05 | | | | | | -1.76E-05 | | | | | | |
| 1206 | 28498.41 | | | | | | | 1.31E-05 | | | | | | | 1.22E-05 |
| 1207 | 28498.93 | 7.71E-06 | | 1.34E-05 | | | | | -2.26E-05 | | -3.82E-06 | | | | |
| 1208 | 28509.96 | | | | 1.21E-05 | | | | | | | 1.60E-05 | | | |
| 1209 | 28554.55 | | | | | 8.83E-06 | | | | | | | -2.70E-07 | | |
| 1210 | 28581.67 | | 1.29E-05 | | | | | | -1.75E-05 | | | | | | |
| 1211 | 28587.27 | | | | | | | 1.35E-05 | | | | | | | 1.26E-05 |
| 1212 | 28589.19 | | | 1.37E-05 | | | | | | | -3.55E-06 | | | | |
| 1213 | 28600.74 | 9.05E-06 | | | 1.28E-05 | 8.90E-06 | 6.51E-06 | | -2.14E-05 | | | 1.66E-05 | -3.11E-07 | 6.96E-06 | |
| 1214 | 28641.23 | | 1.11E-05 | | | | | | -1.92E-05 | | | | | | |
| 1215 | 28647.84 | | | | | | | 1.17E-05 | | | | | | | 1.08E-05 |
| 1216 | 28650.53 | | | 1.19E-05 | | | | | | | -5.00E-06 | | | | |
| 1217 | 28662.72 | 8.00E-06 | | | | | | | -2.30E-05 | | | | | | |
| 1218 | 28684.83 | | | | 1.20E-05 | | | | | | | 1.53E-05 | | | |
| 1219 | 28706.58 | | | | | 7.56E-06 | | | | | | | -2.48E-06 | | |
| 1220 | 28727.8 | | 8.11E-06 | | | | | | -2.19E-05 | | | | | | |
| 1221 | 28731.53 | | | | | | | 8.88E-06 | | | | | | | 8.31E-06 |
| 1222 | 28739.33 | | | 9.21E-06 | | | | | | | -7.70E-06 | | | | |
| 1223 | 28756.8 | 5.58E-06 | | | | | | | -2.52E-05 | | | | | | |
| 1224 | 28769.39 | | | | 8.71E-06 | | | | | | | 1.35E-05 | | | |

| | A | B | C | D | E | F | G | H | I | J | K | L | M | N | O |
|---|---|---|---|---|---|---|---|---|---|---|---|---|---|---|---|
| 1225 | 28809.22 | | | | | 7.80E-06 | | | | | | | -1.86E-06 | | |
| 1226 | 28845.14 | | | | | | 6.46E-06 | | | | | | | 7.09E-06 | |
| 1227 | 28886.42 | | 8.63E-06 | | | | | | | -2.08E-05 | | | | | |
| 1228 | 28887.38 | | | | | | | 9.13E-06 | | | | | | | 8.93E-06 |
| 1229 | 28890.59 | | | 9.78E-06 | | | | | | | -7.05E-06 | | | | |
| 1230 | 28911.31 | 5.63E-06 | | | | | | | -2.45E-05 | | | | | | |
| 1231 | 28925.41 | | | | 8.21E-06 | | | | | | | 1.41E-05 | | | |
| 1232 | 28958.88 | | | | | 7.34E-06 | | | | | | | -1.89E-06 | | |
| 1233 | 29013.82 | | | | | | | 9.63E-06 | | | | | | | 9.89E-06 |
| 1234 | 29014.45 | | 9.78E-06 | | | | | | | -1.97E-05 | | | | | |
| 1235 | 29021.29 | | | 1.04E-05 | | | | | | | -6.08E-06 | | | | |
| 1236 | 29045.24 | 6.07E-06 | | | | | | | -2.42E-05 | | | | | | |
| 1237 | 29054.75 | | | | 8.73E-06 | | | | | | | 1.46E-05 | | | |
| 1238 | 29061.52 | | | | | | 6.70E-06 | | | | | | | 7.46E-06 | |
| 1239 | 29079.23 | | | | | 7.98E-06 | | | | | | | -1.66E-06 | | |
| 1240 | 29105.99 | | | | | | | 9.23E-06 | | | | | | | 8.91E-06 |
| 1241 | 29111.85 | | 9.76E-06 | | | | | | | -2.14E-05 | | | | | |
| 1242 | 29113.58 | | | 9.61E-06 | | | | | | | -7.33E-06 | | | | |
| 1243 | 29137.06 | 7.27E-06 | | | | | | | -2.31E-05 | | | | | | |
| 1244 | 29138.95 | | | | 9.09E-06 | | | | | | | 1.51E-05 | | | |
| 1245 | 29163.36 | | | | | 9.11E-06 | | | | | | | 5.36E-07 | | |
| 1246 | 29212.16 | | | | | | | 1.00E-05 | | | | | | | 1.02E-05 |
| 1247 | 29217.85 | | 1.08E-05 | 1.03E-05 | | | | | | -2.03E-05 | -6.21E-06 | | | | |
| 1248 | 29243.87 | 9.16E-06 | | | 1.24E-05 | 1.36E-05 | | | -2.09E-05 | | | | 1.89E-05 | 8.08E-06 | |
| 1249 | 29286.74 | | | | | | 6.74E-06 | 9.77E-06 | | | | | | 7.47E-06 | 1.05E-05 |
| 1250 | 29289.66 | | 1.02E-05 | 9.23E-06 | | | | | | -1.95E-05 | -6.57E-06 | | | | |
| 1251 | 29309.88 | | | | 1.13E-05 | | | | | | | 1.75E-05 | | | |
| 1252 | 29309.98 | 9.56E-06 | | | | | | | -1.96E-05 | | | | | | |
| 1253 | 29325.63 | | | | | 1.28E-05 | | | | | | | 7.14E-06 | | |
| 1254 | 29366.02 | | | | | | | 7.29E-06 | | | | | | | 8.34E-06 |
| 1255 | 29373.49 | | | 7.62E-06 | | | | | | | -8.16E-06 | | | | |
| 1256 | 29373.74 | | 8.63E-06 | | | | | | | -2.05E-05 | | | | | |
| 1257 | 29379.4 | | | | 1.10E-05 | | | | | | | 1.75E-05 | | | |
| 1258 | 29384.79 | 9.60E-06 | | | | 1.31E-05 | | | -1.88E-05 | | | | 7.36E-06 | | |
| 1259 | 29438.84 | | | | | | 6.75E-06 | | | | | | | 7.30E-06 | |
| 1260 | 29454.06 | | | | | | | 6.63E-06 | | | | | | | 7.74E-06 |
| 1261 | 29464.71 | | | 8.03E-06 | | | | | | | -8.34E-06 | | | | |
| 1262 | 29467.27 | | 1.00E-05 | | | | | | | -1.93E-05 | | | | | |
| 1263 | 29483.25 | | | | 1.20E-05 | | | | | | | 1.78E-05 | | | |
| 1264 | 29493.05 | 1.21E-05 | | | | | | | -1.48E-05 | | | | | | |
| 1265 | 29496.99 | | | | | 1.44E-05 | | | | | | | 1.24E-05 | | |
| 1266 | 29544.53 | | | | | | | 6.54E-06 | | | | | | | 7.54E-06 |
| 1267 | 29554.36 | | | 8.77E-06 | | | | | | | -7.39E-06 | | | | |
| 1268 | 29555.61 | | 1.13E-05 | | | | | | | -1.73E-05 | | | | | |
| 1269 | 29568.23 | | | | 1.23E-05 | | | | | | | 1.89E-05 | | | |
| 1270 | 29590.45 | 1.22E-05 | | | | | | | -1.34E-05 | | | | | | |
| 1271 | 29603.49 | | | | | 1.48E-05 | | | | | | | 1.34E-05 | | |
| 1272 | 29672.23 | | | | | | | 5.68E-06 | | | | | | | 7.44E-06 |
| 1273 | 29694.9 | | | 8.67E-06 | | | | | | | -7.34E-06 | | | | |
| 1274 | 29696.72 | | 1.15E-05 | | | | | | | -1.66E-05 | | | | | |
| 1275 | 29702.65 | | | | 1.17E-05 | | | | | | | 1.92E-05 | | | |
| 1276 | 29726.12 | 1.38E-05 | | | | | | | -1.11E-05 | | | | | | |
| 1277 | 29736.68 | | | | | 1.57E-05 | 6.64E-06 | | | | | | 1.48E-05 | 7.31E-06 | |
| 1278 | 29763.83 | | | | | | | 6.22E-06 | | | | | | | 8.29E-06 |
| 1279 | 29766.46 | | | 8.88E-06 | | | | | | | -7.38E-06 | | | | |
| 1280 | 29770.72 | | 1.14E-05 | | | | | | | -1.74E-05 | | | | | |
| 1281 | 29773.2 | | | | 1.17E-05 | | | | | | | 1.89E-05 | | | |
| 1282 | 29784.75 | 1.37E-05 | | | | | | | -1.15E-05 | | | | | | |
| 1283 | 29794.31 | | | | | 1.56E-05 | | | | | | | 1.47E-05 | | |
| 1284 | 29859.52 | | | | | | | 6.01E-06 | | | | | | | 8.68E-06 |
| 1285 | 29881.96 | | | 9.02E-06 | | | | | | | -5.78E-06 | | | | |
| 1286 | 29891.75 | | 1.21E-05 | | | | | | | -1.39E-05 | | | | | |
| 1287 | 29894.7 | | | | 1.14E-05 | | | | | | | 2.04E-05 | | | |
| 1288 | 29908.48 | 1.47E-05 | | | | | | | -8.23E-06 | | | | | | |
| 1289 | 29923.19 | | | | | 1.67E-05 | | | | | | | 1.66E-05 | | |
| 1290 | 29973.39 | | | | | | | 7.33E-06 | | | | | | | 1.13E-05 |
| 1291 | 29991.16 | | | 9.71E-06 | | | | | | | -4.20E-06 | | | | |
| 1292 | 29999.04 | | 1.37E-05 | | | | | | | -1.20E-05 | | | | | |
| 1293 | 29999.48 | | | | 1.20E-05 | | | | | | | 2.13E-05 | | | |
| 1294 | 30026.67 | 1.50E-05 | | | | | | | -6.62E-06 | | | | | | |
| 1295 | 30035.76 | | | | | | | 6.85E-06 | | | | | | 7.63E-06 | |
| 1296 | 30037.46 | | | | | 1.63E-05 | | | | | | | 1.46E-05 | | |
| 1297 | 30069.48 | | | | | | | 7.07E-06 | | | | | | | 1.06E-05 |
| 1298 | 30102.4 | | | 8.89E-06 | | | | | | | -5.01E-06 | | | | |
| 1299 | 30109.28 | | | | 1.22E-05 | | | | | | | 2.14E-05 | | | |
| 1300 | 30110.16 | | 1.41E-05 | | | | | | | -1.14E-05 | | | | | |
| 1301 | 30128.4 | 1.52E-05 | | | | | | | -5.94E-06 | | | | | | |
| 1302 | 30147.48 | | | | | 1.61E-05 | | | | | | | 1.53E-05 | | |
| 1303 | 30155.16 | | | | | | | 6.47E-06 | | | | | | | 1.02E-05 |
| 1304 | 30177.19 | | | 8.09E-06 | | | | | | | -5.90E-06 | | | | |
| 1305 | 30187.36 | | | | 1.26E-05 | | | | | | | 2.07E-05 | | | |
| 1306 | 30190.52 | | 1.48E-05 | | | | | | | -1.07E-05 | | | | | |
| 1307 | 30211.96 | 1.66E-05 | | | | | | | -3.73E-06 | | | | | | |
| 1308 | 30219.59 | | | | | 1.66E-05 | | | | | | | 1.69E-05 | | |
| 1309 | 30260.36 | | | | | | | 7.49E-06 | | | | | | | 1.12E-05 |
| 1310 | 30281.85 | | | 8.51E-06 | | | | | | | -5.51E-06 | | | | |
| 1311 | 30287.19 | | | | 1.25E-05 | | | | | | | 2.08E-05 | | | |
| 1312 | 30297.86 | | 1.31E-05 | | | | | | | -1.10E-05 | | | | | |
| 1313 | 30324.87 | 1.58E-05 | | | | | | | -4.52E-06 | | | | | | |
| 1314 | 30330.73 | | | | | 1.71E-05 | | | | | | | 1.73E-05 | | |
| 1315 | 30349.42 | | | | | | | 7.61E-06 | | | | | | | 1.14E-05 |
| 1316 | 30374.77 | | | 8.76E-06 | 1.28E-05 | | | | | | -4.58E-06 | 2.13E-05 | | | |
| 1317 | 30390.81 | | 1.36E-05 | | | | | | | -1.02E-05 | | | | | |
| 1318 | 30413.06 | 1.58E-05 | | | | | | | -3.33E-06 | | | | | | |
| 1319 | 30463.68 | | | | | 1.66E-05 | | | | | | | 1.62E-05 | | |
| 1320 | 30484.74 | | | | | | | 7.31E-06 | | | | | | | 1.11E-05 |
| 1321 | 30502.3 | | | | | | 6.87E-06 | | | | | | | 7.73E-06 | |
| 1322 | 30505.77 | | | 7.62E-06 | | | | | | | -4.69E-06 | | | | |
| 1323 | 30507.49 | | | | 1.27E-05 | | | | | | | 2.08E-05 | | | |
| 1324 | 30514.42 | | 1.21E-05 | | | | | | | -1.01E-05 | | | | | |
| 1325 | 30539.16 | 1.57E-05 | | | | | | | -1.62E-06 | | | | | | |
| 1326 | 30556.05 | | | | | 1.68E-05 | | | | | | | 1.60E-05 | | |

| | A | B | C | D | E | F | G | H | I | J | K | L | M | N | O |
|---|---|---|---|---|---|---|---|---|---|---|---|---|---|---|---|
| 1327 | 30575.59 | | | | | | | 8.12E-06 | | | | | | | 1.13E-05 |
| 1328 | 30597.07 | | | 4.50E-06 | 1.03E-05 | | | | | | -6.83E-06 | 1.87E-05 | | | |
| 1329 | 30606.86 | | 7.13E-06 | | | | | | | -1.41E-05 | | | | | |
| 1330 | 30651.31 | 1.06E-05 | | | | | | | -6.20E-06 | | | | | | |
| 1331 | 30692.42 | | | | | 1.68E-05 | | | | | | | 1.54E-05 | | |
| 1332 | 30696.75 | | | | | | | 7.41E-06 | | | | | | | 1.15E-05 |
| 1333 | 30711.72 | | | 5.03E-06 | 1.08E-05 | | | | | | -6.02E-06 | 1.92E-05 | | | |
| 1334 | 30721.56 | | 7.44E-06 | | | | | | | -1.35E-05 | | | | | |
| 1335 | 30752.49 | 1.13E-05 | | | | | | | -5.18E-06 | | | | | | |
| 1336 | 30758.19 | | | | | | 6.73E-06 | | | | | | | 7.54E-06 | |
| 1337 | 30789.75 | | | | | | | 7.24E-06 | | | | | | | 1.22E-05 |
| 1338 | 30806.55 | | | | | 1.65E-05 | | | | | | | 1.57E-05 | | |
| 1339 | 30816.58 | | | | 1.04E-05 | | | | | | | 1.97E-05 | | | |
| 1340 | 30817.94 | | | 4.97E-06 | | | | | | | -5.68E-06 | | | | |
| 1341 | 30835.1 | | 7.72E-06 | | | | | | | -1.31E-05 | | | | | |
| 1342 | 30870.87 | 1.17E-05 | | | | | | | -4.96E-06 | | | | | | |
| 1343 | 30914.86 | | | | | | | 5.51E-06 | | | | | | | 1.05E-05 |
| 1344 | 30928.06 | | | | 8.59E-06 | | | | | | | | 1.63E-05 | | |
| 1345 | 30930.41 | | | | | 1.59E-05 | | | | | | | 1.28E-05 | | |
| 1346 | 30933.19 | | | 3.58E-06 | | | | | | | -6.90E-06 | | | | |
| 1347 | 30948.66 | | 7.14E-06 | | | | | | | -1.40E-05 | | | | | |
| 1348 | 30966.89 | 1.12E-05 | | | | | | | -5.91E-06 | | | | | | |
| 1349 | 30991.48 | | | | | | | 5.28E-06 | | | | | | | 1.02E-05 |
| 1350 | 31013.56 | | | | 8.88E-06 | | | | | | | | 1.60E-05 | | |
| 1351 | 31016.33 | | | | | 1.56E-05 | | | | | | | 1.31E-05 | | |
| 1352 | 31022.04 | | | 2.96E-06 | | | | | | | -7.42E-06 | | | | |
| 1353 | 31042.67 | | 6.26E-06 | | | | | | | -1.40E-05 | | | | | |
| 1354 | 31069.89 | 1.20E-05 | | | | | | | -4.73E-06 | | | | | | |
| 1355 | 31096.3 | | | | | | | 5.15E-06 | | | | | | | 1.07E-05 |
| 1356 | 31113.78 | | | | 9.00E-06 | | | | | | | | 1.62E-05 | | |
| 1357 | 31123.77 | | | 2.76E-06 | | | | | | | -7.75E-06 | | | | |
| 1358 | 31131.02 | | | | | | 6.70E-06 | | | | | | | 7.42E-06 | |
| 1359 | 31132.13 | | | | | 1.50E-05 | | | | | | | 1.33E-05 | | |
| 1360 | 31146.54 | | 7.02E-06 | | | | | | | -1.29E-05 | | | | | |
| 1361 | 31172.62 | 1.33E-05 | | | | | | | -3.81E-06 | | | | | | |
| 1362 | 31199.93 | | | | | | | 5.53E-06 | | | | | | | 1.11E-05 |
| 1363 | 31207.14 | | | | 9.12E-06 | | | | | | | | 1.63E-05 | | |
| 1364 | 31217.15 | | | 2.99E-06 | | | | | | | -7.33E-06 | | | | |
| 1365 | 31225.98 | | | | | 1.47E-05 | | | | | | | 1.23E-05 | | |
| 1366 | 31232.6 | | 6.97E-06 | | | | | | | -1.31E-05 | | | | | |
| 1367 | 31265.4 | 1.26E-05 | | | | | | | -4.37E-06 | | | | | | |
| 1368 | 31299.29 | | | | | | | 5.09E-06 | | | | | | | 1.01E-05 |
| 1369 | 31317.34 | | | | 9.14E-06 | | | | | | | | 1.61E-05 | | |
| 1370 | 31329.68 | | | 2.83E-06 | | | | | | | -7.22E-06 | | | | |
| 1371 | 31349.14 | | 7.26E-06 | | | | | | | -1.24E-05 | | | | | |
| 1372 | 31350.75 | | | | | 1.49E-05 | | | | | | | 1.28E-05 | | |
| 1373 | 31382.3 | 1.46E-05 | | | | | | | -1.98E-06 | | | | | | |
| 1374 | 31416.84 | | | | | | | 5.24E-06 | | | | | | | 1.06E-05 |
| 1375 | 31439.95 | | | | 9.59E-06 | | | | | | | | 1.64E-05 | | |
| 1376 | 31453.92 | | | 3.64E-06 | | | | | | | -6.53E-06 | | | | |
| 1377 | 31459.56 | | | | | | 6.73E-06 | | | | | | | 7.66E-06 | |
| 1378 | 31474.57 | | 8.04E-06 | | | | | | | -1.13E-05 | | | | | |
| 1379 | 31486.13 | | | | | 1.48E-05 | | | | | | | 1.24E-05 | | |
| 1380 | 31501.87 | 1.36E-05 | | | | | | | -2.09E-06 | | | | | | |
| 1381 | 31550.51 | | | | | | | 5.98E-06 | | | | | | | 1.11E-05 |
| 1382 | 31567.54 | | | | 1.13E-05 | | | | | | | 1.70E-05 | | | |
| 1383 | 31594.5 | | | 4.35E-06 | | | | | | | -5.27E-06 | | | | |
| 1384 | 31605.39 | | 8.58E-06 | | | | | | | -1.10E-05 | | | | | |
| 1385 | 31650.03 | 1.48E-05 | | | | | | | -5.78E-07 | | | | | | |
| 1386 | 31657.53 | | | | | 1.51E-05 | | | | | | | 1.20E-05 | | |
| 1387 | 31697.61 | | | | | | | 6.77E-06 | | | | | | | 1.02E-05 |
| 1388 | 31717.81 | | | | 1.13E-05 | | | | | | | | 1.58E-05 | | |
| 1389 | 31746.86 | | | 3.57E-06 | | | | | | | -5.33E-06 | | | | |
| 1390 | 31763.69 | | 8.51E-06 | | | | | | | -1.04E-05 | | | | | |
| 1391 | 31792.52 | 1.50E-05 | | | | | | | 1.28E-07 | | | | | | |
| 1392 | 31807.92 | | | | | 1.55E-05 | | | | | | | 1.17E-05 | | |
| 1393 | 31835.56 | | | | | | | 6.45E-06 | | | | | | | 1.04E-05 |
| 1394 | 31858.35 | | | | 1.12E-05 | | | | | | | | 1.70E-05 | | |
| 1395 | 31863.17 | | | | | | 6.80E-06 | | | | | | | 7.69E-06 | |
| 1396 | 31867.04 | | | 4.26E-06 | | | | | | | -5.11E-06 | | | | |
| 1397 | 31907.79 | | 9.12E-06 | | | | | | | -9.55E-06 | | | | | |
| 1398 | 31933 | 1.53E-05 | | | | | | | 5.12E-07 | | | | | | |
| 1399 | 31961.71 | | | | | 1.60E-05 | | | | | | | 1.16E-05 | | |
| 1400 | 31977.64 | | | | | | | 7.33E-06 | | | | | | | 1.07E-05 |
| 1401 | 31994.84 | | | | 1.17E-05 | | | | | | | | 1.69E-05 | | |
| 1402 | 32008.38 | | | 4.66E-06 | | | | | | | -4.84E-06 | | | | |
| 1403 | 32035.25 | | 1.01E-05 | | | | | | | -8.91E-06 | | | | | |
| 1404 | 32086.18 | | | | | | 6.85E-06 | | | | | | | 7.61E-06 | |
| 1405 | 32090.11 | 1.67E-05 | | | | | | | 1.50E-06 | | | | | | |
| 1406 | 32140.14 | | | | | | | 8.11E-06 | | | | | | | 1.07E-05 |
| 1407 | 32144.89 | | | | | 1.60E-05 | | | | | | | 1.08E-05 | | |
| 1408 | 32157.05 | | | | 1.24E-05 | | | | | | | 1.71E-05 | | | |
| 1409 | 32178.49 | | | 4.42E-06 | | | | | | | -4.89E-06 | | | | |
| 1410 | 32224.04 | | 1.06E-05 | | | | | | | -8.55E-06 | | | | | |
| 1411 | 32249.11 | 1.71E-05 | | | | | | | 1.74E-06 | | | | | | |
| 1412 | 32287.92 | | | | | | | 7.95E-06 | | | | | | | 1.06E-05 |
| 1413 | 32290.63 | | | | | 1.59E-05 | | | | | | | 1.06E-05 | | |
| 1414 | 32296.47 | | | | 1.26E-05 | | | | | | | 1.66E-05 | | | |
| 1415 | 32311.13 | | | 4.55E-06 | | | | | | | -4.20E-06 | | | | |
| 1416 | 32332.21 | | 1.08E-05 | | | | | | | -8.52E-06 | | | | | |
| 1417 | 32375.47 | 1.82E-05 | | | | | | | 2.21E-06 | | | | | | |
| 1418 | 32416.49 | | | | | | | 9.03E-06 | | | | | | | 1.15E-05 |
| 1419 | 32439.25 | | | | 1.34E-05 | | | | | | | 1.75E-05 | | | |
| 1420 | 32443.28 | | | | | 1.58E-05 | | | | | | | 1.11E-05 | | |
| 1421 | 32451.61 | | | 5.18E-06 | | | | | | | -2.32E-06 | | | | |
| 1422 | 32489.83 | | 1.07E-05 | | | | | | | -6.61E-06 | | | | | |
| 1423 | 32523.61 | | | | | | 6.89E-06 | | | | | | | 7.86E-06 | |
| 1424 | 32531.98 | 1.77E-05 | | | | | | | 3.75E-06 | | | | | | |
| 1425 | 32546.04 | | | | | | | 8.60E-06 | | | | | | | 1.09E-05 |
| 1426 | 32562.56 | | | | 1.25E-05 | | | | | | | 1.62E-05 | | | |
| 1427 | 32575.96 | | | | | 1.54E-05 | | | | | | | 1.08E-05 | | |
| 1428 | 32587.04 | | | 5.46E-06 | | | | | | | -1.61E-06 | | | | |

| | A | B | C | D | E | F | G | H | I | J | K | L | M | N | O |
|---|---|---|---|---|---|---|---|---|---|---|---|---|---|---|---|
| 1429 | 32625.47 | | 1.12E-05 | | | | | | | -5.37E-06 | | | | | |
| 1430 | 32666.73 | 1.78E-05 | | | | | | | 4.64E-06 | | | | | | |
| 1431 | 32690.27 | | | | | | | 8.25E-06 | | | | | | | 1.10E-05 |
| 1432 | 32712.14 | | | | 1.18E-05 | | | | | | | 1.51E-05 | | | |
| 1433 | 32729.09 | | | | | 1.56E-05 | | | | | | | 1.06E-05 | | |
| 1434 | 32730.8 | | | 5.92E-06 | | | | | | -1.64E-06 | | | | | |
| 1435 | 32776.56 | | 1.10E-05 | | | | | | | -5.60E-06 | | | | | |
| 1436 | 32809.72 | 1.74E-05 | | | | | | | 4.68E-06 | | | | | | |
| 1437 | 32819.58 | | | | | | 7.00E-06 | | | | | | | 8.05E-06 | |
| 1438 | 32835.19 | | | | | | | 8.00E-06 | | | | | | | 1.17E-05 |
| 1439 | 32839.48 | | | | 1.18E-05 | | | | | | | 1.53E-05 | | | |
| 1440 | 32868.26 | | | 6.58E-06 | | | | | | -1.69E-06 | | | | | |
| 1441 | 32889.01 | | | | | 1.67E-05 | | | | | | | 1.25E-05 | | |
| 1442 | 32902.7 | | 1.18E-05 | | | | | | | -4.47E-06 | | | | | |
| 1443 | 32923.8 | 1.92E-05 | | | | | | | 5.45E-06 | | | | | | |
| 1444 | 32943.06 | | | | | | | 7.87E-06 | | | | | | | 1.14E-05 |
| 1445 | 32963.85 | | | | 1.19E-05 | | | | | | | 1.58E-05 | | | |
| 1446 | 32976.39 | | | 7.02E-06 | | | | | | -1.13E-06 | | | | | |
| 1447 | 33000.64 | | | | | 1.71E-05 | | | | | | | 1.27E-05 | | |
| 1448 | 33025.49 | | 1.28E-05 | | | | | | | -1.97E-06 | | | | | |
| 1449 | 33076.31 | 1.97E-05 | | | | | | | 6.99E-06 | | | | | | |
| 1450 | 33096.74 | | | | | | | 7.71E-06 | | | | | | | 1.10E-05 |
| 1451 | 33126.29 | | | | 1.22E-05 | | | | | | | 1.52E-05 | | | |
| 1452 | 33150.01 | | | 7.41E-06 | | | | | | -5.40E-07 | | | | | |
| 1453 | 33181.65 | | | | | 1.65E-05 | 7.11E-06 | | | | | | 1.22E-05 | 8.03E-06 | |
| 1454 | 33183.22 | | 1.30E-05 | | | | | | | -1.61E-06 | | | | | |
| 1455 | 33207.18 | 2.07E-05 | | | | | | | 8.06E-06 | | | | | | |
| 1456 | 33236.34 | | | | | | | 7.71E-06 | | | | | | | 1.08E-05 |
| 1457 | 33265.99 | | | | 1.18E-05 | | | | | | | 1.34E-05 | | | |
| 1458 | 33292.01 | | | 7.72E-06 | | | | | | -1.31E-07 | | | | | |
| 1459 | 33324.95 | | | | | 1.64E-05 | | | | | | | 9.78E-06 | | |
| 1460 | 33334.01 | | 1.29E-05 | | | | | | | -1.36E-06 | | | | | |
| 1461 | 33388.03 | 2.02E-05 | | | | | | | 6.95E-06 | | | | | | |
| 1462 | 33421.06 | | | | | | | 6.98E-06 | | | | | | | 1.05E-05 |
| 1463 | 33445.09 | | | | 1.09E-05 | | | | | | | 1.23E-05 | | | |
| 1464 | 33465.46 | | | 8.49E-06 | | | | | | -1.10E-08 | | | | | |
| 1465 | 33483.03 | | | | | | 7.24E-06 | | | | | | | 7.88E-06 | |
| 1466 | 33493.76 | | | | | 1.63E-05 | | | | | | | 9.07E-06 | | |
| 1467 | 33516.14 | | 1.39E-05 | | | | | | | -4.15E-07 | | | | | |
| 1468 | 33564.7 | 2.00E-05 | | | | | | | 8.34E-06 | | | | | | |
| 1469 | 33597.53 | | | | | | | 6.32E-06 | | | | | | | 9.95E-06 |
| 1470 | 33622.36 | | | | 1.07E-05 | | | | | | | 1.00E-05 | | | |
| 1471 | 33644.56 | | | 7.71E-06 | | | | | | 4.47E-07 | | | | | |
| 1472 | 33701.78 | | | | | 1.56E-05 | | | | | | | 6.36E-06 | | |
| 1473 | 33705.62 | | 1.23E-05 | | | | | | | -1.82E-06 | | | | | |
| 1474 | 33755.22 | 1.93E-05 | | | | | | | 8.75E-06 | | | | | | |
| 1475 | 33793.63 | | | | | | | 5.67E-06 | | | | | | | 9.22E-06 |
| 1476 | 33841.23 | | | | 1.03E-05 | | | | | | | 1.06E-05 | | | |
| 1477 | 33859.09 | | | 7.52E-06 | | | | | | 8.21E-07 | | | | | |
| 1478 | 33909.7 | | 1.28E-05 | | | | | | | -3.26E-07 | | | | | |
| 1479 | 33926.07 | | | | | 1.58E-05 | | | | | | | 7.51E-06 | | |
| 1480 | 33954.54 | | | | | | | 7.39E-06 | | | | | | 8.09E-06 | |
| 1481 | 33962.87 | 1.90E-05 | | | | | | | 9.53E-06 | | | | | | |
| 1482 | 33990.7 | | | | | | | 5.52E-06 | | | | | | | 9.05E-06 |
| 1483 | 34028.4 | | | | 1.00E-05 | | | | | | | 1.10E-05 | | | |
| 1484 | 34043.27 | | | 7.05E-06 | | | | | | 4.34E-07 | | | | | |
| 1485 | 34077.64 | | 1.19E-05 | | | | | | | -1.61E-07 | | | | | |
| 1486 | 34091.29 | | | | | 1.53E-05 | | | | | | | 6.16E-06 | | |
| 1487 | 34113.34 | 1.82E-05 | | | | | | | 8.97E-06 | | | | | | |
| 1488 | 34134.75 | | | | | | | 5.06E-06 | | | | | | | 9.44E-06 |
| 1489 | 34184.92 | | | | 8.89E-06 | | | | | | | 9.15E-06 | | | |
| 1490 | 34206.83 | | | 6.40E-06 | | | | | | -2.95E-07 | | | | | |
| 1491 | 34282.96 | | 1.20E-05 | | | | | | | 1.93E-07 | | | | | |
| 1492 | 34330.38 | | | | | 1.51E-05 | | | | | | | 5.24E-06 | | |
| 1493 | 34335.77 | 1.80E-05 | | | | | | | 8.96E-06 | | | | | | |
| 1494 | 34377 | | | | | | 7.57E-06 | | | | | | | 8.20E-06 | |
| 1495 | 34382.75 | | | | | | | 4.18E-06 | | | | | | | 8.04E-06 |
| 1496 | 34427.57 | | | | 8.28E-06 | | | | | | | 8.42E-06 | | | |
| 1497 | 34439.12 | | | 5.89E-06 | | | | | | -7.96E-07 | | | | | |
| 1498 | 34518.91 | | 1.21E-05 | | | | | | | 2.12E-07 | | | | | |
| 1499 | 34562.95 | 1.85E-05 | | | | | | | 9.91E-06 | | | | | | |
| 1500 | 34571.99 | | | | | 1.55E-05 | | | | | | | 5.61E-06 | | |
| 1501 | 34612.74 | | | | | | | 4.31E-06 | | | | | | | 7.54E-06 |
| 1502 | 34687.53 | | | | 8.32E-06 | | | | | | | 8.08E-06 | | | |
| 1503 | 34702.57 | | | 5.34E-06 | | | | | | -6.98E-07 | | | | | |
| 1504 | 34762.53 | | 1.15E-05 | | | | | | | -1.52E-07 | | | | | |
| 1505 | 34825.08 | 1.77E-05 | | | | | | | 9.02E-06 | | | | | | |
| 1506 | 34845.73 | | | | | 1.26E-05 | | | | | | | -3.37E-07 | | |
| 1507 | 34861.54 | | | | | | | 3.95E-06 | | | | | | | 6.69E-06 |
| 1508 | 34909.46 | | | | 8.14E-06 | | | | | | | 7.37E-06 | | | |
| 1509 | 34917.66 | | | | | | 7.18E-06 | | | | | | | 7.59E-06 | |
| 1510 | 34918.76 | | | 5.72E-06 | | | | | | 2.89E-07 | | | | | |
| 1511 | 34982.02 | | 1.24E-05 | | | | | | | 1.23E-06 | | | | | |
| 1512 | 35059.98 | 1.88E-05 | | | | | | | 1.13E-05 | | | | | | |
| 1513 | 35120.3 | | | | | 1.22E-05 | | | | | | | 3.53E-07 | | |
| 1514 | 35128.49 | | | | | | | 4.15E-06 | | | | | | | 6.70E-06 |
| 1515 | 35187.55 | | | | 7.96E-06 | | | | | | | 6.52E-06 | | | |
| 1516 | 35188.84 | | | 4.52E-06 | | | | | | -1.21E-07 | | | | | |
| 1517 | 35256.57 | | 1.25E-05 | | | | | | | 1.48E-06 | | | | | |
| 1518 | 35330.28 | 1.85E-05 | | | | | | | 9.93E-06 | | | | | | |
| 1519 | 35393.9 | | | | | | | 5.31E-06 | | | | | | | 8.14E-06 |
| 1520 | 35416.76 | | | | | 1.13E-05 | | | | | | | -4.71E-06 | | |
| 1521 | 35459.66 | | | 4.56E-06 | 8.24E-06 | | | | | | 4.84E-07 | 6.35E-06 | | | |
| 1522 | 35532.5 | | 1.23E-05 | | | | | | | 1.32E-06 | | | | | |
| 1523 | 35614.2 | 1.73E-05 | | | | | | | 9.67E-06 | | | | | | |
| 1524 | 35643.43 | | | | | | 7.13E-06 | | | | | | | 7.49E-06 | |
| 1525 | 35681.49 | | | | | | | 4.58E-06 | | | | | | | 8.55E-06 |
| 1526 | 35710.92 | | | | | 1.07E-05 | | | | | | | -7.20E-06 | | |
| 1527 | 35737.92 | | | 3.90E-06 | | | | | | -2.79E-06 | | | | | |
| 1528 | 35738.51 | | | | 8.35E-06 | | | | | | | 5.59E-06 | | | |
| 1529 | 35808.6 | | 1.11E-05 | | | | | | | 5.39E-07 | | | | | |
| 1530 | 35865.44 | 1.67E-05 | | | | | | | 8.57E-06 | | | | | | |

| | A | B | C | D | E | F | G | H | I | J | K | L | M | N | O |
|---|---|---|---|---|---|---|---|---|---|---|---|---|---|---|---|
| 1531 | 35951.41 | | | | | | | 4.39E-06 | | | | | | | 6.51E-06 |
| 1532 | 35977.88 | | | | | 1.08E-05 | | | | | | | -9.89E-06 | | |
| 1533 | 36021.08 | | | 3.11E-06 | | | | | | | -4.64E-06 | | | | |
| 1534 | 36027.57 | | | | 8.85E-06 | | | | | | | 3.62E-06 | | | |
| 1535 | 36073.86 | | 9.50E-06 | | | | | | | -1.59E-06 | | | | | |
| 1536 | 36142.97 | 1.58E-05 | | | | | | | 6.57E-06 | | | | | | |
| 1537 | 36228.14 | | | | | | | 3.47E-06 | | | | | | | 4.45E-06 |
| 1538 | 36242.13 | | | | | 9.88E-06 | | | | | | | -1.14E-05 | | |
| 1539 | 36243.34 | | | | | | 7.29E-06 | | | | | | | 7.58E-06 | |
| 1540 | 36291.16 | | | 1.83E-06 | | | | | | | -6.07E-06 | | | | |
| 1541 | 36302.21 | | | | 7.50E-06 | | | | | | | 1.60E-06 | | | |
| 1542 | 36346.19 | | 8.51E-06 | | | | | | | -1.89E-06 | | | | | |
| 1543 | 36408.74 | 1.53E-05 | | | | | | | 7.11E-06 | | | | | | |
| 1544 | 36476.72 | | | | | | | 3.28E-06 | | | | | | | 3.32E-06 |
| 1545 | 36487.44 | | | | | 1.04E-05 | | | | | | | -1.14E-05 | | |
| 1546 | 36541.19 | | | 1.46E-06 | | | | | | | -5.62E-06 | | | | |
| 1547 | 36572.59 | | | | 7.91E-06 | | | | | | | 4.52E-07 | | | |
| 1548 | 36600.74 | | 8.70E-06 | | | | | | | -1.05E-06 | | | | | |
| 1549 | 36657.63 | 1.52E-05 | | | | | | | 8.42E-06 | | | | | | |
| 1550 | 36657.78 | | | | | | 7.96E-06 | | | | | | | 8.31E-06 | |
| 1551 | 36733.11 | | | | | 8.97E-06 | | 3.69E-06 | | | | | -1.19E-05 | | -1.29E-06 |
| 1552 | 36758.36 | | | 1.35E-06 | | | | | | | -8.46E-06 | | | | |
| 1553 | 36815.81 | | | | 7.17E-06 | | | | | | | -2.08E-06 | | | |
| 1554 | 36834.05 | | 7.87E-06 | | | | | | | -1.74E-06 | | | | | |
| 1555 | 36898.7 | 1.57E-05 | | | | | | | 1.01E-05 | | | | | | |
| 1556 | 36938.01 | | | | | 8.64E-06 | | | | | | | -1.14E-05 | | |
| 1557 | 36939.3 | | | | | | | 4.27E-06 | | | | | | | 5.66E-06 |
| 1558 | 37012.16 | | | 1.51E-06 | | | | | | | -6.92E-06 | | | | |
| 1559 | 37074.93 | | | | 5.79E-06 | | | | | | | -1.94E-06 | | | |
| 1560 | 37086.06 | | 6.75E-06 | | | | | | | -7.31E-06 | | | | | |
| 1561 | 37137.31 | 1.49E-05 | | | | | | | 5.74E-06 | | | | | | |
| 1562 | 37160.22 | | | | | 8.18E-06 | | | | | | | -1.38E-05 | | |
| 1563 | 37175.15 | | | | | | | 2.08E-06 | | | | | | | -1.54E-05 |
| 1564 | 37191.91 | | | | | | 8.04E-06 | | | | | | | 8.54E-06 | |
| 1565 | 37212.88 | | | 7.73E-07 | | | | | | | -1.61E-05 | | | | |
| 1566 | 37261.08 | | | | 5.84E-06 | | | | | | | -2.53E-06 | | | |
| 1567 | 37264.77 | | 6.92E-06 | | | | | | | -6.74E-06 | | | | | |
| 1568 | 37324.26 | 1.47E-05 | | | | | | | 6.32E-06 | | | | | | |
| 1569 | 37418.21 | | | | | 8.73E-06 | | | | | | | -1.33E-05 | | |
| 1570 | 37431.13 | | | | | | | 1.87E-06 | | | | | | | -1.60E-05 |
| 1571 | 37475.99 | | | | | | 8.98E-06 | | | | | | | 1.03E-05 | |
| 1572 | 37482.98 | | | 2.27E-07 | | | | | | | -1.63E-05 | | | | |
| 1573 | 37542.19 | | 6.27E-06 | | | | | | | -7.38E-06 | | | | | |
| 1574 | 37548.36 | | | | 5.75E-06 | | | | | | | -2.66E-06 | | | |
| 1575 | 37584.35 | 1.41E-05 | | | | | | | 5.82E-06 | | | | | | |
| 1576 | 37633.18 | | | | | 7.72E-06 | | | | | | | -1.37E-05 | | |
| 1577 | 37658.55 | | | | | | | 1.30E-06 | | | | | | | -1.62E-05 |
| 1578 | 37688.88 | | | -8.96E-07 | | | | | | | -1.71E-05 | | | | |
| 1579 | 37748.09 | | 5.77E-06 | | | | | | | -7.48E-06 | | | | | |
| 1580 | 37757.69 | | | | 4.92E-06 | | | | | | | -2.00E-06 | | | |
| 1581 | 37810.81 | 1.35E-05 | | | | | | | 5.86E-06 | | | | | | |
| 1582 | 37870.67 | | | | | 7.18E-06 | 9.10E-06 | | | | | | -1.38E-05 | 9.99E-06 | |
| 1583 | 37898.13 | | | | | | | 9.32E-07 | | | | | | | -1.62E-05 |
| 1584 | 37931.63 | | | -1.05E-06 | | | | | | | -1.67E-05 | | | | |
| 1585 | 37996.73 | | 5.70E-06 | | | | | | | -6.22E-06 | | | | | |
| 1586 | 38020.49 | | | | 5.38E-06 | | | | | | | -8.66E-07 | | | |
| 1587 | 38047.37 | 1.43E-05 | | | | | | | 8.50E-06 | | | | | | |
| 1588 | 38170.05 | | | | | 7.64E-06 | | | | | | | -1.13E-05 | | |
| 1589 | 38203.48 | | | | | | | 1.91E-06 | | | | | | | -1.43E-05 |
| 1590 | 38234.32 | | | -8.95E-07 | | | | | | | -1.67E-05 | | | | |
| 1591 | 38272.51 | | 6.32E-06 | | | | | | | -5.39E-06 | | | | | |
| 1592 | 38287.77 | | | | 5.92E-06 | | | | | | | -3.53E-08 | | | |
| 1593 | 38314.71 | 1.45E-05 | | | | | | | 8.72E-06 | | | | | | |
| 1594 | 38363.52 | | | | | | 9.74E-06 | | | | | | | 1.06E-05 | |
| 1595 | 38406.21 | | | | | 7.97E-06 | | | | | | | -1.07E-05 | | |
| 1596 | 38423.68 | | | | | | | 1.91E-06 | | | | | | | -1.33E-05 |
| 1597 | 38471.01 | | | -6.67E-07 | | | | | | | -1.60E-05 | | | | |
| 1598 | 38505.48 | | 6.75E-06 | | | | | | | -3.78E-06 | | | | | |
| 1599 | 38535.74 | | | | 6.18E-06 | | | | | | | 1.19E-06 | | | |
| 1600 | 38563.78 | 1.50E-05 | | | | | | | 1.12E-05 | | | | | | |
| 1601 | 38656.03 | | | | | 8.90E-06 | | | | | | | -6.90E-06 | | |
| 1602 | 38666.58 | | | | | | | 3.32E-06 | | | | | | | -1.26E-05 |
| 1603 | 38687.44 | | | | | 1.02E-05 | | | | | | | 1.07E-05 | | |
| 1604 | 38695.41 | | | 2.31E-07 | | | | | | | -1.71E-05 | | | | |
| 1605 | 38731.05 | | 7.48E-06 | | | | | | | -3.56E-06 | | | | | |
| 1606 | 38756.66 | | | | 6.73E-06 | | | | | | | 1.66E-06 | | | |
| 1607 | 38760.63 | 1.57E-05 | | | | | | | 1.10E-05 | | | | | | |
| 1608 | 38853.2 | | | | | 8.70E-06 | | | | | | | -6.34E-06 | | |
| 1609 | 38858.75 | | | | | | | 2.81E-06 | | | | | | | -1.25E-05 |
| 1610 | 38902.12 | | | 4.35E-07 | | | | | | | -1.65E-05 | | | | |
| 1611 | 38959.96 | | 7.70E-06 | | | | | | | -2.24E-06 | | | | | |
| 1612 | 38990.06 | | | | 6.70E-06 | | | | | | | 2.63E-06 | | | |
| 1613 | 39014.99 | 1.62E-05 | | | | | | | 1.28E-05 | | | | | | |
| 1614 | 39113.81 | | | | | | | 1.11E-05 | | | | | | 1.13E-05 | |
| 1615 | 39140.4 | | | | | 8.66E-06 | | | | | | | -5.72E-06 | | |
| 1616 | 39153.6 | | | | | | | 2.93E-06 | | | | | | | -1.32E-05 |
| 1617 | 39186.91 | | | -6.29E-09 | | | | | | | -1.66E-05 | | | | |
| 1618 | 39226.68 | | 7.22E-06 | | | | | | | -1.71E-06 | | | | | |
| 1619 | 39282.55 | | | | 7.54E-06 | | | | | | | 2.35E-06 | | | |
| 1620 | 39283.8 | 1.60E-05 | | | | | | | 1.43E-05 | | | | | | |
| 1621 | 39392.35 | | | | | 8.33E-06 | | | | | | | -4.86E-06 | | |
| 1622 | 39411.13 | | | | | | | 3.28E-06 | | | | | | | -1.24E-05 |
| 1623 | 39450.31 | | | 5.42E-07 | | | | | | | -1.65E-05 | | | | |
| 1624 | 39484.53 | | 7.11E-06 | | | | | | | -1.37E-06 | | | | | |
| 1625 | 39547.16 | | | | 7.54E-06 | | | | | | | 2.09E-06 | | | |
| 1626 | 39550.21 | 1.56E-05 | | | | | | | 1.34E-05 | | | | | | |
| 1627 | 39678.24 | | | | | 8.14E-06 | | | | | | | -4.87E-06 | | |
| 1628 | 39738.04 | | | | | | | 3.67E-06 | | | | | | | -1.09E-05 |
| 1629 | 39784.21 | | | 1.69E-06 | | | | | | | -1.47E-05 | | | | |
| 1630 | 39816.48 | | 8.64E-06 | | | | | | | 7.80E-07 | | | | | |
| 1631 | 39855.89 | | | | 7.76E-06 | | | | | | | 5.14E-06 | | | |
| 1632 | 39866.81 | 1.68E-05 | | | | | | | 1.51E-05 | | | | | | |

| | A | B | C | D | E | F | G | H | I | J | K | L | M | N | O |
|---|---|---|---|---|---|---|---|---|---|---|---|---|---|---|---|
| 1633 | 39985.31 | | | | | 7.79E-06 | | | | | | | -8.74E-07 | | |
| 1634 | 40010.45 | | | | | | | 3.39E-06 | | | | | | | -1.03E-05 |
| 1635 | 40049.3 | | | 2.18E-06 | | | | | | | -1.37E-05 | | | | |
| 1636 | 40104.04 | | 9.19E-06 | | | | | | | 2.48E-06 | | | | | |
| 1637 | 40163.02 | | | | 8.95E-06 | | | | | | | 6.32E-06 | | | |
| 1638 | 40169.32 | | | | | | 1.15E-05 | | | | | | | 1.17E-05 | |
| 1639 | 40175.64 | 1.73E-05 | | | | | | | 1.60E-05 | | | | | | |
| 1640 | 40295.22 | | | | | 8.81E-06 | | | | | | | 4.12E-07 | | |
| 1641 | 40300.36 | | | | | | | 6.18E-06 | | | | | | | -9.17E-06 |
| 1642 | 40323.21 | | | 2.46E-06 | | | | | | | -1.36E-05 | | | | |
| 1643 | 40333.62 | | 9.23E-06 | | | | | | | 2.80E-06 | | | | | |
| 1644 | 40370.06 | | | | 1.03E-05 | | | | | | | 6.81E-06 | | | |
| 1645 | 40391.33 | 1.70E-05 | | | | | | | 1.78E-05 | | | | | | |
| 1646 | 40512.18 | | | | | 8.64E-06 | | | | | | | 6.41E-07 | | |
| 1647 | 40532.95 | | | | | | | 5.33E-06 | | | | | | | -8.79E-06 |
| 1648 | 40561.18 | | | 1.44E-06 | | | | | | | -1.38E-05 | | | | |
| 1649 | 40591.73 | | 9.29E-06 | | | | | | | 4.48E-06 | | | | | |
| 1650 | 40675.13 | | | | 9.38E-06 | | | | | | | 7.77E-06 | | | |
| 1651 | 40675.82 | 1.76E-05 | | | | | | | 1.86E-05 | | | | | | |
| 1652 | 40789.24 | | | | | 8.71E-06 | | | | | | | 1.20E-06 | | |
| 1653 | 40798.9 | | | | | | | 5.68E-06 | | | | | | | -7.56E-06 |
| 1654 | 40832.63 | | | 1.70E-06 | | | | | | | -1.27E-05 | | | | |
| 1655 | 40863.68 | | 8.87E-06 | | | | | | | 4.24E-06 | | | | | |
| 1656 | 40926.52 | | | | 9.19E-06 | | | | | | | 6.83E-06 | | | |
| 1657 | 40928.35 | 1.72E-05 | | | | | | | 1.85E-05 | | | | | | |
| 1658 | 41089.25 | | | | | 8.28E-06 | | | | | | | 1.07E-06 | | |
| 1659 | 41098.23 | | | | | | | 5.48E-06 | | | | | | | -8.33E-06 |
| 1660 | 41100.92 | | | | | | 1.17E-05 | | | | | | | 1.19E-05 | |
| 1661 | 41147.08 | | | 8.08E-07 | | | | | | | -1.36E-05 | | | | |
| 1662 | 41194.47 | | 7.12E-06 | | | | | | | 3.35E-06 | | | | | |
| 1663 | 41243.22 | | | | 9.45E-06 | | | | | | | 5.14E-06 | | | |
| 1664 | 41246.77 | 1.63E-05 | | | | | | | 1.81E-05 | | | | | | |
| 1665 | 41356.81 | | | | | 8.59E-06 | | | | | | | -2.37E-06 | | |
| 1666 | 41398.96 | | | | | | | 5.73E-06 | | | | | | | -1.08E-05 |
| 1667 | 41454.45 | | | 8.45E-07 | | | | | | | -1.41E-05 | | | | |
| 1668 | 41474.55 | | 6.44E-06 | | | | | | | 2.07E-06 | | | | | |
| 1669 | 41557.29 | 1.53E-05 | | | 1.01E-05 | | | | 1.78E-05 | | | 2.32E-06 | | | |
| 1670 | 41683.12 | | | | | 8.39E-06 | | | | | | | -5.77E-06 | | |
| 1671 | 41694.2 | | | | | | 1.13E-05 | | | | | | | 1.13E-05 | |
| 1672 | 41744.12 | | | | | | | 6.01E-06 | | | | | | | -1.26E-05 |
| 1673 | 41808.85 | | | 5.36E-07 | | | | | | | -1.55E-05 | | | | |
| 1674 | 41832.05 | | 6.83E-06 | | | | | | | 2.43E-06 | | | | | |
| 1675 | 41875.53 | 1.54E-05 | | | | | | | 1.71E-05 | | | | | | |
| 1676 | 41880.71 | | | | 9.86E-06 | | | | | | | 9.05E-07 | | | |
| 1677 | 41959.67 | | | | | 7.10E-06 | | | | | | | -1.11E-05 | | |
| 1678 | 42055.37 | | | | | | | 4.95E-06 | | | | | | | -1.51E-05 |
| 1679 | 42092.58 | | | -6.21E-08 | | | | | | | -1.67E-05 | | | | |
| 1680 | 42107.89 | | 6.28E-06 | | | | 1.14E-05 | | | 2.98E-06 | | | | 1.15E-05 | |
| 1681 | 42162.49 | 1.45E-05 | | | | | | | 1.73E-05 | | | | | | |
| 1682 | 42165.55 | | | | 9.38E-06 | | | | | | | 3.68E-07 | | | |
| 1683 | 42281.1 | | | | | 6.84E-06 | | | | | | | -9.63E-06 | | |
| 1684 | 42371.51 | | | | | | | 4.55E-06 | | | | | | | -1.59E-05 |
| 1685 | 42418.66 | | | 2.09E-07 | | | | | | | -1.70E-05 | | | | |
| 1686 | 42418.89 | | 7.39E-06 | | | | | | | 4.01E-06 | | | | | |
| 1687 | 42431.41 | | | | | | 1.15E-05 | | | | | | | 1.16E-05 | |
| 1688 | 42464.66 | 1.53E-05 | | | | | | | 1.83E-05 | | | | | | |
| 1689 | 42472.21 | | | | 1.03E-05 | | | | | | | 1.40E-06 | | | |
| 1690 | 42555.82 | | | | | 7.21E-06 | | | | | | | -1.03E-05 | | |
| 1691 | 42640.97 | | | | | | 1.21E-05 | -9.63E-07 | | | | | | 1.25E-05 | -2.21E-05 |
| 1692 | 42689.15 | | | -5.98E-06 | | | | | | | -2.35E-05 | | | | |
| 1693 | 42690.42 | | 1.04E-06 | | | | | | | -2.65E-06 | | | | | |
| 1694 | 42697.41 | | | | | | | 1.20E-05 | | | | | | 1.22E-05 | |
| 1695 | 42719.1 | 2.64E-06 | | | | | | | 4.63E-06 | | | | | | |
| 1696 | 42731.11 | | | | 2.13E-07 | | | | | | | -1.89E-05 | | | |
| 1697 | 42777.5 | | | | | -7.92E-06 | | | | | | | -3.66E-05 | | | |
| 1698 | 42837.68 | | | | | | 1.22E-05 | | | | | | | 1.25E-05 | |
| 1699 | 42901.53 | | | | | | | -1.76E-06 | | | | | | | -2.48E-05 |
| 1700 | 42936.54 | | | -4.52E-06 | | | | | | | -2.29E-05 | | | | |
| 1701 | 42943.71 | | 1.86E-06 | | | | | | | -1.42E-06 | | | | | |
| 1702 | 42989.67 | 2.22E-06 | | | | | | | 4.72E-06 | | | | | | |
| 1703 | 43015.21 | | | | 7.05E-07 | | | | | | | -1.80E-05 | | | |
| 1704 | 43090.48 | | | | | -8.49E-06 | | | | | | | -3.85E-05 | | | |
| 1705 | 43092.44 | | | | | | 1.20E-05 | | | | | | | 1.21E-05 | |
| 1706 | 43118.92 | | | | | | 1.20E-05 | | | | | | | 1.19E-05 | |
| 1707 | 43138.19 | | | | | | | -2.78E-06 | | | | | | | -2.50E-05 |
| 1708 | 43204.81 | | 1.64E-06 | -5.84E-06 | | | | | | -2.59E-06 | -2.38E-05 | | | | |
| 1709 | 43252.82 | | | | | | 1.15E-05 | | | | | | | 1.15E-05 | |
| 1710 | 43256.76 | 2.28E-06 | | | | | | | 4.85E-06 | | | | | | |
| 1711 | 43289.48 | | | | 3.35E-06 | | | | | | | -1.42E-05 | | | |
| 1712 | 43390.26 | | | | | -6.95E-06 | | | | | | | -3.59E-05 | | | |
| 1713 | 43449.54 | | | | | | | -3.97E-07 | | | | | | | -2.30E-05 |
| 1714 | 43518.08 | | 4.84E-06 | | | | 1.17E-05 | | | 2.58E-06 | | | | 1.17E-05 | |
| 1715 | 43520.18 | | | -2.84E-06 | | | | | | | -2.04E-05 | | | | |
| 1716 | 43547.3 | 5.61E-06 | | | | | | | 1.05E-05 | | | | | | |
| 1717 | 43560.47 | | | | 8.45E-06 | | | | | | | -9.61E-06 | | | |
| 1718 | 43708.37 | | | | | -6.45E-06 | | | | | | | -3.44E-05 | | | |
| 1719 | 43768.55 | | | | | | | 1.55E-06 | | | | | | | -2.10E-05 |
| 1720 | 43797.07 | | | | | | 1.12E-05 | | | | | | | 1.08E-05 | |
| 1721 | 43797.83 | | 6.64E-06 | | | | | | | 5.57E-06 | | | | | |
| 1722 | 43800.22 | | | -1.17E-06 | | | | | | | -1.66E-05 | | | | |
| 1723 | 43879.6 | 9.95E-06 | | | | | | | 1.68E-05 | | | | | | |
| 1724 | 43923.71 | | | | 1.13E-05 | | | | | | | -6.73E-06 | | | |
| 1725 | 44079.96 | | | | | -4.76E-06 | | | | | | | -2.81E-05 | | | |
| 1726 | 44093.45 | | | | | | | 5.97E-06 | | | | | | | -1.55E-05 |
| 1727 | 44132.63 | | | | | | 1.07E-05 | | | | | | | 1.04E-05 | |
| 1728 | 44138.14 | | 9.37E-06 | | | | | | | 8.34E-06 | | | | | |
| 1729 | 44156.02 | | | 1.42E-06 | | | | | | | -1.45E-05 | | | | |
| 1730 | 44185.65 | 1.04E-05 | | | | | | | 2.00E-05 | | | | | | |
| 1731 | 44253.87 | | | | 1.21E-05 | | | | | | | -4.62E-06 | | | |
| 1732 | 44455.08 | | | | | | | 6.69E-06 | | | | | | | -1.49E-05 |
| 1733 | 44490.6 | | | | | -4.52E-06 | | | | | | | -2.61E-05 | | | |
| 1734 | 44515.39 | | 1.05E-05 | | | | | | | 1.11E-05 | | | | | |

| | A | B | C | D | E | F | G | H | I | J | K | L | M | N | O |
|---|---|---|---|---|---|---|---|---|---|---|---|---|---|---|---|
| 1735 | 44536.66 | | | 2.52E-06 | | | | | | | -1.30E-05 | | | | |
| 1736 | 44566.73 | 1.19E-05 | | | | | | | | 1.99E-05 | | | | | |
| 1737 | 44631.97 | | | | 1.21E-05 | | | | | | | -3.45E-06 | | | |
| 1738 | 44833.4 | | | | | | | 8.02E-06 | | | | | | | -1.30E-05 |
| 1739 | 44838.46 | | | | | | 1.14E-05 | | | | | | | 1.15E-05 | |
| 1740 | 44863.41 | | 1.20E-05 | | | | | | | | 1.33E-05 | | | | |
| 1741 | 44883.69 | | | | | -2.88E-06 | | | | | | | -2.15E-05 | | |
| 1742 | 44909.56 | | | 4.02E-06 | | | | | | | -9.42E-06 | | | | |
| 1743 | 44918.61 | 1.40E-05 | | | | | | | 2.45E-05 | | | | | | |
| 1744 | 45011.61 | | | | 1.46E-05 | | | | | | | 7.37E-08 | | | |
| 1745 | 45236.1 | | | | | | | | 1.09E-05 | | | | | | -1.16E-05 |
| 1746 | 45236.94 | | 1.31E-05 | | | | | | | | 1.59E-05 | | | | |
| 1747 | 45313.52 | 1.48E-05 | | | | | | | 2.48E-05 | | | | | | |
| 1748 | 45328.64 | | | 5.00E-06 | | | | | | | -6.32E-06 | | | | |
| 1749 | 45338.33 | | | | | -2.77E-06 | | | | | | | -2.25E-05 | | |
| 1750 | 45457.15 | | | | 1.65E-05 | | | | | | | -9.62E-08 | | | |
| 1751 | 45634.07 | | 1.37E-05 | | | | | | | | 1.59E-05 | | | | |
| 1752 | 45647.43 | | | | | | | | 1.06E-05 | | | | | | -1.28E-05 |
| 1753 | 45694.11 | 1.54E-05 | | | | | | | 2.44E-05 | | | | | | |
| 1754 | 45714.58 | | | | | -3.32E-06 | | | | | | | -2.27E-05 | | |
| 1755 | 45730 | | | 5.62E-06 | | | | | | | -5.94E-06 | | | | |
| 1756 | 45748.27 | | | | | | 1.10E-05 | | | | | | | 1.16E-05 | |
| 1757 | 45889.17 | | | | 1.65E-05 | | | | | | | 2.81E-06 | | | |
| 1758 | 46059.66 | | 1.40E-05 | | | | | | | | 1.71E-05 | | | | |
| 1759 | 46090.27 | | | | | | | | 1.16E-05 | | | | | | -1.11E-05 |
| 1760 | 46148.09 | 1.60E-05 | | | | | | | | 2.66E-05 | | | | | |
| 1761 | 46180.13 | | | 6.04E-06 | | | | | | | -5.04E-06 | | | | |
| 1762 | 46196.14 | | | | | -3.21E-06 | | | | | | | -2.16E-05 | | |
| 1763 | 46371.12 | | | | 1.77E-05 | | | | | | | 6.40E-06 | | | |
| 1764 | 46618.1 | | 1.53E-05 | | | | | | | | 2.00E-05 | | | | |
| 1765 | 46656.43 | | | | | | | | 1.42E-05 | | | | | | -7.73E-06 |
| 1766 | 46690.18 | 1.78E-05 | | | | | | | | 3.11E-05 | | | | | |
| 1767 | 46736.91 | | | 8.07E-06 | | | | | | | -7.82E-07 | | | | |
| 1768 | 46801.01 | | | | | -2.40E-06 | | | | | | | -1.96E-05 | | |
| 1769 | 46811.62 | | | | | | 1.04E-05 | | | | | | | 1.12E-05 | |
| 1770 | 46886.33 | | | | 1.91E-05 | | | | | | | 7.58E-06 | | | |
| 1771 | 47167.58 | | 1.84E-05 | | | | | | | | 2.38E-05 | | | | |
| 1772 | 47281.29 | | | | | | | | 1.78E-05 | | | | | | -3.91E-06 |
| 1773 | 47295.65 | 2.01E-05 | | | | | | | | 3.71E-05 | | | | | |
| 1774 | 47396.16 | | | 1.01E-05 | | | | | | | 2.16E-06 | | | | |
| 1775 | 47484.89 | | | | | -1.19E-06 | | | | | | | -1.43E-05 | | |
| 1776 | 47555.79 | | | | 2.01E-05 | | | | | | | 1.05E-05 | | | |
| 1777 | 47822.18 | | 1.87E-05 | | | | | | | | 2.74E-05 | | | | |
| 1778 | 47916.08 | 2.06E-05 | | | | | | | 3.99E-05 | | | | | | |
| 1779 | 47925.44 | | | | | | | | 1.73E-05 | | | | | | -2.65E-06 |
| 1780 | 48068.66 | | | 1.00E-05 | | | | | | | 3.77E-06 | | | | |
| 1781 | 48311.21 | | | | | 1.86E-05 | -3.14E-06 | | | | | | 1.15E-05 | -1.39E-05 | | |
| 1782 | 48598.36 | | 2.00E-05 | | | | | | | | 3.14E-05 | | | | |
| 1783 | 48713.7 | 2.18E-05 | | | | | | | | 4.20E-05 | | | | | |
| 1784 | 48744.95 | | | | | | | | 1.67E-05 | | | | | | -1.55E-06 |
| 1785 | 48867.65 | | | | | | 9.77E-06 | | | | | | | 9.88E-06 | |
| 1786 | 48883.37 | | | 1.06E-05 | | | | | | | 6.68E-06 | | | | |
| 1787 | 49144.72 | | | | | 1.75E-05 | -2.76E-06 | | | | | | 1.39E-05 | -1.20E-05 | | |
| 1788 | 49417.99 | | 2.15E-05 | | | | | | | | 2.34E-05 | | | | |
| 1789 | 49536.23 | 2.24E-05 | | | | | | | 3.32E-05 | | | | | | |
| 1790 | 49599.79 | | | | | | | | 1.82E-05 | | | | | | 1.60E-06 |
| 1791 | 49630.34 | | | 1.06E-05 | | | | | | | 7.10E-06 | | | | |
| 1792 | 49920.86 | | | | | -3.20E-06 | | | | | | | -1.09E-05 | | |
| 1793 | 49958.4 | | | | 1.75E-05 | | | | | | | 1.38E-05 | | | |
| 1794 | 50178.08 | | 2.17E-05 | | | | | | | | 2.21E-05 | | | | |
| 1795 | 50318.06 | 2.36E-05 | | | | | | | 4.49E-05 | | | | | | |
| 1796 | 50390.3 | | | | | | | | 1.76E-05 | | | | | | 3.94E-06 |
| 1797 | 50573.19 | | | 1.20E-05 | | | | | | | 1.40E-05 | | | | |
| 1798 | 50888.57 | | | | | -3.46E-06 | | | | | | | -9.05E-06 | | |
| 1799 | 50889.25 | | | | 1.72E-05 | | | | | | | 1.48E-05 | | | |
| 1800 | 51031.97 | | 2.23E-05 | | | | | | | | 2.67E-05 | | | | |
| 1801 | 51178.39 | 2.29E-05 | | | | | | | 4.32E-05 | | | | | | |
| 1802 | 51316.2 | | | | | | | | 1.86E-05 | | | | | | 3.94E-06 |
| 1803 | 51332.96 | | | | | | 8.64E-06 | | | | | | | 8.32E-06 | |
| 1804 | 51414.64 | | | 9.04E-06 | | | | | | | 1.02E-05 | | | | |
| 1805 | 51855.33 | | | | 1.58E-05 | | | | | | | 1.24E-05 | | | |
| 1806 | 51956.35 | | | | | -6.49E-06 | | | | | | | -1.23E-05 | | |
| 1807 | 52045.65 | | 2.21E-05 | | | | | | | | 2.43E-05 | | | | |
| 1808 | 52146.94 | 2.32E-05 | | | | | | | 4.42E-05 | | | | | | |
| 1809 | 52430.91 | | | | | | | | 1.90E-05 | | | | | | 2.17E-06 |
| 1810 | 52532.25 | | | 9.16E-06 | | | | | | | 8.48E-06 | | | | |
| 1811 | 52595.48 | | | | | | 3.21E-06 | | | | | | | 9.83E-07 | |
| 1812 | 52931.05 | | | | 1.60E-05 | | | | | | | 1.27E-05 | | | |
| 1813 | 52983.56 | | 2.44E-05 | | | -6.29E-06 | | | | | 2.99E-05 | | -8.90E-06 | | |
| 1814 | 53093.22 | 2.56E-05 | | | | | | | 5.59E-05 | | | | | | |
| 1815 | 53406.23 | | | | | | | | 2.03E-05 | | | | | | 5.99E-06 |
| 1816 | 53478.21 | | | 1.34E-05 | | | | | | | 1.32E-05 | | | | |
| 1817 | 53601.08 | | | | | | 2.12E-06 | | | | | | | -1.04E-06 | |
| 1818 | 53773.49 | | 2.94E-05 | | | | | | | | 3.46E-05 | | | | |
| 1819 | 53799.57 | | | | 1.64E-05 | | | | | | | 1.32E-05 | | | |
| 1820 | 53907.85 | 2.99E-05 | | | | | | | 5.22E-05 | | | | | | |
| 1821 | 53936.2 | | | | | -6.53E-06 | | | | | | | -9.15E-06 | | |
| 1822 | 54278.97 | | | | | | | | 3.42E-05 | | | | | | 2.51E-05 |
| 1823 | 54352.08 | | | 1.67E-05 | | | | | | | 2.11E-05 | | | | |
| 1824 | 54577.8 | | 3.52E-05 | | | | | | | | 3.83E-05 | | | | |
| 1825 | 54608.65 | | | | 2.35E-05 | | | | | | | 2.25E-05 | | | |
| 1826 | 54624.86 | 3.49E-05 | | | | | | | 5.47E-05 | | | | | | |
| 1827 | 54706.74 | | | | | | 2.74E-06 | | | | | | | -6.03E-07 | |
| 1828 | 54757.76 | | | | | -6.04E-06 | | | | | | | -3.99E-06 | | |
| 1829 | 54840.06 | | | | | | | | 3.38E-05 | | | | | | 2.76E-05 |
| 1830 | 54888.18 | | | | 1.76E-05 | | | | | | | 2.30E-05 | | | |
| 1831 | 55057.07 | | | 3.78E-05 | | | | | | | 4.06E-05 | | | | |
| 1832 | 55161.65 | 3.76E-05 | | | | | | | 5.39E-05 | | | | | | |
| 1833 | 55167.25 | | | | 2.57E-05 | | | | | | | 2.47E-05 | | | |
| 1834 | 55375.29 | | | | | -2.17E-06 | | | | | | | -4.53E-07 | | |
| 1835 | 55394.28 | | | | | | | | 3.55E-05 | | | | | | 3.10E-05 |
| 1836 | 55402.95 | | | 1.97E-05 | | | | | | | 2.42E-05 | | | | |

| | A | B | C | D | E | F | G | H | I | J | K | L | M | N | O |
|---|---|---|---|---|---|---|---|---|---|---|---|---|---|---|---|
| 1837 | 55588.59 | | 4.04E-05 | | | | | | | 4.40E-05 | | | | | |
| 1838 | 55631.75 | 4.13E-05 | | | | | | | 6.01E-05 | | | | | | |
| 1839 | 55661.9 | | | | 2.97E-05 | | | | | | | 2.69E-05 | | | |
| 1840 | 55881.45 | | | | | | | 3.64E-05 | | | | | | | 3.24E-05 |
| 1841 | 55882.47 | | | 2.15E-05 | | | | | | | | 2.21E-05 | | | |
| 1842 | 56019.54 | | | | | -1.60E-06 | | | | | | | 1.47E-06 | | |
| 1843 | 56102.09 | | 4.49E-05 | | | | | | | 4.54E-05 | | | | | |
| 1844 | 56190.16 | 4.52E-05 | | | | | | | 6.04E-05 | | | | | | |
| 1845 | 56227.3 | | | | 3.06E-05 | | | | | | | 2.98E-05 | | | |
| 1846 | 56493.59 | | | | | | | 3.91E-05 | | | | | | | 3.63E-05 |
| 1847 | 56506.45 | | | 2.32E-05 | | | | | | | | 2.39E-05 | | | |
| 1848 | 56679.92 | | 4.71E-05 | | | | | | | 4.67E-05 | | | | | |
| 1849 | 56745.74 | | | | | -6.44E-07 | | | | | | | 3.11E-06 | | |
| 1850 | 56762.19 | 4.74E-05 | | | | | | | 6.11E-05 | | | | | | |
| 1851 | 56827.6 | | | | 3.12E-05 | | | | | | | 3.10E-05 | | | |
| 1852 | 56838.71 | | | | | | 7.82E-06 | | | | | | | 5.79E-06 | |
| 1853 | 57200.98 | | | | | | | 3.88E-05 | | | | | | | 3.65E-05 |
| 1854 | 57209.25 | | | 2.34E-05 | | | | | | | | 2.58E-05 | | | |
| 1855 | 57405.42 | | 4.74E-05 | | | | | | | 4.85E-05 | | | | | |
| 1856 | 57537.7 | 4.78E-05 | | | | | | | 6.14E-05 | | | | | | |
| 1857 | 57647.14 | | | | 3.16E-05 | | | | | | | 3.10E-05 | | | |
| 1858 | 57665.32 | | | | | -1.43E-06 | | | | | | | 2.70E-06 | | |
| 1859 | 58059.73 | | | | | | | 3.83E-05 | | | | | | | 3.62E-05 |
| 1860 | 58066.97 | | | 2.30E-05 | | | | | | | | 2.60E-05 | | | |
| 1861 | 58395.82 | | 4.83E-05 | | | | | | | 4.79E-05 | | | | | |
| 1862 | 58514.05 | 4.85E-05 | | | | | | | 5.93E-05 | | | | | | |
| 1863 | 58719.81 | | | | 2.98E-05 | | | | | | | 2.85E-05 | | | |
| 1864 | 58938.6 | | | | | -3.81E-06 | | | | | | | 3.02E-06 | | |
| 1865 | 59308.95 | | | | | | | 3.94E-05 | | | | | | | 3.87E-05 |
| 1866 | 59321.7 | | | 2.73E-05 | | | | | | | | 3.05E-05 | | | |
| 1867 | 59468.62 | | | | | | 9.58E-06 | | | | | | | 7.50E-06 | |
| 1868 | 59636.51 | | 5.25E-05 | | | | | | | 4.99E-05 | | | | | |
| 1869 | 59786.05 | 5.14E-05 | | | | | | | 6.03E-05 | | | | | | |
| 1870 | 59994.78 | | | | 2.69E-05 | | | | | | | 2.67E-05 | | | |
| 1871 | 60225.59 | | | | | -8.39E-06 | | | | | | | 2.93E-06 | | |
| 1872 | 60351.97 | | | 2.38E-05 | | | | 3.50E-05 | | | 2.37E-05 | | | | 3.46E-05 |
| 1873 | 60407.01 | | 5.22E-05 | | | | | | | 4.54E-05 | | | | | |
| 1874 | 60609.11 | 5.03E-05 | | | | | | | 5.52E-05 | | | | | | |
| 1875 | 60863.91 | | | | 2.66E-05 | | | | | | | 2.95E-05 | | | |
| 1876 | 61105.07 | | | | | -7.58E-06 | | | | | | | 5.70E-06 | | |
| 1877 | 61164.97 | | | 2.64E-05 | | | | 3.67E-05 | | | 2.76E-05 | | | | 4.04E-05 |
| 1878 | 61315.96 | | 5.67E-05 | | | | | | | 4.99E-05 | | | | | |
| 1879 | 61424.52 | 5.40E-05 | | | | | | | 5.83E-05 | | | | | | |
| 1880 | 61620.53 | | | | 2.67E-05 | | | | | | | 3.17E-05 | | | |
| 1881 | 61876.2 | | | | | -7.25E-06 | | | | | | | 6.78E-06 | | |
| 1882 | 61919.05 | | | | | | | 3.53E-05 | | | | | | | 4.21E-05 |
| 1883 | 61924.74 | | | 2.64E-05 | | | | | | | | 2.86E-05 | | | |
| 1884 | 62063.87 | | 5.66E-05 | | | | | | | 5.26E-05 | | | | | |
| 1885 | 62153.89 | 5.47E-05 | | | | | | | 6.02E-05 | | | | | | |
| 1886 | 62390.66 | | | | 2.51E-05 | | | | | | | 3.48E-05 | | | |
| 1887 | 62753.65 | | | 2.50E-05 | | | | | | | | 3.20E-05 | | | |
| 1888 | 62756.94 | | | | | | | 3.37E-05 | | | | | | | 4.47E-05 |
| 1889 | 62768.09 | | | | | -7.98E-06 | | | | | | | 5.45E-06 | | |
| 1890 | 62842.6 | | | | | | 1.31E-05 | | | | | | | 1.37E-05 | |
| 1891 | 62877.76 | | 5.55E-05 | | | | | | | 5.56E-05 | | | | | |
| 1892 | 63034.83 | 5.41E-05 | | | | | | | 6.17E-05 | | | | | | |
| 1893 | 63283.5 | | | | 2.46E-05 | | | | | | | 3.61E-05 | | | |
| 1894 | 63565.66 | | | | | | | 4.47E-05 | | | | | | | 5.69E-05 |
| 1895 | 63574.98 | | | | | -3.31E-06 | | | | | | | 1.47E-05 | | |
| 1896 | 63578.11 | | | 3.86E-05 | | | | | | | | 4.66E-05 | | | |
| 1897 | 63790.94 | | 6.88E-05 | | | | | | | 7.01E-05 | | | | | |
| 1898 | 63820.82 | 7.91E-05 | | | | | | | 8.26E-05 | | | | | | |
| 1899 | 64125.01 | | | | 3.32E-05 | | | | | | | 4.37E-05 | | | |
| 1900 | 64218.34 | | | | | | 1.32E-05 | | | | | | | 1.38E-05 | |
| 1901 | 64390.31 | | | | | -3.28E-06 | | 4.52E-05 | | | | | 1.61E-05 | | 6.02E-05 |
| 1902 | 64432.69 | | | 3.97E-05 | | | | | | | | 4.95E-05 | | | |
| 1903 | 64639.46 | | 7.25E-05 | | | | | | | 7.72E-05 | | | | | |
| 1904 | 64907.05 | 8.32E-05 | | | | | | | 9.03E-05 | | | | | | |
| 1905 | 65355.46 | | | | 3.43E-05 | | | | | | | 4.80E-05 | | | |
| 1906 | 65953.36 | | | | | | | 4.64E-05 | | | | | | | 6.28E-05 |
| 1907 | 66009.09 | | | 4.26E-05 | | | | | | | | 5.50E-05 | | | |
| 1908 | 66269.96 | | 7.54E-05 | | | -3.50E-06 | | | | 8.20E-05 | | | 1.78E-05 | | |
| 1909 | 66544.74 | 8.71E-05 | | | | | | | 9.40E-05 | | | | | | |
| 1910 | 67372.25 | | | | 3.14E-05 | | | | | | | 5.01E-05 | | | |
| 1911 | 68008.92 | | | | | | | 4.93E-05 | | | | | | | 7.54E-05 |
| 1912 | 68055.07 | | | 4.65E-05 | | | | | | | | 5.58E-05 | | | |
| 1913 | 68249.87 | | 8.40E-05 | | | | | | | 8.91E-05 | | | | | |
| 1914 | 68387.24 | 9.56E-05 | | | | | | | 0.000102 | | | | | | |
| 1915 | 68754.03 | | | | | -1.96E-06 | | | | | | | 1.95E-05 | | |
| 1916 | 68951.23 | | | | 3.12E-05 | | | | | | | 5.20E-05 | | | |
| 1917 | 69153.61 | | | | | | | 4.67E-05 | | | | | | | 6.79E-05 |
| 1918 | 69167.58 | | | 4.64E-05 | | | | | | | | 5.19E-05 | | | |
| 1919 | 69276.97 | | 8.31E-05 | | | | | | | 8.05E-05 | | | | | |
| 1920 | 69410.21 | 9.84E-05 | | | | | | | 9.48E-05 | | | | | | |
| 1921 | 69516.99 | | | | | | 1.38E-05 | | | | | | | 1.40E-05 | |
| 1922 | 69833.15 | | | | | 1.38E-06 | | | | | | | 2.05E-05 | | |
| 1923 | 70034.8 | | | | 3.15E-05 | | | | | | | 5.21E-05 | | | |
| 1924 | 70331.9 | | | | | | | 4.79E-05 | | | | | | | 7.06E-05 |
| 1925 | 70366.1 | | | 4.96E-05 | | | | | | | | 5.66E-05 | | | |
| 1926 | 70418.02 | | 9.02E-05 | | | | | | | 9.00E-05 | | | | | |
| 1927 | 70551.74 | 0.00011 | | | | | | | 0.000118 | | | | | | |
| 1928 | 71306.02 | | | | | 2.99E-06 | | | | | | | 1.69E-05 | | |
| 1929 | 71740.82 | | | | 2.98E-05 | | | | | | | 4.41E-05 | | | |
| 1930 | 72538.32 | | | | | | | 4.74E-05 | | | | | | | 6.60E-05 |
| 1931 | 72558.36 | | | 5.14E-05 | | | | | | | | 5.75E-05 | | | |
| 1932 | 72688.68 | | 9.16E-05 | | | | | | | 9.31E-05 | | | | | |
| 1933 | 72852.51 | 0.000117 | | | | | | | 0.000125 | | | | | | |
| 1934 | 75277.72 | | | | | 4.83E-06 | | | | | | | 1.15E-05 | | |
| 1935 | 75630.09 | | | | 3.05E-05 | | | | | | | 4.17E-05 | | | |
| 1936 | 77000.63 | | | | | | | 4.83E-05 | | | | | | | 6.59E-05 |
| 1937 | 77049.08 | | | 5.52E-05 | | | | | | | | 6.47E-05 | | | |
| 1938 | 77265.24 | | 9.43E-05 | | | | | | | 9.72E-05 | | | | | |

| | A | B | C | D | E | F | G | H | I | J | K | L | M | N | O |
|---|---|---|---|---|---|---|---|---|---|---|---|---|---|---|---|
| 1939 | 77606.26 | 0.000121 | | | | | | 0.000134 | | | | | | | |
| 1940 | 82162.22 | | | | | 6.90E-06 | | | | | | | 6.84E-06 | | |
| 1941 | 82843.51 | | | | 3.30E-05 | | | | | | | 3.94E-05 | | | |
| 1942 | 84970.7 | | | | | | | 6.59E-05 | | | | | | | 0.000156 |
| 1943 | 85065.31 | | | 6.02E-05 | | | | | | | 9.99E-05 | | | | |
| 1944 | 85233.66 | | 9.80E-05 | | | | | | | 0.000132 | | | | | |
| 1945 | 85326.16 | 0.000139 | | | | | | | 0.000179 | | | | | | |
| 1946 | 95605.61 | | | | | | 1.54E-05 | | | | | | | 1.54E-05 | |
| 1947 | 102287.5 | | 0.000113 | 8.66E-05 | 3.82E-05 | 1.27E-05 | | 0.000102 | | 0.00017 | 0.000137 | 3.44E-05 | 7.23E-06 | | 0.000189 |
| 1948 | 102745.3 | 0.00022 | | | | | | | 0.000308 | | | | | | |

|   | A | B | C | D | E | F | G | H | I | J | K | L | M | N | O |
|---|---|---|---|---|---|---|---|---|---|---|---|---|---|---|---|
| 1 | Rank | Inf | 0.002 | 0.0015 | 0.001 | 0.0005 | 0.0001 | 0.0013 | 0.0020 Ran | 0.0015 Ran | 0.0010 Ran | 0.0005 Ran | 0.0001 Ran | 0.0013 Random | |
| 2 | 0 | -2.75E-06 | -7.16E-06 | -1.53E-05 | -1.48E-05 | -1.84E-05 | -8.81E-06 | -1.56E-05 | 4.98E-05 | -7.54E-06 | -1.73E-05 | -1.63E-05 | -1.96E-05 | -7.42E-06 | -1.73E-05 |
| 3 | 144.9339 | | | | | | -7.27E-06 | | | | | | | -5.61E-06 | |
| 4 | 170.1765 | | | | | -1.75E-05 | | | | | | | -1.81E-05 | | |
| 5 | 187.6639 | | | -1.52E-05 | -1.43E-05 | | | -1.54E-05 | | | -1.72E-05 | -1.57E-05 | | | -1.69E-05 |
| 6 | 188.5724 | | -7.45E-06 | | | | | | | | | | | | |
| 7 | 195.6518 | -6.14E-06 | | | | | | | 4.27E-05 | | | | | | |
| 8 | 219.2789 | | | | | | | | | -2.08E-05 | | | | | |
| 9 | 295.1507 | | | | | | -6.60E-06 | | | | | | -1.94E-05 | -5.02E-06 | |
| 10 | 368.8204 | | | | | -1.85E-05 | | | | | | | -1.94E-05 | | |
| 11 | 370.1827 | | | -1.75E-05 | -1.67E-05 | | | -1.77E-05 | | | -2.03E-05 | -1.84E-05 | | | -1.99E-05 |
| 12 | 373.158 | | -1.12E-05 | | | | | | | | | | | | |
| 13 | 373.9716 | -1.10E-05 | | | | | | 2.10E-05 | | | | | | | |
| 14 | 377.8136 | | | | | | -6.19E-06 | | | | | | | -4.26E-06 | |
| 15 | 407.8617 | | | | | | | | | -2.47E-05 | | | | | |
| 16 | 438.6157 | | | | | | -5.40E-06 | | | | | | | -3.65E-06 | |
| 17 | 477.6567 | | | | | -1.92E-05 | | | | | | | -2.02E-05 | | |
| 18 | 478.5557 | | | | -1.84E-05 | | | | | | | -2.07E-05 | | | |
| 19 | 478.6781 | | | -1.95E-05 | | | | -1.96E-05 | | | -2.26E-05 | | | | -2.25E-05 |
| 20 | 482.499 | | -1.32E-05 | | | | | | | | | | | | |
| 21 | 482.7168 | -1.47E-05 | | | | | | | 1.16E-05 | | | | | | |
| 22 | 490.5911 | | | | | | | | | -2.60E-05 | | | | | |
| 23 | 502.6359 | | | | | | -4.98E-06 | | | | | | | -3.04E-06 | |
| 24 | 558.2789 | | | | | | -4.83E-06 | | | | | | | -2.84E-06 | |
| 25 | 581.5944 | | | | | | | | | -2.94E-05 | | | | | |
| 26 | 581.9667 | | | | | -1.99E-05 | | | | | | | -2.08E-05 | | |
| 27 | 583.9893 | | | -2.08E-05 | -1.93E-05 | | | -2.08E-05 | | | -2.40E-05 | -2.17E-05 | | | -2.39E-05 |
| 28 | 586.5086 | | -1.47E-05 | | | | | | | | | | | | |
| 29 | 587.3981 | -1.74E-05 | | | | | | | 3.79E-06 | | | | | | |
| 30 | 597.711 | | | | | | -4.69E-06 | | | | | | | -2.79E-06 | |
| 31 | 639.0041 | | | | | | -4.53E-06 | | | | | | | -2.74E-06 | |
| 32 | 652.1336 | | | | | | | | | | -3.19E-05 | | | | |
| 33 | 658.6859 | | | | -1.99E-05 | -2.00E-05 | | | | | | -2.27E-05 | -2.10E-05 | | | |
| 34 | 659.1005 | | | -2.14E-05 | | | | -2.16E-05 | | | -2.50E-05 | | | | -2.47E-05 |
| 35 | 664.0816 | -2.17E-05 | -1.65E-05 | | | | | | 1.97E-07 | | | | | | |
| 36 | 681.8055 | | | | | | -4.39E-06 | | | | | | | -2.69E-06 | |
| 37 | 728.2594 | | | | | | -4.28E-06 | | | | | | | -2.59E-06 | |
| 38 | 737.7113 | | | | | -1.99E-05 | | | | | | | -2.11E-05 | | |
| 39 | 744.7291 | | | | -2.05E-05 | | | -2.22E-05 | | | | -2.32E-05 | | | | -2.53E-05 |
| 40 | 744.7865 | | | -2.21E-05 | | | | | | | -2.57E-05 | | | | | |
| 41 | 751.2108 | | -1.75E-05 | | | | | | | | | | | | |
| 42 | 752.3151 | -2.17E-05 | | | | | | | -4.56E-06 | | | | | | |
| 43 | 770.9584 | | | | | | | | | | -3.28E-05 | | | | |
| 44 | 776.7122 | | | | | | -4.22E-06 | | | | | | | -2.56E-06 | |
| 45 | 819.4231 | | | | | -2.02E-05 | | | | | | | -2.13E-05 | | | |
| 46 | 819.4403 | | | | | | -4.16E-06 | | | | | | | -2.44E-06 | |
| 47 | 825.4923 | | | | -2.12E-05 | | | | | | | -2.39E-05 | | | | |
| 48 | 826.4174 | | | -2.27E-05 | | | | -2.28E-05 | | | -2.67E-05 | | | | -2.60E-05 |
| 49 | 832.7611 | | | | | | | | | | -2.62E-05 | | | | | |
| 50 | 834.2102 | | -1.80E-05 | | | | | | | | | | | | |
| 51 | 834.6203 | -2.42E-05 | | | | | | | -1.07E-05 | | | | | | |
| 52 | 858.7477 | | | | | | | -4.02E-06 | | | | | | | -2.31E-06 | |
| 53 | 890.8636 | | | | | -2.04E-05 | | | | | | | -2.14E-05 | | | |
| 54 | 895.3621 | | | | -2.14E-05 | | | | | | | -2.43E-05 | | | | |
| 55 | 895.6228 | | | | | | | | -2.31E-05 | | | | | | | -2.66E-05 |
| 56 | 895.7544 | | | -2.31E-05 | | | | | | | -2.73E-05 | | | | | |
| 57 | 902.3289 | | -1.82E-05 | | | | | | | | | | | | |
| 58 | 902.3985 | -2.57E-05 | | | | | | | -1.46E-05 | | | | | | |
| 59 | 904.7207 | | | | | | | -3.75E-06 | | | | | | | -2.07E-06 | |
| 60 | 916.8311 | | | | | | | | | | -2.24E-05 | | | | | |
| 61 | 948.4332 | | | | | | | -3.77E-06 | | | | | | | -2.19E-06 | |
| 62 | 959.6388 | | | | | -2.06E-05 | | | | | | | -2.14E-05 | | | |
| 63 | 963.3911 | | | | -2.17E-05 | | | | | | | -2.44E-05 | | | | |
| 64 | 964.0542 | | | -2.33E-05 | | | | -2.33E-05 | | | -2.74E-05 | | | | | -2.68E-05 |
| 65 | 971.302 | -2.64E-05 | | | | | | | -1.68E-05 | | | | | | |
| 66 | 972.2128 | | -1.89E-05 | | | | | | | | | | | | |
| 67 | 990.3324 | | | | | | | -3.83E-06 | | | | | | | -2.27E-06 | |
| 68 | 1019.581 | | | | | | | -3.64E-06 | | | | | | | -2.14E-06 | |
| 69 | 1024.796 | | | | | | | | | | -2.56E-05 | | | | | |
| 70 | 1027.458 | | | | | -2.04E-05 | | | | | | | -2.10E-05 | | | |
| 71 | 1033.16 | | | | -2.16E-05 | | | | | | | -2.42E-05 | | | | |
| 72 | 1034.717 | | | | | | | | -2.32E-05 | | | | | | | -2.67E-05 |
| 73 | 1035.186 | | | -2.32E-05 | | | | | | | -2.72E-05 | | | | | |
| 74 | 1040.942 | -2.61E-05 | | | | | | | -1.84E-05 | | | | | | |
| 75 | 1041.19 | | -1.90E-05 | | | | | | | | | | | | |
| 76 | 1064.842 | | | | | | | -3.35E-06 | | | | | | | -1.69E-06 | |
| 77 | 1081.141 | | | | | | | | | | -2.13E-05 | | | | | |
| 78 | 1107.281 | | | | | -2.03E-05 | | | | | | | -2.11E-05 | | | |
| 79 | 1113.154 | | | | -2.16E-05 | | -3.08E-06 | | | | | -2.43E-05 | | -1.55E-06 | | |
| 80 | 1114.646 | | | | | | | -2.31E-05 | | | | | | | | -2.69E-05 |
| 81 | 1114.676 | | | -2.32E-05 | | | | | | | -2.74E-05 | | | | | |
| 82 | 1120.718 | -2.82E-05 | | | | | | | -2.27E-05 | | | | | | |
| 83 | 1121.724 | | -1.95E-05 | | | | | | | | | | | | |
| 84 | 1159.708 | | | | | | | -2.99E-06 | | | | | | | -1.51E-06 | |
| 85 | 1174.953 | | | | | | | | | | -2.08E-05 | | | | | |
| 86 | 1177.232 | | | | | -2.06E-05 | | | | | | | -2.13E-05 | | | |
| 87 | 1186.275 | | | | -2.21E-05 | | | | | | | -2.47E-05 | | | | |
| 88 | 1186.833 | | | | | | | | -2.35E-05 | | | | | | | -2.73E-05 |
| 89 | 1187.417 | | | -2.36E-05 | | | | | | | -2.78E-05 | | | | | |
| 90 | 1194.654 | -2.88E-05 | | | | | | | -2.62E-05 | | | | | | |
| 91 | 1195.562 | | -2.00E-05 | | | | | | | | | | | | |
| 92 | 1196.27 | | | | | | | -3.07E-06 | | | | | | | -1.58E-06 | |
| 93 | 1234.649 | | | | | | | | | | -2.14E-05 | | | | | |
| 94 | 1236.609 | | | | | | | -3.06E-06 | | | | | | | -1.56E-06 | |
| 95 | 1244.262 | | | | | -2.07E-05 | | | | | | | -2.13E-05 | | | |
| 96 | 1248.906 | | | | -2.22E-05 | | | | | | | -2.48E-05 | | | | |
| 97 | 1253.18 | | | | | | | | -2.37E-05 | | | | | | | -2.74E-05 |
| 98 | 1253.421 | | | -2.38E-05 | | | | | | | -2.80E-05 | | | | | |
| 99 | 1258.927 | -2.99E-05 | | | | | | | -2.87E-05 | | | | | | |
| 100 | 1259.441 | | -2.05E-05 | | | | | | | | | | | | |
| 101 | 1270.858 | | | | | | | -3.01E-06 | | | | | | | -1.56E-06 | |
| 102 | 1302.176 | | | | | | | | | | -2.26E-05 | | | | | |

| | A | B | C | D | E | F | G | H | I | J | K | L | M | N | O |
|---|---|---|---|---|---|---|---|---|---|---|---|---|---|---|---|
| 103 | 1307.084 | | | | | -2.05E-05 | | | | | | | -2.10E-05 | | |
| 104 | 1316.431 | | | | -2.20E-05 | | | | | | | -2.46E-05 | | | |
| 105 | 1320.016 | | | | | | -2.36E-05 | | | | | | | | -2.72E-05 |
| 106 | 1320.104 | | | -2.38E-05 | | | -2.97E-06 | | | | -2.79E-05 | | | -1.55E-06 | |
| 107 | 1328.313 | -3.05E-05 | | | | | | | -3.00E-05 | | | | | | |
| 108 | 1329.953 | | -2.08E-05 | | | | | | | | | | | | |
| 109 | 1368.977 | | | | | | -2.84E-06 | | | | | | | -1.45E-06 | |
| 110 | 1377.725 | | | | | -2.07E-05 | | | | | | | -2.12E-05 | | |
| 111 | 1390.677 | | | | -2.21E-05 | | | | | | | -2.48E-05 | | | |
| 112 | 1392.658 | | | | | | | | | -2.06E-05 | | | | | |
| 113 | 1393.394 | | | | | | | -2.37E-05 | | | | | | | -2.73E-05 |
| 114 | 1393.476 | | | -2.39E-05 | | | | | | | -2.79E-05 | | | | |
| 115 | 1403.714 | -3.10E-05 | | | | | | | -3.11E-05 | | | | | | |
| 116 | 1404.434 | | -2.08E-05 | | | | | | | | | | | | |
| 117 | 1409.511 | | | | | | -2.66E-06 | | | | | | | -1.19E-06 | |
| 118 | 1451.473 | | | | | -2.06E-05 | | | | | | | -2.10E-05 | | |
| 119 | 1459.325 | | | | -2.20E-05 | | | | | | | -2.48E-05 | | | |
| 120 | 1460.879 | | | | | | -2.37E-05 | | | | | | | | -2.74E-05 |
| 121 | 1461.199 | | | -2.39E-05 | | | | | | | -2.81E-05 | | | | |
| 122 | 1461.408 | | | | | | -2.54E-06 | | | | | | | -1.02E-06 | |
| 123 | 1467.206 | -3.18E-05 | | | | | | | -3.12E-05 | | | | | | |
| 124 | 1467.366 | | -2.05E-05 | | | | | | | | | | | | |
| 125 | 1472.079 | | | | | | | | | -1.81E-05 | | | | | |
| 126 | 1500.337 | | | | | | -2.54E-06 | | | | | | | -1.03E-06 | |
| 127 | 1509.143 | | | | | -2.07E-05 | | | | | | | -2.11E-05 | | |
| 128 | 1520.119 | | | | -2.22E-05 | | | | | | | -2.51E-05 | | | |
| 129 | 1525.794 | | | | | | -2.37E-05 | | | | | | | | -2.77E-05 |
| 130 | 1526.476 | | | -2.39E-05 | | | | | | | -2.83E-05 | | | | |
| 131 | 1533.978 | -3.21E-05 | | | | | | | -3.31E-05 | | | | | | |
| 132 | 1533.983 | | -2.05E-05 | | | | | | | | | | | | |
| 133 | 1541.509 | | | | | | -2.42E-06 | | | | | | | -8.72E-07 | |
| 134 | 1544.519 | | | | | | | | | | -1.91E-05 | | | | |
| 135 | 1568.599 | | | | | -2.05E-05 | | | | | | | -2.08E-05 | | |
| 136 | 1574.852 | | | | | | -2.30E-06 | | | | | | | -7.94E-07 | |
| 137 | 1578.242 | | | | -2.21E-05 | | | | | | | -2.50E-05 | | | |
| 138 | 1579.756 | | | | | | | -2.36E-05 | | | | | | | -2.76E-05 |
| 139 | 1579.807 | | | -2.38E-05 | | | | | | | -2.82E-05 | | | | |
| 140 | 1588.806 | -3.22E-05 | | | | | | | -3.20E-05 | | | | | | |
| 141 | 1589.377 | | -2.04E-05 | | | | | | | | | | | | |
| 142 | 1606.19 | | | | | | | | | -1.99E-05 | | | | | |
| 143 | 1614.524 | | | | | | -2.26E-06 | | | | | | | -7.50E-07 | |
| 144 | 1626.078 | | | | | -2.05E-05 | | | | | | | -2.09E-05 | | |
| 145 | 1632.452 | | | | -2.22E-05 | | | | | | | -2.52E-05 | | | |
| 146 | 1635.599 | | | | | | | -2.37E-05 | | | | | | | -2.78E-05 |
| 147 | 1636.508 | | | -2.40E-05 | | | | | | | -2.84E-05 | | | | |
| 148 | 1645.394 | | | | | | -2.21E-06 | | | | | | | -6.94E-07 | |
| 149 | 1645.89 | -3.24E-05 | -2.06E-05 | | | | | | -3.22E-05 | | | | | | |
| 150 | 1654.887 | | | | | | | | | | -2.01E-05 | | | | |
| 151 | 1682.752 | | | | | -2.05E-05 | | | | | | | -2.09E-05 | | |
| 152 | 1695.352 | | | | | | -2.19E-06 | | | | | | | -6.96E-07 | |
| 153 | 1699.534 | | | | -2.20E-05 | | | | | | | -2.51E-05 | | | |
| 154 | 1702.203 | | | | | | | -2.35E-05 | | | | | | | -2.77E-05 |
| 155 | 1702.621 | | | -2.38E-05 | | | | | | | -2.83E-05 | | | | |
| 156 | 1712.97 | -3.25E-05 | | | | | | | -3.32E-05 | | | | | | |
| 157 | 1714.055 | | -2.00E-05 | | | | | | | | | | | | |
| 158 | 1736.577 | | | | | | -1.89E-06 | | | | | | | -4.33E-07 | |
| 159 | 1737.319 | | | | | | | | | | -2.06E-05 | | | | |
| 160 | 1747.607 | | | | | -2.04E-05 | | | | | | | -2.08E-05 | | |
| 161 | 1758.16 | | | | -2.21E-05 | | | | | | | -2.52E-05 | | | |
| 162 | 1761.288 | | | -2.39E-05 | | | | -2.36E-05 | | | -2.83E-05 | | | | -2.77E-05 |
| 163 | 1769.155 | -3.27E-05 | | | | | | | -3.53E-05 | | | | | | |
| 164 | 1770.694 | | -2.10E-05 | | | | | | | | | | | | |
| 165 | 1774.609 | | | | | | -1.92E-06 | | | | | | | -4.39E-07 | |
| 166 | 1805.783 | | | | | -2.04E-05 | | | | | | | -2.08E-05 | | |
| 167 | 1813.219 | | | | | | | | | | -2.00E-05 | | | | |
| 168 | 1819.161 | | | | -2.21E-05 | | | | | | | -2.53E-05 | | | |
| 169 | 1820.127 | | | | | | -1.79E-06 | | | | | | | -2.40E-07 | |
| 170 | 1822.617 | | | -2.39E-05 | | | | -2.36E-05 | | | -2.84E-05 | | | | -2.78E-05 |
| 171 | 1831.041 | -3.31E-05 | -2.06E-05 | | | | | | -3.51E-05 | | | | | | |
| 172 | 1853.479 | | | | | | -1.75E-06 | | | | | | | -2.25E-07 | |
| 173 | 1858.864 | | | | | -2.04E-05 | | | | | | | -2.08E-05 | | |
| 174 | 1874.469 | | | | | | | | | | -2.08E-05 | | | | |
| 175 | 1874.86 | | | | -2.21E-05 | | | | | | | -2.54E-05 | | | |
| 176 | 1876.884 | | | -2.39E-05 | | | | -2.36E-05 | | | -2.85E-05 | | | | -2.79E-05 |
| 177 | 1883.847 | | | | | | -1.61E-06 | | | | | | | -8.89E-08 | |
| 178 | 1884.917 | -3.38E-05 | | | | | | | -3.63E-05 | | | | | | |
| 179 | 1885.322 | | -2.08E-05 | | | | | | | | | | | | |
| 180 | 1917.808 | | | | | -2.04E-05 | | | | | | | -2.08E-05 | | |
| 181 | 1926.996 | | | | | | -1.65E-06 | | | | | | | -2.00E-07 | |
| 182 | 1932.727 | | | | -2.21E-05 | | | | | | | -2.55E-05 | | | |
| 183 | 1936.535 | | | | | | | -2.37E-05 | | | | | | | -2.79E-05 |
| 184 | 1937.048 | | | -2.39E-05 | | | | | | | -2.87E-05 | | | | |
| 185 | 1940.387 | | | | | | | | | | -1.95E-05 | | | | |
| 186 | 1948.037 | | -2.09E-05 | | | | | | | | | | | | |
| 187 | 1948.994 | -3.37E-05 | | | | | | | -3.67E-05 | | | | | | |
| 188 | 1969.516 | | | | | | -1.45E-06 | | | | | | | 2.92E-08 | |
| 189 | 1978.244 | | | | | -2.03E-05 | | | | | | | -2.07E-05 | | |
| 190 | 1991.703 | | | | -2.19E-05 | | | | | | | -2.52E-05 | | | |
| 191 | 1997.058 | | | | | | | -2.36E-05 | | | | | | | -2.77E-05 |
| 192 | 1997.794 | | | -2.38E-05 | | | | | | | -2.84E-05 | | | | |
| 193 | 2016.121 | | | | | | -1.55E-06 | | | | | | | -4.15E-08 | |
| 194 | 2017.453 | -3.48E-05 | -2.08E-05 | | | | | | -3.77E-05 | -1.96E-05 | | | | | |
| 195 | 2040.002 | | | | | -2.03E-05 | | | | | | | -2.07E-05 | | |
| 196 | 2050.903 | | | | | | -1.58E-06 | | | | | | | -7.82E-08 | |
| 197 | 2050.952 | | | | -2.20E-05 | | | | | | | -2.53E-05 | | | |
| 198 | 2054.362 | | | | | | | -2.37E-05 | | | | | | | -2.78E-05 |
| 199 | 2055.214 | | | -2.40E-05 | | | | | | | -2.86E-05 | | | | |
| 200 | 2070.308 | | | | | | | | | | -1.81E-05 | | | | |
| 201 | 2071.765 | | -2.13E-05 | | | | | | | | | | | | |
| 202 | 2072.674 | -3.52E-05 | | | | | | | -3.79E-05 | | | | | | |
| 203 | 2097.124 | | | | | | -1.54E-06 | | | | | | | 2.99E-09 | |
| 204 | 2100.487 | | | | | -2.03E-05 | | | | | | | -2.07E-05 | | |

| | A | B | C | D | E | F | G | H | I | J | K | L | M | N | O |
|---|---|---|---|---|---|---|---|---|---|---|---|---|---|---|---|
| 205 | 2117.602 | | | | -2.21E-05 | | | | | | | -2.52E-05 | | | |
| 206 | 2119.925 | | | | | | | -2.37E-05 | | | | | | | -2.78E-05 |
| 207 | 2119.926 | | | -2.41E-05 | | | | | | | | -2.86E-05 | | | |
| 208 | 2132.673 | | -2.10E-05 | | | | | | | | | | | | |
| 209 | 2132.801 | -3.60E-05 | | | | | | | -3.85E-05 | | | | | | |
| 210 | 2141.497 | | | | | | -1.59E-06 | | | | | | | -1.42E-07 | |
| 211 | 2143.981 | | | | | | | | | -2.06E-05 | | | | | |
| 212 | 2156.42 | | | | | -2.03E-05 | | | | | | | -2.07E-05 | | |
| 213 | 2170.295 | | | | -2.21E-05 | | | | | | | -2.53E-05 | | | |
| 214 | 2171.201 | | | | | | | -1.43E-06 | | | | | | -4.79E-09 | |
| 215 | 2172.662 | | | | | | | -2.37E-05 | | | | | | | -2.78E-05 |
| 216 | 2173.14 | | | -2.41E-05 | | | | | | | | -2.86E-05 | | | |
| 217 | 2187.156 | | -2.12E-05 | | | | | | | | | | | | |
| 218 | 2188.055 | -3.60E-05 | | | | | | | -3.80E-05 | | | | | | |
| 219 | 2206.532 | | | | | | -1.29E-06 | | | | | | | 1.46E-07 | |
| 220 | 2207.789 | | | | | -2.02E-05 | | | | | | | -2.06E-05 | | |
| 221 | 2218.152 | | | | | | | | | -2.01E-05 | | | | | |
| 222 | 2226.375 | | | | -2.21E-05 | | | | | | | -2.54E-05 | | | |
| 223 | 2229.637 | | | | | | | -2.37E-05 | | | | | | | -2.79E-05 |
| 224 | 2230.852 | | | -2.41E-05 | | | | | | | | -2.87E-05 | | | |
| 225 | 2246.387 | | | | | | -1.26E-06 | | | | | | | 7.62E-08 | |
| 226 | 2248.713 | | -2.15E-05 | | | | | | | | | | | | |
| 227 | 2249.391 | -3.61E-05 | | | | | | | -3.81E-05 | | | | | | |
| 228 | 2271.862 | | | | | -2.02E-05 | | | | | | | -2.06E-05 | | |
| 229 | 2287.295 | | | | -2.20E-05 | | -1.28E-06 | | | | | -2.54E-05 | | 4.06E-08 | |
| 230 | 2288.458 | | | | | | | | | | -1.55E-05 | | | | |
| 231 | 2292.045 | | | -2.41E-05 | | | | -2.37E-05 | | | | -2.87E-05 | | | -2.79E-05 |
| 232 | 2304.699 | -3.61E-05 | -2.16E-05 | | | | | | -3.95E-05 | | | | | | |
| 233 | 2319.583 | | | | | | -1.35E-06 | | | | | | | 9.14E-09 | |
| 234 | 2322.467 | | | | | -2.01E-05 | | | | | | | -2.05E-05 | | |
| 235 | 2333.008 | | | | -2.20E-05 | | | | | | | -2.53E-05 | | | |
| 236 | 2335.76 | | | | | | | -2.37E-05 | | | | | | | -2.79E-05 |
| 237 | 2335.925 | | | -2.42E-05 | | | | | | | | -2.87E-05 | | | |
| 238 | 2345.339 | | | | | | -6.07E-07 | | | | | | | 8.25E-07 | |
| 239 | 2351.322 | | -2.17E-05 | | | | | | | | | | | | |
| 240 | 2351.434 | -3.70E-05 | | | | | | | -3.95E-05 | | | | | | |
| 241 | 2355.002 | | | | | | | | | | -2.10E-05 | | | | |
| 242 | 2367.323 | | | | | -2.02E-05 | | | | | | | -2.05E-05 | | |
| 243 | 2377.592 | | | | | | | -2.54E-07 | | | | | | 1.18E-06 | |
| 244 | 2378.368 | | | | -2.20E-05 | | | | | | | -2.52E-05 | | | |
| 245 | 2381.04 | | | | | | | | -2.36E-05 | | | | | | -2.78E-05 |
| 246 | 2381.049 | | | -2.42E-05 | | | | | | | | -2.87E-05 | | | |
| 247 | 2396.97 | | -2.18E-05 | | | | | | | | | | | | |
| 248 | 2397.274 | -3.77E-05 | | | | | | | -4.05E-05 | | | | | | |
| 249 | 2408.784 | | | | | | -1.17E-07 | | | | | | | 1.27E-06 | |
| 250 | 2410.359 | | | | | | | | | | -2.19E-05 | | | | |
| 251 | 2410.495 | | | | | -2.02E-05 | | | | | | | -2.06E-05 | | |
| 252 | 2427.205 | | | | -2.19E-05 | | | | | | | -2.51E-05 | | | |
| 253 | 2428.789 | | | -2.42E-05 | | | | -2.36E-05 | | | | -2.87E-05 | | | -2.78E-05 |
| 254 | 2436.234 | | | | | | -1.95E-07 | | | | | | | 1.25E-06 | |
| 255 | 2443.38 | | -2.18E-05 | | | | | | | | | | | | |
| 256 | 2446.08 | -3.80E-05 | | | | | | | -4.14E-05 | | | | | | |
| 257 | 2458.389 | | | | | -2.02E-05 | | | | | | | -2.07E-05 | | |
| 258 | 2473.332 | | | | | | -1.22E-07 | | | | | | | 1.27E-06 | |
| 259 | 2476.906 | | | | -2.18E-05 | | | | | | | -2.52E-05 | | | |
| 260 | 2479.826 | | | | | | | -2.38E-05 | | | | | | | -2.79E-05 |
| 261 | 2480.07 | | | -2.42E-05 | | | | | | | | -2.86E-05 | | | |
| 262 | 2482.803 | | | | | | | | | -2.21E-05 | | | | | |
| 263 | 2496.102 | | -2.18E-05 | | | | | | | | | | | | |
| 264 | 2498.213 | -3.77E-05 | | | | | | | -4.09E-05 | | | | | | |
| 265 | 2511.856 | | | | | | -4.08E-08 | | | | | | | 1.35E-06 | |
| 266 | 2512.864 | | | | | -2.01E-05 | | | | | | | -2.06E-05 | | |
| 267 | 2527.541 | | | | -2.16E-05 | | | | | | | -2.48E-05 | | | |
| 268 | 2530.563 | | | | | | | -2.36E-05 | | | | | | | -2.76E-05 |
| 269 | 2530.601 | | | -2.39E-05 | | | | | | | | -2.83E-05 | | | |
| 270 | 2538.423 | | | | | | 1.49E-07 | | | | | | | 1.54E-06 | |
| 271 | 2544.662 | | -2.14E-05 | | | | | | | | | | | | |
| 272 | 2546.178 | -3.66E-05 | | | | | | | -3.97E-05 | | | | | | |
| 273 | 2553.797 | | | | | | | | | | -2.27E-05 | | | | |
| 274 | 2557.147 | | | | | -1.99E-05 | | | | | | | -2.03E-05 | | |
| 275 | 2573.982 | | | | -2.15E-05 | | 2.14E-07 | | | | | -2.47E-05 | | 1.65E-06 | |
| 276 | 2579.596 | | | | | | | -2.34E-05 | | | | | | | -2.75E-05 |
| 277 | 2579.8 | | | -2.38E-05 | | | | | | | | -2.84E-05 | | | |
| 278 | 2596.075 | | -2.11E-05 | | | | | | | | | | | | |
| 279 | 2598.56 | -3.71E-05 | | | | | | | -4.04E-05 | | | | | | |
| 280 | 2605.879 | | | | | -1.98E-05 | | | | | | | -2.02E-05 | | |
| 281 | 2610.494 | | | | | | 5.33E-07 | | | | | | | 2.04E-06 | |
| 282 | 2630.612 | | | | -2.14E-05 | | | | | | | -2.46E-05 | | | |
| 283 | 2632.321 | | | | | | | | | | -2.20E-05 | | | | |
| 284 | 2635.333 | | | | | | | -2.33E-05 | | | | | | | -2.73E-05 |
| 285 | 2635.9 | | | -2.37E-05 | | | | | | | | -2.83E-05 | | | |
| 286 | 2649.055 | | | | | | 6.68E-07 | | | | | | | 2.15E-06 | |
| 287 | 2651.347 | | -2.10E-05 | | | | | | | | | | | | |
| 288 | 2652.499 | -3.71E-05 | | | | | | | -3.95E-05 | | | | | | |
| 289 | 2661.437 | | | | | -1.97E-05 | | | | | | | -2.01E-05 | | |
| 290 | 2681.02 | | | | -2.14E-05 | | | | | | | -2.46E-05 | | | |
| 291 | 2686.614 | | | | | | | -2.33E-05 | | | | | | | -2.74E-05 |
| 292 | 2687.29 | | | -2.38E-05 | | | | | | | | -2.83E-05 | | | |
| 293 | 2688.889 | | | | | | 7.95E-07 | | | | | | | 2.34E-06 | |
| 294 | 2709.29 | | -2.15E-05 | | | | | | | | | | | | |
| 295 | 2710.011 | -3.74E-05 | | | | | | | -4.05E-05 | | | | | | |
| 296 | 2716.596 | | | | | -1.96E-05 | | | | | | | -2.01E-05 | | |
| 297 | 2722.378 | | | | | | | | | | -2.13E-05 | | | | |
| 298 | 2730.532 | | | | | | 8.51E-07 | | | | | | | 2.37E-06 | |
| 299 | 2736.201 | | | | -2.14E-05 | | | | | | | -2.46E-05 | | | |
| 300 | 2740.47 | | | | | | | -2.34E-05 | | | | | | | -2.74E-05 |
| 301 | 2740.951 | | | -2.38E-05 | | | | | | | | -2.84E-05 | | | |
| 302 | 2762.44 | | -2.13E-05 | | | | | | | | | | | | |
| 303 | 2763.578 | -3.73E-05 | | | | | | | -4.07E-05 | | | | | | |
| 304 | 2766.931 | | | | | | 8.77E-07 | | | | | | | 2.30E-06 | |
| 305 | 2768.478 | | | | | -1.95E-05 | | | | | | | -1.99E-05 | | |
| 306 | 2776.666 | | | | | | | | | | -2.02E-05 | | | | |

| | A | B | C | D | E | F | G | H | I | J | K | L | M | N | O |
|---|---|---|---|---|---|---|---|---|---|---|---|---|---|---|---|
| 307 | 2783.054 | | | | -2.13E-05 | | | | | | | -2.45E-05 | | | |
| 308 | 2785.942 | | | | | | | -2.33E-05 | | | | | | | -2.73E-05 |
| 309 | 2786.713 | | | -2.38E-05 | | | | | | | -2.83E-05 | | | | |
| 310 | 2797.639 | | | | | | 8.98E-07 | | | | | | | 2.33E-06 | |
| 311 | 2805.188 | | -2.11E-05 | | | | | | | | | | | | |
| 312 | 2805.846 | -3.75E-05 | | | | | | | -4.13E-05 | | | | | | |
| 313 | 2808.983 | | | | | -1.96E-05 | | | | | | | -2.00E-05 | | |
| 314 | 2828.81 | | | | -2.13E-05 | | | | | | | -2.45E-05 | | | |
| 315 | 2829.1 | | | | | | 9.16E-07 | | | | | | | 2.35E-06 | |
| 316 | 2838.052 | | | | | | | -2.32E-05 | | | | | | | -2.72E-05 |
| 317 | 2838.824 | | | -2.37E-05 | | | | | | | -2.82E-05 | | | | |
| 318 | 2857.111 | | | | | | | | | -2.16E-05 | | | | | |
| 319 | 2861.21 | | -2.11E-05 | | | | | | | | | | | | |
| 320 | 2861.341 | -3.74E-05 | | | | | | | -4.25E-05 | | | | | | |
| 321 | 2863.069 | | | | | -1.95E-05 | | | | | | | -1.99E-05 | | |
| 322 | 2867.16 | | | | | | 9.49E-07 | | | | | | | 2.42E-06 | |
| 323 | 2886.112 | | | | -2.11E-05 | | | | | | | -2.44E-05 | | | |
| 324 | 2891.306 | | | | | | | -2.30E-05 | | | | | | | -2.71E-05 |
| 325 | 2891.631 | | | -2.36E-05 | | | | | | | -2.81E-05 | | | | |
| 326 | 2903.905 | | | | | | 9.19E-07 | | | | | | | 2.41E-06 | |
| 327 | 2910.749 | | -2.10E-05 | | | | | | | | | | | | |
| 328 | 2912.964 | -3.74E-05 | | | | | | | -4.16E-05 | | | | | | |
| 329 | 2913.789 | | | | | -1.95E-05 | | | | | | | -1.99E-05 | | |
| 330 | 2935.355 | | | | -2.11E-05 | | | | | | | -2.45E-05 | | | |
| 331 | 2942.496 | | | | | | 9.88E-07 | | | | | | | 2.51E-06 | |
| 332 | 2942.566 | | | | | | | -2.29E-05 | | | | | | | -2.70E-05 |
| 333 | 2943.5 | | | -2.35E-05 | | | | | | | -2.80E-05 | | | | |
| 334 | 2945.01 | | | | | | | | | -2.24E-05 | | | | | |
| 335 | 2963.767 | -3.72E-05 | | | | | | | -4.19E-05 | | | | | | |
| 336 | 2964.22 | | -2.06E-05 | | | | | | | | | | | | |
| 337 | 2966.471 | | | | | -1.93E-05 | | | | | | | -1.97E-05 | | |
| 338 | 2980.808 | | | | | | 9.76E-07 | | | | | | | 2.47E-06 | |
| 339 | 2994.143 | | | | -2.10E-05 | | | | | | | -2.44E-05 | | | |
| 340 | 2998.929 | | | | | | | -2.29E-05 | | | | | | | -2.70E-05 |
| 341 | 2999.927 | | | -2.34E-05 | | | | | | | -2.81E-05 | | | | |
| 342 | 3006.874 | | | | | | | | | -2.28E-05 | | | | | |
| 343 | 3027.496 | -3.67E-05 | | | | | | | -4.21E-05 | | | | | | |
| 344 | 3027.585 | | | | | | 9.64E-07 | | | | | | | 2.49E-06 | |
| 345 | 3027.618 | | -2.05E-05 | | | -1.93E-05 | | | | | | | -1.97E-05 | | |
| 346 | 3046.456 | | | | -2.09E-05 | | | | | | | -2.44E-05 | | | |
| 347 | 3050.117 | | | | | | | -2.28E-05 | | | | | | | -2.69E-05 |
| 348 | 3050.598 | | | -2.34E-05 | | | | | | | -2.80E-05 | | | | |
| 349 | 3052.941 | | | | | | 1.00E-06 | | | | | | | 2.53E-06 | |
| 350 | 3067.207 | | | | | -1.91E-05 | | | | | | | -1.96E-05 | | |
| 351 | 3067.754 | -3.68E-05 | | | | | | | -4.23E-05 | | | | | | |
| 352 | 3068.133 | | -2.06E-05 | | | | | | | | | | | | |
| 353 | 3080.478 | | | | | | | | | -2.40E-05 | | | | | |
| 354 | 3081.486 | | | | | | 1.04E-06 | | | | | | | 2.57E-06 | |
| 355 | 3094.136 | | | | -2.07E-05 | | | | | | | -2.42E-05 | | | |
| 356 | 3098.952 | | | | | | | -2.27E-05 | | | | | | | -2.66E-05 |
| 357 | 3099.609 | | | -2.32E-05 | | | | | | | -2.77E-05 | | | | |
| 358 | 3117.218 | | | | | | 1.21E-06 | | | | | | | 2.68E-06 | |
| 359 | 3118.2 | | | | | -1.90E-05 | | | | | | | -1.94E-05 | | |
| 360 | 3119.321 | -3.82E-05 | | | | | | | -4.31E-05 | | | | | | |
| 361 | 3120.746 | | -2.04E-05 | | | | | | | | | | | | |
| 362 | 3139.474 | | | | -2.06E-05 | | | | | | | -2.39E-05 | | | |
| 363 | 3143.213 | | | | | | | | | | -2.66E-05 | | | | |
| 364 | 3145.76 | | | | | | | -2.26E-05 | | | | | | | -2.65E-05 |
| 365 | 3146.529 | | | -2.32E-05 | | | | | | | -2.77E-05 | | | | |
| 366 | 3149.379 | | | | | | 1.15E-06 | | | | | | | 2.66E-06 | |
| 367 | 3161.951 | | | | | -1.89E-05 | | | | | | | -1.93E-05 | | |
| 368 | 3164.909 | -3.81E-05 | | | | | | | -4.23E-05 | | | | | | |
| 369 | 3165.954 | | -2.01E-05 | | | | | | | | | | | | |
| 370 | 3188.888 | | | | | | 1.18E-06 | | | | | | | 2.67E-06 | |
| 371 | 3191.264 | | | | | | | | | | -2.78E-05 | | | | |
| 372 | 3192.479 | | | | -2.05E-05 | | | | | | | -2.39E-05 | | | |
| 373 | 3200.88 | | | | | | | -2.24E-05 | | | | | | | -2.65E-05 |
| 374 | 3201.286 | | | -2.31E-05 | | | | | | | -2.76E-05 | | | | |
| 375 | 3221.607 | | | | | -1.89E-05 | | | | | | | -1.94E-05 | | |
| 376 | 3226.987 | -3.79E-05 | | | | | | | -4.31E-05 | | | | | | |
| 377 | 3227.893 | | -2.00E-05 | | | | | | | | | | | | |
| 378 | 3231.615 | | | | | | 1.22E-06 | | | | | | | 2.65E-06 | |
| 379 | 3247.217 | | | | -2.04E-05 | | | | | | | -2.38E-05 | | | |
| 380 | 3254.018 | | | | | | | -2.25E-05 | | | | | | | -2.64E-05 |
| 381 | 3256.142 | | | -2.31E-05 | | | | | | | -2.75E-05 | | | | |
| 382 | 3265.233 | | | | | | | | | -2.90E-05 | | | | | |
| 383 | 3273.4 | | | | | | 1.24E-06 | | | | | | | 2.64E-06 | |
| 384 | 3277.356 | | | | | -1.88E-05 | | | | | | | -1.92E-05 | | |
| 385 | 3283.14 | -3.76E-05 | | | | | | | -4.17E-05 | | | | | | |
| 386 | 3284.12 | | -1.99E-05 | | | | | | | | | | | | |
| 387 | 3300.516 | | | | -2.03E-05 | | | | | | | -2.36E-05 | | | |
| 388 | 3304.596 | | | | | | 1.36E-06 | | | | | | | 2.75E-06 | |
| 389 | 3310.593 | | | | | | | -2.22E-05 | | | | | | | -2.60E-05 |
| 390 | 3311.775 | | | -2.28E-05 | | | | | | | -2.72E-05 | | | | |
| 391 | 3328.002 | | | | | -1.86E-05 | | | | | | | -1.91E-05 | | |
| 392 | 3336.817 | -3.66E-05 | | | | | | | -4.16E-05 | | | | | | |
| 393 | 3338.83 | | -1.98E-05 | | | | | | | | | | | | |
| 394 | 3349.315 | | | | | | 1.40E-06 | | | | | | | 2.76E-06 | |
| 395 | 3358.09 | | | | | | | | | -2.96E-05 | | | | | |
| 396 | 3364.628 | | | | -2.02E-05 | | | | | | | -2.35E-05 | | | |
| 397 | 3370.399 | | | | | | | -2.23E-05 | | | | | | | -2.60E-05 |
| 398 | 3372.486 | | | -2.28E-05 | | | | | | | -2.71E-05 | | | | |
| 399 | 3390.695 | | | | | | 1.35E-06 | | | | | | | 2.67E-06 | |
| 400 | 3392.057 | | | | | -1.86E-05 | | | | | | | -1.91E-05 | | |
| 401 | 3396.791 | -3.69E-05 | | | | | | | -4.08E-05 | | | | | | |
| 402 | 3399.22 | | -2.00E-05 | | | | | | | | | | | | |
| 403 | 3415.186 | | | | | | | | | -3.03E-05 | | | | | |
| 404 | 3417.276 | | | | -2.01E-05 | | | | | | | -2.36E-05 | | | |
| 405 | 3424.477 | | | -2.27E-05 | | | 1.33E-06 | -2.22E-05 | | | -2.70E-05 | | | 2.66E-06 | -2.59E-05 |
| 406 | 3438.996 | | | | | -1.85E-05 | | | | | | | -1.90E-05 | | |
| 407 | 3446.7 | -3.72E-05 | | | | | | | -4.21E-05 | | | | | | |
| 408 | 3450.943 | | -1.99E-05 | | | | | | | | | | | | |

| | A | B | C | D | E | F | G | H | I | J | K | L | M | N | O |
|---|---|---|---|---|---|---|---|---|---|---|---|---|---|---|---|
| 409 | 3461.898 | | | | | | 1.36E-06 | | | | | | | 2.68E-06 | |
| 410 | 3471.129 | | | | -1.99E-05 | | | | | | | -2.34E-05 | | | |
| 411 | 3477.201 | | | | | | | | -2.20E-05 | | | | | | -2.57E-05 |
| 412 | 3478.932 | | | -2.25E-05 | | | | | | | -2.68E-05 | | | | |
| 413 | 3484.017 | | | | | | | | | -3.02E-05 | | | | | |
| 414 | 3493.785 | | | | | -1.85E-05 | | | | | | | -1.90E-05 | | |
| 415 | 3499.883 | -3.69E-05 | | | | | | | -4.21E-05 | | | | | | |
| 416 | 3501.961 | | | | | | 1.42E-06 | | | | | | | 2.75E-06 | |
| 417 | 3504.497 | | -1.99E-05 | | | | | | | | | | | | |
| 418 | 3526.131 | | | | -1.98E-05 | | | | | | | -2.34E-05 | | | |
| 419 | 3534.178 | | | | | | | | -2.19E-05 | | | | | | -2.56E-05 |
| 420 | 3535.302 | | | -2.25E-05 | | | | | | | -2.68E-05 | | | | |
| 421 | 3551.723 | | | | | | 1.39E-06 | | | | | | | 2.77E-06 | |
| 422 | 3551.788 | | | | | -1.84E-05 | | | | | | | -1.88E-05 | | |
| 423 | 3561.471 | -3.69E-05 | | | | | | | -4.14E-05 | | | | | | |
| 424 | 3566.475 | | -1.96E-05 | | | | | | | | | | | | |
| 425 | 3567.088 | | | | | | | | | -3.23E-05 | | | | | |
| 426 | 3584.968 | | | | -1.96E-05 | | | | | | | -2.31E-05 | | | |
| 427 | 3597.274 | | | | | | | | -2.17E-05 | | | | | | -2.54E-05 |
| 428 | 3598.757 | | | -2.22E-05 | | | | | | | -2.63E-05 | | | | |
| 429 | 3599.274 | | | | | | 1.48E-06 | | | | | | | 2.87E-06 | |
| 430 | 3618.577 | | | | | -1.82E-05 | | | | | | | -1.87E-05 | | |
| 431 | 3627.075 | -3.67E-05 | | | | | | | -4.17E-05 | | | | | | |
| 432 | 3631.022 | | -1.95E-05 | | | | | | | | | | | | |
| 433 | 3633.766 | | | | | | | | | | -3.41E-05 | | | | |
| 434 | 3635.794 | | | | | | 1.45E-06 | | | | | | | 2.79E-06 | |
| 435 | 3647.26 | | | | -1.94E-05 | | | | | | | -2.29E-05 | | | |
| 436 | 3655.669 | | | | | | | | -2.15E-05 | | | | | | -2.52E-05 |
| 437 | 3656.926 | | | -2.20E-05 | | | | | | | -2.62E-05 | | | | |
| 438 | 3670.949 | | | | | -1.80E-05 | | | | | | | -1.85E-05 | | |
| 439 | 3682.726 | -3.55E-05 | | | | | | | -3.92E-05 | | | | | | |
| 440 | 3683.592 | | | | | | 1.55E-06 | | | | | | | 2.86E-06 | |
| 441 | 3687.728 | | -1.92E-05 | | | | | | | | | | | | |
| 442 | 3703.541 | | | | -1.93E-05 | | | | | | | -2.28E-05 | | | |
| 443 | 3709.17 | | | | | | | | -2.15E-05 | | | | | | -2.49E-05 |
| 444 | 3711.328 | | | -2.20E-05 | | | | | | | -2.59E-05 | | | | |
| 445 | 3717.141 | | | | | | | | | | -3.52E-05 | | | | |
| 446 | 3723.015 | | | | | | 1.51E-06 | | | | | | | 2.79E-06 | |
| 447 | 3725.145 | | | | | -1.79E-05 | | | | | | | -1.84E-05 | | |
| 448 | 3737.635 | -3.64E-05 | | | | | | | -4.00E-05 | | | | | | |
| 449 | 3747.051 | | -1.92E-05 | | | | | | | | | | | | |
| 450 | 3767.648 | | | | -1.95E-05 | | | | | | | -2.30E-05 | | | |
| 451 | 3778.267 | | | | | | | | -2.17E-05 | | | | | | -2.54E-05 |
| 452 | 3779.448 | | | -2.23E-05 | | | | | | | -2.64E-05 | | | | |
| 453 | 3782.658 | | | | | | 1.49E-06 | | | | | | | 2.82E-06 | |
| 454 | 3786.227 | | | | | | | | | | -3.53E-05 | | | | |
| 455 | 3791.202 | | | | | -1.79E-05 | | | | | | | -1.84E-05 | | |
| 456 | 3800.647 | -3.66E-05 | | | | | | | -3.96E-05 | | | | | | |
| 457 | 3809.543 | | -1.92E-05 | | | | | | | | | | | | |
| 458 | 3826.401 | | | | -1.96E-05 | | | | | | | -2.29E-05 | | | |
| 459 | 3826.692 | | | | | | 1.47E-06 | | | | | | | 2.75E-06 | |
| 460 | 3830.962 | | | | | | | | -2.16E-05 | | | | | | -2.52E-05 |
| 461 | 3831.642 | | | -2.22E-05 | | | | | | | -2.63E-05 | | | | |
| 462 | 3844.31 | | | | | -1.78E-05 | | | | | | | -1.82E-05 | | |
| 463 | 3861.275 | -3.65E-05 | | | | | | | -3.95E-05 | | | | | | |
| 464 | 3866.106 | | | | | | | | | | -3.50E-05 | | | | |
| 465 | 3867.188 | | -1.89E-05 | | | | | | | | | | | | |
| 466 | 3870.218 | | | | | | 1.51E-06 | | | | | | | 2.76E-06 | |
| 467 | 3889.095 | | | | -1.95E-05 | | | | | | | -2.28E-05 | | | |
| 468 | 3894.266 | | | | | | | | -2.16E-05 | | | | | | -2.52E-05 |
| 469 | 3897.039 | | | -2.21E-05 | | | | | | | -2.61E-05 | | | | |
| 470 | 3908.278 | | | | | -1.78E-05 | | | | | | | -1.82E-05 | | |
| 471 | 3913.24 | | | | | | 1.57E-06 | | | | | | | 2.84E-06 | |
| 472 | 3919.029 | -3.65E-05 | | | | | | | -3.93E-05 | | | | | | |
| 473 | 3929.478 | | -1.85E-05 | | | | | | | | | | | | |
| 474 | 3936.833 | | | | | | | | | | -3.63E-05 | | | | |
| 475 | 3942.627 | | | | -1.94E-05 | | | | | | | -2.25E-05 | | | |
| 476 | 3949.624 | | | | | | 1.48E-06 | -2.14E-05 | | | | | | 2.59E-06 | -2.50E-05 |
| 477 | 3951.331 | | | -2.19E-05 | | | | | | | -2.58E-05 | | | | |
| 478 | 3961.712 | | | | | -1.78E-05 | | | | | | | -1.83E-05 | | |
| 479 | 3980.257 | -3.68E-05 | | | | | | | -3.87E-05 | | | | | | |
| 480 | 3990.911 | | -1.84E-05 | | | | | | | | | | | | |
| 481 | 4003.874 | | | | | | 1.34E-06 | | | | | | | 2.39E-06 | |
| 482 | 4007.089 | | | | -1.93E-05 | | | | | | | -2.23E-05 | | | |
| 483 | 4011.424 | | | | | | | | | | -3.63E-05 | | | | |
| 484 | 4017.448 | | | | | | | | -2.13E-05 | | | | | | -2.49E-05 |
| 485 | 4018.483 | | | -2.18E-05 | | | | | | | -2.58E-05 | | | | |
| 486 | 4031.369 | | | | | -1.77E-05 | | | | | | | -1.83E-05 | | |
| 487 | 4041.03 | -3.60E-05 | | | | | | | -3.83E-05 | | | | | | |
| 488 | 4050.492 | | | | | | 1.42E-06 | | | | | | | 2.45E-06 | |
| 489 | 4052.088 | | -1.80E-05 | | | | | | | | | | | | |
| 490 | 4066.836 | | | | -1.93E-05 | | | | | | | -2.23E-05 | | | |
| 491 | 4070.172 | | | | | | | | | | -3.66E-05 | | | | |
| 492 | 4074.009 | | | | | | | | -2.12E-05 | | | | | | -2.47E-05 |
| 493 | 4078.215 | | | -2.18E-05 | | | | | | | -2.59E-05 | | | | |
| 494 | 4088.023 | | | | | -1.75E-05 | | | | | | | -1.81E-05 | | |
| 495 | 4103.338 | | | | | | 1.37E-06 | | | | | | | 2.36E-06 | |
| 496 | 4108.812 | -3.68E-05 | | | | | | | -3.82E-05 | | | | | | |
| 497 | 4117.369 | | -1.80E-05 | | | | | | | | | | | | |
| 498 | 4133.382 | | | | -1.92E-05 | | | | | | | -2.23E-05 | | | |
| 499 | 4141.904 | | | | | | | | -2.10E-05 | | | | | | -2.45E-05 |
| 500 | 4143.538 | | | -2.17E-05 | | | | | | | -2.58E-05 | | | | |
| 501 | 4146.157 | | | | | | 1.38E-06 | | | | | | | 2.37E-06 | |
| 502 | 4149.956 | | | | | | | | | | -3.76E-05 | | | | |
| 503 | 4151.91 | | | | | -1.75E-05 | | | | | | | -1.80E-05 | | |
| 504 | 4164.784 | -3.72E-05 | | | | | | | -3.70E-05 | | | | | | |
| 505 | 4176.78 | | -1.77E-05 | | | | | | | | | | | | |
| 506 | 4181.137 | | | | | | 1.39E-06 | | | | | | | 2.38E-06 | |
| 507 | 4188.02 | | | | -1.91E-05 | | | | | | | -2.20E-05 | | | |
| 508 | 4196.165 | | | | | | | | -2.08E-05 | | | | | | -2.42E-05 |
| 509 | 4196.542 | | | -2.15E-05 | | | | | | | -2.54E-05 | | | | |
| 510 | 4208.726 | | | | | -1.73E-05 | | | | | | | -1.78E-05 | | |

| | A | B | C | D | E | F | G | H | I | J | K | L | M | N | O |
|---|---|---|---|---|---|---|---|---|---|---|---|---|---|---|---|
| 511 | 4225.08 | | | | | | | | | -3.84E-05 | | | | | |
| 512 | 4226.628 | -3.71E-05 | | | | | | | -3.75E-05 | | | | | | |
| 513 | 4231.254 | | | | | | 1.48E-06 | | | | | | | 2.46E-06 | |
| 514 | 4238.304 | | -1.75E-05 | | | | | | | | | | | | |
| 515 | 4251.622 | | | | -1.90E-05 | | | | | | | -2.20E-05 | | | |
| 516 | 4263.137 | | | | | | | -2.05E-05 | | | | | | | -2.40E-05 |
| 517 | 4264.842 | | | -2.12E-05 | | | | | | | -2.52E-05 | | | | |
| 518 | 4269.307 | | | | | -1.72E-05 | | | | | | | -1.77E-05 | | |
| 519 | 4273.204 | | | | | | 1.45E-06 | | | | | | | 2.44E-06 | |
| 520 | 4289.023 | -3.68E-05 | | | | | | | -3.67E-05 | | | | | | |
| 521 | 4305.747 | | -1.74E-05 | | | | | | | | | | | | |
| 522 | 4314.792 | | | | | | | | | -3.71E-05 | | | | | |
| 523 | 4320.002 | | | | -1.89E-05 | | | | | | | -2.20E-05 | | | |
| 524 | 4328.785 | | | | | | | -2.03E-05 | | | | | | | -2.38E-05 |
| 525 | 4329.476 | | | | | | 1.37E-06 | | | | | | | 2.30E-06 | |
| 526 | 4330.205 | | | -2.10E-05 | | | | | | | -2.51E-05 | | | | |
| 527 | 4339.136 | | | | | -1.73E-05 | | | | | | | -1.78E-05 | | |
| 528 | 4367.242 | -3.68E-05 | | | | | | | -3.63E-05 | | | | | | |
| 529 | 4381.716 | | -1.70E-05 | | | | | | | | | | | | |
| 530 | 4392.348 | | | | | | 1.27E-06 | | | | | | | 2.20E-06 | |
| 531 | 4395.346 | | | | -1.88E-05 | | | | | | | -2.16E-05 | | | |
| 532 | 4401.987 | | | | | | | | | -3.74E-05 | | | | | |
| 533 | 4403.649 | | | | | | | | -1.99E-05 | | | | | | -2.33E-05 |
| 534 | 4406.539 | | | -2.07E-05 | | | | | | | -2.47E-05 | | | | |
| 535 | 4412.712 | | | | | -1.73E-05 | | | | | | | -1.77E-05 | | |
| 536 | 4429.531 | -3.68E-05 | | | | | | | -3.66E-05 | | | | | | |
| 537 | 4438.112 | | | | | | 1.23E-06 | | | | | | | 2.13E-06 | |
| 538 | 4445.176 | | -1.69E-05 | | | | | | | | | | | | |
| 539 | 4457.171 | | | | -1.87E-05 | | | | | | | -2.15E-05 | | | |
| 540 | 4467.145 | | | | | | | | -1.95E-05 | | | | | | -2.27E-05 |
| 541 | 4468.949 | | | -2.04E-05 | | | | | | | -2.42E-05 | | | | |
| 542 | 4473.719 | | | | | -1.72E-05 | | | | | | | -1.76E-05 | | |
| 543 | 4480.883 | | | | | | | | | -3.72E-05 | | | | | |
| 544 | 4486.423 | | | | | | 1.24E-06 | | | | | | | 2.20E-06 | |
| 545 | 4497.969 | -3.60E-05 | | | | | | | -3.62E-05 | | | | | | |
| 546 | 4519.173 | | -1.66E-05 | | | | | | | | | | | | |
| 547 | 4531.715 | | | | -1.84E-05 | | | | | | | -2.13E-05 | | | |
| 548 | 4544.124 | | | | | | | | -1.92E-05 | | | | | | -2.26E-05 |
| 549 | 4546.321 | | | -2.01E-05 | | | | | | | -2.40E-05 | | | | |
| 550 | 4547.489 | | | | | | 1.27E-06 | | | | | | | 2.27E-06 | |
| 551 | 4550.611 | | | | | -1.71E-05 | | | | | | | -1.76E-05 | | |
| 552 | 4578.757 | -3.58E-05 | | | | | | | -3.69E-05 | | | | | | |
| 553 | 4584.18 | | | | | | | | | | -3.66E-05 | | | | |
| 554 | 4591.966 | | -1.60E-05 | | | | | | | | | | | | |
| 555 | 4600.521 | | | | -1.82E-05 | | | | | | | -2.12E-05 | | | |
| 556 | 4603.546 | | | | | | 1.29E-06 | | | | | | | 2.31E-06 | |
| 557 | 4610.118 | | | | | | | | -1.93E-05 | | | | | | -2.25E-05 |
| 558 | 4610.462 | | | -2.02E-05 | | | | | | | -2.39E-05 | | | | |
| 559 | 4614.123 | | | | | -1.70E-05 | | | | | | | -1.74E-05 | | |
| 560 | 4639.85 | -3.59E-05 | | | | | | | -3.70E-05 | | | | | | |
| 561 | 4656.185 | | | | | | 1.27E-06 | | | | | | | 2.32E-06 | |
| 562 | 4658.058 | | -1.56E-05 | | | | | | | | | | | | |
| 563 | 4661.511 | | | | | | | | | -3.67E-05 | | | | | |
| 564 | 4667.853 | | | | -1.82E-05 | | | | | | | -2.11E-05 | | | |
| 565 | 4679.493 | | | | | | | | -1.93E-05 | | | | | | -2.25E-05 |
| 566 | 4681.691 | | | -2.02E-05 | | | | | | | -2.38E-05 | | | | |
| 567 | 4684.109 | | | | | -1.70E-05 | | | | | | | -1.74E-05 | | |
| 568 | 4709.881 | | | | | | 1.25E-06 | | | | | | | 2.33E-06 | |
| 569 | 4710.014 | -3.62E-05 | | | | | | | -3.75E-05 | | | | | | |
| 570 | 4725.021 | | -1.57E-05 | | | | | | | | | | | | |
| 571 | 4739.726 | | | | -1.78E-05 | | | | | | | -2.06E-05 | | | |
| 572 | 4748.253 | | | | | | | | -1.91E-05 | | | | | | -2.21E-05 |
| 573 | 4750.742 | | | -1.99E-05 | | | | | | | -2.32E-05 | | | | |
| 574 | 4752.263 | | | | | -1.67E-05 | | | | | | | -1.69E-05 | | |
| 575 | 4753.373 | | | | | | 1.52E-06 | | | | | | | 2.50E-06 | |
| 576 | 4786.35 | -3.59E-05 | | | | | | | -3.72E-05 | | | | | | |
| 577 | 4786.984 | | | | | | | | | | -3.63E-05 | | | | |
| 578 | 4803.155 | | -1.50E-05 | | | | | | | | | | | | |
| 579 | 4819.101 | | | | -1.78E-05 | | | | | | | -2.04E-05 | | | |
| 580 | 4825.641 | | | | | | 1.58E-06 | | | | | | | 2.62E-06 | |
| 581 | 4829.208 | | | | | | | | -1.91E-05 | | | | | | -2.21E-05 |
| 582 | 4831.342 | | | -1.98E-05 | | | | | | | -2.30E-05 | | | | |
| 583 | 4832.421 | | | | | -1.66E-05 | | | | | | | -1.67E-05 | | |
| 584 | 4865.633 | -3.58E-05 | | | | | | | -3.66E-05 | | | | | | |
| 585 | 4877.778 | | | | | | 1.64E-06 | | | | | | | 2.77E-06 | |
| 586 | 4878.49 | | -1.42E-05 | | | | | | | | | | | | |
| 587 | 4890.612 | | | | -1.74E-05 | | | | | | | -2.02E-05 | | | |
| 588 | 4898.988 | | | | | | | | | -3.53E-05 | | | | | |
| 589 | 4900.819 | | | | | | | | -1.87E-05 | | | | | | -2.18E-05 |
| 590 | 4903.08 | | | -1.95E-05 | | | | | | | -2.26E-05 | | | | |
| 591 | 4903.415 | | | | | -1.62E-05 | | | | | | | -1.64E-05 | | |
| 592 | 4924.778 | | | | | | 1.81E-06 | | | | | | | 2.90E-06 | |
| 593 | 4933.846 | -3.45E-05 | | | | | | | -3.55E-05 | | | | | | |
| 594 | 4947.974 | | -1.21E-05 | | | | | | | | | | | | |
| 595 | 4958.299 | | | | -1.71E-05 | | | | | | | -1.98E-05 | | | |
| 596 | 4969.322 | | | | | | | | -1.82E-05 | | | | | | -2.10E-05 |
| 597 | 4969.801 | | | | | -1.60E-05 | | | | | | | -1.61E-05 | | |
| 598 | 4970.072 | | | -1.88E-05 | | | | | | | -2.17E-05 | | | | |
| 599 | 4977.33 | | | | | | | | | -3.54E-05 | | | | | |
| 600 | 4980.422 | | | | | | 1.82E-06 | | | | | | | 2.92E-06 | |
| 601 | 5020.617 | -3.39E-05 | | | | | | | -3.50E-05 | | | | | | |
| 602 | 5034.884 | | -1.09E-05 | | | | | | | | | | | | |
| 603 | 5044.723 | | | | -1.69E-05 | | | | | | | -1.96E-05 | | | |
| 604 | 5049.423 | | | | | | 1.85E-06 | | | | | | | 2.96E-06 | |
| 605 | 5060.625 | | | | | | | | -1.79E-05 | | | | | | -2.06E-05 |
| 606 | 5060.955 | | | | | -1.58E-05 | | | | | | | -1.58E-05 | | |
| 607 | 5062.414 | | | -1.86E-05 | | | | | | | -2.12E-05 | | | | |
| 608 | 5084.876 | | | | | | | | | | -4.06E-05 | | | | |
| 609 | 5106.548 | -3.38E-05 | | | | | | | -3.46E-05 | | | | | | |
| 610 | 5123.209 | | -1.06E-05 | | | | | | | | | | | | |
| 611 | 5123.439 | | | | | | 1.83E-06 | | | | | | | 2.86E-06 | |
| 612 | 5133.146 | | | | -1.63E-05 | | | | | | | -1.90E-05 | | | |

| | A | B | C | D | E | F | G | H | I | J | K | L | M | N | O |
|---|---|---|---|---|---|---|---|---|---|---|---|---|---|---|---|
| 613 | 5147.389 | | | | | | | -1.76E-05 | | | | | | | -2.04E-05 |
| 614 | 5147.403 | | | | | -1.56E-05 | | | | | | | -1.55E-05 | | |
| 615 | 5148.966 | | | -1.86E-05 | | | | | | | -2.09E-05 | | | | |
| 616 | 5174.975 | | | | | | 1.97E-06 | | | | | | | 2.99E-06 | |
| 617 | 5175.513 | | | | | | | | | -4.51E-05 | | | | | |
| 618 | 5189.657 | -3.29E-05 | | | | | | | -3.42E-05 | | | | | | |
| 619 | 5210.285 | | -9.68E-06 | | | | | | | | | | | | |
| 620 | 5219.256 | | | | -1.57E-05 | | | | | | | -1.82E-05 | | | |
| 621 | 5227.825 | | | | | -1.51E-05 | | | | | | | -1.51E-05 | | |
| 622 | 5229.954 | | | | | | | | -1.71E-05 | | | | | | -1.93E-05 |
| 623 | 5231.986 | | | -1.81E-05 | | | | | | | -2.02E-05 | | | | |
| 624 | 5247.001 | | | | | | 2.01E-06 | | | | | | | 3.04E-06 | |
| 625 | 5283.878 | -3.27E-05 | | | | | | | -3.42E-05 | | | | | | |
| 626 | 5299.94 | | -6.88E-06 | | | | | | | | | | | | |
| 627 | 5309.289 | | | | -1.57E-05 | | | | | | | -1.82E-05 | | | |
| 628 | 5321.957 | | | | | | | | | | -4.59E-05 | | | | |
| 629 | 5325.341 | | | | | -1.49E-05 | | | | | | | -1.50E-05 | | |
| 630 | 5325.345 | | | | | | | -1.67E-05 | | | | | | | -1.94E-05 |
| 631 | 5327.425 | | | -1.80E-05 | | | | | | | -2.01E-05 | | | | |
| 632 | 5337.035 | | | | | | 2.03E-06 | | | | | | | 2.98E-06 | |
| 633 | 5380.067 | -3.27E-05 | | | | | | | -3.35E-05 | | | | | | |
| 634 | 5397.685 | | -8.55E-06 | | | | | | | | | | | | |
| 635 | 5405.434 | | | | -1.59E-05 | | | | | | | -1.84E-05 | | | |
| 636 | 5410.843 | | | | | | | | | | -4.81E-05 | | | | |
| 637 | 5412.877 | | | | | -1.48E-05 | | | | | | | -1.50E-05 | | |
| 638 | 5417.653 | | | | | | | | -1.69E-05 | | | | | | -1.98E-05 |
| 639 | 5420.287 | | | -1.82E-05 | | | 1.99E-06 | | | | | -2.06E-05 | | | 2.91E-06 | |
| 640 | 5462.95 | -3.25E-05 | | | | | | | -3.29E-05 | | | | | | |
| 641 | 5475.502 | | | | | | 2.16E-06 | | | | | | | 3.14E-06 | |
| 642 | 5478.875 | | -9.10E-06 | | | | | | | | | | | | |
| 643 | 5489.049 | | | | -1.57E-05 | | | | | | | -1.83E-05 | | | |
| 644 | 5499.807 | | | | | -1.44E-05 | | | | | | | -1.45E-05 | | |
| 645 | 5500.736 | | | | | | | -1.67E-05 | | | | | | | -1.97E-05 |
| 646 | 5504.384 | | | -1.78E-05 | | | | | | | -2.05E-05 | | | | |
| 647 | 5555.858 | | | | | | 2.13E-06 | | | | | | | 3.07E-06 | |
| 648 | 5564.523 | | | | | | | | | -4.78E-05 | | | | | |
| 649 | 5564.589 | -3.16E-05 | | | | | | | -3.24E-05 | | | | | | |
| 650 | 5583.713 | | -9.18E-06 | | | | | | | | | | | | |
| 651 | 5592.803 | | | | -1.55E-05 | | | | | | | -1.78E-05 | | | |
| 652 | 5597.015 | | | | | -1.41E-05 | | | | | | | -1.42E-05 | | |
| 653 | 5603.962 | | | | | | | -1.60E-05 | | | | | | | -1.90E-05 |
| 654 | 5607.092 | | | -1.71E-05 | | | | | | | -2.01E-05 | | | | |
| 655 | 5642.615 | | | | | | 2.22E-06 | | | | | | | 3.20E-06 | |
| 656 | 5671.302 | -3.14E-05 | | | | | | | -3.17E-05 | | | | | | |
| 657 | 5685.635 | | -9.59E-06 | | | | | | | | | | | | |
| 658 | 5697.92 | | | | -1.57E-05 | | | | | | | -1.79E-05 | | | |
| 659 | 5707.361 | | | | | -1.42E-05 | | | | | | | -1.42E-05 | | |
| 660 | 5713.655 | | | | | | | -1.64E-05 | | | | | | | -1.94E-05 |
| 661 | 5714.513 | | | -1.76E-05 | | | | | | | -2.02E-05 | | | | |
| 662 | 5716.028 | | | | | | | | | -4.84E-05 | | | | | |
| 663 | 5730.869 | | | | | | 2.22E-06 | | | | | | | 3.23E-06 | |
| 664 | 5777.103 | -3.08E-05 | | | | | | | -3.15E-05 | | | | | | |
| 665 | 5793.301 | | -9.63E-06 | | | | | | | | | | | | |
| 666 | 5801.367 | | | | -1.53E-05 | | | | | | | -1.74E-05 | | | |
| 667 | 5813.795 | | | | | -1.36E-05 | | | | | | | -1.38E-05 | | |
| 668 | 5822.43 | | | | | | | -1.62E-05 | | | | | | | -1.90E-05 |
| 669 | 5825.747 | | | -1.71E-05 | | | | | | | -2.00E-05 | | | | |
| 670 | 5829.61 | | | | | | 2.24E-06 | | | | | | | 3.21E-06 | |
| 671 | 5848.783 | | | | | | | | | -4.71E-05 | | | | | |
| 672 | 5885.66 | -3.08E-05 | | | | | | | -3.09E-05 | | | | | | |
| 673 | 5900.515 | | -9.58E-06 | | | | | | | | | | | | |
| 674 | 5906.884 | | | | -1.49E-05 | | | | | | | -1.71E-05 | | | |
| 675 | 5915.629 | | | | | -1.34E-05 | | | | | | | -1.36E-05 | | |
| 676 | 5919.917 | | | | | | 2.35E-06 | | | | | | | 3.40E-06 | |
| 677 | 5925.612 | | | | | | | -1.60E-05 | | | | | | | -1.87E-05 |
| 678 | 5926.765 | | | -1.70E-05 | | | | | | | -1.97E-05 | | | | |
| 679 | 5992.925 | -2.99E-05 | | | | | | | -3.02E-05 | | | | | | |
| 680 | 6005.878 | | | | | | | | | -4.66E-05 | | | | | |
| 681 | 6011.555 | | -8.83E-06 | | | | | | | | | | | | |
| 682 | 6017.807 | | | | | | 2.38E-06 | | | | | | | 3.44E-06 | |
| 683 | 6019.711 | | | | -1.47E-05 | | | | | | | -1.69E-05 | | | |
| 684 | 6022.236 | | | | | -1.34E-05 | | | | | | | -1.37E-05 | | |
| 685 | 6034.861 | | | | | | | -1.57E-05 | | | | | | | -1.82E-05 |
| 686 | 6038.58 | | | -1.68E-05 | | | | | | | -1.96E-05 | | | | |
| 687 | 6107.193 | -3.02E-05 | | | | | | | -3.02E-05 | | | | | | |
| 688 | 6120.266 | | | | | | 2.34E-06 | | | | | | | 3.38E-06 | |
| 689 | 6122.307 | | -9.80E-06 | | | | | | | | | | | | |
| 690 | 6132.987 | | | | -1.48E-05 | | | | | | | -1.73E-05 | | | |
| 691 | 6139.119 | | | | | -1.35E-05 | | | | | | | -1.38E-05 | | |
| 692 | 6145.884 | | | | | | | -1.60E-05 | | | | | | | -1.87E-05 |
| 693 | 6148.031 | | | -1.70E-05 | | | | | | | -2.01E-05 | | | | |
| 694 | 6158.33 | | | | | | | | | -4.60E-05 | | | | | |
| 695 | 6218.028 | -2.96E-05 | | | | | | | -2.91E-05 | | | | | | |
| 696 | 6218.548 | | | | | | 2.34E-06 | | | | | | | 3.37E-06 | |
| 697 | 6240.472 | | -9.74E-06 | | | | | | | | | | | | |
| 698 | 6249.876 | | | | -1.47E-05 | | | | | | | -1.72E-05 | | | |
| 699 | 6254.219 | | | | | -1.35E-05 | | | | | | | -1.37E-05 | | |
| 700 | 6268.239 | | | | | | | -1.61E-05 | | | | | | | -1.88E-05 |
| 701 | 6271.899 | | | -1.70E-05 | | | | | | | -2.01E-05 | | | | |
| 702 | 6332.827 | | | | | | 2.42E-06 | | | | | | | 3.35E-06 | |
| 703 | 6333.664 | | | | | | | | | -4.47E-05 | | | | | |
| 704 | 6337.472 | -3.00E-05 | | | | | | | -2.97E-05 | | | | | | |
| 705 | 6352.259 | | -9.81E-06 | | | | | | | | | | | | |
| 706 | 6360.004 | | | | -1.47E-05 | | | | | | | -1.73E-05 | | | |
| 707 | 6366.14 | | | | | -1.36E-05 | | | | | | | -1.38E-05 | | |
| 708 | 6378.411 | | | | | | | -1.59E-05 | | | | | | | -1.87E-05 |
| 709 | 6382.066 | | | -1.69E-05 | | | | | | | -2.01E-05 | | | | |
| 710 | 6425.012 | | | | | | 2.49E-06 | | | | | | | 3.31E-06 | |
| 711 | 6452.584 | -2.94E-05 | | | | | | | -2.87E-05 | | | | | | |
| 712 | 6473.052 | | -9.52E-06 | | | | | | | | | | | | |
| 713 | 6486.615 | | | | -1.51E-05 | | | | | | | -1.75E-05 | | | |
| 714 | 6490.249 | | | | | -1.37E-05 | | | | | | | -1.38E-05 | | |

| | A | B | C | D | E | F | G | H | I | J | K | L | M | N | O |
|---|---|---|---|---|---|---|---|---|---|---|---|---|---|---|---|
| 715 | 6506.464 | | | | | | | -1.62E-05 | | | | | | | -1.90E-05 |
| 716 | 6507.802 | | -1.71E-05 | | | | | | | | -2.04E-05 | | | | |
| 717 | 6517.099 | | | | | | | | | -4.40E-05 | | | | | |
| 718 | 6529.396 | | | | | 2.41E-06 | | | | | | | | 3.21E-06 | |
| 719 | 6571.355 | -2.89E-05 | | | | | | | -2.84E-05 | | | | | | |
| 720 | 6593.441 | | -9.77E-06 | | | | | | | | | | | | |
| 721 | 6606.497 | | | | -1.51E-05 | | | | | | | -1.75E-05 | | | |
| 722 | 6607.026 | | | | | -1.37E-05 | | | | | | | -1.39E-05 | | |
| 723 | 6624.824 | | | | | | | -1.62E-05 | | | | | | | -1.90E-05 |
| 724 | 6626.952 | | | -1.71E-05 | | | | | | | -2.04E-05 | | | | |
| 725 | 6630.854 | | | | | | 2.46E-06 | | | | | | | 3.29E-06 | |
| 726 | 6693.465 | | | | | | | | | | -4.32E-05 | | | | |
| 727 | 6708.399 | -2.90E-05 | | | | | | | -2.83E-05 | | | | | | |
| 728 | 6729.474 | | -9.61E-06 | | | | | | | | | | | | |
| 729 | 6742.084 | | | | | -1.38E-05 | | | | | | | -1.40E-05 | | |
| 730 | 6742.213 | | | | -1.51E-05 | | | | | | | -1.76E-05 | | | |
| 731 | 6757.198 | | | | | | 2.42E-06 | | | | | | | 3.19E-06 | |
| 732 | 6758.272 | | | | | | | | -1.61E-05 | | | | | | | -1.89E-05 |
| 733 | 6760.426 | | | -1.71E-05 | | | | | | | -2.03E-05 | | | | |
| 734 | 6848.417 | -2.84E-05 | | | | | | | -2.78E-05 | | | | | | |
| 735 | 6859.761 | | -8.70E-06 | | | | | | | | | | | | |
| 736 | 6866.637 | | | | | | | | | -3.97E-05 | | | | | |
| 737 | 6868.566 | | | | | -1.39E-05 | 2.37E-06 | | | | | | | -1.41E-05 | 3.13E-06 | |
| 738 | 6871.86 | | | | -1.52E-05 | | | | | | | -1.77E-05 | | | | |
| 739 | 6893.442 | | | | | | | | -1.62E-05 | | | | | | | -1.90E-05 |
| 740 | 6896.551 | | | -1.73E-05 | | | | | | | -2.04E-05 | | | | | |
| 741 | 6988.048 | -2.72E-05 | | | | | | | -2.81E-05 | | | | | | | |
| 742 | 6999.035 | | -8.84E-06 | | | | | 2.25E-06 | | | | | | | 3.03E-06 | |
| 743 | 7005.92 | | | | | -1.41E-05 | | | | | | | -1.42E-05 | | | |
| 744 | 7010.747 | | | | -1.55E-05 | | | | | | | -1.78E-05 | | | | |
| 745 | 7027.185 | | | | | | | | -1.65E-05 | | | | | | | -1.92E-05 |
| 746 | 7030.703 | | | -1.74E-05 | | | | | | | -2.05E-05 | | | | | |
| 747 | 7064.473 | | | | | | | | | -3.91E-05 | | | | | | |
| 748 | 7105.198 | | | | | | 2.26E-06 | | | | | | | 2.96E-06 | | |
| 749 | 7128.762 | -2.69E-05 | | | | | | | -2.74E-05 | | | | | | | |
| 750 | 7140.272 | | -9.06E-06 | | | | | | | | | | | | | |
| 751 | 7148.722 | | | | | -1.41E-05 | | | | | | | -1.42E-05 | | | |
| 752 | 7153.315 | | | | -1.56E-05 | | | | | | | -1.77E-05 | | | | |
| 753 | 7179.053 | | | | | | | | -1.65E-05 | | | | | | | -1.92E-05 |
| 754 | 7183.299 | | | -1.75E-05 | | | | | | | -2.04E-05 | | | | | |
| 755 | 7241.16 | | | | | | 2.25E-06 | | | | | | | 2.97E-06 | | |
| 756 | 7282.176 | | | | | | | | | | -3.67E-05 | | | | | |
| 757 | 7287.181 | -2.68E-05 | | | | | | | -2.74E-05 | | | | | | | |
| 758 | 7295.387 | | -9.83E-06 | | | | | | | | | | | | | |
| 759 | 7304.289 | | | | | -1.42E-05 | | | | | | | -1.43E-05 | | | |
| 760 | 7308.163 | | | | -1.58E-05 | | | | | | | -1.78E-05 | | | | |
| 761 | 7319.56 | | | | | | | | -1.66E-05 | | | | | | | -1.95E-05 |
| 762 | 7324.027 | | | -1.75E-05 | | | | | | | -2.06E-05 | | | | | |
| 763 | 7337.201 | | | | | | 2.20E-06 | | | | | | | 2.85E-06 | | |
| 764 | 7422.313 | -2.61E-05 | | | | | | | -2.70E-05 | | | | | | | |
| 765 | 7424.662 | | -9.88E-06 | | | | | | | | | | | | | |
| 766 | 7435.192 | | | | | -1.42E-05 | | | | | | | -1.43E-05 | | | |
| 767 | 7443.157 | | | | -1.55E-05 | | | | | | | -1.76E-05 | | | | |
| 768 | 7463.784 | | | | | | 2.29E-06 | | | | | | | 2.86E-06 | | |
| 769 | 7464.023 | | | | | | | | -1.64E-05 | | | | | | | -1.93E-05 |
| 770 | 7470.589 | | | -1.73E-05 | | | | | | | -2.05E-05 | | | | | |
| 771 | 7503.601 | | | | | | | | | -3.53E-05 | | | | | | |
| 772 | 7566.723 | -2.52E-05 | | | | | | | -2.61E-05 | | | | | | | |
| 773 | 7568.346 | | -1.00E-05 | | | | | | | | | | | | | |
| 774 | 7574.219 | | | | | -1.42E-05 | | | | | | | -1.43E-05 | | | |
| 775 | 7581.885 | | | | | | 2.28E-06 | | | | | | | 2.84E-06 | | |
| 776 | 7583.451 | | | | -1.54E-05 | | | | | | | -1.75E-05 | | | | |
| 777 | 7604.468 | | | | | | | | -1.63E-05 | | | | | | | -1.92E-05 |
| 778 | 7608.904 | | | -1.72E-05 | | | | | | | -2.03E-05 | | | | | |
| 779 | 7707.45 | | | | | | 2.19E-06 | | | | | | | 2.71E-06 | | |
| 780 | 7720.402 | | -9.63E-06 | | | | | | | | | | | | | |
| 781 | 7723.487 | -2.48E-05 | | | | | | | -2.49E-05 | | | | | | | |
| 782 | 7727.676 | | | | | -1.40E-05 | | | | | | | -1.41E-05 | | | |
| 783 | 7739.77 | | | | -1.52E-05 | | | | | | | -1.71E-05 | | | | |
| 784 | 7747.187 | | | | | | | | | -3.06E-05 | | | | | | |
| 785 | 7759.805 | | | | | | | | -1.61E-05 | | | | | | | -1.88E-05 |
| 786 | 7763.323 | | | -1.69E-05 | | | | | | | -2.00E-05 | | | | | |
| 787 | 7831.321 | | | | | | 2.23E-06 | | | | | | | 2.65E-06 | | |
| 788 | 7874.433 | | -8.05E-06 | | | | | | | | | | | | | |
| 789 | 7876.981 | -2.35E-05 | | | | | | | -2.30E-05 | | | | | | | |
| 790 | 7881.325 | | | | | -1.38E-05 | | | | | | | -1.39E-05 | | | |
| 791 | 7891.499 | | | | -1.49E-05 | | | | | | | -1.68E-05 | | | | |
| 792 | 7920.895 | | | | | | | | -1.57E-05 | | | | | | | -1.83E-05 |
| 793 | 7924.344 | | | -1.65E-05 | | | | | | | -1.95E-05 | | | | | |
| 794 | 7944.511 | | | | | | 2.18E-06 | | | | | | | 2.53E-06 | | |
| 795 | 7953.447 | | | | | | | | | | -2.48E-05 | | | | | |
| 796 | 8021.143 | | -8.34E-06 | | | | | | | | | | | | | |
| 797 | 8024.528 | | | | | -1.35E-05 | | | | | | | -1.35E-05 | | | |
| 798 | 8026.594 | -2.34E-05 | | | | | | | -2.41E-05 | | | | | | | |
| 799 | 8044.126 | | | | -1.41E-05 | | 2.15E-06 | | | | | | -1.58E-05 | | 2.44E-06 | |
| 800 | 8066.939 | | | | | | | | -1.50E-05 | | | | | | | -1.74E-05 |
| 801 | 8070.147 | | | -1.56E-05 | | | | | | | -1.85E-05 | | | | | |
| 802 | 8151.241 | | | | | | 2.27E-06 | | | | | | | 2.58E-06 | | |
| 803 | 8163.861 | | | | | | | | | | -2.47E-05 | | | | | |
| 804 | 8177.566 | | -7.73E-06 | | | | | | | | | | | | | |
| 805 | 8178.364 | | | | | -1.34E-05 | | | | | | | -1.34E-05 | | | |
| 806 | 8193.091 | -2.24E-05 | | | | | | | -2.23E-05 | | | | | | | |
| 807 | 8205.38 | | | | -1.29E-05 | | | | | | | -1.36E-05 | | | | |
| 808 | 8226.392 | | | | | | | | -1.34E-05 | | | | | | | -1.54E-05 |
| 809 | 8230.637 | | | -1.41E-05 | | | | | | | -1.69E-05 | | | | | |
| 810 | 8271.287 | | | | | | 2.39E-06 | | | | | | | 2.80E-06 | | |
| 811 | 8340.939 | | -6.16E-06 | | | | | | | | | | | | | |
| 812 | 8345.053 | | | | | -1.28E-05 | | | | | | | -1.27E-05 | | | |
| 813 | 8351.512 | -2.16E-05 | | | | | | | -2.03E-05 | | | | | | | |
| 814 | 8363.525 | | | | -1.26E-05 | | | | | | | -1.41E-05 | | | | |
| 815 | 8398.73 | | | | | | | | -1.37E-05 | | | | | | | -1.58E-05 |
| 816 | 8406.336 | | | | | | | | | | -2.11E-05 | | | | | |

| | A | B | C | D | E | F | G | H | I | J | K | L | M | N | O |
|---|---|---|---|---|---|---|---|---|---|---|---|---|---|---|---|
| 817 | 8408.331 | | | -1.42E-05 | | | 2.44E-06 | | | | -1.69E-05 | | | 2.81E-06 | |
| 818 | 8507.681 | | -5.39E-06 | | | | | | | | | | | | |
| 819 | 8512.606 | | | | | -1.24E-05 | | | | | | | -1.24E-05 | | |
| 820 | 8518.922 | | | | | | 2.49E-06 | | | | | | | 2.74E-06 | |
| 821 | 8523.973 | -2.09E-05 | | | | | | | -1.94E-05 | | | | | | |
| 822 | 8542.828 | | | | -1.22E-05 | | | | | | | -1.36E-05 | | | |
| 823 | 8571.784 | | | | | | | -1.31E-05 | | | | | | | -1.54E-05 |
| 824 | 8579.613 | | | -1.33E-05 | | | | | | | -1.60E-05 | | | | |
| 825 | 8647.799 | | | | | | 2.46E-06 | | | | | | | 2.69E-06 | |
| 826 | 8660.105 | | | | | | | | | -1.98E-05 | | | | | |
| 827 | 8686.462 | | -4.54E-06 | | | | | | | | | | | | |
| 828 | 8701.666 | | | | | -1.15E-05 | | | | | | | -1.16E-05 | | |
| 829 | 8710.987 | -2.01E-05 | | | | | | | -1.94E-05 | | | | | | |
| 830 | 8725.354 | | | | -1.01E-05 | | | | | | | | -1.19E-05 | | |
| 831 | 8765.839 | | | | | | | | -1.13E-05 | | | | | | -1.29E-05 |
| 832 | 8768.931 | | | | -1.11E-05 | | | | | | | -1.32E-05 | | | |
| 833 | 8801.033 | | | | | | | 2.37E-06 | | | | | | | 2.59E-06 | |
| 834 | 8880.492 | | -4.04E-06 | | | | | | | | | | | | |
| 835 | 8886.937 | | | | | | -1.11E-05 | | | | | | | -1.11E-05 | | |
| 836 | 8900.296 | -1.99E-05 | | | | | | | | -1.81E-05 | | | | | | |
| 837 | 8907.659 | | | | | | | | | | -1.81E-05 | | | | | |
| 838 | 8915.997 | | | | | -9.42E-06 | | | | | | | | -1.14E-05 | | |
| 839 | 8950.165 | | | | | | | | -1.03E-05 | | | | | | | -1.18E-05 |
| 840 | 8952.681 | | | | -1.04E-05 | | | | | | | -1.23E-05 | | | | |
| 841 | 8961.594 | | | | | | | 2.42E-06 | | | | | | | 2.58E-06 | |
| 842 | 9068.611 | | -3.16E-06 | | | | | | | | | | | | |
| 843 | 9081.236 | | | | | | -1.08E-05 | | | | | | | -1.07E-05 | | |
| 844 | 9091.769 | -1.90E-05 | | | | | | | | -1.62E-05 | | | | | | |
| 845 | 9110.773 | | | | | -9.10E-06 | | | | | | | | -1.10E-05 | | |
| 846 | 9133.323 | | | | | | | 2.46E-06 | | | | | | | 2.64E-06 | |
| 847 | 9142.904 | | | | | | | | -9.97E-06 | | | | | | | -1.12E-05 |
| 848 | 9143.934 | | | | | | | | | | -1.61E-05 | | | | | |
| 849 | 9147.638 | | | | -1.02E-05 | | | | | | | | -1.17E-05 | | | |
| 850 | 9283.216 | | -2.68E-06 | | | | | | | | | | | | |
| 851 | 9301.822 | | | | | | -1.05E-05 | | | | | | | -1.02E-05 | | |
| 852 | 9313.474 | -1.84E-05 | | | | | | | | -1.60E-05 | | | | | | |
| 853 | 9320.099 | | | | | | | 2.46E-06 | | | | | | | 2.59E-06 | |
| 854 | 9324.464 | | | | | -9.16E-06 | | | | | | | | -1.09E-05 | | |
| 855 | 9353.571 | | | | | | | | -1.03E-05 | | | | | | | -1.17E-05 |
| 856 | 9357.644 | | | | -1.06E-05 | | | | | | | -1.20E-05 | | | | |
| 857 | 9425.589 | | | | | | | | | | -6.53E-06 | | | | | |
| 858 | 9490.477 | | -2.68E-06 | | | | | | | | | | | | |
| 859 | 9503.959 | | | | | | -1.06E-05 | | | | | | | -1.05E-05 | | |
| 860 | 9519.996 | | | | | | | 2.45E-06 | | | | | | | 2.48E-06 | |
| 861 | 9522.776 | -1.84E-05 | | | | | | | | -1.58E-05 | | | | | | |
| 862 | 9538.846 | | | | | -9.24E-06 | | | | | | | | -1.12E-05 | | |
| 863 | 9575.071 | | | | | | | | -1.03E-05 | | | | | | | -1.19E-05 |
| 864 | 9582.355 | | | | -1.08E-05 | | | | | | | -1.21E-05 | | | | |
| 865 | 9703.731 | | | | | | | | | | -2.42E-06 | | | | | |
| 866 | 9722.89 | | -2.23E-06 | | | | | | | | | | | | |
| 867 | 9741.49 | | | | | | | 2.56E-06 | | | | | | | 2.54E-06 | |
| 868 | 9748.489 | | | | | | -1.02E-05 | | | | | | | -9.95E-06 | | |
| 869 | 9767.177 | -1.79E-05 | | | | | | | | -1.54E-05 | | | | | | |
| 870 | 9795.754 | | | | | -9.03E-06 | | | | | | | | -1.08E-05 | | |
| 871 | 9850.681 | | | | | | | | -9.98E-06 | | | | | | | -1.14E-05 |
| 872 | 9856.069 | | | | -1.05E-05 | | | | | | | -1.15E-05 | | | | |
| 873 | 9983.656 | | -2.74E-06 | | | | | | | | | | | | |
| 874 | 9994.72 | | | | | | | | | | 1.25E-05 | | | | | |
| 875 | 9996.202 | | | | | | | 2.50E-06 | | | | | | | 2.48E-06 | |
| 876 | 10014.25 | | | | | | -9.92E-06 | | | | | | | -9.61E-06 | | |
| 877 | 10023.73 | -1.72E-05 | | | | | | | | -1.39E-05 | | | | | | |
| 878 | 10048.74 | | | | | -8.80E-06 | | | | | | | | -1.06E-05 | | |
| 879 | 10090.19 | | | | | | | | -9.37E-06 | | | | | | | -1.06E-05 |
| 880 | 10096.82 | | | | -9.91E-06 | | | | | | | -1.08E-05 | | | | |
| 881 | 10230.58 | | -2.45E-06 | | | | | | | | | | | | |
| 882 | 10236.56 | | | | | | | 2.42E-06 | | | | | | | 2.35E-06 | |
| 883 | 10254.72 | | | | | | | | | | 1.82E-05 | | | | | |
| 884 | 10270.42 | | | | | | -9.47E-06 | | | | | | | -9.12E-06 | | |
| 885 | 10282.16 | -1.70E-05 | | | | | | | | -1.29E-05 | | | | | | |
| 886 | 10323.21 | | | | | -8.61E-06 | | | | | | | | -1.05E-05 | | |
| 887 | 10359.37 | | | | | | | | -9.44E-06 | | | | | | | -1.05E-05 |
| 888 | 10372.11 | | | | -9.77E-06 | | | | | | | -1.07E-05 | | | | |
| 889 | 10504.91 | | -2.14E-06 | | | | | | | | | | | | |
| 890 | 10511.36 | | | | | | | 2.41E-06 | | | | | | | 2.29E-06 | |
| 891 | 10559.87 | | | | | | -8.79E-06 | | | | | | | -8.21E-06 | | |
| 892 | 10571.04 | -1.75E-05 | | | | | | | | -1.25E-05 | | | | | | |
| 893 | 10575.59 | | | | | | | | | | 2.10E-05 | | | | | |
| 894 | 10603.94 | | | | | -8.30E-06 | | | | | | | | -1.03E-05 | | |
| 895 | 10642.08 | | | | | | | | -8.99E-06 | | | | | | | -1.03E-05 |
| 896 | 10649.85 | | | | -9.44E-06 | | | | | | | -1.04E-05 | | | | |
| 897 | 10789.56 | | -1.94E-06 | | | | | | | | | | | | |
| 898 | 10800.25 | | | | | | | 2.50E-06 | | | | | | | 2.34E-06 | |
| 899 | 10830.24 | | | | | | -8.42E-06 | | | | | | | -7.87E-06 | | |
| 900 | 10840.04 | -1.70E-05 | | | | | | | | -1.24E-05 | | | | | | |
| 901 | 10891.13 | | | | | -8.01E-06 | | | | | | | | -1.01E-05 | | |
| 902 | 10938.86 | | | | | | | | -8.86E-06 | | | | | | | -1.04E-05 |
| 903 | 10944.48 | | | | -9.24E-06 | | | | | | | -1.05E-05 | | | | |
| 904 | 10968.49 | | | | | | | | | | 1.67E-05 | | | | | |
| 905 | 11068.8 | | -1.69E-06 | | | | | | | | | | | | |
| 906 | 11109.58 | | | | | | | 2.53E-06 | | | | | | | 2.30E-06 | |
| 907 | 11120.98 | | | | | | -7.85E-06 | | | | | | | -7.00E-06 | | |
| 908 | 11127.13 | -1.73E-05 | | | | | | | | -1.28E-05 | | | | | | |
| 909 | 11164.87 | | | | | -7.18E-06 | | | | | | | | -9.37E-06 | | |
| 910 | 11215.8 | | | | | | | | -7.95E-06 | | | | | | | -9.38E-06 |
| 911 | 11227.37 | | | | -8.55E-06 | | | | | | | -9.68E-06 | | | | |
| 912 | 11369.19 | | -1.21E-06 | | | | | | | | | | | | |
| 913 | 11402.54 | | | | | | | | | | 1.67E-05 | | | | | |
| 914 | 11425.71 | | | | | | -7.52E-06 | | | | | | | -6.82E-06 | | |
| 915 | 11427.48 | | | | | | | 2.53E-06 | | | | | | | 2.34E-06 | |
| 916 | 11428.51 | -1.72E-05 | | | | | | | | -1.32E-05 | | | | | | |
| 917 | 11467.94 | | | | | -7.15E-06 | | | | | | | | -9.29E-06 | | |
| 918 | 11530.38 | | | | | | | | -8.10E-06 | | | | | | | -9.48E-06 |

| | A | B | C | D | E | F | G | H | I | J | K | L | M | N | O |
|---|---|---|---|---|---|---|---|---|---|---|---|---|---|---|---|
| 919 | 11540.5 | | | -8.84E-06 | | | | | | | -9.83E-06 | | | | |
| 920 | 11666.25 | | -1.47E-06 | | | | | | | | | | | | |
| 921 | 11735.6 | | | | | -7.13E-06 | | | | | | | -6.45E-06 | | |
| 922 | 11743.28 | -1.71E-05 | | | | | | | -1.44E-05 | | | | | | |
| 923 | 11748.06 | | | | | | 2.52E-06 | | | | | | | 2.36E-06 | |
| 924 | 11787.47 | | | | -6.79E-06 | | | | | | | | -8.83E-06 | | |
| 925 | 11834.55 | | | | | | | -7.62E-06 | | | | | | | -9.26E-06 |
| 926 | 11840.22 | | | | | | | | | 1.49E-05 | | | | | |
| 927 | 11845.31 | | | -8.33E-06 | | | | | | | -9.64E-06 | | | | |
| 928 | 11974.04 | | -1.47E-06 | | | | | | | | | | | | |
| 929 | 12054.31 | | | | | -6.95E-06 | | | | | | | -6.46E-06 | | |
| 930 | 12054.87 | -1.68E-05 | | | | | | | -1.35E-05 | | | | | | |
| 931 | 12125.35 | | | | | | 2.55E-06 | | | | | | | 2.37E-06 | |
| 932 | 12125.48 | | | | -6.51E-06 | | | | | | | | -8.59E-06 | | |
| 933 | 12172.26 | | | | | | | -7.57E-06 | | | | | | | -9.20E-06 |
| 934 | 12181.65 | | | -8.17E-06 | | | | | | | -9.71E-06 | | | | |
| 935 | 12315.98 | | -1.96E-06 | | | | | | | | | | | | |
| 936 | 12371.57 | | | | | | | | | 1.30E-05 | | | | | |
| 937 | 12396.18 | | | | | -7.07E-06 | | | | | | | -6.54E-06 | | |
| 938 | 12399.82 | -1.66E-05 | | | | | | | -1.32E-05 | | | | | | |
| 939 | 12446.06 | | | | -6.81E-06 | | | | | | | | -8.91E-06 | | |
| 940 | 12507.44 | | | | | | | -7.70E-06 | | | | | | | -9.35E-06 |
| 941 | 12511.94 | | | | | | 2.64E-06 | | | | | | | 2.43E-06 | |
| 942 | 12520.77 | | | -8.20E-06 | | | | | | | -9.82E-06 | | | | |
| 943 | 12638.05 | | -2.34E-06 | | | | | | | | | | | | |
| 944 | 12723.59 | -1.65E-05 | | | | -7.17E-06 | | | -1.37E-05 | | | | -6.67E-06 | | |
| 945 | 12795.17 | | | | -6.84E-06 | | | | | | | | -9.08E-06 | | |
| 946 | 12838.14 | | | | | | | -7.66E-06 | | | | | | | -9.37E-06 |
| 947 | 12841.81 | | | -8.06E-06 | | | | | | | -9.72E-06 | | | | |
| 948 | 12880.49 | | | | | | | | | 1.59E-05 | | | | | |
| 949 | 12880.65 | | | | | | 2.71E-06 | | | | | | | 2.49E-06 | |
| 950 | 12981.63 | | -3.32E-06 | | | | | | | | | | | | |
| 951 | 13103.44 | | | | | -7.09E-06 | | | | | | | -6.94E-06 | | |
| 952 | 13109.34 | -1.65E-05 | | | | | | | -1.47E-05 | | | | | | |
| 953 | 13175.77 | | | | -6.67E-06 | | | | | | | | -9.01E-06 | | |
| 954 | 13239.89 | | | | | | | -7.42E-06 | | | | | | | -9.17E-06 |
| 955 | 13249.4 | | | -7.94E-06 | | | | | | | -9.76E-06 | | | | |
| 956 | 13304.35 | | | | | | 2.80E-06 | | | | | | | 2.52E-06 | |
| 957 | 13346.94 | | -2.40E-06 | | | | | | | | | | | | |
| 958 | 13473.03 | | | | | -7.11E-06 | | | | | | | -6.78E-06 | | |
| 959 | 13484.79 | -1.66E-05 | | | | | | | -1.44E-05 | | | | | | |
| 960 | 13497.88 | | | | | | | | | 1.69E-05 | | | | | |
| 961 | 13550.15 | | | | -6.63E-06 | | | | | | | | -8.91E-06 | | |
| 962 | 13601.06 | | | | | | | | -6.99E-06 | | | | | | | -8.69E-06 |
| 963 | 13612.12 | | | -7.52E-06 | | | | | | | -9.26E-06 | | | | |
| 964 | 13705.07 | | | | | | | 2.86E-06 | | | | | | | 2.54E-06 | |
| 965 | 13742.76 | | -2.95E-06 | | | | | | | | | | | | |
| 966 | 13880.58 | | | | | -6.88E-06 | | | | | | | -6.34E-06 | | |
| 967 | 13891.78 | -1.78E-05 | | | | | | | -1.38E-05 | | | | | | |
| 968 | 13972.19 | | | | -6.47E-06 | | | | | | | | -8.68E-06 | | |
| 969 | 14041.84 | | | | | | | -6.82E-06 | | | | | | | -8.43E-06 |
| 970 | 14048.39 | | | -7.43E-06 | | | | | | | -9.05E-06 | | | | |
| 971 | 14144.96 | | | | | | | | | 2.05E-05 | | | | | |
| 972 | 14152.41 | | | | | | 2.63E-06 | | | | | | | 2.30E-06 | |
| 973 | 14153.14 | | -2.97E-06 | | | | | | | | | | | | |
| 974 | 14279.54 | | | | | -7.55E-06 | | | | | | | -7.15E-06 | | |
| 975 | 14296.46 | -1.78E-05 | | | | | | | -1.43E-05 | | | | | | |
| 976 | 14362.88 | | | | -7.00E-06 | | | | | | | | -9.21E-06 | | |
| 977 | 14439.08 | | | | | | | -7.02E-06 | | | | | | | -8.36E-06 |
| 978 | 14449.82 | | | -7.65E-06 | | | | | | | -9.05E-06 | | | | |
| 979 | 14533.87 | | -3.39E-06 | | | | | | | | | | | | |
| 980 | 14557.24 | | | | | | | 2.56E-06 | | | | | | | 2.16E-06 | |
| 981 | 14679.13 | | | | | -7.84E-06 | | | | | | | -7.44E-06 | | |
| 982 | 14696.9 | -1.79E-05 | | | | | | | -1.46E-05 | | | | | | |
| 983 | 14771.99 | | | | -7.50E-06 | | | | | | | | -9.52E-06 | | |
| 984 | 14812.67 | | | | | | | | | | 2.30E-05 | | | | | |
| 985 | 14822.47 | | | | | | | | -7.69E-06 | | | | | | | -9.14E-06 |
| 986 | 14831.78 | | | -8.29E-06 | | | | | | | -9.71E-06 | | | | |
| 987 | 14896.21 | | | | | | | 2.44E-06 | | | | | | | 1.90E-06 | |
| 988 | 14921.4 | | -3.76E-06 | | | | | | | | | | | | |
| 989 | 15027.75 | | | | | -8.09E-06 | | | | | | | -7.78E-06 | | |
| 990 | 15041.84 | -1.83E-05 | | | | | | | -1.43E-05 | | | | | | |
| 991 | 15103.13 | | | | -7.90E-06 | | | | | | | | -9.82E-06 | | |
| 992 | 15152.54 | | | | | | | | -7.95E-06 | | | | | | | -9.51E-06 |
| 993 | 15165.97 | | | -8.57E-06 | | | | | | | -1.01E-05 | | | | |
| 994 | 15244.73 | | | | | | | 2.38E-06 | | | | | | | 1.83E-06 | |
| 995 | 15296.07 | | -4.44E-06 | | | | | | | | | | | | |
| 996 | 15419.95 | | | | | | -8.74E-06 | | | | | | | -8.36E-06 | | |
| 997 | 15457.9 | -1.91E-05 | | | | | | | -1.44E-05 | | | | | | |
| 998 | 15483.98 | | | | | | | | | | 2.50E-05 | | | | | |
| 999 | 15492.89 | | | | | -8.45E-06 | | | | | | | -1.02E-05 | | | |
| 1000 | 15532.61 | | | | | | | | -8.19E-06 | | | | | | | -9.51E-06 |
| 1001 | 15537.06 | | | | -8.81E-06 | | | | | | | -1.01E-05 | | | | |
| 1002 | 15590.74 | | | | | | | 2.28E-06 | | | | | | | 1.70E-06 | |
| 1003 | 15653.4 | | -4.11E-06 | | | | | | | | | | | | |
| 1004 | 15783.52 | | | | | | -9.08E-06 | | | | | | | -8.69E-06 | | |
| 1005 | 15813.24 | -1.96E-05 | | | | | | | -1.39E-05 | | | | | | |
| 1006 | 15863.65 | | | | | -8.73E-06 | | | | | | | -1.05E-05 | | | |
| 1007 | 15925.89 | | | | | | | | -8.05E-06 | | | | | | | -9.42E-06 |
| 1008 | 15930.08 | | | -8.71E-06 | | | | | | | -1.00E-05 | | | | |
| 1009 | 15931.06 | | | | | | | 2.17E-06 | | | | | | | 1.64E-06 | |
| 1010 | 16006.48 | | -5.37E-06 | | | | | | | | | | | | |
| 1011 | 16095.39 | | | | | | | | | | 2.65E-05 | | | | | |
| 1012 | 16118.45 | | | | | -9.10E-06 | | | | | | | -8.88E-06 | | | |
| 1013 | 16156.48 | -2.08E-05 | | | | | | | -1.42E-05 | | | | | | |
| 1014 | 16205.5 | | | | -8.98E-06 | | | | | | | | -1.09E-05 | | | |
| 1015 | 16259.51 | | | | | | | | -8.24E-06 | | | | | | | -9.76E-06 |
| 1016 | 16269.48 | | | -9.00E-06 | | | | | | | -1.02E-05 | | | | |
| 1017 | 16323.46 | | | | | | | 2.18E-06 | | | | | | | 1.66E-06 | |
| 1018 | 16350.96 | | -5.37E-06 | | | | | | | | | | | | |
| 1019 | 16445.96 | | | | | -9.28E-06 | | | | | | | -9.17E-06 | | | |
| 1020 | 16490.78 | -2.15E-05 | | | | | | | -1.39E-05 | | | | | | |

| | A | B | C | D | E | F | G | H | I | J | K | L | M | N | O |
|---|---|---|---|---|---|---|---|---|---|---|---|---|---|---|---|
| 1021 | 16520.98 | | | | -9.40E-06 | | | | | 2.87E-05 | | -1.15E-05 | | | |
| 1022 | 16556.75 | | | | | | | | | 2.87E-05 | | | | | |
| 1023 | 16570.63 | | | | | | | -8.94E-06 | | | | | | | -1.04E-05 |
| 1024 | 16573.31 | | | -9.76E-06 | | | | | | | | -1.08E-05 | | | |
| 1025 | 16643.36 | | -5.39E-06 | | | | | | | | | | | | |
| 1026 | 16653.78 | | | | | | 2.21E-06 | | | | | | | 1.61E-06 | |
| 1027 | 16748.7 | | | | | -9.99E-06 | | | | | | | -9.81E-06 | | |
| 1028 | 16800.65 | -2.17E-05 | | | | | | | -1.30E-05 | | | | | | |
| 1029 | 16820.71 | | | | -9.89E-06 | | | | | | | -1.20E-05 | | | |
| 1030 | 16857.84 | | | | | | | -9.39E-06 | | | | | | | -1.07E-05 |
| 1031 | 16859.96 | | | -1.02E-05 | | | | | | | | -1.11E-05 | | | |
| 1032 | 16862.38 | | | | | | | | | 2.84E-05 | | | | | |
| 1033 | 16945.28 | | -5.82E-06 | | | | | | | | | | | | |
| 1034 | 17022.11 | | | | | | 2.07E-06 | | | | | | | 1.48E-06 | |
| 1035 | 17052.38 | | | | | -1.05E-05 | | | | | | -1.05E-05 | | | |
| 1036 | 17103.47 | -2.28E-05 | | | | | | | -1.33E-05 | | | | | | |
| 1037 | 17117.69 | | | | -1.06E-05 | | | | | | | -1.30E-05 | | | |
| 1038 | 17128.35 | | | | | | | | | 3.08E-05 | | | | | |
| 1039 | 17148.9 | | | | | | | -1.02E-05 | | | | | | | -1.18E-05 |
| 1040 | 17150.32 | | | -1.09E-05 | | | | | | | | -1.19E-05 | | | |
| 1041 | 17227.71 | | -6.77E-06 | | | | | | | | | | | | |
| 1042 | 17289.38 | | | | | | 1.95E-06 | | | | | | | 1.34E-06 | |
| 1043 | 17295.7 | | | | | -1.12E-05 | | | | | | | -1.11E-05 | | |
| 1044 | 17355.35 | -2.36E-05 | | | | | | | -1.45E-05 | | | | | | |
| 1045 | 17360.79 | | | | -1.15E-05 | | | | | | | -1.36E-05 | | | |
| 1046 | 17366.09 | | | | | | | | | | 2.97E-05 | | | | |
| 1047 | 17388.63 | | | | | | | -1.10E-05 | | | | | | | -1.25E-05 |
| 1048 | 17388.74 | | | -1.18E-05 | | | | | | | | -1.27E-05 | | | |
| 1049 | 17458.03 | | -7.44E-06 | | | | | | | | | | | | |
| 1050 | 17517.91 | | | | | | 1.80E-06 | | | | | | | 1.27E-06 | |
| 1051 | 17529.05 | | | | | -1.15E-05 | | | | | | -1.16E-05 | | | |
| 1052 | 17575 | -2.45E-05 | | | | | | | -1.59E-05 | | | | | | |
| 1053 | 17576.88 | | | | | | | | | | 3.02E-05 | | | | |
| 1054 | 17577.96 | | | | -1.18E-05 | | | | | | | -1.40E-05 | | | |
| 1055 | 17601.68 | | | -1.22E-05 | | | | -1.15E-05 | | | | -1.33E-05 | | | -1.29E-05 |
| 1056 | 17674.3 | | -7.58E-06 | | | | | | | | | | | | |
| 1057 | 17735.19 | | | | | -1.18E-05 | | | | | | | -1.19E-05 | | |
| 1058 | 17751.77 | | | | | | 1.79E-06 | | | | | | | 1.35E-06 | |
| 1059 | 17796.76 | | | | | | | | | | 3.16E-05 | | | | |
| 1060 | 17799.78 | -2.45E-05 | | | | | | | -1.68E-05 | | | | | | |
| 1061 | 17799.93 | | | | -1.26E-05 | | | | | | | -1.47E-05 | | | |
| 1062 | 17824.13 | | | | | | | | -1.22E-05 | | | | | | -1.36E-05 |
| 1063 | 17824.15 | | | -1.29E-05 | | | | | | | | -1.40E-05 | | | |
| 1064 | 17891.4 | | -8.63E-06 | | | | | | | | | | | | |
| 1065 | 17942.23 | | | | | -1.23E-05 | | | | | | | -1.23E-05 | | |
| 1066 | 17956.78 | | | | | | 1.82E-06 | | | | | | | 1.40E-06 | |
| 1067 | 17973.02 | | | | | | | | | | 3.69E-05 | | | | |
| 1068 | 18001.51 | | | | -1.29E-05 | | | | | | | -1.49E-05 | | | |
| 1069 | 18004 | -2.42E-05 | | | | | | | -1.40E-05 | | | | | | |
| 1070 | 18019.7 | | | | | | | -1.24E-05 | | | | | | | -1.37E-05 |
| 1071 | 18020.07 | | | -1.31E-05 | | | | | | | | -1.42E-05 | | | |
| 1072 | 18086.3 | | -8.79E-06 | | | | | | | | | | | | |
| 1073 | 18145.18 | | | | | | 1.72E-06 | | | | | | | 1.30E-06 | |
| 1074 | 18150.61 | | | | | -1.27E-05 | | | | | | | -1.28E-05 | | |
| 1075 | 18195.29 | | | | | | | | | | 3.58E-05 | | | | |
| 1076 | 18210.35 | | | | -1.34E-05 | | | | | | | -1.54E-05 | | | |
| 1077 | 18217.33 | -2.41E-05 | | | | | | | -1.17E-05 | | | | | | |
| 1078 | 18240.32 | | | | | | | | -1.29E-05 | | | | | | -1.44E-05 |
| 1079 | 18241 | | | -1.34E-05 | | | | | | | | -1.46E-05 | | | |
| 1080 | 18289.59 | | -8.12E-06 | | | | | | | | | | | | |
| 1081 | 18342.13 | | | | | | 1.74E-06 | | | | | | | 1.31E-06 | |
| 1082 | 18351.05 | | | | | -1.31E-05 | | | | | | | -1.32E-05 | | |
| 1083 | 18379.61 | | | | | | | | | | 3.85E-05 | | | | |
| 1084 | 18396.29 | | | | -1.39E-05 | | | | | | | -1.58E-05 | | | |
| 1085 | 18408.99 | -2.43E-05 | | | | | | | -2.23E-05 | | | | | | |
| 1086 | 18428.71 | | | -1.42E-05 | | | | -1.35E-05 | | | | -1.51E-05 | | | -1.50E-05 |
| 1087 | 18488.89 | | -8.35E-06 | | | | | | | | | | | | |
| 1088 | 18513.48 | | | | | | 1.71E-06 | | | | | | | 1.28E-06 | |
| 1089 | 18539.12 | | | | | -1.35E-05 | | | | | | | -1.36E-05 | | |
| 1090 | 18566.77 | | | | | | | | | | 3.92E-05 | | | | |
| 1091 | 18592.46 | | | | -1.38E-05 | | | | | | | -1.59E-05 | | | |
| 1092 | 18613.33 | -2.32E-05 | | | | | | | -2.29E-05 | | | | | | |
| 1093 | 18621.52 | | | -1.42E-05 | | | | -1.36E-05 | | | | -1.52E-05 | | | -1.49E-05 |
| 1094 | 18675.32 | | | | | | 1.74E-06 | | | | | | | 1.34E-06 | |
| 1095 | 18683.98 | | -8.66E-06 | | | | | | | | | | | | |
| 1096 | 18734.89 | | | | | -1.38E-05 | | | | | | | -1.40E-05 | | |
| 1097 | 18763.94 | | | | | | | | | | 3.81E-05 | | | | |
| 1098 | 18790.3 | | | | -1.42E-05 | | | | | | | -1.61E-05 | | | |
| 1099 | 18802.21 | -2.34E-05 | | | | | | | -2.55E-05 | | | | | | |
| 1100 | 18804.54 | | | -1.43E-05 | | | | | | | | -1.53E-05 | | | |
| 1101 | 18805.58 | | | | | | | -1.39E-05 | | | | | | | -1.51E-05 |
| 1102 | 18812.32 | | | | | | 1.73E-06 | | | | | | | 1.33E-06 | |
| 1103 | 18854.91 | | -9.27E-06 | | | | | | | | | | | | |
| 1104 | 18903 | | | | | -1.41E-05 | | | | | | | -1.44E-05 | | |
| 1105 | 18926.08 | | | | | | | | | | 3.81E-05 | | | | |
| 1106 | 18957.75 | | | | | | 1.57E-06 | | | | | | | 1.18E-06 | |
| 1107 | 18957.81 | | | | -1.47E-05 | | | | | | | -1.66E-05 | | | |
| 1108 | 18973.69 | -2.31E-05 | | | | | | | -2.70E-05 | | | | | | |
| 1109 | 18974.05 | | | -1.48E-05 | | | | -1.45E-05 | | | | -1.62E-05 | | | -1.58E-05 |
| 1110 | 19029.14 | | -9.53E-06 | | | | | | | | | | | | |
| 1111 | 19064.99 | | | | | -1.43E-05 | | | | | | | -1.47E-05 | | |
| 1112 | 19078.23 | | | | | | 1.50E-06 | | | | | | | 1.11E-06 | |
| 1113 | 19109.95 | | | | | | | | | | 3.29E-05 | | | | |
| 1114 | 19115.69 | | | | -1.49E-05 | | | | | | | -1.69E-05 | | | |
| 1115 | 19131.03 | | | -1.49E-05 | | | | | | | | -1.63E-05 | | | |
| 1116 | 19132 | | | | | | | -1.46E-05 | | | | | | | -1.61E-05 |
| 1117 | 19135.21 | -2.29E-05 | | | | | | | -2.91E-05 | | | | | | |
| 1118 | 19189.14 | | -9.45E-06 | | | | | | | | | | | | |
| 1119 | 19218.25 | | | | | | 1.47E-06 | | | | | | | 1.09E-06 | |
| 1120 | 19234.49 | | | | | -1.49E-05 | | | | | | | -1.52E-05 | | |
| 1121 | 19283.39 | | | | | | | | | | 3.26E-05 | | | | |
| 1122 | 19296.49 | | | | -1.54E-05 | | | | | | | -1.73E-05 | | | |

| | A | B | C | D | E | F | G | H | I | J | K | L | M | N | O |
|---|---|---|---|---|---|---|---|---|---|---|---|---|---|---|---|
| 1123 | 19315.26 | | | -1.54E-05 | | | | -1.50E-05 | | | -1.69E-05 | | | | -1.65E-05 |
| 1124 | 19324.66 | -2.22E-05 | | | | | | | -2.88E-05 | | | | | | |
| 1125 | 19352.89 | | | | | | 1.45E-06 | | | | | | | 1.02E-06 | |
| 1126 | 19380.3 | | -1.03E-05 | | | | | | | | | | | | |
| 1127 | 19413.57 | | | | | -1.51E-05 | | | | | | | -1.53E-05 | | |
| 1128 | 19461.7 | | | | | | | | | 3.23E-05 | | | | | |
| 1129 | 19470.54 | | | | -1.53E-05 | | | | | | | | -1.72E-05 | | |
| 1130 | 19495.48 | | | -1.54E-05 | | | | | | | -1.68E-05 | | | | |
| 1131 | 19495.77 | | | | | | | -1.49E-05 | | | | | | | -1.65E-05 |
| 1132 | 19497.02 | | | | | | 1.36E-06 | | | | | | | 9.40E-07 | |
| 1133 | 19505.71 | -2.11E-05 | | | | | | | -2.86E-05 | | | | | | |
| 1134 | 19545.28 | | -1.09E-05 | | | | | | | | | | | | |
| 1135 | 19580.47 | | | | | -1.57E-05 | | | | | | | -1.62E-05 | | |
| 1136 | 19591.45 | | | | | | 1.30E-06 | | | | | | | 7.95E-07 | |
| 1137 | 19621.47 | | | | | | | | | 2.92E-05 | | | | | |
| 1138 | 19625.2 | | | | -1.57E-05 | | | | | | | | -1.76E-05 | | |
| 1139 | 19646.05 | | | -1.57E-05 | | | | -1.52E-05 | | | -1.71E-05 | | | | -1.68E-05 |
| 1140 | 19659.82 | -2.04E-05 | | | | | | | -2.71E-05 | | | | | | |
| 1141 | 19714.12 | | -1.04E-05 | | | | | | | | | | | | |
| 1142 | 19719.78 | | | | | | 1.18E-06 | | | | | | | 6.77E-07 | |
| 1143 | 19753.04 | | | | | -1.56E-05 | | | | | | | -1.62E-05 | | |
| 1144 | 19791.69 | | | | | | | | | 2.82E-05 | | | | | |
| 1145 | 19805.03 | | | | -1.58E-05 | | | | | | | | -1.79E-05 | | |
| 1146 | 19835.02 | | | -1.59E-05 | | | | -1.55E-05 | | | -1.77E-05 | | | | -1.72E-05 |
| 1147 | 19847.94 | -2.02E-05 | | | | | | | -2.71E-05 | | | | | | |
| 1148 | 19857.3 | | | | | | 1.14E-06 | | | | | | | 6.96E-07 | |
| 1149 | 19885.86 | | -9.78E-06 | | | | | | | | | | | | |
| 1150 | 19928.83 | | | | | -1.58E-05 | | | | | | | -1.66E-05 | | |
| 1151 | 19956.74 | | | | | | | | | | 2.56E-05 | | | | |
| 1152 | 19968.42 | | | | | | 1.14E-06 | | | | | | | 7.16E-07 | |
| 1153 | 19978.52 | | | | -1.55E-05 | | | | | | | | -1.79E-05 | | |
| 1154 | 19995.84 | | | -1.58E-05 | | | | -1.52E-05 | | | -1.74E-05 | | | | -1.68E-05 |
| 1155 | 20011.07 | -1.96E-05 | | | | | | | -2.98E-05 | | | | | | |
| 1156 | 20042.06 | | -1.09E-05 | | | | | | | | | | | | |
| 1157 | 20068.9 | | | | | | | | | 2.21E-05 | | | | | |
| 1158 | 20070.97 | | | | | | 1.10E-06 | | | | | | | 6.83E-07 | |
| 1159 | 20080.21 | | | | | -1.60E-05 | | | | | | | -1.67E-05 | | |
| 1160 | 20134.47 | | | | -1.58E-05 | | | | | | | | -1.80E-05 | | |
| 1161 | 20163.84 | | | -1.60E-05 | | | | | | | -1.77E-05 | | | | |
| 1162 | 20166.19 | | | | | | | -1.54E-05 | | | | | | | -1.73E-05 |
| 1163 | 20180.41 | -1.92E-05 | | | | | | | -3.03E-05 | | | | | | |
| 1164 | 20193.34 | | | | | | 1.06E-06 | | | | | | | 6.11E-07 | |
| 1165 | 20211.53 | | -1.10E-05 | | | | | | | | | | | | |
| 1166 | 20240.38 | | | | | | | | | 2.09E-05 | | | | | |
| 1167 | 20245.73 | | | | | -1.57E-05 | | | | | | | -1.68E-05 | | |
| 1168 | 20294.38 | | | | -1.58E-05 | | | | | | | -1.84E-05 | | | |
| 1169 | 20311.85 | | | | | | 1.03E-06 | | | | | | | 5.85E-07 | |
| 1170 | 20315.77 | | | -1.60E-05 | | | | -1.55E-05 | | | -1.78E-05 | | | | -1.73E-05 |
| 1171 | 20336.52 | -1.86E-05 | | | | | | | -3.09E-05 | | | | | | |
| 1172 | 20363.08 | | -1.10E-05 | | | | | | | | | | | | |
| 1173 | 20389.41 | | | | | -1.60E-05 | | | | | | | -1.71E-05 | | |
| 1174 | 20396.31 | | | | | | | | | 1.83E-05 | | | | | |
| 1175 | 20402 | | | | | | 1.01E-06 | | | | | | | 3.89E-07 | |
| 1176 | 20442.1 | | | | -1.63E-05 | | | | | | | | -1.89E-05 | | |
| 1177 | 20463.02 | | | -1.65E-05 | | | | | | | -1.84E-05 | | | | |
| 1178 | 20463.6 | | | | | | | -1.60E-05 | | | | | | | -1.80E-05 |
| 1179 | 20492.33 | -1.82E-05 | | | | | | | -2.96E-05 | | | | | | |
| 1180 | 20516.13 | | | | | | 9.53E-07 | | | | | | | 3.09E-07 | |
| 1181 | 20517.16 | | -1.09E-05 | | | | | | | | | | | | |
| 1182 | 20552.29 | | | | | -1.59E-05 | | | | | | | -1.71E-05 | | |
| 1183 | 20563.11 | | | | | | | | | 1.79E-05 | | | | | |
| 1184 | 20603.73 | | | | -1.65E-05 | | | | | | | -1.90E-05 | | | |
| 1185 | 20614.32 | | | | | | 8.88E-07 | | | | | | | 1.52E-07 | |
| 1186 | 20618.96 | | | -1.67E-05 | | | | | | | -1.86E-05 | | | | |
| 1187 | 20619.05 | | | | | | | -1.62E-05 | | | | | | | -1.83E-05 |
| 1188 | 20639.68 | -1.75E-05 | | | | | | | -2.89E-05 | | | | | | |
| 1189 | 20672.41 | | -1.04E-05 | | | | | | | | | | | | |
| 1190 | 20684.15 | | | | | | | | | 1.79E-05 | | | | | |
| 1191 | 20695.88 | | | | | -1.61E-05 | | | | | | | -1.73E-05 | | |
| 1192 | 20723.05 | | | | | | 8.87E-07 | | | | | | | 2.08E-07 | |
| 1193 | 20747.75 | | | | -1.65E-05 | | | | | | | -1.90E-05 | | | |
| 1194 | 20765.87 | | | -1.70E-05 | | | | | | | -1.86E-05 | | | | |
| 1195 | 20766.48 | | | | | | | -1.65E-05 | | | | | | | -1.85E-05 |
| 1196 | 20794.31 | -1.66E-05 | | | | | | | -2.68E-05 | | | | | | |
| 1197 | 20819.38 | | -9.99E-06 | | | | | | | | | | | | |
| 1198 | 20840.37 | | | | | | | | | 1.71E-05 | | | | | |
| 1199 | 20840.64 | | | | | | 9.04E-07 | | | | | | | 2.05E-07 | |
| 1200 | 20857.81 | | | | | -1.61E-05 | | | | | | | -1.72E-05 | | |
| 1201 | 20913.78 | | | | -1.65E-05 | | | | | | | -1.89E-05 | | | |
| 1202 | 20938.53 | | | -1.68E-05 | | | | | | | -1.85E-05 | | | | |
| 1203 | 20938.71 | | | | | | | -1.64E-05 | | | | | | | -1.84E-05 |
| 1204 | 20964.49 | | | | | | 8.48E-07 | | | | | | | 1.31E-07 | |
| 1205 | 20968.31 | -1.58E-05 | | | | | | | -2.51E-05 | | | | | | |
| 1206 | 20988.14 | | -9.76E-06 | | | | | | | | | | | | |
| 1207 | 21010.02 | | | | | | | | | 1.41E-05 | | | | | |
| 1208 | 21019.62 | | | | | -1.59E-05 | | | | | | | -1.73E-05 | | |
| 1209 | 21079.96 | | | | -1.64E-05 | | | | | | | -1.86E-05 | | | |
| 1210 | 21081.75 | | | | | | 8.51E-07 | | | | | | | 1.32E-07 | |
| 1211 | 21099.52 | | | -1.66E-05 | | | | -1.62E-05 | | | -1.82E-05 | | | | -1.79E-05 |
| 1212 | 21127.74 | -1.52E-05 | | | | | | | -2.25E-05 | | | | | | |
| 1213 | 21136.62 | | | | | | | | | 1.41E-05 | | | | | |
| 1214 | 21148.57 | | -1.01E-05 | | | | | | | | | | | | |
| 1215 | 21180.21 | | | | | -1.59E-05 | | | | | | | -1.73E-05 | | |
| 1216 | 21183.52 | | | | | | 8.10E-07 | | | | | | | 3.31E-08 | |
| 1217 | 21236.01 | | | | -1.64E-05 | | | | | | | -1.85E-05 | | | |
| 1218 | 21258.02 | | | -1.66E-05 | | | | -1.61E-05 | | | -1.80E-05 | | | | -1.75E-05 |
| 1219 | 21289.73 | -1.53E-05 | | | | | | | -2.34E-05 | | | | | | |
| 1220 | 21299.5 | | | | | | 8.05E-07 | | | | | | | -2.88E-08 | |
| 1221 | 21311.48 | | -1.00E-05 | | | | | | | | | | | | |
| 1222 | 21312.18 | | | | | | | | | 1.52E-05 | | | | | |
| 1223 | 21340.18 | | | | | -1.58E-05 | | | | | | | -1.71E-05 | | |
| 1224 | 21404.3 | | | | -1.66E-05 | | | | | | | -1.85E-05 | | | |

| | A | B | C | D | E | F | G | H | I | J | K | L | M | N | O |
|---|---|---|---|---|---|---|---|---|---|---|---|---|---|---|---|
| 1225 | 21422.82 | | | | | | 7.85E-07 | | | | | | | -3.12E-08 | |
| 1226 | 21428.36 | | | -1.69E-05 | | | | -1.64E-05 | | | -1.85E-05 | | | | -1.80E-05 |
| 1227 | 21449.69 | -1.52E-05 | | | | | | | -2.67E-05 | | | | | | |
| 1228 | 21467.78 | | -9.98E-06 | | | | | | | | | | | | |
| 1229 | 21480.84 | | | | | | | | | 1.48E-05 | | | | | |
| 1230 | 21498.68 | | | | | -1.58E-05 | | | | | | | -1.72E-05 | | |
| 1231 | 21524.19 | | | | | | 7.08E-07 | | | | | | | -1.08E-07 | |
| 1232 | 21558.19 | | | | -1.62E-05 | | | | | | | | -1.82E-05 | | |
| 1233 | 21577.35 | | | -1.67E-05 | | | | -1.63E-05 | | | -1.80E-05 | | | | -1.78E-05 |
| 1234 | 21609.16 | -1.43E-05 | | | | | | | -2.81E-05 | | | | | | |
| 1235 | 21620.06 | | | | | | 5.21E-07 | | | | | | | -5.38E-07 | |
| 1236 | 21626.18 | | -1.13E-05 | | | | | | | | | | | | |
| 1237 | 21638.35 | | | | | | | | | 1.30E-05 | | | | | |
| 1238 | 21652.69 | | | | | -1.64E-05 | | | | | | | -1.78E-05 | | |
| 1239 | 21708.53 | | | | -1.68E-05 | | | | | | | -1.85E-05 | | | |
| 1240 | 21721.54 | | | | | | 5.13E-07 | | | | | | | -5.73E-07 | |
| 1241 | 21730.22 | | | -1.68E-05 | | | | | | | -1.83E-05 | | | | |
| 1242 | 21731.23 | | | | | | | -1.65E-05 | | | | | | | -1.80E-05 |
| 1243 | 21754.14 | -1.40E-05 | | | | | | | -2.30E-05 | | | | | | |
| 1244 | 21762.75 | | | | | | | | | 1.33E-05 | | | | | |
| 1245 | 21767.63 | | -1.21E-05 | | | | | | | | | | | | |
| 1246 | 21799.56 | | | | | -1.66E-05 | | | | | | | -1.80E-05 | | |
| 1247 | 21817.14 | | | | | | 4.45E-07 | | | | | | | -6.24E-07 | |
| 1248 | 21848.57 | | | | -1.69E-05 | | | | | | | -1.88E-05 | | | |
| 1249 | 21870.74 | | | -1.70E-05 | | | | | | | -1.86E-05 | | | | |
| 1250 | 21871.24 | | | | | | | | | 9.92E-06 | | | | | |
| 1251 | 21872.52 | | | | | | | -1.67E-05 | | | | | | | -1.82E-05 |
| 1252 | 21896.91 | -1.17E-05 | | | | | | | -2.31E-05 | | | | | | |
| 1253 | 21898.94 | | | | | | 4.42E-07 | | | | | | | -6.31E-07 | |
| 1254 | 21908.65 | | -1.12E-05 | | | | | | | | | | | | |
| 1255 | 21924.96 | | | | | -1.65E-05 | | | | | | | -1.79E-05 | | |
| 1256 | 21984.8 | | | | | | 4.60E-07 | | | | | | | -5.95E-07 | |
| 1257 | 21992.6 | | | | -1.64E-05 | | | | | | | -1.81E-05 | | | |
| 1258 | 22018.59 | | | -1.65E-05 | | | | | | | -1.78E-05 | | | | |
| 1259 | 22019.94 | | | | | | | -1.62E-05 | | | | | | | -1.73E-05 |
| 1260 | 22057.42 | | | | | | | | | 8.91E-06 | | | | | |
| 1261 | 22057.45 | -1.08E-05 | | | | | | | -2.26E-05 | | | | | | |
| 1262 | 22067.32 | | -1.12E-05 | | | | | | | | | | | | |
| 1263 | 22090.8 | | | | | -1.63E-05 | | | | | | | -1.77E-05 | | |
| 1264 | 22096.68 | | | | | | 4.27E-07 | | | | | | | -6.15E-07 | |
| 1265 | 22149.19 | | | | -1.62E-05 | | | | | | | -1.80E-05 | | | |
| 1266 | 22174.25 | | | -1.64E-05 | | | | | | | -1.74E-05 | | | | |
| 1267 | 22174.32 | | | | | | | -1.60E-05 | | | | | | | -1.70E-05 |
| 1268 | 22190.34 | | | | | | 3.02E-07 | | | | | | | -7.29E-07 | |
| 1269 | 22191.92 | | | | | | | | | 1.05E-05 | | | | | |
| 1270 | 22205.59 | -1.04E-05 | | | | | | | -2.08E-05 | | | | | | |
| 1271 | 22212.78 | | -1.15E-05 | | | | | | | | | | | | |
| 1272 | 22235.86 | | | | | -1.62E-05 | | | | | | | -1.75E-05 | | |
| 1273 | 22277.46 | | | | | | 2.35E-07 | | | | | | | -8.12E-07 | |
| 1274 | 22290.66 | | | | -1.58E-05 | | | | | | | -1.78E-05 | | | |
| 1275 | 22313.87 | | | -1.59E-05 | | | | | | | -1.64E-05 | | | | |
| 1276 | 22314.43 | | | | | | | -1.54E-05 | | | | | | | -1.65E-05 |
| 1277 | 22336.97 | -9.17E-06 | | | | | | | -2.29E-05 | | | | | | |
| 1278 | 22345.12 | | -1.13E-05 | | | | | | | | | | | | |
| 1279 | 22360.27 | | | | | | 2.57E-07 | | | | | | | -7.88E-07 | |
| 1280 | 22365.72 | | | | | | | | | 1.06E-05 | | | | | |
| 1281 | 22366.32 | | | | | -1.69E-05 | | | | | | | -1.83E-05 | | |
| 1282 | 22432.52 | | | | -1.58E-05 | | | | | | | -1.77E-05 | | | |
| 1283 | 22451.76 | | | | | | 1.68E-07 | | | | | | | -8.95E-07 | |
| 1284 | 22453.77 | | | -1.59E-05 | | | | -1.56E-05 | | | -1.70E-05 | | | | -1.66E-05 |
| 1285 | 22498.13 | -9.18E-06 | | | | | | | -2.35E-05 | | | | | | |
| 1286 | 22500.78 | | -1.07E-05 | | | | | | | | | | | | |
| 1287 | 22524.03 | | | | | -1.69E-05 | | | | | | | -1.84E-05 | | |
| 1288 | 22546.74 | | | | | | 1.19E-07 | | | | | | | -9.21E-07 | |
| 1289 | 22566.69 | | | | | | | | | 1.14E-05 | | | | | |
| 1290 | 22589.06 | | | | -1.44E-05 | | | | | | | -1.64E-05 | | | |
| 1291 | 22609.81 | | | -1.52E-05 | | | | | | | -1.63E-05 | | | | |
| 1292 | 22609.91 | | | | | | | -1.49E-05 | | | | | | | -1.58E-05 |
| 1293 | 22649.15 | | | | | | -8.49E-08 | | | | | | | -1.04E-06 | |
| 1294 | 22651.85 | -7.62E-06 | | | | | | | -2.15E-05 | | | | | | |
| 1295 | 22652.12 | | -1.01E-05 | | | | | | | | | | | | |
| 1296 | 22671.52 | | | | | -1.63E-05 | | | | | | | -1.77E-05 | | |
| 1297 | 22714.28 | | | | -1.41E-05 | | | | | | | -1.59E-05 | | | |
| 1298 | 22723.96 | | | | | | | | | 1.08E-05 | | | | | |
| 1299 | 22726.68 | | | | | | -8.44E-08 | | | | | | | -1.08E-06 | |
| 1300 | 22744.72 | | | -1.48E-05 | | | | -1.46E-05 | | | -1.57E-05 | | | | -1.54E-05 |
| 1301 | 22793.41 | | -9.86E-06 | | | | | | | | | | | | |
| 1302 | 22797.57 | -5.88E-06 | | | | | | | -2.15E-05 | | | | | | |
| 1303 | 22811.48 | | | | | -1.63E-05 | | | | | | | -1.76E-05 | | |
| 1304 | 22824.36 | | | | | | -9.58E-08 | | | | | | | -1.16E-06 | |
| 1305 | 22864.08 | | | | -1.38E-05 | | | | | | | -1.53E-05 | | | |
| 1306 | 22895.94 | | | -1.42E-05 | | | | -1.40E-05 | | | -1.50E-05 | | | | -1.48E-05 |
| 1307 | 22905.26 | | | | | | | | | 1.08E-05 | | | | | |
| 1308 | 22942.4 | | | | | | -1.28E-07 | | | | | | | -1.18E-06 | |
| 1309 | 22946.51 | | -8.81E-06 | | | | | | | | | | | | |
| 1310 | 22948.54 | -4.42E-06 | | | | | | | -2.00E-05 | | | | | | |
| 1311 | 22954.99 | | | | | -1.56E-05 | | | | | | | -1.62E-05 | | |
| 1312 | 23009.81 | | | | -1.35E-05 | | | | | | | -1.51E-05 | | | |
| 1313 | 23016.75 | | | | | | -2.10E-07 | | | | | | | -1.20E-06 | |
| 1314 | 23032.81 | | | -1.39E-05 | | | | -1.36E-05 | | | -1.49E-05 | | | | -1.46E-05 |
| 1315 | 23082.37 | | -7.93E-06 | | | | | | | | | | | | |
| 1316 | 23083.56 | | | | | | | | | 9.95E-06 | | | | | |
| 1317 | 23094.17 | -2.42E-06 | | | | | | | -1.47E-05 | | | | | | |
| 1318 | 23100.11 | | | | | -1.55E-05 | | | | | | | -1.62E-05 | | |
| 1319 | 23121.08 | | | | | | -2.76E-07 | | | | | | | -1.21E-06 | |
| 1320 | 23166.09 | | | | -1.31E-05 | | | | | | | -1.47E-05 | | | |
| 1321 | 23198.14 | | | -1.34E-05 | | | | -1.32E-05 | | | -1.45E-05 | | | | -1.41E-05 |
| 1322 | 23229.9 | | | | | | -3.79E-07 | | | | | | | -1.31E-06 | |
| 1323 | 23234.16 | | -6.94E-06 | | | | | | | | | | | | |
| 1324 | 23241.34 | -4.26E-07 | | | | | | | -1.37E-05 | | | | | | |
| 1325 | 23246.71 | | | | | -1.51E-05 | | | | | | | -1.59E-05 | | |
| 1326 | 23261.59 | | | | | | | | | 6.42E-06 | | | | | |

| | A | B | C | D | E | F | G | H | I | J | K | L | M | N | O |
|---|---|---|---|---|---|---|---|---|---|---|---|---|---|---|---|
| 1327 | 23310.4 | | | | -1.28E-05 | | | | | | | -1.42E-05 | | | |
| 1328 | 23314.61 | | | | | | -4.50E-07 | | | | | | | -1.43E-06 | |
| 1329 | 23330.25 | | | -1.30E-05 | | | | | -1.28E-05 | | | -1.41E-05 | | | -1.37E-05 |
| 1330 | 23381.35 | | -7.48E-06 | | | | | | | | | | | | |
| 1331 | 23396.39 | | | | | -1.45E-05 | | | | | | | -1.53E-05 | | |
| 1332 | 23397.44 | 4.33E-06 | | | | | | | -1.19E-05 | | | | | | |
| 1333 | 23418.63 | | | | | | -5.11E-07 | | | | | | | -1.52E-06 | |
| 1334 | 23467.97 | | | | -1.17E-05 | | | | | | | -1.26E-05 | | | |
| 1335 | 23469.35 | | | | | | | | | 6.07E-06 | | | | | |
| 1336 | 23495.54 | | | -1.21E-05 | | | | -1.19E-05 | | | -1.30E-05 | | | | -1.27E-05 |
| 1337 | 23530.64 | | | | | | -5.78E-07 | | | | | | | -1.64E-06 | |
| 1338 | 23541.13 | | -6.29E-06 | | | | | | | | | | | | |
| 1339 | 23547.76 | | | | | -1.34E-05 | | | | | | | -1.44E-05 | | |
| 1340 | 23550.56 | 6.70E-06 | | | | | | | -7.48E-06 | | | | | | |
| 1341 | 23622.62 | | | | -1.09E-05 | | | | | | | -1.20E-05 | | | |
| 1342 | 23624.16 | | | | | | -5.39E-07 | | | | | | | -1.56E-06 | |
| 1343 | 23635.23 | | | | | | | | | 4.57E-06 | | | | | |
| 1344 | 23648.28 | | | | | | -1.05E-05 | | | | | | | | -1.14E-05 |
| 1345 | 23648.66 | | | -1.07E-05 | | | | | | | -1.15E-05 | | | | |
| 1346 | 23696.59 | | -3.69E-06 | | | | | | | | | | | | |
| 1347 | 23702.54 | | | | | -1.24E-05 | | | | | | | -1.31E-05 | | |
| 1348 | 23711.14 | 1.12E-05 | | | | | | | -6.35E-07 | | | | | | |
| 1349 | 23726.12 | | | | | | | -4.38E-07 | | | | | | -1.51E-06 | |
| 1350 | 23775.6 | | | | -9.60E-06 | | | | | | | -1.02E-05 | | | |
| 1351 | 23817.45 | | | | | | | | -9.16E-06 | | | | | | -9.57E-06 |
| 1352 | 23817.64 | | | -9.20E-06 | | | | | | | -9.67E-06 | | | | |
| 1353 | 23822.98 | | | | | | | | | 5.74E-06 | | | | | |
| 1354 | 23847.51 | | | | | | -3.70E-07 | | | | | | | -1.43E-06 | |
| 1355 | 23853.85 | | -2.93E-06 | | | | | | | | | | | | |
| 1356 | 23858.53 | | | | | -1.17E-05 | | | | | | | -1.22E-05 | | |
| 1357 | 23870.02 | 1.31E-05 | | | | | | | 5.46E-06 | | | | | | |
| 1358 | 23952.93 | | | | -8.93E-06 | | | | | | | -9.48E-06 | | | |
| 1359 | 23973.9 | | | | | | -5.22E-07 | | | | | | | -1.55E-06 | |
| 1360 | 23980.46 | | | -8.26E-06 | | | | -8.38E-06 | | | -8.67E-06 | | | | -8.50E-06 |
| 1361 | 24003.09 | | | | | | | | | 5.10E-06 | | | | | |
| 1362 | 24037.17 | | -9.55E-07 | | | | | | | | | | | | |
| 1363 | 24046.03 | | | | | -1.05E-05 | | | | | | | -1.11E-05 | | |
| 1364 | 24054.66 | 1.56E-05 | | | | | | | 1.15E-05 | | | | | | |
| 1365 | 24112.53 | | | | | | -7.36E-07 | | | | | | | -1.79E-06 | |
| 1366 | 24129.19 | | | | -7.10E-06 | | | | | | | -7.67E-06 | | | |
| 1367 | 24166.04 | | | | | | | -6.78E-06 | | | | | | | -6.91E-06 |
| 1368 | 24167.76 | | | -6.54E-06 | | | | | | | -6.97E-06 | | | | |
| 1369 | 24210.33 | | | | | | | | | 5.72E-06 | | | | | |
| 1370 | 24221.45 | | 1.48E-06 | | | | | | | | | | | | |
| 1371 | 24228.43 | | | | | -9.66E-06 | | | | | | | -1.01E-05 | | |
| 1372 | 24255.92 | 2.00E-05 | | | | | | | 1.73E-05 | | | | | | |
| 1373 | 24259.91 | | | | | | -8.19E-07 | | | | | | | -2.06E-06 | |
| 1374 | 24323.26 | | | | -5.36E-06 | | | | | | | -5.89E-06 | | | |
| 1375 | 24354.74 | | | | | | | | | 5.53E-06 | | | | | |
| 1376 | 24356.04 | | | | | | | -4.57E-06 | | | | | | | -4.23E-06 |
| 1377 | 24358.05 | | | -3.93E-06 | | | | | | | -4.06E-06 | | | | |
| 1378 | 24389 | | | | | | -7.94E-07 | | | | | | | -2.09E-06 | |
| 1379 | 24416.86 | | 1.04E-05 | | | | | | | | | | | | |
| 1380 | 24425.03 | | | | | -8.23E-06 | | | | | | | -8.37E-06 | | |
| 1381 | 24452.35 | 2.96E-05 | | | | | | | 4.31E-05 | | | | | | |
| 1382 | 24535.04 | | | | -2.58E-06 | | | | | | | -3.26E-06 | | | |
| 1383 | 24581.36 | | | | | | | | -1.40E-06 | | | | | | -4.83E-07 |
| 1384 | 24581.57 | | | -3.58E-07 | | | | | | | -3.43E-07 | | | | |
| 1385 | 24609.18 | | | | | | -7.24E-07 | | | | | | | -2.04E-06 | |
| 1386 | 24610.69 | | | | | | | | | 4.70E-06 | | | | | |
| 1387 | 24652.49 | | 1.55E-05 | | | | | | | | | | | | |
| 1388 | 24664.03 | | | | | -6.03E-06 | | | | | | | -6.55E-06 | | |
| 1389 | 24677.65 | 3.81E-05 | | | | | | | 5.51E-05 | | | | | | |
| 1390 | 24781.62 | | | | 1.10E-07 | | | | | | | -5.77E-07 | | | |
| 1391 | 24791.65 | | | | | | | | | | -1.37E-06 | | | | |
| 1392 | 24804.34 | | | | | | | -8.38E-07 | | | | | | -2.25E-06 | |
| 1393 | 24824.54 | | | | 1.66E-06 | | | | 8.49E-07 | | | 1.67E-06 | | | | 1.33E-06 |
| 1394 | 24880.2 | | 1.60E-05 | | | | | | | | | | | | |
| 1395 | 24890.69 | | | | | -3.19E-06 | | | | | | | -3.68E-06 | | |
| 1396 | 24902.75 | 3.31E-05 | | | | | | | 3.91E-05 | | | | | | |
| 1397 | 24936.83 | | | | | | | | | -1.60E-07 | | | | | |
| 1398 | 24981.02 | | | | 4.33E-06 | | | | | | | 4.91E-06 | | | |
| 1399 | 25004.84 | | | | | | -9.32E-07 | | | | | | | -2.35E-06 | |
| 1400 | 25027.8 | | | | | | | 5.00E-06 | | | | | | | 5.56E-06 |
| 1401 | 25028.09 | | | 6.44E-06 | | | | | | | 6.47E-06 | | | | |
| 1402 | 25086.73 | | 1.35E-05 | | | | | | | | | | | | |
| 1403 | 25095.13 | | | | | -1.46E-06 | | | | | | | -1.56E-06 | | |
| 1404 | 25104.01 | 3.06E-05 | | | | | | | 3.43E-05 | | | | | | |
| 1405 | 25130.58 | | | | | | | | | -2.31E-06 | | | | | |
| 1406 | 25184.81 | | | | 1.51E-05 | | | | | | | 1.91E-05 | | | |
| 1407 | 25216.49 | | | | | | -1.02E-06 | | | | | | | -2.44E-06 | |
| 1408 | 25219.72 | | | | | | | 1.60E-05 | | | | | | | 1.69E-05 |
| 1409 | 25219.96 | | | 1.69E-05 | | | | | | | 1.85E-05 | | | | |
| 1410 | 25243.48 | | | | | | | | | -3.36E-06 | | | | | |
| 1411 | 25264.88 | | 1.59E-05 | | | | | | | | | | | | |
| 1412 | 25272.57 | | | | | 6.59E-06 | | | | | | | 6.17E-06 | | |
| 1413 | 25283.44 | 3.50E-05 | | | | | | | 4.26E-05 | | | | | | |
| 1414 | 25373.98 | | | | 1.31E-05 | | | | | | | 1.59E-05 | | | |
| 1415 | 25383.69 | | | | | | | | | -6.54E-06 | | | | | |
| 1416 | 25412.8 | | | | 1.38E-05 | | | | 1.41E-05 | | | 1.42E-05 | | | | 1.46E-05 |
| 1417 | 25428.06 | | | | | | -8.29E-07 | | | | | | | -2.32E-06 | |
| 1418 | 25465.01 | | 1.35E-05 | | | | | | | | | | | | |
| 1419 | 25479.11 | | | | | 5.77E-06 | | | | | | | 5.13E-06 | | |
| 1420 | 25480.03 | 3.06E-05 | | | | | | | 3.43E-05 | | | | | | |
| 1421 | 25531.11 | | | | | | | | | -8.56E-06 | | | | | |
| 1422 | 25552.13 | | | | 1.16E-05 | | | | | | | 1.47E-05 | | | |
| 1423 | 25597.96 | | | | | | | 1.26E-05 | | | | | | | 1.46E-05 |
| 1424 | 25599.43 | | | 1.28E-05 | | | | | | | 1.46E-05 | | | | |
| 1425 | 25632.03 | | | | | | -8.05E-07 | | | | | | | -2.27E-06 | |
| 1426 | 25642.48 | | 1.24E-05 | | | | | | | | | | | | |
| 1427 | 25650.77 | 3.10E-05 | | | | | | | 4.12E-05 | | | | | | |
| 1428 | 25653.41 | | | | | 6.49E-06 | | | | | | | 6.03E-06 | | |

| | A | B | C | D | E | F | G | H | I | J | K | L | M | N | O |
|---|---|---|---|---|---|---|---|---|---|---|---|---|---|---|---|
| 1429 | 25659.79 | | | | | | | | | -1.00E-05 | | | | | |
| 1430 | 25726.88 | | | | 1.10E-05 | | | | | | | 1.41E-05 | | | |
| 1431 | 25767.96 | | | 1.15E-05 | | | | 1.14E-05 | | | 1.25E-05 | | | | 1.31E-05 |
| 1432 | 25773.81 | | | | | | | | | -1.02E-05 | | | | | |
| 1433 | 25803.36 | | 1.16E-05 | | | | | | | | | | | | |
| 1434 | 25806.76 | 2.80E-05 | | | | | | | 3.11E-05 | | | | | | |
| 1435 | 25825.45 | | | | | 6.44E-06 | | | | | | | 6.00E-06 | | |
| 1436 | 25843.23 | | | | | | -9.02E-07 | | | | | | | -2.36E-06 | |
| 1437 | 25900.66 | | | | 1.09E-05 | | | | | | | | 1.34E-05 | | |
| 1438 | 25912.7 | | | | | | | | | -1.01E-05 | | | | | |
| 1439 | 25937.41 | | | | | | | 1.16E-05 | | | | | | | 1.26E-05 |
| 1440 | 25938.01 | | | 1.16E-05 | | | | | | | 1.23E-05 | | | | |
| 1441 | 25971.26 | | 1.12E-05 | | | | | | | | | | | | |
| 1442 | 25971.67 | 2.63E-05 | | | | | | | 2.63E-05 | | | | | | |
| 1443 | 25984.7 | | | | | 7.37E-06 | | | | | | | 6.86E-06 | | |
| 1444 | 25986.1 | | | | | | | | | -1.02E-05 | | | | | |
| 1445 | 26031.66 | | | | | | -9.14E-07 | | | | | | | -2.36E-06 | |
| 1446 | 26069.91 | | | | 1.12E-05 | | | | | | | 1.38E-05 | | | |
| 1447 | 26099.64 | | | | | | | 1.13E-05 | | | | | | | 1.25E-05 |
| 1448 | 26099.93 | | | 1.17E-05 | | | | | | | 1.24E-05 | | | | |
| 1449 | 26129.75 | | | | | | | | | -8.28E-06 | | | | | |
| 1450 | 26139.66 | 2.66E-05 | | | | | | | 2.16E-05 | | | | | | |
| 1451 | 26142.13 | | 1.09E-05 | | | | | | | | | | | | |
| 1452 | 26156.03 | | | | | 7.64E-06 | | | | | | | 7.07E-06 | | |
| 1453 | 26226.59 | | | | 1.13E-05 | | | | | | | 1.37E-05 | | | |
| 1454 | 26234.73 | | | | | | -9.98E-07 | | | | | | | -2.42E-06 | |
| 1455 | 26238.37 | | | | | | | | | -8.29E-06 | | | | | |
| 1456 | 26249.43 | | | | | | | 1.11E-05 | | | | | | | 1.24E-05 |
| 1457 | 26250.93 | | | 1.14E-05 | | | | | | | 1.21E-05 | | | | |
| 1458 | 26280.53 | 2.69E-05 | | | | | | | 1.82E-05 | | | | | | |
| 1459 | 26282.65 | | 1.00E-05 | | | | | | | | | | | | |
| 1460 | 26298.2 | | | | | 7.77E-06 | | | | | | | 7.06E-06 | | |
| 1461 | 26339.87 | | | | | | | | | -7.97E-06 | | | | | |
| 1462 | 26380.99 | | | | 1.11E-05 | | | | | | | 1.33E-05 | | | |
| 1463 | 26414.94 | | | 1.18E-05 | | | | 1.13E-05 | | | 1.19E-05 | | | | 1.22E-05 |
| 1464 | 26429.32 | | | | | | -1.10E-06 | | | | | | | -2.50E-06 | |
| 1465 | 26442 | 2.78E-05 | | | | | | | 1.66E-05 | | | | | | |
| 1466 | 26451.69 | | 1.07E-05 | | | | | | | | | | | | |
| 1467 | 26468.54 | | | | | 7.99E-06 | | | | | | | 7.16E-06 | | |
| 1468 | 26483.42 | | | | | | | | | -7.45E-06 | | | | | |
| 1469 | 26539 | | | | 1.12E-05 | | | | | | | 1.34E-05 | | | |
| 1470 | 26571.67 | | | | | | | 1.17E-05 | | | | | | | 1.24E-05 |
| 1471 | 26572.96 | | | 1.21E-05 | | | | | | | 1.21E-05 | | | | |
| 1472 | 26583.35 | | | | | | | | | -7.58E-06 | | | | | |
| 1473 | 26594.21 | 2.62E-05 | | | | | | | 1.37E-05 | | | | | | |
| 1474 | 26608.28 | | 1.03E-05 | | | | | | | | | | | | |
| 1475 | 26622.75 | | | | | 8.27E-06 | | | | | | | 7.38E-06 | | |
| 1476 | 26628.95 | | | | | | -1.25E-06 | | | | | | | -2.68E-06 | |
| 1477 | 26680.23 | | | | | | | | | -9.57E-06 | | | | | |
| 1478 | 26686.2 | | | | 1.15E-05 | | | | | | | 1.35E-05 | | | |
| 1479 | 26719.25 | | | | | | | 1.20E-05 | | | | | | | 1.25E-05 |
| 1480 | 26719.61 | | | 1.25E-05 | | | | | | | 1.23E-05 | | | | |
| 1481 | 26736.18 | 2.67E-05 | | | | | | | 1.57E-05 | | | | | | |
| 1482 | 26751.33 | | 1.06E-05 | | | | | | | | | | | | |
| 1483 | 26775.66 | | | | | 9.09E-06 | | | | | | | 7.86E-06 | | |
| 1484 | 26809.18 | | | | | | -1.30E-06 | | | | | | | -2.71E-06 | |
| 1485 | 26811.74 | | | | | | | | | -8.22E-06 | | | | | |
| 1486 | 26831.36 | | | | 1.23E-05 | | | | | | | 1.42E-05 | | | |
| 1487 | 26857.55 | | | | | | | 1.26E-05 | | | | | | | 1.30E-05 |
| 1488 | 26858.42 | | | 1.31E-05 | | | | | | | 1.32E-05 | | | | |
| 1489 | 26875.22 | 2.66E-05 | | | | | | | 1.46E-05 | | | | | | |
| 1490 | 26894.78 | | 1.07E-05 | | | | | | | | | | | | |
| 1491 | 26912.85 | | | | | 9.86E-06 | | | | | | | 8.66E-06 | | |
| 1492 | 26937.46 | | | | | | | | | -8.59E-06 | | | | | |
| 1493 | 26977.88 | | | | 1.24E-05 | | | | | | | 1.41E-05 | | | |
| 1494 | 27011.98 | | | | | | | 1.29E-05 | | | | | | | 1.30E-05 |
| 1495 | 27012.28 | | | 1.32E-05 | | | | | | | 1.32E-05 | | | | |
| 1496 | 27030.43 | 2.68E-05 | | | | | | | 1.39E-05 | | | | | | |
| 1497 | 27030.63 | | | | | | -1.30E-06 | | | | | | | -2.74E-06 | |
| 1498 | 27046.05 | | | | | | | | | -1.04E-05 | | | | | |
| 1499 | 27046.72 | | 1.11E-05 | | | | | | | | | | | | |
| 1500 | 27065.1 | | | | | 1.02E-05 | | | | | | | 8.85E-06 | | |
| 1501 | 27125.25 | | | | 1.23E-05 | | | | | | | 1.38E-05 | | | |
| 1502 | 27155.6 | | | | | | | 1.22E-05 | | | | | | | 1.22E-05 |
| 1503 | 27157.81 | | | 1.24E-05 | | | | | | | 1.25E-05 | | | | |
| 1504 | 27159.93 | | | | | | | | | -1.03E-05 | | | | | |
| 1505 | 27169.42 | 2.65E-05 | | | | | | | 1.19E-05 | | | | | | |
| 1506 | 27191.63 | | 9.87E-06 | | | | | | | | | | | | |
| 1507 | 27210.4 | | | | | 1.04E-05 | | | | | | | 8.89E-06 | | |
| 1508 | 27216.88 | | | | | | -1.30E-06 | | | | | | | -2.80E-06 | |
| 1509 | 27266.25 | | | | 1.24E-05 | | | | | | | 1.39E-05 | | | |
| 1510 | 27281.19 | | | | | | | | | -1.08E-05 | | | | | |
| 1511 | 27299.2 | | | | | | | 1.26E-05 | | | | | | | 1.25E-05 |
| 1512 | 27299.9 | | | 1.29E-05 | | | | | | | 1.27E-05 | | | | |
| 1513 | 27309.24 | 2.65E-05 | | | | | | | 1.12E-05 | | | | | | |
| 1514 | 27331.61 | | 1.01E-05 | | | | | | | | | | | | |
| 1515 | 27353.59 | | | | | 1.10E-05 | | | | | | | 9.36E-06 | | |
| 1516 | 27414.49 | | | | | | -1.25E-06 | | | | | | | -2.71E-06 | |
| 1517 | 27417.79 | | | | 1.36E-05 | | | | | | | 1.49E-05 | | | |
| 1518 | 27418.54 | | | | | | | | | -1.22E-05 | | | | | |
| 1519 | 27444.52 | | | | | | | 1.35E-05 | | | | | | | 1.36E-05 |
| 1520 | 27446.2 | | | 1.43E-05 | | | | | | | 1.39E-05 | | | | |
| 1521 | 27456.74 | 2.78E-05 | | | | | | | 1.26E-05 | | | | | | |
| 1522 | 27479.88 | | 1.09E-05 | | | | | | | | | | | | |
| 1523 | 27507.17 | | | | | 1.25E-05 | | | | | | | 1.07E-05 | | |
| 1524 | 27523.33 | | | | | | | | | -1.21E-05 | | | | | |
| 1525 | 27561.29 | | | | 1.39E-05 | | | | | | | 1.54E-05 | | | |
| 1526 | 27594.06 | | | | | | | 1.41E-05 | | | | | | | 1.43E-05 |
| 1527 | 27594.38 | | | 1.48E-05 | | | | | | | 1.46E-05 | | | | |
| 1528 | 27603.39 | 2.82E-05 | | | | | | | 1.29E-05 | | | | | | |
| 1529 | 27605.82 | | | | | | -1.33E-06 | | | | | | | -2.85E-06 | |
| 1530 | 27627.43 | | 1.15E-05 | | | | | | | | | | | | |

| | A | B | C | D | E | F | G | H | I | J | K | L | M | N | O |
|---|---|---|---|---|---|---|---|---|---|---|---|---|---|---|---|
| 1531 | 27645.13 | | | | | 1.32E-05 | | | | | | | 1.15E-05 | | |
| 1532 | 27657.43 | | | | | | | | | -1.08E-05 | | | | | |
| 1533 | 27697.21 | | | | 1.47E-05 | | | | | | | 1.62E-05 | | | |
| 1534 | 27729.54 | | | | | | | 1.48E-05 | | | | | | | 1.52E-05 |
| 1535 | 27730.66 | | | 1.54E-05 | | | | | | | | 1.57E-05 | | | |
| 1536 | 27737.98 | 2.80E-05 | | | | | | | 1.26E-05 | | | | | | |
| 1537 | 27769.48 | | 1.21E-05 | | | | | | | | | | | | |
| 1538 | 27770.55 | | | | | | | | | -1.35E-05 | | | | | |
| 1539 | 27788.63 | | | | | 1.40E-05 | | | | | | | 1.23E-05 | | |
| 1540 | 27799.63 | | | | | | -1.34E-06 | | | | | | | -2.82E-06 | |
| 1541 | 27844.27 | | | | 1.50E-05 | | | | | | | 1.71E-05 | | | |
| 1542 | 27867.81 | | | | | | | | | -1.41E-05 | | | | | |
| 1543 | 27876.1 | | | | | | | 1.52E-05 | | | | | | | 1.62E-05 |
| 1544 | 27876.46 | | | 1.62E-05 | | | | | | | | 1.76E-05 | | | |
| 1545 | 27883.06 | 2.95E-05 | | | | | | | 1.37E-05 | | | | | | |
| 1546 | 27905.31 | | 1.39E-05 | | | | | | | | | | | | |
| 1547 | 27922.82 | | | | | 1.45E-05 | | | | | | | 1.33E-05 | | |
| 1548 | 27959.49 | | | | | | | | | -1.54E-05 | | | | | |
| 1549 | 27969.85 | | | | 1.56E-05 | | | | | | | 1.80E-05 | | | |
| 1550 | 27984.68 | | | | | | -1.32E-06 | | | | | | | -2.84E-06 | |
| 1551 | 28003.97 | | | | | | | 1.60E-05 | | | | | | | 1.73E-05 |
| 1552 | 28004.12 | | | 1.74E-05 | | | | | | | 1.85E-05 | | | | |
| 1553 | 28008.14 | 2.95E-05 | | | | | | | 1.45E-05 | | | | | | |
| 1554 | 28037.92 | | 1.38E-05 | | | | | | | | | | | | |
| 1555 | 28057.89 | | | | | 1.56E-05 | | | | | | | 1.44E-05 | | |
| 1556 | 28065.5 | | | | | | | | | -1.59E-05 | | | | | |
| 1557 | 28115.13 | | | | 1.63E-05 | | | | | | | 1.85E-05 | | | |
| 1558 | 28152.09 | | | | | | | 1.66E-05 | | | | | | | 1.76E-05 |
| 1559 | 28152.44 | | | 1.80E-05 | | | | | | | 1.89E-05 | | | | |
| 1560 | 28156.37 | 2.94E-05 | | | | | | | 1.46E-05 | | | | | | |
| 1561 | 28181.55 | | | | | | | | | -1.65E-05 | | | | | |
| 1562 | 28186.77 | | 1.39E-05 | | | | | | | | | | | | |
| 1563 | 28201.28 | | | | | 1.71E-05 | | | | | | | 1.60E-05 | | |
| 1564 | 28208.33 | | | | | | -1.21E-06 | | | | | | | -2.77E-06 | |
| 1565 | 28252.77 | | | | 1.70E-05 | | | | | | | 1.90E-05 | | | |
| 1566 | 28284.39 | | | | | | | 1.73E-05 | | | | | | | 1.79E-05 |
| 1567 | 28284.43 | | | 1.85E-05 | | | | | | | 1.90E-05 | | | | |
| 1568 | 28285.92 | 3.02E-05 | | | | | | | 1.50E-05 | | | | | | |
| 1569 | 28304.48 | | | | | | | | | -1.71E-05 | | | | | |
| 1570 | 28325.39 | | 1.47E-05 | | | | | | | | | | | | |
| 1571 | 28341.31 | | | | | 1.78E-05 | | | | | | | 1.66E-05 | | |
| 1572 | 28384.66 | | | | 1.72E-05 | | | | | | | 1.93E-05 | | | |
| 1573 | 28391.59 | | | | | | -1.26E-06 | | | | | | | -2.78E-06 | |
| 1574 | 28398.76 | | | | | | | | | -1.61E-05 | | | | | |
| 1575 | 28411.35 | | | | | | | 1.78E-05 | | | | | | | 1.87E-05 |
| 1576 | 28411.51 | | | 1.92E-05 | | | | | | | 1.97E-05 | | | | |
| 1577 | 28412.3 | 3.08E-05 | | | | | | | 1.49E-05 | | | | | | |
| 1578 | 28445.04 | | 1.51E-05 | | | | | | | | | | | | |
| 1579 | 28459.82 | | | | | 1.85E-05 | | | | | | | 1.76E-05 | | |
| 1580 | 28488.99 | | | | | | | | | -1.70E-05 | | | | | |
| 1581 | 28510.11 | | | | 1.83E-05 | | | | | | | 2.03E-05 | | | |
| 1582 | 28540.14 | 3.26E-05 | | 2.00E-05 | | | | 1.84E-05 | 1.60E-05 | | 2.09E-05 | | | | 1.95E-05 |
| 1583 | 28573 | | 1.91E-05 | | | | | | | | | | | | |
| 1584 | 28581.67 | | | | | | | | | -1.68E-05 | | | | | |
| 1585 | 28588.09 | | | | | 1.98E-05 | | | | | | | 1.85E-05 | | |
| 1586 | 28606.63 | | | | | | -1.23E-06 | | | | | | | -2.81E-06 | |
| 1587 | 28641.23 | | | | | | | | | -1.77E-05 | | | | | |
| 1588 | 28651.93 | | | | 1.97E-05 | | | | | | | 2.18E-05 | | | |
| 1589 | 28679.22 | 3.29E-05 | | | | | | | 1.61E-05 | | | | | | |
| 1590 | 28683.07 | | | 2.04E-05 | | | | 1.93E-05 | | | 2.13E-05 | | | | 2.04E-05 |
| 1591 | 28711.17 | | 2.00E-05 | | | | | | | | | | | | |
| 1592 | 28724.56 | | | | | 2.00E-05 | | | | | | | 1.89E-05 | | |
| 1593 | 28727.8 | | | | | | | | | -2.04E-05 | | | | | |
| 1594 | 28774.49 | | | | 2.06E-05 | | | | | | | 2.31E-05 | | | |
| 1595 | 28793.53 | 3.34E-05 | | | | | | | 1.67E-05 | | | | | | |
| 1596 | 28801.25 | | | 2.23E-05 | | | | 2.09E-05 | | | 2.40E-05 | | | | 2.24E-05 |
| 1597 | 28833.25 | | | | | | -1.28E-06 | | | | | | | -2.86E-06 | |
| 1598 | 28842.41 | | 1.80E-05 | | | | | | | | | | | | |
| 1599 | 28857.77 | | | | | 2.05E-05 | | | | | | | 1.92E-05 | | |
| 1600 | 28886.42 | | | | | | | | | -1.92E-05 | | | | | |
| 1601 | 28900.68 | | | | 2.16E-05 | | | | | | | 2.37E-05 | | | |
| 1602 | 28924.88 | 3.31E-05 | | | | | | | 1.56E-05 | | | | | | |
| 1603 | 28932.69 | | | 2.51E-05 | | | | 2.41E-05 | | | 2.77E-05 | | | | 2.58E-05 |
| 1604 | 28977.87 | | 1.91E-05 | | | | | | | | | | | | |
| 1605 | 28989.99 | | | | | 2.27E-05 | | | | | | | 2.22E-05 | | |
| 1606 | 29014.45 | | | | | | | | | -1.81E-05 | | | | | |
| 1607 | 29046.86 | | | | 2.25E-05 | | | | | | | 2.47E-05 | | | |
| 1608 | 29058.1 | 3.49E-05 | | | | | | | 1.64E-05 | | | | | | |
| 1609 | 29073.77 | | | | | | | 2.38E-05 | | | | | | | 2.58E-05 |
| 1610 | 29073.86 | | | 2.47E-05 | | | | | | | 2.74E-05 | | | | |
| 1611 | 29077.49 | | | | | | -8.89E-07 | | | | | | | -2.51E-06 | |
| 1612 | 29109.92 | | 1.69E-05 | | | | | | | | | | | | |
| 1613 | 29111.85 | | | | | | | | | -1.96E-05 | | | | | |
| 1614 | 29129.42 | | | | | 2.30E-05 | | | | | | | 2.25E-05 | | |
| 1615 | 29171.69 | | | | 2.26E-05 | | | | | | | 2.47E-05 | | | |
| 1616 | 29184.78 | 3.57E-05 | | | | | | | 1.67E-05 | | | | | | |
| 1617 | 29201.93 | | | 2.51E-05 | | | | 2.40E-05 | | | 2.75E-05 | | | | 2.59E-05 |
| 1618 | 29217.85 | | | | | | | | | -1.79E-05 | | | | | |
| 1619 | 29245.77 | | 1.73E-05 | | | | | | | | | | | | |
| 1620 | 29260.33 | | | | | 2.36E-05 | | | | | | | 2.30E-05 | | |
| 1621 | 29288.48 | | | | | | -8.47E-07 | | | | | | | -2.49E-06 | |
| 1622 | 29289.66 | | | | | | | | | -1.72E-05 | | | | | |
| 1623 | 29298.6 | | | | 2.26E-05 | | | | | | | 2.47E-05 | | | |
| 1624 | 29304.97 | 3.65E-05 | | | | | | | 1.70E-05 | | | | | | |
| 1625 | 29330.87 | | | 2.53E-05 | | | | 2.43E-05 | | | 2.73E-05 | | | | 2.58E-05 |
| 1626 | 29365.55 | | 1.71E-05 | | | | | | | | | | | | |
| 1627 | 29373.74 | | | | | | | | | -1.82E-05 | | | | | |
| 1628 | 29375.47 | | | | | 2.40E-05 | | | | | | | 2.36E-05 | | |
| 1629 | 29425.6 | | | | 2.27E-05 | | | | | | | 2.48E-05 | | | |
| 1630 | 29430 | 3.73E-05 | | | | | | | 1.79E-05 | | | | | | |
| 1631 | 29453.02 | | | 2.51E-05 | | | | 2.41E-05 | | | 2.70E-05 | | | | 2.56E-05 |
| 1632 | 29467.27 | | | | | | | | | -1.69E-05 | | | | | |

| | A | B | C | D | E | F | G | H | I | J | K | L | M | N | O |
|---|---|---|---|---|---|---|---|---|---|---|---|---|---|---|---|
| 1633 | 29485.85 | | 1.73E-05 | | | | | | | | | | | | |
| 1634 | 29498.43 | | | | | 2.43E-05 | | | | | | | 2.39E-05 | | |
| 1635 | 29539.43 | | | | | | -9.27E-07 | | | | | | | -2.58E-06 | |
| 1636 | 29540.26 | 3.80E-05 | | | | | | | 1.69E-05 | | | | | | |
| 1637 | 29541.03 | | | | 2.31E-05 | | | | | | | 2.55E-05 | | | |
| 1638 | 29555.61 | | | | | | | | | -1.46E-05 | | | | | |
| 1639 | 29563.31 | | | | | | | 2.46E-05 | | | | | | | 2.66E-05 |
| 1640 | 29563.51 | | | 2.59E-05 | | | | | | | | 2.78E-05 | | | |
| 1641 | 29596.24 | | 1.75E-05 | | | | | | | | | | | | |
| 1642 | 29614.88 | | | | | 2.48E-05 | | | | | | | 2.46E-05 | | |
| 1643 | 29666.8 | 3.68E-05 | | | | | | | 1.61E-05 | | | | | | |
| 1644 | 29675.77 | | | | 2.35E-05 | | | | | | | 2.62E-05 | | | |
| 1645 | 29696.72 | | | | | | | | | -1.37E-05 | | | | | |
| 1646 | 29708.07 | | | 2.60E-05 | | | | 2.47E-05 | | | | 2.77E-05 | | | 2.66E-05 |
| 1647 | 29743.43 | | 1.75E-05 | | | | | | | | | | | | |
| 1648 | 29751.33 | | | | | 2.49E-05 | | | | | | | 2.49E-05 | | |
| 1649 | 29770.72 | | | | | | | | | -1.40E-05 | | | | | |
| 1650 | 29787.43 | 3.69E-05 | | | | | | | 1.62E-05 | | | | | | |
| 1651 | 29795.67 | | | | 2.37E-05 | | | | | | | 2.64E-05 | | | |
| 1652 | 29812.71 | | | | | | -9.67E-07 | | | | | | | -2.71E-06 | |
| 1653 | 29822.2 | | | 2.60E-05 | | | | 2.47E-05 | | | | 2.77E-05 | | | 2.67E-05 |
| 1654 | 29859.52 | | 1.67E-05 | | | | | | | | | | | | |
| 1655 | 29874.31 | | | | | 2.51E-05 | | | | | | | 2.50E-05 | | |
| 1656 | 29891.75 | | | | | | | | | -1.07E-05 | | | | | |
| 1657 | 29903.64 | 3.80E-05 | | | | | | | 1.79E-05 | | | | | | |
| 1658 | 29911.11 | | | | 2.40E-05 | | | | | | | 2.69E-05 | | | |
| 1659 | 29940.79 | | | 2.59E-05 | | | | 2.47E-05 | | | | 2.77E-05 | | | 2.69E-05 |
| 1660 | 29976.92 | | 1.67E-05 | | | | | | | | | | | | |
| 1661 | 29989.66 | | | | | 2.52E-05 | | | | | | | 2.52E-05 | | |
| 1662 | 29999.04 | | | | | | | | | -8.66E-06 | | | | | |
| 1663 | 30016.26 | 3.80E-05 | | | | | | | 1.85E-05 | | | | | | |
| 1664 | 30029.91 | | | | 2.39E-05 | | | | | | | 2.68E-05 | | | |
| 1665 | 30057.69 | | | | | | | 2.48E-05 | | | | | | | 2.70E-05 |
| 1666 | 30057.7 | | | 2.59E-05 | | | | | | | | 2.77E-05 | | | |
| 1667 | 30102.57 | | 1.66E-05 | | | | | | | | | | | | |
| 1668 | 30104.35 | | | | | | -9.85E-07 | | | | | | | -2.73E-06 | |
| 1669 | 30110.16 | | | | | | | | | -7.95E-06 | | | | | |
| 1670 | 30110.39 | | | | | 2.53E-05 | | | | | | | 2.54E-05 | | |
| 1671 | 30140.64 | 3.86E-05 | | | | | | | 1.96E-05 | | | | | | |
| 1672 | 30149.7 | | | | 2.41E-05 | | | | | | | 2.71E-05 | | | |
| 1673 | 30185.75 | | | 2.60E-05 | | | | 2.49E-05 | | | | 2.79E-05 | | | 2.71E-05 |
| 1674 | 30190.52 | | | | | | | | | -6.25E-06 | | | | | |
| 1675 | 30211.05 | | 1.71E-05 | | | | | | | | | | | | |
| 1676 | 30221.9 | | | | | 2.56E-05 | | | | | | | 2.58E-05 | | |
| 1677 | 30251.79 | 3.90E-05 | | | | | | | 1.90E-05 | | | | | | |
| 1678 | 30273.07 | | | | 2.43E-05 | | | | | | | 2.73E-05 | | | |
| 1679 | 30297.86 | | | | | | | | | -6.70E-06 | | | | | |
| 1680 | 30300.98 | | | | | | | 2.50E-05 | | | | | | | 2.70E-05 |
| 1681 | 30301.17 | | | 2.60E-05 | | | | | | | | 2.80E-05 | | | |
| 1682 | 30336.19 | | 1.62E-05 | | | | | | | | | | | | |
| 1683 | 30345.4 | | | | | 2.57E-05 | | | | | | | 2.59E-05 | | |
| 1684 | 30364.41 | 3.93E-05 | | | | | | | 1.98E-05 | | | | | | |
| 1685 | 30386.5 | | | | 2.43E-05 | | | | | | | 2.73E-05 | | | |
| 1686 | 30390.81 | | | | | | | | | -5.95E-06 | | | | | |
| 1687 | 30404.58 | | | | | | -1.02E-06 | | | | | | | -2.78E-06 | |
| 1688 | 30423.29 | | | | | | | 2.49E-05 | | | | | | | 2.71E-05 |
| 1689 | 30423.93 | | | 2.58E-05 | | | | | | | | 2.78E-05 | | | |
| 1690 | 30469.53 | | 1.65E-05 | | | | | | | | | | | | |
| 1691 | 30484 | | | | | 2.60E-05 | | | | | | | 2.62E-05 | | |
| 1692 | 30501.57 | 3.99E-05 | | | | | | | 2.04E-05 | | | | | | |
| 1693 | 30514.42 | | | | | | | | | -6.25E-06 | | | | | |
| 1694 | 30526.73 | | | | 2.45E-05 | | | | | | | 2.77E-05 | | | |
| 1695 | 30552.29 | | | | | | | 2.51E-05 | | | | | | | 2.74E-05 |
| 1696 | 30552.69 | | | 2.61E-05 | | | | | | | | 2.82E-05 | | | |
| 1697 | 30584.82 | | 1.64E-05 | | | | | | | | | | | | |
| 1698 | 30593.71 | | | | | 2.60E-05 | | | | | | | 2.63E-05 | | |
| 1699 | 30604.5 | 3.96E-05 | | | | | | | 2.00E-05 | | | | | | |
| 1700 | 30606.86 | | | | | | | | | -1.02E-05 | | | | | |
| 1701 | 30652.4 | | | | 2.46E-05 | | | | | | | 2.77E-05 | | | |
| 1702 | 30670.74 | | | 2.62E-05 | | | | 2.52E-05 | | | | 2.82E-05 | | | 2.74E-05 |
| 1703 | 30710.6 | | 1.63E-05 | | | | | | | | | | | | |
| 1704 | 30717.34 | | | | | 2.61E-05 | | | | | | | 2.64E-05 | | |
| 1705 | 30721.56 | | | | | | | | | -9.67E-06 | | | | | |
| 1706 | 30723.82 | | | | | | -9.79E-07 | | | | | | | -2.84E-06 | |
| 1707 | 30727.07 | 3.84E-05 | | | | | | | 1.98E-05 | | | | | | |
| 1708 | 30758.22 | | | | 2.49E-05 | | | | | | | 2.78E-05 | | | |
| 1709 | 30781.39 | | | | | | | 2.53E-05 | | | | | | | 2.75E-05 |
| 1710 | 30782.23 | | | 2.61E-05 | | | | | | | | 2.82E-05 | | | |
| 1711 | 30833.69 | | 1.57E-05 | | | | | | | | | | | | |
| 1712 | 30835.1 | | | | | | | | | -9.26E-06 | | | | | |
| 1713 | 30846.66 | | | | | 2.62E-05 | | | | | | | 2.66E-05 | | |
| 1714 | 30856.12 | 3.71E-05 | | | | | | | 1.86E-05 | | | | | | |
| 1715 | 30888.52 | | | | 2.47E-05 | | | | | | | 2.76E-05 | | | |
| 1716 | 30914.86 | | | | | | | 2.51E-05 | | | | | | | 2.76E-05 |
| 1717 | 30915.92 | | | 2.59E-05 | | | | | | | | 2.82E-05 | | | |
| 1718 | 30948.66 | | | | | | | | | -9.86E-06 | | | | | |
| 1719 | 30954.47 | | 1.58E-05 | | | | | | | | | | | | |
| 1720 | 30966.76 | | | | | 2.64E-05 | | | | | | | 2.69E-05 | | |
| 1721 | 30972.33 | 3.72E-05 | | | | | | | 1.95E-05 | | | | | | |
| 1722 | 31018.5 | | | | 2.46E-05 | | | | | | | 2.78E-05 | | | |
| 1723 | 31042.67 | | | | | | | | | -9.80E-06 | | | | | |
| 1724 | 31044.82 | | | | | | -1.01E-06 | | | | | | | -2.87E-06 | |
| 1725 | 31046.89 | | | | | | | 2.49E-05 | | | | | | | 2.76E-05 |
| 1726 | 31048.2 | | | 2.57E-05 | | | | | | | | 2.83E-05 | | | |
| 1727 | 31091.5 | | 1.56E-05 | | | | | | | | | | | | |
| 1728 | 31101.8 | | | | | 2.66E-05 | | | | | | | 2.71E-05 | | |
| 1729 | 31110.66 | 3.55E-05 | | | | | | | 1.89E-05 | | | | | | |
| 1730 | 31146.54 | | | | | | | | | -8.63E-06 | | | | | |
| 1731 | 31156.7 | | | | 2.46E-05 | | | | | | | 2.78E-05 | | | |
| 1732 | 31184.78 | | | | | | | 2.49E-05 | | | | | | | 2.77E-05 |
| 1733 | 31186.33 | | | 2.57E-05 | | | | | | | | 2.83E-05 | | | |
| 1734 | 31224.81 | | 1.41E-05 | | | | | | | | | | | | |

| | A | B | C | D | E | F | G | H | I | J | K | L | M | N | O |
|---|---|---|---|---|---|---|---|---|---|---|---|---|---|---|---|
| 1735 | 31232.6 | | | | | | | | | -8.76E-06 | | | | | |
| 1736 | 31235.12 | | | | | 2.67E-05 | | | | | | | 2.71E-05 | | |
| 1737 | 31236.57 | 3.55E-05 | | | | | | | 2.00E-05 | | | | | | |
| 1738 | 31281.77 | | | | 2.44E-05 | | | | | | | 2.77E-05 | | | |
| 1739 | 31306.22 | | | | | | | 2.47E-05 | | | | | | | 2.74E-05 |
| 1740 | 31310.85 | | | 2.54E-05 | | | | | | | 2.79E-05 | | | | |
| 1741 | 31349.14 | | | | | | | | | -7.90E-06 | | | | | |
| 1742 | 31359.72 | | 1.36E-05 | | | | | | | | | | | | |
| 1743 | 31368.28 | 3.49E-05 | | | | 2.66E-05 | | | 1.98E-05 | | | | 2.70E-05 | | |
| 1744 | 31371.92 | | | | | | -1.15E-06 | | | | | | | -3.00E-06 | |
| 1745 | 31405.5 | | | | 2.43E-05 | | | | | | | 2.77E-05 | | | |
| 1746 | 31442.9 | | | | | | | 2.47E-05 | | | | | | | 2.75E-05 |
| 1747 | 31444.16 | | | 2.54E-05 | | | | | | | 2.80E-05 | | | | |
| 1748 | 31474.57 | | | | | | | | | -7.21E-06 | | | | | |
| 1749 | 31482.36 | | 1.37E-05 | | | | | | | | | | | | |
| 1750 | 31488.08 | | | | | 2.66E-05 | | | | | | | 2.70E-05 | | |
| 1751 | 31489.12 | 3.49E-05 | | | | | | | 2.11E-05 | | | | | | |
| 1752 | 31547.3 | | | | 2.43E-05 | | | | | | | 2.76E-05 | | | |
| 1753 | 31580.25 | | | | | | | 2.47E-05 | | | | | | | 2.75E-05 |
| 1754 | 31582.68 | | | 2.54E-05 | | | | | | | 2.81E-05 | | | | |
| 1755 | 31605.39 | | | | | | | | | -6.56E-06 | | | | | |
| 1756 | 31627.29 | | 1.36E-05 | | | | | | | | | | | | |
| 1757 | 31634.52 | 3.48E-05 | | | | | | | 2.17E-05 | | | | | | |
| 1758 | 31634.83 | | | | | 2.66E-05 | | | | | | | 2.71E-05 | | |
| 1759 | 31693.5 | | | | 2.42E-05 | | | | | | | 2.75E-05 | | | |
| 1760 | 31715.23 | | | | | | | 2.45E-05 | | | | | | | 2.73E-05 |
| 1761 | 31716.76 | | | 2.53E-05 | | | | | | | 2.78E-05 | | | | |
| 1762 | 31725.94 | | | | | | -1.34E-06 | | | | | | | -3.15E-06 | |
| 1763 | 31763.69 | | | | | | | | | -6.14E-06 | | | | | |
| 1764 | 31776.72 | | 1.35E-05 | | | | | | | | | | | | |
| 1765 | 31777.36 | 3.42E-05 | | | | | | | 2.06E-05 | | | | | | |
| 1766 | 31779.8 | | | | | 2.66E-05 | | | | | | | 2.69E-05 | | |
| 1767 | 31833.11 | | | | 2.42E-05 | | | | | | | 2.74E-05 | | | |
| 1768 | 31863.83 | | | | | | | 2.44E-05 | | | | | | | 2.71E-05 |
| 1769 | 31864.02 | | | 2.52E-05 | | | | | | | 2.77E-05 | | | | |
| 1770 | 31907.79 | | | | | | | | | -5.30E-06 | | | | | |
| 1771 | 31920.01 | 3.32E-05 | 1.32E-05 | | | | | | 1.94E-05 | | | | | | |
| 1772 | 31921.96 | | | | | 2.65E-05 | | | | | | | 2.70E-05 | | |
| 1773 | 31980.49 | | | | 2.42E-05 | | | | | | | 2.79E-05 | | | |
| 1774 | 32008.38 | | | | | | | 2.45E-05 | | | | | | | 2.75E-05 |
| 1775 | 32010.63 | | | 2.52E-05 | | | | | | | 2.81E-05 | | | | |
| 1776 | 32035.25 | | | | | | | | | -4.42E-06 | | | | | |
| 1777 | 32073.63 | 3.16E-05 | | | | | | | 2.18E-05 | | | | | | |
| 1778 | 32073.73 | | 1.36E-05 | | | | | | | | | | | | |
| 1779 | 32075.93 | | | | | 2.68E-05 | | | | | | | 2.76E-05 | | |
| 1780 | 32082.55 | | | | | | -1.40E-06 | | | | | | | -3.27E-06 | |
| 1781 | 32139.93 | | | | 2.44E-05 | | | | | | | 2.79E-05 | | | |
| 1782 | 32176.06 | | | | | | | 2.46E-05 | | | | | | | 2.78E-05 |
| 1783 | 32177.26 | | | 2.52E-05 | | | | | | | 2.83E-05 | | | | |
| 1784 | 32224.04 | | | | | | | | | -4.10E-06 | | | | | |
| 1785 | 32226.67 | 3.12E-05 | | | | | | | 2.15E-05 | | | | | | |
| 1786 | 32227.52 | | 1.34E-05 | | | | | | | | | | | | |
| 1787 | 32230.2 | | | | | 2.67E-05 | | | | | | | 2.76E-05 | | |
| 1788 | 32289.75 | | | | 2.41E-05 | | | | | | | 2.78E-05 | | | |
| 1789 | 32315.83 | | | | | | | 2.46E-05 | | | | | | | 2.78E-05 |
| 1790 | 32317.58 | | | 2.52E-05 | | | | | | | 2.85E-05 | | | | |
| 1791 | 32332.21 | | | | | | | | | -4.35E-06 | | | | | |
| 1792 | 32363.84 | 3.10E-05 | | | | | | | 2.13E-05 | | | | | | |
| 1793 | 32369.16 | | 1.34E-05 | | | | | | | | | | | | |
| 1794 | 32370.81 | | | | | 2.66E-05 | | | | | | | 2.74E-05 | | |
| 1795 | 32413.74 | | | | | | -1.41E-06 | | | | | | | -3.35E-06 | |
| 1796 | 32423.04 | | | | 2.41E-05 | | | | | | | 2.77E-05 | | | |
| 1797 | 32450.57 | | | | | | | 2.45E-05 | | | | | | | 2.78E-05 |
| 1798 | 32451.66 | | | 2.52E-05 | | | | | | | 2.84E-05 | | | | |
| 1799 | 32489.83 | | | | | | | | | -2.20E-06 | | | | | |
| 1800 | 32517.93 | 3.08E-05 | | | | | | | 2.17E-05 | | | | | | |
| 1801 | 32522.91 | | 1.36E-05 | | | | | | | | | | | | |
| 1802 | 32523.93 | | | | | 2.68E-05 | | | | | | | 2.75E-05 | | |
| 1803 | 32589.17 | | | | 2.40E-05 | | | | | | | 2.76E-05 | | | |
| 1804 | 32622.79 | | | | | | | 2.46E-05 | | | | | | | 2.78E-05 |
| 1805 | 32623.66 | | | 2.53E-05 | | | | | | | 2.85E-05 | | | | |
| 1806 | 32625.47 | | | | | | | | | -1.03E-06 | | | | | |
| 1807 | 32681.23 | 3.11E-05 | | | | | | | 2.19E-05 | | | | | | |
| 1808 | 32683.25 | | 1.40E-05 | | | | | | | | | | | | |
| 1809 | 32686.58 | | | | | 2.69E-05 | | | | | | | 2.76E-05 | | |
| 1810 | 32749.24 | | | | 2.40E-05 | | | | | | | 2.75E-05 | | | |
| 1811 | 32776.56 | | | | | | | | | -9.68E-07 | | | | | |
| 1812 | 32786 | | | | | | | 2.46E-05 | | | | | | | 2.76E-05 |
| 1813 | 32787.34 | | | 2.52E-05 | | | | | | | 2.82E-05 | | | | |
| 1814 | 32827.49 | | | | | | -1.47E-06 | | | | | | | -3.44E-06 | |
| 1815 | 32844.22 | 3.03E-05 | | | | | | | 2.16E-05 | | | | | | |
| 1816 | 32849.62 | | 1.35E-05 | | | | | | | | | | | | |
| 1817 | 32851.16 | | | | | 2.69E-05 | | | | | | | 2.76E-05 | | |
| 1818 | 32902.7 | | | | | | | | | 1.02E-06 | | | | | |
| 1819 | 32910.28 | | | | 2.36E-05 | | | | | | | 2.73E-05 | | | |
| 1820 | 32926.89 | | | | | | | 2.44E-05 | | | | | | | 2.75E-05 |
| 1821 | 32930.27 | | | 2.48E-05 | | | | | | | 2.80E-05 | | | | |
| 1822 | 32988.77 | 3.08E-05 | | | | | | | 2.17E-05 | | | | | | |
| 1823 | 32994.08 | | 1.39E-05 | | | 2.69E-05 | | | | | | | 2.77E-05 | | |
| 1824 | 33025.49 | | | | | | | | | 3.20E-06 | | | | | |
| 1825 | 33055.95 | | | | 2.35E-05 | | | | | | | 2.71E-05 | | | |
| 1826 | 33093.73 | | | | | | | 2.45E-05 | | | | | | | 2.78E-05 |
| 1827 | 33094.13 | | | 2.50E-05 | | | | | | | 2.83E-05 | | | | |
| 1828 | 33124.88 | | | | | | -1.48E-06 | | | | | | | -3.48E-06 | |
| 1829 | 33153.74 | 3.07E-05 | | | | | | | 2.28E-05 | | | | | | |
| 1830 | 33158.44 | | 1.43E-05 | | | | | | | | | | | | |
| 1831 | 33158.7 | | | | | 2.71E-05 | | | | | | | 2.82E-05 | | |
| 1832 | 33183.22 | | | | | | | | | 3.45E-06 | | | | | |
| 1833 | 33228.48 | | | | 2.38E-05 | | | | | | | 2.75E-05 | | | |
| 1834 | 33257.13 | | | | | | | 2.46E-05 | | | | | | | 2.80E-05 |
| 1835 | 33261.39 | | | 2.52E-05 | | | | | | | 2.85E-05 | | | | |
| 1836 | 33319.71 | 3.07E-05 | | | | | | | 2.35E-05 | | | | | | |

| | A | B | C | D | E | F | G | H | I | J | K | L | M | N | O |
|---|---|---|---|---|---|---|---|---|---|---|---|---|---|---|---|
| 1837 | 33322.2 | | | | | 2.69E-05 | | | | | | | 2.80E-05 | | |
| 1838 | 33323.86 | | 1.41E-05 | | | | | | | | | | | | |
| 1839 | 33334.01 | | | | | | | | | 3.72E-06 | | | | | |
| 1840 | 33395.72 | | | | 2.37E-05 | | | | | | | 2.75E-05 | | | |
| 1841 | 33431.01 | | | | | | | 2.46E-05 | | | | | | | 2.81E-05 |
| 1842 | 33431.73 | | | 2.52E-05 | | | | | | | 2.86E-05 | | | | |
| 1843 | 33478.74 | | | | | | -1.41E-06 | | | | | | | -3.51E-06 | |
| 1844 | 33484.8 | 3.07E-05 | | | | | | | 2.38E-05 | | | | | | |
| 1845 | 33488.86 | | 1.45E-05 | | | | | | | | | | | | |
| 1846 | 33489.47 | | | | | 2.68E-05 | | | | | | | 2.79E-05 | | |
| 1847 | 33516.14 | | | | | | | | | 5.13E-06 | | | | | |
| 1848 | 33560.46 | | | 2.38E-05 | | | | | | | | 2.79E-05 | | | |
| 1849 | 33581.91 | | | | | | | 2.51E-05 | | | | | | | 2.88E-05 |
| 1850 | 33583.65 | | 2.55E-05 | | | | | | | | | 2.93E-05 | | | |
| 1851 | 33663.64 | 3.07E-05 | | | | | | | 2.38E-05 | | | | | | |
| 1852 | 33664.17 | | | | | 2.70E-05 | | | | | | | 2.83E-05 | | |
| 1853 | 33665.55 | | 1.52E-05 | | | | | | | | | | | | |
| 1854 | 33705.62 | | | | | | | | | 3.15E-06 | | | | | |
| 1855 | 33739.22 | | | | 2.42E-05 | | | | | | | 2.84E-05 | | | |
| 1856 | 33769.31 | | | | | | | 2.56E-05 | | | | | | | 2.95E-05 |
| 1857 | 33772.71 | | | 2.61E-05 | | | | | | | 2.99E-05 | | | | |
| 1858 | 33857.26 | | | | | 2.70E-05 | | | | | | | 2.83E-05 | | |
| 1859 | 33859.09 | | 1.55E-05 | | | | | | | | | | | | |
| 1860 | 33862.15 | 3.12E-05 | | | | | | | 2.43E-05 | | | | | | |
| 1861 | 33909.7 | | | | | | | | | 4.57E-06 | | | | | |
| 1862 | 33924.69 | | | | | | -1.53E-06 | | | | | | | -3.66E-06 | |
| 1863 | 33929.97 | | | | 2.40E-05 | | | | | | | 2.81E-05 | | | |
| 1864 | 33965.47 | | | | | | | 2.54E-05 | | | | | | | 2.94E-05 |
| 1865 | 33967.17 | | | 2.59E-05 | | | | | | | 2.99E-05 | | | | |
| 1866 | 34030.82 | | | | | 2.67E-05 | | | | | | | 2.81E-05 | | |
| 1867 | 34033.64 | | 1.48E-05 | | | | | | | | | | | | |
| 1868 | 34033.74 | 2.97E-05 | | | | | | | 2.35E-05 | | | | | | |
| 1869 | 34077.64 | | | | | | | | | 4.90E-06 | | | | | |
| 1870 | 34090.52 | | | | 2.37E-05 | | | | | | | 2.75E-05 | | | |
| 1871 | 34110.12 | | | | | | | 2.50E-05 | | | | | | | 2.89E-05 |
| 1872 | 34112.57 | | | 2.55E-05 | | | | | | | 2.93E-05 | | | | |
| 1873 | 34194.17 | | | | | 2.66E-05 | | | | | | | 2.80E-05 | | |
| 1874 | 34203.11 | | 1.49E-05 | | | | | | | | | | | | |
| 1875 | 34209.15 | 2.97E-05 | | | | | | | 2.45E-05 | | | | | | |
| 1876 | 34274.38 | | | | 2.36E-05 | | | | | | | 2.76E-05 | | | |
| 1877 | 34282.96 | | | | | | | | | 4.75E-06 | | | | | |
| 1878 | 34295.21 | | | | | | -1.89E-06 | | | | | | | -4.14E-06 | |
| 1879 | 34304.85 | | | | | | | 2.51E-05 | | | | | | | 2.91E-05 |
| 1880 | 34307.96 | | | 2.56E-05 | | | | | | | | 2.94E-05 | | | |
| 1881 | 34405.71 | | | | | 2.65E-05 | | | | | | | 2.79E-05 | | |
| 1882 | 34410.36 | | 1.51E-05 | | | | | | | | | | | | |
| 1883 | 34420.95 | 2.96E-05 | | | | | | | 2.45E-05 | | | | | | |
| 1884 | 34490.79 | | | | 2.35E-05 | | | | | | | 2.76E-05 | | | |
| 1885 | 34514.8 | | | 2.65E-05 | | | | 2.56E-05 | | | 3.03E-05 | | | | 2.96E-05 |
| 1886 | 34518.91 | | | | | | | | | 5.16E-06 | | | | | |
| 1887 | 34592.88 | | | | | 2.66E-05 | | | | | | | 2.80E-05 | | |
| 1888 | 34598.32 | | 1.58E-05 | | | | | | | | | | | | |
| 1889 | 34603.2 | 2.85E-05 | | | | | | | 2.58E-05 | | | | | | |
| 1890 | 34669.63 | | | | 2.35E-05 | | | | | | | 2.77E-05 | | | |
| 1891 | 34695.66 | | | | | | | 2.57E-05 | | | | | | | 2.95E-05 |
| 1892 | 34697.38 | | | 2.61E-05 | | | | | | | 3.00E-05 | | | | |
| 1893 | 34711.5 | | | | | | -1.90E-06 | | | | | | | -4.23E-06 | |
| 1894 | 34762.53 | | | | | | | | | 4.71E-06 | | | | | |
| 1895 | 34789.07 | | | | | 2.66E-05 | | | | | | | 2.79E-05 | | |
| 1896 | 34793.46 | | 1.58E-05 | | | | | | | | | | | | |
| 1897 | 34798.34 | 2.86E-05 | | | | | | | 2.50E-05 | | | | | | |
| 1898 | 34877.36 | | | | 2.35E-05 | | | | | | | 2.78E-05 | | | |
| 1899 | 34905.21 | | | | | | | 2.57E-05 | | | | | | | 2.94E-05 |
| 1900 | 34909.77 | | | 2.61E-05 | | | | | | | 2.99E-05 | | | | |
| 1901 | 34982.02 | | | | | | | | | 6.24E-06 | | | | | |
| 1902 | 35019.93 | | | | | 2.68E-05 | | | | | | | 2.79E-05 | | |
| 1903 | 35021.62 | | 1.68E-05 | | | | | | | | | | | | |
| 1904 | 35033.95 | 2.86E-05 | | | | | | | 2.61E-05 | | | | | | |
| 1905 | 35113.62 | | | | 2.35E-05 | | | | | | | 2.78E-05 | | | |
| 1906 | 35141.84 | | | | | | | 2.60E-05 | | | | | | | 2.96E-05 |
| 1907 | 35146.33 | | | 2.64E-05 | | | | | | | 3.00E-05 | | | | |
| 1908 | 35232.17 | | | | | | -2.02E-06 | | | | | | | -4.38E-06 | |
| 1909 | 35240.86 | | | | | 2.66E-05 | | | | | | | 2.77E-05 | | |
| 1910 | 35244.8 | | 1.74E-05 | | | | | | | | | | | | |
| 1911 | 35251.8 | 2.83E-05 | | | | | | | 2.64E-05 | | | | | | |
| 1912 | 35256.57 | | | | | | | | | 6.93E-06 | | | | | |
| 1913 | 35338.53 | | | | 2.36E-05 | | | | | | | 2.78E-05 | | | |
| 1914 | 35373.12 | | | | | | | 2.60E-05 | | | | | | | 2.96E-05 |
| 1915 | 35374.29 | | | 2.67E-05 | | | | | | | 3.01E-05 | | | | |
| 1916 | 35501.33 | | | | | 2.66E-05 | | | | | | | 2.76E-05 | | |
| 1917 | 35506.87 | | 1.84E-05 | | | | | | | | | | | | |
| 1918 | 35523.18 | 2.86E-05 | | | | | | | 2.68E-05 | | | | | | |
| 1919 | 35532.5 | | | | | | | | | 6.57E-06 | | | | | |
| 1920 | 35601.17 | | | | 2.34E-05 | | | | | | | 2.76E-05 | | | |
| 1921 | 35634.35 | | | | | | | 2.56E-05 | | | | | | | 2.93E-05 |
| 1922 | 35635.51 | | | 2.60E-05 | | | | | | | | 2.97E-05 | | | |
| 1923 | 35746.27 | | 1.87E-05 | | | | | | | | | | | | |
| 1924 | 35750.02 | | | | | 2.63E-05 | | | | | | | 2.72E-05 | | |
| 1925 | 35768.74 | 2.94E-05 | | | | | | | 2.75E-05 | | | | | | |
| 1926 | 35808.6 | | | | | | | | | 5.27E-06 | | | | | |
| 1927 | 35855.91 | | | | 2.33E-05 | | | | | | | 2.72E-05 | | | |
| 1928 | 35865.44 | | | | | | -2.38E-06 | | | | | | | -4.84E-06 | |
| 1929 | 35894.35 | | | | | | | 2.53E-05 | | | | | | | 2.91E-05 |
| 1930 | 35905.77 | | | 2.58E-05 | | | | | | | | 2.95E-05 | | | |
| 1931 | 36043.81 | | 1.96E-05 | | | | | | | | | | | | |
| 1932 | 36046.09 | | | | | 2.62E-05 | | | | | | | 2.71E-05 | | |
| 1933 | 36064.34 | 2.96E-05 | | | | | | | 2.81E-05 | | | | | | |
| 1934 | 36073.86 | | | | | | | | | 3.06E-06 | | | | | |
| 1935 | 36143.98 | | | | 2.34E-05 | | | | | | | 2.73E-05 | | | |
| 1936 | 36174.86 | | | | | | | 2.54E-05 | | | | | | | 2.91E-05 |
| 1937 | 36177.18 | | | 2.59E-05 | | | | | | | 2.98E-05 | | | | |
| 1938 | 36324.42 | | 2.01E-05 | | | | | | | | | | | | |

|  | A | B | C | D | E | F | G | H | I | J | K | L | M | N | O |
|---|---|---|---|---|---|---|---|---|---|---|---|---|---|---|---|
| 1939 | 36329.83 |  |  |  |  | 2.59E-05 |  |  |  |  |  |  | 2.68E-05 |  |  |
| 1940 | 36342.27 | 3.13E-05 |  |  |  |  |  |  |  | 3.04E-05 |  |  |  |  |  |
| 1941 | 36346.19 |  |  |  |  |  |  |  |  |  | 2.87E-06 |  |  |  |  |
| 1942 | 36442.46 |  |  |  | 2.32E-05 |  |  |  |  |  |  | 2.69E-05 |  |  |  |
| 1943 | 36487.44 |  |  |  |  |  | 2.50E-05 |  |  |  |  |  |  |  | 2.87E-05 |
| 1944 | 36488.09 |  |  | 2.55E-05 |  |  |  |  |  |  | 2.97E-05 |  |  |  |  |
| 1945 | 36600.74 |  |  |  |  |  |  |  |  |  | 3.87E-06 |  |  |  |  |
| 1946 | 36620.8 |  |  |  |  |  | -2.39E-06 |  |  |  |  |  |  | -4.90E-06 |  |
| 1947 | 36627.86 |  | 2.07E-05 |  |  |  |  |  |  |  |  |  |  |  |  |
| 1948 | 36646.07 |  |  |  |  | 2.56E-05 |  |  |  |  |  |  | 2.62E-05 |  |  |
| 1949 | 36664.12 | 3.13E-05 |  |  |  |  |  |  |  | 3.02E-05 |  |  |  |  |  |
| 1950 | 36736.01 |  |  |  | 2.31E-05 |  |  |  |  |  |  | 2.69E-05 |  |  |  |
| 1951 | 36766.18 |  |  |  |  |  |  |  | 2.51E-05 |  |  |  |  |  | 2.93E-05 |
| 1952 | 36775.06 |  |  | 2.57E-05 |  |  |  |  |  |  |  | 3.00E-05 |  |  |  |
| 1953 | 36834.05 |  |  |  |  |  |  |  |  |  | 2.34E-06 |  |  |  |  |
| 1954 | 36919.41 |  | 1.99E-05 |  |  |  |  |  |  |  |  |  |  |  |  |
| 1955 | 36942.59 |  |  |  |  | 2.54E-05 |  |  |  |  |  |  | 2.60E-05 |  |  |
| 1956 | 36961.4 | 3.22E-05 |  |  |  |  |  |  |  | 3.12E-05 |  |  |  |  |  |
| 1957 | 37047.21 |  |  |  | 2.38E-05 |  |  |  |  |  |  | 2.74E-05 |  |  |  |
| 1958 | 37080.15 |  |  |  |  |  |  |  | 2.60E-05 |  |  |  |  |  | 2.92E-05 |
| 1959 | 37085.99 |  |  | 2.63E-05 |  |  |  |  |  |  |  | 2.98E-05 |  |  |  |
| 1960 | 37086.06 |  |  |  |  |  |  |  |  |  | -2.12E-06 |  |  |  |  |
| 1961 | 37224.83 |  | 1.99E-05 |  |  |  |  |  |  |  |  |  |  |  |  |
| 1962 | 37261.08 |  |  |  |  | 2.49E-05 |  |  |  |  |  |  | 2.55E-05 |  |  |
| 1963 | 37264.77 |  |  |  |  |  |  |  |  |  | -1.59E-06 |  |  |  |  |
| 1964 | 37281.23 | 3.20E-05 |  |  |  |  |  |  |  | 3.24E-05 |  |  |  |  |  |
| 1965 | 37352.94 |  |  |  |  |  |  | -2.31E-06 |  |  |  |  |  |  | -4.92E-06 |  |
| 1966 | 37355.93 |  |  |  | 2.44E-05 |  |  |  |  |  |  | 2.70E-05 |  |  |  |
| 1967 | 37391.07 |  |  |  |  |  |  |  | 2.65E-05 |  |  |  |  |  | 2.92E-05 |
| 1968 | 37392.09 |  |  | 2.67E-05 |  |  |  |  |  |  |  | 2.96E-05 |  |  |  |
| 1969 | 37536 |  | 1.97E-05 |  |  |  |  |  |  |  |  |  |  |  |  |
| 1970 | 37542.19 |  |  |  |  |  |  |  |  |  | -2.38E-06 |  |  |  |  |
| 1971 | 37574.34 |  |  |  |  | 2.46E-05 |  |  |  |  |  |  | 2.50E-05 |  |  |
| 1972 | 37582.88 | 3.31E-05 |  |  |  |  |  |  |  | 3.50E-05 |  |  |  |  |  |
| 1973 | 37666.19 |  |  |  | 2.41E-05 |  |  |  |  |  |  | 2.67E-05 |  |  |  |
| 1974 | 37698.43 |  |  |  |  |  |  |  | 2.58E-05 |  |  |  |  |  | 2.87E-05 |
| 1975 | 37698.56 |  |  | 2.60E-05 |  |  |  |  |  |  |  | 2.90E-05 |  |  |  |
| 1976 | 37748.09 |  |  |  |  |  |  |  |  |  | -2.31E-06 |  |  |  |  |
| 1977 | 37859.43 |  | 1.81E-05 |  |  |  |  |  |  |  |  |  |  |  |  |
| 1978 | 37912.98 | 3.20E-05 |  |  |  | 2.44E-05 |  |  |  | 3.43E-05 |  |  |  | 2.47E-05 |  |  |
| 1979 | 37993.11 |  |  |  | 2.38E-05 |  |  |  |  |  |  | 2.63E-05 |  |  |  |
| 1980 | 37996.73 |  |  |  |  |  |  |  |  |  | -1.37E-06 |  |  |  |  |
| 1981 | 38020.62 |  |  |  |  |  |  |  | 2.55E-05 |  |  |  |  |  | 2.84E-05 |
| 1982 | 38023.9 |  |  | 2.58E-05 |  |  |  |  |  |  |  | 2.87E-05 |  |  |  |
| 1983 | 38087.38 |  |  |  |  |  |  | -2.48E-06 |  |  |  |  |  |  | -5.24E-06 |  |
| 1984 | 38163.97 |  | 1.81E-05 |  |  |  |  |  |  |  |  |  |  |  |  |
| 1985 | 38222.58 | 3.27E-05 |  |  |  |  |  |  |  | 3.51E-05 |  |  |  |  |  |
| 1986 | 38237.08 |  |  |  |  | 2.46E-05 |  |  |  |  |  |  | 2.48E-05 |  |  |
| 1987 | 38272.51 |  |  |  |  |  |  |  |  |  | -6.37E-07 |  |  |  |  |
| 1988 | 38294.3 |  |  |  | 2.39E-05 |  |  |  |  |  |  | 2.61E-05 |  |  |  |
| 1989 | 38319.43 |  |  |  |  |  |  |  | 2.59E-05 |  |  |  |  |  | 2.86E-05 |
| 1990 | 38324.35 |  |  | 2.63E-05 |  |  |  |  |  |  |  | 2.91E-05 |  |  |  |
| 1991 | 38470.25 |  | 1.87E-05 |  |  |  |  |  |  |  |  |  |  |  |  |
| 1992 | 38505.48 |  |  |  |  |  |  |  |  |  | 1.08E-06 |  |  |  |  |
| 1993 | 38506.51 | 3.42E-05 |  |  |  |  |  |  |  | 3.73E-05 |  |  |  |  |  |
| 1994 | 38527.29 |  |  |  |  | 2.42E-05 |  |  |  |  |  |  | 2.44E-05 |  |  |
| 1995 | 38597.76 |  |  |  | 2.37E-05 |  |  |  |  |  |  | 2.58E-05 |  |  |  |
| 1996 | 38632.52 |  |  |  |  |  |  |  | 2.58E-05 |  |  |  |  |  | 2.82E-05 |
| 1997 | 38634.77 |  |  | 2.61E-05 |  |  |  |  |  |  |  | 2.87E-05 |  |  |  |
| 1998 | 38731.05 |  |  |  |  |  |  |  |  |  | 1.46E-06 |  |  |  |  |
| 1999 | 38760.25 |  | 1.79E-05 |  |  |  |  |  |  |  |  |  |  |  |  |
| 2000 | 38804.81 | 3.50E-05 |  |  |  |  |  |  |  | 3.87E-05 |  |  |  |  |  |
| 2001 | 38831.88 |  |  |  |  |  |  | -2.26E-06 |  |  |  |  |  |  | -4.90E-06 |  |
| 2002 | 38861.24 |  |  |  |  | 2.38E-05 |  |  |  |  |  |  | 2.43E-05 |  |  |
| 2003 | 38910.94 |  |  |  | 2.31E-05 |  |  |  |  |  |  | 2.54E-05 |  |  |  |
| 2004 | 38928.92 |  |  |  |  |  |  |  | 2.50E-05 |  |  |  |  |  | 2.75E-05 |
| 2005 | 38937.65 |  |  | 2.55E-05 |  |  |  |  |  |  |  | 2.82E-05 |  |  |  |
| 2006 | 38959.96 |  |  |  |  |  |  |  |  |  | 2.78E-06 |  |  |  |  |
| 2007 | 39074.46 |  | 1.81E-05 |  |  |  |  |  |  |  |  |  |  |  |  |
| 2008 | 39127.28 | 3.61E-05 |  |  |  |  |  |  |  | 4.06E-05 |  |  |  |  |  |
| 2009 | 39185.09 |  |  |  |  | 2.38E-05 |  |  |  |  |  |  | 2.42E-05 |  |  |
| 2010 | 39226.68 |  |  |  |  |  |  |  |  |  | 3.05E-06 |  |  |  |  |
| 2011 | 39242.88 |  |  |  | 2.30E-05 |  |  |  |  |  |  | 2.53E-05 |  |  |  |
| 2012 | 39268.49 |  |  |  |  |  |  |  | 2.48E-05 |  |  |  |  |  | 2.75E-05 |
| 2013 | 39273.94 |  |  | 2.53E-05 |  |  |  |  |  |  |  | 2.81E-05 |  |  |  |
| 2014 | 39370.26 |  | 1.80E-05 |  |  |  |  |  |  |  |  |  |  |  |  |
| 2015 | 39417.96 | 3.66E-05 |  |  |  |  |  |  |  | 4.11E-05 |  |  |  |  |  |
| 2016 | 39484.53 |  |  |  |  |  |  |  |  |  | 2.88E-06 |  |  |  |  |
| 2017 | 39492.99 |  |  |  |  | 2.39E-05 |  |  |  |  |  |  | 2.43E-05 |  |  |
| 2018 | 39543.65 |  |  |  | 2.28E-05 |  | -1.97E-06 |  |  |  |  |  | 2.52E-05 |  | -4.68E-06 |  |
| 2019 | 39565.83 |  |  |  |  |  |  |  | 2.48E-05 |  |  |  |  |  | 2.75E-05 |
| 2020 | 39570.11 |  |  | 2.53E-05 |  |  |  |  |  |  |  | 2.81E-05 |  |  |  |
| 2021 | 39700.58 |  | 1.70E-05 |  |  |  |  |  |  |  |  |  |  |  |  |
| 2022 | 39742.93 | 3.73E-05 |  |  |  |  |  |  |  | 4.32E-05 |  |  |  |  |  |
| 2023 | 39816.48 |  |  |  |  |  |  |  |  |  | 5.09E-06 |  |  |  |  |
| 2024 | 39845.91 |  |  |  |  | 2.39E-05 |  |  |  |  |  |  | 2.42E-05 |  |  |
| 2025 | 39876.23 |  |  |  |  | 2.25E-05 |  |  |  |  |  |  | 2.48E-05 |  |  |
| 2026 | 39899.87 |  |  |  |  |  |  |  | 2.44E-05 |  |  |  |  |  | 2.70E-05 |
| 2027 | 39905.1 |  |  | 2.49E-05 |  |  |  |  |  |  |  | 2.77E-05 |  |  |  |
| 2028 | 40012.8 |  | 1.78E-05 |  |  |  |  |  |  |  |  |  |  |  |  |
| 2029 | 40035.7 | 3.79E-05 |  |  |  |  |  |  |  | 4.53E-05 |  |  |  |  |  |
| 2030 | 40104.04 |  |  |  |  |  |  |  |  |  | 6.59E-06 |  |  |  |  |
| 2031 | 40174.34 |  |  |  |  | 2.42E-05 |  |  |  |  |  |  | 2.46E-05 |  |  |
| 2032 | 40192.1 |  |  |  | 2.31E-05 |  |  |  |  |  |  | 2.54E-05 |  |  |  |
| 2033 | 40221.88 |  |  |  |  |  |  |  | 2.47E-05 |  |  |  |  |  | 2.74E-05 |
| 2034 | 40227.56 |  |  | 2.51E-05 |  |  |  |  |  |  |  | 2.83E-05 |  |  |  |
| 2035 | 40227.78 |  |  |  |  |  |  | -1.63E-06 |  |  |  |  |  |  | -4.52E-06 |  |
| 2036 | 40325.47 |  | 1.74E-05 |  |  |  |  |  |  |  |  |  |  |  |  |
| 2037 | 40333.62 |  |  |  |  |  |  |  |  |  | 6.98E-06 |  |  |  |  |
| 2038 | 40343.04 | 3.77E-05 |  |  |  |  |  |  |  | 4.55E-05 |  |  |  |  |  |
| 2039 | 40468.21 |  |  |  |  | 2.45E-05 |  |  |  |  |  |  | 2.46E-05 |  |  |
| 2040 | 40493.75 |  |  |  | 2.36E-05 |  |  |  |  |  |  | 2.56E-05 |  |  |  |

| | A | B | C | D | E | F | G | H | I | J | K | L | M | N | O |
|---|---|---|---|---|---|---|---|---|---|---|---|---|---|---|---|
| 2041 | 40526.77 | | | | | | | 2.51E-05 | | | | | | | 2.76E-05 |
| 2042 | 40530.14 | | | 2.57E-05 | | | | | | | 2.85E-05 | | | | |
| 2043 | 40591.73 | | | | | | | | | 8.63E-06 | | | | | |
| 2044 | 40600.65 | | 1.83E-05 | | | | | | | | | | | | |
| 2045 | 40630.9 | 3.76E-05 | | | | | | | 4.55E-05 | | | | | | |
| 2046 | 40776.93 | | | | | 2.44E-05 | | | | | | | 2.45E-05 | | |
| 2047 | 40786.81 | | | 2.36E-05 | | | | | | | | 2.59E-05 | | | |
| 2048 | 40803.55 | | | | | | | 2.48E-05 | | | | | | | 2.77E-05 |
| 2049 | 40812.14 | | | 2.56E-05 | | | | | | | 2.90E-05 | | | | |
| 2050 | 40863.68 | | | | | | | | | 8.98E-06 | | | | | |
| 2051 | 40912.48 | | 1.80E-05 | | | | | | | | | | | | |
| 2052 | 40931.61 | 3.88E-05 | | | | | | | 4.66E-05 | | | | | | |
| 2053 | 40965.91 | | | | | | -1.38E-06 | | | | | | | -3.94E-06 | |
| 2054 | 41101.73 | | | | | 2.46E-05 | | | | | | | 2.49E-05 | | |
| 2055 | 41101.74 | | | 2.38E-05 | | | | | | | | 2.66E-05 | | | |
| 2056 | 41131.81 | | | | | | | 2.49E-05 | | | | | | | 2.81E-05 |
| 2057 | 41138.09 | | | 2.58E-05 | | | | | | | 2.93E-05 | | | | |
| 2058 | 41194.47 | | | | | | | | | 7.60E-06 | | | | | |
| 2059 | 41208.76 | | 1.77E-05 | | | | | | | | | | | | |
| 2060 | 41212.15 | 3.88E-05 | | | | | | | 4.63E-05 | | | | | | |
| 2061 | 41408.8 | | | | 2.41E-05 | | | | | | | 2.69E-05 | | | |
| 2062 | 41418.32 | | | | | 2.45E-05 | | | | | | | 2.49E-05 | | |
| 2063 | 41440.98 | | | | | | | 2.48E-05 | | | | | | | 2.82E-05 |
| 2064 | 41444.35 | | | 2.56E-05 | | | | | | | 2.92E-05 | | | | |
| 2065 | 41474.55 | | | | | | | | | 6.75E-06 | | | | | |
| 2066 | 41532.3 | | 1.75E-05 | | | | | | | | | | | | |
| 2067 | 41536.08 | 3.92E-05 | | | | | | | 5.03E-05 | | | | | | |
| 2068 | 41753.06 | | | | 2.38E-05 | | | | | | | 2.64E-05 | | | |
| 2069 | 41770.87 | | | | | 2.45E-05 | | | | | | | 2.49E-05 | | |
| 2070 | 41790.33 | | | | | | | 2.48E-05 | | | | | | | 2.80E-05 |
| 2071 | 41799.96 | | | 2.57E-05 | | | | | | | 2.93E-05 | | | | |
| 2072 | 41811.13 | | | | | | -1.29E-06 | | | | | | | -3.90E-06 | |
| 2073 | 41832.05 | | | | | | | | | 7.44E-06 | | | | | |
| 2074 | 41850.5 | 3.86E-05 | | | | | | | 4.82E-05 | | | | | | |
| 2075 | 41856.6 | | 1.75E-05 | | | | | | | | | | | | |
| 2076 | 42103.02 | | | | 2.41E-05 | | | | | | | 2.68E-05 | | | |
| 2077 | 42107.89 | | | | | | | | | 7.74E-06 | | | | | |
| 2078 | 42122.57 | | | | | 2.47E-05 | | | | | | | 2.52E-05 | | |
| 2079 | 42129.44 | | | | | | | 2.51E-05 | | | | | | | 2.88E-05 |
| 2080 | 42138.42 | | | 2.61E-05 | | | | | | | 3.00E-05 | | | | |
| 2081 | 42189.07 | 3.81E-05 | | | | | | | 4.97E-05 | | | | | | |
| 2082 | 42199.65 | | 1.79E-05 | | | | | | | | | | | | |
| 2083 | 42418.89 | | | | | | | | | 9.48E-06 | | | | | |
| 2084 | 42470.75 | | | | 2.42E-05 | | | | | | | 2.69E-05 | | | |
| 2085 | 42528.87 | | | | | | | 2.53E-05 | | | | | | | 2.92E-05 |
| 2086 | 42529.64 | | | | | 2.49E-05 | | | | | | | 2.55E-05 | | |
| 2087 | 42532.19 | | | 2.63E-05 | | | | | | | 3.06E-05 | | | | |
| 2088 | 42569.49 | 3.74E-05 | | | | | | | 5.06E-05 | | | | | | |
| 2089 | 42593 | | 1.74E-05 | | | | | | | | | | | | |
| 2090 | 42688.31 | | | | | | -1.16E-06 | | | | | | | -3.87E-06 | |
| 2091 | 42690.42 | | | | | | | | | 3.25E-06 | | | | | |
| 2092 | 42887.59 | | | | 2.45E-05 | | | | | | | 2.76E-05 | | | |
| 2093 | 42925.52 | | | | | | | 2.62E-05 | | | | | | | 3.07E-05 |
| 2094 | 42934.19 | | | 2.73E-05 | | | | | | | 3.28E-05 | | | | |
| 2095 | 42940.54 | | | | | 2.50E-05 | | | | | | | 2.55E-05 | | |
| 2096 | 42943.71 | | | | | | | | | 4.10E-06 | | | | | |
| 2097 | 42961.34 | 3.82E-05 | | | | | | | 5.14E-05 | | | | | | |
| 2098 | 42988.61 | | 1.76E-05 | | | | | | | | | | | | |
| 2099 | 43204.81 | | | | | | | | | 3.09E-06 | | | | | |
| 2100 | 43316.87 | | | | 2.44E-05 | | | | | | | 2.79E-05 | | | |
| 2101 | 43376.79 | | | | | | | 2.60E-05 | | | | | | | 3.09E-05 |
| 2102 | 43383.73 | | | 2.72E-05 | | | | | | | 3.24E-05 | | | | |
| 2103 | 43392.37 | 3.76E-05 | | | | | | | 5.18E-05 | | | | | | |
| 2104 | 43396.33 | | | | | 2.50E-05 | | | | | | | 2.58E-05 | | |
| 2105 | 43425.18 | | 1.74E-05 | | | | | | | | | | | | |
| 2106 | 43518.08 | | | | | | | | | 7.95E-06 | | | | | |
| 2107 | 43691.84 | | | | | | -1.31E-06 | | | | | | | -4.04E-06 | |
| 2108 | 43797.83 | | | | | | | | | 1.13E-05 | | | | | |
| 2109 | 43799.23 | | | | 2.50E-05 | | | | | | | 2.87E-05 | | | |
| 2110 | 43858.18 | | | | | | | 2.70E-05 | | | | | | | 3.29E-05 |
| 2111 | 43863.82 | | | 2.83E-05 | | | | | | | 3.52E-05 | | | | |
| 2112 | 43871.62 | 3.71E-05 | | | | | | | 5.39E-05 | | | | | | |
| 2113 | 43904.05 | | | | | 2.56E-05 | | | | | | | 2.64E-05 | | |
| 2114 | 43924.51 | | 1.68E-05 | | | | | | | | | | | | |
| 2115 | 44138.14 | | | | | | | | | 1.39E-05 | | | | | |
| 2116 | 44300.53 | | | | 2.55E-05 | | | | | | | 2.95E-05 | | | |
| 2117 | 44345.91 | 3.90E-05 | | | | | | 2.81E-05 | 5.66E-05 | | | | | | 3.42E-05 |
| 2118 | 44357.23 | | | 2.91E-05 | | | | | | | 3.60E-05 | | | | |
| 2119 | 44395.85 | | | | | 2.57E-05 | | | | | | | 2.66E-05 | | |
| 2120 | 44410.5 | | 1.78E-05 | | | | | | | | | | | | |
| 2121 | 44515.39 | | | | | | | | | 1.65E-05 | | | | | |
| 2122 | 44759.66 | | | | | | -1.57E-06 | | | | | | | -4.50E-06 | |
| 2123 | 44766.72 | | | | 2.58E-05 | | | | | | | 2.98E-05 | | | |
| 2124 | 44812.84 | 3.89E-05 | | | | | | | 5.75E-05 | | | | | | |
| 2125 | 44819.24 | | | | | | | 2.92E-05 | | | | | | | 3.54E-05 |
| 2126 | 44828.07 | | | 3.07E-05 | | | | | | | 3.75E-05 | | | | |
| 2127 | 44863.41 | | | | | | | | | 1.86E-05 | | | | | |
| 2128 | 44883.38 | | 1.83E-05 | | | | | | | | | | | | |
| 2129 | 44887.86 | | | | | 2.59E-05 | | | | | | | 2.71E-05 | | |
| 2130 | 45236.94 | | | | | | | | | 2.11E-05 | | | | | |
| 2131 | 45288.61 | | | | 2.61E-05 | | | | | | | 3.04E-05 | | | |
| 2132 | 45319.85 | 3.93E-05 | | | | | | | 5.94E-05 | | | | | | |
| 2133 | 45325.83 | | | | | | | 3.01E-05 | | | | | | | 3.60E-05 |
| 2134 | 45334.45 | | | 3.11E-05 | | | | | | | 3.78E-05 | | | | |
| 2135 | 45386.01 | | 1.93E-05 | | | | | | | | | | | | |
| 2136 | 45408.6 | | | | | 2.56E-05 | | | | | | | 2.71E-05 | | |
| 2137 | 45634.07 | | | | | | | | | 2.03E-05 | | | | | |
| 2138 | 45748.99 | | | | | | -1.79E-06 | | | | | | | -5.11E-06 | |
| 2139 | 45800.11 | | | | 2.69E-05 | | | | | | | 3.08E-05 | | | |
| 2140 | 45821.28 | 4.05E-05 | | | | | | | 6.18E-05 | | | | | | |
| 2141 | 45831.16 | | | | | | | 3.17E-05 | | | | | | | 3.68E-05 |
| 2142 | 45835.77 | | | 3.28E-05 | | | | | | | 3.82E-05 | | | | |

| | A | B | C | D | E | F | G | H | I | J | K | L | M | N | O |
|---|---|---|---|---|---|---|---|---|---|---|---|---|---|---|---|
| 2143 | 45909.35 | | 2.03E-05 | | | | | | | | | | | | |
| 2144 | 45954.23 | | | | | 2.53E-05 | | | | | | | 2.69E-05 | | |
| 2145 | 46059.66 | | | | | | | | | 2.16E-05 | | | | | |
| 2146 | 46388.7 | 3.89E-05 | | | 2.64E-05 | | | | 6.12E-05 | | | 3.06E-05 | | | |
| 2147 | 46425.22 | | | | | | | 3.02E-05 | | | | | | | 3.64E-05 |
| 2148 | 46430.82 | | | 3.14E-05 | | | | | | | 3.77E-05 | | | | |
| 2149 | 46500.53 | | 2.02E-05 | | | | | | | | | | | | |
| 2150 | 46601.51 | | | | | 2.54E-05 | | | | | | | 2.69E-05 | | |
| 2151 | 46618.1 | | | | | | | | | 2.55E-05 | | | | | |
| 2152 | 46971.02 | | | | | | -1.73E-06 | | | | | | | -4.98E-06 | |
| 2153 | 47043.23 | 3.85E-05 | | | | | | | 6.25E-05 | | | | | | |
| 2154 | 47056.81 | | | | 2.61E-05 | | | | | | | 3.05E-05 | | | |
| 2155 | 47107.94 | | | | | | | 2.99E-05 | | | | | | | 3.63E-05 |
| 2156 | 47111.35 | | | 3.10E-05 | | | | | | | 3.80E-05 | | | | |
| 2157 | 47167.58 | | | | | | | | | 2.79E-05 | | | | | |
| 2158 | 47169.05 | | 1.92E-05 | | | | | | | | | | | | |
| 2159 | 47295.66 | | | | | 2.55E-05 | | | | | | | 2.68E-05 | | |
| 2160 | 47822.18 | | | | | | | | | 3.09E-05 | | | | | |
| 2161 | 47829.2 | 3.77E-05 | | | | | | | 6.37E-05 | | | | | | |
| 2162 | 47831.67 | | | | 2.73E-05 | | | | | | | 3.15E-05 | | | |
| 2163 | 47867.85 | | | | | | | 3.03E-05 | | | | | | | 3.72E-05 |
| 2164 | 47876.03 | | | 3.11E-05 | | | | | | | 3.93E-05 | | | | |
| 2165 | 47941.69 | | 2.01E-05 | | | | | | | | | | | | |
| 2166 | 48147.33 | | | | | 2.56E-05 | | | | | | | 2.69E-05 | | |
| 2167 | 48423.28 | | | | | | -1.43E-06 | | | | | | | -4.44E-06 | |
| 2168 | 48598.36 | | | | | | | | | 3.38E-05 | | | | | |
| 2169 | 48747.59 | | | | 2.81E-05 | | | | | | | 3.30E-05 | | | |
| 2170 | 48755.75 | 3.82E-05 | | | | | | | 6.43E-05 | | | | | | |
| 2171 | 48767.53 | | | | | | | 3.03E-05 | | | | | | | 3.72E-05 |
| 2172 | 48783.16 | | | 3.09E-05 | | | | | | | 3.87E-05 | | | | |
| 2173 | 48878.92 | | 1.94E-05 | | | | | | | | | | | | |
| 2174 | 49140.09 | | | | | 2.64E-05 | | | | | | | 2.78E-05 | | |
| 2175 | 49417.99 | | | | | | | | | 2.79E-05 | | | | | |
| 2176 | 49838.05 | | | | 2.94E-05 | | | | | | | 3.46E-05 | | | |
| 2177 | 49861.69 | 3.96E-05 | | | | | | | 6.48E-05 | | | | | | |
| 2178 | 49865.18 | | | | | | | 3.15E-05 | | | | | | | 3.87E-05 |
| 2179 | 49881.48 | | | 3.21E-05 | | | | | | | 4.03E-05 | | | | |
| 2180 | 50012.58 | | 2.06E-05 | | | | | | | | | | | | |
| 2181 | 50178.08 | | | | | | | | | 2.62E-05 | | | | | |
| 2182 | 50360.65 | | | | | 2.77E-05 | | | | | | | 2.90E-05 | | |
| 2183 | 50456.52 | | | | | | -1.47E-06 | | | | | | | -4.36E-06 | |
| 2184 | 51019.59 | | | | 3.00E-05 | | | | | | | 3.49E-05 | | | |
| 2185 | 51031.97 | | | | | | | | | 3.11E-05 | | | | | |
| 2186 | 51039.8 | | | | | | | 3.19E-05 | | | | | | | 3.89E-05 |
| 2187 | 51054.67 | | | 3.25E-05 | | | | | | | 4.05E-05 | | | | |
| 2188 | 51057.56 | 3.97E-05 | | | | | | | 6.59E-05 | | | | | | |
| 2189 | 51206.05 | | 2.16E-05 | | | | | | | | | | | | |
| 2190 | 51669.57 | | | | | 2.82E-05 | | | | | | | 2.98E-05 | | |
| 2191 | 52045.65 | | | | | | | | | 2.87E-05 | | | | | |
| 2192 | 52249.99 | | | | 3.06E-05 | | | | | | | 3.60E-05 | | | |
| 2193 | 52261.93 | | | | | | | 3.29E-05 | | | | | | | 4.03E-05 |
| 2194 | 52271.59 | | | 3.33E-05 | | | | | | | 4.19E-05 | | | | |
| 2195 | 52308.78 | 3.98E-05 | | | | | | | 6.63E-05 | | | | | | |
| 2196 | 52378.73 | | 2.24E-05 | | | | | | | | | | | | |
| 2197 | 52917.59 | | | | | 2.96E-05 | | | | | | | 3.16E-05 | | |
| 2198 | 52983.56 | | | | | | | | | 3.40E-05 | | | | | |
| 2199 | 53319.95 | | | | 3.20E-05 | | | | | | | 3.77E-05 | | | |
| 2200 | 53330.59 | | | | | | | 3.41E-05 | | | | | | | 4.19E-05 |
| 2201 | 53339.4 | | | 3.46E-05 | | | | | | | 4.38E-05 | | | | |
| 2202 | 53380.36 | 4.00E-05 | | | | | | | 6.68E-05 | | | | | | |
| 2203 | 53426.11 | | | | | | -1.68E-06 | | | | | | | -4.59E-06 | |
| 2204 | 53450.35 | | 2.35E-05 | | | | | | | | | | | | |
| 2205 | 53773.49 | | | | | | | | | 3.96E-05 | | | | | |
| 2206 | 53909.8 | | | | | 3.01E-05 | | | | | | | 3.22E-05 | | |
| 2207 | 54234.18 | | | | 3.24E-05 | | | | | | | 3.79E-05 | | | |
| 2208 | 54237.43 | | | 3.47E-05 | | | | 3.44E-05 | | | 4.44E-05 | | | | 4.24E-05 |
| 2209 | 54283.99 | 4.09E-05 | | | | | | | 6.82E-05 | | | | | | |
| 2210 | 54332.97 | | 2.48E-05 | | | | | | | | | | | | |
| 2211 | 54577.8 | | | | | | | | | 4.32E-05 | | | | | |
| 2212 | 54767.51 | | | | | 3.05E-05 | | | | | | | 3.27E-05 | | |
| 2213 | 55023.1 | | | | 3.32E-05 | | | | | | | 3.92E-05 | | | |
| 2214 | 55032.57 | | | | | | | 3.50E-05 | | | | | | | 4.32E-05 |
| 2215 | 55033.57 | | | 3.54E-05 | | | | | | | 4.52E-05 | | | | |
| 2216 | 55057.07 | | | | | | | | | 4.58E-05 | | | | | |
| 2217 | 55076.72 | 4.18E-05 | | | | | | | 7.08E-05 | | | | | | |
| 2218 | 55102.19 | | 2.60E-05 | | | | | | | | | | | | |
| 2219 | 55499.18 | | | | | 2.99E-05 | | | | | | | 3.25E-05 | | |
| 2220 | 55588.59 | | | | | | | | | 4.89E-05 | | | | | |
| 2221 | 55591.24 | | | | | | -2.09E-06 | | | | | | | -5.03E-06 | |
| 2222 | 55747.11 | | | | 3.27E-05 | | | | | | | 3.93E-05 | | | |
| 2223 | 55748.59 | | | 3.53E-05 | | | | 3.47E-05 | | | 4.51E-05 | | | | 4.33E-05 |
| 2224 | 55782.45 | 4.15E-05 | | | | | | | 7.22E-05 | | | | | | |
| 2225 | 55805.48 | | 2.60E-05 | | | | | | | | | | | | |
| 2226 | 56102.09 | | | | | | | | | 5.03E-05 | | | | | |
| 2227 | 56234.81 | | | | | 2.93E-05 | | | | | | | 3.20E-05 | | |
| 2228 | 56439.45 | | | | 3.26E-05 | | | | | | | 3.94E-05 | | | |
| 2229 | 56441.39 | | | 3.51E-05 | | | | 3.42E-05 | | | 4.51E-05 | | | | 4.32E-05 |
| 2230 | 56443.53 | 4.02E-05 | | | | | | | 7.12E-05 | | | | | | |
| 2231 | 56496.33 | | 2.45E-05 | | | | | | | | | | | | |
| 2232 | 56679.92 | | | | | | | | | 5.22E-05 | | | | | |
| 2233 | 56994.8 | | | | | 2.92E-05 | | | | | | | 3.19E-05 | | |
| 2234 | 57007.59 | | | | | | -1.79E-06 | | | | | | | -4.62E-06 | |
| 2235 | 57191.07 | 4.01E-05 | | | | | | | 7.19E-05 | | | | | | |
| 2236 | 57206.73 | | | | 3.22E-05 | | | | | | | 3.93E-05 | | | |
| 2237 | 57216.04 | | | 3.47E-05 | | | | 3.38E-05 | | | 4.50E-05 | | | | 4.31E-05 |
| 2238 | 57266.28 | | 2.40E-05 | | | | | | | | | | | | |
| 2239 | 57405.42 | | | | | | | | | 5.40E-05 | | | | | |
| 2240 | 57849.9 | | | | | 2.85E-05 | | | | | | | 3.14E-05 | | |
| 2241 | 58079.22 | 4.02E-05 | | | | | | | 7.30E-05 | | | | | | |
| 2242 | 58140.33 | | | | 3.19E-05 | | | | | | | 3.93E-05 | | | |
| 2243 | 58152.54 | | | 3.42E-05 | | | | 3.34E-05 | | | 4.53E-05 | | | | 4.31E-05 |
| 2244 | 58214.62 | | 2.30E-05 | | | | | | | | | | | | |

| | A | B | C | D | E | F | G | H | I | J | K | L | M | N | O |
|---|---|---|---|---|---|---|---|---|---|---|---|---|---|---|---|
| 2245 | 58395.82 | | | | | | | | | 5.35E-05 | | | | | |
| 2246 | 59062.56 | | | | | 2.87E-05 | | | | | | | 3.13E-05 | | |
| 2247 | 59282.93 | 4.09E-05 | | | | | | | 7.55E-05 | | | | | | |
| 2248 | 59444.57 | | | | | | -1.91E-06 | | | | | | | -4.76E-06 | |
| 2249 | 59456.98 | | | | 3.24E-05 | | | | | | | 3.99E-05 | | | |
| 2250 | 59462.65 | | | 3.48E-05 | | | | 3.41E-05 | | | 4.62E-05 | | | | 4.40E-05 |
| 2251 | 59580.91 | | 2.36E-05 | | | | | | | | | | | | |
| 2252 | 59636.51 | | | | | | | | | 5.56E-05 | | | | | |
| 2253 | 60407.01 | | | | | | | | | 5.23E-05 | | | | | |
| 2254 | 60723.25 | | | | | 2.92E-05 | | | | | | | 3.22E-05 | | |
| 2255 | 60967.58 | 4.22E-05 | | | | | | | 7.60E-05 | | | | | | |
| 2256 | 61070.05 | | | | 3.32E-05 | | | | | | | 4.12E-05 | | | |
| 2257 | 61083.35 | | | | | | | 3.52E-05 | | | | | | | 4.53E-05 |
| 2258 | 61085.87 | | | 3.59E-05 | | | | | | | 4.76E-05 | | | | |
| 2259 | 61178 | | 2.60E-05 | | | | | | | | | | | | |
| 2260 | 61315.96 | | | | | | | | | 5.49E-05 | | | | | |
| 2261 | 62063.87 | | | | | | | | | 5.64E-05 | | | | | |
| 2262 | 62310.59 | | | | | 2.93E-05 | | | | | | | 3.22E-05 | | |
| 2263 | 62557.8 | | | | | | -2.07E-06 | | | | | | | -5.24E-06 | |
| 2264 | 62622.93 | 4.45E-05 | | | | | | | 7.92E-05 | | | | | | |
| 2265 | 62720.79 | | | | 3.43E-05 | | | | | | | 4.25E-05 | | | |
| 2266 | 62736.51 | | | | | | | 3.64E-05 | | | | | | | 4.70E-05 |
| 2267 | 62742.38 | | | 3.70E-05 | | | | | | | 4.93E-05 | | | | |
| 2268 | 62874.25 | | 2.68E-05 | | | | | | | | | | | | |
| 2269 | 62877.76 | | | | | | | | | 6.04E-05 | | | | | |
| 2270 | 63790.94 | | | | | | | | | 7.32E-05 | | | | | |
| 2271 | 64403.78 | | | | | 3.02E-05 | | | | | | | 3.31E-05 | | |
| 2272 | 64639.46 | | | | | | | | | 8.12E-05 | | | | | |
| 2273 | 64658.51 | 4.89E-05 | | | | | | | 8.31E-05 | | | | | | |
| 2274 | 64840.67 | | | | 3.73E-05 | | | | | | | 4.66E-05 | | | |
| 2275 | 64857.06 | | | 4.14E-05 | | | | 4.05E-05 | | | 5.74E-05 | | | | 5.43E-05 |
| 2276 | 64986.32 | | 3.14E-05 | | | | | | | | | | | | |
| 2277 | 66269.96 | | | | | | | | | 8.59E-05 | | | | | |
| 2278 | 67372.25 | | | | | 3.44E-05 | | | | | | | 3.88E-05 | | |
| 2279 | 67643.52 | 5.00E-05 | | | | | | | 8.68E-05 | | | | | | |
| 2280 | 67932.67 | | | | 4.13E-05 | | | | | | | 4.89E-05 | | | |
| 2281 | 67953.14 | | | 4.30E-05 | | | | 4.31E-05 | | | 5.72E-05 | | | | 5.34E-05 |
| 2282 | 68002.68 | | | | | | -3.14E-06 | | | | | | | -6.84E-06 | |
| 2283 | 68050.85 | | 3.26E-05 | | | | | | | | | | | | |
| 2284 | 68249.87 | | | | | | | | | 9.47E-05 | | | | | |
| 2285 | 69276.97 | | | | | | | | | 8.61E-05 | | | | | |
| 2286 | 70163.55 | | | | | 3.49E-05 | | | | | | | 3.94E-05 | | |
| 2287 | 70270.85 | 4.94E-05 | | | | | | | 8.91E-05 | | | | | | |
| 2288 | 70408.23 | | | | 3.93E-05 | | | | | | | 4.59E-05 | | | |
| 2289 | 70418.02 | | | | | | | | | 9.66E-05 | | | | | |
| 2290 | 70420.92 | | 3.17E-05 | | | | | 4.09E-05 | | | | | | | 5.13E-05 |
| 2291 | 70421.17 | | | 4.15E-05 | | | | | | | 5.53E-05 | | | | |
| 2292 | 72688.68 | | | | | | | | | 9.91E-05 | | | | | |
| 2293 | 75009.44 | | | | | 3.41E-05 | | | | | | | 3.83E-05 | | |
| 2294 | 75371.93 | 5.56E-05 | | | | | | | 9.97E-05 | | | | | | |
| 2295 | 75597.27 | | 3.65E-05 | | | | | | | | | | | | |
| 2296 | 75650.4 | | | | 4.05E-05 | | | | | | | 4.89E-05 | | | |
| 2297 | 75673.42 | | | | | | | 4.30E-05 | | | | | | | 5.68E-05 |
| 2298 | 75685.64 | | | 4.33E-05 | | | | | | | 5.97E-05 | | | | |
| 2299 | 77265.24 | | | | | | | | | 0.000103 | | | | | |
| 2300 | 85233.66 | | | | | | | | | 0.000137 | | | | | |
| 2301 | 100841 | | | | | | -4.71E-06 | | | | | | | -9.18E-06 | |
| 2302 | 101835.9 | 6.86E-05 | 5.56E-05 | 6.22E-05 | 6.44E-05 | 5.44E-05 | | 6.32E-05 | 0.000125 | | 8.37E-05 | 7.51E-05 | 6.23E-05 | | 8.27E-05 |
| 2303 | 102287.5 | | | | | | | | | 0.000171 | | | | | |

## Supplementary Dataset 2

| | | | Included |
|---|---|---|---|
| | | 1 | Included |
| | | 0 | Excluded |

| Model code | Model purpose | Data period | Genetic cutoff | R2 | ΔR2 | Age | Cases_Ant | Cases_PCR | Cases_Total | Connections | Flow | Geographical Distance | Population | Sex | Deprivation (SIMD) | Temporal Distance | Random | Containment Health Index | Government Response Index | Stringency Index | Summary variables |
|---|---|---|---|---|---|---|---|---|---|---|---|---|---|---|---|---|---|---|---|---|---|
| RF1Y1 | | Whole | 0.0013 | 0.78 | | 1 | 1 | 1 | 1 | 1 | 1 | 1 | 1 | 1 | 1 | 1 | 0 | 0 | 0 | 0 | 0 |
| RF2Y1 | Commuting Flow variables (both Seq 1 and Seq 2)- testing importance and influence on R2 | Whole | 0.0013 | 0.78 | 0 | 1 | 1 | 1 | 1 | 1 | 0 | 1 | 1 | 1 | 1 | 1 | 0 | 0 | 0 | 0 | 0 |
| RF3Y1 | Sex (both Seq 1 and Seq 2)- testing importance and influence on R2 | Whole | 0.0013 | 0.77 | 0.01 | 1 | 1 | 1 | 1 | 1 | 0 | 1 | 1 | 0 | 1 | 1 | 0 | 0 | 0 | 0 | 0 |
| RF4Y1 | Age (both Seq 1 and Seq 2)- testing importance and influence on R2 | Whole | 0.0013 | 0.71 | 0.06 | 1 | 1 | 1 | 1 | 1 | 0 | 1 | 1 | 0 | 1 | 1 | 0 | 0 | 0 | 0 | 0 |
| RF5Y1 | Additional variant without Age variable | Whole | 0.0013 | 0.71 | | 0 | 0 | 0 | 1 | 1 | 0 | 1 | 1 | 0 | 1 | 1 | 0 | 0 | 0 | 0 | 0 |
| **RF6Y1** | **Cases_PCR & Cases_Ant (both Seq 1 and Seq 2)- testing importance and influence on R2** | Whole | 0.0013 | **0.76** | **0.01** | **1** | **0** | **0** | **1** | **1** | **0** | **1** | **1** | **0** | **1** | **1** | **0** | **0** | **0** | **0** | **0** |
| RF7Y1 | Cases_PCR & Cases_Total (both Seq 1 and Seq 2)- testing importance and influence on R2 | Whole | 0.0013 | 0.75 | 0.02 | 1 | 0 | 1 | 0 | 1 | 0 | 1 | 1 | 0 | 1 | 1 | 0 | 0 | 0 | 0 | 0 |
| RF8Y1 | Cases_Total & Cases_Ant (both Seq 1 and Seq 2)- testing importance and influence on R2 | Whole | 0.0013 | 0.72 | 0.05 | 1 | 1 | 0 | 0 | 1 | 0 | 1 | 1 | 0 | 1 | 1 | 0 | 0 | 0 | 0 | 0 |
| RF9Y1 | Temporal Distance (both Seq 1 and Seq 2)- testing importance and influence on R2 | Whole | 0.0013 | 0.68 | 0.08 | 1 | 0 | 0 | 1 | 1 | 0 | 1 | 1 | 0 | 1 | 0 | 0 | 0 | 0 | 0 | 0 |
| RF10Y1 | Population (both Seq 1 and Seq 2)- testing importance and influence on R2 | Whole | 0.0013 | 0.74 | 0.02 | 1 | 0 | 0 | 1 | 1 | 0 | 1 | 0 | 0 | 1 | 1 | 0 | 0 | 0 | 0 | 0 |
| RF11Y1 | Deprivation (SIMD)- (both Seq 1 and Seq 2)- testing importance and influence on R2 | Whole | 0.0013 | 0.73 | 0.03 | 1 | 0 | 0 | 1 | 1 | 0 | 1 | 1 | 0 | 0 | 1 | 0 | 0 | 0 | 0 | 0 |
| RF12Y1 | Commuting connections (both Seq 1 and Seq 2)- testing importance and influence on R2 | Whole | 0.0013 | 0.77 | -0.01 | 1 | 0 | 0 | 1 | 0 | 0 | 1 | 1 | 0 | 1 | 1 | 0 | 0 | 0 | 0 | 0 |
| RF13Y1 | Testing effect of Random Variable | Whole | 0.0013 | 0.75 | NA | 1 | 0 | 0 | 1 | 1 | 0 | 1 | 1 | 0 | 1 | 1 | 1 | 0 | 0 | 0 | 0 |
| RF14Y1 | Geographical Distance (both Seq 1 and Seq 2)- testing importance and influence on R2 | Whole | 0.0013 | 0.76 | 0 | 1 | 0 | 0 | 1 | 1 | 0 | 0 | 1 | 0 | 1 | 1 | 0 | 0 | 0 | 0 | 0 |
| RF15Y1 | All varibales with Cases- testing importance and influence on R2 | Whole | 0.0013 | 0.57 | 0.19 | 1 | 0 | 0 | 0 | 1 | 0 | 1 | 1 | 0 | 1 | 1 | 0 | 0 | 0 | 0 | 0 |
| RF16Y1 | Additional variant (no cases & connections) | Whole | 0.0013 | 0.64 | NA | 1 | 0 | 0 | 0 | 0 | 0 | 1 | 1 | 1 | 1 | 1 | 0 | 0 | 0 | 0 | 0 |
| RF17Y1 | Aditional variant (no population and daprivation) | Whole | 0.0013 | 0.57 | NA | 1 | 0 | 0 | 1 | 1 | 0 | 1 | 0 | 0 | 0 | 1 | 0 | 0 | 0 | 0 | 0 |
| RF18Y1 | Additional variant (no population) | Whole | 0.0013 | 0.74 | NA | 1 | 0 | 0 | 1 | 1 | 0 | 1 | 0 | 0 | 1 | 1 | 0 | 0 | 0 | 0 | 0 |
| RF20Y1 | Added summary varinables (means and differences) | Whole | 0.0013 | 0.72 | NA | 1 | 0 | 0 | 1 | 1 | 0 | 1 | 1 | 0 | 1 | 1 | 0 | 0 | 0 | 0 | 1 |
| RFR1Y1 | Added average Stringency Index, average Government Response Index, and average Containment Health Index | Whole | 0.0013 | 0.8 | NA | 1 | 0 | 0 | 1 | 1 | 0 | 1 | 1 | 0 | 1 | 1 | 0 | 1 | 1 | 1 | 0 |
| RFR2Y1 | Added average Containment Health Index | Whole | 0.0013 | 0.81 | NA | 1 | 0 | 0 | 1 | 1 | 0 | 1 | 1 | 0 | 1 | 1 | 0 | 1 | 0 | 0 | 0 |
| RF6Y1Train | RF6Y1 model train only on 75% of randomly selected data rows | | 0.0013 | 0.73 | | 1 | 0 | 0 | 1 | 1 | 0 | 1 | 1 | 0 | 1 | 1 | 0 | 0 | 0 | 0 | 0 |
| RFSLY1 | RF6Y1 variables for lockdown periods data only | Lockdown | 0.0013 | | NA | 1 | 0 | 0 | 1 | 1 | 0 | 1 | 1 | 0 | 1 | 1 | 0 | 0 | 0 | 0 | 0 |
| RFSnLY1 | RF6Y1 variables for out of lockdown periods data only | Out of lockdown | 0.0013 | | NA | 1 | 0 | 0 | 1 | 1 | 0 | 1 | 1 | 0 | 1 | 1 | 0 | 0 | 0 | 0 | 0 |
| RFNTSLY1 | No ganetic cutoff;  out-of-lockdown period; similarity to wild type ordering | Lockdown | None | 0.81 | NA | 1 | 0 | 0 | 1 | 1 | 0 | 1 | 1 | 0 | 1 | 1 | 0 | 0 | 0 | 0 | 0 |
| RFNTSnLY1 | No ganetic cutoff;  out-of-lockdown period; similarity to wild type ordering | Out of lockdown | None | 0.78 | NA | 1 | 0 | 0 | 1 | 1 | 0 | 1 | 1 | 0 | 1 | 1 | 0 | 0 | 0 | 0 | 0 |
| RFNTSLY1Ran | No ganetic cutoff;  lockdown period; random ordering | Lockdown | None | 0.65 | NA | 1 | 0 | 0 | 1 | 1 | 0 | 1 | 1 | 0 | 1 | 1 | 0 | 0 | 0 | 0 | 0 |
| RFNTSnLY1Ran | No ganetic cutoff;  out-of-lockdown period; random ordering | Out of lockdown | None | 0.66 | NA | 1 | 0 | 0 | 1 | 1 | 0 | 1 | 1 | 0 | 1 | 1 | 0 | 0 | 0 | 0 | 0 |
| RFMTSLY1 | Identical to RFSLY1 | Lockdown | 0.0013 | 0.8 | NA | 1 | 0 | 0 | 1 | 1 | 0 | 1 | 1 | 0 | 1 | 1 | 0 | 0 | 0 | 0 | 0 |
| RFMTSnLY1 | Identical to RFSnLY2 | Out of lockdown | 0.0013 | 0.74 | NA | 1 | 0 | 0 | 1 | 1 | 0 | 1 | 1 | 0 | 1 | 1 | 0 | 0 | 0 | 0 | 0 |
| RFMTSLY1Ran | Ganetic cutoff= 0.0013 (approx. 40 SNPs); lockdown period; random ordering | Lockdown | 0.0013 | 0.67 | NA | 1 | 0 | 0 | 1 | 1 | 0 | 1 | 1 | 0 | 1 | 1 | 0 | 0 | 0 | 0 | 0 |
| RFMTSnLY1Ran | Ganetic cutoff= 0.0013 (approx. 40 SNPs);  out-of-lockdown period; random ordering | Out of lockdown | 0.0013 | 0.64 | NA | 1 | 0 | 0 | 1 | 1 | 0 | 1 | 1 | 0 | 1 | 1 | 0 | 0 | 0 | 0 | 0 |
| RFGT1SLY1 | Ganetic cutoff= 0.0020 (approx. 60 SNPs);  out-of-lockdown period; similarity to wild type ordering | Lockdown | 0.002 | 0.81 | NA | 1 | 0 | 0 | 1 | 1 | 0 | 1 | 1 | 0 | 1 | 1 | 0 | 0 | 0 | 0 | 0 |
| RFGT1SnLY1 | Ganetic cutoff=0.0020;  out-of-lockdown period; similarity to wild type ordering | Out of lockdown | 0.002 | 0.75 | NA | 1 | 0 | 0 | 1 | 1 | 0 | 1 | 1 | 0 | 1 | 1 | 0 | 0 | 0 | 0 | 0 |
| RFGT1SLY1Ran | Ganetic cutoff=0.0020;  lockdown period; random ordering | Lockdown | 0.002 | 0.63 | NA | 1 | 0 | 0 | 1 | 1 | 0 | 1 | 1 | 0 | 1 | 1 | 0 | 0 | 0 | 0 | 0 |
| RFGT1SnLY1Ran | Ganetic cutoff=0.0020;  out-of-lockdown period; random ordering | Out of lockdown | 0.002 | 0.63 | NA | 1 | 0 | 0 | 1 | 1 | 0 | 1 | 1 | 0 | 1 | 1 | 0 | 0 | 0 | 0 | 0 |
| RFGT2SLY1 | Ganetic cutoff=0.0015;  out-of-lockdown period; similarity to wild type ordering | Lockdown | 0.0015 | 0.79 | NA | 1 | 0 | 0 | 1 | 1 | 0 | 1 | 1 | 0 | 1 | 1 | 0 | 0 | 0 | 0 | 0 |
| RFGT2SnLY1 | Ganetic cutoff=0.0015;  out-of-lockdown period; similarity to wild type ordering | Out of lockdown | 0.0015 | 0.74 | NA | 1 | 0 | 0 | 1 | 1 | 0 | 1 | 1 | 0 | 1 | 1 | 0 | 0 | 0 | 0 | 0 |
| RFGT2SLY1Ran | Ganetic cutoff=0.0015;  lockdown period; random ordering | Lockdown | 0.0015 | 0.61 | NA | 1 | 0 | 0 | 1 | 1 | 0 | 1 | 1 | 0 | 1 | 1 | 0 | 0 | 0 | 0 | 0 |
| RFGT2SnLY1Ran | Ganetic cutoff=0.0015;  out-of-lockdown period; random ordering | Out of lockdown | 0.0015 | 0.64 | NA | 1 | 0 | 0 | 1 | 1 | 0 | 1 | 1 | 0 | 1 | 1 | 0 | 0 | 0 | 0 | 0 |
| RFGT3SLY1 | Ganetic cutoff=0.001;  out-of-lockdown period; similarity to wild type ordering | Lockdown | 0.001 | 0.82 | NA | 1 | 0 | 0 | 1 | 1 | 0 | 1 | 1 | 0 | 1 | 1 | 0 | 0 | 0 | 0 | 0 |
| RFGT3SnLY1 | Ganetic cutoff= 0.001;  out-of-lockdown period; similarity to wild type ordering | Out of lockdown | 0.001 | 0.73 | NA | 1 | 0 | 0 | 1 | 1 | 0 | 1 | 1 | 0 | 1 | 1 | 0 | 0 | 0 | 0 | 0 |
| RFGT3SLY1Ran | Ganetic cutoff=0.001;  lockdown period; random ordering | Lockdown | 0.001 | 0.71 | NA | 1 | 0 | 0 | 1 | 1 | 0 | 1 | 1 | 0 | 1 | 1 | 0 | 0 | 0 | 0 | 0 |
| RFGT3SnLY1Ran | Ganetic cutoff=0.001;  out-of-lockdown period; random ordering | Out of lockdown | 0.001 | 0.64 | NA | 1 | 0 | 0 | 1 | 1 | 0 | 1 | 1 | 0 | 1 | 1 | 0 | 0 | 0 | 0 | 0 |
| RFGT4SLY1 | Ganetic cutoff= 0.0013 (approx. 40 SNPs); out-of-lockdown period; similarity to wild type ordering | Lockdown | 0.0005 | 0.81 | NA | 1 | 0 | 0 | 1 | 1 | 0 | 1 | 1 | 0 | 1 | 1 | 0 | 0 | 0 | 0 | 0 |
| RFGT4SnLY1 | Ganetic cutoff= 0.0005;  out-of-lockdown period; similarity to wild type ordering | Out of lockdown | 0.0005 | 0.69 | NA | 1 | 0 | 0 | 1 | 1 | 0 | 1 | 1 | 0 | 1 | 1 | 0 | 0 | 0 | 0 | 0 |
| RFGT4SLY1Ran | Ganetic cutoff= 0.0005;  lockdown period; random ordering | Lockdown | 0.0005 | 0.64 | NA | 1 | 0 | 0 | 1 | 1 | 0 | 1 | 1 | 0 | 1 | 1 | 0 | 0 | 0 | 0 | 0 |
| RFGT4SnLY1Ran | Ganetic cutoff=0.0005;  out-of-lockdown period; random ordering | Out of lockdown | 0.0005 | 0.61 | NA | 1 | 0 | 0 | 1 | 1 | 0 | 1 | 1 | 0 | 1 | 1 | 0 | 0 | 0 | 0 | 0 |
| RFGT5SLY1 | Ganetic cutoff=0.0001;  out-of-lockdown period; similarity to wild type ordering | Lockdown | 0.0001 | 0.64 | NA | 1 | 0 | 0 | 1 | 1 | 0 | 1 | 1 | 0 | 1 | 1 | 0 | 0 | 0 | 0 | 0 |
| RFGT5SnLY1 | Ganetic cutoff=0.0001;  out-of-lockdown period; similarity to wild type ordering | Out of lockdown | 0.0001 | 0.52 | NA | 1 | 0 | 0 | 1 | 1 | 0 | 1 | 1 | 0 | 1 | 1 | 0 | 0 | 0 | 0 | 0 |
| RFGT5SLY1Ran | Ganetic cutoff= 0.0001;  lockdown period; random ordering | Lockdown | 0.0001 | 0.47 | NA | 1 | 0 | 0 | 1 | 1 | 0 | 1 | 1 | 0 | 1 | 1 | 0 | 0 | 0 | 0 | 0 |
| RFGT5SnLY1Ran | Ganetic cutoff=0.0001;  out-of-lockdown period; random ordering | Out of lockdown | 0.0001 | 0.38 | NA | 1 | 0 | 0 | 1 | 1 | 0 | 1 | 1 | 0 | 1 | 1 | 0 | 0 | 0 | 0 | 0 |